\documentclass[aps,twocolumn,prb,superscriptaddress,preprintnumbers, eqsecnum]{revtex4-2}

\usepackage{amsmath}
\usepackage{amsthm}
\usepackage{amssymb}
\usepackage{mathtools}
\usepackage{multirow}
\usepackage{graphicx}
\usepackage{bm} % bold math
\usepackage{slashed} 
\usepackage{color} 
\usepackage{xcolor}
\usepackage{amsfonts}
\usepackage[breaklinks=true,colorlinks,citecolor=blue,linkcolor=blue,urlcolor=blue]{hyperref}
\usepackage{cases}
\usepackage{url}
\usepackage{framed}
\usepackage{dsfont}  % Either change all \mathbb to \mathds
\usepackage{cancel,soul} % to strike things out
\usepackage[normalem]{ulem}
\usepackage{appendix}
\usepackage{hhline}
\usepackage{subfigure}
\usepackage{physics}
\usepackage{tikz}
\usepackage{pgfplots}
\pgfplotsset{compat=1.16}
\usepgfplotslibrary{fillbetween}
\usetikzlibrary{arrows.meta}
\usepackage{simplewick} % Wick normal ordering

\definecolor{color}{HTML}{F2B361}

\def\=={\raisebox{0.35pt}{$\mathrm{:}$}\!\!=}

\DeclareMathOperator*{\argmax}{arg\,max}
\DeclareMathOperator{\sgn}{sgn}

\theoremstyle{definition}

\theoremstyle{remark}

\allowdisplaybreaks

\usepackage{subfigure}
\usepackage{booktabs}
\usepackage[colorlinks,
			breaklinks=true,
			linkcolor=blue,
			citecolor=blue,
			urlcolor=blue]{hyperref}

\def \ii {\mathrm{i}}

\begin{document}
	
\preprint{MIT-CTP/6130}

\title{Competing symmetry breaking and topology in quantum spin chains}

\author{Anthony Rey}
\affiliation{
Condensed Matter Theory Group,
PSI Center for Scientific Computing,
Theory and Data,
5232 Villigen PSI, Switzerland
            }

\author{\"{O}mer M. Aksoy}
\affiliation{
Center for Theoretical Physics — a Leinweber Institute, Massachusetts Institute of Technology, Cambridge, MA 02139, USA
            }

\author{Claudio Chamon}
\affiliation{
Department of Physics and Astronomy, Purdue University, West Lafayette,
Indiana 47907, USA
}
\affiliation{
Purdue Quantum Science and Engineering Institute,
Purdue University, West Lafayette, Indiana, 47907, USA
}

\author{Akira Furusaki}
\affiliation{
RIKEN Center for Emergent Matter Science, Wako, Saitama 351-0198, Japan
            }

\author{Christopher Mudry}
\affiliation{
Condensed Matter Theory Group,
PSI Center for Scientific Computing,
Theory and Data,
5232 Villigen PSI, Switzerland
            }
\affiliation{
Institut de Physique, EPF Lausanne, CH-1015 Lausanne, Switzerland
            }  

\date{\today}

\begin{abstract}
We investigate how spontaneous symmetry breaking and
symmetry-protected topology compete and cooperate in
one-dimensional quantum spin-1/2 systems.
We introduce two families of local Hamiltonians with
$\mathbb{Z}^{\,}_{2}\times\mathbb{Z}^{\,}_{2}\times\mathbb{Z}^{T}_{2}$
symmetry which host a zoo of broken-symmetry phases with and
without gapless boundary spins, namely, six long-range-ordered gapped phases
that differ by their $\mathbb{Z}^{\,}_{2}$-symmetry-breaking patterns
($\mathrm{FM}^{\,}_{x,y,z}$ and $\mathrm{Neel}^{\,}_{x,y,z}$),
their partner symmetry-protected topological phases
($\mathrm{FM}^{\mathrm{SPT}}_{x,y}$ and $\mathrm{Neel}^{\mathrm{SPT}}_{x,y}$),
and two gapless phases
($\mathrm{gFM}^{\,}_{z}$ and $\mathrm{gNeel}^{\,}_{z}$)
with incommensurate $\mathbb{Z}^{T}_{2}$-breaking magnetic long-range order
and central charge $\mathsf{c}=1$.
Direct transitions between competing gapped phases can
be continuous or weakly first order.
We demonstrate that continuous phase transition lines
have exotic critical properties.
For example, a line of deconfined quantum critical points
(between the $\mathrm{Neel}^{\,}_{x}$ and $\mathrm{Neel}^{\,}_{y}$ phases)
described by Tomonaga-Luttinger liquid theory is terminated by the
$\mathrm{gFM}^{\,}_{z}$ phase at a Lifshitz tricritical point
which exhibits spontaneous breaking of the continuous $\mathrm{U}(1)$
symmetry despite the locality and one-dimensionality of the Hamiltonian,
contrary to a naive expectation from the
Hohenberg-Mermin-Wagner-Coleman theorem.
Furthermore, critical lines between symmetry-protected
topological phases with different $\mathbb{Z}^{\,}_{2}$-symmetry breaking
(e.g., $\mathrm{Neel}^{\mathrm{SPT}}_{x}$ and $\mathrm{Neel}^{\mathrm{SPT}}_{y}$)
additionally retain low-energy boundary degrees of freedom,
realizing symmetry-enriched criticality
(or a gapless SPT phase).
Our results reveal unconventional mechanisms by which
competing symmetry breaking and topology generate ordered critical phases
and multicriticality.
\end{abstract}

\maketitle

\tableofcontents

\section{Introduction}
\label{sec:Introduction}

\subsection{Motivation}

Landau's paradigm classifies phases of matter through the symmetries
that are broken spontaneously in the ground state.  According to this
paradigm, two phases that break distinct symmetries cannot be
connected adiabatically without encountering a phase transition.
The discovery of symmetry-protected topological (SPT) phases has enlarged
this perspective by showing that even phases with identical
symmetry-breaking patterns can remain distinct when they differ by
topological attributes protected by unbroken symmetries%
~\cite{Gu09,Pollmann10,Chen11a,Chen11b,Fidkowski11a,Pollmann12}.
Consequently, a complete characterization of a gapped phase may
require specifying both its pattern of spontaneous symmetry breaking
and its SPT content.

An interesting question then arises when several gapped phases compete
simultaneously.  When two competing phases are separated in coupling
space, their distinction can originate from different mechanisms.  On
the one hand, they can realize different patterns of symmetry breaking.
On the other hand, they can realize identical broken
symmetries while differing by their SPT character.  More generally,
some phase transitions may involve only symmetry-breaking
distinctions, while others may involve topological distinctions, or a
combination of both.  Understanding the outcome of such competing
tendencies is one of the central motivations for the present work.

To address this question, we introduce two families of one-dimensional
quantum spin-$1/2$ lattice Hamiltonians.  Each family realizes four
gapped phases arranged at the corners of a two-dimensional coupling
space.  Every phase supports long-range order, but the nature of the
distinction between any pair of neighboring corner phases differs
between the two Hamiltonian families.  In one family, neighboring
phases can differ simultaneously by their symmetry-breaking pattern
and by their SPT character.  In the other family, neighboring phases
may differ only by their SPT character or only by their
symmetry-breaking pattern. The purpose of introducing these two
families is therefore to disentangle the role played by conventional
order and by topological distinctions in determining the global
structure of the phase diagram.

A priori, one might expect that the competition between four gapped
phases is resolved solely through phase boundaries separating
neighboring pairs of phases, without introducing additional phases of
matter. Instead, we find that this expectation can fail.
As the competition between the four gapped phases becomes most
intense, an additional phase emerges in the center of the phase
diagram.  This emergent phase is either gapless or gapped.
In both cases, the central phase occupies a
finite region of coupling space and separates the four corner phases
from one another. Its existence implies that the collective competition
between several gapped phases can generate phases of matter
that are not anticipated from considering any pair of competing phases
in isolation.

\subsection{Results}

\subsubsection{Phase diagrams}
\label{subsubsec:Phase diagrams}

\begin{figure}
\centering
\includegraphics[width=1\columnwidth]{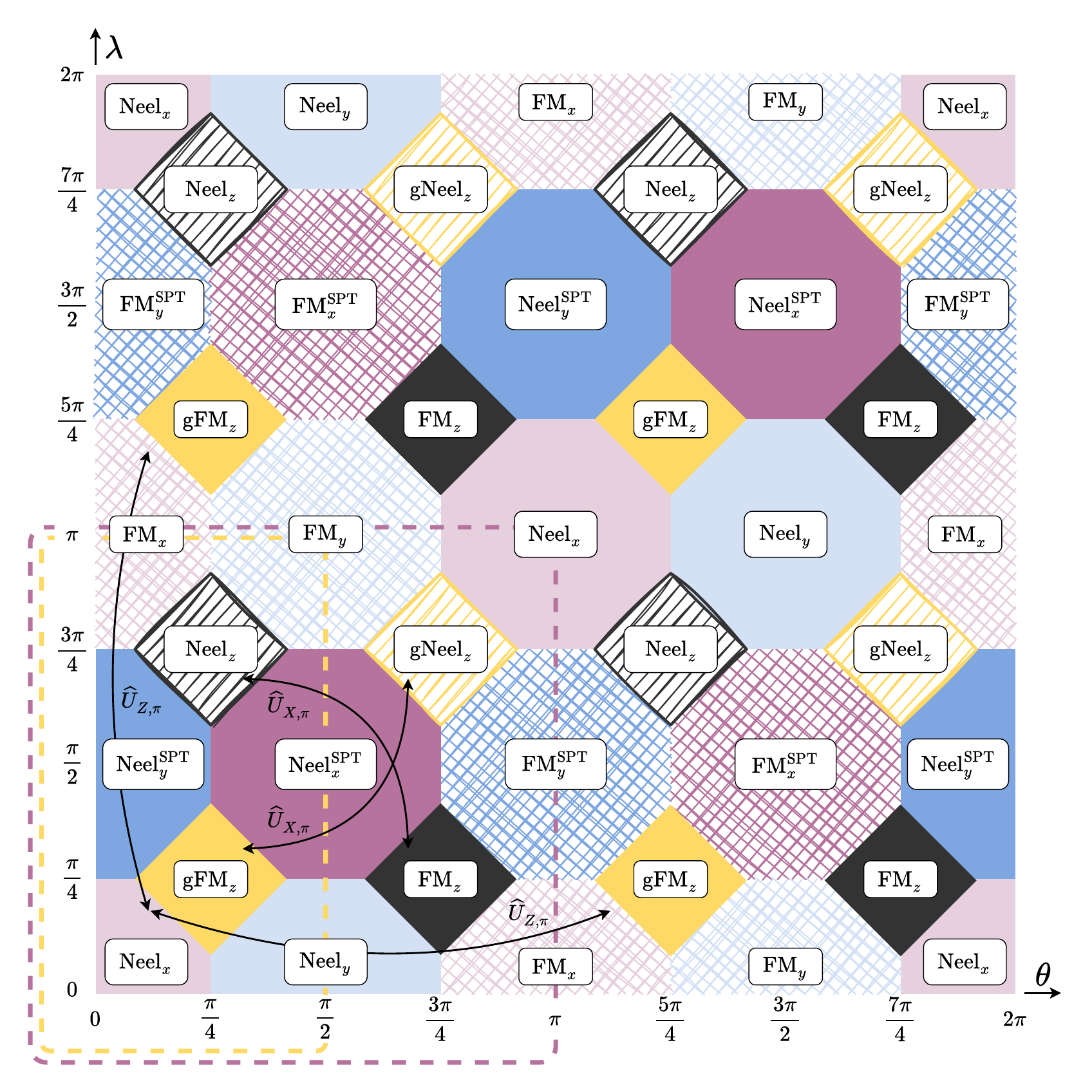}
\caption{%
Phase diagram of Hamiltonian
$\widehat{H}^{(\mathsf{b})}_{\mathrm{I}}(\theta,\lambda)$
in Eq.~\eqref{eq:def HA and HB c}.
Phase boundaries are not drawn to scale.
		}
\label{fig:phase diagram A}
\end{figure}

\begin{figure}
\centering
\includegraphics[width=1\columnwidth]{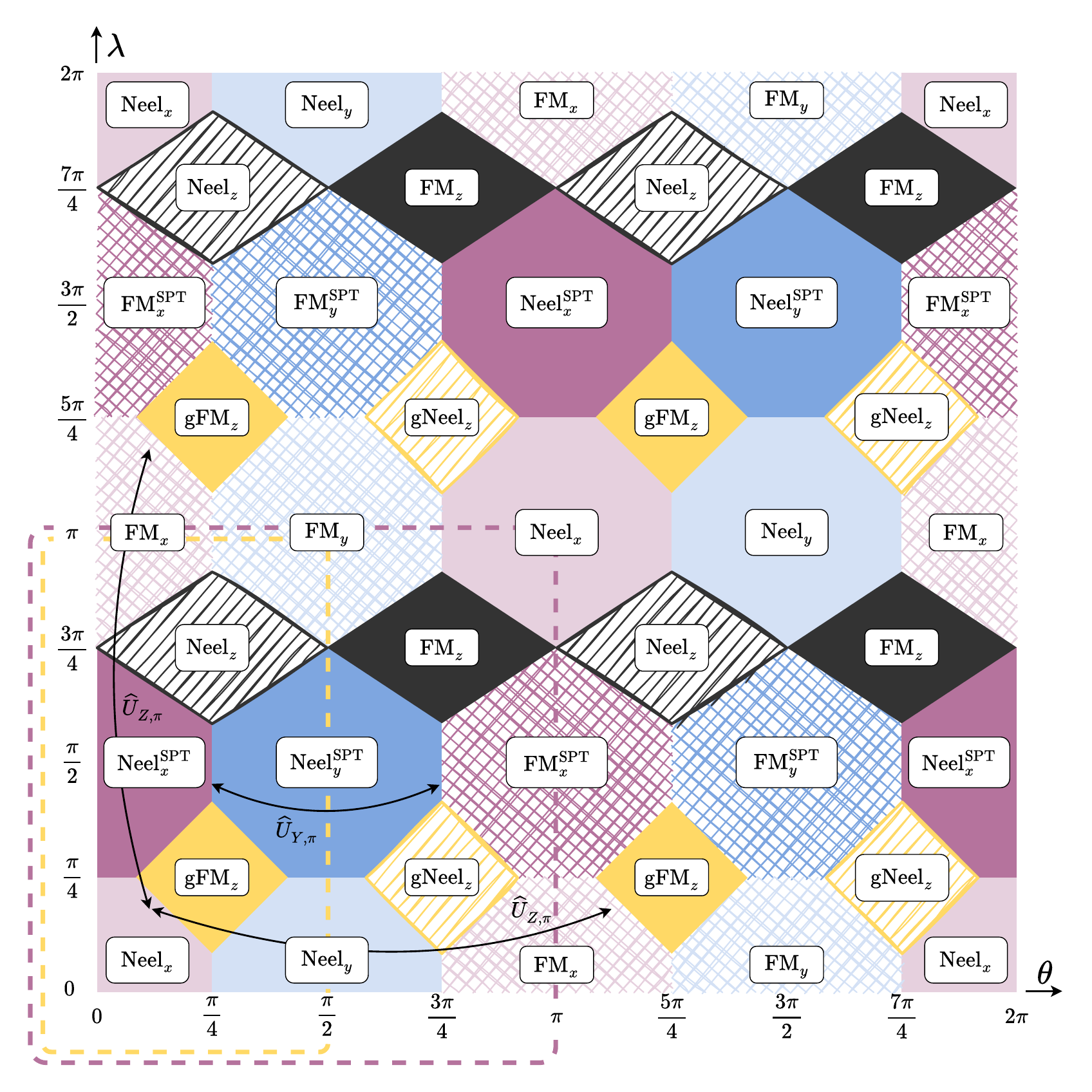}
\caption{%
Phase diagram of Hamiltonian
$\widehat{H}^{(\mathsf{b})}_{\mathrm{II}}(\theta,\lambda)$
in Eq.~\eqref{eq:def HA and HB d}.
Phase boundaries are not drawn to scale.
		}
\label{fig:phase diagram B}
\end{figure}

The main results of this paper are
summarized by the phase diagram
\ref{fig:phase diagram A}
for the family of one-dimensional quantum spin-$1/2$ lattice Hamiltonians
\eqref{eq:def HA and HB c}
and the phase diagram
\ref{fig:phase diagram B}
for the family of one-dimensional quantum spin-$1/2$ lattice Hamiltonians
\eqref{eq:def HA and HB d}.
Phase diagrams \ref{fig:phase diagram A} and \ref{fig:phase diagram B}
were deduced from exact diagonalization (ED) and
density matrix renormalization group (DMRG).

For both Hamiltonians
\eqref{eq:def HA and HB c}
and
\eqref{eq:def HA and HB d},
the coupling space is the torus $\mathbb{T}^{2}$ parametrized by the
pair of angles $\theta\in[0,2\pi]$ and $\lambda\in[0,2\pi]$.
Any pair of points in $\mathbb{T}^{2}$
that are related by the transformation
$\theta\mapsto\theta+\pi$
or
$\lambda\mapsto\lambda+\pi$
or their composition
are assigned a pair of unitarily equivalent
Hamiltonians in either one of Figs.\
\ref{fig:phase diagram A}
or
\ref{fig:phase diagram B}.
This is why we may reduce the coupling space to the region
(\ref{eq:coupling space purple})
that is encircled by the dashed purple frame in
Figs.\
\ref{fig:phase diagram A}
or
\ref{fig:phase diagram B}.
The same argument repeats itself
in the dashed purple frame of
Fig.\ \ref{fig:phase diagram A}
(Fig.\ \ref{fig:phase diagram B})
with the help of a rotation by $\pi$ centered
about the fixed point $(\pi/2,\pi/2)$
(with the help of a mirror transformation
about the fixed vertical line $\theta=\pi/4$).
Hence, the phase diagram for the torus
$\mathbb{T}^{2}$
can be deduced from the phase diagram of the region
(\ref{eq:coupling space yellow})
that is encircled by the dashed yellow frame in either one of
Figs.\
\ref{fig:phase diagram A}
or
\ref{fig:phase diagram B}.

By construction,
either one of Hamiltonians
\eqref{eq:def HA and HB c}
and
\eqref{eq:def HA and HB d}
becomes a classical Hamiltonian at the point
$(m\,\pi/2,n\,\pi/2)\in\mathbb{T}^{2}$
with $m$ and $n$ integers.
These classical Hamiltonians, when periodic boundary conditions (PBCs)
are imposed, are unitarily equivalent to either the ferromagnetic
or antiferromagnetic nearest-neighbor
one-dimensional classical Ising model.
Hence, their ground states are gapped and
break spontaneously an internal $\mathbb{Z}^{\,}_{2}$ discrete symmetry.
The ordered moments are aligned along the axis $x$ ($y$) in spin-$1/2$ space 
\begin{enumerate}
\item[I.] 
when $m+n$ is even (odd) for Hamiltonian 
\eqref{eq:def HA and HB c}, and
\item[II.]
when $m$ is even (odd) for Hamiltonian 
\eqref{eq:def HA and HB d}.
\end{enumerate}
When the integer $n$ is odd and open boundary conditions (OBCs)
are imposed, the ground-state degeneracy of these classical Hamiltonians
jumps from two to eight for any finite size of the chain. This sensitivity
to the choice of boundary conditions is the signature of a symmetry protected
topological (SPT) phase. Accordingly, we label the points
$(m\,\pi/2,n\,\pi/2)\in\mathbb{T}^{2}$
with $m$ and $n$ integers
by
\begin{enumerate}
\item
the ordering vector of the long-range colinear order, i.e.,
antiferromagnetic (Neel) or ferromagnetic (FM),
\item
the directions $x$ or $y$ of the ordered moments in spin-$1/2$ space, 
\item
and the presence or absence of SPT degeneracies when OBCs are selected.
\end{enumerate}
This gives a total of eight distinct fixed-point classical Hamiltonians
that we tag with the names
\begin{subequations}
\label{eq:eight gapped phases}
\begin{equation}
\mathrm{Neel}^{\,}_{x},\qquad
\mathrm{Neel}^{\,}_{y},\qquad
\mathrm{FM}^{\,}_{x},\qquad
\mathrm{FM}^{\,}_{y},
\end{equation}
when their ground states remain twofold degenerate with OBCs
and
\begin{equation}
\mathrm{Neel}^{\mathrm{SPT}}_{x},\quad
\mathrm{Neel}^{\mathrm{SPT}}_{y},\quad
\mathrm{FM}^{\mathrm{SPT}}_{x},\quad
\mathrm{FM}^{\mathrm{SPT}}_{y},
\end{equation}
\end{subequations}
when their ground states become eightfold degenerate with OBCs.

On the torus $\mathbb{T}^{2}$, each fixed point
$(m\,\pi/2,n\,\pi/2)\in\mathbb{T}^{2}$
with $m$ and $n$ a pair of integers
is the center of a gapped phase
(i) that has the (approximate) shape of a polygon
and (ii) that is named according to the terminology
\eqref{eq:eight gapped phases}.

For Hamiltonian \eqref{eq:def HA and HB c},
the polygon is an 8-gon. Four of its eight edges are boundaries
along which a direct quantum phase transition
takes place to another gapped phase controlled
by the fixed point
$(m'\,\pi/2,n'\,\pi/2)\in\mathbb{T}^{2}$
with $m'\neq m$ and $n'\neq n$ integers.
The other four edges are boundaries
along which a direct quantum phase transition
takes place
to an extended phase supporting long-range-colinear order 
along the $z$ axis in quantum spin-$1/2$ space with the following properties.
\begin{enumerate}
\item
They are either ferromagnetic (FM) or antiferromagnetic (Neel).
\item
They are either gapless or gapful.
\item
They have the approximate shape of a diamond (a 4-gon).
\item
They are centered at the positions
$(p\,\pi/4,q\,\pi/4)\in\mathbb{T}^{2}$
with $p$ and $q$ a pair of odd integers.
\item
Each diamond has four vertices, each of which is a quantum
tricritical point.
\end{enumerate}
This gives a total of four distinct diamond-shaped extended
long-range-ordered phases that we tag with the names
\begin{subequations}
\label{eq:four extended diamond phase}
\begin{equation}
\hbox{gNeel}^{\,}_{z},
\qquad
\hbox{gFM}^{\,}_{z},
\end{equation}
when they are gapless and
\begin{equation}
\hbox{Neel}^{\,}_{z},
\qquad
\hbox{FM}^{\,}_{z},
\end{equation}
\end{subequations}
when they are gapful.
Symmetry arguments dictate that
the eight distinct gapped ordered phases
\eqref{eq:eight gapped phases} supporting
a colinear order along the $x$ or $y$ direction in spin-$1/2$ space
must have vertical and horizontal boundaries.
Vertical boundaries realize deconfined quantum critical points
(DQCPs) with
central charge $\mathsf{c}=1$.
Horizontal boundaries realize weakly first-order
topological phase transitions between distinct long-range ordered phases.
Our numerics lack the precision needed to characterize the nature of the
quantum tricritical points.

For Hamiltonian \eqref{eq:def HA and HB d},
the polygon is a 7-gon. Three of its seven edges are boundaries
along which a direct quantum phase transition
takes place to another gapped phase controlled
by the fixed point
$(m'\,\pi/2,n'\,\pi/2)\in\mathbb{T}^{2}$
with $m'\neq m$ and $n'\neq n$ integers.
The other four edges are boundaries
along which a direct quantum phase transition
takes place to an extended phase supporting long-range-colinear order
along the $z$ axis in quantum spin-$1/2$ space with the following properties.
\begin{enumerate}
\item
They are either ferromagnetic (FM) or antiferromagnetic (Neel).
\item
They are either gapless or gapful.
\item
They have the approximate shape of a diamond (a 4-gon).
\item
They are centered at the positions
$(p\,\pi/4,q\,\pi/4)\in\mathbb{T}^{2}$
with $p$ and $q$ a pair of odd integers.
\item
Each diamond has four vertices, two of which are tricritical points
and the other two of which are quantum tetracritical points.
\end{enumerate}
This gives a total of four distinct diamond-shaped extended
long-range-ordered phases that we tag with the names
(\ref{eq:four extended diamond phase}).
Hereto, symmetry arguments dictate that
the eight distinct gapped ordered phases
\eqref{eq:eight gapped phases} supporting
a colinear order along the $x$ or $y$ direction in spin-$1/2$ space
must have vertical and horizontal boundaries.
Vertical boundaries realize deconfined quantum critical points with
central charge $\mathsf{c}=1$.
Horizontal boundaries realize topological quantum critical points
with central charge $\mathsf{c}=1$ except at the critical points
$(\pi/4,0)\hbox{ mod }(\pi/2,\pi/2)$ with central charge
$\mathsf{c}=2$.
Our numerics lack the precision needed to characterize the nature of the
quantum tricritical points.

\begin{figure}
\centering
\includegraphics[width=1\columnwidth]{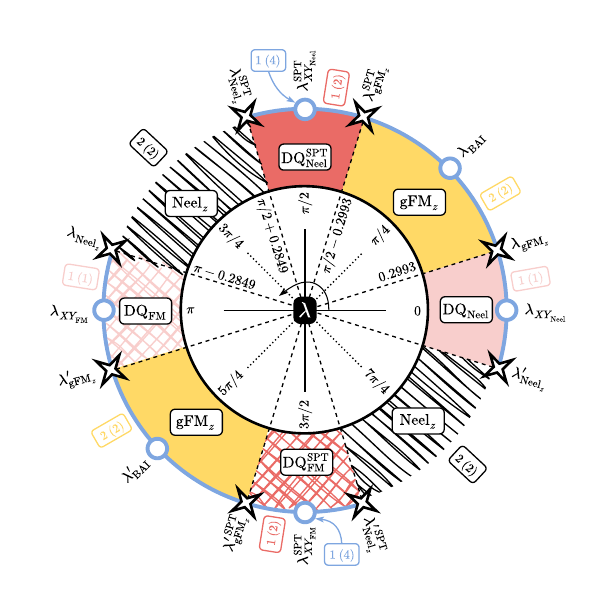}
\caption{
Phase diagram of Hamiltonian 
$\widehat{H}^{(\mathsf{b})}_{\mathrm{U}(1)}(\lambda)$
defined by Eq.\ (\ref{eq:def H U1})
as a function of the angle $\lambda\in[0,2\pi]$
along the unit circle colored in black.
We use a colored pie chart with eight sectors
to characterize the four long-range-ordered phases
(\ref{eq:four extended diamond phase})
and the four lines of deconfined-quantum-critical points (DQCPs)
(\ref{eq:four critical lines between the two diamond phases})
along the cut $\theta=\pi/4$
in Fig.~\ref{fig:phase diagram A}
or \ref{fig:phase diagram B}.
Integrable points
(tricritical points)
are denoted by circles (stars)
on the outer perimeter of the pie chart.
Intervals colored in blue
on the outer perimeter of the pie chart
refer to gapless phases of matter described by
a conformal field theory (CFT) with central charge $\mathsf{c}=1$.
There are six such intervals, each of which is bounded by
a pair of tricritical points and centered about an integrable point.
All six integrable and all eight tricritical points
can be arranged into unitarily equivalent pairs differing by the
angle $\pi$. Finally, we give the degeneracies with PBCs (OBCs) of 
(i) the four long-range-ordered phases,
(ii) the four lines of DQCPs, and
(iii) the two integrable points at $\lambda=\pi/2,3\pi/2$.
		}
\label{fig:phase diagram U1}
\end{figure}

At $\theta=\pi/4$, for any $\lambda\in[0,2\pi]$
and for any boundary conditions,
Hamiltonian
\eqref{eq:def HA and HB c}
equals Hamiltonian
\eqref{eq:def HA and HB d}, i.e.,
\begin{equation}
\widehat{H}^{(\mathsf{b})}_{\mathrm{U}(1)}(\lambda)\equiv
\widehat{H}^{(\mathsf{b})}_{\mathrm{I}}(\pi/4,\lambda)=
\widehat{H}^{(\mathsf{b})}_{\mathrm{II}}(\pi/4,\lambda).
\label{eq:def H U1}
\end{equation}
The subscript $\mathrm{U}(1)$ is a reference to the following
two facts. First, Hamiltonian (\ref{eq:def H U1})
commutes with the generator of an
$\mathrm{SO}(2)\cong\mathrm{U}(1)$
rotation in spin-$1/2$ space
that acts globally on every site. Second,
Hamiltonian (\ref{eq:def H U1})
can be fermionized, in which case the total fermionic charge
is conserved and, as such, is the generator of a global
$\mathrm{U}(1)$ on-site symmetry. 

We represent 
the cut along the line $\theta=\pi/4$ of the torus $\mathbb{T}^{2}$
in Fig.\ \ref{fig:phase diagram A}
with the unit circle colored in black
in Fig.\ \ref{fig:phase diagram U1}.
We use a colored pie chart divided into eight sectors
to describe the two long-range-ordered phases
\begin{equation}
\hbox{gFM}^{\,}_{z},
\qquad
\hbox{Neel}^{\,}_{z},
\label{eq:two intervening diamond phases for theta=piover4}
\end{equation}
out of the four defined in Eq.\
(\ref{eq:four extended diamond phase})
and the four lines of DQCPs,
which we tag as
\begin{equation}
\mathrm{DQ}^{\,}_{\mathrm{Neel}},
\qquad
\mathrm{DQ}^{\mathrm{SPT}}_{\mathrm{Neel}},
\qquad
\mathrm{DQ}^{\,}_{\mathrm{FM}},
\qquad
\mathrm{DQ}^{\mathrm{SPT}}_{\mathrm{FM}},
\label{eq:four critical lines between the two diamond phases}
\end{equation}
that are realized along the unit circle colored in black in
Fig.\ \ref{fig:phase diagram U1}. Here,
for each critical line, the subscript (superscript)
refers to the wave vector of the
competing colinear long-range order of the adjacent gapped phases
in Fig.\ \ref{fig:phase diagram A} or \ref{fig:phase diagram B}
(refers to the presence of an SPT attribute in each of the
adjacent gapped phases
in Fig.\ \ref{fig:phase diagram A} or \ref{fig:phase diagram B}).

\begin{table}
\caption{
Minimal ground-state degeneracies with PBCs (OBCs)
for the two long-range-ordered phases
and the two free-fermion points with SPT attributes. 
		 }
% %%%%%%%%%%%
% We need to put the \label AFTER \caption to avoid
% the Table to be referenced using \ref by the section number instead
% of the true Table number.
% %%%%%%%%%%%
\label{tab:degeneracies for LRO phases and free fermion SPT's}
\begin{ruledtabular}
\begin{tabular}{cccc}
$\mathrm{gFM}^{\,}_{z}$&
$\mathrm{Neel}^{\,}_{z}$&
$\lambda^{\mathrm{SPT}}_{XY^{\,}_{\mathrm{Neel}}}=\pi/2$&
$\lambda^{\mathrm{SPT}}_{XY^{\,}_{\mathrm{FM}}}=3\pi/2$\\
2\,(2)&
2\,(2)&
1\,(4)&
1\,(4)\\
\end{tabular}
\end{ruledtabular}
\end{table}

\begin{table}
\caption{
Minimal ground-state degeneracies with PBCs (OBCs)
for the four lines of DQCPs
(excluding $\lambda=\pi/2,3\pi/2$). 
		 }
% %%%%%%%%%%%
% We need to put the \label AFTER \caption to avoid
% the Table to be referenced using \ref by the section number instead
% of the true Table number.
% %%%%%%%%%%%
\label{tab:degeneracies for DQCPs}
\begin{ruledtabular}
\begin{tabular}{cccc}
$\mathrm{DQ}^{\,}_{\mathrm{Neel}}$&
$\mathrm{DQ}^{\mathrm{SPT}}_{\mathrm{Neel}}$&
$\mathrm{DQ}^{\,}_{\mathrm{FM}}$&
$\mathrm{DQ}^{\mathrm{SPT}}_{\mathrm{FM}}$\\
1\,(1)&
1\,(2)&
1\,(1)&
1\,(2)\\
\end{tabular}
\end{ruledtabular}
\end{table}

The two long-range-ordered phases
(\ref{eq:two intervening diamond phases for theta=piover4})
terminate at the tricritical points
\begin{subequations}
\begin{equation}
\lambda^{\,}_{\mathrm{gFM}^{\,}_{z}},\qquad
\lambda^{\mathrm{SPT}}_{\mathrm{gFM}^{\,}_{z}},\qquad
\lambda^{\mathrm{SPT}}_{\mathrm{Neel}^{\,}_{z}},\qquad
\lambda^{\,}_{\mathrm{Neel}^{\,}_{z}},
\end{equation}
when $\lambda\in[0,\pi]$ and
\begin{equation}
\lambda^{\prime}_{\mathrm{gFM}^{\,}_{z}},\qquad
\lambda^{\prime\,\mathrm{SPT}}_{\mathrm{gFM}^{\,}_{z}},\qquad
\lambda^{\prime\,\mathrm{SPT}}_{\mathrm{Neel}^{\,}_{z}},\qquad
\lambda^{\prime}_{\mathrm{Neel}^{\,}_{z}},
\end{equation}
when $\lambda'=\pi+\lambda\in[\pi,2\pi]$.
Our numerical estimates for the positions
\begin{equation}
\begin{footnotesize}
\begin{array}{llll}
0.2993,&
\frac{\pi}{2}-0.2993,&
\frac{\pi}{2}+0.2849,&
\pi-0.2849,\\&&&\\
\pi+0.2993,&
\frac{3\pi}{2}-0.2993,&
\frac{3\pi}{2}+0.2849,&
2\pi-0.2849,
\end{array}
\end{footnotesize}
\end{equation}
\end{subequations}
of these eight tricritical points
are approximately to scale in Fig.\ \ref{fig:phase diagram U1}.
They fix the boundaries of the four long-range-ordered phases
and of the four lines of DQCPs.

All six gapless phases
\begin{equation}
\begin{aligned}
&\mathrm{DQ}^{\,}_{\mathrm{Neel}},
&&\mathrm{gFM}^{\,}_{z}
&&&\mathrm{DQ}^{\mathrm{SPT}}_{\mathrm{Neel}},
\\
&\mathrm{DQ}^{\,}_{\mathrm{FM}},
&&\mathrm{gFM}^{\,}_{z},
&&&\mathrm{DQ}^{\mathrm{SPT}}_{\mathrm{FM}},
\label{eq:six critical lines at theta=piover4}
\end{aligned}
\end{equation}
in Fig.\ \ref{fig:phase diagram U1}
realize a conformal field theory (CFT)
with central charge $\mathsf{c}=1$.
This is indicated in Fig.\ \ref{fig:phase diagram U1} by a blue coloring
of the outer perimeter of the pie chart.

Of the six gapless phases (\ref{eq:six critical lines at theta=piover4}),
the four lines of DQCPs
(\ref{eq:four critical lines between the two diamond phases})
are approximately centered about the four
free-fermion quantum critical points
\begin{equation}
\begin{split}
&
\lambda^{\,}_{XY^{\,}_{\mathrm{Neel}}}=0,\qquad
\lambda^{\mathrm{SPT}}_{XY^{\,}_{\mathrm{Neel}}}=\frac{\pi}{2},\\
&
\lambda^{\,}_{XY^{\,}_{\mathrm{FM}}}=\pi,\qquad
\lambda^{\mathrm{SPT}}_{XY^{\,}_{\mathrm{FM}}}=\frac{3\pi}{2},
\end{split}
\end{equation}
each one of which 
realizes a conformal field theory (CFT)
with central charge $\mathsf{c}=1$.
The free-fermion DQCPs at
$\lambda^{\,}_{XY^{\,}_{\mathrm{Neel}}}$
and
$\lambda^{\,}_{XY^{\,}_{\mathrm{FM}}}$
are assigned a minimal ground-state degeneracy of one for both PBCs and OBCs,
as the finite-size energy separation between the first excited state above
the non-degenerate ground state scales like
$(2N)^{-1}$, with $2N$ the number of sites,
for either choice of boundary conditions.
This is not so for the free-fermion DQCPs at
$\lambda^{\mathrm{SPT}}_{XY^{\,}_{\mathrm{Neel}}}$
and
$\lambda^{\mathrm{SPT}}_{XY^{\,}_{\mathrm{FM}}}$,
as the degeneracies of their ground states
jump from one to four for any number $2N$ of sites
when selecting OBCs instead of PBCs.
Moreover, the ground-state degeneracy of four at
$\lambda^{\mathrm{SPT}}_{XY^{\,}_{\mathrm{Neel}}}$
and
$\lambda^{\mathrm{SPT}}_{XY^{\,}_{\mathrm{FM}}}$
when OBCs are selected,
is reduced to two along the critical lines
$\mathrm{DQ}^{\mathrm{SPT}}_{\mathrm{Neel}}$
and
$\mathrm{DQ}^{\mathrm{SPT}}_{\mathrm{FM}}$, respectively.
Indeed, our numerics for the scaling with $2N$ of
the energy separation between two
consecutive energy eigenstates along the critical lines
$\mathrm{DQ}^{\mathrm{SPT}}_{\mathrm{Neel}}$
and
$\mathrm{DQ}^{\mathrm{SPT}}_{\mathrm{FM}}$
establish that
(i) the energy separation between the first excited state above
the non-degenerate ground state scales like $(2N)^{-\kappa}$
with $\kappa=1$ for PBCs and $\kappa>1$ for OBCs,
(ii) while the energy separation between the second and first excited
states scales like $(2N)^{-1}$ for both PBCs and OBCs.
Hereto, we declare that the ground state and the first excited state
become degenerate in the limit $2N\uparrow\infty$ when their
finite-size level splitting decays faster than $(2N)^{-1}$.
This sensitivity to the choice of boundary conditions
of the ground-state degeneracy in a quantum critical phase or
at a quantum critical point
is an instance of symmetry-enriched gaplessness
\cite{Kestner11,Sau11,Fidkowski11b,Keselman150202037,
Ruhman17,Scaffidi170501557,Jiang18,Verresen18,Parker18,
Keselman18,Jones19,Ben-Zion20,Wang20,verresen2020topology,Duque21,Thorngren21,
Borla21,Verresen21,Hidaka22,Yu22,chang2022absence,prembabu2022boundary,Yang23,
Wen23,Mondal23,Yang23b,wang2023stability,li2023intrinsicallypurely,
huang2023topological,wen2023classification,Su24,yu2024universal,
ando2024gauging,zhong2024topological,yu2024quantum,li2024decorated,
zhang2024quantum,yang2024gifts}.
This is one of the main results of this paper.

Another main result of this paper is the existence of
two long-range-ordered and gapless phases
$\mathrm{gFM}^{\,}_{z}$
that are centered at the points
\begin{equation}
\lambda^{\,}_{\mathrm{BAI}}=\frac{\pi}{4},
\qquad
\lambda^{\prime}_{\mathrm{BAI}}=\frac{5\pi}{4},
\end{equation}
in Fig.\ \ref{fig:phase diagram U1}.
They were shown numerically to realize a
CFT with central charge $\mathsf{c}=1$
in Ref.\ \cite{Rey2025}. Moreover, both central points were
studied in their own right in Refs.\
\cite{Yang20,Zadnik21a,Zadnik21b,Pozsgay21,Jones22}.
Remarkably, it was shown in
Refs.\
\cite{Zadnik21a,Zadnik21b,Pozsgay21}
that these points are Bethe Ansatz integrable (BAI).

We report in
Table
\ref{tab:degeneracies for LRO phases and free fermion SPT's}
the degeneracies with PBCs (OBCs)
of the two distinct long-range-ordered phases together with
the two free-fermion points with SPT attributes.
We report in Table \ref{tab:degeneracies for DQCPs}
the degeneracies with PBCs (OBCs)
of the four lines of DQCPs, excluding $\lambda=\pi/2,3\pi/2$.

\subsubsection{Analytical results}
\label{subsubsec:Analytical results}

The analytical results that we present in this paper are all obtained
by studying, either perturbatively or non-perturbatively,
the vicinity of free-fermion critical points.
Without loss of generality, we only consider the regions
of the phase diagrams in Figs.\
\ref{fig:phase diagram A}
or
\ref{fig:phase diagram B}
that are enclosed by the dashed yellow frames.

As stated in Eq.\
(\ref{eq:def H U1}),
Hamiltonian
\eqref{eq:def HA and HB c}
equals Hamiltonian
\eqref{eq:def HA and HB d}
along the vertical line $\theta=\pi/4$.
This line hosts two critical points
at $(\pi/4,0)$ and $(\pi/4,\pi/2)$
that are unitarily related if and only if PBCs are selected.
The point $(\pi/4,0)$ is an example of a deconfined quantum critical point,
here a conformal field theory with the central charge $\mathsf{c}=1$
realized by a free Dirac Hamiltonian with
$\mathfrak{u}(1)$ current algebra.

We show that the effect of $|\lambda|\ll1$
along the interval $(\pi/4,\lambda)$
is captured by a Luttinger liquid with critical exponents that are
smooth functions of $\lambda$. This Luttinger liquid phase is equivalent
to that of the nearest-neighbor XXZ quantum spin-1/2 chain
with a dominant antiferromagnetic
XX exchange and a subdominant ZZ exchange that is ferromagnetic
when $0<\lambda\ll1$ and antiferromagnetic when $-1\ll\lambda<0$.

On the one hand,
we predict that the tricritical point
$(\pi/4,\lambda^{\prime}_{\mathrm{Neel}^{\,}_{z}})$
in Fig.\ \ref{fig:phase diagram U1}
is the $\mathrm{SU}(2)$ Heisenberg point of 
the nearest-neighbor antiferromagnetic XXZ quantum spin-1/2 chain,
i.e., a conformal field theory with central charge $\mathsf{c}=1$
and
$\mathfrak{su}(2)_{1}$ current algebra. Correspondingly, the
$\mathrm{Neel}^{\,}_{z}$ phase is the counterpart of the
ordered phase of the nearest-neighbor antiferromagnetic XXZ
quantum spin-1/2 chain when the ZZ exchange dominates over the XX
exchange.

On the other hand,
we do not predict that the tricritical point
$(\pi/4,\lambda^{\,}_{\mathrm{gFM}^{\,}_{z}})$
in Fig.\ \ref{fig:phase diagram U1}
is the $\mathrm{SU}(2)$ Heisenberg point of 
the nearest-neighbor ferromagnetic XXZ quantum spin-1/2 chain
at which a first-order phase transition to a ferromagnetic phase takes place.
Instead, we predict that the tricritical point
$(\pi/4,\lambda^{\,}_{\mathrm{gFM}^{\,}_{z}})$
is a continuous quantum critical point with Gaussian dynamical exponent
$\mathsf{z}=2$
at which a commensurate to incommensurate phase transition takes place.
The tricritical point
$(\pi/4,\lambda^{\,}_{\mathrm{gFM}^{\,}_{z}})$
distinguishes a commensurate Luttinger liquid 
from an incommensurate Luttinger liquid with the incommensuration
signaling the smooth onset of a non-vanishing uniform magnetization along the
$z$ axis in quantum spin-$1/2$ (a gradient-ordered Luttinger liquid).
This tricritical point is an example of a Lifshitz transition in a
quantum spin-$1/2$ chain at which the conventional Mermin-Wagner
restriction is evaded as follows.
The two-point correlation functions of the $x$
and $y$ components of the quantum spin-$1/2$ develop long-range order,
thereby spontaneously breaking the continuous $\mathrm{U}(1)$
symmetry, while the two-point correlation function of the $z$
component remains algebraic (see
Refs.~\cite{Watanabe24,Fava_PNAS_24,Nahum2506.21540,Andriyakhina2026,
chepiga2026,shimomura2026} for related
examples of continuous symmetry breaking in one dimension).

With OBCs, the exactly solvable critical points
$(\pi/4,0)$ and $(\pi/4,\pi/2)$ are not unitarily related anymore.
The critical point at $(\pi/4,\pi/2)$ inherits a four-fold degeneracy
from the SPT character of the gapped phases that it separates
along the line $(\theta,\pi/2)$.
By performing degenerate perturbation theory, we quantify the
lifting of this four-fold degeneracy as a function
of the number $2N$ of sites in the chain. For OBCs,
the agreement between perturbation theory and numerics is good
and explains why the Luttinger liquid phase is augmented
by a two-fold degeneracy originating from boundary zero modes
along the interval $(\pi/4,\pi/2+\lambda)$, $|\lambda|\ll1$.

The point $(0,\pi/4)$ [$(0,3\pi/4)$] behaves very differently for
Hamiltonians
\eqref{eq:def HA and HB c}
and
\eqref{eq:def HA and HB d}.
For both models,
the point $(0,\pi/4)$ [$(0,3\pi/4)$]
separates an SPT trivial from an SPT non-trivial
gapped phase,
whereby each gapped phase also breaks spontaneously a discrete symmetry.
However, it is only for Hamiltonian
\eqref{eq:def HA and HB c}
that the point $(0,\pi/4)$ [$(0,3\pi/4)$]
separates two gapped phases supporting competing long-range order.

A signature of this difference is that
the Jordan-Wigner transformation that turns Hamiltonian
\eqref{eq:def HA and HB d}
into a non-interacting Majorana chain at $(0,\pi/4)$ [$(0,3\pi/4)$],
turns Hamiltonian
\eqref{eq:def HA and HB c}
into a strongly interacting Majorana chain at $(0,\pi/4)$ [$(0,3\pi/4)$].

The point $(0,\pi/4)$ [$(0,3\pi/4)$] of Hamiltonian
\eqref{eq:def HA and HB c}
is predicted by mean-field theory to realize a first-order phase transition,
very much in the same way that the point 
$(\pi/4,0)$
is also predicted by mean-field theory
to realize a first-order phase transition.
However, the $\mathrm{U}(1)$ microscopic symmetry at the point
$(\pi/4,0)$ invalidates the prediction of mean-field theory because
the Mermin-Wagner theorem is operative. As there is no such
microscopic $\mathrm{U}(1)$ symmetry at the point $(0,\pi/4)$
[$(0,3\pi/4)$], an emergent $\mathrm{U}(1)$
symmetry would be needed to produce
a deconfined quantum critical point.
Numerics are consistent with a weakly first-order phase transition
at the point $(0,\pi/4)$.

The non-interacting Majorana chain that represents Hamiltonian
\eqref{eq:def HA and HB d}
at both $(0,\pi/4)$ and $(0,3\pi/4)$ 
is gapless in the continuum limit and realizes a
continuous quantum critical point whose universal properties are
captured by the $\mathsf{c}=2$ conformal field theory with
$\mathfrak{su}(2)^{\,}_{1}\oplus\mathfrak{su}(2)^{\,}_{1}$
current algebra. This point is a gapless superconductor in the
symmetry class BDI that separates two gapped superconducting phases,
one of which is a trivial while the other is a non-trivial
SPT phase in the symmetry class BDI.
The stability to a small perturbation
$0<|\theta|\ll1$
is dramatically different at $(0,\pi/4)$ than it is at $(0,3\pi/4)$.
In the latter case, the leading effect of normal ordering
of the interaction induced by $0<|\theta|\ll1$ is to produce
a Majorana mass, i.e., the point  $(0,3\pi/4)$ is a tetracritical point
at which four long-ranged-ordered gapped phases meet. In the former case,
the leading effect of normal ordering of the interaction induced by
$0<|\theta|\ll1$ is to shift the positions of the nodes of the
gapless superconductors, while the subleading effect is that of a
marginally relevant current-current interaction that gaps one and only one
of the two sectors with $\mathfrak{su}(2)^{\,}_{1}$ current algebra at
$(0,\pi/4)$. Hence, the neighborhood
$\{(\theta,\pi/4)\,|\,0<|\theta|\ll1\},$
of
$(0,\pi/4)$ is predicted to realize an incommensurate Luttinger liquid
with central charge $\mathsf{c}=1$ and $\mathfrak{su}(2)^{\,}_{1}$
current algebra that terminates in a pair of tricritical points.
We could not elucidate the nature of this pair of tricritical points,
but it is tempting to conjecture that they also realize a Lifshitz transition
to the $\mathrm{gFM}^{\,}_{z}$ phase.

\subsection{Outline}

The two families of Hamiltonians
\eqref{eq:def HA and HB c}  
and
\eqref{eq:def HA and HB d}  
are defined in Sec.\ \ref{sec:Two families of Hamiltonians}.
Their symmetries, dualities, and the special cases that have been studied
in the literature are also presented in Sec.\
\ref{sec:Two families of Hamiltonians}.
The main results of this paper are summarized by
the phase diagrams
\ref{fig:phase diagram A}
and
\ref{fig:phase diagram B}.
The shapes of all competing phases
and the nature of their boundaries
in coupling space are obtained by numerical means in
Sec.\ \ref{sec:Numerical exploration of the phase diagrams}.
Analytical results are derived in Sec.\
\ref{sec:analytical_exploration_of_the_phase_diagrams}.
The Appendices provide technical details and additional numerical
and analytical results. Appendix
\ref{appsec:iDMRG}
reviews iDMRG \cite{schollwock2011,hauschild2018}
and the transfer-matrix formalism
\cite{Ornstein1914,Herviou2025}
used throughout the numerical analysis.
Appendix~\ref{sec:Additional evidence gFMz}
provides additional numerical evidence for the stability of the extended
$\mathrm{gFM}^{\,}_{z}$ phase.
Appendix
\ref{sec:effective_interactions_between_boundary_spin_1_2_degrees_of_freedom}
derives the effective interaction between
the boundary spin-$1/2$ degrees of freedom.
Appendices
\ref{appsec:Degeneracies of the ground states H1II piover4 or 3piover3}
and
\ref{appsec:Degeneracies of the ground states H0II piover4 or 3piover3}
derive the ground-state degeneracies at the free-Majorana critical points
for OBCs and PBCs, respectively.

\section{Two families of Hamiltonians}
\label{sec:Two families of Hamiltonians}

\subsection{Hilbert space}
\label{subsec:Hilbert space}

Let $2N$ be an even integer and let
\begin{subequations}
\label{eq:def Hilbert space}
\begin{equation}
\Lambda\coloneqq
\left\{j\left|\,j=1,\ldots,2N\right.\right\}
\label{eq:def Hilbert space a}
\end{equation}
denote a chain with
lattice spacing $\mathfrak{a}$
of cardinality $2N$.
For any site $j\in\Lambda$,
define the operators
$\widehat{X}^{\,}_{j}$,
$\widehat{Y}^{\,}_{j}$,
and
$\widehat{Z}^{\,}_{j}$
by postulating that they 
obey the Pauli algebra
\begin{align}
&
\widehat{X}^{\,}_{j}\,
\widehat{Y}^{\,}_{j}=
-\widehat{Y}^{\,}_{j}\,
\widehat{X}^{\,}_{j}=
\ii\widehat{Z}^{\,}_{j},
\label{eq:def Hilbert space b}
\\
&
\widehat{X}^{2}_{j}=
\mathds{1}^{\,}_{\mathfrak{H}},
\label{eq:def Hilbert space c}
\end{align}
up to cyclic permutations of the letters $X$, $Y$, and $Z$,
while any two of
$\widehat{X}^{\,}_{j}$,
$\widehat{Y}^{\,}_{j}$,
and
$\widehat{Z}^{\,}_{j}$
commute when indexed by distinct site labels.
Here, $\mathds{1}^{\,}_{\mathfrak{H}}$
is the identity on their domain of definition,
namely the Hilbert space
\begin{equation}
\mathfrak{H}\cong
\mathbb{C}^{2^{|\Lambda|}}.
\label{eq:def Hilbert space d}
\end{equation}
We shall also make the identification
\begin{equation}
\widehat{X}^{\,}_{j+2N}=
\widehat{X}^{\,}_{j},
\quad
\widehat{Y}^{\,}_{j+2N}=
\widehat{Y}^{\,}_{j},
\quad
\widehat{Z}^{\,}_{j+2N}=
\widehat{Z}^{\,}_{j},
\label{eq:def Hilbert space f}
\end{equation}
\end{subequations}
for any $j\in\Lambda$
in order to accommodate (PBCs) only.

\begin{widetext}
\subsection{Lattice Hamiltonians}
\label{subsec:Lattice Hamiltonians}

We define the two-dimensional torus
\begin{subequations}
\label{eq:def HA and HB}
\begin{equation}
\mathbb{T}^{2}\coloneqq
\Big\{
(\theta,\lambda)\in\mathbb{R}^{2}\, \Big|\,
\theta\in[0,2\pi],
\
\lambda\in[0,2\pi]
\Big\}
\label{eq:def HA and HB a}
\end{equation}
as coupling space and
the integer $\mathsf{b}=0,1$ shall select PBCs
when $\mathsf{b}=0$ and open boundary conditions (OBCs)
when $\mathsf{b}=1$.
For every directed nearest-neighbor bond $\langle j,j+1\rangle$
of the lattice,
we define the Hermitian bond operator
\begin{equation}
\widehat{h}^{\,}_{j}(\theta)\coloneqq
\cos(\theta)\,
\widehat{X}^{\,}_{j}\,
\widehat{X}^{\,}_{j+1}
+
\sin(\theta)\,
\widehat{Y}^{\,}_{j}\,
\widehat{Y}^{\,}_{j+1}.
\label{eq:def HA and HB b}
\end{equation}
\noindent
The first family of Hamiltonians is defined by
\begin{align}
\mathrlap{\widehat{H}^{(\mathsf{b})}_{\mathrm{I}}(\theta,\lambda)\coloneqq
\cos(\lambda)
\sum_{j=1}^{2N-\mathsf{b}}
\widehat{h}^{\,}_{j}(\theta)
+
\sin(\lambda)
\sum_{j=1}^{2N-3\mathsf{b}}
\widehat{Z}^{\,}_{j}\,
\widehat{h}^{\,}_{j+1}(\theta)\,
\widehat{Z}^{\,}_{j+3}.}
\hphantom{\widehat{H}^{(\mathsf{b})}_{\mathrm{II}}(\theta,\lambda)\coloneqq
\cos(\lambda)
\sum_{j=1}^{2N-\mathsf{b}}
\widehat{h}^{\,}_{j}(\theta)
+
\sin(\lambda)
\sum_{j=1}^{2N-3\mathsf{b}}
\widehat{Z}^{\,}_{j}\,
\widehat{h}^{\,}_{j+1}\left(\frac{\pi}{2}-\theta\right)
\widehat{Z}^{\,}_{j+3}.}
\label{eq:def HA and HB c}
\end{align}
The second family of Hamiltonians is defined by
\begin{align}
\widehat{H}^{(\mathsf{b})}_{\mathrm{II}}(\theta,\lambda)\coloneqq
\cos(\lambda)
\sum_{j=1}^{2N-\mathsf{b}}
\widehat{h}^{\,}_{j}(\theta)
+
\sin(\lambda)
\sum_{j=1}^{2N-3\mathsf{b}}
\widehat{Z}^{\,}_{j}\,
\widehat{h}^{\,}_{j+1}\left(\frac{\pi}{2}-\theta\right)
\widehat{Z}^{\,}_{j+3}.
\label{eq:def HA and HB d}
\end{align}
\end{subequations}

The labels in the region defined by the dashed purple boundary in 
Figs.\
\ref{fig:phase diagram A}
and
\ref{fig:phase diagram B}
at the points
\begin{subequations}
\begin{equation}
\begin{array}{lll}
(0,0),\qquad
& 
(\pi/2,0),\qquad
& 
(\pi,0),\\
(0,\pi/2),\qquad
&
(\pi/2,\pi/2),\qquad
&
(\pi,\pi/2),\\
(0,\pi),\qquad
&
(\pi/2,\pi),\qquad
&
(\pi,\pi),
\end{array}
\end{equation}
follow from
$\widehat{H}^{(\mathsf{b})}_{\mathrm{I}}(\theta,\lambda)$
and
$\widehat{H}^{(\mathsf{b})}_{\mathrm{II}}(\theta,\lambda)$
simplifying at these points to the classical Hamiltonians
\begin{equation}
\begin{array}{lll}
+
\sum\limits_{j=1}^{2N-\mathsf{b}\hphantom{3}}
\widehat{X}^{\,}_{j}\,\widehat{X}^{\,}_{j+1},\qquad
&
+
\sum\limits_{j=1}^{2N-\mathsf{b}\hphantom{3}}
\widehat{Y}^{\,}_{j}\,\widehat{Y}^{\,}_{j+1},\qquad
& 
-
\sum\limits_{j=1}^{2N-\mathsf{b}\hphantom{3}}
\widehat{X}^{\,}_{j}\,\widehat{X}^{\,}_{j+1},
\\
+
\sum\limits_{j=1}^{2N-3\mathsf{b}}
\widehat{Z}^{\,}_{j}\,
\widehat{X}^{\,}_{j+1}\,\widehat{X}^{\,}_{j+2}\,
\widehat{Z}^{\,}_{j+3},\qquad
&
+
\sum\limits_{j=1}^{2N-3\mathsf{b}}
\widehat{Z}^{\,}_{j}\,
\widehat{Y}^{\,}_{j+1}\,\widehat{Y}^{\,}_{j+2}\,
\widehat{Z}^{\,}_{j+3},\qquad
& 
-
\sum\limits_{j=1}^{2N-3\mathsf{b}}
\widehat{Z}^{\,}_{j}\,
\widehat{X}^{\,}_{j+1}\,\widehat{X}^{\,}_{j+2}\,
\widehat{Z}^{\,}_{j+3}\,,
\\
-
\sum\limits_{j=1}^{2N-\mathsf{b}\hphantom{3}}
\widehat{X}^{\,}_{j}\,\widehat{X}^{\,}_{j+1},\qquad
& 
-
\sum\limits_{j=1}^{2N-\mathsf{b}\hphantom{3}}
\widehat{Y}^{\,}_{j}\,\widehat{Y}^{\,}_{j+1},\qquad
& 
+
\sum\limits_{j=1}^{2N-\mathsf{b}\hphantom{3}}
\widehat{X}^{\,}_{j}\,\widehat{X}^{\,}_{j+1},
\end{array}
\end{equation}
and
\begin{equation}
\begin{array}{lll}
+
\sum\limits_{j=1}^{2N-\mathsf{b}\hphantom{3}}
\widehat{X}^{\,}_{j}\,\widehat{X}^{\,}_{j+1},\qquad
&
+
\sum\limits_{j=1}^{2N-\mathsf{b}\hphantom{3}}
\widehat{Y}^{\,}_{j}\,\widehat{Y}^{\,}_{j+1},\qquad
& 
-
\sum\limits_{j=1}^{2N-\mathsf{b}\hphantom{3}}
\widehat{X}^{\,}_{j}\,\widehat{X}^{\,}_{j+1},
\\
+
\sum\limits_{j=1}^{2N-3\mathsf{b}}
\widehat{Z}^{\,}_{j}\,
\widehat{Y}^{\,}_{j+1}\,\widehat{Y}^{\,}_{j+2}\,
\widehat{Z}^{\,}_{j+3},\qquad
&
+
\sum\limits_{j=1}^{2N-3\mathsf{b}}
\widehat{Z}^{\,}_{j}\,
\widehat{X}^{\,}_{j+1}\,\widehat{X}^{\,}_{j+2}\,
\widehat{Z}^{\,}_{j+3},\qquad
& 
-
\sum\limits_{j=1}^{2N-3\mathsf{b}}
\widehat{Z}^{\,}_{j}\,
\widehat{Y}^{\,}_{j+1}\,\widehat{Y}^{\,}_{j+2}\,
\widehat{Z}^{\,}_{j+3}\,,
\\
-
\sum\limits_{j=1}^{2N-\mathsf{b}\hphantom{3}}
\widehat{X}^{\,}_{j}\,\widehat{X}^{\,}_{j+1},\qquad
& 
-
\sum\limits_{j=1}^{2N-\mathsf{b}\hphantom{3}}
\widehat{Y}^{\,}_{j}\,\widehat{Y}^{\,}_{j+1},\qquad
& 
+
\sum\limits_{j=1}^{2N-\mathsf{b}\hphantom{3}}
\widehat{X}^{\,}_{j}\,\widehat{X}^{\,}_{j+1},
\end{array}
\end{equation}
\end{subequations}
respectively. These are classical Hamiltonians as they are the
sum of pairwise commuting local Hermitian operators labeled
by site index $j\in\Lambda$.
\end{widetext}

\subsection{Relations in coupling space}
\label{sec:relations_in_coupling_space}

\subsubsection{Reduced coupling space}

For any choice $\mathsf{b}=0,1$ of boundary conditions,
the coupling space (\ref{eq:def HA and HB a})
can be reduced to the region
\begin{equation}
\Omega^{\,}_{\mathrm{P}}\coloneqq
\Big\{
(\theta,\lambda)\in\mathbb{T}^{2}\ \Big|\ 
\theta\in[0,\pi],
\
\lambda\in[0,\pi]
\Big\}\subset\mathbb{T}^{2}
\label{eq:coupling space purple}
\end{equation}
highlighted in purple in
Figs.~\ref{fig:phase diagram A}
and \ref{fig:phase diagram B}.
\begin{proof}
To see this, observe that,
for any choice of the boundary conditions
$\mathsf{b}=0,1$,
\begin{equation}
\widehat{H}^{(\mathsf{b})}(\theta+\pi,\lambda)=
\widehat{H}^{(\mathsf{b})}(\theta,\lambda+\pi)=
-
\widehat{H}^{(\mathsf{b})}(\theta,\lambda),
\label{eq:sign change by shift of pi on 2-torus}
\end{equation}
where 
$\widehat{H}^{(\mathsf{b})}\equiv
\widehat{H}^{(\mathsf{b})}_{\mathrm{I}},
\widehat{H}^{(\mathsf{b})}_{\mathrm{II}}$.
Now, one verifies that the action
\begin{subequations}
\label{eq:def UeZ}
\begin{equation}
\widehat{U}^{\,}_{Z,\pi}\,
\widehat{H}^{(\mathsf{b})}(\theta,\lambda)\,
\widehat{U}^{\dag}_{Z,\pi}=
-
\widehat{H}^{(\mathsf{b})}(\theta,\lambda)
\label{eq:def UeZ a}
\end{equation}
for
$\widehat{H}^{(\mathsf{b})}\equiv
\widehat{H}^{(\mathsf{b})}_{\mathrm{I}},
\widehat{H}^{(\mathsf{b})}_{\mathrm{II}}$
by conjugation with the unitary operator
\begin{equation}
\widehat{U}^{\,}_{Z,\pi}\coloneqq
\prod\limits_{j=1}^{N}\,
\widehat{Z}^{\,}_{2j},
\label{eq:def UeZ b}
\end{equation}
\end{subequations}
realizes the transformation
(\ref{eq:sign change by shift of pi on 2-torus}).
By combining
Eqs.\ (\ref{eq:sign change by shift of pi on 2-torus})
and  (\ref{eq:def UeZ}),
we conclude that
\begin{subequations}
\label{eq:conjugation by UeZ}
\begin{equation}
\widehat{U}^{\,}_{Z,\pi}\,
\widehat{H}^{(\mathsf{b})}(\theta,\lambda)\,
\widehat{U}^{\dag}_{Z,\pi}=
\widehat{H}^{(\mathsf{b})}(\theta+\pi,\lambda)
\label{eq:conjugation by UeZ shifts theta by pi}
\end{equation}
and
\begin{equation}
\widehat{U}^{\,}_{Z,\pi}\,
\widehat{H}^{(\mathsf{b})}(\theta,\lambda)\,
\widehat{U}^{\dag}_{Z,\pi}=
\widehat{H}^{(\mathsf{b})}(\theta,\lambda+\pi),
\label{eq:conjugation by UeZ shifts lambda by pi}
\end{equation}
\end{subequations}
for
$\widehat{H}^{(\mathsf{b})}\equiv
\widehat{H}^{(\mathsf{b})}_{\mathrm{I}},
\widehat{H}^{(\mathsf{b})}_{\mathrm{II}}$.
As any two Hamiltonians related on the 2-torus
(\ref{eq:def HA and HB a})
by the transformation
$\theta\mapsto\theta+\pi$
or
$\lambda\mapsto\lambda+\pi$
are unitarily equivalent, they possess the same
energy eigenvalues and their eigenstates are unitarily related.
\end{proof}

The action of $\widehat{U}^{\,}_{Z,\pi}$
on the coupling space is pictured by arrows in
Figs.\
\ref{fig:phase diagram A}
and
\ref{fig:phase diagram B}.

We focus on Hamiltonian
$\widehat{H}^{(\mathsf{b})}_{\mathrm{I}}$
with $\mathsf{b}=0,1$.
For any choice $\mathsf{b}=0,1$ of boundary conditions,
we claim that we can restrict the coupling space
$\Omega^{\,}_{\mathrm{P}}$ to the region
\begin{equation}
\Omega^{\,}_{\mathrm{Y}}\coloneqq
\Big\{
(\theta,\lambda)\in\Omega^{\,}_{\mathrm{P}}\ \Big|\ 
\theta\in[0,\pi/2],
\
\lambda\in[0,\pi]
\Big\}\subset\Omega^{\,}_{\mathrm{P}}
\label{eq:coupling space yellow}
\end{equation}
for $\widehat{H}^{(\mathsf{b})}_{\mathrm{I}}(\theta,\lambda)$.
Region \eqref{eq:coupling space yellow}
is highlighted in yellow in
Fig.~\ref{fig:phase diagram A}.
\begin{proof}
Consider the bijective transformation
\begin{subequations}
\begin{equation}
\begin{split}
\Omega^{\,}_{\mathrm{P}}&\to\Omega^{\,}_{\mathrm{P}},
\\
(\theta,\lambda)&\mapsto(\pi-\theta,\pi-\lambda),
\end{split}
\end{equation}
with the fixed point $(\pi/2,\pi/2)$,
see Fig.~\ref{fig:phase diagram A}.
This is a counterclockwise rotation by $\pi$
of the region $\Omega^{\,}_{\mathrm{P}}$
about the fixed point $(\pi/2,\pi/2)$.
The reduction of coupling space from region 
\eqref{eq:coupling space purple} to region
\eqref{eq:coupling space yellow} 
follows from the existence of a unitary transformation
$\widehat{U}$ acting by conjugation as
\begin{equation}
\widehat{U}\,
\widehat{H}^{(\mathsf{b})}_{\mathrm{I}}(\theta,\lambda)\,
\widehat{U}^{\dag}=
\widehat{H}^{(\mathsf{b})}_{\mathrm{I}}
(\pi-\theta,\pi-\lambda).
\label{eq:unitary U_uparrow A}
\end{equation}
One verifies that the choice
\begin{equation}
\widehat{U}^{\,}_{X,\pi}\coloneqq
\prod\limits_{j=1}^{N}\,
\widehat{X}^{\,}_{2j}
\label{eq:def U_up}
\end{equation}
\end{subequations}
realizes the unitary transformation
\eqref{eq:unitary U_uparrow A}.
\end{proof}

We focus on Hamiltonian
$\widehat{H}^{(\mathsf{b})}_{\mathrm{II}}$
with $\mathsf{b}=0,1$.
For any choice $\mathsf{b}=0,1$ of boundary conditions,
we claim that we can restrict the coupling space
$\Omega^{\,}_{\mathrm{P}}$ to the region
$\Omega^{\,}_{\mathrm{Y}}$ defined by Eq.\
(\ref{eq:coupling space yellow})
for $\widehat{H}^{(\mathsf{b})}_{\mathrm{II}}(\theta,\lambda)$.
Region \eqref{eq:coupling space yellow}
is highlighted in yellow in
Fig.\ \ref{fig:phase diagram B}.
\begin{proof}
Consider the bijective transformation
\begin{subequations}
\begin{equation}
\begin{split}
\Omega^{\,}_{\mathrm{P}}&\to\Omega^{\,}_{\mathrm{P}},
\\
(\theta,\lambda)&\mapsto(\pi-\theta,\lambda),
\end{split}
\end{equation}
with the vertical fixed line $(\pi/2,\lambda)$,
see Fig.~\ref{fig:phase diagram B}.
This is a reflection of region $\Omega^{\,}_{\mathrm{P}}$
about the vertical line
$(\pi/2,\lambda)$.
The reduction of coupling space from region
\eqref{eq:coupling space purple} to region
\eqref{eq:coupling space yellow}
follows from the existence of a unitary transformation
$\widehat{U}$ acting by conjugation as
\begin{equation}
\widehat{U}\,
\widehat{H}^{(\mathsf{b})}_{\mathrm{II}}(\theta,\lambda)\,
\widehat{U}^{\dag}=
\widehat{H}^{(\mathsf{b})}_{\mathrm{II}}
(\pi-\theta,\lambda).
\label{eq:unitary U_rightarrow B}
\end{equation}
One verifies that the choice
\begin{equation}
\widehat{U}^{\,}_{Y,\pi}\coloneqq
\prod\limits_{j=1}^{N}\,
\widehat{Y}^{\,}_{2j}
\label{eq:U_Y even a}
\end{equation}
\end{subequations}
realizes the unitary transformation
\eqref{eq:unitary U_rightarrow B}.
\end{proof}

\begin{widetext}
We decompose region $\Omega^{\,}_{\mathrm{Y}}$
defined by Eq.\ \eqref{eq:coupling space yellow}
as either the union of the bottom (B) and top (T) regions
\begin{subequations}
\label{eq:coupling quadrants}
\begin{align}
&
\Omega^{\,}_{\mathrm{Y,B}}\coloneqq
\Big\{
(\theta,
\lambda)\in\Omega^{\,}_{\mathrm{Y}}
\,\Big|\,
\theta\in
[0,\pi/2],\,
\lambda\in
[0,\pi/2]
\Big\}\subset\Omega^{\,}_{\mathrm{Y}},
\label{eq:coupling quadrants a}
\\
&
\Omega^{\,}_{\mathrm{Y,T}}\coloneqq
\Big\{
(\theta,
\lambda)\in\Omega^{\,}_{\mathrm{Y}}
\,\Big|\,
\theta\in
[0,\pi/2],\,
\lambda\in[\pi/2,\pi]
\Big\}\subset\Omega^{\,}_{\mathrm{Y}},
\label{eq:coupling quadrants b}
\end{align}
respectively, or the union of the left (L) and right (R) regions
\begin{align}
&
\Omega^{\,}_{\mathrm{Y,L}}\coloneqq
\Big\{
(\theta,
\lambda)\in\Omega^{\,}_{\mathrm{Y}}
\,\Big|\,
\theta\in
[0,\pi/4],\,
\lambda\in
[0,\pi]
\Big\}\subset\Omega^{\,}_{\mathrm{Y}},
\label{eq:coupling quadrants c}
\\
&
\Omega^{\,}_{\mathrm{Y,R}}\coloneqq
\Big\{
(\theta,
\lambda)\in\Omega^{\,}_{\mathrm{Y}}
\,\Big|\,
\theta\in
[\pi/4,\pi/2],\,
\lambda\in
[0,\pi]
\Big\}\subset\Omega^{\,}_{\mathrm{Y}},
\label{eq:coupling quadrants d}
\end{align}
\end{subequations}
respectively.
\end{widetext}
No unitary transformation could be found to
relate the regions
$\Omega^{\,}_{\mathrm{Y,B}}$
and
$\Omega^{\,}_{\mathrm{Y,T}}$ 
for Hamiltonians
\eqref{eq:def HA and HB c}
and
\eqref{eq:def HA and HB d}
and for any boundary conditions $\mathsf{b}=0,1$.
However, we claim that we can halve the range of
$\theta$
in region
$\Omega^{\,}_{\mathrm{Y}}$
to the region
$\Omega^{\,}_{\mathrm{Y,L}}$
for Hamiltonians
\eqref{eq:def HA and HB c}
and
\eqref{eq:def HA and HB d}
and for any boundary conditions $\mathsf{b}=0,1$.
\begin{proof}
Introduce the unitary transformation
\begin{subequations}
\begin{equation}
\widehat{U}^{\,}_{\mathrm{R}}\coloneqq
\left(
\prod\limits_{j=1}^{2N}\,
e^{-\ii\frac{\pi}{4}\,
\widehat{Z}^{\,}_{j}}
\right)
\widehat{U}^{\,}_{X},
\qquad
\widehat{U}^{\,}_{X}\coloneqq
\prod\limits_{j=1}^{2N}\,
\widehat{X}^{\,}_{j}.
\label{eq:U_R}
\end{equation}
The operator $\widehat{U}^{\,}_{\mathrm{R}}$
defined by Eq.~\eqref{eq:U_R} is the composition
of the global rotation of $\pi$ about the
$x$ axis (through $\widehat{U}^{\,}_{\mathrm{X}}$)
with the global clockwise rotation of $\pi/2$
about the $z$ axis
[through $\widehat{U}^{\dag}_{Z}(\pi/2)$
defined in Eq.\ (\ref{eq:def GU(1) a})].
Its action by conjugation on the Pauli operators is
\begin{equation}
\widehat{U}^{\,}_{\mathrm{R}}:
\bigl(
\widehat{X}^{\,}_{j},\,
\widehat{Y}^{\,}_{j},\,
\widehat{Z}^{\,}_{j}
\bigr)
\mapsto
\bigl(
\widehat{Y}^{\,}_{j},\,
\widehat{X}^{\,}_{j},\,
-\widehat{Z}^{\,}_{j}
\bigr).
\label{eq:action U_R on 3 Paulis}
\end{equation}
\end{subequations}
For any choice $\mathsf{b}=0,1$ of boundary conditions
and for any $\theta\in[0,\pi/4]$ and $\lambda\in[0,\pi]$,
one verifies the identities
\begin{align}
&
\widehat{U}^{\,}_{\mathrm{R}}\,
\widehat{H}^{(\mathsf{b})}(\theta,\lambda)\,
\widehat{U}^{\dag}_{\mathrm{R}}=
\widehat{H}^{(\mathsf{b})}(\pi/2-\theta,\lambda),
\label{eq:action U_R on Hamiltonians}
\end{align}
where
$\widehat{H}^{(\mathsf{b})}\equiv
\widehat{H}^{(\mathsf{b})}_{\mathrm{I}},
\widehat{H}^{(\mathsf{b})}_{\mathrm{II}}$.
\end{proof}

Therefore, we can restrict
the coupling space $\Omega^{\,}_{\mathrm{Y}}$ in
Eq.~\eqref{eq:coupling space yellow}
of the Hamiltonians
\eqref{eq:def HA and HB c}
and
\eqref{eq:def HA and HB d}
to the union of the regions
\begin{subequations}
\label{eq:coupling quadrants left}
\begin{align}
&
\Omega^{\,}_{\mathrm{Y,B,L}}\coloneqq
\left(	
\Omega^{\,}_{\mathrm{Y,B}}\cap
\Omega^{\,}_{\mathrm{Y,L}}
\right)\subset
\Omega^{\,}_{\mathrm{Y,B}}\subset
\Omega^{\,}_{\mathrm{Y}},
\label{eq:coupling quadrants left a}
\\
&
\Omega^{\,}_{\mathrm{Y,T,L}}\coloneqq
\left(	
\Omega^{\,}_{\mathrm{Y,T}}\cap
\Omega^{\,}_{\mathrm{Y,L}}
\right)\subset
\Omega^{\,}_{\mathrm{Y,T}}\subset
\Omega^{\,}_{\mathrm{Y}},
\label{eq:coupling quadrants left b}
\end{align}
\end{subequations}
for any boundary conditions $\mathsf{b}=0,1$.

\subsubsection{Reduced coupling space for PBCs only}

For any boundary condition ($\mathsf{b}=0,1$),
introduce the unitary transformations
\begin{subequations}	
\begin{align}
\widehat{U}^{(\mathsf{b})}_{\mathrm{CZ}}\coloneqq&\,
\prod\limits_{j=1}^{2N-\mathsf{b}}\,
\frac{1}{2}
\left(
\mathds{1}^{\,}_{\mathfrak{H}}+
\widehat{Z}^{\,}_{j}+	
\widehat{Z}^{\,}_{j+1}-	
\widehat{Z}^{\,}_{j}\,\widehat{Z}^{\,}_{j+1}	
\right),
\label{eq:U_CZ}
\\
\widehat{U}^{(\mathsf{b})}_{\mathrm{E}}\coloneqq&\,
\widehat{U}^{(\mathsf{b})}_{\mathrm{CZ}}\,
\widehat{U}^{\,}_{\mathrm{R}}.
\label{eq:U_E}
\end{align}
\end{subequations}

The operator $\widehat{U}^{(\mathsf{b})}_{\mathrm{E}}$
defined by Eq.~\eqref{eq:U_E} is the composition of
the operator $\widehat{U}^{\,}_{\mathrm{R}}$ with
the control-$Z$ (CZ) operator
$\widehat{U}^{(\mathsf{b})}_{\mathrm{CZ}}$.
The action of $\widehat{U}^{(\mathsf{b})}_{\mathrm{CZ}}$
by conjugation on the Pauli operators is
\begin{subequations}
\label{eq:action U_CZ on 3 Paulis}
\begin{widetext}
\begin{equation}
\widehat{U}^{(\mathsf{b})}_{\mathrm{CZ}}:
\bigl(
\widehat{X}^{\,}_{j},\,
\widehat{Y}^{\,}_{j},\,
\widehat{Z}^{\,}_{j}
\bigr)
\mapsto
\bigl(
\widehat{Z}^{\,}_{j-1}
\widehat{X}^{\,}_{j}
\widehat{Z}^{\,}_{j+1},\,
\widehat{Z}^{\,}_{j-1}
\widehat{Y}^{\,}_{j}
\widehat{Z}^{\,}_{j+1},\,
\widehat{Z}^{\,}_{j}
\bigr),
\label{eq:action U_CZ on 3 Paulis in bulk}
\end{equation}
\end{widetext}
for $j=1+\mathsf{b},\cdots,2N-\mathsf{b}$.
For completeness,
the action of
$\widehat{U}^{(1)}_{\mathrm{CZ}}$ on the Pauli operators
at $j=1$ and $j=2N$ when $\mathsf{b}=1$ is given by
\begin{equation}
\widehat{U}^{(1)}_{\mathrm{CZ}}:
\bigl(
\widehat{X}^{\,}_{j},\,
\widehat{Y}^{\,}_{j},\,
\widehat{Z}^{\,}_{j}
\bigr)
\mapsto
\bigl(
\widehat{X}^{\,}_{j}
\widehat{Z}^{\,}_{j+1},\,
\widehat{Y}^{\,}_{j}
\widehat{Z}^{\,}_{j+1},\,
\widehat{Z}^{\,}_{j}
\bigr),
\label{eq:action U_CZ on 3 Paulis j=1}
\end{equation}
and
\begin{equation}
\widehat{U}^{(1)}_{\mathrm{CZ}}:
\bigl(
\widehat{X}^{\,}_{j},\,
\widehat{Y}^{\,}_{j},\,
\widehat{Z}^{\,}_{j}
\bigr)
\mapsto
\bigl(
\widehat{Z}^{\,}_{j-1}
\widehat{X}^{\,}_{j},\,
\widehat{Z}^{\,}_{j-1}
\widehat{Y}^{\,}_{j},\,
\widehat{Z}^{\,}_{j}
\bigr).
\label{eq:action U_CZ on 3 Paulis j=2N}
\end{equation}
\end{subequations}
respectively.

On the one hand, it follows that
we can reduce the region
$\Omega^{\,}_{\mathrm{Y,B}}$
to its bottom region
\begin{equation}
\Omega^{\,}_{\mathrm{Y,B^{\downarrow}}}\coloneqq
\Big\{
(\theta,
\lambda)\in\Omega^{\,}_{\mathrm{Y}}
\,\Big|\,
\theta\in
[0,\pi/2],\,
\lambda\in
[0,\pi/4]
\Big\}
\label{eq:coupling space PBCs bottom}
\end{equation}
when selecting PBCs for both
Hamiltonians
\eqref{eq:def HA and HB c}
and
\eqref{eq:def HA and HB d}.
Indeed, for any $(\theta,\lambda)\in\Omega^{\,}_{\mathrm{Y,B}}$, i.e.,
for any point from the bottom quadrant of $\Omega^{\,}_{\mathrm{Y}}$,
one verifies the pair of identities
\begin{subequations}
\label{eq:reduce coupling space PBC bottom}
\begin{equation}	
\widehat{U}^{(0)}_{\mathrm{E}}\,
\widehat{H}^{(0)}_{\mathrm{I}}(\theta,\lambda)\,
\widehat{U}^{(0)\,\dag}_{\mathrm{E}}=
\widehat{H}^{(0)}_{\mathrm{I}}(\theta,\pi/2-\lambda),
\label{eq:reduce coupling space PBC bottom a}
\end{equation}
and
\begin{equation}	
\widehat{U}^{(0)}_{\mathrm{CZ}}\,
\widehat{H}^{(0)}_{\mathrm{II}}(\theta,\lambda)\,
\widehat{U}^{(0)\,\dag}_{\mathrm{CZ}}=
\widehat{H}^{(0)}_{\mathrm{II}}(\theta,\pi/2-\lambda).
\label{eq:reduce coupling space PBC bottom b}
\end{equation}
\end{subequations}

On the other hand,
we can also reduce the region
$\Omega^{\,}_{\mathrm{Y,T}}$
to its top region 
\begin{equation}
\Omega^{\,}_{\mathrm{Y,T^{\uparrow}}}\coloneqq
\Big\{
(\theta,
\lambda)\in\Omega^{\,}_{\mathrm{Y}}
\,\Big|\,
\theta\in
[0,\pi/2],\,
\lambda\in
[3\pi/4,\pi]
\Big\}
\label{eq:coupling space PBCs top}
\end{equation}
when selecting PBCs for both
Hamiltonians
\eqref{eq:def HA and HB c}
and
\eqref{eq:def HA and HB d}.
Indeed, for any $(\theta,\lambda)\in\Omega^{\,}_{\mathrm{Y,T}}$, i.e.,
for any point from the top quadrant of $\Omega_{\mathrm{Y}}$,
one verifies the pair of identities
\begin{subequations}
\label{eq:reduce coupling space PBC top}
\begin{equation}	
\left(
\widehat{U}^{\,}_{Z,\pi}\,
\widehat{U}^{(0)}_{\mathrm{E}}
\right)
\widehat{H}^{(0)}_{\mathrm{I}}(\theta,\lambda)
\left(
\widehat{U}^{\,}_{Z,\pi}\,
\widehat{U}^{(0)}_{\mathrm{E}}
\right)^{\dag}=
\widehat{H}^{(0)}_{\mathrm{I}}(\theta,3\pi/2-\lambda),
\end{equation}
and
\begin{equation}	
\left(
\widehat{U}^{\,}_{Z,\pi}\,
\widehat{U}^{(0)}_{\mathrm{CZ}}
\right)
\widehat{H}^{(0)}_{\mathrm{II}}(\theta,\lambda)
\left(
\widehat{U}^{\,}_{Z,\pi}\,
\widehat{U}^{(0)}_{\mathrm{CZ}}
\right)^{\dag}=
\widehat{H}^{(0)}_{\mathrm{II}}(\theta,3\pi/2-\lambda).
\end{equation}
\end{subequations}

Consequently, we can restrict
the coupling space $\Omega^{\,}_{\mathrm{Y}}$ in
Eq.~\eqref{eq:coupling space yellow}
of the Hamiltonians
\eqref{eq:def HA and HB c}
and
\eqref{eq:def HA and HB d}
to the union of the regions
\begin{subequations}
\label{eq:coupling quadrants left PBCs}
\begin{align}
&
\Omega^{\,}_{\mathrm{Y,B^{\downarrow},L}}\coloneqq
\left(	
\Omega^{\,}_{\mathrm{Y,B^{\downarrow}}}\cap
\Omega^{\,}_{\mathrm{Y,L}}
\right)\subset
\Omega^{\,}_{\mathrm{Y,B,L}}\subset
\Omega^{\,}_{\mathrm{Y}},
\label{eq:coupling quadrants left PBCs a}
\\
&
\Omega^{\,}_{\mathrm{Y,T^{\uparrow},L}}\coloneqq
\left(
\Omega^{\,}_{\mathrm{Y,T^{\uparrow}}}\cap
\Omega^{\,}_{\mathrm{Y,L}}	
\right)\subset
\Omega^{\,}_{\mathrm{Y,T,L}}\subset
\Omega^{\,}_{\mathrm{Y}},
\label{eq:coupling quadrants left PBCs b}
\end{align}
\end{subequations}
upon choosing PBCs with $\mathsf{b}=0$.
This union is a halving of the union of the regions
(\ref{eq:coupling quadrants left a})
and
(\ref{eq:coupling quadrants left b}).

\subsection{Symmetries}
\label{subsec:Symmetries}

\subsubsection{Discrete symmetries}

For any choice $\mathsf{b}=0,1$
of boundary conditions
and for any $\theta,\lambda\in[0,2\pi]$,
Hamiltonians
\eqref{eq:def HA and HB c} and \eqref{eq:def HA and HB d}
are both invariant under conjugation by
the elements of the group
\begin{subequations}
\label{eq:symmetries}
\begin{equation}
\mathrm{G}\coloneqq
\mathbb{Z}^{X}_{2}\times
\mathbb{Z}^{Y}_{2}\times
\mathbb{Z}^{T}_{2}
\end{equation}
generated by the unitary operators
\begin{equation}
\widehat{U}^{\,}_{X}\coloneqq
\prod\limits_{j\in\Lambda}
\widehat{X}^{\,}_{j},
\qquad
\widehat{U}^{\,}_{Y}\coloneqq
\prod\limits_{j\in\Lambda}
\widehat{Y}^{\,}_{j},
\label{eq:symmetries a}
\end{equation}
and the antiunitary operator
\begin{equation}
\widehat{U}^{\,}_{T}\coloneqq
\mathrlap{\widehat{U}^{\,}_{Y}\,
\mathsf{K},}
\hphantom{\prod\limits_{j\in\Lambda}\widehat{Y}^{\,}_{j},}
\qquad
\left(\widehat{U}^{\,}_{T}\right)^{2}=
(-1)^{2N}\,
\mathds{1}^{\,}_{\mathfrak{H}},
\label{eq:symmetries b}
\end{equation}
\end{subequations}
respectively.
Here, $\mathsf{K}$ denotes complex conjugation.
The subgroup
$\mathbb{Z}^{X}_{2}\times
\mathbb{Z}^{Y}_{2}$
is realized by any two of the three global $\pi$ rotations about
the $x$, $y$, and $z$ axis in spin space,
while the subgroup
$\mathbb{Z}^{T}_{2}$ is realized by the
reversal of time.

We have explicitly demanded that the
system size $2N$ be
an even integer for $\widehat{U}^{\,}_{T}$
to square to the identity and for the
symmetry generators in 
Eq.~\eqref{eq:symmetries} to commute pairwise.

Upon choosing PBCs,
the discrete translation group $\mathbb{Z}^{\,}_{2N}$,
generated by the translation
by one lattice spacing,
holds for any $\theta,\lambda\in[0,2\pi]$.

\subsubsection{Continuous symmetry when $\theta=\pi/4$}

At $\theta=\pi/4$ in region $\Omega^{\,}_{\mathrm{Y}}$,
Hamiltonians
\eqref{eq:def HA and HB c}
and
\eqref{eq:def HA and HB d}
are both invariant under conjugation by
the elements of the continuous group
\begin{subequations}
\label{eq:def GU(1)}
\begin{equation}
\mathrm{G}^{\,}_{\mathrm{U}(1)}\cong\mathrm{U}(1)
\label{eq:def GU(1) a} 
\end{equation} 
generated by the unitary operator
\begin{equation}
\widehat{U}^{\,}_{Z}(\alpha)\coloneqq
e^{+\ii\frac{\alpha}{2}
\sum\limits_{j=1}^{2N}\,
\widehat{Z}^{\,}_{j}},
\qquad
\alpha\in[0,2\pi[.
\label{eq:def GU(1) b}
\end{equation}
We observe that
$\widehat{U}^{\,}_{\mathrm{R}}$
defined in Eq.~\eqref{eq:U_R}
is nothing but
\begin{equation}
\widehat{U}^{\,}_{\mathrm{R}}=
\widehat{U}^{\,}_{Z}(\pi/2)\,
\widehat{U}^{\,}_{X}.
\label{eq:def GU(1) c}
\end{equation}
\end{subequations}

\subsubsection{Continuous symmetry when $\theta=3\pi/4$ }

At $\theta=3\pi/4$ in region $\Omega^{\,}_{\mathrm{Y}}$,
Hamiltonians
\eqref{eq:def HA and HB c}
and
\eqref{eq:def HA and HB d}
are both invariant under conjugation by
the elements of the continuous group
\begin{subequations}
\label{eq:def GU(1)e}
\begin{equation}
\mathrm{G}^{\,}_{\mathrm{U}(1),\pi}\cong\mathrm{U}(1)
\label{eq:def GU(1)e a}
\end{equation} 
generated by the unitary operator
\begin{equation}
\widehat{U}^{\prime}_{Z,\pi}(\alpha)\coloneqq
e^{-\ii\frac{\alpha}{2}
\sum\limits_{j=1}^{2N}\,
(-1)^{j}\,\widehat{Z}^{\,}_{j}},
\qquad
\alpha\in[0,2\pi[.
\label{eq:def GU(1)e b}
\end{equation}
We observe that
\begin{equation}
\widehat{U}^{\prime}_{Z,\pi}(\alpha)=
\widehat{U}^{\,}_{X,\pi}\,
\widehat{U}^{\prime}_{Z}(\alpha)\,
\widehat{U}^{\dag}_{X,\pi},
\qquad
\alpha\in[0,2\pi[.
\label{eq:def GU(1)e c}
\end{equation}
\end{subequations}

\subsection{Limiting cases studied in the literature}
\label{subsec:Limiting cases studied in the literature}

Without loss of generality,
we work in region $\Omega^{\,}_{\mathrm{Y,B}}$
defined in Eq.\ (\ref{eq:coupling quadrants a})
of the coupling space $\mathbb{T}^{2}$
defined in Eq.\ (\ref{eq:def HA and HB a}).

Lieb, Schultz, and Mattis (LSM)
in Ref.\ \cite{Lieb61}
solved exactly Hamiltonians
\eqref{eq:def HA and HB c}
and
\eqref{eq:def HA and HB d}
along the boundary $\lambda=0$ of
the region $\Omega^{\,}_{\mathrm{Y,B}}$
by means of the Jordan-Wigner transformation \cite{Jordan28}. 

Hamiltonians
\eqref{eq:def HA and HB c}
and
\eqref{eq:def HA and HB d}
are equal along the line $\theta=\pi/4$
in region $\Omega^{\,}_{\mathrm{Y,B}}$.
At the point
\begin{subequations}
\label{eq:def HBAI}
\begin{equation}
\theta=\lambda=\pi/4
\end{equation} 
with PBCs,
Hamiltonians
\eqref{eq:def HA and HB c}
and
\eqref{eq:def HA and HB d}
simplify to
\begin{align}
&
\widehat{H}^{(0)}_{\mathrm{BAI}}\coloneqq
\sum_{j=1}^{2N}
\widehat{P}^{\,}_{j-1,j+2}\,
\left(
\widehat{X}^{\,}_{j}\,\widehat{X}^{\,}_{j+1}
+
\widehat{Y}^{\,}_{j}\,\widehat{Y}^{\,}_{j+1}
\right),
\\
&
\widehat{P}^{\,}_{j-1,j+2}\coloneqq
\frac{1+\widehat{Z}^{\,}_{j-1}\,\widehat{Z}^{\,}_{j+2}}{2}.
\end{align}
\end{subequations}
Here, the projector
$\widehat{P}^{\,}_{j-1,j+2}$
projects out configurations
with opposite spins on sites $j-1$ and $j+2$.
Upon fermionization \cite{Jordan28},
Hamiltonian (\ref{eq:def HBAI}) turns into
a correlated hopping for spinless electrons.
Hamiltonian $\widehat{H}^{(\mathsf{b})}_{\mathrm{BAI}}$
with PBCs $\mathsf{b}=0$ or OBCs $\mathsf{b}=1$
was studied in Refs.\
\cite{Yang20,Zadnik21a,Zadnik21b,Pozsgay21,Jones22}
on its own right. Remarkably, it was shown in
Refs.\
\cite{Zadnik21a,Zadnik21b}
that $\widehat{H}^{(0)}_{\mathrm{BAI}}$
is Bethe Ansatz integrable (BAI).
Reference \cite{Pozsgay21}
showed that $\widehat{H}^{(1)}_{\mathrm{BAI}}$
is also BAI.

Suzuki in Ref.\ \cite{Suzuki1971}
showed that Hamiltonian
\eqref{eq:def HA and HB d}
with PBCs 
along the boundary $\theta=0$ ($\theta=\pi/2$)
of region $\Omega^{\,}_{\mathrm{Y,B}}$
is integrable in that it maps to
a quadratic chain of spinless fermions.
Suzuki was unaware that the point 
at $(0,\pi/4)$ [$(\pi/2,\pi/4)$]
realizes a continuous quantum topological phase transition
with central charge $\mathsf{c}=2$ that
separates two gapped phases such that,
as shown in Ref.\
\cite{Verresen17}
(see also Ref.\ \cite{MudryChamon2025}),
\begin{enumerate}
\item[(i)]
each gapped phase breaks spontaneously the symmetry group
(\ref{eq:symmetries})
down to the same subgroup
$\mathbb{Z}^{X}_{2}$ ($\mathbb{Z}^{Y}_{2}$),
\item[(ii)]
while one and only one of the two gapped phases
acquires an additional
topologically protected degeneracy of four
when OBCs are selected.
\end{enumerate}

Rey et al.\ in Ref.\ \cite{Rey2025} introduced
Hamiltonian
\eqref{eq:def HA and HB c}
in the region $\Omega^{\,}_{\mathrm{Y,B}}\subset\mathbb{T}^{2}$
and explored its phase diagram numerically.

\section{Numerical exploration of the phase diagrams}
\label{sec:Numerical exploration of the phase diagrams}

We numerically explore the phase diagrams of Hamiltonians
\eqref{eq:def HA and HB c} and
\eqref{eq:def HA and HB d}
that are given in
Figs.~\ref{fig:phase diagram A} and
\ref{fig:phase diagram B}, respectively.
The study of Hamiltonians
\eqref{eq:def HA and HB c} and
\eqref{eq:def HA and HB d}
was performed through ED and DMRG with either OBCs or PBCs
and through infinite-size DMRG (iDMRG)
\footnote{
The ED study was done with our own code.
The DMRG study was done with our own code
as well as the Julia library ITensors.
The iDMRG study was done with our own code
		 }.

Without loss of generality, we restrict the study
of Hamiltonians
\eqref{eq:def HA and HB c} and
\eqref{eq:def HA and HB d}
from the coupling space defined by
the torus $\mathbb{T}^{2}$ in Eq.~\eqref{eq:def HA and HB a}
to the coupling space defined by the
region $\Omega^{\,}_{\mathrm{Y}}$ in
Eq.~\eqref{eq:coupling space yellow}.

Finally, we refer the reader to Ref.~\cite{Rey2025},
in which the phase diagram of
Hamiltonian \eqref{eq:def HA and HB c}
was extensively studied numerically in the region
$\Omega^{\,}_{\mathrm{Y,B}}\subset
\Omega^{\,}_{\mathrm{Y}}\subset
\mathbb{T}^{2}$.

\subsection{Extended phases}
\label{subsec:Extended phases with non-vanishing area}

Within the torus \eqref{eq:def HA and HB a}, each Hamiltonian
\eqref{eq:def HA and HB c} or
\eqref{eq:def HA and HB d}
realizes eight distinct gapped phases that
are labeled in Eq.~\eqref{eq:eight gapped phases},
together with four different
diamond-shaped phases that support
colinear magnetic long-range order
along the $z$ axis in spin space
which are labeled in Eq.~\eqref{eq:four extended diamond phase}.
These 12 phases are extended phases
separated by phase boundaries.
Extended phases also meet at multicritical points.

In this section we explore the phases labeled in
Eqs.~\eqref{eq:eight gapped phases} and
\eqref{eq:four extended diamond phase}
that appear in the region $\Omega^{\,}_{\mathrm{Y}}$ defined by
Eq.~\eqref{eq:coupling space yellow}.
We hereby define a set of observables that
characterize the phase diagrams.

First, for each $\alpha=x,y,z$
labeling the $x$, $y$, $z$ axis in space, respectively,
we define on the site $j\in\Lambda$ the Pauli operator
\begin{equation}
\widehat{\sigma}^{\alpha}_{j}\coloneqq
\begin{cases}
\widehat{X}^{\,}_{j},&\alpha=x,\\
\widehat{Y}^{\,}_{j},&\alpha=y,\\
\widehat{Z}^{\,}_{j},&\alpha=z.\\
\end{cases}
\end{equation}

Second, any state from $\mathfrak{H}$
defined in Eq.~\eqref{eq:def Hilbert space a}
is denoted
$\ket*{\Psi}\in\mathfrak{H}$.

Third, for any $\alpha=x,y,z$,
the uniform (FM) magnetization
$m^{\alpha}_{\mathrm{uni}}$,
the staggered (AFM) magnetization
$m^{\alpha}_{\mathrm{sta}}$,
the uniform (FM) SPT magnetization
$m^{\mathrm{SPT},\alpha}_{\mathrm{uni}}$,
and
the staggered (AFM) SPT magnetization
$m^{\mathrm{SPT},\alpha}_{\mathrm{sta}}$
in the state $\ket*{\Psi}$
are defined by the expectation values
\begin{widetext}
\begin{subequations}
\label{eq:def order parameters}
\begin{align}
&
m^{\alpha}_{\mathrm{uni}}[\Psi]\coloneqq
\frac{1}{2N}\,
\sum_{j=1}^{2N} \ev*{\,\widehat{\sigma}^{\alpha}_{j}}{\Psi},
&&
m^{\mathrm{SPT},\alpha}_{\mathrm{uni}}[\Psi]\coloneqq
\frac{1}{2N-2\mathsf{b}}\,
\sum_{j=1+\mathsf{b}}^{2N-\mathsf{b}}
\ev*{\,
\widehat{\sigma}^{z}_{j-1}\,
\widehat{\sigma}^{\alpha}_{j}\,
\widehat{\sigma}^{z}_{j+1}
}{\Psi},
\label{eq:muni + spt}
\\
&
m^{\alpha}_{\mathrm{sta}}[\Psi]\coloneqq
\frac{1}{2N}\,
\sum_{j=1}^{2N}
(-1)^{j}\,
\ev*{\,\widehat{\sigma}^{\alpha}_{j}}{\Psi},
&&
m^{\mathrm{SPT},\alpha}_{\mathrm{sta}}[\Psi]\coloneqq
\frac{1}{2N-2\mathsf{b}}\,
\sum_{j=1+\mathsf{b}}^{2N-\mathsf{b}}
(-1)^{j}\,
\ev*{\,
\widehat{\sigma}^{z}_{j-1}\,
\widehat{\sigma}^{\alpha}_{j}\,
\widehat{\sigma}^{z}_{j+1}
}{\Psi},
\label{eq:msta + spt}
\end{align}
\end{subequations}
\end{widetext}
respectively. We recall that
the parameter $\mathsf{b}=0\ (1)$ dictates
the choice of PBCs (OBCs).
The magnetizations defined in Eq.\ (\ref{eq:def order parameters})
are the order parameters of the extended phases.
That is, any one of the eight distinct
gapped phases \eqref{eq:eight gapped phases}
or any one of the four different diamond-shaped phases
\eqref{eq:four extended diamond phase} 
is uniquely determined by the set of
non-vanishing expectation values among the
twelve defined in Eq.~\eqref{eq:def order parameters}
that is acquired in the thermodynamic limit $2N\to\infty$.

Fourth, for any $\alpha=x,y,z$,
for any $s^{\,}_{0}=2,\cdots,2N-2$
and for any $s=1,\ldots,2N-s^{\,}_{0}-1$,
the two-point correlation function
$C^{\alpha}$,
the connected component
of the two-point correlation function
$C^{\alpha,\mathrm{c}}$,
the dressed two-point correlation function
$C^{\mathrm{SPT},\alpha}$,
and the connected dressed two-point correlation function
$C^{\mathrm{SPT},\alpha,\mathrm{c}}$
in the state $\ket*{\Psi}\in\mathfrak{H}$
are defined by their expectation values
\begin{widetext}
\begin{subequations}
\label{eq:Cs}
\begin{align}
&
\mathrlap{C^{\alpha}_{s^{\,}_{0},s^{\,}_{0}+s}[\Psi]}
\hphantom{C^{\mathrm{SPT},\alpha,\mathrm{c}}_{s^{\,}_{0},s^{\,}_{0}+s}[\Psi]}\coloneqq
\ev*{\, \widehat{\sigma}^{\alpha}_{s^{\,}_{0}}\,
\widehat{\sigma}^{\alpha}_{s^{\,}_{0}+s}}{\Psi},
\label{eq:Calpha}
\\
&
\mathrlap{C^{\alpha,\mathrm{c}}_{s^{\,}_{0},s^{\,}_{0}+s}[\Psi]}
\hphantom{C^{\mathrm{SPT},\alpha,\mathrm{c}}_{s^{\,}_{0},s^{\,}_{0}+s}[\Psi]}\coloneqq
\ev*{\,\widehat{\sigma}^{\alpha}_{s^{\,}_{0}}\,
\widehat{\sigma}^{\alpha}_{s^{\,}_{0}+s}}{\Psi}
-\!
\ev*{\,\widehat{\sigma}^{\alpha}_{s^{\,}_{0}}}{\Psi}
\ev*{\,\widehat{\sigma}^{\alpha}_{s^{\,}_{0}+s}}{\Psi},
\label{eq:CalphaConn}
\\
&
\mathrlap{C^{\mathrm{SPT},\alpha}_{s^{\,}_{0},s^{\,}_{0}+s}[\Psi]}
\hphantom{C^{\mathrm{SPT},\alpha,\mathrm{c}}_{s^{\,}_{0},s^{\,}_{0}+s}[\Psi]}\coloneqq
\ev*{\,
(
\widehat{\sigma}^{z}_{s^{\,}_{0}-1}\,
\widehat{\sigma}^{\alpha}_{s^{\,}_{0}}\,
\widehat{\sigma}^{z}_{s^{\,}_{0}+1}
)\,
(
\widehat{\sigma}^{z}_{s^{\,}_{0}-1+s}\,
\widehat{\sigma}^{\alpha}_{s^{\,}_{0}+s}\,
\widehat{\sigma}^{z}_{s^{\,}_{0}+1+s}
)\,
}{\Psi},
\label{eq:CalphaSPT}
\\
&
C^{\mathrm{SPT},\alpha,\mathrm{c}}_{s^{\,}_{0},s^{\,}_{0}+s}[\Psi]\coloneqq
\ev*{\,
(
\widehat{\sigma}^{z}_{s^{\,}_{0}-1}\,
\widehat{\sigma}^{\alpha}_{s^{\,}_{0}}\,
\widehat{\sigma}^{z}_{s^{\,}_{0}+1}
)\,
(
\widehat{\sigma}^{z}_{s^{\,}_{0}-1+s}\,
\widehat{\sigma}^{\alpha}_{s^{\,}_{0}+s}\,
\widehat{\sigma}^{z}_{s^{\,}_{0}+1+s}
)\,
}{\Psi}
\nonumber\\
&
\hphantom{C^{\mathrm{SPT},\alpha,\mathrm{c}}_{s^{\,}_{0},s^{\,}_{0}+s}[\Psi]\coloneqq\,}
-\!
\ev*{\,
(
\widehat{\sigma}^{z}_{s^{\,}_{0}-1}\,
\widehat{\sigma}^{\alpha}_{s^{\,}_{0}}\,
\widehat{\sigma}^{z}_{s^{\,}_{0}+1}
)
}{\Psi}
\ev*{\,
(
\widehat{\sigma}^{z}_{s^{\,}_{0}-1+s}\,
\widehat{\sigma}^{\alpha}_{s^{\,}_{0}+s}\,
\widehat{\sigma}^{z}_{s^{\,}_{0}+1+s}
)\,
}{\Psi}, 
\label{eq:CalphaSPTConn}
\end{align}
\end{subequations}
\end{widetext}
respectively. When OBCs are selected, these two-point correlation
functions depend on both $s^{\,}_{0}$ and $s^{\,}_{0}+s$.
Notice that $C^{\mathrm{SPT},\alpha}
_{s^{\,}_{0},s^{\,}_{0}+s}$ is the
correlation between two monomials of order three. 
For $\alpha=x,y$, the correlation $C^{\mathrm{SPT},\alpha}
_{s^{\,}_{0},s^{\,}_{0}+s}$
is the relevant correlation function for one of the SPT phases.
The correlation function $C^{\mathrm{SPT},\alpha}
_{s^{\,}_{0},s^{\,}_{0}+s}$
in Eq.\ (\ref{eq:CalphaSPT})
for $\alpha=x,y$
is obtained by conjugation of the spin operators
on the right-hand side of Eq.~\eqref{eq:Calpha}
by $\widehat{U}^{\,}_{\mathrm{CZ}}$
defined in Eq.~\eqref{eq:U_CZ}.
Any one of the eight distinct gapped phases
\eqref{eq:eight gapped phases}
or any one of the four different diamond-shaped phases
\eqref{eq:four extended diamond phase} 
is also uniquely determined by
the long-range order (LRO) established when a subset from
the six correlation functions
\eqref{eq:Calpha} and \eqref{eq:CalphaSPT}
is non-vanishing in the limit $s\to\infty$.
 
Finally, we shall use the short-hand notation
\begin{equation}
\ev{\,\bm{\cdot}\,}^{\,}_{\Psi} \equiv \ev{\,\bm{\cdot}\,}{\Psi}
\label{eq:bracket psi}
\end{equation}
for any expectation value in the state $|\Psi\rangle\in\mathfrak{H}$.

\subsubsection{The \texorpdfstring{$\mathrm{Neel}^{\,}_{x}$}{NeelX} phase}

We begin by investigating the $\mathrm{Neel}^{\,}_{x}$ phase
that extends
from the classical fixed point located at
$(0,0)\in\Omega^{\,}_{\mathrm{Y}}$
to a finite region of coupling space.

At any point in this phase, the ground state of Hamiltonians
\eqref{eq:def HA and HB c} and
\eqref{eq:def HA and HB d}
breaks spontaneously the symmetry group
$\mathrm{G}=\mathbb{Z}^{X}_{2}\times
\mathbb{Z}^{Y}_{2}\times
\mathbb{Z}^{T}_{2}$
defined in Eq.~\eqref{eq:symmetries} down to the subgroup
\begin{equation}
\mathrm{G}^{\,}_{\mathrm{Neel}^{\,}_{x}}\coloneqq
\mathbb{Z}^{X}_{2}\times
\mathbb{Z}^{YT}_{2},
\quad
\mathbb{Z}^{YT}_{2}\coloneqq
\left\{
\mathds{1}^{\,}_{\mathfrak{H}},\ 
\widehat{U}^{\,}_{Y}\,
\widehat{U}^{\,}_{T}
\right\},
\end{equation}
which is represented linearly on the ground state's manifold
for both PBCs and OBCs.

At the point $(0,0)\in\Omega^{\,}_{\mathrm{Y}}$
the Hamiltonian \eqref{eq:def HA and HB c}
equals Hamiltonian \eqref{eq:def HA and HB d}, i.e.,
\begin{subequations}
\begin{align}
\widehat{H}^{(\mathsf{b})}_{\mathrm{Neel}^{\,}_{x}}
&\coloneqq
\sum\limits_{j=1}^{2N-\mathsf{b}}
\widehat{X}^{\,}_{j}\,\widehat{X}^{\,}_{j+1}
\\
&\hphantom{:}\equiv
\widehat{H}^{(\mathsf{b})}_{\mathrm{I}}(0,0)\equiv
\widehat{H}^{(\mathsf{b})}_{\mathrm{II}}(0,0)
\label{eq:def H NeelX}
\end{align}
\end{subequations}
and is equivalent to the antiferromagnetic nearest-neighbor
classical Ising chain.
If we label the eigenstate of Hamiltonian
$\widehat{H}^{(\mathsf{b})}_{\mathrm{Neel}^{\,}_{x}}$ 
with eigenenergy $E^{\,}_{n}$ by
$\ket*{\Psi^{\,}_{n}}$, its two ground states
can be expressed as
\begin{subequations}
\begin{align}
&
\ket*{\Psi^{\,}_{0}}\coloneqq
|
\leftarrow\rightarrow\cdots
\leftarrow\rightarrow\
\leftarrow\rightarrow\cdots
\leftarrow\rightarrow
\rangle,
\label{eq:NeelX states a}
\\
&
\ket*{\Psi^{\,}_{1}}\coloneqq
\widehat{U}^{\,}_{T}\ket*{\Psi^{\,}_{0}},
\label{eq:NeelX states b}
\end{align}
in the basis for which
\begin{equation}
\widehat{X}^{\,}_{j}\,|\rightarrow\rangle^{\,}_{j}=
+|\rightarrow\rangle^{\,}_{j},
\qquad
\widehat{X}^{\,}_{j}\,|\leftarrow\rangle^{\,}_{j}=
-|\leftarrow\rangle^{\,}_{j}.
\label{eq:NeelX states c}
\end{equation}
\label{eq:NeelX states}
\end{subequations}
All ordered energy eigenvalues
\begin{equation}
E^{\,}_{0}\leq
E^{\,}_{1}\leq
\cdots\leq
E^{\,}_{n}\leq
E^{\,}_{n+1}\leq
\cdots
\end{equation}
of Hamiltonian
$\widehat{H}^{(\mathsf{b})}_{\mathrm{Neel}^{\,}_{x}}$
are at least two-fold degenerate, for any spin configuration must share the
same energy as the one obtained by reversing the direction of all
spins through the operation of time reversal $\widehat{U}^{\,}_{T}$
defined in Eq.~\eqref{eq:symmetries b}.

When moving slightly away from the corner
$(0,0)\in\Omega^{\,}_{\mathrm{Y}}$ in
Figs.~\ref{fig:phase diagram A}
and \ref{fig:phase diagram B},
the $\mathrm{Neel}^{\,}_{x}$ states \eqref{eq:NeelX states}
are no longer exact ground states.
Moreover, for any finite number $2N$ of sites, the energy
$E^{\,}_{0}$ is strictly lower than the energy
$E^{\,}_{1}$ of the first excited state.
However, the energy spacing $E^{\,}_{1}-E^{\,}_{0}$
between the ground state energy $E^{\,}_{0}$ and the first
excited state energy $E^{\,}_{1}$
is observed numerically to vanish exponentially fast with $2N$
so that a two-fold degeneracy of the ground state is recovered
in the thermodynamic limit $2N\to\infty$.
The energy gap $E^{\,}_{2}-E^{\,}_{1}$ is non-vanishing
and closes at the phase boundaries in the thermodynamic limit.

At any point in the $\mathrm{Neel}^{\,}_{x}$ phase,
we have verified using the DMRG method that
the ground state of Hamiltonians
\eqref{eq:def HA and HB c} and
\eqref{eq:def HA and HB d}
supports a non-vanishing magnetization
\begin{equation}
0<\bigl|m^{x}_{\mathrm{sta}}\bigr|\le1,
\end{equation}
where the equality $\bigl|m^{x}_{\mathrm{sta}}\bigr|=1$
holds only at the
classical point $(0,0)\in\Omega^{\,}_{\mathrm{Y}}$.

\begin{figure}
\subfigure{\includegraphics[width=1\columnwidth]{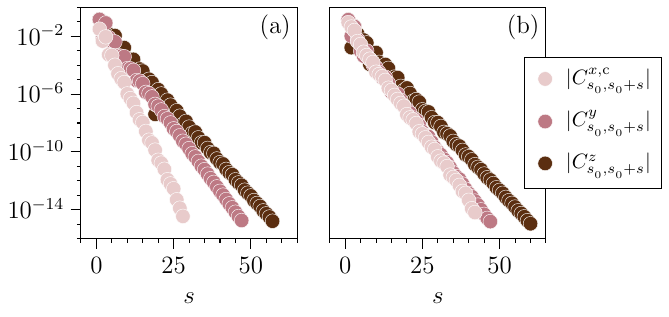} \label{fig:NeelX OBCs Correlations a}}%
\subfigure{\label{fig:NeelX OBCs Correlations b}}%
\caption{
The case of OBCs.
(a) Dependence on $s$ given $s^{\,}_{0}$ of the absolute value of the
spin-spin correlation functions
$C^{x,\mathrm{c}}_{s^{\,}_{0},s^{\,}_{0}+s}[\Psi^{\,}_{0}]$,
$C^{y}_{s^{\,}_{0},s^{\,}_{0}+s}[\Psi^{\,}_{0}]$,
and
$C^{z}_{s^{\,}_{0},s^{\,}_{0}+s}[\Psi^{\,}_{0}]$, 
defined in Eq.~\eqref{eq:Cs},
computed using the DMRG ($\chi=128$) 
for the ground state
$|\Psi^{\,}_{0}\rangle$
of Hamiltonian
$\widehat{H}^{(1)}_{\mathrm{I}}$ in
Eq.~\eqref{eq:def HA and HB c}
at $(\theta,\lambda)=(\pi/10,\pi/10)$,
with $s^{\,}_{0}=2N/2-2N/4$ and $2N=256$.
(b) Same as (a)
for Hamiltonian $\widehat{H}^{(1)}_{\mathrm{II}}$.
}
\label{fig:NeelX OBCs Correlations}
\end{figure}

We have also verified numerically,
see Fig.~\ref{fig:NeelX OBCs Correlations},
that the dependence on the separation $s$
of the connected two-point correlation
$C^{x,\mathrm{c}}_{s^{\,}_{0},s^{\,}_{0}+s}$
and the two-point correlations
$C^{y}_{s^{\,}_{0},s^{\,}_{0}+s}$ and
$C^{z}_{s^{\,}_{0},s^{\,}_{0}+s}$ defined in
Eq.~\eqref{eq:Cs} of the Hamiltonians
\eqref{eq:def HA and HB c} and
\eqref{eq:def HA and HB d} at the point
$(\pi/10,\pi/10)$,
is that of an exponential decay.
More precisely, we use the Ansatz 
\begin{subequations}
\label{eq:NeelX OBCs exponential decay}
\begin{equation}
C^{\alpha}_{s^{\,}_{0},s^{\,}_{0}+s}
\propto
e^{-s/\xi^{\alpha}},
\qquad
C^{\alpha,\mathrm{c}}_{s^{\,}_{0},s^{\,}_{0}+s}
\propto
e^{-s/\xi^{\alpha,\mathrm{c}}},
\end{equation}
\end{subequations}
to fit the DMRG data.
As shown in Fig.~\ref{fig:NeelX OBCs Correlations a}
for Hamiltonian \eqref{eq:def HA and HB c}
at $(\theta,\lambda)=(\pi/10,\pi/10)$,
the two-point functions
$C^{x,\mathrm{c}}_{s^{\,}_{0},s^{\,}_{0}+s}$,
$C^{y}_{s^{\,}_{0},s^{\,}_{0}+s}$ and
$C^{z}_{s^{\,}_{0},s^{\,}_{0}+s}$
decay exponentially in $s$ with the correlations lengths
\begin{subequations}
\begin{align}
&\xi^{x,\mathrm{c}}=0.94\pm0.01,
\\
&\mathrlap{\xi^{y}}
\hphantom{\xi^{x,\mathrm{c}}}=1.493\pm0.007,
\\
&\mathrlap{\xi^{z}}
\hphantom{\xi^{x,\mathrm{c}}}=1.83\pm0.03,
\end{align}
\end{subequations}
respectively.
Similarly, Fig.~\ref{fig:NeelX OBCs Correlations b}
for Hamiltonian \eqref{eq:def HA and HB d}
at $(\theta,\lambda)=(\pi/10,\pi/10)$ shows that
the two-point functions
$C^{x,\mathrm{c}}_{s^{\,}_{0},s^{\,}_{0}+s}$,
$C^{y}_{s^{\,}_{0},s^{\,}_{0}+s}$ and
$C^{z}_{s^{\,}_{0},s^{\,}_{0}+s}$
decay exponentially in $s$ with the correlations lengths
\begin{subequations}
\begin{align}
&\xi^{x,\mathrm{c}}=1.41\pm0.01,
\\
&\mathrlap{\xi^{y}}
\hphantom{\xi^{x,\mathrm{c}}}=1.502\pm0.007,
\\
&\mathrlap{\xi^{z}}
\hphantom{\xi^{x,\mathrm{c}}}=1.90\pm0.01,
\end{align}
\end{subequations}
respectively.

Three additional gapped phases are unitarily
related to the $\mathrm{Neel}^{\,}_{x}$ phase through
unitary transformations introduced in
Sec.~\ref{sec:relations_in_coupling_space}.
Any one of these three phases can be characterized
by the order parameter obtained by the conjugation of the 
order parameter $m^{x}_{\mathrm{sta}}$
used to characterize the $\mathrm{Neel}^{\,}_{x}$
phase under the corresponding unitary transformation,
as we shall summarize.
In all cases, the numerical signatures
obtained from ED and DMRG agree with
the predictions implied by these relations.

The $\mathrm{Neel}^{\,}_{y}$ phase extends
from the classical fixed point located at
$(\pi/2,0)\in\Omega^{\,}_{\mathrm{Y}}$
to a finite region of
coupling space.
The unitary transformation $\widehat{U}^{\,}_{\mathrm{R}}$
defined in Eq.~\eqref{eq:U_R}
realizes the reflection $\theta\mapsto\pi/2-\theta$
in the phase diagram
in Figs.~\ref{fig:phase diagram A}
and \ref{fig:phase diagram B}.
Consequently, the numerical characterization
of the $\mathrm{Neel}^{\,}_{x}$ phase
made in this section must also hold for
the $\mathrm{Neel}^{\,}_{y}$ phase if the replacements 
\begin{equation}
(\theta,\lambda) \mapsto (\pi/2-\theta,\lambda),\quad
\widehat{X}^{\,}_{j} \mapsto \widehat{Y}^{\,}_{j}, \quad
m^{x}_{\mathrm{sta}} \mapsto m^{y}_{\mathrm{sta}},
\label{eq:changes NeelX to NeelY}
\end{equation}
are made.

The $\mathrm{FM}^{\,}_{x}$ phase extends
from the classical fixed point located at
$(0,\pi)\in\Omega^{\,}_{\mathrm{Y}}$
to a finite region of
coupling space.
The unitary transformation $\widehat{U}^{\,}_{Z,\pi}$
defined in Eq.~\eqref{eq:def UeZ b}
realizes the translation $\lambda\mapsto\lambda+\pi$
in the phase diagram
in Figs.~\ref{fig:phase diagram A}
and \ref{fig:phase diagram B}.
Consequently, the numerical characterization
of the $\mathrm{Neel}^{\,}_{x}$ phase
made in this section must also hold for
the $\mathrm{FM}^{\,}_{x}$ phase if the replacements 
\begin{equation}
(\theta,\lambda) \mapsto (\theta,\lambda+\pi),\quad
m^{x}_{\mathrm{sta}} \mapsto m^{x}_{\mathrm{uni}},
\label{eq:changes NeelX to FMX}
\end{equation}
are made.

The $\mathrm{FM}^{\,}_{y}$ phase extends
from the classical fixed point located at
$(\pi/2,\pi)\in\Omega^{\,}_{\mathrm{Y}}$
to a finite region of
coupling space.
The product of the unitary transformations
$\widehat{U}^{\,}_{Z,\pi}$ and
$\widehat{U}^{\,}_{\mathrm{R}}$
defined in Eqs.~\eqref{eq:def UeZ b} and
\eqref{eq:U_R}, respectively,
realizes the translation $\lambda\mapsto\lambda+\pi$
together with the reflection $\theta\mapsto\pi/2-\theta$
in the phase diagram
in Figs.~\ref{fig:phase diagram A}
and \ref{fig:phase diagram B}.
Consequently, the numerical characterization
of the $\mathrm{Neel}^{\,}_{x}$ phase
made in this section must also hold for
the $\mathrm{FM}^{\,}_{y}$ phase if the replacements 
\begin{equation}
(\theta,\lambda) \mapsto (\pi/2-\theta,\lambda+\pi),\quad
\widehat{X}^{\,}_{j} \mapsto \widehat{Y}^{\,}_{j}, \quad
m^{x}_{\mathrm{sta}} \mapsto m^{y}_{\mathrm{uni}},
\label{eq:changes NeelX to FMY}
\end{equation}
are made.

\subsubsection{The \texorpdfstring{$\mathrm{Neel}^{\mathrm{SPT}}_{x}$}{NeelX-SPT} phase}
\label{sec:the_texorpdfstring_mathrm_neel}

The $\mathrm{Neel}^{\mathrm{SPT}}_{x}$ phase extends
from the 
classical fixed point located at
(i)
$(\pi/2,\pi/2)\in\Omega^{\,}_{\mathrm{Y}}$
for Hamiltonian $\widehat{H}^{(\mathsf{b})}_{\mathrm{I}}$ and
(ii)
$(0,\pi/2)\in\Omega^{\,}_{\mathrm{Y}}$
for Hamiltonian $\widehat{H}^{(\mathsf{b})}_{\mathrm{II}}$
to a finite region of
coupling space.

In this phase, the ground state of Hamiltonians
\eqref{eq:def HA and HB c} and
\eqref{eq:def HA and HB d}
breaks spontaneously the symmetry group
$\mathrm{G}=\mathbb{Z}^{X}_{2}\times
\mathbb{Z}^{Y}_{2}\times
\mathbb{Z}^{T}_{2}$
defined in Eq.~\eqref{eq:symmetries} down to the subgroup
\begin{equation}
\mathrm{G}^{\,}_{\mathrm{Neel}^{\mathrm{SPT}}_{x}}\coloneqq
\mathbb{Z}^{X}_{2}\times
\mathbb{Z}^{YT}_{2},
\quad
\mathbb{Z}^{YT}_{2}\coloneqq
\left\{
\mathds{1}^{\,}_{\mathfrak{H}},\ 
\widehat{U}^{\,}_{Y}\,
\widehat{U}^{\,}_{T}
\right\},
\label{eq:NeelXSPT unbroken group}
\end{equation}
which is represented
(i)
linearly on the ground-state manifold
if PBCs are chosen, or
(ii)
projectively on the boundaries of the chain
if OBCs are chosen.
The group $\mathrm{G}^{\,}_{\mathrm{Neel}^{\mathrm{SPT}}_{x}}$
thus acts as a protecting symmetry upon choosing OBCs.

At the point
$(\pi/2,\pi/2)\in\Omega^{\,}_{\mathrm{Y}}$
the Hamiltonian \eqref{eq:def HA and HB c}
equals Hamiltonian \eqref{eq:def HA and HB d}
at the point
$(0,\pi/2)\in\Omega^{\,}_{\mathrm{Y}}$, i.e.,
\begin{subequations}
\begin{align}
\widehat{H}^{(\mathsf{b})}_{\mathrm{Neel}^{\mathrm{SPT}}_{x}}
&\coloneqq
\sum\limits_{j=1}^{2N-3\mathsf{b}}
\widehat{Z}^{\,}_{j}\,
\widehat{Y}^{\,}_{j+1}\,
\widehat{Y}^{\,}_{j+2}\,
\widehat{Z}^{\,}_{j+3}
\\
&\hphantom{:}\equiv
\widehat{H}^{(\mathsf{b})}_{\mathrm{I}}(\pi/2,\pi/2)\equiv
\widehat{H}^{(\mathsf{b})}_{\mathrm{II}}(0,\pi/2)
\label{eq:def H NeelX SPT}
\end{align}
\end{subequations}
and is unitarily equivalent to the classical antiferromagnetic
Ising chain in Eq.~(\ref{eq:def H NeelX})
if we impose PBCs ($\mathsf{b}=0$),
owing to the existence of the unitary
$\widehat{U}^{(0)}_{\mathrm{CZ}}$ defined in
Eq.~\eqref{eq:U_CZ}.
The ground state of
$\widehat{H}^{(0)}_{\mathrm{Neel}^{\mathrm{SPT}}_{x}}$
is thus two-fold degenerate and separated
from the excited states by a non-vanishing energy gap.
In contrast, when OBCs are chosen, the ground state of
$\widehat{H}^{(1)}_{\mathrm{Neel}^{\mathrm{SPT}}_{x}}$
is eight-fold degenerate and separated
from the excited states by a non-vanishing energy gap.
The jump of the ground-state degeneracy from two to eight 
when changing the PBCs to OBCs can be understood in the following way
(see Ref.~\cite{Rey2025}
for more detail).
\begin{itemize}
\item[(i)] The unbroken symmetry
$\mathrm{G}^{\,}_{\mathrm{Neel}^{\mathrm{SPT}}_{x}}$
in Eq.~\eqref{eq:NeelXSPT unbroken group} is represented
projectively at each boundary of the chain.
\item[(ii)] The unbroken symmetry
$\mathrm{G}^{\,}_{\mathrm{Neel}^{\mathrm{SPT}}_{x}}$
in Eq.~\eqref{eq:NeelXSPT unbroken group} fractionalizes
at each boundary of the chain.
\item[(iii)] The ground state of
$\widehat{H}^{(1)}_{\mathrm{Neel}^{\mathrm{SPT}}_{x}}$
supports two decoupled quantum spin-$1/2$ degrees of freedom
localized on each boundary.
\end{itemize}

In the $\mathrm{Neel}^{\mathrm{SPT}}_{x}$ phase,
we have verified numerically using the DMRG method that
the ground state of Hamiltonians
\eqref{eq:def HA and HB c} and
\eqref{eq:def HA and HB d}
supports a non-vanishing magnetization
\begin{equation}
0< \bigl|m^{\mathrm{SPT},x}_{\mathrm{sta}}\bigr|\le1,
\end{equation}
where the equality 
$\bigl|m^{\mathrm{SPT},x}_\mathrm{sta}\bigr|=1$ holds only at the
classical point $(\pi/2,\pi/2)\in\Omega^{\,}_{\mathrm{Y}}$
for Hamiltonian \eqref{eq:def HA and HB c}
and the classical point $(0,\pi/2)\in\Omega^{\,}_{\mathrm{Y}}$
for Hamiltonian \eqref{eq:def HA and HB d}.
The reason for which $m^{\mathrm{SPT},x}_{\mathrm{sta}}$
is the relevant order parameter for the
$\mathrm{Neel}^{\mathrm{SPT}}_{x}$ phase follows from the factorization
$\widehat{Z}^{\,}_{j}\,
\widehat{Y}^{\,}_{j+1}\,
\widehat{Y}^{\,}_{j+2}\,
\widehat{Z}^{\,}_{j+3}
=
\left(
\widehat{Z}^{\,}_{j}\,
\widehat{X}^{\,}_{j+1}\,
\widehat{Z}^{\,}_{j+2}
\right)
\left(
\widehat{Z}^{\,}_{j+1}\,
\widehat{X}^{\,}_{j+2}\,
\widehat{Z}^{\,}_{j+3}
\right)$,
a consequence of
the Pauli algebra \eqref{eq:def Hilbert space}.

In the $\mathrm{Neel}^{\mathrm{SPT}}_{x}$ phase
and upon choosing PBCs (OBCs),
the energy spacing $E^{\,}_{1}-E^{\,}_{0}$
($E^{\,}_{7}-E^{\,}_{0}$)
between the ground state energy $E^{\,}_{0}$ and the first (seventh)
excited state energy $E^{\,}_{1}$ ($E^{\,}_{7}$)
is observed numerically to vanish exponentially fast with $2N$
so that a two-fold
(eight-fold) degeneracy of the ground state is recovered
in the thermodynamic limit $2N\to\infty$.
The energy gap $E^{\,}_{2}-E^{\,}_{1}$
($E^{\,}_{8}-E^{\,}_{7}$) is non-vanishing
and only closes at the phase boundaries in the
thermodynamic limit.

Finally, we have verified numerically
that the dependence on the separation $s$
of the connected two-point correlation
$C^{\mathrm{SPT},x,\mathrm{c}}_{s^{\,}_{0},s^{\,}_{0}+s}$ and the two-point correlations
$C^{\mathrm{SPT},y}_{s^{\,}_{0},s^{\,}_{0}+s}$ and
$C^{\mathrm{SPT},z}_{s^{\,}_{0},s^{\,}_{0}+s}$ defined in
Eq.~\eqref{eq:Cs} of the Hamiltonians
\eqref{eq:def HA and HB c} and
\eqref{eq:def HA and HB d} at any point
in the $\mathrm{Neel}^{\mathrm{SPT}}_{x}$ phase,
are decaying exponentially.

The $\mathrm{Neel}^{\mathrm{SPT}}_{y}$ phase extends
from the 
classical fixed point located at
(i)
$(0,\pi/2)\in\Omega^{\,}_{\mathrm{Y}}$
for Hamiltonian $\widehat{H}^{(\mathsf{b})}_{\mathrm{I}}$ and
(ii)
$(\pi/2,\pi/2)\in\Omega^{\,}_{\mathrm{Y}}$
for Hamiltonian $\widehat{H}^{(\mathsf{b})}_{\mathrm{II}}$
to a finite region of the
coupling space.
The unitary transformation $\widehat{U}^{\,}_{\mathrm{R}}$
defined in Eq.~\eqref{eq:U_R}
realizes the reflection $\theta\mapsto\pi/2-\theta$
in the phase diagram
in Figs.~\ref{fig:phase diagram A}
and \ref{fig:phase diagram B}.
Consequently, the numerical characterization
of the $\mathrm{Neel}^{\mathrm{SPT}}_{x}$ phase
made in this section must also hold for
the $\mathrm{Neel}^{\mathrm{SPT}}_{y}$ phase if the replacements 
\begin{equation}
(\theta,\lambda) \mapsto (\pi/2-\theta,\lambda),\quad
\widehat{X}^{\,}_{j} \mapsto \widehat{Y}^{\,}_{j}, \quad
m^{\mathrm{SPT},x}_{\mathrm{sta}} \mapsto m^{\mathrm{SPT},y}_{\mathrm{sta}},
\label{eq:changes NeelX SPT to NeelY SPT}
\end{equation}
are made.

The $\mathrm{FM}^{\mathrm{SPT}}_{x}$ phase extends
from the classical fixed point located at
(i) $(3\pi/2,\pi/2)\in\mathbb{T}^{2}$
for Hamiltonian $\widehat{H}^{(\mathsf{b})}_{\mathrm{I}}$
and
(ii) $(\pi,\pi/2)\in\Omega^{\,}_{\mathrm{P}}$
for Hamiltonian $\widehat{H}^{(\mathsf{b})}_{\mathrm{II}}$
to a finite region of
the coupling space.
The unitary transformation $\widehat{U}^{\,}_{Z,\pi}$
defined in Eq.~\eqref{eq:def UeZ b}
realizes the translation $\theta\mapsto\theta+\pi$
in the phase diagram
in Figs.~\ref{fig:phase diagram A}
and \ref{fig:phase diagram B}.
Consequently, the numerical characterization
of the $\mathrm{Neel}^{\mathrm{SPT}}_{x}$ phase
made in this section must also hold for
the $\mathrm{FM}^{\mathrm{SPT}}_{x}$ phase if the replacements 
\begin{equation}
(\theta,\lambda) \mapsto (\theta+\pi,\lambda),\quad
m^{\mathrm{SPT},x}_{\mathrm{sta}} \mapsto m^{\mathrm{SPT},x}_{\mathrm{uni}},
\label{eq:changes NeelXSPT to FMXSPT}
\end{equation}
are made.

The $\mathrm{FM}^{\mathrm{SPT}}_{y}$ phase extends
from the classical fixed point located at
(i) $(\pi,\pi/2)\in\Omega^{\,}_{\mathrm{P}}$
for Hamiltonian $\widehat{H}^{(\mathsf{b})}_{\mathrm{I}}$
and
(ii) $(3\pi/2,\pi/2)\in\mathbb{T}^{2}$
for Hamiltonian $\widehat{H}^{(\mathsf{b})}_{\mathrm{II}}$
to a finite region of
the coupling space.
The product of the unitary transformations
$\widehat{U}^{\,}_{Z,\pi}$ and
$\widehat{U}^{\,}_{\mathrm{R}}$
defined in Eqs.~\eqref{eq:def UeZ b} and
\eqref{eq:U_R}, respectively,
realizes the translation translation $\theta\mapsto\theta+\pi$
together with the reflection $\theta\mapsto\pi/2-\theta$
in the phase diagram
in Figs.~\ref{fig:phase diagram A}
and \ref{fig:phase diagram B}.
Consequently, the numerical characterization
of the $\mathrm{Neel}^{\mathrm{SPT}}_{x}$ phase
made in this section must also hold for
the $\mathrm{FM}^{\mathrm{SPT}}_{y}$ phase if the replacements 
\begin{equation}
(\theta,\lambda) \mapsto (3\pi/2-\theta,\lambda),\quad
\widehat{X}^{\,}_{j} \mapsto \widehat{Y}^{\,}_{j}, \quad
m^{\mathrm{SPT},x}_{\mathrm{sta}} \mapsto m^{\mathrm{SPT},y}_{\mathrm{uni}},
\label{eq:changes NeelX to FMY SPT}
\end{equation}
are made.

\subsubsection{The \texorpdfstring{$\mathrm{gFM}^{\,}_{z}$}{gFMz} phase}
\label{ssub:the_texorpdfstring_mathrm_gfm}

Any one of the phases~\eqref{eq:eight gapped phases}
occupies an extended region of the
coupling space $\mathbb{T}^{2}$ defined in Eq.~\eqref{eq:def HA and HB a}.
Since any two phases in \eqref{eq:eight gapped phases} differ by
(i) their long-range order,
(ii) their SPT order, or
(iii) a combination of both,
they must be separated by a quantum phase transition.
A first question is therefore what is the fate of
the phases \eqref{eq:eight gapped phases}
when they meet.

In Ref.~\cite{Rey2025}, the Hamiltonian
$\widehat{H}^{(\mathsf{b})}_{\mathrm{I}}$ defined in
Eq.~\eqref{eq:def HA and HB c} was extensively studied within
$\Omega^{\,}_{\mathrm{Y,B}}
\subset\Omega^{\,}_{\mathrm{Y}}
\subset\mathbb{T}^{2}$
in Eq.~\eqref{eq:coupling quadrants a}.
Reference \cite{Rey2025}
shows the existence of an extended phase,
approximately centered at $(\pi/4,\pi/4)$ that is
shaped like a diamond,
with a two-fold ground-state degeneracy for both OBCs and PBCs.
This phase, denoted
$\mathrm{gFM}_{z}$ in Figs.~\ref{fig:phase diagram A} and
\ref{fig:phase diagram B}, is gapless and
supports long-range ferromagnetic
order along the $z$ axis in spin space,
with incommensurate two-point correlations.
The $\mathrm{gFM}^{\,}_{z}$ phase is centered about the point
$(\pi/4,\pi/4)\in\Omega^{\,}_{\mathrm{Y,B}}$
and connects any two of the gapped
$\mathrm{Neel}^{\,}_{x}$,
$\mathrm{Neel}^{\,}_{y}$,
$\mathrm{Neel}^{\mathrm{SPT}}_{x}$, and
$\mathrm{Neel}^{\mathrm{SPT}}_{y}$ phases in
Figs.~\ref{fig:phase diagram A}
and \ref{fig:phase diagram B}.

The $\mathrm{gFM}^{\,}_{z}$ phase breaks
spontaneously the symmetry group
$\mathrm{G}=\mathbb{Z}^{X}_{2}\times
\mathbb{Z}^{Y}_{2}\times
\mathbb{Z}^{T}_{2}$
defined in Eq.~\eqref{eq:symmetries} down to the subgroup
\begin{subequations}
\label{eq:gFMz SSB Pattern}
\begin{equation}
\mathrm{G}^{\,}_{\mathrm{gFM}^{\,}_{z}}\coloneqq
\mathbb{Z}^{XY}_{2}\times
\mathbb{Z}^{YT}_{2},
\end{equation}
where
\begin{align}
&\mathbb{Z}^{XY}_{2}\coloneqq
\left\{
\mathds{1}^{\,}_{\mathfrak{H}},\ 
\widehat{U}^{\,}_{X}\,
\widehat{U}^{\,}_{Y}
\right\},
\\
&\mathrlap{\mathbb{Z}^{YT}_{2}}
\hphantom{\mathbb{Z}^{XY}_{2}}\coloneqq
\left\{
\mathds{1}^{\,}_{\mathfrak{H}},\ 
\widehat{U}^{\,}_{Y}\,
\widehat{U}^{\,}_{T}
\right\},
\end{align}
\end{subequations}
which is represented linearly on the ground-state
manifold for both PBCs and OBCs.

Here, we generalize the study of Ref.~\cite{Rey2025}
to decide if such an extended phase exists for
$\widehat{H}^{(\mathsf{b})}_{\mathrm{I}}$
[$\widehat{H}^{(\mathsf{b})}_{\mathrm{II}}$]
elsewhere within the reduced coupling space
$\Omega^{\,}_{\mathrm{Y,T}}\subset
\Omega^{\,}_{\mathrm{Y}}\subset
\mathbb{T}^{2}$ defined in Eq.~\eqref{eq:coupling quadrants b}
[within $\Omega^{\,}_{\mathrm{Y}}\subset
\mathbb{T}^{2}$ defined in Eq.~\eqref{eq:coupling space yellow}].

Because Hamiltonian
\eqref{eq:def HA and HB c}
equals Hamiltonian
\eqref{eq:def HA and HB d}
at $\theta=\pi/4$ for any $\lambda\in[0,2\pi]$
[see Eq.~\eqref{eq:def H U1}],
we already know that the
$\mathrm{gFM}^{\,}_{z}$ phase
exists for Hamiltonian \eqref{eq:def HA and HB d}
within the reduced coupling space
$\Omega^{\,}_{\mathrm{Y,B}}\subset
\Omega^{\,}_{\mathrm{Y}}\subset
\mathbb{T}^{2}$
along the same vertical interval $(\pi/4,\lambda)$
centered about $\lambda=\pi/4$
as it does for Hamiltonian \eqref{eq:def HA and HB c}. 
We are going to show numerically that
(i)
the
$\mathrm{gFM}^{\,}_{z}$ phase
for Hamiltonian \eqref{eq:def HA and HB d}
is also extended about $\theta=\pi/4$ along the $\theta$ direction
and that (ii) there is no $\mathrm{gFM}^{\,}_{z}$ phase when
$\lambda>\pi/2$ in $\Omega^{\,}_{\mathrm{Y}}$ for both Hamiltonians
\eqref{eq:def HA and HB c} and \eqref{eq:def HA and HB d}.
The location of the boundaries of the $\mathrm{gFM}^{\,}_{z}$ phase
of Hamiltonian \eqref{eq:def HA and HB d} is given
in Sec.~\ref{subsubsec:The Neelx-gFMz phase boundary}.

\begin{figure}
\subfigure{\includegraphics[width=1\columnwidth]{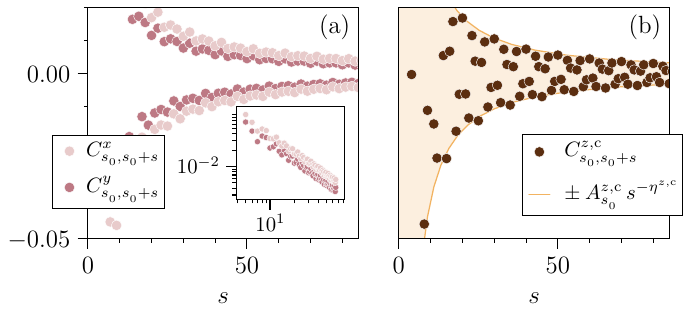} \label{fig:gFMz OBCs Correlations a}}%
\subfigure{\label{fig:gFMz OBCs Correlations b}}
\caption{
The case of OBCs.
(a)
Dependence on $s$ given $s^{\,}_{0}$ for the
spin-spin correlations
$C^{x}_{s^{\,}_{0},s^{\,}_{0}+s}$
and
$C^{y}_{s^{\,}_{0},s^{\,}_{0}+s}$,
defined in Eq.~\eqref{eq:Calpha},
computed using DMRG ($\chi=128$)
for the ground state
$|\Psi^{\,}_{0}\rangle$
of Hamiltonian
$\widehat{H}^{(1)}_{\mathrm{II}}$ in
Eq.~\eqref{eq:def HA and HB d}
at $(\theta,\lambda)=(\pi/4-\pi/40,\pi/4-\pi/40)$,
with $s^{\,}_{0}=2N/2-2N/4$ and
$2N=256$.
The inset illustrates the absolute values of
$C^{x}_{s^{\,}_{0},s^{\,}_{0}+s}$
and
$C^{y}_{s^{\,}_{0},s^{\,}_{0}+s}$
in a log-log scale to extract the exponents
$\eta^{x}=1.07\pm0.01$ and
$\eta^{y}=1.08\pm0.02$, respectively,
using Eq.~\eqref{eq:gFMz OBCs algebraic decay}.
(b)
Dependence on $s$ given $s^{\,}_{0}$ for the
spin-spin correlations
$C^{z,\mathrm{c}}_{s^{\,}_{0},s^{\,}_{0}+s}$, 
defined in Eq.~\eqref{eq:CalphaConn},
computed using DMRG ($\chi=128$)
for the ground state
$|\Psi^{\,}_{0}\rangle$
of Hamiltonian
$\widehat{H}^{(1)}_{\mathrm{II}}$ in
Eq.~\eqref{eq:def HA and HB d}
at $(\theta,\lambda)=(\pi/4-\pi/40,\pi/4-\pi/40)$,
with $s^{\,}_{0}=2N/2-2N/4$ and
$2N=256$.
We use
Eq.~\eqref{eq:gFMz OBCs algebraic decay cosine}
to compute the exponent $\eta^{z,\mathrm{c}}$ for the connected
two-point function.
The shaded yellow region serves as a visual guide to observe the
algebraic decay $s^{-\eta^{z,\mathrm{c}}}$.
}
\label{fig:gFMz OBCs Correlations}
\end{figure}

We illustrate in Fig.~\ref{fig:gFMz OBCs Correlations a}
the dependence on $s$ given $s^{\,}_{0}$ for the
spin-spin correlation functions
$C^{x}_{s^{\,}_{0},s^{\,}_{0}+s}$,
$C^{y}_{s^{\,}_{0},s^{\,}_{0}+s}$,
and
$C^{z,\mathrm{c}}_{s^{\,}_{0},s^{\,}_{0}+s}$,
defined in Eq.~\eqref{eq:Cs}.
Here, the ground state is obtained by the DMRG
for Hamiltonian
$\widehat{H}^{(1)}_{\mathrm{II}}(\pi/4-\pi/40,\pi/4-\pi/40)$
in Eq.~\eqref{eq:def HA and HB d} under OBCs.
We find using the Ansatz
\begin{equation}
C^{\alpha}_{s^{\,}_{0},s^{\,}_{0}+s}=
A^{\alpha}_{s}\,
s^{-\eta^{\alpha}},
\label{eq:gFMz OBCs algebraic decay}
\end{equation}
that $C^{x}_{s^{\,}_{0},s^{\,}_{0}+s}$ and $C^{y}_{s^{\,}_{0},s^{\,}_{0}+s}$
decay algebraically with the exponents
\begin{equation}
\eta^{x}=1.04\pm0.02,
\quad
\eta^{y}=1.04\pm0.02,
\label{eq:gFMz OBCs HB Fitting Paramters XY}
\end{equation}
and that the function $A^{\alpha}_{s}\propto (-1)^{s}$.
This shows that the point $(\pi/4-\pi/40,\pi/4-\pi/40)$
of Hamiltonian $\widehat{H}^{(1)}_{\mathrm{II}}$
realizes a critical phase.
This critical phase is conjectured to be the
$\mathrm{gFM}^{\,}_{z}$ phase in Fig.~\ref{fig:phase diagram B}.
To confirm this conjecture, we find
that the correlation function
$|C^{z}_{s^{\,}_{0},s^{\,}_{0}+s}|$
defined in Eq.~\eqref{eq:Calpha}
approaches a constant value $\left(m^{z}_{\mathrm{uni}}\right)^{2}$
in the limit $s\gg1$,
where $|m^{z}_{\mathrm{uni}}|\simeq0.392$.
This provides evidence that the Hamiltonian
$\widehat{H}^{(1)}_{\mathrm{II}}(\pi/4-\pi/40,\pi/4-\pi/40)$
in Eq.~\eqref{eq:def HA and HB d} under OBCs realizes
a critical phase supporting ferromagnetic
long-range order along the $z$ axis in spin space.
We have also computed $C^{z,\mathrm{c}}_{s^{\,}_{0},s^{\,}_{0}+s}$
in Fig.~\ref{fig:gFMz OBCs Correlations b}.
We find using the Ansatz
\begin{equation}
C^{z,\mathrm{c}}_{s^{\,}_{0},s^{\,}_{0}+s}=
A^{z,\mathrm{c}}_{s^{\,}_{0}}\,
s^{-\eta^{z,\mathrm{c}}}
\cos\big(q^{z,\mathrm{c}}\,s\big),
\label{eq:gFMz OBCs algebraic decay cosine}
\end{equation}
for $1\ll s^{\,}_{0},s,s^{\,}_{0}+s\ll2N$,
where the exponent of the algebraic decay
and the wave number of the oscillation
of the connected two-point function
take the values
\begin{subequations}
\label{eq:gFMz OBCs HB Fitting Paramters Z}
\begin{align}
&\eta^{z,\mathrm{c}}=1.16\pm0.01,
\\
&
q^{z,\mathrm{c}}=2\pi\times\left(0.30002\pm 4\cdot 10^{-5}\right),
\end{align}
\end{subequations}
respectively.
The value of $q^{z,\mathrm{c}}$ throughout the line
$\lambda^{\,}_{\mathrm{gFM}^{\,}_{z}}<\lambda\leq\pi/4$ is in good agreement with
the one expected for a Luttinger liquid phase with $\mathsf{c}=1$, i.e.,
\begin{equation}
q^{z,\mathrm{c}}=
\pi(1-|m^z_{\mathrm{uni}}|).
\end{equation}

We list below the value of the exponents and the wavevectors of oscillations
obtained for Hamiltonian
$\widehat{H}^{(1)}_{\mathrm{I}}(\pi/4-\pi/40,\pi/4-\pi/40)$
in Eq.~\eqref{eq:def HA and HB c} under OBCs, where it realizes the
$\mathrm{gFM}^{\,}_{z}$ phase \cite{Rey2025}
[compare to Eqs.~\eqref{eq:gFMz OBCs HB Fitting Paramters XY}
and \eqref{eq:gFMz OBCs HB Fitting Paramters Z}]
\begin{subequations}
\begin{align}
&\eta^{x}=0.98\pm0.01,
\\
&\eta^{y}=1.07\pm0.02,
\\
&\eta^{z,\mathrm{c}}=1.18\pm0.02,
\\
&
q^{z,\mathrm{c}}=2\pi\times\left(0.30010\pm 5\cdot 10^{-5}\right),
\end{align}
\end{subequations}
and where $|m^{z}_{\mathrm{uni}}|\simeq0.397$.

\begin{figure}
\centering
\subfigure{\includegraphics[width=1\columnwidth]{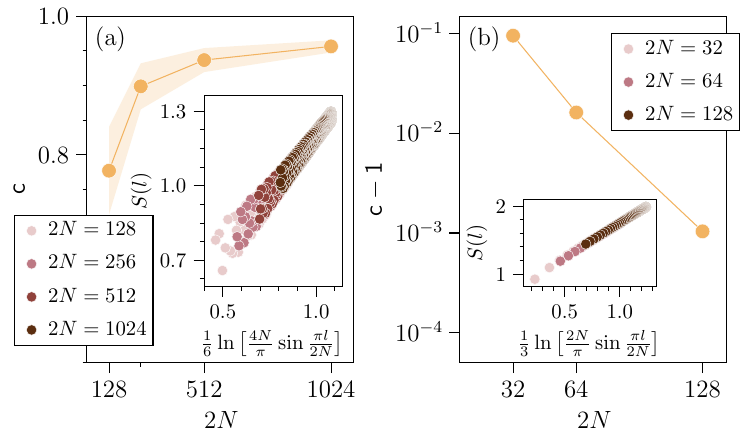} \label{fig:gFMz HB Central Charge a}}%
\subfigure{\label{fig:gFMz HB Central Charge b}}%
\caption{
Dependence of central charge $\mathsf{c}$
on the number $2N$ of sites
of the ground state of
Hamiltonian $\widehat{H}^{(\mathsf{b})}_{\mathrm{II}}$
in Eq.~\eqref{eq:def HA and HB d}
at the point $(\pi/4-\pi/40,\pi/4-\pi/40)$,
obtained using DMRG with
(a) OBCs and $\chi=256$ and
(b) PBCs and $\chi=512$.
The inset of each panel shows the 
dependence on the number $2N$ of sites of the
entanglement entropy $S^{\,}_{\mathrm{MPS},2N}({l})$. 
The entanglement entropy $S^{\,}_{\mathrm{MPS},2N}({l})$
is obtained from the approximation of
the ground state in terms of
a matrix product state.
The slope of the linear fit performed for each system
size $2N$ gives
the central charge $\mathsf{c}$.
The confidence interval is indicated
as a shaded yellow region in (a).
}
\label{fig:gFMz HB Central Charge}
\end{figure}

The $\mathrm{gFM}^{\,}_{z}$ phase of Hamiltonian
$\widehat{H}^{(\mathsf{b})}_{\mathrm{I}}$
in Eq.~\eqref{eq:def HA and HB c} 
realizes a CFT with central charge $\mathsf{c}=1$ \cite{Rey2025}.
Moreover, the equality \eqref{eq:def H U1}
of Hamiltonian
$\widehat{H}^{(\mathsf{b})}_{\mathrm{II}}$
to Hamiltonian
$\widehat{H}^{(\mathsf{b})}_{\mathrm{I}}$
along $\theta=\pi/4$ for any $\lambda\in[0,2\pi]$,
implies that
any point in the $\mathrm{gFM}^{\,}_{z}$ phase of Hamiltonian
$\widehat{H}^{(\mathsf{b})}_{\mathrm{II}}$
in Eq.~\eqref{eq:def HA and HB d}
realizes a CFT with central charge $\mathsf{c}=1$.
Accordingly, we report in
Fig.~\ref{fig:gFMz HB Central Charge}
the value of the central charge of
Hamiltonian \eqref{eq:def HA and HB d}
under both PBCs and OBCs
at the point $(\pi/4-\pi/40,\pi/4-\pi/40)$.
The value of the central charge
\begin{subequations}
\label{eq:CFT prediction for S2N(l)}
\begin{equation}
\mathsf{c}\coloneqq\lim_{2N\to\infty}\mathsf{c}^{\,}_{2N}
\label{eq:CFT prediction for S2N(l) a}
\end{equation}
has been determined by comparing the CFT prediction
\cite{calabrese2004}
\begin{equation}
S^{\,}_{2N}({l})=
\begin{cases}
\displaystyle
\frac{\mathsf{c}^{\,}_{2N}}{3}
\ln\!
\left[
\frac{2N}{\pi}
\sin\left(\frac{\pi{l}}{2N}\right)
\right]
+
\hbox{const},
&
\mathrm{PBCs},
\\&\\
\displaystyle
\frac{\mathsf{c}^{\,}_{2N}}{6}
\ln\!
\left[
\frac{4N}{\pi}
\sin\left(\frac{\pi{l}}{2N}\right)
\right]
+
\hbox{const},
&
\mathrm{OBCs},
\end{cases}
\label{eq:CFT prediction for S2N(l) b}
\end{equation}
\end{subequations}
for the bipartite entanglement entropy $S^{\,}_{2N}({l})$
against its numerical evaluation $S^{\,}_{\mathrm{MPS},2N}({l})$, i.e.,
the bipartite entanglement
entropy with respect to
the bond ${l}$ of the matrix product state that
approximates the ground state of Hamiltonian
\eqref{eq:def HA and HB c};
see Ref.~\cite{Rey2025}.

In Fig.~\ref{fig:gFMz HB Central Charge a},
the bipartite entanglement entropy
$S^{\,}_{\mathrm{MPS},2N}({l})$
shows oscillations as a function of ${l}$.
This oscillatory behavior is interpreted as the
signature of the incommensuration in the
$\mathrm{gFM}^{\,}_{z}$ phase~\cite{Ye150205049},
see Ref.~\cite{Rey2025}. 
Extracting the central charge in finite systems
is more accurate with PBCs, as finite-size effects are strongly
suppressed compared to OBCs. 
However, choosing PBC significantly
increases the computational cost of the DMRG algorithm
(see Ref.\ \cite{schollwock2011}).
Using PBCs in Fig.~\ref{fig:gFMz HB Central Charge b},
we find that the point $(\pi/4-\pi/40,\pi/4-\pi/40)$
of Hamiltonian $\widehat{H}^{(\mathsf{b})}_{\mathrm{II}}$
in Eq.~\eqref{eq:def HA and HB d}
realizes a CFT with central charge $\mathsf{c}=1$.

Additional numerical evidence for the stability of the
$\mathrm{gFM}^{\,}_{z}$ phase away from the $\mathrm{U}(1)$-symmetric
lines is presented in Appendix~\ref{sec:Additional evidence gFMz}.
There, we use finite-entanglement scaling, the scale
dependence of the transverse spin correlations, and the
momentum-resolved correlations of the microscopic operators that break
the $\mathrm{U}(1)$ symmetry to test the alternative scenario in which
the apparently extended $\mathrm{gFM}^{\,}_{z}$ region is instead a
crossover regime governed by a very large but finite physical
correlation length. Within the system sizes and MPS correlation
lengths accessible numerically, these diagnostics support the
interpretation of the $\mathrm{gFM}^{\,}_{z}$ region as a genuine
extended critical phase.

\subsubsection{The \texorpdfstring{$\mathrm{Neel}^{\,}_{z}$}{NeelZ} phase}

We now turn our attention to the determination of the
properties of the $\mathrm{Neel}^{\,}_{z}$ phase 
centered about the point
$(\pi/4,3\pi/4)\in\Omega^{\,}_{\mathrm{Y,T}}$
that is surrounded by the gapped
$\mathrm{FM}^{\,}_{x}$,
$\mathrm{FM}^{\,}_{y}$,
$\mathrm{Neel}^{\mathrm{SPT}}_{x}$, and
$\mathrm{Neel}^{\mathrm{SPT}}_{y}$ phases in
Figs.~\ref{fig:phase diagram A}
and \ref{fig:phase diagram B}.
Without loss of generality, we restrict the study
of Hamiltonians \eqref{eq:def HA and HB c}
and \eqref{eq:def HA and HB d} to the region
$\Omega^{\,}_{\mathrm{Y,T}}\subset
\Omega^{\,}_{\mathrm{Y}}\subset
\mathbb{T}^{2}$ defined in Eq.~\eqref{eq:coupling quadrants b}.
The labeling $\mathrm{Neel}^{\,}_{z}$ is chosen because this phase is
an extended gapped phase supporting antiferromagnetic
long-range order
with the staggered magnetization pointing along the
$z$ axis in spin space, as we now explain.

\begin{figure}
\centering
\subfigure{\includegraphics[width=1\columnwidth]{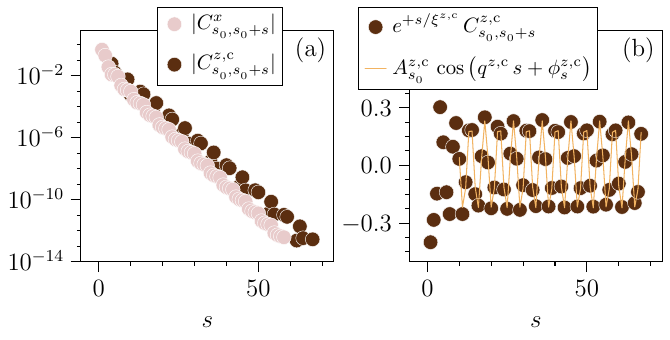} \label{fig:NeelZ OBCs Correlations and Spins a}}%
\subfigure{\label{fig:NeelZ OBCs Correlations and Spins b}}%
\caption{
The case of OBCs.
(a)
Dependence on $s$ given $s^{\,}_{0}$ of the
absolute value of the
spin-spin correlation functions
$C^{x}_{s^{\,}_{0},s^{\,}_{0}+s}$ and
$C^{z,\mathrm{c}}_{s^{\,}_{0},s^{\,}_{0}+s}$
defined in Eq.~\eqref{eq:Cs},
computed using DMRG
($\chi=128$ and $2N=1024$) 
for the ground state $|\Psi\rangle$
of Hamiltonian $\widehat{H}^{(1)}_{\mathrm{U}(1)}$
in Eq.~\eqref{eq:def H U1} at $\lambda=3\pi/4$,
with $s^{\,}_{0}=2N/2-2N/4$.
We extract the correlation lengths
$\xi^{x}=2.25\pm0.03$ and
$\xi^{z,\mathrm{c}}=2.47\pm0.04$ using
Eq.~\eqref{eq:NeelX OBCs exponential decay}.
(b) Dependence on $s$ given $s^{\,}_{0}$ of the
absolute value of the
rescaled spin-spin correlation function
$e^{+s/\xi^{z,\mathrm{c}}}\,
C^{z,\mathrm{c}}_{s^{\,}_{0},s^{\,}_{0}+s}$,
that is fitted using
Eq.~\eqref{eq:NeelZ OBCs exponential decay cosine}.
}
\label{fig:NeelZ OBCs Correlations and Spins}
\end{figure}

By computing the spin expectation values on each site
$j=1,\ldots,2N$ of the chain under OBCs at the point $(\pi/4,3\pi/4)$
in Figs.~\ref{fig:phase diagram A} 
and \ref{fig:phase diagram B},
we observed that the only non-vanishing order parameter is
\begin{equation}
\bigl|m^{z}_{\mathrm{sta}}\bigl|\simeq 0.486.
\end{equation}
Unlike for the $\mathrm{gFM}^{\,}_{z}$ phase under OBCs,
there are no visible oscillations in the local expectation
values $\ev*{\widehat{Z}^{\,}_{j}}$
in the bulk of the chain of sites $j=1,\ldots,2N$.
However, the oscillations are present in the
spin-spin correlation function
$C^{z,\mathrm{c}}_{s^{\,}_{0},s^{\,}_{0}+s}$
at the point $(\pi/4,3\pi/4)$, as is shown
in Fig.~\ref{fig:NeelZ OBCs Correlations and Spins}.
Determining the wavevector of oscillations was possible using the Ansatz
\begin{equation}
C^{z,\mathrm{c}}_{s^{\,}_{0},s^{\,}_{0}+s}=
A^{z,\mathrm{c}}_{s^{\,}_{0}}\,
e^{-s/\xi^{z,\mathrm{c}}}\,
\cos\big(q^{z,\mathrm{c}}\,s\big)
\label{eq:NeelZ OBCs exponential decay cosine}
\end{equation}
for $1\ll s^{\,}_{0},s,s^{\,}_{0}+s\ll2N$.
We find the correlation length
$\xi^{z,\mathrm{c}}=2.47\pm0.04$,
as in Fig.~\ref{fig:NeelZ OBCs Correlations and Spins a},
and the incommensurate wave number
\begin{equation}
q^{z,\mathrm{c}}=2\pi\times\left(0.22213\pm 4\cdot 10^{-5}\right).
\end{equation}
We have verified that the correlation function
$|C^{z}_{s^{\,}_{0},s^{\,}_{0}+s}|$
defined in Eq.~\eqref{eq:Calpha}
approaches a constant value $\left(m^{z}_{\mathrm{sta}}\right)^{2}$
in the limit $s\gg1$.
This provides evidence that the point $(\pi/4,3\pi/4)$
for both Hamiltonians
\eqref{eq:def HA and HB c} and \eqref{eq:def HA and HB d}
realizes a gapped phase with long-range antiferromagnetic
order along the $z$ axis in spin space,
hence the labeling $\mathrm{Neel}^{\,}_{z}$.
Using the Ansatz \eqref{eq:NeelX OBCs exponential decay},
we find that $C^{x}_{s^{\,}_{0},s^{\,}_{0}+s}=C^{y}_{s^{\,}_{0},s^{\,}_{0}+s}$
decay exponentially with the correlation length
$\xi^{x}=\xi^{y}=2.25\pm0.03$.
The correlation function $C^{x}_{s^{\,}_{0},s^{\,}_{0}+s}$
also oscillates with a periodicity of $4$ in the separation $s$.

\begin{figure}
\centering
\subfigure{\includegraphics[width=1\columnwidth]{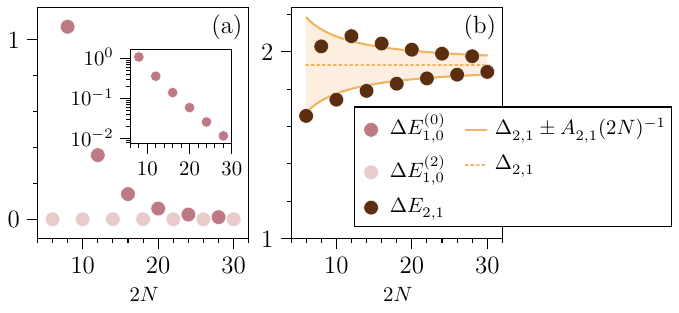} \label{fig:NeelZ OBCs Gaps a}}%
\subfigure{\label{fig:NeelZ OBCs Gaps b}}%
\caption{
The case of PBCs. 
Dependence on the system size $2N$ of the energy differences
$\Delta E^{\,}_{n+1,n}\coloneqq
E^{\,}_{n+1}-E^{\,}_{n}$, with
(a) $n=0$ and
(b) $n=1$,
above the ground state energy $E^{\,}_{0}$
of Hamiltonian \eqref{eq:def H U1} at $\lambda=3\pi/4$ 
obtained with ED.
The superscript $(p)$, with $p=0,2$, indicates the restriction
to the system sizes $2N\equiv p \bmod 4$.
The inset of (a) highlights the exponential decay
of $E^{\,}_{1}-E^{\,}_{0}$ as a function of $2N$
as \eqref{eq:NeelZ PBCs Exponential}.
We show in (b) visual guides to the behavior
\eqref{eq:NeelZ PBCs Gap a}
of $E^{\,}_{2}-E^{\,}_{1}$ as a function of $2N$.
We infer the value
$\Delta^{\,}_{2,1}=1.929\pm0.004$
for the spectral gap between the two-degenerate ground states
and all excitations.
}
\label{fig:NeelZ OBCs Gaps}
\end{figure}

The exponential decay of the two-point functions
\eqref{eq:Cs} indicates that the point $(\pi/4,3\pi/4)$
realizes a gapped phase.
The existence of long-range antiferromagnetic
order along the $z$ axis in spin space
at the point $(\pi/4,3\pi/4)$
indicates that the symmetry group
$\mathrm{G}=\mathbb{Z}^{X}_{2}\times
\mathbb{Z}^{Y}_{2}\times
\mathbb{Z}^{T}_{2}$
defined in Eq.~\eqref{eq:symmetries} is broken down to the subgroup
\eqref{eq:gFMz SSB Pattern}.
Hence, a two-fold degeneracy of the ground state
under both OBCs and PBCs is expected 
in the thermodynamic limit.

We show in Fig.~\ref{fig:NeelZ OBCs Gaps a}
that the ground state at $(\pi/4,3\pi/4)$ under PBCs is
(i) exactly two-fold degenerate
for any system size $2N$ such that $2N=2\bmod 4$, in which case
$E^{\,}_{1}-E^{\,}_{0}=0$ and
(ii) two-fold degenerate in the thermodynamic limit $2N\to\infty$
when $2N=0\bmod 4$, in which case
the spacing $E^{\,}_{1}-E^{\,}_{0}$ vanishes
exponentially with the system size as
\begin{subequations}
\label{eq:NeelZ PBCs Exponential}
\begin{equation}
E^{\,}_{1}-E^{\,}_{0}=
A^{\,}_{1,0}\,e^{-2N/\xi^{\,}_{1,0}},
\end{equation}
where
\begin{equation}
\xi^{\,}_{1,0}=4.77\pm0.06.
\end{equation}
\end{subequations}
We have also found that the state
with energy $E^{\,}_{0}$ has lattice momentum $k=0$, while the state
with energy $E^{\,}_{1}$ has lattice momentum $k=\pi$.

Additionally, the energy spacing
$E^{\,}_{2}-E^{\,}_{1}$
in Fig.~\ref{fig:NeelZ OBCs Gaps b}
is best fitted with the Ansatz
\begin{subequations}
\label{eq:NeelZ PBCs Gap}
\begin{equation}
E^{\,}_{2}-E^{\,}_{1}= 
\Delta^{\,}_{2,1}
+
(-1)^{N}
A^{\,}_{2,1}\,
(2N)^{-1},
\label{eq:NeelZ PBCs Gap a}
\end{equation}
where
\begin{equation}
\Delta^{\,}_{2,1}=1.929\pm0.004.
\label{eq:NeelZ PBCs Gap b}
\end{equation}
\end{subequations}
Correspondingly, $\Delta^{\,}_{2,1}$ is an estimate of the spectral gap
in the thermodynamic limit $2N\to\infty$ between a two-fold
degenerate ground state and all excitations at the point $(\pi/4,3\pi/4)$
in the coupling space. The finite-size corrections to the
spectral gap decays algebraically as is expected from the
presence a continuum of excitations above the gap
\cite{Colome-Tache2010}.

We have also verified that the properties of the point
$(\pi/4,3\pi/4)$ extends to any point in the
$\mathrm{Neel}^{\,}_{z}$ phase in
Figs.~\ref{fig:phase diagram A}
and
\ref{fig:phase diagram B},
for both Hamiltonians
\eqref{eq:def HA and HB c}
and
\eqref{eq:def HA and HB d}.
In particular,
we have found that the staggered magnetization 
$m^{z}_{\mathrm{sta}}$
reaches the maximum value of $|m^{z}_{\mathrm{sta}}|\approx0.5$
at the point $(\pi/4,3\pi/4)$ in the
$\mathrm{Neel}^{\,}_{z}$ phase after
the thermodynamic limit has been taken.

\begin{figure}
\centering
\subfigure{\includegraphics[width=1\columnwidth]{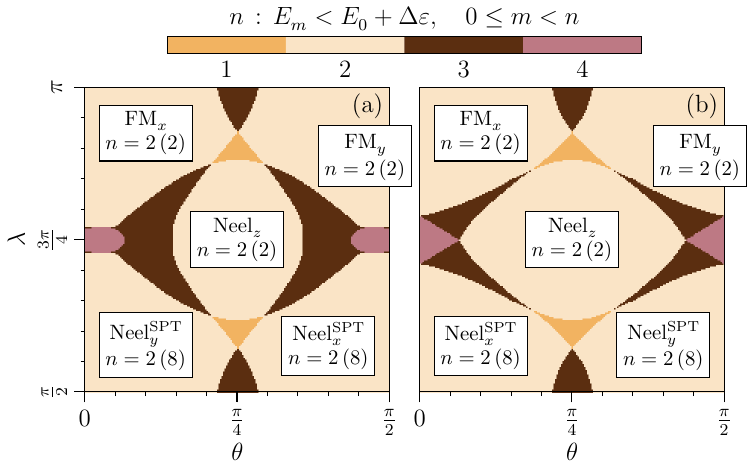} \label{fig:NeelZ PBCs Energy Spacing a}}%
\subfigure{\label{fig:NeelZ PBCs Energy Spacing b}}%
\caption{
Number $n$ of eigenenergies $E^{\,}_{m}$ 
with an energy smaller than
$E^{\,}_{0}+\Delta\varepsilon$, where $E^{\,}_{0}$ is the ground
state energy, calculated using ED by diagonalizing
the Hamiltonian
(a)
$\widehat{H}^{(0)}_{\mathrm{I}}$
in Eq.~\eqref{eq:def HA and HB c} and
(b)
$\widehat{H}^{(0)}_{\mathrm{II}}$
in Eq.~\eqref{eq:def HA and HB d},
under PBCs for $2N=14$ sites,
$\Delta\varepsilon=0.7$,
and the integer $0\leq m < n$.
The numbers in parentheses are those for OBCs.
}
\label{fig:NeelZ PBCs Energy Spacing}
\end{figure}

Finally, we briefly discuss the location of
the boundaries of the $\mathrm{Neel}^{\,}_{z}$ phase
in terms of the ground-state degeneracy.
We postpone the quantitative study of
it to Sec.~\ref{subsubsec:The FMx-Neelz phase boundary}.
Each of the classical fixed point
$(0, \pi/2)$,
$(\pi/2, \pi/2)$,
$(0, \pi)$ and
$(\pi/2, \pi)$
in the phase diagrams in Figs.~\ref{fig:phase diagram A}
and \ref{fig:phase diagram B}
realizes a gapped phase supporting
colinear magnetic long-range order.
This long-range order implies a minimal
degeneracy of $2$ that originates from the
spontaneous breaking of the time-reversal symmetry and, as such
is independent of the choice between PBCs and OBCs.
Additionally, the degeneracy at the points $(0, \pi/2)$ and
$(\pi/2, \pi/2)$ increases by a factor of $4$ upon choosing OBCs.
The locus of points in the coupling
space along which such a degeneracy changes defines the phase boundary
of each one of these four phases. An estimate of the location of
these phase boundaries is achieved by diagonalizing Hamiltonians
\eqref{eq:def HA and HB c} and
\eqref{eq:def HA and HB d} for $2N=14$ sites and counting the number
$n$ of eigenenergies $E^{\,}_{m}$ within a distance of
$\Delta\varepsilon=0.7>1/(2N)$ from the ground-state energy.
We illustrate in
Figs.~\ref{fig:NeelZ PBCs Energy Spacing a}
and \ref{fig:NeelZ PBCs Energy Spacing b}
such a procedure for Hamiltonians
\eqref{eq:def HA and HB c} and \eqref{eq:def HA and HB d},
respectively, where PBCs and OBCs have been imposed.
In doing so, we find a phase centered around the point
$(\pi/4,3\pi/4)$
that is shaped like a diamond with a degeneracy of two for both PBCs and OBCs.
Remarkably, comparing Figs.~\ref{fig:NeelZ PBCs Energy Spacing a}
and \ref{fig:NeelZ PBCs Energy Spacing b},
in the vicinity of $(0,3\pi/4)$ and $(\pi/2,3\pi/4)$,
we observe a discrepancy
that we shall interpret below as the
$\mathrm{Neel}^{\,}_{z}$ phase for Hamiltonian
\eqref{eq:def HA and HB d} ending
in two tetracritical points, one
at $(0,3\pi/4)$ and one at $(\pi/2,3\pi/4)$.

\subsection{Phase transitions}
\label{sec:phase_transitions}

We have constructed numerically the phase diagrams of
Hamiltonians
\eqref{eq:def HA and HB c}
and
\eqref{eq:def HA and HB d}
in Figs.\
\ref{fig:phase diagram A}
and
\ref{fig:phase diagram B},
respectively.
Each phase diagram hosts
twelve distinct phases, which
were reported in Eqs.\
\eqref{eq:eight gapped phases}
and
\eqref{eq:four extended diamond phase}.
All twelve phases support colinear magnetic long-range order,
four of them support SPT order,
and two of them are gapless.
We now turn our attention to the study of the transitions
between those twelve phases that share a common boundary in
Figs.\
\ref{fig:phase diagram A}
and
\ref{fig:phase diagram B}.
We have studied numerically
(i)
the two vertical phase boundaries 
$\mathrm{Neel}^{\,}_{x}-\mathrm{Neel}^{\,}_{y}$,
$\mathrm{Neel}^{\mathrm{SPT}}_{x}-\mathrm{Neel}^{\mathrm{SPT}}_{y}$,
(ii)
the two horizontal phase boundaries
$\mathrm{Neel}^{\,}_{x}-\mathrm{Neel}^{\mathrm{SPT}}_{y}$,
$\mathrm{Neel}^{\,}_{x}-\mathrm{Neel}^{\mathrm{SPT}}_{x}$,
(iii)
the four diagonal phase boundaries
$\mathrm{Neel}^{\,}_{x}-\mathrm{gFM}^{\,}_{z}$,
$\mathrm{FM}^{\,}_{x}-\mathrm{Neel}^{\,}_{z}$
from Fig.\
\ref{fig:phase diagram A}
and from Fig.\
\ref{fig:phase diagram B}, respectively,
and (iv)
one tetracritical point in Fig.\
\ref{fig:phase diagram B}. 
Our system sizes are too small to reliably study numerically
the properties of
any one of the tricritical points
in Figs.\
\ref{fig:phase diagram A}
and
\ref{fig:phase diagram B}.
The bulk properties of all remaining phase boundaries
in Figs.\
\ref{fig:phase diagram A}
and
\ref{fig:phase diagram B}
and tetracritical points
in Fig.\
\ref{fig:phase diagram B} 
are unitarily equivalent to those we have studied.

\begin{table}
\caption{
Summary of the critical exponents defined in
Eq.~\eqref{eq:def critical exponents}
of Hamiltonian \eqref{eq:def H XY}
at the free-fermion critical point
$\theta^{\,}_{\mathrm{c}}=\pi/4=0.78538\ldots$
and $\lambda=0$
obtained using iDMRG with $\chi=256$
and ED with $2N\leq26$.
We assumed $\theta^{\,}_{\mathrm{c}}$ to be a fitting
parameter in the Ansatz
Eqs.~\eqref{eq:nu critical exponent}
and \eqref{eq:beta critical exponent}.
We provide the theoretical values of the exponents for comparison.
    }
% %%%%%%%%%%%
% We need to put the \label AFTER \caption to avoid
% the Table to be referenced using \ref by the section number instead
% of the true Table number.
% %%%%%%%%%%%
\label{tab:DQ Neel Critical Exponents}
\centering
\setlength{\tabcolsep}{2pt}
\renewcommand{\arraystretch}{1.12}
\begin{tabular}{c c c c}
\hline\hline
\textbf{Exponent}
& \textbf{Numerics}
& $\theta^{\,}_{\mathrm{c}}$
& \textbf{Theory}
\\ \hline
$\nu$ 
& $1.01\pm0.02$ 
& $0.7866\pm0.0003$
& $1$
\\
$\beta$ 
& $0.248\pm0.003$ 
& $0.78539\pm0.00005$
& $1/4$
\\
$\eta^{x}$ 
& $0.5001\pm0.005$ 
& --
& $1/2$ 
\\
$\eta^{y}$ 
& $0.5001\pm0.005$ 
& --
& $1/2$ 
\\
$\eta^{z}$ 
& $2.0008\pm0.0002$ 
& --
& $2$ 
\\
$\mathsf{z}$ 
& $1.0029\pm0.0001$ 
& --
& $1$ 
\\
\hline\hline
\end{tabular}
\end{table}

\subsubsection{The $\mathrm{Neel}^{\,}_{x}-\mathrm{Neel}^{\,}_{y}$
			  phase boundary}
\label{subsubsec:The Neelx-Neely phase boundary}

\begin{figure*}
\centering
\subfigure{\includegraphics[width=0.525\textwidth]{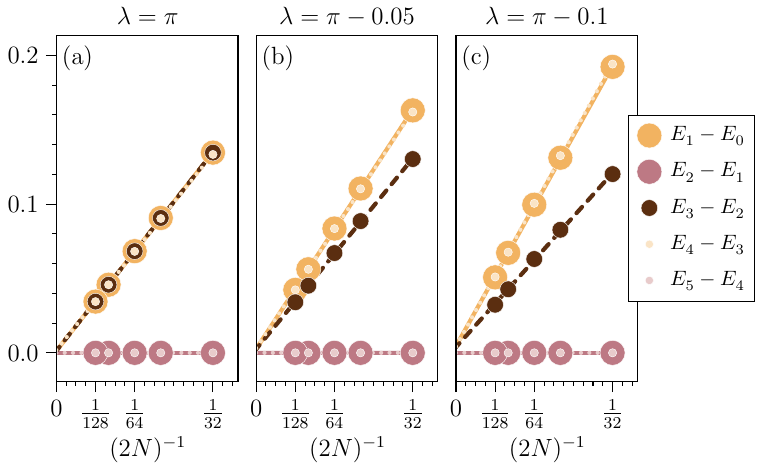} \label{fig:DQ FM Gaps + Central Charge + Exponent a}}%
\subfigure{\includegraphics[width=0.242\textwidth]{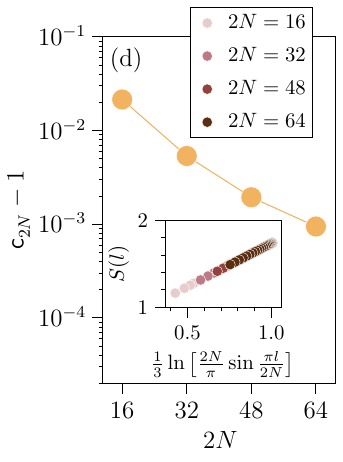} \label{fig:DQ FM Gaps + Central Charge + Exponent b}}%
\subfigure{\includegraphics[width=0.221\textwidth]{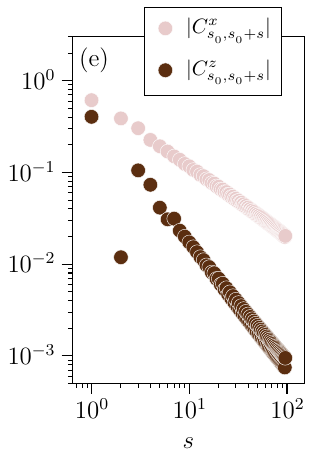} \label{fig:DQ FM Gaps + Central Charge + Exponent c}}%
\subfigure{\label{fig:DQ FM Gaps + Central Charge + Exponent d}}%
\subfigure{\label{fig:DQ FM Gaps + Central Charge + Exponent e}}%
\caption{
Dependence on 
$(2N)^{-1}$ of the energy differences
$E^{\,}_{n}-E^{\,}_{n-1}$, $n=1,2,3,4,5$, above 
the ground state energy $E^{\,}_{0}$
of Hamiltonian \eqref{eq:def H U1}
obtained with DMRG ($\chi = 128$) with OBCs
for different
(a) $\lambda=\pi$,
(b) $\lambda=\pi-0.05$ and
(c) $\lambda=\pi-0.1$. 
Linear fits for each $n$ are represented as lines whose 
color corresponds to the color of the data
points in the legend.
(d)
Dependence of central charge $\mathsf{c}$
on the number $2N$ of sites
of the ground state of
Hamiltonian \eqref{eq:def H U1}
obtained with DMRG ($\chi=512$)
under PBCs at $\lambda=\pi-0.1$.
The inset shows the 
dependence on the number $2N$ of sites of the
entanglement entropy $S^{\,}_{\mathrm{MPS},2N}({l})$.
The slope of the linear fit performed for each system
size $2N$ gives
the central charge $\mathsf{c}$.
(e)
Dependence on $s$ given $s^{\,}_{0}$ for the
spin-spin correlators
$C^{x}_{s^{\,}_{0},s^{\,}_{0}+s}$,
$C^{z}_{s^{\,}_{0},s^{\,}_{0}+s}$
defined in Eq.~\eqref{eq:Cs},
computed using the DMRG ($\chi=256$
and $2N=256$)
under OBCs at $\lambda=\pi-0.2$.
We extract the decay exponents
$\eta^{x}=0.757\pm0.005$ and
$\eta^{z}=1.36\pm0.02$ using
Eq.~\eqref{eq:eta critical exponent}.
}
\label{fig:DQ FM Gaps + Central Charge + Exponent}
\end{figure*}

The
$\mathrm{Neel}^{\,}_{x}$
and
$\mathrm{Neel}^{\,}_{y}$
phases are
long-range-ordered gapped phases
that are centered about
the classical fixed points
$(0,0)$ and $(\pi/2,0)$, respectively,
for both Hamiltonians 
\eqref{eq:def HA and HB c}
and
\eqref{eq:def HA and HB d}.
They are separated by a vertical boundary
along the $\mathrm{U}(1)$-symmetric
$\theta=\pi/4$ line.
The
$\mathrm{Neel}^{\,}_{x}$
and
$\mathrm{Neel}^{\,}_{y}$ phases
meet along the vertical open set
\begin{subequations}
\begin{equation}
\mathrm{DQ}^{\,}_{\mathrm{Neel}}\coloneqq
\left\{
(\pi/4,\lambda)
\ \left| \
\lambda^{\prime}_{\mathrm{Neel}^{\,}_{z}}
<
\lambda
<
\lambda^{\,}_{\mathrm{gFM}^{\,}_{z}}
\right.
\right\}.
\end{equation}
Here, D stands for deconfined and Q stands for quantum.
The end points labeled by
\begin{equation}
\lambda^{\prime}_{\mathrm{Neel}^{\,}_{z}}\coloneqq-0.2849\pm0.0005
\end{equation}
and
\begin{equation}
\lambda^{\,}_{\mathrm{gFM}^{\,}_{z}}\coloneqq0.2993\pm0.0009
\end{equation}
\end{subequations}
are tricritical points.
The set $\mathrm{DQ}^{\,}_{\mathrm{Neel}}$
realizes a line of quantum critical points,
each of which is a CFT with the central charge
$\mathsf{c}=1$,
for both Hamiltonians 
\eqref{eq:def HA and HB c}
and
\eqref{eq:def HA and HB d}.
Here we consider the $\lambda=0$ case as a prototypical case.

At $\lambda=0$, for any $\theta\in[0,\pi/2]$
Hamiltonian
\eqref{eq:def HA and HB c}
equals Hamiltonian
\eqref{eq:def HA and HB d}, i.e.,
\begin{equation}
\widehat{H}^{(\mathsf{b})}_{XY}(\theta)\equiv
\widehat{H}^{(\mathsf{b})}_{\mathrm{I}}(\theta,0)=
\widehat{H}^{(\mathsf{b})}_{\mathrm{II}}(\theta,0).
\label{eq:def H XY}
\end{equation}
The subscript $XY$ is a reference to the fact that
$\widehat{H}^{(\mathsf{b})}_{XY}(\theta)$ is
the antiferromagnetic quantum spin-$1/2$ $XY$ chain,
which realizes a critical point at the
$\mathrm{U}(1)$-symmetric point $\theta=\pi/4$.
The point $(\pi/4,0)$ is therefore denoted by
\begin{equation}
\lambda^{\,}_{\mathrm{Neel}^{\,}_{XY}}=0
\label{eq:def lambdaNeelXY}
\end{equation}
in Fig.~\ref{fig:phase diagram U1}.

We report in Table \ref{tab:DQ Neel Critical Exponents}
the critical exponents
$\nu$, $\beta$, and $\eta^{\alpha}$ with $\alpha=x,y,z$
and the dynamical critical exponent $\mathsf{z}$
obtained from iDMRG and ED simulations
for the Hamiltonian
\eqref{eq:def H XY}
at $\theta_{\mathrm{c}}^{\,}\equiv\pi/4$ and $\lambda=0$.
These critical exponents compare well with the exact ones. 
Hence, these preliminary results serve as
a benchmark for our numerical methods and
as a reference for those quantum phase transitions with $\lambda\ne0$
in Figs.~\ref{fig:phase diagram A}
and
\ref{fig:phase diagram B}
that are not exactly solvable.

We define the critical exponents in the general case
where the coupling $g\in\{\theta,\lambda\}$ drives the
quantum phase transition.
Here we study the $\mathrm{Neel}^{\,}_{x}$ to $\mathrm{Neel}^{\,}_{y}$
phase transition where $g=\theta$ is understood.

The exponent $\nu$ governs the divergence of the
correlation length $\xi$ when the coupling
$g$ approaches 
its critical value $g^{\,}_{\mathrm{c}}$.
Close to criticality, we assume the standard scaling form
\begin{subequations}
\label{eq:def critical exponents}
\begin{equation}
\xi(g)\sim
|g-g^{\,}_{\mathrm{c}}|^{-\nu}.
\label{eq:nu critical exponent}
\end{equation}
In practice, we fit $\ln\xi$
versus $\ln|g-g^{\,}_{\mathrm{c}}|$ to obtain the slope $-\nu$.

The critical scaling exponent $\beta$ controls the change of
the order parameter in the long-range ordered phase as
$g$ approaches $g^{\,}_{\mathrm{c}}$,
which is here given by
$m^{x}_{\mathrm{sta}}$ 
for $g<g^{\,}_{\mathrm{c}}$
and
$m^{y}_{\mathrm{sta}}$
for
$g>g^{\,}_{\mathrm{c}}$.
If we reserve the notation $M(g)$ for the absolute value of
either one of these two order parameters, we expect the scaling
\begin{equation}
M(g)\sim
|g-g^{\,}_{\mathrm{c}}|^{\beta}.
\label{eq:beta critical exponent}
\end{equation}
The exponent $\beta$ can be measured from the linear dependence
of $\ln M$ on $\ln|g-g^{\,}_{\mathrm{c}}|$.

At criticality, $g=g^{\,}_{\mathrm{c}}$,
all correlation functions are algebraic. Correspondingly,
the universal scaling exponent $\eta^{\alpha}$ is defined
at criticality by
\begin{equation}
C^{\alpha}_{s^{\,}_{0},s^{\,}_{0}+s}=
A^{\alpha}_{s}\,
s^{-\eta^{\alpha}},
\label{eq:eta critical exponent}
\end{equation}
\end{subequations}
where $A^{\alpha}_{s}$ is a nonuniversal amplitude.
For a finite chain of length $2N$, to minimize finite-size corrections,
we use the right-hand side of
Eq.\ (\ref{eq:eta critical exponent})
as a fitting Ansatz when evaluating numerically the
dependence on $s$ of the left-hand side within the window
$s^{\,}_{\mathrm{min}}\leq s\leq s^{\,}_{\mathrm{max}}$
chosen such that the decay is unambiguously algebraic.
We have measured the
nonuniversal amplitudes $A^{\alpha}_{s}$ in 
Eq.~\eqref{eq:eta critical exponent}.
At $\lambda=0$, we find
\begin{subequations}
\label{eq:measured amplitudes As at (0,thetac)}
\begin{align}
&
A^{x}_{s}\propto(-1)^{s},
\\
&
A^{y}_{s}\propto(-1)^{s},
\\
&
A^{z}_{s}\propto
\begin{cases}
0 & \hbox{if } s\equiv0\bmod 2,
\\
1 & \hbox{if } s\not\equiv0\bmod 2.
\end{cases}
\end{align}
\end{subequations}
As is well known, when $\lambda=0$,
the $z$ component of the quantum spin-$1/2$ has a
non-vanishing algebraic correlation
if and only if $s$ is an integer multiple of $2$.

The relation
(\ref{eq:conjugation by UeZ shifts lambda by pi})
implies that the phases
$\mathrm{FM}^{\,}_{x}$
and
$\mathrm{FM}^{\,}_{y}$,
which are centered about
the classical fixed points
$(0,\pi)$ and $(\pi/2,\pi)$
for both Hamiltonians 
\eqref{eq:def HA and HB c}
and
\eqref{eq:def HA and HB d},
meet along the unitarily equivalent phase boundary
\begin{subequations}
\begin{equation}
\begin{split}
\mathrm{DQ}^{\,}_{\mathrm{FM}}\coloneqq&\,
\left\{
(\pi/4,\lambda)
\ \left| \
\pi+\lambda^{\prime}_{\mathrm{Neel}^{\,}_{z}}
<
\lambda
<
\pi+\lambda^{\,}_{\mathrm{gFM}^{\,}_{z}}
\right.
\right\}
\\
\equiv&\,
\left\{
(\pi/4,\lambda)
\ \left| \
\lambda^{\,}_{\mathrm{Neel}^{\,}_{z}}
<
\lambda
<
\lambda^{\prime}_{\mathrm{gFM}^{\,}_{z}}
\right.
\right\},
\end{split}
\end{equation}
%\end{subequations}
where
(see
Sec.~\ref{subsubsec:Tricritical points gFMz})
\begin{align}
&
\lambda^{\,}_{\mathrm{Neel}^{\,}_{z}}=\pi-0.2849\pm0.0005\\
&
\lambda^{\prime}_{\mathrm{gFM}^{\,}_{z}}=\pi+0.2993\pm0.0009
\end{align}
\end{subequations}
are tricritical points.
The phase boundaries
$\mathrm{DQ}^{\,}_{\mathrm{Neel}}$
and
$\mathrm{DQ}^{\,}_{\mathrm{FM}}$
are reported in Fig.\ \ref{fig:phase diagram U1}.
Any point labeled by $\lambda$
along the phase boundary $\mathrm{DQ}^{\,}_{\mathrm{FM}}$
shares the same critical properties as the point labeled by $\lambda-\pi$
along the phase boundary $\mathrm{DQ}^{\,}_{\mathrm{Neel}}$.

Without loss of generality,
we study numerically the interval
\begin{equation}
\mathrm{DQ}^{-}_{\mathrm{FM}}\coloneqq
\left\{
(\pi/4,\lambda)
\ \big|\
\lambda^{\,}_{\mathrm{Neel}^{\,}_{z}}<\lambda\leq\pi
\right\}
\end{equation}
and use three criteria to establish that it is an interval of
critical points with central charge $\mathsf{c}=1$ that ends with
the tricritical point
$\lambda^{\,}_{\mathrm{Neel}^{\,}_{z}}$.

First, we know that the ground-state energy
$E^{\,}_{0}$ is nondegenerate under OBCs at the
DQCP $(\pi/4,\pi)$, an integrable (free-fermion) point.
The energy difference between
the first excited state $E^{\,}_{1}$
and the ground state $E^{\,}_{0}$ decreases like
$(2N)^{-1}$ as opposed to being exponentially small in $2N$.
We have verified in
Figs.~\ref{fig:DQ FM Gaps + Central Charge + Exponent a},
\ref{fig:DQ FM Gaps + Central Charge + Exponent b}, and
\ref{fig:DQ FM Gaps + Central Charge + Exponent c}
that the uniqueness of the ground state persists for 
all points on the line $(\pi/4,\lambda)$
with $\lambda\leq\pi$
until the gapped $\mathrm{Neel}^{\,}_{z}$ phase is reached. 

The rate at which two consecutive energy eigenvalues
approach each other
as the system size is increased is a useful
characterization of excited states. To this end, we define
the finite-size level-spacing amplitude $v^{\,}_{n,n-1}$ by
\begin{subequations}
\label{eq:velocities}
\begin{equation}
v^{\,}_{n,n-1}\coloneqq
\lim\limits_{2N\to\infty}
2N(E^{\,}_{n}-E^{\,}_{n-1}),
\label{eq:velocities a}
\end{equation}
assuming the partial ordering
\begin{equation}
E^{\,}_{n-1}\leq E^{\,}_{n},
\qquad
n=1,2,3,\cdots,
\label{eq:velocities b}
\end{equation}
\end{subequations}
of the exact energy eigenvalues.
An amplitude $v^{\,}_{n,n-1}$ vanishes in the thermodynamic limit
if $E^{\,}_{n}$ and $E^{\,}_{n-1}$
are degenerate for any choice of $2N$, or
if $E^{\,}_{n}-E^{\,}_{n-1}$ decays faster than $(2N)^{-1}$.
At $(\pi/4,\pi)$, it is observed that all the 
non-vanishing amplitudes are independent of $2N$, equal, and given by
\begin{equation}
v^{\,}_{1,0}= 
v^{\,}_{3,2}=
v^{\,}_{4,3}=
4.27\pm0.02,
\end{equation} 
which should be compared with the exact value in the large-$N$ limit, $\sqrt{2}\pi$.

Moving away from $(\pi/4,\pi)$
in the interval $\mathrm{DQ}^{-}_{\mathrm{FM}}$,
we deduce from Figs.\
\ref{fig:DQ FM Gaps + Central Charge + Exponent a},
\ref{fig:DQ FM Gaps + Central Charge + Exponent b},
and
\ref{fig:DQ FM Gaps + Central Charge + Exponent c}
that the amplitudes 
$v^{\,}_{1,0}$ and $v^{\,}_{4,3}$
remain equal within numerical precision,
reach their maxima at $\lambda^{\,}_{\mathrm{Neel}^{\,}_{z}}$,
and decrease as a function of increasing $\lambda$
on the interval $\mathrm{DQ}^{-}_{\mathrm{FM}}$.
However, the amplitudes
$v^{\,}_{1,0}$ and $v^{\,}_{4,3}$
differ from $v^{\,}_{3,2}$ both in their values and their monotonicity on
$\lambda$. Indeed, $v^{\,}_{3,2}$ is maximal at $\lambda=\pi$
and decreases as a function of
$|\pi-\lambda|$ on the interval $\mathrm{DQ}^{-}_{\mathrm{FM}}$.
The strength of the finite-size effects
close to the tricritical point at
$(\pi/4,\lambda^{\,}_{\mathrm{Neel}^{\,}_{z}})$
prevent a reliable numerical estimate of these amplitudes
at and in the vicinity of the tricritical point.
If we combine this numerical study of the amplitudes
$v^{\,}_{1,0}$, $v^{\,}_{3,2}$, and $v^{\,}_{4,3}$
to the one made on the interval
\begin{equation}
\mathrm{DQ}^{+}_{\mathrm{Neel}}\coloneqq
\left\{
(\pi/4,\lambda)
\ \big|\
0<\lambda\leq\lambda^{\,}_{\mathrm{gFM}^{\,}_{z}}
\right\}
\end{equation}
in Ref.\ \cite{Rey2025},
we conclude that
the amplitudes 
$v^{\,}_{1,0}=v^{\,}_{4,3}$
reach their maxima at
$\lambda^{\,}_{\mathrm{Neel}^{\,}_{z}}$
($\lambda^{\prime}_{\mathrm{Neel}^{\,}_{z}}$)
and decrease as a function of increasing $\lambda$
on the phase boundary
$\mathrm{DQ}^{\,}_{\mathrm{FM}}$
($\mathrm{DQ}^{\,}_{\mathrm{Neel}}$),
while the amplitude
$v^{\,}_{3,2}$ is maximal at
$\lambda=\pi$
($\lambda=0$)
and decreases as a function of
$|\pi-\lambda|$
($|\lambda|$)
on the phase boundary
$\mathrm{DQ}^{\,}_{\mathrm{FM}}$
($\mathrm{DQ}^{\,}_{\mathrm{Neel}}$).

The finite-size spectra discussed above are understood from
a $\mathsf{c}=1$ CFT, the free-boson theory with the Lagrangian
(see also Ref.~\cite{Oshikawa2006})
\begin{equation}
\mathcal{L}=\frac{g}{4\pi}(\partial^{\,}_{\mu}\varphi)^{2},
\label{eq: free-boson lagrangian}
\end{equation}
where the bosonic field $\varphi$ obeys $\varphi\equiv\varphi+2\pi n$,
$n\in\mathbb{Z}$.  When the Dirichlet boundary conditions are imposed
on $\varphi$ at $x=0$ and $x=L$, the first five excitation energies
measured from the ground-state energy are given by
\begin{equation}
(\varepsilon^{\,}_{1},
\varepsilon^{\,}_{2},
\varepsilon^{\,}_{3},
\varepsilon^{\,}_{4},
\varepsilon^{\,}_{5})
=\frac{\pi}{L}(g, g, 1, 1+g, 1+g),
\label{eq: spectra of free bosons under OBCs}
\end{equation}
where the speed of bosonic excitations is set equal to unity.
We have assumed that $0<g<1$, and $g=\frac{1}{2}$ at the free-fermion point.
We see that the pattern of the energy differences
$E^{\,}_{n}-E^{\,}_{n-1}\propto \varepsilon^{\,}_{n}-\varepsilon^{\,}_{n-1}$
in Fig.~\ref{fig:DQ FM Gaps + Central Charge + Exponent} is consistent with
Eq.\ (\ref{eq: spectra of free bosons under OBCs}).
The parameter $g$ can be obtained from $v^{\,}_{n,n-1}$ by
\begin{equation}
g=\frac{v^{\,}_{1,0}}{v^{\,}_{1,0}+v^{\,}_{3,2}}.
\end{equation}

Second, any point in the interval $\mathrm{DQ}^{-}_{\mathrm{FM}}$
realizes a CFT with central charge
$\mathsf{c}=1$ \cite{Rey2025}.
For instance, we report in
Fig.~\ref{fig:DQ FM Gaps + Central Charge + Exponent d}
the value of the central charge of
Hamiltonian \eqref{eq:def H U1}
under PBCs at the point $(\pi/4,\pi-0.1)$.
The value of the central charge $\mathsf{c}$
has been determined using
Eq.~\eqref{eq:CFT prediction for S2N(l)}.

Third, we have verified numerically that
all three correlation functions
$C^{x}_{s^{\,}_{0},s^{\,}_{0}+s}$,
$C^{y}_{s^{\,}_{0},s^{\,}_{0}+s}$,
and
$C^{z}_{s^{\,}_{0},s^{\,}_{0}+s}$
decay algebraically
for selected points in the interval
$\mathrm{DQ}^{-}_{\mathrm{FM}}$.
We report this behavior 
in Fig.~\ref{fig:DQ FM Gaps + Central Charge + Exponent e}
at the point $(\pi/4,\pi-0.2)$.
Using the Ansatz \eqref{eq:eta critical exponent},
we find that both
$C^{x}_{s^{\,}_{0},s^{\,}_{0}+s}=
C^{y}_{s^{\,}_{0},s^{\,}_{0}+s}$
and
$C^{z}_{s^{\,}_{0},s^{\,}_{0}+s}$
decay algebraically with the exponents
$\eta^{x}=0.757\pm0.005$
and
$\eta^{z}=1.36\pm0.02$, respectively.
Moreover, $A^{z}_{s}\propto (-1)^{s}$.
If we assume that the leading contribution to $C^{x}_{s^{\,}_{0},s^{\,}_{0}+s}$ comes
from a primary field in the $\mathsf{c}=1$ CFT in Eq.\ (\ref{eq: free-boson lagrangian}),
the exponent $\eta^{x}$ is related to the parameter $g$
by $\eta^{x}=2g^{2}$.

\begin{figure}
\centering
\includegraphics[width=1\columnwidth]{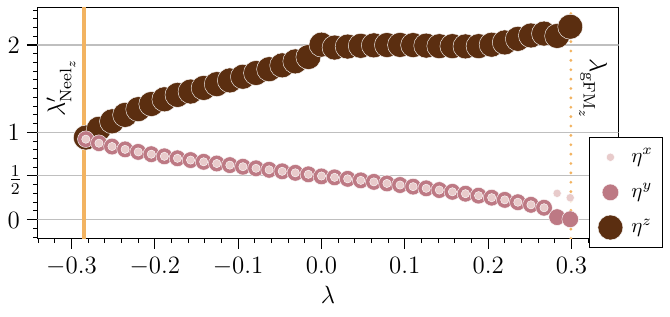}
\caption{
Dependence on $\lambda$ of the decay exponents
$\eta^{x}$, $\eta^{y}$ and $\eta^{z}$
of the spin-spin correlators
$C^{x}_{s^{\,}_{0},s^{\,}_{0}+s}$,
$C^{y}_{s^{\,}_{0},s^{\,}_{0}+s}$ and
$C^{z}_{s^{\,}_{0},s^{\,}_{0}+s}$
defined in Eq.~\eqref{eq:Cs},
computed using the iDMRG ($\chi=256$ and $L=4$)
for the ground state
$|\Psi^{\,}_{0}\rangle$
of Hamiltonian \eqref{eq:def H U1}.
The decay exponents
$\eta^{x}$, $\eta^{y}$ and $\eta^{z}$
are extracted using Eq.~\eqref{eq:eta critical exponent}.
The interval of the coupling space shown corresponds to the
$\mathrm{DQ}^{\,}_{\mathrm{Neel}}$ line.
We extrapolate this result to the
$\mathrm{DQ}^{\,}_{\mathrm{FM}}$ line
upon doing the replacement
$\lambda\mapsto\lambda+\pi$.
We indicate by the vertical yellow lines the locations
of the tricritical points.
}
\label{fig:DQ Neel Eta}
\end{figure}

In combination with the numerical results from Ref.\
\cite{Rey2025}, we have obtained that,
along the phase boundary $\mathrm{DQ}^{\,}_{\mathrm{FM}}$
($\mathrm{DQ}^{\,}_{\mathrm{Neel}}$), 
$\eta^{x}=\eta^{y}$
is a monotonically decreasing function of increasing $\lambda$
that reaches
\begin{itemize}
\item[(i)] a maximum value of
$1$ at $\lambda=\lambda^{\,}_{\mathrm{Neel}^{\,}_{z}}\ 
(\lambda^{\prime}_{\mathrm{Neel}^{\,}_{z}})$,
\item[(ii)] the free-fermion value of
$\frac{1}{2}$ at $\lambda=\pi\ (0)$, and
\item[(iii)] a minimum value of
$0$ at $\lambda=\lambda^{\prime}_{\mathrm{gFM}^{\,}_{z}}\ 
(\lambda^{\,}_{\mathrm{gFM}^{\,}_{z}})$,
\end{itemize}
while
$\eta^{z}$
is an increasing function of $\lambda\leq\pi\ (0)$
and a constant function of $\lambda\geq\pi\ (0)$
that reaches
\begin{itemize}
\item[(i)] a minimum value of
$1$ at $\lambda=\lambda^{\,}_{\mathrm{Neel}^{\,}_{z}}\ 
(\lambda^{\prime}_{\mathrm{Neel}^{\,}_{z}})$,
\item[(ii)] the free-fermion value of
$2$ at $\lambda=\pi\ (0)$.
\end{itemize}
This is illustrated in Fig.~\ref{fig:DQ Neel Eta}.
Remarkably, the exponents $\eta^{\alpha}$ with $\alpha=x,y,z$
along the phase boundary $\mathrm{DQ}^{\,}_{\mathrm{Neel}}$
follow those of the critical spin-$1/2$ XXZ
chain in the following way.

We define the spin-$1/2$ XXZ
chain as
\begin{equation}
\widehat{H}^{\,}_{\mathrm{XXZ}}(\Delta)\coloneqq
\sum_{j=1}^{2N}
\left(
\widehat{X}^{\,}_{j}
\widehat{X}^{\,}_{j+1}
+
\widehat{Y}^{\,}_{j}
\widehat{Y}^{\,}_{j+1}
+
\Delta\,
\widehat{Z}^{\,}_{j}
\widehat{Z}^{\,}_{j+1}
\right).
\label{eq:def H XXZ}
\end{equation}
It is known that the Hamiltonian
\eqref{eq:def H XXZ}
realizes a critical phase in the interval
\begin{subequations}
\begin{equation}
-1<\Delta\leq1,
\end{equation}
with the two-point functions
defined in Eq.~\eqref{eq:Calpha} decaying algebraically
with the leading terms
\cite{Luther1975, Lukyanov1999, Hikihara1998}
\begin{align}
&C^{x}_{s^{\,}_{0},s^{\,}_{0}+s}=
B^{x}_{s}\,s^{-\eta},
&&B^{x}_{s}\propto(-1)^{s},
\\
&C^{z}_{s^{\,}_{0},s^{\,}_{0}+s}=
B^{z}_{s}\,s^{-\frac{1}{\eta}}
-
\frac{1}{\pi^{2}\,\eta}\,s^{-2},
&&B^{z}_{s}\propto(-1)^{s}.
\label{eq: <ZZ> in XXZ}
\end{align}
Here, the exponent $\eta$ obeys
\cite{Luther1975, Lukyanov1999, Hikihara1998}
\begin{equation}
\eta=1-\frac{1}{\pi}\cos^{-1}\Delta.
\end{equation}
\end{subequations}

Our results on the $\mathrm{DQ}^{\,}_{\mathrm{Neel}}$ phase boundary
thus compare to the critical spin-$1/2$ XXZ
chain as follows.
The end points $\lambda^{\prime}_{\mathrm{Neel}^{\,}_{z}}$
and $\lambda^{\,}_{\mathrm{gFM}^{\,}_{z}}$ of the critical $\mathrm{DQ}^{\,}_{\mathrm{Neel}}$ line
correspond to $\Delta=+1$ and $\Delta=-1$ in the XXZ chain,
respectively.
The exponent $\eta^{x}$ is identified with $\eta$.
In the interval $\lambda^{\prime}_{\mathrm{Neel}^{\,}_{z}}<\lambda\leq0$,
the relation $\eta^{x}\eta^{z}\simeq1$ holds, as in the XXZ chain
when $0\leq\Delta<1$.
In the interval $-1<\Delta<0$,
the leading term of $C^{z}_{s^{\,}_{0},s^{\,}_{0}+s}$ in the XXZ chain is
the uniform term $\propto s^{-2}$ in Eq.\ (\ref{eq: <ZZ> in XXZ}).
In accordance with this, we have observed $\eta^{z}\simeq2$ in Fig.~\ref{fig:DQ Neel Eta}
in the interval $0<\lambda<\lambda^{\,}_{\mathrm{gFM}^{\,}_{z}}$.

\begin{figure}
\centering
\subfigure{\includegraphics[width=1\columnwidth]{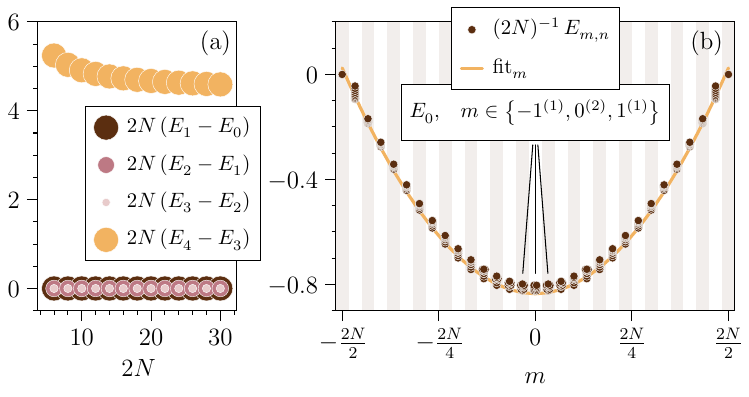} \label{fig:DQ Neel SPT OBCs pi/2 Decay Energies a}}%
\subfigure{\label{fig:DQ Neel SPT OBCs pi/2 Decay Energies b}}%
\caption{
The case of OBCs.
(a)
Dependence on
$2N$ of the energy differences
$2N\,(E^{\,}_{n}-E^{\,}_{n-1})$, $n=1,2,3,4$, above 
the ground state energy $E^{\,}_{0}$
of Hamiltonian \eqref{eq:def H U1}
obtained with ED for $\lambda=\pi/2$.
A linear fit is performed to obtain the amplitude
$v^{\,}_{4,3}=4.91\pm0.02$, defined in Eq.~\eqref{eq:velocities}.
The energy differences $E^{\,}_{n}-E^{\,}_{n-1}$, $n=1,2,3$
vanish for any $2N$ within numerical precision.
(b)
Lowest energies per site $(2N)^{-1}\,E^{\,}_{m,n}$
per magnetization sector $m$ of the
Hamiltonian \eqref{eq:def H U1}
obtained with ED for $\lambda=\pi/2$.
The lowest energy $E^{\,}_{m,0}$ per magnetization $m$
is well fitted by the function
$\mathrm{fit}^{\,}_{m}=(0.005\pm 3\cdot 10^{-4})\times
|m|^{2.00\pm 0.03}-0.824\pm 0.003$.
We highlight the fact that the
ground-state energy $E^{\,}_{0}$
is four-fold degenerate and indicate
the value of the magnetization $m$ of the ground state
associated to each of the four ground-state energies.
The superscript indicates the degeneracy of
each of the four ground-state energies
within the magnetization sector $m$.
}
\label{fig:DQ Neel SPT OBCs pi/2 Decay Energies}
\end{figure}

\begin{figure*}
\subfigure{\includegraphics[width=0.222\textwidth]{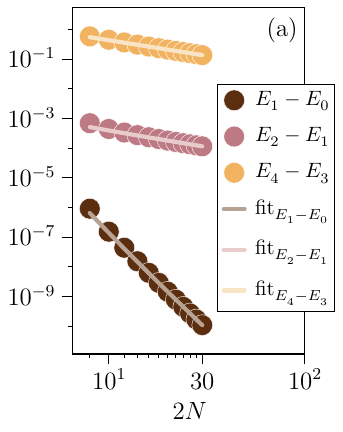} \label{fig:DQ Neel SPT OBCs Boundary a}}%
\subfigure{\includegraphics[width=0.2\textwidth]{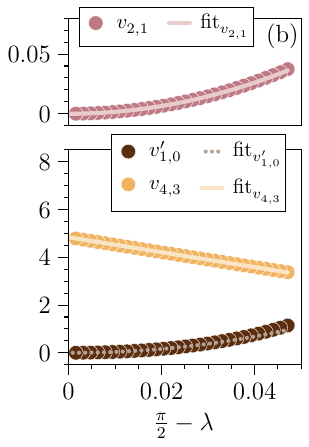} \label{fig:DQ Neel SPT OBCs Boundary b}}%
\subfigure{\includegraphics[width=0.292\textwidth]{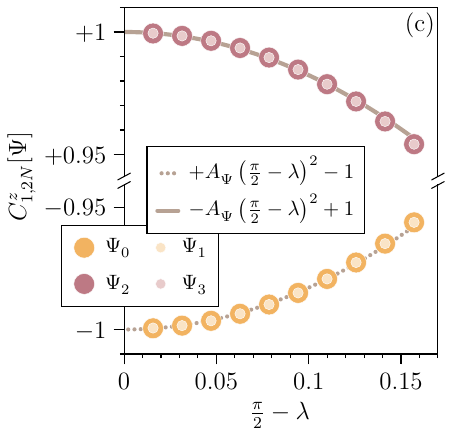} \label{fig:DQ Neel SPT OBCs Boundary c}}%
\subfigure{\includegraphics[width=0.285\textwidth]{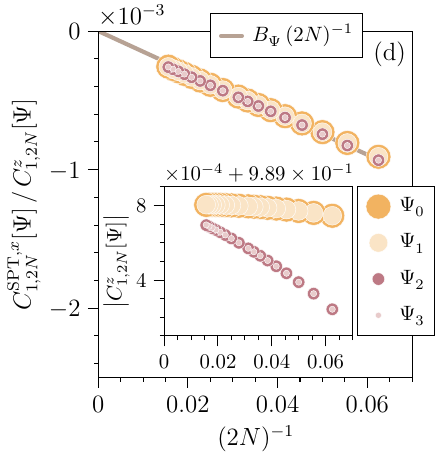} \label{fig:DQ Neel SPT OBCs Boundary d}}%
\caption{
The case of OBCs.
(a)
Dependence on $2N\in[8,30]$ of the energy differences
$E^{\,}_{n}-E^{\,}_{n-1}$, $n=1,2,4$
above the ground-state energy $E^{\,}_{0}$
obtained by diagonalizing Hamiltonian
\eqref{eq:def H U1} with ED at
$\lambda=\pi/2\left(1-\tfrac{1}{100}\right)$.
The fits used are explicitly given in
Table \ref{tab:Fits DQ Neel SPT OBCs}.
(b)
Dependence on $\lambda$ of the finite-size level-spacing amplitude
$v^{\,}_{n,n-1}$, $n=2,4$, defined in Eq.~\eqref{eq:velocities}
and $A^{\,}_{\mathrm{p}}$, defined in Eq.~\eqref{eq:velocities prime}
obtained by computing the dependence of
$E^{\,}_{n}-E^{\,}_{n-1}$, $n=1,2,4$
on $2N$ obtained by diagonalizing Hamiltonian
\eqref{eq:def H U1} with ED.
The fits used are explicitly given in
Table \ref{tab:Fits DQ Neel SPT OBCs}.
(c)
Two-point correlation function
$C^{z}_{1,2N}[\Psi]\equiv
\langle\Psi|\,\widehat{Z}^{\,}_{1}\,
\widehat{Z}^{\,}_{2N}\,|\Psi\rangle$
as a function of $\lambda$ for $2N=32$.
The symbol $\Psi$ labels any one of the first four energy eigenstates
$\Psi^{\,}_{0}$, $\Psi^{\,}_{1}$, $\Psi^{\,}_{2}$,
and $\Psi^{\,}_{3}$
approximated by DMRG with $\chi=128$
of Hamiltonian \eqref{eq:def H U1}.
We have fitted $C^{z}_{1,2N}[\Psi]$ with the Ansatz
$A^{\,}_{\Psi}\,\bigl(\frac{\pi}{2}-\lambda\bigr)^{2}-1$
where $A^{\,}_{\Psi}=1.71\pm0.01$ for $\Psi=\Psi^{\,}_{0}$.
The dependence of $A^{\,}_{\Psi}$ on the choice of $\Psi$
is negligible.
(d)
Anisotropy ratio
$C^{\mathrm{SPT},x}_{1,2N}[\Psi]/C^{z}_{1,2N}[\Psi]$
as a function of $(2N)^{-1}$ at $\lambda=\pi/2
\left(1-\tfrac{5}{100}\right)$
with $2N\in[16,64]$,
where 
$C^{\mathrm{SPT},x}_{1,2N}[\Psi]\equiv
\langle\Psi|
(\widehat{X}^{\,}_{1}\,\widehat{Z}^{\,}_{2})\,
(\widehat{Z}^{\,}_{2N-1}\,\widehat{X}^{\,}_{2N})
|\Psi\rangle$.
We have fitted
$C^{\mathrm{SPT},x}_{1,2N}[\Psi]/C^{z}_{1,2N}[\Psi]$
with the Ansatz $B^{\,}_{\Psi}\,(2N)^{-1}$
where $B^{\,}_{\Psi}=-0.0149\pm0.0001$ for $\Psi=\Psi^{\,}_{0}$.
The dependence of $B^{\,}_{\Psi}$ on the choice of $\Psi$
is negligible.
The inset shows the dependence of
$\big|C^{z}_{1,2N}[\Psi]\big|$ on $(2N)^{-1}$.
The symbol $\Psi$ labels any one of the first four energy eigenstates
$\Psi^{\,}_{0}$, $\Psi^{\,}_{1}$, $\Psi^{\,}_{2}$,
and $\Psi^{\,}_{3}$
approximated by DMRG with $\chi=128$
of Hamiltonian \eqref{eq:def H U1}.
}
\label{fig:DQ Neel SPT OBCs Boundary}
\end{figure*}

\begin{table*}
\caption{
Equations for the fits used in
Figs.~\ref{fig:DQ Neel SPT OBCs Boundary a}
and \ref{fig:DQ Neel SPT OBCs Boundary b}.
    }
% %%%%%%%%%%%
% We need to put the \label AFTER \caption to avoid
% the Table to be referenced using \ref by the section number instead
% of the true Table number.
% %%%%%%%%%%%
\label{tab:Fits DQ Neel SPT OBCs}
\centering
\renewcommand{\arraystretch}{1.5}
\setlength{\tabcolsep}{1pt}
\begin{tabular}{lcl c lcl}
\hline\hline
\multicolumn{3}{c}{Fig.~\ref{fig:DQ Neel SPT OBCs Boundary a}} &
\phantom{abcd}&
\multicolumn{3}{c}{Fig.~\ref{fig:DQ Neel SPT OBCs Boundary b}} \\
\cmidrule(lr){1-3}\cmidrule(lr){5-7}
$\mathrm{fit}^{\,}_{E^{\,}_{1}-E^{\,}_{0}}$ &
$=$ &
$0.097(2)\times(2N-1)^{-3.119(5)}\times(2N-1)^{-3}$ &
&
$\mathrm{fit}^{\,}_{A^{\,}_{\mathrm{p}}}$ &
$=$ &
$55(4)\cdot10^{1}\times\bigl(\frac{\pi}{2}-\lambda\bigr)^{2.08(2)}+2(3)\cdot10^{-4}$ \\
$\mathrm{fit}^{\,}_{E^{\,}_{2}-E^{\,}_{1}}$ &
$=$ &
$0.00394(4)\times(2N-1)^{2.950(3)}\times(2N-1)^{-4}$ &
&
$\mathrm{fit}^{\,}_{v^{\,}_{2,1}}$ &
$=$ &
$16.90(6)\times\bigl(\frac{\pi}{2}-\lambda\bigr)^{2.0114(8)}-(14\pm5)\cdot10^{-7}$ \\
$\mathrm{fit}^{\,}_{E^{\,}_{4}-E^{\,}_{3}}$ &
$=$ &
$4.87(3)\times(2N)^{-1.054(2)}$ &
&
$\mathrm{fit}^{\,}_{v^{\,}_{4,3}}$ &
$=$ &
$-30.92(3)\times\bigl(\frac{\pi}{2}-\lambda\bigr)^{0.9967(3)}+4.83837(6)$ \\
\hline\hline
\end{tabular}
\end{table*}

\subsubsection{The $\mathrm{Neel}^{\mathrm{SPT}}_{x}-\mathrm{Neel}^{\mathrm{SPT}}_{y}$ phase boundary}
\label{subsubsec:The NeelSPTx-NeelSPTy phase boundary}

We turn our attention to the study of
the phase boundary
$\mathrm{DQ}^{\mathrm{SPT}}_{\mathrm{Neel}}$
between the $\mathrm{Neel}^{\mathrm{SPT}}_{x}$ and
$\mathrm{Neel}^{\mathrm{SPT}}_{y}$ phases.
The main results are summarized in
Figs.~\ref{fig:DQ Neel SPT OBCs pi/2 Decay Energies},
and \ref{fig:DQ Neel SPT OBCs Boundary}.

The
$\mathrm{Neel}^{\mathrm{SPT}}_{x}$
and
$\mathrm{Neel}^{\mathrm{SPT}}_{y}$
phases are
long-range-ordered and SPT gapped phases
that are centered about
the classical fixed points
$(\pi/2,\pi/2)$ and $(0,\pi/2)$ 
for Hamiltonian
\eqref{eq:def HA and HB c}
and
$(0,\pi/2)$ and $(\pi/2,\pi/2)$
for Hamiltonian
\eqref{eq:def HA and HB d}.
They are separated by a vertical boundary
along the $\mathrm{U}(1)$-symmetric
$\theta=\pi/4$ line. It follows from
the combination of the relations
(\ref{eq:reduce coupling space PBC bottom a})
and
(\ref{eq:conjugation by UeZ shifts lambda by pi})
for Hamiltonian
\eqref{eq:def HA and HB c},
and from the combination of the relations
(\ref{eq:reduce coupling space PBC bottom b})
and
(\ref{eq:conjugation by UeZ shifts lambda by pi})
for Hamiltonian
\eqref{eq:def HA and HB d},
that the
$\mathrm{Neel}^{\mathrm{SPT}}_{x}$
and
$\mathrm{Neel}^{\mathrm{SPT}}_{y}$
phases meet at the phase boundary
\begin{subequations}
\begin{equation}
\begin{split}
\mathrm{DQ}^{\mathrm{SPT}}_{\mathrm{Neel}}\coloneqq&\,
\left\{
(\pi/4,\lambda)
\ \left| \
\frac{\pi}{2}-\lambda^{\,}_{\mathrm{gFM}^{\,}_{z}}
<
\lambda
<
\frac{\pi}{2}-\lambda^{\prime}_{\mathrm{Neel}^{\,}_{z}}
\right.
\right\}
\\
\equiv&\,
\left\{
(\pi/4,\lambda)
\ \left| \
\lambda^{\mathrm{SPT}}_{\mathrm{gFM}^{\,}_{z}}
<
\lambda
<
\lambda^{\mathrm{SPT}}_{\mathrm{Neel}^{\,}_{z}}
\right.
\right\},
\end{split}
\end{equation}
where the end points
\begin{equation}
\lambda^{\mathrm{SPT}}_{\mathrm{gFM}^{\,}_{z}}\coloneqq\frac{\pi}{2}-0.2993\pm0.0009
\end{equation}
and
\begin{equation}
\lambda^{\mathrm{SPT}}_{\mathrm{Neel}^{\,}_{z}}\coloneqq\frac{\pi}{2}+0.2849\pm0.0005
\end{equation}
are tricritical points
\footnote{
Our numerics are consistent with this prediction
		 }.
As a corollary, the relation
(\ref{eq:conjugation by UeZ shifts lambda by pi})
implies that the phases
$\mathrm{FM}^{\mathrm{SPT}}_{x}$
and
$\mathrm{FM}^{\mathrm{SPT}}_{y}$,
which are centered about
the classical fixed points
$(\pi/2,3\pi/2)$ and $(0,3\pi/2)$ 
for Hamiltonian
\eqref{eq:def HA and HB c}
and
$(0,3\pi/2)$ and $(\pi/2,3\pi/2)$
for Hamiltonian
\eqref{eq:def HA and HB d},
meet along the phase boundary
\begin{equation}
\begin{split}
\mathrm{DQ}^{\mathrm{SPT}}_{\mathrm{FM}}\coloneqq&\,
\left\{
(\pi/4,\lambda)
\ \left| \
\frac{3\pi}{2}-\lambda^{\,}_{\mathrm{gFM}^{\,}_{z}}
<
\lambda
<
\frac{3\pi}{2}-\lambda^{\prime}_{\mathrm{Neel}^{\,}_{z}}
\right.
\right\}
\\
\equiv&\,
\left\{
(\pi/4,\lambda)
\ \left| \
\lambda^{\prime\mathrm{SPT}}_{\mathrm{gFM}^{\,}_{z}}
<
\lambda
<
\lambda^{\prime\mathrm{SPT}}_{\mathrm{Neel}^{\,}_{z}}
\right.
\right\}
\end{split}
\end{equation} 
\end{subequations}
for both Hamiltonians
\eqref{eq:def HA and HB c}
and
\eqref{eq:def HA and HB d}
\footnote{
Our numerics are consistent with this prediction
		 }.
All bulk critical properties for the point
labeled by $\lambda$ along the phase boundary
$\mathrm{DQ}^{\mathrm{SPT}}_{\mathrm{FM}}$
are identical to the bulk critical properties of the points
labeled
$\lambda-\pi/2$,
$\lambda-\pi$,
and
$\lambda-3\pi/2$
along the phase boundaries
$\mathrm{DQ}^{\,}_{\mathrm{FM}}$,
$\mathrm{DQ}^{\mathrm{SPT}}_{\mathrm{Neel}}$,
and
$\mathrm{DQ}^{\,}_{\mathrm{Neel}}$,
respectively
\footnote{
Our numerics are consistent with this prediction
		 }. However, the phase boundaries
$\mathrm{DQ}^{\mathrm{SPT}}_{\mathrm{FM}}$
and
$\mathrm{DQ}^{\mathrm{SPT}}_{\mathrm{Neel}}$
differ from the phase boundaries
$\mathrm{DQ}^{\,}_{\mathrm{FM}}$
and
$\mathrm{DQ}^{\,}_{\mathrm{Neel}}$
through the sensitivity of the phase boundaries
$\mathrm{DQ}^{\mathrm{SPT}}_{\mathrm{FM}}$
and
$\mathrm{DQ}^{\mathrm{SPT}}_{\mathrm{Neel}}$
to the change from PBCs to OBCs,
as we now demonstrate numerically
through Figs.
\ref{fig:DQ Neel SPT OBCs pi/2 Decay Energies}
and
\ref{fig:DQ Neel SPT OBCs Boundary}.

Without loss of generality, we only need to
consider the phase boundary
$\mathrm{DQ}^{\mathrm{SPT}}_{\mathrm{Neel}}$.
We have studied the sensitivity to the choice of boundary conditions
of a point $\lambda$ along $\mathrm{DQ}^{\mathrm{SPT}}_{\mathrm{Neel}}$
by computing numerically the degeneracy of the ground state
as a function of the system size $2N$.
The point
\begin{equation}
(\pi/4,\pi/2)\in\mathrm{DQ}^{\mathrm{SPT}}_{\mathrm{Neel}},
\end{equation}
denoted
$\lambda^{\mathrm{SPT}}_{XY^{\,}_{\mathrm{Neel}}}$
in Fig.\ \ref{fig:phase diagram U1},
is a fixed point of the transformation
(\ref{eq:action U_R on Hamiltonians}).
As was shown analytically in Ref.\
\cite{Rey2025},
this fixed point is non-degenerate under PBCs but inherits
a degeneracy of four under OBCs from the neighboring
eight-fold degenerate SPT phases
$\mathrm{Neel}^{\mathrm{SPT}}_{x}$
and
$\mathrm{Neel}^{\mathrm{SPT}}_{y}$.
Indeed, the Hamiltonian equivalent to Hamiltonian
(\ref{eq:def H XY})
when $\lambda=\pi/2$, i.e.,
\begin{equation}
\widehat{H}^{(\mathsf{b})}_{\mathrm{U}(1)}(\pi/2)\equiv
\widehat{H}^{(\mathsf{b})}_{\mathrm{I}}(\pi/4,\pi/2)=
\widehat{H}^{(\mathsf{b})}_{\mathrm{II}}(\pi/4,\pi/2),
\end{equation}
is mapped by the Jordan-Wigner transformation to a free-fermion chain
with one spinless fermion localized on the first site and one
spinless fermion localized on the last site dropping from
$\widehat{H}^{(1)}_{\mathrm{U}(1)}(\pi/2)$
when OBCs are selected with the choice $\mathsf{b}=1$.
It was shown numerically in Ref.\
\cite{Rey2025} that this degeneracy of four is lifted to a degeneracy of
two on the interval
\begin{equation}
\begin{split}
\mathrm{DQ}^{\mathrm{SPT},<}_{\mathrm{Neel}}\coloneqq
\left\{
(\pi/4,\lambda)
\ \left| \
\lambda^{\mathrm{SPT}}_{\mathrm{gFM}^{\,}_{z}}
<
\lambda
<
\pi/2
\right.
\right\}.
\end{split}
\end{equation}
We have verified numerically that this is also so on the interval
\begin{equation}
\begin{split}
\mathrm{DQ}^{\mathrm{SPT},>}_{\mathrm{Neel}}\coloneqq
\left\{
(\pi/4,\lambda)
\ \left| \
\pi/2
<
\lambda
<
\lambda^{\mathrm{SPT}}_{\mathrm{Neel}^{\,}_{z}}
\right.
\right\}.
\end{split}
\end{equation}
Moreover,
Figs.\
\ref{fig:DQ Neel SPT OBCs pi/2 Decay Energies}
and
\ref{fig:DQ Neel SPT OBCs Boundary}
show that the change of the degeneracy from four to two as
$\lambda$ deviates from $\pi/2$ along the phase boundary
$\mathrm{DQ}^{\mathrm{SPT}}_{\mathrm{Neel}}$
is tied to boundary magnetism.
We conclude that the phase boundary
$\mathrm{DQ}^{\mathrm{SPT}}_{\mathrm{Neel}}$
is qualitatively distinct from the phase boundary
$\mathrm{DQ}^{\,}_{\mathrm{Neel}}$,
for the former, unlike the latter,
supports boundary zero modes.

Along the horizontal line $\lambda=\pi/2$
it is convenient for benchmarking our numerics to study numerically
Hamiltonian
\begin{equation}
\widehat{H}^{(\mathsf{b})}_{XY^{\mathrm{SPT}}}(\theta)\equiv
\widehat{H}^{(\mathsf{b})}_{\mathrm{I}}(\theta,\pi/2)\equiv
\widehat{H}^{(\mathsf{b})}_{\mathrm{II}}(\pi/2-\theta,\pi/2),
\label{eq:def H XY SPT}
\end{equation}
where $\mathsf{b}=0\,(1)$ selects PBCs (OBCs) and
$\theta\in[0,\pi/2]$.

The following behavior has been verified numerically.
On the one hand,
under PBCs and for any $0<\theta<\pi/4$, 
the exact two-fold degeneracy of the ground state of
$\widehat{H}^{(0)}_{XY^{\mathrm{SPT}}}(\theta)$
present at the end point $\theta=0$
is lifted for any finite size $2N$. 
More precisely, the energy difference $E^{\,}_{1}-E^{\,}_{0}$
vanishes exponentially fast with increasing $2N$.
It is only in the thermodynamic limit $2N\to\infty$ that the two-fold
degeneracy of the ground state is recovered for $0<\theta<\pi/4$.
At the critical point $\theta=\pi/4$, 
the difference between the energy $E^{\,}_{1}$
and the ground-state energy $E^{\,}_{0}$ of
$\widehat{H}^{(0)}_{XY^{\mathrm{SPT}}}(\pi/4)$
vanishes as $(2N)^{-1}$ with increasing $2N$.
Hence, the ground state of Hamiltonian
\begin{subequations}
\begin{equation}
\widehat{H}^{(0)}_{XY^{\mathrm{SPT}}}(\pi/4)\equiv
\widehat{H}^{(0)}_{\mathrm{U}(1)}(\pi/2)
\label{eq:degeneracy H0XYSPT(piover4)}
\end{equation}
is gapless and
non-degenerate in the thermodynamic limit $2N\to\infty$.
On the other hand,
under OBCs and for any $0<\theta<\pi/4$,
the exact eight-fold degeneracy of the ground state of
$\widehat{H}^{(1)}_{XY^{\mathrm{SPT}}}(\theta)$
present at the end point $\theta=0$ is lifted
for any finite size $2N$. 
More precisely, the energy difference
$E^{\,}_{7}-E^{\,}_{0}$
vanishes exponentially fast with increasing $2N$.
It is only in the thermodynamic limit $2N\to\infty$ that the eight-fold
degeneracy of the ground state is recovered.
At the critical point $\theta=\pi/4$,
the difference between the energy $E^{\,}_{7}$
and the energy $E^{\,}_{6}$ of
$\widehat{H}^{(1)}_{XY^{\mathrm{SPT}}}(\pi/4)$
vanishes as $(2N)^{-1}$ with increasing $2N$.
Hence, the ground state of Hamiltonian
$\widehat{H}^{(1)}_{XY^{\mathrm{SPT}}}(\pi/4)$
is not eight-fold degenerate anymore. Instead,
Fig.~\ref{fig:DQ Neel SPT OBCs pi/2 Decay Energies}
shows that the energy difference $E^{\,}_{4}-E^{\,}_{3}$ of
$\widehat{H}^{(1)}_{XY^{\mathrm{SPT}}}(\pi/4)$
vanishes as $(2N)^{-1}$ with increasing $2N$, while we
have verified that the energy differences
$
E^{\,}_{n}-E^{\,}_{n-1},
$
with
$n=1,2,3,$
are always of order $10^{-15}$, i.e.,
they vanish for any system size $2N$ within numerical precision.
Hence, the ground state of Hamiltonian
\begin{equation}
\widehat{H}^{(1)}_{XY^{\mathrm{SPT}}}(\pi/4)\equiv
\widehat{H}^{(1)}_{\mathrm{U}(1)}(\pi/2)
\label{eq:degeneracy H1XYSPT(piover4)}
\end{equation}
\end{subequations}
is four-fold degenerate, a degeneracy smaller
by a factor of two as compared to the degeneracy
of $\widehat{H}^{(1)}_{XY^{\mathrm{SPT}}}(\theta)$
at $\theta=0$.
As with PBCs, this decrease of the degeneracy by a factor of two
is caused by the absence of the spontaneous
breaking of the time-reversal symmetry at criticality.

The dependence of the degeneracy of the ground state
upon the choice of boundary conditions
is a signature of the SPT character of the point
$(\pi/4,\pi/2)$ of Hamiltonians
\eqref{eq:def HA and HB c} and \eqref{eq:def HA and HB d}.
This SPT attribute extends to any point of the
$\mathrm{DQ}^{\mathrm{SPT}}_{\mathrm{Neel}}$ phase boundary
as we now show.

We have computed the lowest four energy eigenvalues
$E^{\,}_{0}\leq E^{\,}_{1}\leq E^{\,}_{2}\leq E^{\,}_{3}$
of Hamiltonian \eqref{eq:def H U1}
at $\theta=\pi/4$ as a function of
$\lambda^{\mathrm{SPT}}_{\mathrm{gFM}^{\,}_{z}}\leq\lambda\leq\pi/2$
and the system size
$2N\in[8,30]$ with ED.
The generic dependence on $2N$ that we find is reported in
Fig.~\ref{fig:DQ Neel SPT OBCs Boundary a}
for the values of $E^{\,}_{n}-E^{\,}_{n-1}$, with $n=1,2,4,$
for the illustrative case of
$\lambda=\pi/2\,(1-1/100)$.
We make the following observations.
\begin{itemize}
\item[(i)]
The energy difference $E^{\,}_{3}-E^{\,}_{2}$
vanishes for any $2N$ within numerical precision.
\item[(ii)]
The energy differences $E^{\,}_{n}-E^{\,}_{n-1}$
for $n=2,4$ vanish in the thermodynamic limit $2N\to\infty$ 
as $(2N)^{-1}$,
as is expected for a continuum with dynamical exponent $\mathsf{z}=1$.
\item[(iii)]
The energy difference $E^{\,}_{1}-E^{\,}_{0}$
vanishes in the thermodynamic limit as the power law
(with the goodness of fit $R^{2}=0.99994$)
\footnote{
The scaling Ansatz
(\ref{eq:DQ Neel SPT Power Law E1-E0})
is a better fit than the exponential scaling Ansatz
(with the goodness of fit $R^{2}=0.988$)
$E^{\,}_{1}-E^{\,}_{0}=
A^{\,}_{\mathrm{e}}\,e^{-2N/\xi},
$
where
$
A^{\,}_{\mathrm{e}}=(10\pm3)\cdot 10^{-7}
$
and
$
\xi=3.1\pm0.1
$
for $2N\in[14,28]$
		 }
\begin{subequations}
\label{eq:DQ Neel SPT Power Law E1-E0}
\begin{equation}
A^{\,}_{\mathrm{p}}\,(2N-1)^{-\kappa}\,(2N-1)^{-3},
\end{equation}
where
\begin{equation}
A^{\,}_{\mathrm{p}}=0.097\pm0.02,
\qquad
\kappa=3.119\pm0.005,
\end{equation}
\end{subequations}
for $2N\in[20,30]$.
Here, the scaling Ansatz
(\ref{eq:DQ Neel SPT Power Law E1-E0})
is deduced analytically in Sec.\
\ref{subsubsec:effective boundary interaction around lambda piover2}.
We interpret a power law decay of the energy spacing $E^{\,}_{1}-E^{\,}_{0}$
faster than that of the continuum as a degeneracy
in the thermodynamic limit.
\item[(iv)]
On the one hand,
the use of DMRG to access larger values of $2N$ is consistent with
(i) and (ii). On the other hand, DMRG predicts that
$E^{\,}_{1}-E^{\,}_{0}$
vanishes for any $2N$ within numerical precision
instead of the scaling Ansatz (\ref{eq:DQ Neel SPT Power Law E1-E0}).
We interpret the latter as an artifact of DMRG.
\end{itemize}

Figure \ref{fig:DQ Neel SPT OBCs Boundary b}
quantifies the dependence on $\lambda$ of the parameters entering
the scaling Ansatz (\ref{eq:DQ Neel SPT Power Law E1-E0}).
We find that the dependence on
$\lambda$ of the scaling exponent
$\kappa$ in Eq.~\eqref{eq:DQ Neel SPT Power Law E1-E0}
is weak, contrary to that of $A^{\,}_{\mathrm{p}}$.
To study the dependence on $\lambda$ of the lowest four energy differences,
we make use of the finite-size level-spacing amplitude
$v^{\,}_{n,n-1}$ for $n=2,4$
defined in Eq.~\eqref{eq:velocities a} and we introduce
the finite-size level-spacing amplitude
\begin{equation}
A^{\,}_{\mathrm{p}}\coloneqq
\lim\limits_{2N\to\infty}
(2N-1)^{-\kappa}
(2N-1)^{-3}
(E^{\,}_{1}-E^{\,}_{0}).
\label{eq:velocities prime}
\end{equation}
We fit the dependence of these amplitudes using the Ansatz
\begin{equation}
A\times(\pi/2-\lambda)^{\sigma}+B,
\qquad
\pi/2-\lambda^{\,}_{\mathrm{gFM}^{\,}_{z}}\leq\lambda<\pi/2,
\label{eq:DQ Neel SPT Ansatz pi/2-lambda}
\end{equation} 
holding $\kappa=3.119$ fixed.
We deduce from Fig.\
\ref{fig:DQ Neel SPT OBCs Boundary b}
that, for the values
$\pi/2\,(1-3/100)\leq\lambda<\pi/2$,
the amplitudes
$A^{\,}_{\mathrm{p}}$
and
$v^{\,}_{2,1}$
are quadratic functions ($\sigma\simeq2$) of $\pi/2-\lambda$,
while the speed $v^{\,}_{4,3}$ is the linear function
($\sigma\simeq1$) of $\pi/2-\lambda$.
More precisely,
the Ansatz \eqref{eq:DQ Neel SPT Ansatz pi/2-lambda}
for the dependence of $A^{\,}_{\mathrm{p}}$
on $\lambda$ delivers
\begin{subequations}
\label{eq:DQ Neel SPT Ansatz E1-E0 Paramters pi/2-lambda}
\begin{align}
&
\sigma=2.00\pm0.07,
\\
&
A=400\pm100,
\\
&
B=-(1\pm4)\cdot10^{-4},
\end{align}
\end{subequations}
The large uncertainty on the parameters
\eqref{eq:DQ Neel SPT Ansatz E1-E0 Paramters pi/2-lambda}
is interpreted as the presence of strong finite-size effects.

The $\mathrm{U}(1)$-symmetric family of Hamiltonian
$\widehat{H}^{(1)}_{\mathrm{U}(1)}(\lambda)$
obeying OBCs defined in Eq.\
(\ref{eq:def H U1})
has, in the thermodynamic limit,
a ground state that is four-fold degenerate
when $\lambda=\pi/2$ and two-fold degenerate on
$\mathrm{DQ}^{\mathrm{SPT}}_{\mathrm{Neel}}\setminus\{(\pi/4,\pi/2)\}$.
The four-fold degeneracy when $\lambda=\pi/2$
is a consequence of the fact
\cite{Rey2025}
that the pair of quantum spin-$1/2$ operators
\begin{subequations}
\label{eq:DQ Neel SPT Free Spins}
\begin{align}
&\widehat{\vb*{S}}^{\,}_{\mathrm{L}}\coloneqq
\begin{pmatrix}
\widehat{X}^{\,}_{1}\,\widehat{Z}^{\,}_{2}
\\
\widehat{Y}^{\,}_{1}\,\widehat{Z}^{\,}_{2}
\\
\widehat{Z}^{\,}_{1}
\end{pmatrix}
\equiv
\begin{pmatrix}
\widehat{S}^{x}_{\mathrm{L}}
\\
\widehat{S}^{y}_{\mathrm{L}}
\\
\widehat{S}^{z}_{\mathrm{L}}
\end{pmatrix},
\label{eq:DQ Neel SPT Free Spins L}
\end{align}
and
\begin{align}
&
\widehat{\vb*{S}}_{\mathrm{R}}\coloneqq
\begin{pmatrix}
\widehat{Z}^{\,}_{2N-1}\,
\widehat{X}^{\,}_{2N}
\\
\widehat{Z}^{\,}_{2N-1}\,
\widehat{Y}^{\,}_{2N}
\\
\widehat{Z}^{\,}_{2N}
\end{pmatrix}
\equiv
\begin{pmatrix}
\widehat{S}^{x}_{\mathrm{R}}
\\
\widehat{S}^{y}_{\mathrm{R}}
\\
\widehat{S}^{z}_{\mathrm{R}}
\end{pmatrix},
\label{eq:DQ Neel SPT Free Spins R}
\end{align}
\end{subequations}
commute pairwise and with
$\widehat{H}^{(1)}_{\mathrm{U}(1)}(\pi/2)$.
We may thus label the four degenerate ground states of
$\widehat{H}^{(1)}_{\mathrm{U}(1)}(\pi/2)$
by the eigenvalues of
$\widehat{S}^{z}_{\mathrm{L}}$
and
$\widehat{S}^{z}_{\mathrm{R}}$.
We have computed the two-point correlation functions 
between the components of
$\widehat{\vb*{S}}_{\mathrm{L}}$
and
$\widehat{\vb*{S}}_{\mathrm{R}}$
along the interval
$(\pi/4,\lambda)$ with $\pi/2\,(1-1/10)\leq\lambda<\pi/2$.
Remarkably, we find that
$\widehat{S}^{z}_{\mathrm{L}}$ and $\widehat{S}^{z}_{\mathrm{R}}$
are correlated until the tricritical point is reached in any one of the four
lowest-lying-energy eigenstates, as is inferred from
Fig.~\ref{fig:DQ Neel SPT OBCs Boundary c} when $2N=32$.
Indeed, according to
Fig.~\ref{fig:DQ Neel SPT OBCs Boundary c},
it is an empirical observation that
(non-vanishing antiferromagnetic Ising correlation)
\begin{subequations}
\begin{equation}
\langle\Psi^{\,}_{0}|\,
\widehat{S}^{z}_{\mathrm{L}}\,
\widehat{S}^{z}_{\mathrm{R}}\,
|\Psi^{\,}_{0}\rangle=
\langle\Psi^{\,}_{1}|\,
\widehat{S}^{z}_{\mathrm{L}}\,
\widehat{S}^{z}_{\mathrm{R}}\,
|\Psi^{\,}_{1}\rangle<0,
\label{eq:AF ZZ bd correlation}
\end{equation}
while (non-vanishing ferromagnetic Ising correlation)
\begin{align}
\langle\Psi^{\,}_{2}|\,
\widehat{S}^{z}_{\mathrm{L}}\,
\widehat{S}^{z}_{\mathrm{R}}\,
|\Psi^{\,}_{2}\rangle=
\langle\Psi^{\,}_{3}|\,
\widehat{S}^{z}_{\mathrm{L}}\,
\widehat{S}^{z}_{\mathrm{R}}\,
|\Psi^{\,}_{3}\rangle>0,
\label{eq:F ZZ bd correlation}
\end{align}
along the interval
$(\pi/4,\lambda)$ with $\pi/2\,(1-1/10)\leq\lambda<\pi/2$ when $2N=32$.
Moreover, it is observed in
Fig.~\ref{fig:DQ Neel SPT OBCs Boundary c} that the bounds
\begin{equation}
0.95<
\big|
\langle\Psi^{\,}_{i}|\,
\widehat{S}^{z}_{\mathrm{L}}\,
\widehat{S}^{z}_{\mathrm{R}}\,
|\Psi^{\,}_{i}\rangle
\big|<1,
\qquad
i=0,1,2,3
\end{equation}
and in the inset of
Fig.~\ref{fig:DQ Neel SPT OBCs Boundary d} that
\begin{equation}
\big|\langle\Psi^{\,}_{0}|\,
\widehat{S}^{z}_{\mathrm{L}}\,
\widehat{S}^{z}_{\mathrm{R}}\,
|\Psi^{\,}_{0}\rangle\big|
\simeq
\big|\langle\Psi^{\,}_{2}|\,
\widehat{S}^{z}_{\mathrm{L}}\,
\widehat{S}^{z}_{\mathrm{R}}\,
|\Psi^{\,}_{2}\rangle\big|
\end{equation}
\end{subequations}
become independent of $2N$ for large $2N$
along the interval $(\pi/4,\lambda)$ with $\pi/2\,(1-1/10)\leq\lambda<\pi/2$.
In contrast, the limit
\begin{equation}
\lim_{2N\to\infty}
\frac{
\big|\langle\Psi^{\,}_{i}|\,
\widehat{S}^{x(y)}_{\mathrm{L}}\,
\widehat{S}^{x(y)}_{\mathrm{R}}\,
|\Psi^{\,}_{i}\rangle\big|
	 }
	 {
\big|\langle\Psi^{\,}_{i}|\,
\widehat{S}^{z}_{\mathrm{L}}\,
\widehat{S}^{z}_{\mathrm{R}}\,
|\Psi^{\,}_{i}\rangle\big|
	 }=0,
\qquad
i=0,1,2,3
\end{equation}
is reached through the power-law decay $(2N)^{-1}$.

In summary, for any point in
$\mathrm{DQ}^{\mathrm{SPT}}_{\mathrm{Neel}}\setminus\{(\pi/4,\pi/2)\}$,
numerics on an open chain are consistent with the existence of
boundary magnetism in the thermodynamic limit. The ground state is
two-fold degenerate and exhibits antiferromagnetic long-range correlations
along the $z$ axis in spin-$1/2$ space.
The first excited state is two-fold degenerate and exhibits
ferromagnetic long-range correlations
along the $z$ axis in spin-$1/2$ space.
In contrast, 
there is no spontaneous magnetism for a closed ring in the
thermodynamic limit for any point in
$\mathrm{DQ}^{\mathrm{SPT}}_{\mathrm{Neel}}\setminus\{(\pi/4,\pi/2)\}$.

In Sec.~\ref{subsubsec:effective boundary interaction around lambda piover2}
we derive the effective Hamiltonian for the edge spin operators
$\widehat{\vb*{S}}_{\mathrm{L}}$ and
$\widehat{\vb*{S}}_{\mathrm{R}}$
for the low-energy spectra and spin correlations.

\subsubsection{The $\mathrm{Neel}^{\,}_{x}-\mathrm{Neel}^{\mathrm{SPT}}_{y}$
			  phase boundary}
\label{subsubsec:The Neelx-NeelSPTy phase boundary}

The phase boundary between the 
gapped long-range ordered phase
$\mathrm{Neel}^{\,}_{x}$
centered at $(0,0)$ and the gapped SPT phase 
$\mathrm{Neel}^{\mathrm{SPT}}_{y}$
centered at $(0,\pi/2)$ in Fig.\ \ref{fig:phase diagram A}
is only of relevance to Hamiltonian
(\ref{eq:def HA and HB c}).
This horizontal phase boundary in Fig.\
\ref{fig:phase diagram A}
is the open set
\begin{subequations}
\begin{equation}
\mathrm{TQ}^{\mathrm{I}}_{\mathrm{gFM}^{\,}_{z}}\coloneqq
\left\{
(\theta,\pi/4)
\ \left| \
\theta^{\,}_{\mathrm{FM}^{\mathrm{I}}_{z}}
<
\theta
<
\theta^{\,}_{\mathrm{gFM}^{\mathrm{I}}_{z}}
\right.
\right\}
\end{equation}
with the pair of tricritical points
(see Sec.~\ref{subsubsec:Tricritical points gFMz})
labeled by
\begin{equation}
\theta^{\,}_{\mathrm{FM}^{\mathrm{I}}_{z}}\coloneqq-0.233\pm0.003
\end{equation}
and
\begin{equation}
\theta^{\,}_{\mathrm{gFM}^{\mathrm{I}}_{z}}\coloneqq+0.272\pm0.003
\end{equation}
\end{subequations}
as end points. We shall infer the critical properties of
$\mathrm{TQ}^{\mathrm{I}}_{\mathrm{gFM}^{\,}_{z}}$
from approaching the critical point
$(0,\pi/4)\in\mathrm{TQ}^{\mathrm{I}}_{\mathrm{gFM}^{\,}_{z}}$
along the vertical cut
$(0,\lambda)$ with $\lambda$ sufficiently close to $\pi/4$.
We are going to show that,
$(0,\pi/4)\in\mathrm{TQ}^{\mathrm{I}}_{\mathrm{gFM}^{\,}_{z}}$
realizes a weakly first-order quantum topological phase transition.
This property extends to all the points in
$\mathrm{TQ}^{\mathrm{I}}_{\mathrm{gFM}^{\,}_{z}}$
that we have studied.
Accordingly, we conjecture that the phase boundary
$\mathrm{TQ}^{\mathrm{I}}_{\mathrm{gFM}^{\,}_{z}}$
is a line of weakly first-order quantum topological phase transition
ending in a pair of tricritical points.

A continuous quantum phase transition between two gapped ground states
is a phase transition such that the continuum of excitations above the
ground state on both sides of the transition collapses to the ground
state at the transition point. In contrast, a discontinuous
quantum phase transition between two gapped ground states is a level
crossing at the transition point without collapse of the
continuum of excitations. If the gap at the first-order
quantum phase transition is very small,
the first-order quantum phase transition is said to be weak.
A weak first-order phase transition is one that is formally first-order
but behaves ``almost'' like a continuous one does
due to very small discontinuities
\cite{Gleiser1994a,Gleiser1994b,Gleiser1994c}.

We are going to provide numerical support
for the existence of a spectral gap above
a four-fold degenerate ground state of
$\widehat{H}^{(\mathsf{b})}_{\mathrm{I}}(0,\pi/4)$
in the thermodynamic limit.
We are then going to compute the dependence on
$\lambda$ in the vicinity of $\lambda=\pi/4$
for the Binder cumulants and show that this dependence is the one
expected of a discontinuous quantum phase transition at $\lambda=\pi/4$.

\begin{figure}
\centering
\includegraphics[width=1\columnwidth]{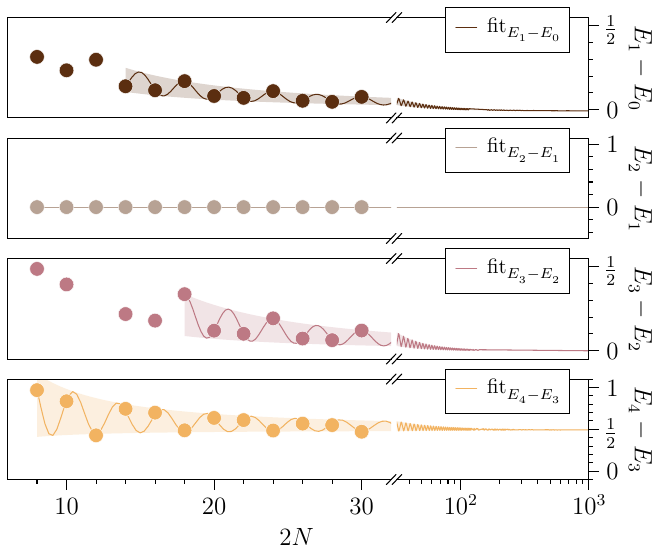}
\caption{
The case of PBCs.
Dependence on $2N\in[8,30]$
of the relevant energy differences
(\ref{eq:def Delta En+1,n (0,piover4)})
obtained from ED of
$\widehat{H}^{(0)}_{\mathrm{I}}(0,\pi/4)$.
Here, the unit of energy has been set to one.
We have used the functional form
\eqref{eq:fit enegy spacings with log corrections and oscillations}
to fit $\Delta E^{\,}_{n,n-1}(2N)$ with $n=1,2,3,4$.
The fitting parameters are given in
Table \ref{tab:TQ gFMz HA Gap}.
We show the extrapolation of the fits to values
of $2N$ up to $10^{3}$.
		}
\label{fig:TQ gFMz HA Gap}
\end{figure}

Define the energy difference
\begin{equation}
\Delta E^{\,}_{n,n-1}(2N)\coloneqq
E^{\,}_{n}(2N)-E^{\,}_{n-1}(2N)
\label{eq:def Delta En+1,n (0,piover4)}
\end{equation}
between two consecutive many-body eigenenergies of
Hamiltonian
$\widehat{H}^{(0)}_{\mathrm{I}}(0,\pi/4)$
obeying PBCs with $2N$ sites. Here
$n=1$ labels the ground state,
$n=2$ labels the first excited state,
and so on.
A preliminary study of the dependence on
$2N\leq20$ of $\Delta E^{\,}_{n,n-1}(2N)$
was done in Ref.\ \cite{Rey2025} using ED,
where it was shown that $\Delta E^{\,}_{n,n-1}(2N)$
decreases in an oscillatory way with $2N$.
Here, we find that all ED values of
$\Delta E^{\,}_{n,n-1}(2N)$ for $2N\in[8,30]$,
shown in Fig.\ \ref{fig:TQ gFMz HA Gap},
can be fitted as a function of $2N$ by the Ansatz
\begin{align}
\Delta E^{\,}_{n,n-1}(2N)=&\,
a^{\,}_{n}
+
b^{\,}_{n}\,(2N\,\ln 2N)^{-c^{\,}_{n}}
\nonumber \\
&\,
\times\left[
1
+
d^{\,}_{n}\,
\cos\left(
\frac{2\pi}{e^{\,}_{n}}\,2N+f^{\,}_{n}
\right)
\right]
\label{eq:fit enegy spacings with log corrections and oscillations}
\end{align}
with the fitting parameters
$a^{\,}_{n}$,
$b^{\,}_{n}$,
$c^{\,}_{n}$,
$d^{\,}_{n}$,
$e^{\,}_{n}$,
and
$f^{\,}_{n}$,
listed in Table \ref{tab:TQ gFMz HA Gap}
for $n=1,3,4$.

According to Fig.\ \ref{fig:TQ gFMz HA Gap},
the energy spacing
$\Delta E^{\,}_{2,1}(2N)$
vanishes within numerical accuracy for all values of $2N$,
while the energy spacings
$\Delta E^{\,}_{1,0}(2N)$
and
$\Delta E^{\,}_{3,2}(2N)$
vanish in the thermodynamic limit $2N\to\infty$
as $(2N\,\ln 2N)^{-1}$ and $(2N\,\ln 2N)^{-3/2}$,
respectively.
By contrast,
the energy spacing
$\Delta E^{\,}_{4,3}(2N)$
extrapolates in the thermodynamic limit $2N\to\infty$
to the non-vanishing value $a^{\,}_{4}=0.49\pm0.03$.
Under PBCs and if the scaling Ansatz
(\ref{eq:fit enegy spacings with log corrections and oscillations})
holds in the thermodynamic limit $2N\to\infty$, 
then
\begin{itemize}
\item[(i)] the energy gap $\Delta^{\,}_{2,1}(0,\lambda)$, that is finite in the
$\mathrm{Neel}^{\,}_{x}$ phase for $\lambda<\pi/4$ and the
$\mathrm{Neel}^{\mathrm{SPT}}_{y}$ phase for $\lambda>\pi/4$,
closes at $\lambda=\pi/4$,
\item[(ii)] the typical non-vanishing energy spacing vanishes as either
$(2N\,\ln 2N)^{-1}$ or $(2N\,\ln 2N)^{-3/2}$
as a function of $2N\to\infty$, i.e.,
as a power law with a multiplicative logarithmic correction, and
\item[(iii)] there exists the spectral energy gap
$\lim\limits_{2N\to\infty}\Delta E^{\,}_{4,3}(2N)=0.49\pm0.03$
at the transition point.
\end{itemize}

\begin{figure}
\centering
\includegraphics[width=1\columnwidth]{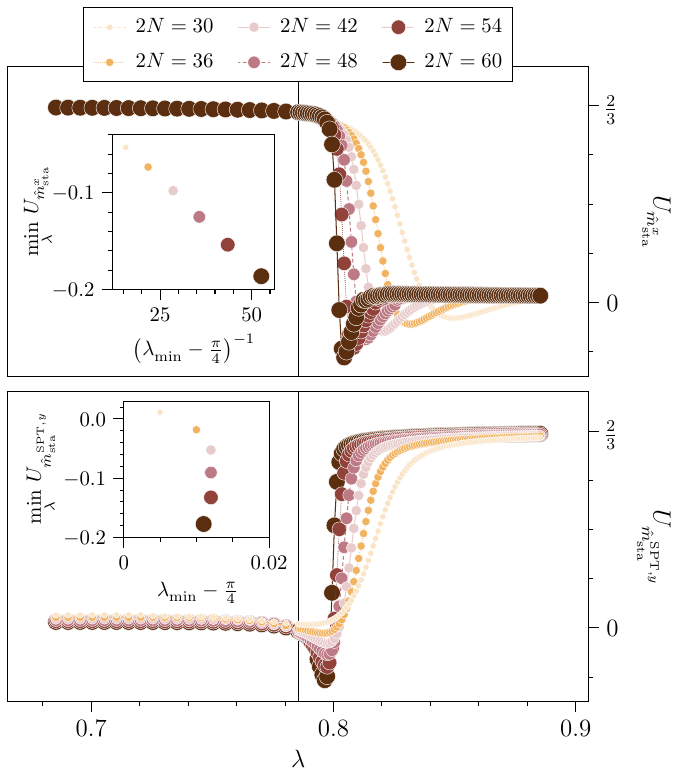}
\caption{
Case of OBCs for Hamiltonian \eqref{eq:def HA and HB c}.
Dependence on
$\pi/4-0.1\leq\lambda\leq\pi/4+0.1$ at fixed $\theta=0$
for the pair of fourth-order Binder cumulants \eqref{eq:def Binder cumulant}
associated to the pair of competing order parameters
$m^{x}_{\mathrm{sta}}$
and
$m^{\mathrm{SPT},y}_{\mathrm{sta}}$
defined in Eqs.\ \eqref{eq:def order parameters}.
The fourth-order Binder cumulants are computed
using DMRG with $2N\in[30,60]$ and $\chi=128$.
The black line indicates the location of $\lambda=\pi/4$.
Each inset shows how the position $\lambda^{\,}_{\mathrm{min}}$ and value
$\min\limits_{\lambda}U$ at which
the fourth-order Binder cumulant is minimal changes with increasing $2N$.
		}
\label{fig:Transition Neel Neel SPT HA Binder}
\end{figure}

The main message from Fig.\ \ref{fig:TQ gFMz HA Gap}
is that the existence of a spectral gap rules out the interpretation
the transition point $(0,\pi/4)$ as a continuous quantum critical point.
The study of the Binder cumulants in
Fig.\ \ref{fig:Transition Neel Neel SPT HA Binder},
in addition to
the spectral gap separating a four-fold degenerate ground state from
the continuum of excitations
(Fig.\ \ref{fig:TQ gFMz HA Gap}),
also suggests that
the transition point $(0,\pi/4)$ is a weakly first-order one.

\begin{table*}
\caption{
Fitting parameters entering Ansatz
(\ref{eq:fit enegy spacings with log corrections and oscillations}).
The values
$-8\cdot10^{3}\pm8\cdot10^{3}$
and
$-1\cdot10^{3}\pm14\cdot10^{3}$
for the fitting parameters
$a_{1}^{\,}$ and $a_{3}^{\,}$, respectively,
are zero within numerical accuracy.
The fitting parameter
$a_{4}^{\,}=0.49\pm0.03$
is non-vanishing.
	}
% %%%%%%%%%%%
% We need to put the \label AFTER \caption to avoid
% the Table to be referenced using \ref by the section number instead
% of the true Table number.
% %%%%%%%%%%%
\label{tab:TQ gFMz HA Gap}
\centering
\renewcommand{\arraystretch}{2}
\setlength{\tabcolsep}{1pt}
\begin{tabular}{lcl}
\hline\hline
$\mathrm{fit}^{\,}_{E^{\,}_{1}-E^{\,}_{0}}$ &
$=$ &
$-(8\pm8)\cdot10^{-3}+
(9\pm2)\times(2N\,\ln 2N)^{-1.08\pm0.08}\times
\left[
1+(0.39\pm0.03)\,\cos\left(
\frac{2\pi}{2.995\pm0.005}\,2N-0.08\pm0.06
\right)\right]$ \\
$\mathrm{fit}^{\,}_{E^{\,}_{3}-E^{\,}_{2}}$ &
$=$ &
$-(1\pm14)\cdot10^{-3}+
(80\pm30)\times(2N\,\ln 2N)^{-1.5\pm0.1}\times
\left[
1+(0.59\pm0.06)\,\cos\left(
\frac{2\pi}{3.00\pm0.02}\,2N-0.1\pm0.2
\right)\right]$ \\
$\mathrm{fit}^{\,}_{E^{\,}_{4}-E^{\,}_{3}}$ &
$=$ &
$0.49\pm0.03+
(5\pm1)\times(2N\,\ln 2N)^{-1.0\pm0.1}\times
\left[
1+(1.3\pm0.3)\,\cos\left(
\frac{2\pi}{3.01\pm0.02}\,2N-3.2\pm0.2
\right)\right]$ \\
\hline\hline
\end{tabular}
\end{table*}

A Binder cumulant is a dimensionless quantity used to analyze phase
transitions and critical phenomena. It provides a way to locate
critical points and study universality classes without requiring prior
knowledge of critical exponents. Let $\widehat{O}$ denote
the operator whose expectation value in the ground state
defines the order parameter at vanishing temperature.
The fourth-order Binder cumulant is defined by
\begin{equation}
U^{\,}_{\widehat{O}}\coloneqq
1
-
\frac{
\langle\widehat{O}^{4}\rangle
	 }
	 {
3
\langle\widehat{O}^{2}\rangle^{2}
	 },
\label{eq:def Binder cumulant}
\end{equation}
where the angular brackets denote the quantum statistical average,
here the ground-state expectation value. The fourth-order
Binder cumulant measures the shape (kurtosis)
of the distribution for the order parameter.
Deep in an Ising-like ordered phase, the distribution of the
order parameter is an even function of that order parameter
with two distinct peaks that converges to a double delta function
in the thermodynamic limit,
in which case
\begin{equation}
\langle\widehat{O}^{4}\rangle\approx
\langle\widehat{O}^{2}\rangle^{2}
\ \Longleftrightarrow\
U^{\,}_{\widehat{O}}\approx\frac{2}{3}.
\end{equation}
Deep in a disordered phase, the distribution of the
order parameter is Gaussian of vanishing mean, in which case
\begin{equation}
\langle\widehat{O}^{4}\rangle\approx
3\,\langle\widehat{O}^{2}\rangle^{2}
\ \Longleftrightarrow\
U^{\,}_{\widehat{O}}\approx0.
\end{equation}

If the transition point between these two limits is a continuous one,
the distribution of the order parameter
at the transition point is scale invariant in the thermodynamic limit.
It is then anticipated that
the fourth-order Binder cumulant
is a smooth and nearly monotonically decreasing function of the tuning parameter
that drives the transition from the ordered phase to the disordered phase.
At the transition point,
the fourth-order Binder cumulant is a universal number.
This position and value of the universal number can be 
obtained from a finite-size scaling analysis as follows.
For any finite system size,
the dependence on the fourth-order Binder cumulant
as a function of the tuning parameter that selects the ordered
phase for small values and the disordered phase for large values is
a smooth and nearly monotonically decreasing function.
Upon increasing the system size,
the fourth-order Binder cumulant approaches the asymptotic values
$2/3$ ($0$) for given small (large) values of the tuning parameter.
Curves for different sizes cross at one value of the tuning parameter
that defines the critical point and the associated critical value of 
the fourth-order Binder cumulant.

If the transition between these two limits is discontinuous,
the distribution of the order parameter
at the transition point is not scale invariant,
has two maxima, owing to the coexistence of the two phases
neighboring the transition point,
whose separation in probability space diverges with the system size.
In other words, the distribution broadens with system sizes, in which case
\begin{equation}
\langle\widehat{O}^{4}\rangle\gg
\langle\widehat{O}^{2}\rangle^{2}
\ \Longleftrightarrow\
U^{\,}_{\widehat{O}}\ll 0.
\end{equation}
It is then anticipated that the fourth-order Binder cumulant
is a strongly non-monotonic function of the tuning parameter
that drives the transition from the ordered phase to the disordered phase.
Close to the transition point,
the fourth-order Binder cumulant is negative with a minimum that diverges
in the thermodynamic limit.

We present in
Fig.~\ref{fig:Transition Neel Neel SPT HA Binder}
the dependence on $\lambda$ of the fourth-order Binder cumulant
(\ref{eq:def Binder cumulant})
for the competing order-parameters
\begin{equation}
\widehat{O}=
\begin{cases}
\frac{1}{2N}
\sum\limits_{j=1}^{2N}
(-1)^{j}\,
\widehat{X}^{\,}_{j},
&
\hbox{if }\mathrm{Neel}^{\,}_{x},
\\&\\
\frac{1}{2N}
\sum\limits_{j=1+\mathsf{b}}^{2N-\mathsf{b}}
(-1)^{j}\,
\widehat{Z}^{\,}_{j-1}\,
\widehat{Y}^{\,}_{j}\,
\widehat{Z}^{\,}_{j+1},
&
\hbox{if }\mathrm{Neel}^{\mathrm{SPT}}_{y},
\end{cases}
\end{equation}
along the line $(0,\lambda)$ with $\lambda$ close to $\pi/4$.
The fourth-order Binder cumulants are estimated for OBCs
($\mathsf{b=1}$) using DMRG with $2N\in[30,60]$.
We use the short-hand
notation
$U^{\,}_{\widehat{O},2N}(\lambda)$
with $2N\in[30,60]$
for any one of the two fourth-order Binder cumulants
labeled by
$\widehat{O}$.
\begin{enumerate}
\item[(i)] For $2N\lesssim48$, $U^{\,}_{\widehat{O},2N}(\lambda)$
is mostly positive and nearly monotonic.
For $2N\gtrsim48$, a pronounced negative minimum develops
and deepens with increasing $2N$.
\item[(ii)]
The interval of $\lambda$ with $U^{\,}_{\widehat{O},2N}(\lambda)<0$
narrows with increasing $2N$
and moves toward the expected critical coupling
$\lambda^{\,}_{\mathrm{c}}=\pi/4$. For example, when $2N=30$, the 
interval for which $U^{\,}_{\widehat{O},2N}(\lambda)<0$ is $[0.839, 0.875]$,
while it is $[0.802, 0.811]$ when $2N=60$.
\item[(iii)] Unlike continuous transitions, where Binder curves
intersect at a nearly size-independent point,
$U^{\,}_{\widehat{O},2N}(\lambda)$
for different values of $2N$ exhibit multiple crossings that drift with $2N$.
\item[(iv)] The value of
$U^{\,}_{\widehat{O}2N}(\lambda)$
at its absolute mimima
grows in absolute value with $2N$.
\end{enumerate}
These features point to a weakly first-order transition at
$(0,\lambda^{\,}_{\mathrm{c}}\simeq\pi/4)$.

Oscillations with the wave vector
\begin{equation}
q^{\,}_{n+1,n}\approx
\frac{2\pi}{3}
\end{equation} 
are seen as a function of the length $2N$
of the ring when computing the energy differences
$\Delta E^{\,}_{n,n-1}(2N)$ for $n=1,3,4$ 
in Fig.\ \ref{fig:TQ gFMz HA Gap}.
We have observed that the dependence on the 
separation $s$ in the two point correlation functions
$C^{\alpha}_{s^{\,}_{0},s^{\,}_{0}+s}$ with $\alpha=x,y,z$
defined in Eq.\ (\ref{eq:Calpha})
at the transition $(0,\pi/4)$ is an
exponential decay multiplied by the periodic oscillation with the wave number
\begin{align}
q^{\,}_{\widehat{X}}=\frac{2\pi}{2},
\qquad
q^{\,}_{\widehat{Y}}=\frac{2\pi}{6},
\qquad
q^{\,}_{\widehat{Z}}=\frac{2\pi}{3},
\end{align}
respectively. Here, we computed $q^{\,}_{\widehat{O}}$
by making use of iDMRG (see Appendix \ref{appsec:iDMRG})
with the identification
$\widehat{O}\equiv\widehat{X},\widehat{Y},\widehat{Z}$
in Eq.\ (\ref{eq:qratio}).

Our final interpretation of Figs.\
\ref{fig:TQ gFMz HA Gap} and
\ref{fig:Transition Neel Neel SPT HA Binder}
is that the point $(0,\pi/4)$ of Hamiltonian
\eqref{eq:def HA and HB c}
is a weakly first-order quantum phase transition,
in contrast to the pair of points
$(\pi/4,0)$
and
$(\pi/4,\pi/2)$
that were shown in Secs.\
\ref{subsubsec:The Neelx-Neely phase boundary}
and
\ref{subsubsec:The NeelSPTx-NeelSPTy phase boundary}
to each realize a continuous quantum phase transition
and the point $(0,\pi/4)$
of Hamiltonian \eqref{eq:def HA and HB d}
that will be shown in Sec.\
\ref{subsubsec:Neelx-NeelSPTx phase boundary}
to also realize a continuous quantum phase transition.
It is as if smoothness at the transition point is
enforced by either deconfined quantum criticality within the same
topological sector of SPT order
at
$(\pi/4,0)$
and
$(\pi/4,\pi/2)$
or by SPT order when there is no competing
spontaneous symmetry breaking at $(0,\pi/4)$
for Hamiltonian \eqref{eq:def HA and HB d}.
At the point $(0,\pi/4)$ of Hamiltonian
\eqref{eq:def HA and HB c}
coexistence of two competing long-range order phases of Ising character
brings about a degeneracy of
\begin{equation}
2+2=4
\end{equation}
at the point $(0,\pi/4)$ of Hamiltonian
\eqref{eq:def HA and HB c} under PBCs
in the thermodynamic limit. Finite-size corrections were too
strong to establish if this degeneracy jumped to 
\begin{equation}
4+8=12
\end{equation}
when selecting OBCs.
We shall give analytic support to this weakly first-order transition scenario
in Sec.\ \ref{sec:analytical_exploration_of_the_phase_diagrams}.

\subsubsection{The $\mathrm{Neel}^{\,}_{x}-\mathrm{Neel}^{\mathrm{SPT}}_{x}$ phase boundary}
\label{subsubsec:Neelx-NeelSPTx phase boundary}

\begin{figure*}
\centering
\includegraphics[width=1\textwidth]{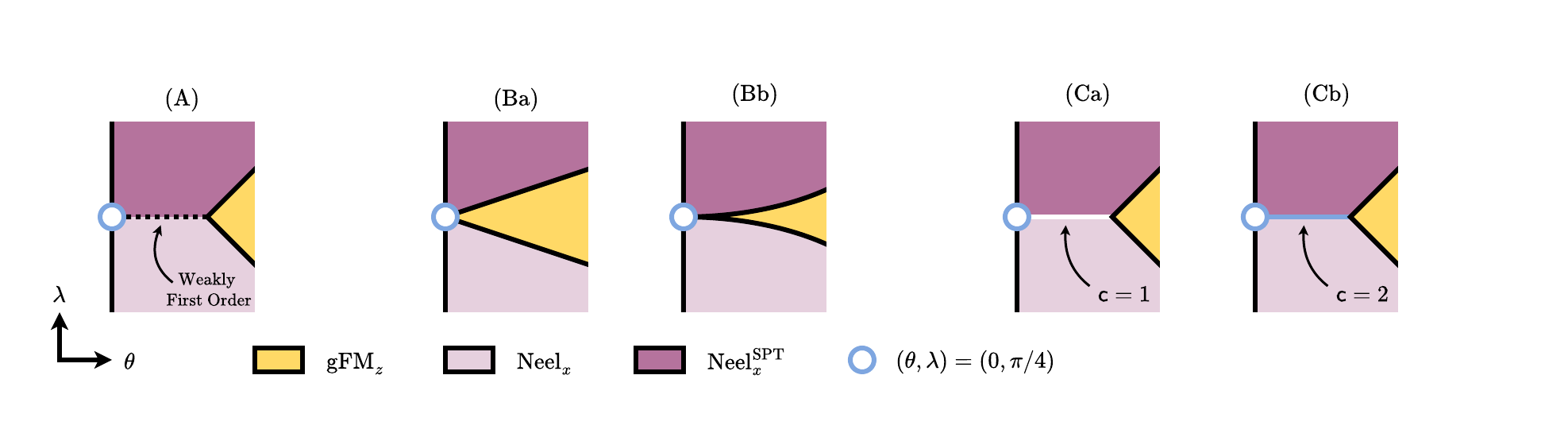}
\caption{
Schematic representation of the five scenarios considered
close to the integrable critical point $(0,\pi/4)$.
In scenario (A), the segment
$0<\theta<\theta^{\,}_{\mathrm{gFM}^{\mathrm{II}}_{z}}$ at $\lambda=\pi/4$
is a line of weak first-order transitions.
In scenario (Ba), increasing $\theta$ at fixed $\lambda=\pi/4$
takes the system directly into the $\mathrm{gFM}^{\,}_{z}$ phase.
In scenario (Bb), the same is true, but the $\mathrm{gFM}^{\,}_{z}$ phase
is bounded by two phase boundaries whose separation from $\lambda=\pi/4$
is exponentially small close to $\theta=0$ and increases with $\theta$.
In scenario (Ca), the point $(0,\pi/4)$ realizes a $\mathsf{c}=2$ CFT,
while the segment with $\theta\neq0$ is a line of
$\mathsf{c}=1$ critical points.
In scenario (Cb), the entire segment,
including the point $(0,\pi/4)$, is a line of
$\mathsf{c}=2$ critical points.
The sketches are schematic and are not meant to represent
the relative sizes of the different regions.
        }
\label{fig:TQ gFMz HB scenarios}
\end{figure*}

The phase boundary between the 
gapped long-range ordered phase
$\mathrm{Neel}^{\,}_{x}$
centered at $(0,0)$ in Fig.\ \ref{fig:phase diagram B}
and the gapped SPT phase 
$\mathrm{Neel}^{\mathrm{SPT}}_{x}$
centered at $(0,\pi/2)$ in Fig.\ \ref{fig:phase diagram B}
is only of relevance to Hamiltonian
$\widehat{H}^{(\mathsf{b})}_{\mathrm{II}}$ defined in
Eq.~\eqref{eq:def HA and HB d}.
This horizontal phase boundary in
Fig.~\ref{fig:phase diagram B}
is the open set
\begin{subequations}
\begin{equation}
\mathrm{TQ}^{\mathrm{II}}_{\mathrm{gFM}^{\,}_{z}}\coloneqq
\left\{
(\theta,\pi/4)
\ \left| \
-\theta^{\,}_{\mathrm{gFM}^{\mathrm{II}}_{z}}
<
\theta
<
\theta^{\,}_{\mathrm{gFM}^{\mathrm{II}}_{z}}
\right.
\right\}
\label{eq:Def TQ gFMz II}
\end{equation}
with the pair of tricritical points
$\pm\theta^{\,}_{\mathrm{gFM}^{\mathrm{II}}_{z}}$
(see Sec.~\ref{subsubsec:Tricritical points gFMz}),
whereby
\begin{equation}
\theta^{\,}_{\mathrm{gFM}^{\mathrm{II}}_{z}}\coloneqq0.32\pm0.1,
\end{equation}
\end{subequations}
as end points.
The fact that $\mathrm{TQ}^{\mathrm{II}}_{\mathrm{gFM}^{\,}_{z}}$ also extends
to values of $\theta<0$ on the line
$-\theta^{\,}_{\mathrm{gFM}^{\mathrm{II}}_{z}}<
\theta<\theta^{\,}_{\mathrm{gFM}^{\mathrm{II}}_{z}}$
is observed as follows.
The sequential conjugation of the Hamiltonian
\begin{subequations}
\label{eq:TQ II NeelZ Transformation}
\begin{align}
\widehat{H}^{(\mathsf{b})}_{\mathrm{II}}(\theta,\lambda)
&\xrightarrow{\widehat{U}^{\,}_{Z,\pi}}
\widehat{H}^{(\mathsf{b})}_{\mathrm{II}}(\theta+\pi,\lambda)
\\
&\xrightarrow{\widehat{U}^{\,}_{Y,\pi}}
\widehat{H}^{(\mathsf{b})}_{\mathrm{II}}(-\theta,\lambda),
\end{align}
under the unitary transformations
\eqref{eq:def UeZ} and \eqref{eq:U_Y even a}
respectively, holds for any
\begin{equation}
\theta\in[-\pi/4,0],
\qquad
\lambda\in[0,\pi/2].
\end{equation}
\end{subequations}
Consequently, the segment
\begin{equation}
(\theta,\pi/4)\in
\mathrm{TQ}^{\mathrm{II}}_{\mathrm{gFM}^{\,}_{z}},
\quad
-\theta^{\,}_{\mathrm{gFM}^{\mathrm{II}}_{z}}<\theta\leq0
\end{equation}
is unitarily equivalent to the segment
\begin{equation}
(\theta,\pi/4)\in
\mathrm{TQ}^{\mathrm{II}}_{\mathrm{gFM}^{\,}_{z}},
\quad
0\leq\theta<\theta^{\,}_{\mathrm{gFM}^{\mathrm{II}}_{z}}.
\end{equation}
It is therefore sufficient to discuss the phase diagram for $\theta\geq0$.

Hamiltonian $\widehat{H}^{(\mathsf{b})}_{\mathrm{II}}(0,\lambda)$
with $\lambda\in[0,\pi]$ can be mapped by a Jordan-Wigner
transformation to a fermionic quadratic form for any number
$2N$ of sites \cite{Verresen17,MudryChamon2025}.
It is therefore integrable.
In the thermodynamic limit, the point
\begin{equation}
(\theta,\lambda)=(0,\pi/4)
\end{equation}
realizes a critical theory with central charge
\begin{equation} \mathsf{c}=2.
\end{equation}
The question addressed in this subsection is what
becomes of this integrable $\mathsf{c}=2$ critical point
when the perturbation $\theta\neq0$
is introduced at fixed $\lambda=\pi/4$.

The five logically distinct scenarios that we consider are
summarized schematically in Fig.~\ref{fig:TQ gFMz HB scenarios}.
They can be separated into three classes.

First, in scenario (A), the segment
\begin{equation}
0<\theta<
\theta^{\,}_{\mathrm{gFM}^{\mathrm{II}}_{z}},
\qquad
\lambda=\pi/4,
\end{equation}
is a line of weak first-order transitions between
$\mathrm{Neel}^{\,}_{x}$ and $\mathrm{Neel}^{\mathrm{SPT}}_{x}$.
Such a scenario can be difficult to distinguish numerically
from a continuous transition when the correlation length
at the transition is very large.

Second, in scenarios (Ba) and (Bb),
the line $\lambda=\pi/4$ does not remain a direct
phase boundary between $\mathrm{Neel}^{\,}_{x}$ and
$\mathrm{Neel}^{\mathrm{SPT}}_{x}$ once $\theta$ is nonzero.
Instead, we enter the $\mathrm{gFM}^{\,}_{z}$ phase
for $\theta\neq0$, which implies
$\theta^{\,}_{\mathrm{gFM}^{\mathrm{II}}_{z}}=0$.
In scenario (Ba), the $\mathrm{gFM}^{\,}_{z}$ region
has a numerically resolvable extent around $\lambda=\pi/4$.
In scenario (Bb), it is bounded by two phase boundaries
whose separation from $\lambda=\pi/4$ is exponentially small
close to $\theta=0$ and grows upon increasing $\theta$.
The latter scenario is particularly difficult to test numerically.
Indeed, if the width of the intervening phase is smaller
than the resolution in $\lambda$, the two phase boundaries
may appear as a single direct transition.

Third, in scenarios (Ca) and (Cb), the line $\lambda=\pi/4$
remains a continuous phase boundary.
In scenario (Ca), the integrable point realizes a
$\mathsf{c}=2$ CFT, but any nonzero $\theta$ realizes
a $\mathsf{c}=1$ CFT in the thermodynamic limit.
In scenario (Cb), the entire line remains
described by a $\mathsf{c}=2$ CFT.

We are going to argue that the
numerical and analytical results presented below
are most naturally consistent with scenario (Ca).

\paragraph{Test of scenario (A).}

\begin{figure}
\centering
\subfigure{\includegraphics[width=1\columnwidth]{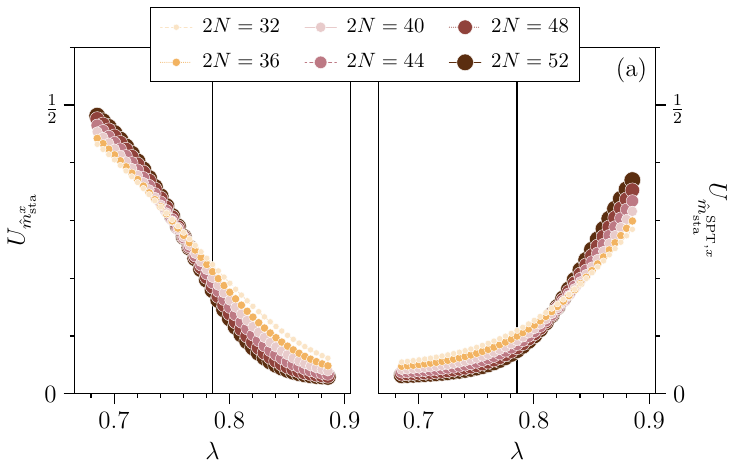} \label{fig:TQ gFMz HB Binder a}}\\
\subfigure{\includegraphics[width=1\columnwidth]{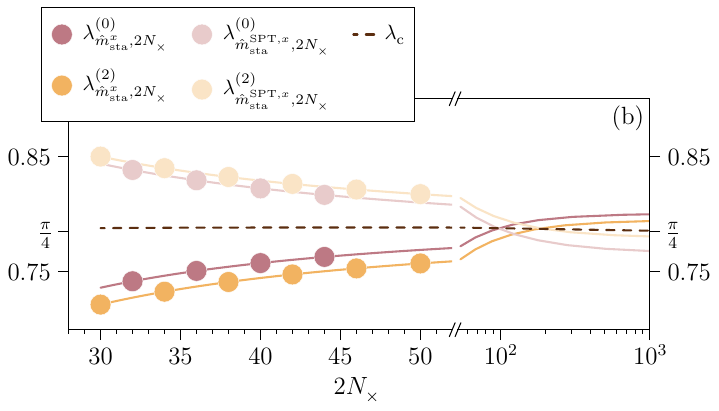} \label{fig:TQ gFMz HB Binder b}}
\caption{
Case of OBCs for Hamiltonian \eqref{eq:def HA and HB d}.
(a)
Dependence on
$\pi/4-1/10\leq\lambda\leq\pi/4+1/10$ at fixed $\theta=1/10$
for the pair of fourth-order Binder cumulants \eqref{eq:def Binder cumulant}
associated to the pair of competing order parameters
$m^{x}_{\mathrm{sta}}$
and
$m^{\mathrm{SPT},x}_{\mathrm{sta}}$
defined in Eqs.\ \eqref{eq:def order parameters}.
Presented are only (to avoid crowding)
the fourth-order Binder cumulants computed
when the number $2N$ of sites is an integer multiple of 4
using DMRG with $2N\in[32,60]$ and $\chi=128$.
The black line indicates the location of $\lambda=\pi/4$.
(b)
Dependence of the crossing points
$\lambda^{(p^{\,}_{\times})}_{\widehat{O},2N^{\,}_{\times}}$
on $2N^{\,}_{\times}$ as defined in Eq.\
(\ref{eq:def Nx and its crossing point}).
It is apparent that
(i)
$\lambda^{(p^{\,}_{\times})}_{\widehat{O},2N^{\,}_{\times}}$
is smaller (larger) than $\pi/4$ when
$\langle\widehat{O}\rangle=m^{x}_{\mathrm{sta}}$
$\big(\langle\widehat{O}\rangle=m^{\mathrm{SPT},x}_{\mathrm{sta}}\big)$
and (ii)
the dependence on $2N^{\,}_{\times}$ of
$\lambda^{(p^{\,}_{\times})}_{\widehat{O},2N^{\,}_{\times}}$
is strictly monotonic when either
$2N^{\,}_{\times}\ \mathrm{mod}\ 4=0$
or
$2N^{\,}_{\times}\ \mathrm{mod}\ 4=2$,
while it is oscillatory with a period of 4 otherwise.
We have fitted the dependence on $2N^{\,}_{\times}$
of our data for $\lambda^{(p^{\,}_{\times})}_{\widehat{O},2N^{\,}_{\times}}$
with the Ansatz
$c+a\times(2N^{\,}_{\times}\ln 2N^{\,}_{\times})^{b}$.
The extrapolation to the limit $2N^{\,}_{\times}\uparrow\infty$ gives
the estimate
$\lambda^{\,}_{\mathrm{c}}=0.785\pm0.003$
for the critical point defined in Eq.\
(\ref{eq:def lambdac from crossing of Binder}).
}
\label{fig:TQ gFMz HB Binder}
\end{figure}

A weak first-order transition is the first
alternative that must be examined.
We do so at the representative point
$(1/10,\pi/4)$
by studying the fourth-order Binder cumulants for
$m^{x}_{\mathrm{sta}}$ and $m^{\mathrm{SPT},x}_{\mathrm{sta}}$.

In this paragraph, we fix the value of
$\theta=1/10$ from now on.
Let
\begin{subequations}
\label{eq:def Nx and its crossing point}
\begin{equation}
U^{\,}_{\widehat{O},2N}(\lambda)
\end{equation}
denote the fourth-order Binder cumulant
\eqref{eq:def Binder cumulant} as a function of $\lambda$
close to $\pi/4$, for either one of
$\langle\widehat{O}\rangle=m^{x}_{\mathrm{sta}}$
or
$\langle\widehat{O}\rangle=m^{\mathrm{SPT},x}_{\mathrm{sta}}$,
when the number of sites in the chain is held fixed to the value $2N$.
Let $2N^{\,}_{\times}$ be an integer such that
\begin{equation}
2N^{\,}_{\times}\,\mathrm{mod}\
4=p^{\,}_{\times},
\qquad
p^{\,}_{\times}=0,2.
\end{equation}
We denote by
\begin{equation}
\lambda^{(p^{\,}_{\times})}_{\widehat{O},2N^{\,}_{\times}}
\end{equation}
\end{subequations}
the crossing point of
$U^{\,}_{\widehat{O},2N^{\,}_{\times}}(\lambda)$ and
$U^{\,}_{\widehat{O},2N^{\,}_{\times}+4}(\lambda)$
for the order parameter $\widehat{O}$.

Figure \ref{fig:TQ gFMz HB Binder a} shows that
(i)
$U^{\,}_{\widehat{O},2N}(\lambda)$
is a positive function of $\lambda$
when holding $2N$ fixed
and (ii)
all the functions
$U^{\,}_{\widehat{O},2N}(\lambda)$ 
cross pairwise in one point upon increasing $2N$.
We claim that these crossing points converge from below (above)
to the value $\lambda_{\mathrm{c}}^{\,}\equiv\pi/4$
when considering the order parameter
$\langle\widehat{O}\rangle=m^{x}_{\mathrm{sta}}$
$\big(\langle\widehat{O}\rangle=m^{\mathrm{SPT},x}_{\mathrm{sta}}\big)$.
We use their average,
\begin{subequations}
\label{eq:def lambdac from crossing of Binder}
\begin{equation}
\lambda^{\,}_{\mathrm{c},2N^{\,}_{\times}}
\coloneqq
\frac{1}{4}
\left[
\sum\limits_{p^{\,}_{\times}}
\sum\limits_{\widehat{O}}
\lambda^{(p^{\,}_{\times})}_{\widehat{O},2N^{\,}_{\times}}
\right],
\end{equation}
to define
\begin{equation}
\lambda^{\,}_{\mathrm{c}}
\equiv
\lim_{2N^{\,}_{\times}\uparrow\infty}
\lambda^{\,}_{\mathrm{c},2N^{\,}_{\times}},
\end{equation}
provided that this limit exists.
The extrapolation in Fig.~\ref{fig:TQ gFMz HB Binder b} gives
\begin{equation}
\lambda^{\,}_{\mathrm{c}}
=
0.785\pm0.003
\approx
\frac{\pi}{4}.
\end{equation}
\end{subequations}

We do not observe the behavior usually associated
with a first-order transition
(compare with Hamiltonian $\widehat{H}^{(\mathsf{b})}_{\mathrm{I}}$
in Fig.~\ref{fig:Transition Neel Neel SPT HA Binder}).
The Binder-cumulant analysis is instead consistent with a
single continuous transition at $\lambda=\pi/4$.
It therefore disfavors scenario (A) at $\theta=1/10$.
Because a sufficiently weak first-order transition can
remain hidden when its associated correlation length
exceeds the accessible system sizes,
this result does not constitute a strict
exclusion of scenario (A).

\paragraph{Tests of scenarios (Ba) and (Bb).}

\begin{figure}
\centering
\subfigure{\includegraphics[width=0.5\columnwidth]{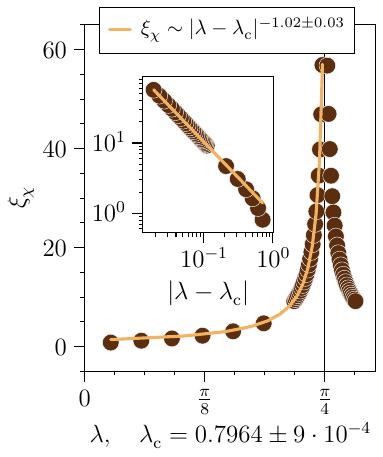} \label{fig:TQ gFMz HB Critical Exponents a}}%
\subfigure{\includegraphics[width=0.5\columnwidth]{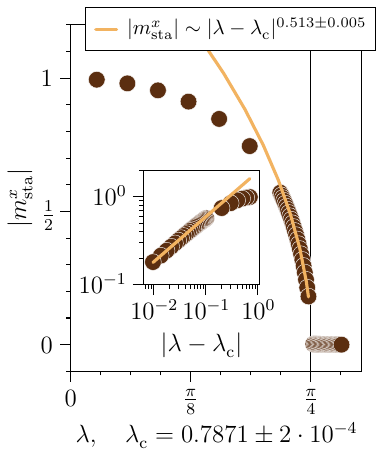} \label{fig:TQ gFMz HB Critical Exponents b}}%
\caption{
(a)
Dependence on $\lambda$ at fixed $\theta=0.1$ of
the correlation length $\xi^{\,}_{\chi}$ defined
in Eq.~\eqref{eq:xibetter} of the ground state of
Hamiltonian \eqref{eq:def HA and HB d}
obtained with iDMRG ($\chi=256$) and a unit cell of size $L=10$.
We fit $\xi^{\,}_{\chi}$ with Eq.~\eqref{eq:nu critical exponent}
to find the exponent $\nu=1.02(3)$ and the fitted location of
the phase transition $\lambda^{\,}_{\mathrm{c}}=0.7964(9)$.
The inset shows $\xi^{\,}_{\chi}$ against $\lambda-\lambda^{\,}_{\mathrm{c}}$
in a log-log plot.
(b)
Dependence on $\lambda$ at fixed $\theta=0.1$ of
the absolute value of $m^{x}_{\mathrm{sta}}$ defined
in Eq.~\eqref{eq:msta + spt} of the ground state of
Hamiltonian \eqref{eq:def HA and HB d}
obtained with iDMRG ($\chi=256$) and a unit cell of size $L=10$.
We fit $|m^{x}_{\mathrm{sta}}|$ with Eq.~\eqref{eq:beta critical exponent}
to find the exponent $\beta=0.513(5)$ and the fitted location of
the phase transition $\lambda^{\,}_{\mathrm{c}}=0.7871(2)$.
The inset shows $|m^{x}_{\mathrm{sta}}|$ against $\lambda-\lambda^{\,}_{\mathrm{c}}$
in a log-log plot.
In all panels, we indicate $\lambda=\pi/4$ by a vertical black line.
}
\label{fig:TQ gFMz HB Critical Exponents}
\end{figure}

\begin{table}
\caption{
Central charge $\mathsf{c}$,
critical exponents
$\nu$ (diverging correlation length),
$\beta$ (order parameter),
$\eta^{\alpha}$ (two-point correlation for
quantum spin $\hat{\sigma}^{\alpha}$),
$\mathsf{z}$ (dynamical exponent)
defined in
Eq.~\eqref{eq:def critical exponents}
for Hamiltonian \eqref{eq:def HA and HB d}
at the two points
$(0,\pi/4)$ and $(1/10,\pi/4)$
obtained using iDMRG with $\chi\geq256$
(ED with PBCs and $2N\leq30$ for $\mathsf{z}$).
}
\label{tab:TQ HB Critical Exponents}
\centering
\renewcommand{\arraystretch}{1.1}
\setlength{\tabcolsep}{5pt}
\begin{tabular}{c c c c c}
\hline\hline
$\theta$
&& $0$
&& $1/10$ \\
\midrule
$\mathsf{c}$
& 
& $1.99\pm0.01$
&
& $1.945\pm0.008$\\
$\nu$
& 
& $1.00\pm0.02$
&
& $1.02\pm0.03$\\
$\beta$
& 
& $0.505\pm0.006$
&
& $0.513\pm0.005$\\
$\eta^{x}$
& 
& $1.004\pm0.003$
&
& $0.921\pm0.005$\\
$\eta^{y}$
& 
& $\infty$
&
& $1.8\pm0.2$\\
$\eta^{z}$
& 
& $1.007\pm0.009$
&
& $1\pm0.1$\\
$\mathsf{z}$
& 
& $1.018\pm0.002$
&
& $1.04\pm0.02$\\
\hline\hline
\end{tabular}
\end{table}

We next test whether the point $(1/10,\pi/4)$
lies inside the $\mathrm{gFM}^{\,}_{z}$ phase,
as proposed in scenarios (Ba) and (Bb).
In both of these scenarios,
$\theta^{\,}_{\mathrm{gFM}^{\mathrm{II}}_{z}}\equiv0$.

As a first test, we have computed the
$\mathrm{gFM}^{\,}_{z}$ order parameter
$m^{z}_{\mathrm{uni}}$ at $(1/10,\pi/4)$
using DMRG with PBCs for several system sizes $2N$
and bond dimensions $\chi$,
as well as iDMRG for several bond dimensions $\chi$.
In both cases, $m^{z}_{\mathrm{uni}}$ vanishes
within numerical precision.
This observation is inconsistent with a numerically
resolved realization of either scenario (Ba) or scenario (Bb).
By itself, however, it is not sufficient to exclude
these scenarios, and scenario (Bb) in particular.

For this purpose, we study the dependence on $\lambda$
of the correlation length and
of the $\mathrm{Neel}^{\,}_{x}$ order parameter $m^{x}_{\mathrm{sta}}$.
Figure~\ref{fig:TQ gFMz HB Critical Exponents a}
shows an enhancement of the correlation length
close to $\lambda=\pi/4$ along $\theta=1/10$.
Using the Ansatz \eqref{eq:nu critical exponent}
on the $\lambda<\pi/4$ side gives
\begin{equation}
\nu=1.02\pm0.03,
\qquad
\lambda^{\,}_{\mathrm{c}}=0.7964\pm0.0009.
\end{equation}
The exponent is consistent with $\nu=1.00\pm0.02$
found in the same way at the integrable point $(0,\pi/4)$.
The fitted value of $\lambda^{\,}_{\mathrm{c}}$
differs from $\pi/4$ by approximately $1\%$.
A displacement in the same direction is also found
when the same analysis is performed at $\theta=0$.

We therefore do not interpret this fitted displacement
by itself as evidence for a phase boundary distinct from $\lambda=\pi/4$.
Similarly, fitting the order parameter in
Fig.~\ref{fig:TQ gFMz HB Critical Exponents b}
with Eq.~\eqref{eq:beta critical exponent} gives
\begin{equation}
\beta=0.513\pm0.005,
\qquad
\lambda^{\,}_{\mathrm{c}}=0.7871\pm0.0002.
\end{equation}
The exponent is consistent with $\beta=0.505(6)$
at the integrable point, while the fitted transition point
lies within approximately $0.2\%$ of $\pi/4$.

Taken together with the Binder-cumulant crossings
in Fig.~\ref{fig:TQ gFMz HB Binder},
these results are consistent with a single direct
transition close to $\lambda=\pi/4$ at $\theta=1/10$.
They disfavor scenario (Ba), in which a resolvable interval
of the $\mathrm{gFM}^{\,}_{z}$ phase would
separate the two gapped phases.

The distinction between a direct transition and
scenario (Bb) is more delicate.
In scenario (Bb), the two boundaries of the
$\mathrm{gFM}^{\,}_{z}$ phase can
be exponentially close to $\lambda=\pi/4$.
Resolving them would require a resolution in $\lambda$
smaller than the separation between the two phase boundaries,
together with correlation lengths $\xi^{\,}_{\chi}$
sufficiently large to distinguish the intervening
phase from a single critical point.
The present data show no resolved splitting of the transition,
but cannot exclude a $\mathrm{gFM}^{\,}_{z}$ interval
narrower than our numerical resolution.
However, this result taken with the numerically observed
absence of a $\mathrm{gFM}^{\,}_{z}$ order parameter
argues against the realization of scenario (Bb).

\paragraph{Tests of scenarios (Ca) and (Cb).}

\begin{figure}
\centering
\subfigure[]{\includegraphics[width=0.52\columnwidth]{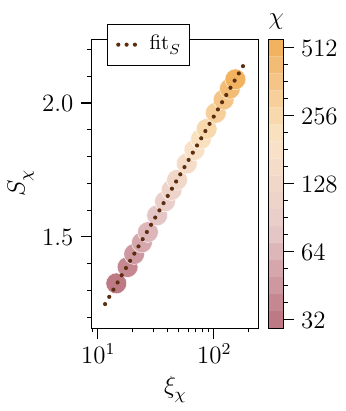}\label{fig:TQ gFMz HB iDMRG Central Charge a}}%
\subfigure[]{\includegraphics[width=0.48\columnwidth]{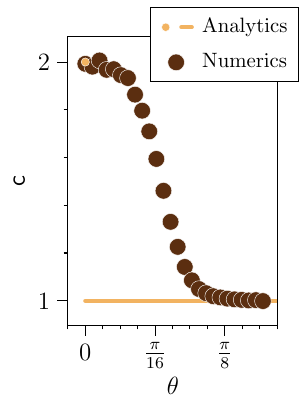} \label{fig:TQ gFMz HB iDMRG Central Charge b}}
\caption{
(a)
Dependence of the finite-entanglement entropy $S^{\,}_{\chi}$
on the MPS correlation length $\xi^{\,}_{\chi}$ at
$(\theta,\lambda)=(1/10,\pi/4)$.
The ground states are obtained with iDMRG for several
bond dimensions $\chi$, represented by different colors.
The correlation length $\xi^{\,}_{\chi}$ is defined in
Eq.~\eqref{eq:xibetter} with $L=10$.
The entropy is evaluated across the first bond of
the repeated unit cell.
The slope of the linear fit $\mathrm{fit}^{\,}_{S}$
over the range $\chi\in[256,512]$,
together with Eq.~\eqref{eq:Entropy From Xi},
gives $\mathsf{c}=1.945\pm0.008$.
(b)
Effective central charges as a function of $\theta$
obtained from the same finite-entanglement procedure at points
$(\theta,\pi/4)$ with $\theta\in[0,0.4]$.
They have converged to the value one (the yellow horizontal line)
for $\theta\gtrsim\pi/8$.
        }
\label{fig:TQ gFMz HB iDMRG Central Charge}
\end{figure}

We finally compare scenarios (Ca) and (Cb).
We argue that the combined numerical results of
Fig.~\ref{fig:TQ gFMz HB iDMRG Central Charge}
and analytical results of
Sec.~\ref{Effect of the normal-ordered quartic term in H-}
favor scenario (Ca).
We do so by studying the finite-entanglement
scaling along $\lambda=\pi/4$.
The interpretation of this scaling requires the comparison
of two length scales.

The first is the finite-entanglement correlation length
$\xi^{\,}_{\chi}$ of the ground state obtained
with iDMRG at finite bond dimension $\chi$.
Even if the exact ground state is critical,
an MPS with finite $\chi$ has a finite correlation
length $\xi^{\,}_{\chi}$.
When this correlation length provides the dominant
infrared cutoff, the finite-entanglement entropy satisfies
\begin{equation}
S^{\,}_{\chi}
=
\frac{\mathsf{c}^{\,}_{\chi}}{6}
\ln\xi^{\,}_{\chi}
+
\gamma,
\label{eq:Entropy From Xi}
\end{equation}
where $\gamma$ is a nonuniversal constant.
We denote by $\mathsf{c}^{\,}_{\chi}$ the effective central charge
extracted from the relation \eqref{eq:Entropy From Xi} because the
finite-entanglement scaling can probe a crossover regime
rather than the asymptotic infrared CFT.

The second correlation length, denoted by $\xi^{\,}_{\infty}$,
is intrinsic to the exact ground state in the thermodynamic limit.
According to scenario (Ca), the perturbation away
from the integrable point gaps one of the two
degrees of freedom of the $\mathsf{c}=2$ theory,
while the other remains gapless.
We denote by $\xi^{\,}_{\star}$ the correlation length
associated with the dynamically generated energy gap
$\Delta^{\,}_{\star}$ of the gapped degree of freedom.
It scales as
\begin{equation}
\xi^{\,}_{\star}
\sim\frac{1}{\Delta^{\,}_{\star}}
\sim\exp\left(\frac{\kappa}{|\theta|}\right),
\qquad
|\theta|\ll1,
\label{eq:Xi Infinity}
\end{equation}
with a nonuniversal constant $\kappa>0$.
Equivalently, $\xi^{\,}_{\star}$ is the length scale
beyond which the gapped degree of freedom ceases to
contribute to the critical behavior.
The exact ground state nevertheless remains critical
because of the remaining gapless degree of freedom,
whose asymptotic contribution to the entanglement
scaling is characterized by $\mathsf{c}=1$.

The one-loop RG analysis of
Sec.~\ref{Effect of the normal-ordered quartic term in H-}
is consistent with the
generation of such a length scale.
For small $\theta$, the scale at which the perturbative
flow reaches strong coupling is very large.
The corresponding length scale behaves nonperturbatively as
\eqref{eq:Xi Infinity}.
We do not attempt to determine quantitatively
the multiplicative factor in
Eq.~\eqref{eq:Xi Infinity}.
Accordingly, the comparison between $\xi^{\,}_{\chi}$ and
$\xi^{\,}_{\star}$ is used only to identify the
expected scaling regimes, and not to make a
prediction for the precise value of $\theta$
at which the numerical crossover occurs.

Within scenario (Ca), the effective central charge
measured with iDMRG depends on the relative magnitude
of $\xi^{\,}_{\chi}$ and $\xi^{\,}_{\star}$.
Three regimes can be distinguished.

When
\begin{equation}
\xi^{\,}_{\chi}\ll\xi^{\,}_{\star},
\label{eq:Regime XiChi < Xi8}
\end{equation}
the numerical ground state probes distances shorter than the scale
at which the dynamically generated gap becomes visible.
Both degrees of freedom then appear effectively gapless.
The finite-entanglement scaling is consequently governed
by the vicinity of the integrable fixed point and gives
\begin{equation}
\mathsf{c}^{\,}_{\chi}\simeq2.
\end{equation}
Thus, an effective central charge close to $2$
at small but nonzero $\theta$ does not necessarily imply
that the asymptotic theory is a $\mathsf{c}=2$ CFT.
It can instead result from an extended preasymptotic regime
in which the accessible correlation lengths $\xi^{\,}_{\chi}$ remain
smaller than $\xi^{\,}_{\star}$.

When
\begin{equation}
\xi^{\,}_{\chi}\sim\xi^{\,}_{\star},
\label{eq:Regime XiChi = Xi8}
\end{equation}
the numerical ground state begins to resolve
the gapped degree of freedom.
The finite-entanglement entropy then probes
the crossover between the short-distance regime,
in which both degrees of freedom contribute,
and the long-distance regime,
in which only one degree of freedom remains gapless.
A fit to Eq.~\eqref{eq:Entropy From Xi} can consequently
yield an effective central charge between $1$ and $2$.
Such an intermediate value should not be interpreted as
the central charge of an additional CFT.

Finally, when
\begin{equation}
\xi^{\,}_{\chi}\gg\xi^{\,}_{\star},
\label{eq:Regime XiChi > Xi8}
\end{equation}
the numerical ground state resolves the dynamically generated gap.
The gapped degree of freedom no longer contributes
to the long-distance entanglement scaling.
Only the remaining gapless degree of freedom contributes,
and the extracted central charge is expected to approach
\begin{equation}
\mathsf{c}^{\,}_{\chi}\simeq1.
\end{equation}

Scenario (Cb) also allows the effective central charge
to evolve smoothly from a value close to $2$ to a value close to $1$,
but for a different physical reason.
In this scenario, both degrees of freedom remain gapless
along the phase boundary for
\begin{equation}
0\leq|\theta|<
\theta^{\,}_{\mathrm{gFM}^{\mathrm{II}}_{z}},
\end{equation}
and this entire segment is asymptotically
described by a $\mathsf{c}=2$ CFT.
There is therefore no finite correlation length
$\xi^{\,}_{\star}$ associated with a dynamically
generated gap within this segment.
Once the finite-entanglement scaling regime is reached,
increasing $\xi^{\,}_{\chi}$ should continue to probe
two gapless degrees of freedom and yield $\mathsf{c}^{\,}_{\chi}\simeq2$.

At $\theta=\theta^{\,}_{\mathrm{gFM}^{\mathrm{II}}_{z}}\approx0.32$,
the system enters the $\mathrm{gFM}^{\,}_{z}$ phase.
This phase contains one gapless degree of freedom and
is characterized by an asymptotic central charge $\mathsf{c}=1$.
Hence, in scenario (Cb), the change from
$\mathsf{c}=2$ to $\mathsf{c}=1$ occurs upon crossing
a phase transition at a nonzero value
$\theta^{\,}_{\mathrm{gFM}^{\mathrm{II}}_{z}}\approx0.32$
of $\theta$.
At finite bond dimension, this transition is rounded
over a finite interval of $\theta$, producing an apparently
smooth drift of $\mathsf{c}^{\,}_{\chi}$ between $2$ and $1$.

Distinguishing numerically scenarios (Ca) and (Cb)
is facilitated if regime \eqref{eq:Regime XiChi > Xi8}
occurs for values
of $0<\theta\ll\theta^{\,}_{\mathrm{gFM}^{\mathrm{II}}_{z}}$,
which is not what we observe numerically.

Figure~\ref{fig:TQ gFMz HB iDMRG Central Charge a}
shows the finite-entanglement scaling at the representative point
$(\theta,\lambda)=(1/10,\pi/4)$.
The dependence of $S^{\,}_{\chi}$ on $\ln\xi^{\,}_{\chi}$
is accurately described by Eq.~\eqref{eq:Entropy From Xi}.
The extracted effective central charge remains close to $2$.
Within scenario (Ca), this result is consistent with 
the regime \eqref{eq:Regime XiChi < Xi8} at $\theta=1/10$.
The numerical simulation would then probe primarily
the preasymptotic regime governed by the vicinity of
the integrable $\mathsf{c}=2$ fixed point.

Figure~\ref{fig:TQ gFMz HB iDMRG Central Charge b}
shows the evolution of the effective central charge
$\mathsf{c}^{\,}_{\chi}$ upon increasing $\theta$.
Close to the integrable point, the estimate remains close to $\mathsf{c}=2$
[regime \eqref{eq:Regime XiChi < Xi8} in scenario (Ca)].
It then departs progressively from $2$ and drifts toward $\mathsf{c}=1$
[regime \eqref{eq:Regime XiChi = Xi8} in scenario (Ca)].
For sufficiently large $\theta$, here $\theta\gtrsim0.39$,
the estimate becomes numerically close to $1$
[regime \eqref{eq:Regime XiChi > Xi8} in scenario (Ca)].
However, as regime \eqref{eq:Regime XiChi > Xi8}
in scenario (Ca) overlaps with the rounding of a step function
at $\theta^{\,}_{\mathrm{gFM}^{\mathrm{II}}_{z}}$
in scenario (Cb) ($0.39>\theta^{\,}_{\mathrm{gFM}^{\mathrm{II}}_{z}}\approx0.32$)
the drift observed in
Fig.~\ref{fig:TQ gFMz HB iDMRG Central Charge b}
does not distinguish unambiguously
scenarios (Ca) and (Cb) by itself.
We favor scenario (Ca) over scenario (Cb)
because the observed crossover in 
Fig.~\ref{fig:TQ gFMz HB iDMRG Central Charge b}
is not centered about $\theta^{\,}_{\mathrm{gFM}^{\mathrm{II}}_{z}}\approx0.32$
and because we found that $\theta>0$ induces marginally
relevant perturbations to the $\mathsf{c}=2$ free-Majorana CFT at
$\theta=0$ within a one-loop RG analysis.

In summary,
combining the finite-entanglement scaling with the other
numerical diagnostics and the RG analysis favors scenario (Ca).
The Binder cumulants and critical scaling are consistent
with a direct transition along $\lambda=\pi/4$,
while the $\mathrm{gFM}^{\,}_{z}$ order parameter vanishes
within numerical precision at $(1/10,\pi/4)$.
Moreover, the RG flow is marginally relevant for any $0<|\theta|\ll1$
and generates a nonperturbative strong-coupling scale.
It therefore implies that the integrable $\mathsf{c}=2$
fixed point does not control the asymptotic infrared behavior
for $0<|\theta|\ll1$, even though an apparent $\mathsf{c}=2$
regime can persist over very large but finite length scales.
Together with the numerical drift toward $\mathsf{c}^{\,}_{\chi}\simeq1$,
this instability is consistent with an asymptotic theory
containing one remaining gapless degree of freedom,
as proposed in scenario (Ca).
The one-loop RG analysis reveals the marginal instability
of the point $(0,\pi/4)$ towards a strong-coupling theory
where half of the critical modes are dynamically gapped,
reducing the central charge from $\mathsf{c}=2$ to $\mathsf{c}=1$.
This is scenario (Ca).

\paragraph{Critical attributes.}

The critical attributes obtained at
$(0,\pi/4)$ and $(0.1,\pi/4)$
are summarized in Table~\ref{tab:TQ HB Critical Exponents}.
At the integrable point, the measured values are consistent
with their exact values.
In particular, the two-point functions behave as
\eqref{eq:eta critical exponent} with nonuniversal amplitudes
\begin{subequations}
\begin{align}
A^{x}_{s}[\Psi]
&\propto(-1)^{s},
\\
A^{y}_{s}[\Psi]
&=0,
\\
A^{z}_{s}[\Psi]
&\propto
\begin{cases}
1,
&
s\equiv0\bmod 4,
\\
0,
&
s\not\equiv0\bmod 4.
\end{cases}
\end{align}
\end{subequations}
Compared with Eqs.~\eqref{eq:eta critical exponent}
and \eqref{eq:measured amplitudes As at (0,thetac)},
the symmetry under the exchange of $x$ and $y$ is maximally broken.
Moreover, the algebraic correlations of the $z$ component
are nonvanishing only when the separation is an
integer multiple of four lattice sites.

At $(1/10,\pi/4)$, the estimates of $\nu$, $\beta$, and $\mathsf{z}$
remain close to those at the integrable point $(0,\pi/4)$.
By contrast, the change of the correlation-function exponents
is more difficult to quantify.
In particular, the uncertainties in $\eta^{y}$ and $\eta^{z}$
are approximately $10\%$.
We therefore do not use the apparent proximity of the
measured exponents at $\theta=0$ and $\theta=1/10$ as evidence
that both points belong to the same $\mathsf{c}=2$ fixed line.
Such a proximity can also occur when the accessible distances
$\xi^{\,}_{\chi}$ remain shorter than the crossover length
$\xi^{\,}_{\infty}$ away from the integrable point.

\paragraph{Wave vectors along $\lambda=\pi/4$.}

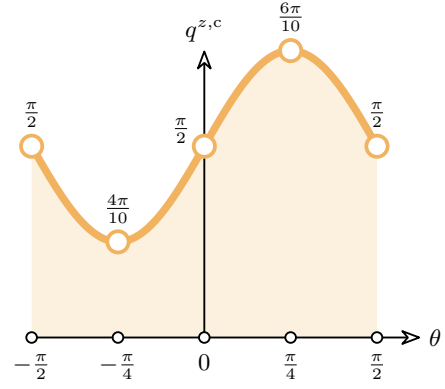
\begin{figure}
\centering
\begin{tikzpicture}
\begin{axis}[
	scale=0.8,
    domain=-pi:pi,
    samples=200,
    axis lines=none,
    ticks=none,
    clip=false
]

% --- filled region under the sine ---
\addplot [
	name path=sin,
	draw=none
] {sin(deg(x))};

\addplot [
	name path=bottom,
	draw=none
] {-2};

\addplot [
	fill=color!20
] fill between[
	of=sin and bottom
];

% --- vertical axis theta = 0 ---
\draw[-{Stealth[length=3mm, open, round]}, line width=0.25mm] (axis cs:0,-2) -- (axis cs: 0, 1)
	node[above] {$q^{z,\mathrm{c}}$};

% --- horizontal axis at q = -2 ---
\draw[-{Stealth[length=3mm, open, round]}, line width=0.25mm] (axis cs:-pi,-2) -- (axis cs: pi+pi/4, -2)
	node[right] {$\theta$};

% --- manual x labels ---
\filldraw[white, draw=black, line width=0.25mm]
    (axis cs: -pi, -2) circle[radius=2pt];
\filldraw[white, draw=black, line width=0.25mm]
    (axis cs: -pi/2, -2) circle[radius=2pt];
\filldraw[white, draw=black, line width=0.25mm]
    (axis cs: 0, -2) circle[radius=2pt];
\filldraw[white, draw=black, line width=0.25mm]
    (axis cs: pi/2, -2) circle[radius=2pt];
\filldraw[white, draw=black, line width=0.25mm]
    (axis cs: pi, -2) circle[radius=2pt];

\node[below=3pt] at (axis cs: -pi, -2) {$-\frac{\pi}{2}$};
\node[below=3pt] at (axis cs: -pi/2, -2) {$-\frac{\pi}{4}$};
\node[below=3pt] at (axis cs: 0, -2) {$0$};
\node[below=3pt] at (axis cs: pi/2, -2) {$\frac{\pi}{4}$};
\node[below=3pt] at (axis cs: pi, -2) {$\frac{\pi}{2}$};

% --- sine curve (draw almost LAST, on top of almost everything) ---
\addplot [
	line width=1mm,
	color=color
] {sin(deg(x))};

% --- labels ---
\filldraw[white, draw=color, line width=0.5mm]
    (axis cs: -pi, 0) circle[radius=4pt];
\filldraw[white, draw=color, line width=0.5mm]
    (axis cs: -pi/2, -1) circle[radius=4pt];
\filldraw[white, draw=color, line width=0.5mm]
    (axis cs: 0, 0) circle[radius=4pt];
\filldraw[white, draw=color, line width=0.5mm]
    (axis cs: pi/2, 1) circle[radius=4pt];
\filldraw[white, draw=color, line width=0.5mm]
    (axis cs: pi, 0) circle[radius=4pt];

\node[above=4pt]  at (axis cs:-pi, 0) {$\frac{\pi}{2}$};
\node[above=4pt]  at (axis cs:-pi/2, -1) {$\frac{4\pi}{10}$};
\node[above=6pt, left=2pt] at (axis cs: 0, 0) {$\frac{\pi}{2}$};
\node[above=4pt] at (axis cs: pi/2, 1) {$\frac{6\pi}{10}$};
\node[above=4pt] at (axis cs: pi, 0) {$\frac{\pi}{2}$};

\end{axis}
\end{tikzpicture}
\caption{\label{fig:TQ gFMz HB q}
Wavevector $q^{z,\mathrm{c}}$ that encodes the oscillations of the
two-point function
$C^{z,\mathrm{c}}_{s^{\,}_{0},s^{\,}_{0}+s}$
defined Eq.\ (\ref{eq:CalphaConn}), obtained numerically
with iDMRG for Hamiltonian
\eqref{eq:def HA and HB d}
at $\lambda=\pi/4$.
The lattice spacing $\mathfrak{a}$ has been set to unity.
        }
\end{figure}

Independently of the preceding identification of the infrared CFT,
iDMRG allows us to resolve the oscillatory behavior of
the two-point functions $C^{x}_{s^{\,}_{0},s^{\,}_{0}+s}[\Psi]$
and $C^{z,\mathrm{c}}_{s^{\,}_{0},s^{\,}_{0}+s}[\Psi]$.
Our numerical results indicate that their wave vectors
are best described by
\begin{subequations}
\label{eq:wave vector for Cx and Cz,c fixed lambda piover4}
\begin{align}
q^{x}(\theta)
&=
\frac{2\pi}{2\mathfrak{a}},
\qquad
\theta\in[-\pi/4,\pi/4],
\label{eq:wave vector for Cx and Cz,c fixed lambda piover4 a}
\\
q^{z,\mathrm{c}}(\theta)
&=
\frac{2\pi}{4\mathfrak{a}}
+
\frac{2\pi}{20\mathfrak{a}}\,
\sin2\theta,
\qquad
\theta\in[-\pi/2,\pi/2].
\label{eq:wave vector for Cx and Cz,c fixed lambda piover4 b}
\end{align}
\end{subequations}
The wave vector $q^{x}$ remains commensurate.
By contrast, $q^{z,\mathrm{c}}$ varies continuously with $\theta$.
Indeed, its dependence is accurately fitted by a single sine harmonic.
Equation \eqref{eq:wave vector for Cx and Cz,c fixed lambda piover4 b}
is inferred from the commensurate values marked by circles in
Fig.~\ref{fig:TQ gFMz HB q} and from the unitary relations discussed below.
Moreover, for $0\leq|\theta|\ll1$, the dependence inferred numerically
\eqref{eq:wave vector for Cx and Cz,c fixed lambda piover4 b}
strongly agrees with the analytical result as discussed in
Sec.~\ref{ssub:comparison_with_numerics}.

At $\theta=\pi/4$,
Eq.~\eqref{eq:wave vector for Cx and Cz,c fixed lambda piover4 b}
gives
\begin{equation}
q^{z,\mathrm{c}}\left(\frac{\pi}{4}\right)
=
2\pi\times\frac{3}{10}
=
\frac{6\pi}{10},
\end{equation}
in units such that $\mathfrak{a}=1$.
This agrees with the wave vector at
$(\pi/4,\pi/4)\in\mathrm{gFM}^{\,}_{z}$ discussed in
Sec.~\ref{ssub:the_texorpdfstring_mathrm_gfm}, as required by
\begin{equation}
\widehat{H}^{(\mathsf{b})}_{\mathrm{II}}
(\pi/4,\pi/4)
=
\widehat{H}^{(\mathsf{b})}_{\mathrm{I}}
(\pi/4,\pi/4).
\end{equation}

It is sufficient to determine $q^{z,\mathrm{c}}(\theta)$
on the interval $\theta\in[0,\pi/4]$.
The remaining intervals follow from the unitary transformations
of the Hamiltonian.
The segment $\theta\in[\pi/4,\pi/2]$ is related to
$\theta\in[0,\pi/4]$ by Eq.~\eqref{eq:action U_R on Hamiltonians}.
Similarly, the segment $\theta\in[-\pi/2,-\pi/4]$ is related to
$\theta\in[-\pi/4,0]$ by Eqs.~\eqref{eq:action U_R on Hamiltonians}
and \eqref{eq:conjugation by UeZ shifts theta by pi}.
Finally, the action of $\widehat{U}^{\,}_{Y,\pi}$ on
$C^{z,\mathrm{c}}_{s^{\,}_{0},s^{\,}_{0}+s}$ introduces
a factor $(-1)^{s}$.
It follows that
\begin{equation}
q^{z,\mathrm{c}}\left(+\frac{\pi}{4}\right)
=
\frac{6\pi}{10}
\quad\Longrightarrow\quad
q^{z,\mathrm{c}}\left(-\frac{\pi}{4}\right)
=
\frac{4\pi}{10}.
\label{eq:symmetry implicatons for q wave vector}
\end{equation}
Indeed, according to the Ansatz \eqref{eq:gFMz OBCs algebraic decay cosine},
the oscillatory part is encoded by
\begin{equation}
\cos\left(
q^{z,\mathrm{c}}s+\phi^{z,\mathrm{c}}
\right).
\end{equation}
The factor $(-1)^{s}$ can be absorbed into the wave vector since
\begin{equation}
(-1)^{s}
\cos\left(q^{z,\mathrm{c}}s\right)
=
\cos\left(q^{z,\mathrm{c}}s+\pi
s\right).
\end{equation}
Hence,
\begin{align}
q^{z,\mathrm{c}}\left(-\frac{\pi}{4}\right)
&\mapsto
q^{z,\mathrm{c}}\left(+\frac{\pi}{4}\right)
+\pi
\mod 2\pi
\nonumber\\
&=
2\pi\times\frac{8}{10}
\mod 2\pi
\nonumber\\
&=
-2\pi\times\frac{2}{10}
\mod 2\pi.
\end{align}
Because the cosine is even, the sign of its wave vector
can be reversed, giving
\begin{equation}
q^{z,\mathrm{c}}\left(-\frac{\pi}{4}\right)
=
\frac{4\pi}{10}
\mod
2\pi,
\end{equation}
which proves Eq.~\eqref{eq:symmetry implicatons for q wave vector}.

\begin{figure}
\centering
\subfigure{\includegraphics[width=0.5\columnwidth]{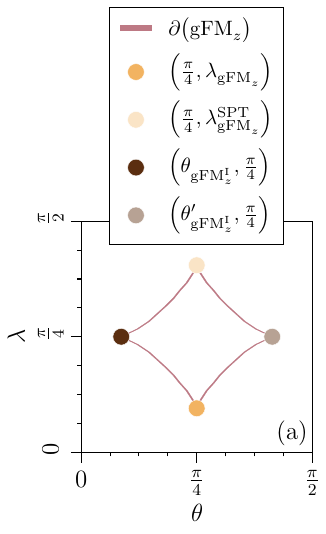} \label{fig:gFMz HA+HB Boundary Location a}}%
\subfigure{\includegraphics[width=0.5\columnwidth]{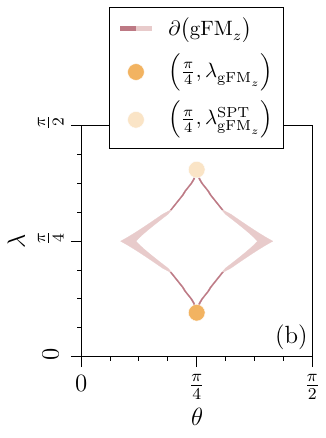} \label{fig:gFMz HA+HB Boundary Location b}}
\caption{
(a)
The case of OBC.
Dependence on $\theta$ and $\lambda$ of the boundary of the
$\mathrm{gFM}^{\,}_{z}$ phase of Hamiltonian
\eqref{eq:def HA and HB c}
within the region \eqref{eq:coupling quadrants b}
of the coupling space.
This phase boundary is obtained using DMRG
($\chi=128$) by
locating the points in \eqref{eq:coupling quadrants b}
for which the dependence on $\theta$ holding $\lambda$ fixed
of the binder cumulants
$U^{\,}_{\hat{m}^{z}_{\mathrm{uni}},2N}$ for $2N=64$ and $2N=128$ cross.
The width of the shaded area represents
sizes of the windows in $\theta$ in which
the curves for smaller systems sizes $2N<64$
crosses the curves for $2N=128$.
(b)
The case of OBC.
Dependence on $\theta$ and $\lambda$ of the boundary of the
$\mathrm{gFM}^{\,}_{z}$ phase of Hamiltonian
\eqref{eq:def HA and HB d}
within the region \eqref{eq:coupling quadrants d}
of the coupling space.
This phase boundary is obtained using DMRG
($\chi=32$ and $2N=128$) by
locating the points in \eqref{eq:coupling quadrants b}
for which the dependence on $\theta$ holding $\lambda$ fixed
of the binder cumulants
$U^{\,}_{\hat{m}^{z}_{\mathrm{uni}},2N}$ for $2N=32$ and $2N=64$ cross.
The width of the shaded area represents
sizes of the windows in $\theta$ in which
the curves for smaller systems sizes $2N<32$
crosses the curves for $2N=64$.
}
\label{fig:gFMz HA+HB Boundary Location}
\end{figure}

\begin{figure}
\centering
\includegraphics[width=1\columnwidth]{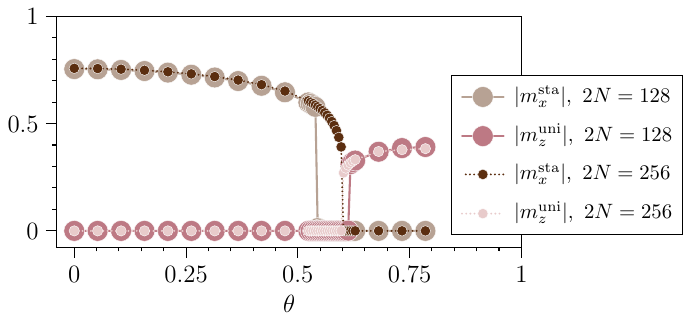}
\caption{
The case of OBCs.
Magnitudes of the relevant order parameters in
Eq.~\eqref{eq:def order parameters}
of Hamiltonian \eqref{eq:def HA and HB d}
obtained using DMRG with
$\lambda\simeq0.576$, $\chi=128$ for $2N=128$ and $2N=256$ sites.
The sharp suppression
of the order parameters on both sides of the quantum phase transition
around $\theta=0.58$ indicates
that the correlation lengths in the gapped phases are
larger than the length of the chain
sufficiently close to the phase boundary.
}
\label{fig:gFMz + NeelZ HB Boundary Order Parameters}
\end{figure}

\subsubsection{The $\mathrm{gFM}^{\,}_{z}$ phase boundary}
\label{subsubsec:The Neelx-gFMz phase boundary}

The numerical estimate for the
boundary of the $\mathrm{gFM}^{\,}_{z}$ phase of
Hamiltonian $\widehat{H}^{(\mathsf{b})}_{\mathrm{I}}$
is reported in
Fig.~\ref{fig:gFMz HA+HB Boundary Location a}.
Similarly, the numerical estimate for the
boundary of the $\mathrm{gFM}^{\,}_{z}$ phase of
Hamiltonian $\widehat{H}^{(\mathsf{b})}_{\mathrm{II}}$
is reported in
Fig.~\ref{fig:gFMz HA+HB Boundary Location b}.
The boundary of the
$\mathrm{gFM}^{\,}_{z}$ phase
for Hamiltonian $\widehat{H}^{(\mathsf{b})}_{\mathrm{I}}$
is obtained from exiting it into
the $\mathrm{Neel}^{\,}_{x}$ or $\mathrm{Neel}^{\mathrm{SPT}}_{y}$ phases,
without loss of generality.
The boundary of the
$\mathrm{gFM}^{\,}_{z}$ phase
for Hamiltonian $\widehat{H}^{(\mathsf{b})}_{\mathrm{II}}$
is obtained from exiting it into
the $\mathrm{Neel}^{\,}_{x}$ or $\mathrm{Neel}^{\mathrm{SPT}}_{x}$ phases,
without loss of generality.

The transition point upon exiting the gapless
$\mathrm{gFM}^{\,}_{z}$ phase into a gapful one is located with
either one of the two following procedures,
as they agree within their numerical accuracies.
Without loss of generality, we consider exiting
the $\mathrm{gFM}^{\,}_{z}$ phase into the $\mathrm{Neel}^{\,}_{x}$ phase.
\begin{itemize}
\item[(I)] We locate the points in the region
$\Omega^{\,}_{\mathrm{Y,B}}$ for which the
$\mathrm{Neel}^{\,}_{x}$ order parameter $m^{x}_{\mathrm{sta}}[\Psi]$
and the $\mathrm{gFM}^{\,}_{z}$ order parameter $m^{z}_{\mathrm{uni}}[\Psi]$
vanish simultaneously
for the MPS state $|\Psi\rangle$ of bond length $\chi$
approximating the ground state,
holding both the number $2N$ of sites in the chain
and the coupling $\lambda$ fixed.
This criterion delivers the interval
\begin{equation}
\theta\in
\left[
\theta^{\,}_{\mathrm{c}}-(\delta\theta^{\,}_{\mathrm{c}})^{\,}_{2N,\chi},
\theta^{\,}_{\mathrm{c}}+(\delta\theta^{\,}_{\mathrm{c}})^{\,}_{2N,\chi}
\right]
\end{equation}
of size $2(\delta\theta^{\,}_{\mathrm{c}})^{\,}_{2N,\chi}$
along the horizontal cut fixed by $\lambda$
for which the
order parameters
$m^{x}_{\mathrm{sta}}[\Psi]$
and 
$m^{z}_{\mathrm{uni}}[\psi]$
both vanish. Provided $\chi$ is chosen sufficiently large
so that (i) the size $2(\delta\theta^{\,}_{\mathrm{c}})^{\,}_{2N,\chi}$
of this interval decreases monotonically with
increasing $2N$ and (ii) the extrapolated limiting value
\begin{equation}
\lim_{2N\to\infty} 2(\delta\theta^{\,}_{\mathrm{c}})^{\,}_{2N,\chi}=0,
\end{equation}
we have identified the continuous quantum critical point
$\theta^{\,}_{\mathrm{c}}$.
\item[(II)] We locate the points in the region
$\Omega^{\,}_{\mathrm{Y,B}}$ for which the
Binder curves $U^{\,}_{\widehat{O},2N}$, defined 
in Eq.~\eqref{eq:def Binder cumulant},
cross for different system sizes $2N$.
Here, the relevant operator $\widehat{O}$
could a priori be chosen to be either the one whose expectation value is
the $\mathrm{Neel}^{\,}_{x}$ order parameter 
$m^{x}_{\mathrm{sta}}[\Psi]\equiv\langle\Psi|\hat{m}^{x}_{\mathrm{sta}}|\Psi\rangle$
or the one whose expectation value is
the $\mathrm{gFM}^{\,}_{z}$ order parameter
$m^{z}_{\mathrm{uni}}[\Psi]\equiv\langle\Psi|\hat{m}^{z}_{\mathrm{uni}}|\Psi\rangle$.
However, due to the fact that the $\mathrm{gFM}^{\,}_{z}$ phase
is critical, we only use the fourth-order Binder cumulant
$U^{\,}_{\hat{m}^{z}_{\mathrm{uni}},2N}$
to detect the boundary of the $\mathrm{gFM}^{\,}_{z}$ phase.
\end{itemize}

An example of the procedure to locate the boundary of the
$\mathrm{gFM}^{\,}_{z}$ phase
of Hamiltonian $\widehat{H}^{(\mathsf{b})}_{\mathrm{II}}$
along $\lambda\approx0.576<\pi/4$ is given in
Fig.~\ref{fig:gFMz + NeelZ HB Boundary Order Parameters}.

In Fig.~\ref{fig:gFMz HA+HB Boundary Location b},
we color the boundary of the $\mathrm{gFM}^{\,}_{z}$ phase
of Hamiltonian $\widehat{H}^{(\mathsf{b})}_{\mathrm{II}}$
as dark and light purple segments.
The dark purple segment are obtained for values of
$\lambda$ such that
(i) incommensuration effects are negligible and
(ii) the correlation length $\xi\ll 2N$ in the
$\mathrm{Neel}^{\,}_{x}$ phase.
The light purple segment was obtained for values of
$\lambda$ where (i) and (ii) do not hold
\footnote{
Two Binder curves for different $2N$ can then
cross at two or more points
due to the presence of strong incommensuration effects.
     }.

It is known from Ref.~\cite{Rey2025} that any point in the boundary
of the $\mathrm{gFM}^{\,}_{z}$ phase of Hamiltonian
$\widehat{H}^{(\mathsf{b})}_{\mathrm{I}}$
realizes a CFT with central charge $\mathsf{c}=1$.
We have verified that this is also so along the boundary
of the $\mathrm{gFM}^{\,}_{z}$ phase of Hamiltonian
$\widehat{H}^{(\mathsf{b})}_{\mathrm{II}}$.

\begin{figure}
\centering
\subfigure{\includegraphics[width=0.5\columnwidth]{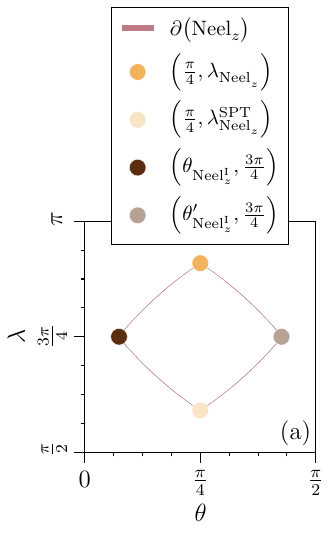} \label{fig:NeelZ HA+HB Boundary Location a}}%
\subfigure{\includegraphics[width=0.5\columnwidth]{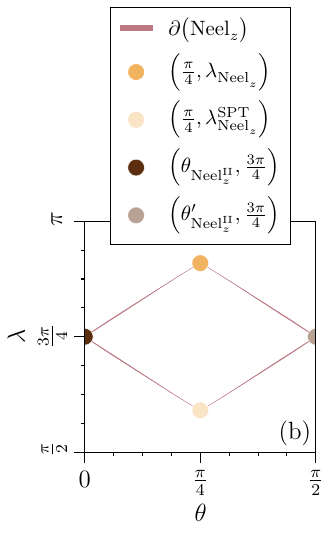} \label{fig:NeelZ HA+HB Boundary Location b}}
\caption{\label{fig:NeelZ HA+HB Boundary Location}
The case of PBCs. 
(a)
Dependence on $\theta$ and $\lambda$ of the boundary of the
$\mathrm{Neel}^{\,}_{z}$ phase of Hamiltonian
\eqref{eq:def HA and HB c}
within the region \eqref{eq:coupling quadrants b}
of the coupling space.
The boundary is found for each fixed value of $\lambda$
as outlined in Eq.\ (\ref{eq:procedure to find boundary Neelz}).
The width of the shaded area represents the
uncertainty in the extracted value
of $\theta^{\,}_{\star}(\lambda)$.
The dots locate
the tricritical points, as computed in
Sec.~\ref{subsubsec:Tricritical points gFMz}.
(b)
Same as (a) but for Hamiltonian
\eqref{eq:def HA and HB d}.
Two of the four tricritical point in (a) become
tetracritical points.
     }
\end{figure}

\begin{figure}
\centering
\subfigure{\includegraphics[width=0.5\columnwidth]{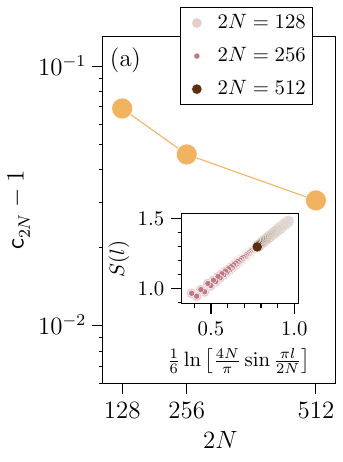} \label{fig:NeelZ HA HB OBCs Boundary Central Charge a}}%
\subfigure{\includegraphics[width=0.5\columnwidth]{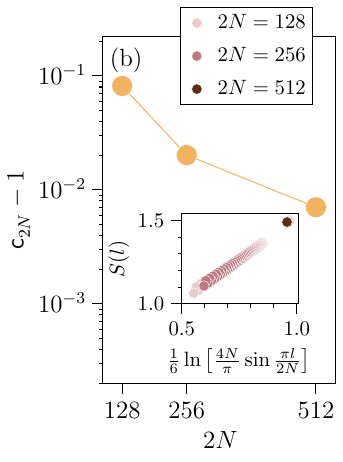} \label{fig:NeelZ HA HB OBCs Boundary Central Charge b}}
\caption{
The case of OBCs.
Dependence of the central charge $\mathsf{c}^{\,}_{2N}$
defined in Eq.\ (\ref{eq:CFT prediction for S2N(l) b})
on the number $2N$ of sites
in the ground state
(a)
of Hamiltonian
\eqref{eq:def HA and HB c}
and
(b)
of Hamiltonian \eqref{eq:def HA and HB d}
at $\lambda\approx2.56>3\pi/4$
when approximated by the matrix product state (MPS)
obtained using DMRG ($\chi=256$).
The inset shows the 
dependence on the number $2N$ of sites of the bipartite
entanglement entropy $S^{\,}_{\mathrm{MPS},2N}({l})$
in the MPS state.
The slope of the linear fit performed for each system
size $2N$ gives the central charge $\mathsf{c}$
defined in Eq.\ (\ref{eq:CFT prediction for S2N(l) a}).
}
\label{fig:NeelZ HA HB OBCs Boundary Central Charge}
\end{figure}

\subsubsection{The $\mathrm{Neel}^{\,}_{z}$ phase boundary}
\label{subsubsec:The FMx-Neelz phase boundary}

The numerical estimate for the
boundary of the $\mathrm{Neel}^{\,}_{z}$ phase of
Hamiltonian $\widehat{H}^{(\mathsf{b})}_{\mathrm{I}}$
is reported in
Fig.~\ref{fig:NeelZ HA+HB Boundary Location a}.
Similarly, the numerical estimate for the
boundary of the $\mathrm{Neel}^{\,}_{z}$ phase of
Hamiltonian $\widehat{H}^{(\mathsf{b})}_{\mathrm{II}}$
is reported in
Fig.~\ref{fig:NeelZ HA+HB Boundary Location b}.
The boundary of the
$\mathrm{Neel}^{\,}_{z}$ phase
for Hamiltonian $\widehat{H}^{(\mathsf{b})}_{\mathrm{I}}$
is obtained from exiting it into
the $\mathrm{FM}^{\,}_{z}$ or $\mathrm{Neel}^{\mathrm{SPT}}_{y}$ phases,
without loss of generality.
The boundary of the
$\mathrm{Neel}^{\,}_{z}$ phase
for Hamiltonian $\widehat{H}^{(\mathsf{b})}_{\mathrm{II}}$
is obtained from exiting it into
the $\mathrm{FM}^{\,}_{x}$ or $\mathrm{Neel}^{\mathrm{SPT}}_{x}$ phases,
without loss of generality.

The boundaries of the
$\mathrm{Neel}^{\,}_{z}$ phase
for both Hamiltonians
$\widehat{H}^{(\mathsf{b})}_{\mathrm{I}}$
and
$\widehat{H}^{(\mathsf{b})}_{\mathrm{II}}$
are obtained through the same procedure.
This procedure relies on ED as the phase boundaries always separate
gapped phases. Without loss of generality, we treat the case of
the boundary of the
$\mathrm{Neel}^{\,}_{z}$ phase
for Hamiltonian $\widehat{H}^{(\mathsf{b})}_{\mathrm{I}}$.

First, we impose PBCs and compute the low-energy
spectrum of $\widehat{H}^{(0)}_{\mathrm{I}}(\theta,\lambda)$
defined in Eq.~\eqref{eq:def HA and HB c},
for system sizes $2N\in[6,20]$,
for $0\leq\theta\leq\pi/4$,
and for $\lambda^{\,}_{\mathrm{Neel}^{\,}_{z}}<\lambda<3\pi/4$.
Here, $\lambda^{\,}_{\mathrm{Neel}^{\,}_{z}}=
0.2849\pm0.0004$ (see Sec.~\ref{subsubsec:Tricritical points gFMz}).
Second, for each $2N\in[6,20]$ and for each $\lambda$ held fixed,
we define
\begin{subequations}
\label{eq:procedure to find boundary Neelz}
\begin{equation}
\theta^{\,}_{\star,2N}(\lambda)
\end{equation}
to be the value of $\theta\in[0,\pi/4]$ 
for which the energy spacing
$E^{\,}_{2}-E^{\,}_{1}$ is the smallest.
Third, we extrapolate the value of
$\theta^{\,}_{\star,2N}(\lambda)$
by taking the thermodynamic limit $2N\to\infty$,
which, if it exists, gives the transition point
\begin{equation}
\theta^{\,}_{\star}(\lambda)\coloneqq
\lim\limits_{2N\to\infty}
\theta^{\,}_{\star,2N}(\lambda).
\end{equation}
Such $\theta^{\,}_{\star}(\lambda)$ for
$\lambda^{\,}_{\mathrm{Neel}^{\,}_{z}}<\lambda<3\pi/4$
are shown as the boundary of the $\mathrm{Neel}^{\,}_{z}$ phase
in Fig.~\ref{fig:NeelZ HA+HB Boundary Location a}.
Finally, we have verified for selected points along
the boundary of the $\mathrm{Neel}^{\,}_{z}$ phase of Hamiltonians
$\widehat{H}^{(\mathsf{b})}_{\mathrm{I}}$
and
$\widehat{H}^{(\mathsf{b})}_{\mathrm{II}}$
that the transition points each
realizes a CFT with central charge
\begin{equation}
\mathsf{c}=1.
\end{equation}
\end{subequations}

For instance, we report in
Fig.~\ref{fig:NeelZ HA HB OBCs Boundary Central Charge a}
the value of the central charge of Hamiltonian
\eqref{eq:def HA and HB c}
and in 
Fig.~\ref{fig:NeelZ HA HB OBCs Boundary Central Charge b}
the value of the central charge of Hamiltonian
\eqref{eq:def HA and HB d}
at the intersection between the horizontal cut
at $\lambda\approx2.566$ and the boundary of the
$\mathrm{Neel}^{\,}_{z}$ phase.
The value of the central charge $\mathsf{c}$
has been determined using
Eq.~\eqref{eq:CFT prediction for S2N(l)}.

\begin{figure}
\centering
\subfigure[]{\includegraphics[width=0.5\columnwidth]{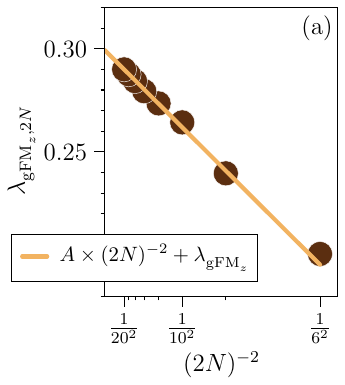} \label{fig:gFMz + NeelZ ED PBCs Tricritical Location a}}%
\subfigure[]{\includegraphics[width=0.5\columnwidth]{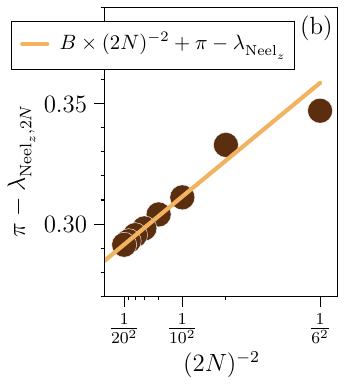} \label{fig:gFMz + NeelZ ED PBCs Tricritical Location b}}
\caption{
Case with PBCs.
(a) 
Dependence on $(2N)^{-2}$
with $2N$ taking the values $2N=6,\cdots,20$
of $\lambda^{\,}_{\mathrm{gFM}^{\,}_{z},2N}$ obtained with ED under PBCs at 
$\theta=\pi/4$. We define $\lambda^{\,}_{\mathrm{gFM}^{\,}_{z},2N}$ for each 
$2N$ as the value of $\lambda$ for which
the energy difference $E^{\,}_{1}-E^{\,}_{0}$ changes from 
$E^{\,}_{1}-E^{\,}_{0}\sim10^{-1}$ to $E^{\,}_{1}-E^{\,}_{0}<10^{-10}$. 
We extrapolate the value of
$\lambda^{\,}_{\mathrm{gFM}^{\,}_{z},2N}$
in the thermodynamic limit $2N\to\infty$,
where $\lambda^{\,}_{\mathrm{gFM}^{\,}_{z}}=0.2993\pm0.0009$.
(b)
Dependence on $(2N)^{-2}$
with $2N$ taking the values $2N=6,\cdots,20$
of $\lambda^{\,}_{\mathrm{Neel}^{\,}_{z},2N}$ obtained with ED under PBCs at 
$\theta=\pi/4$. Here, $\lambda^{\,}_{\mathrm{Neel}^{\,}_{z}}$
is defined to be the value of $\lambda$
along $\theta=\pi/4$ that minimizes $\Delta\varepsilon^{\,}_{*}(\lambda)$.
For each $\lambda$, we define $\Delta\varepsilon^{\,}_{*}(\lambda)$
the energy window
for which the number of energy eigenvalues in
above the ground-state energy jumps from
one for $\Delta\varepsilon<\Delta\varepsilon^{\,}_{*}(\lambda)$,
to four for $\Delta\varepsilon\geq\Delta\varepsilon^{\,}_{*}(\lambda)$.
We extrapolate the value of
$\lambda^{\,}_{\mathrm{Neel}^{\,}_{z},2N}$
to the thermodynamic limit $2N\to\infty$. This extrapolation gives
$\lambda^{\,}_{\mathrm{Neel}^{\,}_{z}}=\pi-0.2849\pm0.0004$.
}
\label{fig:gFMz + NeelZ ED PBCs Tricritical Location}
\end{figure}

\begin{table*}
\caption{
Locations of all
multicritical points for Hamiltonians 
$\widehat{H}^{(0)}_{\mathrm{I}}$ and $\widehat{H}^{(0)}_{\mathrm{II}}$.
    }
% %%%%%%%%%%%
% We need to put the \label AFTER \caption to avoid
% the Table to be referenced using \ref by the section number instead
% of the true Table number.
% %%%%%%%%%%%
\label{tab:Summary Multicritical Location}
\centering
\begin{tabular}{c|c|c|c}
\hline\hline
\textbf{Phase} &
\textbf{Hamiltonian} &
\textbf{Location $(\theta,\lambda)$} &
\textbf{Value}
\\ \hline\hline

\multirow{6}{*}{$\mathrm{gFM}_{z}$}

& I, II
& $(\pi/4,\lambda^{\,}_{\mathrm{gFM}_{z}})$
& $\lambda^{\,}_{\mathrm{gFM}^{\,}_{z}}
=0.2993(9)$
\\ \cline{2-4}

& I, II
& $(\pi/4,\lambda^{\mathrm{SPT}}_{\mathrm{gFM}^{\,}_{z}})$
& $\lambda^{\mathrm{SPT}}_{\mathrm{gFM}^{\,}_{z}}
=\tfrac{\pi}{2}-\lambda^{\,}_{\mathrm{gFM}^{\,}_{z}}$
\\ \cline{2-4}

& I
& $(\theta_{\mathrm{gFM}^{\mathrm{I}}_{z}},\pi/4)$
& $\theta_{\mathrm{gFM}^{\mathrm{I}}_{z}}
=0.272(3)$
\\ \cline{2-4}

& I
& $(\theta^{\prime}_{\mathrm{gFM}^{\mathrm{I}}_{z}},\pi/4)$
& $\theta^{\prime}_{\mathrm{gFM}^{\mathrm{I}}_{z}}
=\tfrac{\pi}{2}-\theta_{\mathrm{gFM}^{\mathrm{I}}_{z}}$
\\ \cline{2-4}

& II
& $(\theta^{\,}_{\mathrm{gFM}^{\mathrm{II}}_{z}},\pi/4)$
& $\theta^{\,}_{\mathrm{gFM}^{\mathrm{II}}_{z}}
=0.32\pm0.1$
\\ \cline{2-4}

& II
& $(\theta^{\prime}_{\mathrm{gFM}^{\mathrm{II}}_{z}},\pi/4)$
& $\theta^{\prime}_{\mathrm{gFM}^{\mathrm{II}}_{z}}
=\tfrac{\pi}{2}-\theta_{\mathrm{gFM}^{\mathrm{II}}_{z}}$
\\ \hline\hline

\multirow{6}{*}{$\mathrm{Neel}^{\,}_{z}$}

& I, II
& $(\pi/4,\lambda^{\,}_{\mathrm{Neel}^{\,}_{z}})$
& $\lambda^{\,}_{\mathrm{Neel}^{\,}_{z}}
=\pi-0.2849(4)$
\\ \cline{2-4}

& I, II
& $(\pi/4,\lambda^{\mathrm{SPT}}_{\mathrm{Neel}^{\,}_{z}})$
& $\lambda^{\mathrm{SPT}}_{\mathrm{Neel}^{\,}_{z}}
=\tfrac{3\pi}{2}-\lambda^{\,}_{\mathrm{Neel}^{\,}_{z}}$
\\ \cline{2-4}

& I
& $(\theta^{\,}_{\mathrm{Neel}^{\mathrm{I}}_{z}},3\pi/4)$
& $\theta^{\,}_{\mathrm{Neel}^{\mathrm{I}}_{z}}
=0.233(3)$
\\ \cline{2-4}

& I
& $(\theta^{\prime}_{\mathrm{Neel}^{\mathrm{I}}_{z}},3\pi/4)$
& $\theta^{\prime}_{\mathrm{Neel}^{\mathrm{I}}_{z}}
=\tfrac{\pi}{2}-\theta_{\mathrm{Neel}^{\mathrm{I}}_{z}}$
\\ \cline{2-4}

& II
& $(\theta_{\mathrm{Neel}^{\mathrm{II}}_{z}},3\pi/4)$
& $\theta_{\mathrm{Neel}^{\mathrm{II}}_{z}}=0$
\\ \cline{2-4}

& II
& $(\theta^{\prime}_{\mathrm{Neel}^{\mathrm{II}}_{z}},3\pi/4)$
& $\theta^{\prime}_{\mathrm{Neel}^{\mathrm{II}}_{z}}
=\tfrac{\pi}{2}-\theta_{\mathrm{Neel}^{\mathrm{II}}_{z}}$
\\ \hline\hline
\end{tabular}
\end{table*}

\subsection{Multicritical points}
\label{subsec:Multicritical points}

Each of the phases $\mathrm{gFM}^{\,}_{z}$
and $\mathrm{Neel}^{\,}_{z}$ 
for Hamiltonians
\eqref{eq:def HA and HB c} and \eqref{eq:def HA and HB d}
are $4$-gons whose vertices are multicritical points.
Their positions are reported in
Table \ref{tab:Summary Multicritical Location},
as we now explain.

\subsubsection{Tricritical points of the $\mathrm{gFM}^{\,}_{z}$ phase}
\label{subsubsec:Tricritical points gFMz}

Along $\theta=\pi/4$, for any $\lambda\in[0,2\pi]$
and for any boundary conditions $\mathsf{b}=0,1$,
Hamiltonian
\eqref{eq:def HA and HB c}
equals Hamiltonian
\eqref{eq:def HA and HB d}.
In agreement with Ref.~\cite{Rey2025},
we have found numerically that
the $\mathrm{gFM}^{\,}_{z}$ phase
terminates at two tricritical points along the line $\theta=\pi/4$,
as we now explain.

When PBCs are chosen, we can reduce the coupling space
\eqref{eq:coupling quadrants a}
of the Hamiltonians
\eqref{eq:def HA and HB c} and \eqref{eq:def HA and HB d}
to the region \eqref{eq:coupling space PBCs bottom}.
In this region,
we have found the location of the tricritical point
\begin{subequations}
\label{eq:tricritical gFMz lambda}
\begin{equation}
\bigl(\tfrac{\pi}{4},\lambda^{\,}_{\mathrm{gFM}^{\,}_{z}}\bigr)\in
\Omega^{\,}_{\mathrm{Y,B^{\downarrow}}},
\end{equation}
that separates the
$\mathrm{Neel}^{\,}_{x}$,
$\mathrm{Neel}^{\,}_{y}$,
and $\mathrm{gFM}^{\,}_{z}$ phases at
\footnote{
If we denote by $\lambda^{\,}_{0}$ the value of
the parameter $\lambda$ of the Hamiltonian
in Ref.~\cite{Rey2025} where the tricritical
point is located, its value can be found to be 
$$
\cot\lambda^{\,}_{\mathrm{gFM}^{\,}_{z}}=
(1-\lambda^{\,}_{0})/\lambda^{\,}_{0}
\quad\iff\quad
\lambda^{\,}_{0}\simeq0.235.
$$
    }
\begin{equation}
\lambda^{\,}_{\mathrm{gFM}^{\,}_{z}}=
0.2993\pm0.0009,
\label{eq:lambda gFMz}
\end{equation}
\end{subequations}
as is indicated in Fig.~\ref{fig:phase diagram U1}.
The numerical value \eqref{eq:lambda gFMz}
is found through the scaling analysis with the system size $2N$
that is reported in
Fig.~\ref{fig:gFMz + NeelZ ED PBCs Tricritical Location a}.
This scaling analysis entails four steps.

First, we compute the low-energy
spectrum of $\widehat{H}^{(0)}_{\mathrm{U}(1)}(\lambda)$
defined in Eq.~\eqref{eq:def H U1} under PBCs
for $0\leq\lambda\leq\pi/4$ and for system sizes $2N\in[6,20]$.
The energy eigenvalues are ordered and denoted
$E_{0}^{\,}\leq E_{1}^{\,}\leq\cdots$.

Second, the integer
$\mathcal{N}^{\,}_{2N}(\Delta\varepsilon)$
counts the number of eigenstates in the window of energy
$[E^{\,}_{0},E^{\,}_{0}+\Delta\varepsilon]$ for a chain made of $2N$ sites.

Third, setting $\Delta\varepsilon=10^{-10}$, we search for
$0\leq\lambda^{\,}_{\mathrm{gFM}^{\,}_{z},2N}\leq\pi/4$ such that
\begin{equation}
\mathcal{N}^{\,}_{2N}(10^{-10})=
\begin{cases}
1, & \hbox{ if } \lambda<\lambda^{\,}_{\mathrm{gFM}^{\,}_{z},2N},
\\
2, & \hbox{ if } \lambda\geq\lambda^{\,}_{\mathrm{gFM}^{\,}_{z},2N}.
\end{cases}
\end{equation}

Finally, we extrapolate the values of
$\lambda^{\,}_{\mathrm{gFM}^{\,}_{z},2N}$
to the thermodynamic limit $2N\to\infty$,
\begin{equation}
\lambda^{\,}_{\mathrm{gFM}^{\,}_{z}}\coloneqq
\lim\limits_{2N\to\infty}
\lambda^{\,}_{\mathrm{gFM}^{\,}_{z},2N}.
\end{equation}

Owing to the existence of the unitary transformation
$\widehat{U}^{(0)}_{\mathrm{CZ}}$ defined in Eq.~\eqref{eq:U_CZ},
one finds that the second vertex of the $\mathrm{gFM}^{\,}_{z}$ phase
is located at
$(\pi/4,\lambda^{\mathrm{SPT}}_{\mathrm{gFM}^{\,}_{z}})$
where
\begin{equation}
\lambda^{\mathrm{SPT}}_{\mathrm{gFM}^{\,}_{z}}=
\tfrac{\pi}{2}-
\lambda^{\,}_{\mathrm{gFM}^{\,}_{z}},
\end{equation}
as is shown in Fig.~\ref{fig:phase diagram U1}.

We now locate the remaining two tricritical points
for the $\mathrm{gFM}^{\,}_{z}$ phase of Hamiltonians
\eqref{eq:def HA and HB c} and
\eqref{eq:def HA and HB d}.
Independently of the choice between OBCs and PBCs,
we can reduce the coupling space
\eqref{eq:coupling quadrants a}
of the Hamiltonians
\eqref{eq:def HA and HB c} and \eqref{eq:def HA and HB d}
to the region \eqref{eq:coupling quadrants left a}.
In this region,
we have found the location of the pair of tricritical points
\begin{subequations}
\begin{align}
&\bigl(\theta^{\,}_{\mathrm{gFM}^{\mathrm{I}}_{z}},\tfrac{\pi}{4}\bigr)\in
\Omega^{\,}_{\mathrm{Y,B,L}},
\\ 
&\bigl(\theta^{\,}_{\mathrm{gFM}^{\mathrm{II}}_{z}},\tfrac{\pi}{4}\bigr)\in
\Omega^{\,}_{\mathrm{Y,B,L}},
\end{align}
where
\begin{align}
&\theta^{\,}_{\mathrm{gFM}^{\mathrm{I}}_{z}}=
0.272\pm0.003,
\label{eq:theta gFMz HA}
\\
&\theta^{\,}_{\mathrm{gFM}^{\mathrm{II}}_{z}}=
0.32\pm0.1,
\label{eq:theta gFMz HB}
\end{align}
for Hamiltonian
$\widehat{H}^{(\mathsf{b})}_{\mathrm{I}}$ and
$\widehat{H}^{(\mathsf{b})}_{\mathrm{II}}$, respectively.
\end{subequations}

The numerical values
\eqref{eq:theta gFMz HA}
and
\eqref{eq:theta gFMz HB}
were obtained by computing the fourth-order Binder cumulants
as was done in 
Figs.
\ref{fig:gFMz HA+HB Boundary Location a}
and
\ref{fig:gFMz HA+HB Boundary Location b},
respectively.
The uncertainties of these
values are larger than those along the vertical line $\theta=\pi/4$
as the finite-size corrections and incommensuration effects
are stronger along the
horizontal line $\lambda=\pi/4$ than along
the vertical line $\theta=\pi/4$,
especially for Hamiltonian $\widehat{H}^{(\mathsf{b})}_{\mathrm{II}}$.

\begin{figure}
\centering
\includegraphics[width=1\columnwidth]{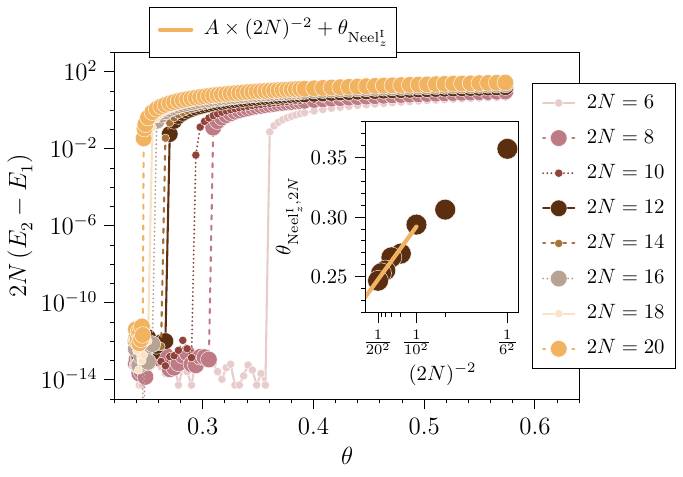}
\caption{\label{fig:NeelZ HA ED PBCs Tricritical Location}
The case of PBCs.
Dependence on $\theta$
of the energy difference $E^{\,}_{2}-E^{\,}_{1}$
above the ground-state energy $E^{\,}_{0}$
of the Hamiltonian $\widehat{H}^{(0)}_{\mathrm{I}}$
along $\lambda=3\pi/4$ obtained with ED
for different values of $2N=6,\cdots,20$.
The inset shows the dependence on $(2N)^{-2}$ of
$\theta^{\,}_{\mathrm{Neel}^{\mathrm{I}}_{z},2N}$
defined for each $2N$ as the value of $\theta$ for which
the energy difference $E^{\,}_{2}-E^{\,}_{1}$ changes from 
$E^{\,}_{2}-E^{\,}_{1}<10^{-10}$ to $E^{\,}_{2}-E^{\,}_{1}\sim10^{-1}$.
We extrapolate the value of
$\theta^{\,}_{\mathrm{Neel}^{\mathrm{I}}_{z},2N}$
to the thermodynamic limit $2N\to\infty$,
where $\theta^{\,}_{\mathrm{Neel}^{\mathrm{I}}_{z}}=0.233\pm0.003$.
     }
\end{figure}

\subsubsection{Tricritical points of the $\mathrm{Neel}^{\,}_{z}$ phase}
\label{subsubsec:Tricritical points NeelZ}

Along $\theta=\pi/4$, for any $\lambda\in[0,2\pi]$
and for any boundary conditions $\mathsf{b}=0,1$,
Hamiltonian
\eqref{eq:def HA and HB c}
equals Hamiltonian
\eqref{eq:def HA and HB d}.
We have found numerically that the $\mathrm{Neel}^{\,}_{z}$ phase
terminates at two tricritical points along the line $\theta=\pi/4$,
as we now explain.

When PBCs are chosen, we can reduce the coupling space
\eqref{eq:coupling quadrants b}
of the Hamiltonians
\eqref{eq:def HA and HB c} and \eqref{eq:def HA and HB d}
to the region \eqref{eq:coupling space PBCs top}.
In this region,
we have found the location of the tricritical point
\begin{subequations}
\begin{equation}
\bigl(\tfrac{\pi}{4},\lambda^{\,}_{\mathrm{Neel}^{\,}_{z}}\bigr)\in
\Omega^{\,}_{\mathrm{Y,T^{\uparrow}}},
\end{equation}
that separates the
$\mathrm{FM}^{\,}_{x}$,
$\mathrm{FM}^{\,}_{y}$ and
$\mathrm{Neel}^{\,}_{z}$ phases at
\begin{equation}
\lambda^{\,}_{\mathrm{Neel}^{\,}_{z}}=
\pi-0.2849\pm0.0004.
\label{eq:lambda NeelZ}
\end{equation}
\end{subequations}
as is indicated in Fig.~\ref{fig:phase diagram U1}.
The numerical value (\ref{eq:lambda NeelZ})
is found through the scaling analysis with the system size $2N$ that
is reported in Fig.\
\ref{fig:gFMz + NeelZ ED PBCs Tricritical Location b}.
This scaling analysis entails four steps.

First, we impose PBCs and compute the low-energy
spectrum of $\widehat{H}^{(0)}_{\mathrm{U}(1)}(\lambda)$
defined in Eq.~\eqref{eq:def H U1},
for $3\pi/4\leq\lambda\leq\pi$ and for system sizes $2N\in[6,20]$.
The energy eigenvalues are ordered and denoted
$E_{0}^{\,}\leq E_{1}^{\,}\leq\cdots$.

Second, the integer
$\mathcal{N}^{\,}_{2N}(\Delta\varepsilon)$
counts the number of eigenstates in the window of energy
$[E^{\,}_{0},E^{\,}_{0}+\Delta\varepsilon]$ for a chain made of $2N$ sites.

Third, we define for each $\lambda\in[3\pi/4,\pi]$
the window of energy
$\Delta\varepsilon^{\,}_{\star}(\lambda)$ that obeys
\begin{equation}
\mathcal{N}^{\,}_{2N}(\Delta\varepsilon)=
\begin{cases}
1, & \hbox{ if } \Delta\varepsilon<\Delta\varepsilon^{\,}_{\star}(\lambda),
\\
4, & \hbox{ if } \Delta\varepsilon\geq\Delta\varepsilon^{\,}_{\star}(\lambda).
\end{cases}
\end{equation}
We then define $\lambda^{\,}_{\mathrm{Neel}^{\,}_{z},2N}$
to be the value of $\lambda$ such that
\begin{equation}
\Delta\varepsilon^{\,}_{\star}(\lambda^{\,}_{\mathrm{Neel}^{\,}_{z},2N})
\leq
\Delta\varepsilon^{\,}_{\star}(\lambda),
\quad
3\pi/4\leq\lambda\leq\pi.
\end{equation}

Finally, we extrapolate the values of
$\lambda^{\,}_{\mathrm{Neel}^{\,}_{z},2N}$
to the thermodynamic limit $2N\to\infty$,
\begin{equation}
\lambda^{\,}_{\mathrm{Neel}^{\,}_{z}}\coloneqq
\lim\limits_{2N\to\infty}
\lambda^{\,}_{\mathrm{Neel}^{\,}_{z},2N}.
\end{equation}

Owing to the existence of the unitary transformation
$\widehat{U}^{(0)}_{\mathrm{CZ}}$ defined in Eq.~\eqref{eq:U_CZ},
one finds that the second vertex of the $\mathrm{Neel}^{\,}_{z}$ phase
is located at
$\bigl(\tfrac{\pi}{4},\lambda^{\mathrm{SPT}}_{\mathrm{Neel}^{\,}_{z}}\bigr)$
where
\begin{equation}
\lambda^{\mathrm{SPT}}_{\mathrm{Neel}^{\,}_{z}}=
\tfrac{3\pi}{2}-
\lambda^{\,}_{\mathrm{Neel}^{\,}_{z}},
\end{equation}
as is shown in Fig.~\ref{fig:phase diagram U1}.

We now locate the remaining two tricritical points
for the $\mathrm{Neel}^{\,}_{z}$ phase of Hamiltonians
\eqref{eq:def HA and HB c}.
Independently of the choice between OBCs and PBCs,
we can reduce the coupling space
\eqref{eq:coupling quadrants b}
of the Hamiltonian
\eqref{eq:def HA and HB c}
to the region \eqref{eq:coupling quadrants left b}.
In this region,
we have found the location of the tricritical point
\begin{subequations}
\begin{equation}
\bigl(\theta^{\,}_{\mathrm{Neel}^{\mathrm{I}}_{z}},\tfrac{3\pi}{4}\bigr)\in
\Omega^{\,}_{\mathrm{Y,T,L}},
\end{equation}
that separates the
$\mathrm{Neel}^{\mathrm{SPT}}_{y}$,
$\mathrm{FM}^{\,}_{z}$,
and $\mathrm{Neel}^{\,}_{z}$ phases at
\begin{equation}
\theta^{\,}_{\mathrm{Neel}^{\mathrm{I}}_{z}}=
0.233\pm0.003.
\label{eq:theta NeelZ HA}
\end{equation}
for Hamiltonian
$\widehat{H}^{(\mathsf{b})}_{\mathrm{I}}$.
\end{subequations}
The numerical value (\ref{eq:theta NeelZ HA})
is found through the scaling analysis with the system size $2N$ that
is reported in Fig.\
\ref{fig:NeelZ HA ED PBCs Tricritical Location}.
This scaling analysis entails the same four steps
as the ones below Eq.\ (\ref{eq:lambda NeelZ}).

Owing to the existence of the unitary transformation
$\widehat{U}^{\,}_{\mathrm{R}}$ defined in Eq.~\eqref{eq:U_R},
one finds that the last vertex of the $\mathrm{Neel}^{\,}_{z}$ phase
of Hamiltonian \eqref{eq:def HA and HB c} is located at
$\bigl(\theta^{\prime}_{\mathrm{Neel}^{\mathrm{I}}_{z}},\tfrac{3\pi}{4}\bigr)$
where
\begin{equation}
\theta^{\prime}_{\mathrm{Neel}^{\mathrm{I}}_{z}}=
\tfrac{\pi}{2}-
\theta^{\,}_{\mathrm{Neel}^{\mathrm{I}}_{z}}.
\end{equation}

\begin{figure}
\centering
\includegraphics[width=1\columnwidth]{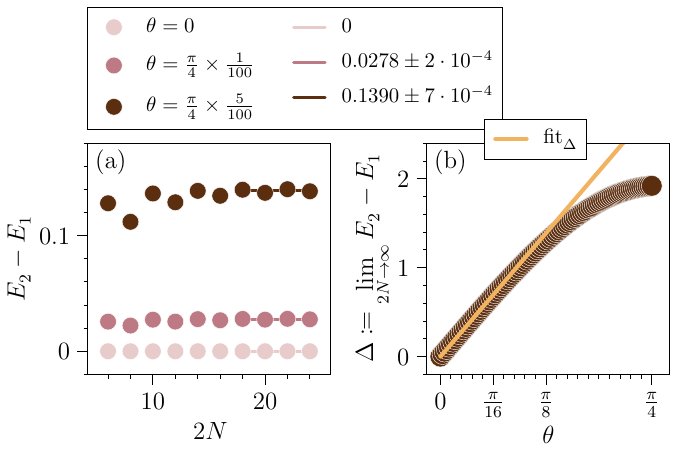}
\caption{
The case of PBCs.
(a)
Dependence on $2N=6,\cdots,24$
of the energy difference $E^{\,}_{2}-E^{\,}_{1}$
for different values of $\theta$
of the Hamiltonian $\widehat{H}^{(0)}_{\mathrm{II}}$
along $\lambda=3\pi/4$ obtained with ED.
We perform a fit that is constant as a function of $2N$
to extrapolate the value of
$E^{\,}_{2}-E^{\,}_{1}$ in the limit $2N\to\infty$.
(b)
Dependence on $\theta$
of the energy difference $E^{\,}_{2}-E^{\,}_{1}$
above the ground-state energy $E^{\,}_{0}$
of the Hamiltonian $\widehat{H}^{(0)}_{\mathrm{II}}$
along $\lambda=3\pi/4$ obtained with ED
in the limit $2N\to\infty$.
We perform a fit to extract the value of
$\theta^{\,}_{\mathrm{Neel}^{\mathrm{II}}_{z}}$
where $\Delta\coloneqq\lim\limits_{2N\to\infty}E^{\,}_{2}-E^{\,}_{1}$ vanishes.
The equation of $\mathrm{fit}^{\,}_{\Delta}$ is given
is Eq.~\eqref{eq:Ansatz theta NeelZ HB}.
}
\label{fig:NeelZ HB ED PBCs Tricritical Location}
\end{figure}

\subsubsection{The tetracritical points from Fig.\ \ref{fig:phase diagram B}}
\label{sec:Quadricritical point}

Owing to the existence of the unitary transformation
$\widehat{U}^{\,}_{\mathrm{R}}$ defined in Eq.~\eqref{eq:U_R},
the $\mathrm{Neel}^{\,}_{z}$ phase
from Fig.\ \ref{fig:phase diagram B}
must terminate at the two multicritical points located at
$(\theta^{\,}_{\mathrm{Neel}^{\mathrm{II}}_{z}},3\pi/4)$
and
$(\theta^{\prime}_{\mathrm{Neel}^{\mathrm{II}}_{z}},3\pi/4)$,
whereby
\begin{subequations}
\label{eq:instabilities tetracritical points}
\begin{equation}
\theta^{\prime}_{\mathrm{Neel}^{\mathrm{II}}_{z}}=
\pi/2-
\theta^{\,}_{\mathrm{Neel}^{\mathrm{II}}_{z}}.
\label{eq:instabilities tetracritical points a}
\end{equation}
We are going to argue numerically that
\begin{equation}
\theta^{\,}_{\mathrm{Neel}^{\mathrm{II}}_{z}}=0.
\label{eq:instabilities tetracritical points b}
\end{equation} 
In other words and in contrast to Hamiltonian
$\widehat{H}^{(\mathsf{b})}_{\mathrm{II}}$
at the integrable point $(0,\pi/4)$,
the integrable point $(0,3\pi/4)$
of Hamiltonian $\widehat{H}^{(\mathsf{b})}_{\mathrm{II}}$
is unstable to both the perturbation
\begin{equation}
\hbox{$\theta\neq0$ with fixed $\lambda=3\pi/4$}
\label{eq:instabilities tetracritical points c}
\end{equation}
and the perturbation
\begin{equation}
\hbox{$\lambda\neq3\pi/4$ with fixed $\theta=0$}.
\label{eq:instabilities tetracritical points d}
\end{equation}
\end{subequations}
This difference between the integrable points
$(0,\pi/4)$
and
$(0,3\pi/4)$
will be interpreted in Sec.\
\ref{sec:analytical_exploration_of_the_phase_diagrams}
as originating from a CFT with central charge
$\mathsf{c}=2$
being perturbed by one and two relevant perturbations,
respectively.

The effect of the perturbation
(\ref{eq:instabilities tetracritical points d})
can be treated as in Sec.\
\ref{subsubsec:Neelx-NeelSPTx phase boundary}
and shown to destroy criticality by opening a spectral gap.
We shall thus only consider the perturbation
(\ref{eq:instabilities tetracritical points c}).
We report in Fig.\
\ref{fig:NeelZ HB ED PBCs Tricritical Location}
the numerical evidences that support the claim
that the integrable point $(0,3\pi/4)$
is a tetracritical point at which the
$\mathrm{Neel}^{\mathrm{SPT}}_{x}$,
$\mathrm{Neel}^{\,}_{z}$,
$\mathrm{FM}^{\,}_{x}$,
and
$\mathrm{FM}^{\,}_{z}$
phases meet.

First, we impose PBCs and compute the low-energy
spectrum of $\widehat{H}^{(0)}_{\mathrm{II}}(\theta,3\pi/4)$
defined in Eq.~\eqref{eq:def HA and HB d}
for $0\leq\theta\leq\pi/4$ and for system sizes $2N\in[6,24]$
using ED. We thus obtain the eigenenergies $E^{\,}_{n,2N}(\theta)$.

Second, for each $\theta$ we do the extrapolation
to the thermodynamic limit
\begin{equation}
\Delta(\theta)\coloneqq
\lim\limits_{2N\to\infty}\left[
E^{\,}_{2,2N}(\theta)-E^{\,}_{1,2N}(\theta)\right].
\end{equation}

Third, making the Ansatz
\begin{subequations}
\label{eq:Ansatz theta NeelZ HB}
\begin{equation}
\Delta(\theta)=
D
+
A\times\bigl(\theta-B\big)^{C},
\end{equation}
with $A,B,C$ and $D$ constant parameters,
we find that
\begin{align}
&A=3.542\pm0.001,
&&B=-(3\pm4)\cdot10^{-6},
\\
&C=1.0003\pm0.0001,
&&D=(1\pm1)\cdot10^{-6},
\end{align}
\end{subequations}
where $\theta$ is chosen in the vicinity of $\theta=0$.

Finally, we deduce from Eq.\
\eqref{eq:Ansatz theta NeelZ HB}
that the estimate for the energy gap $\Delta(\theta)$
above the two-fold degenerate ground-state energy from the
$\mathrm{Neel}^{\,}_{z}$ phase
closes (since $D\simeq0$) linearly in $\theta$ (since $C\simeq1$) at
$\theta\simeq0$ (since $B\simeq0$).
It is the estimate $B\simeq0$
that allows us to locate the tetracritical point at
$(0,3\pi/4)$.

\begin{table}
\caption{
Central charge $\mathsf{c}$
(obtained with DMRG and PBCs),
critical exponents
$\nu$ (diverging correlation length),
$\beta$ (order parameter),
$\eta^{\alpha}$ (two-point correlation for
quantum spin $\hat{\sigma}^{\alpha}$),
$\mathsf{z}$ (dynamical exponent)
defined in
Eq.~\eqref{eq:def critical exponents}
for Hamiltonian \eqref{eq:def HA and HB d}
at the critical point
$(0,3\pi/4)$
obtained using iDMRG with $\chi=256$
(ED with PBCs and $2N\leq30$ for $\mathsf{z}$).
}
\label{tab:TQ HB Critical Exponents 3pi/4}
\centering
\renewcommand{\arraystretch}{1.1}
\setlength{\tabcolsep}{5pt}
\begin{tabular}{c c cc}
\midrule\midrule

& 
& $(0,\lambda)$
& $\bigl(\theta,\tfrac{3\pi}{4}\bigr)$ \\
\cmidrule(lr){3-3}\cmidrule(lr){4-4}

$\nu$
&
& $0.997\pm0.007$
& $1.003\pm0.004$ \\

$\beta$
&
& $0.50\pm0.01$
& $0.498\pm0.007$ \\
\cmidrule(lr){1-4}

& 
& \multicolumn{2}{c}{$(0,\tfrac{3\pi}{4})$} \\
\cmidrule(lr){3-4}

$\mathsf{c}$
&
& \multicolumn{2}{c}{$1.993\pm0.001$} \\

$\eta^{x}$
&
& \multicolumn{2}{c}{$1.013\pm0.009$} \\

$\eta^{z}$
&
& \multicolumn{2}{c}{$1.014\pm0.009$} \\

$\mathsf{z}$
&
& \multicolumn{2}{c}{$0.993\pm0.001$} \\
\midrule\midrule
\end{tabular}
\end{table}

The measured critical attributes at $(0,3\pi/4)$
given in Table \ref{tab:TQ HB Critical Exponents 3pi/4}
are consistent with their exact values.
As it should be,
at the integrable critical point $(0,3\pi/4)$,
the computed two-point function
$C^{y}_{s^{\,}_{0},s^{\,}_{0}+s}[\Psi]$
in the matrix product state $\Psi$ that approximates the ground state
behaves like \eqref{eq:eta critical exponent},
whereby the non-universal amplitudes are measured to be
\begin{subequations}
\label{eq:amplitude functions HB (0,3pi/4)}
\begin{align}
&A^{x}_{s}\propto(-1)^{s},
\\
&A^{y}_{s}=0,
\\
&A^{z}_{s}\propto
\begin{cases}
1 & \hbox{if } s\equiv0\bmod 4,
\\
0 & \hbox{if } s\not\equiv0\bmod 4.
\end{cases}
\end{align}
\end{subequations}
The two-point function
$C^{y}_{s^{\,}_{0},s^{\,}_{0}+s}$
defined in Eq.~\eqref{eq:Cs}
vanish for any $s$.

\section{Analytical exploration of the phase diagrams}
\label{sec:analytical_exploration_of_the_phase_diagrams}

Section
\ref{sec:analytical_exploration_of_the_phase_diagrams}
complements the numerical results of Sec.\
\ref{sec:Numerical exploration of the phase diagrams}
with a set of analytical approaches that elucidate
the origin of the phase diagrams
and clarify the nature of the critical and multicritical points. The
strategy throughout is to identify exactly solvable or weakly
interacting limits, derive effective low-energy descriptions, and
connect them to the numerical observations of long-range order,
topology, and quantum criticality.

The first part of Sec.\
\ref{sec:analytical_exploration_of_the_phase_diagrams}
is devoted to the $\mathrm{U}(1)$-symmetric
line $\theta=\pi/4$, along which Hamiltonians
\eqref{eq:def HA and HB c}
and
\eqref{eq:def HA and HB d}
coincide. The analysis begins with a Jordan-Wigner transformation
that maps the spin degrees of freedom to interacting spinless
fermions. This mapping provides a unified description of the
deconfined critical lines, the incommensurate ferromagnetic phase
$\mathrm{gFM}^{\,}_{z}$, and the gapped antiferromagnetic phase
$\mathrm{Neel}^{\,}_{z}$. A mean-field treatment of the fermionic
interactions already identifies the onset of uniform and staggered
magnetization as potential instabilities of the half-filled free fermion state.
However, to account for the important role of quantum fluctuations
in one-dimensional space, we investigate the stability of the
half-filled free fermion state from the viewpoint of bosonization.
The continuum description obtained from the interacting
fermion model is shown to be equivalent to a Luttinger liquid over an
extended parameter regime. This framework explains the numerical
observation of critical phases with central charge $\mathsf{c}=1$ and
establishes the connection with the low-energy physics of the
nearest-neighbor quantum spin-1/2 XXZ chain.
The bosonized theory also identifies the
leading perturbations that destabilize the critical state and provides
a continuum language in which the tricritical points can be understood.

A central result follows from the analysis of the tricritical point
bordering the $\mathrm{gFM}_z$ phase. The bosonic theory predicts a
Lifshitz instability at which the stiffness associated with the
critical mode vanishes and changes sign. On one side of this
instability the system realizes a conventional Luttinger liquid with
commensurate correlations. On the other side, the ground state
develops a finite gradient of the bosonic field, producing
incommensurate correlations together with long-range ferromagnetic
order along the $z$ axis. The tricritical point is therefore
simultaneously a Lifshitz point, a disorder point, and the endpoint of
a deconfined critical line. Upon approaching this point from the
Luttinger-liquid side, numerics show that the transverse scaling
exponent $\eta^{x}=\eta^{y}$ and the characteristic velocity both
vanish. At the $\mathsf{z}=2$ Gaussian Lifshitz point, the
corresponding transverse vertex correlations exhibit long-range
order. The agreement between these numerical and analytical results
supports spontaneous breaking of the continuous $\mathrm{U}(1)$
symmetry at the tricritical point, while the longitudinal correlations
remain algebraic. This realizes critical continuous symmetry breaking
in a local one-dimensional quantum spin chain
\cite{Watanabe24,Fava_PNAS_24,Nahum2506.21540,Andriyakhina2026}.

The tricritical point bordering the $\mathrm{Neel}^{\,}_{z}$ phase is
described differently. Here, the bosonized theory approaches the
$\mathrm{SU}(2)$-symmetric point of the nearest-neighbor antiferromagnetic
quantum spin-1/2 XXZ chain. The critical line
terminates at a Berezinskii-Kosterlitz-Thouless (BKT) transition beyond
which a marginal perturbation becomes relevant,
pins the bosonic field, opens a spectral gap,
and generates long-range staggered order along the $z$ axis.

The last part of the analysis along the $\mathrm{U}(1)$-symmetric line
concerns the boundary physics near $\lambda=\pi/2$ when open boundary
conditions are imposed. Degenerate perturbation theory reveals that
the reduction of the ground-state degeneracy from four to two
originates from an effective interaction between fractionalized
spin-$1/2$ degrees of freedom localized at opposite ends of the
chain. The resulting effective boundary Hamiltonian accurately
reproduces the finite-size spectrum observed numerically and explains
the persistence of symmetry-enriched criticality along the deconfined
critical lines adjacent to the symmetry-protected topological phases.

The second major theme of Sec.\
\ref{sec:analytical_exploration_of_the_phase_diagrams}
concerns the free-Majorana line $\theta=\pi/2$
for Hamiltonian
\eqref{eq:def HA and HB d}.
Along this line, the Hamiltonian can be mapped exactly
onto non-interacting Majorana chains. This mapping explains the
structure of the gapped phases, the sensitivity of the ground-state
degeneracy to boundary conditions, and the appearance of critical
points with central charge $\mathsf{c}=2$. It also provides a
convenient starting point from which to investigate the stability of
these critical points.

To this end, the third part of Sec.\
\ref{sec:analytical_exploration_of_the_phase_diagrams}
studies perturbations away from the free-Majorana line.
Normal-ordering of the interaction generated
by departing from $\theta=\pi/2$ produces effective quadratic and
quartic terms whose impact depends strongly on which critical point is
considered. One critical point is shown to be unstable because a mass
term is generated immediately, producing a spectral gap. The other
critical point remains gapless at quadratic order, and its fate is
controlled by marginal interactions. A one-loop renormalization-group
analysis reveals that these interactions dynamically gap half of the
critical modes, reducing the central charge from $\mathsf{c}=2$ to
$\mathsf{c}=1$. The resulting phase is interpreted as an
incommensurate critical phase described by an $\mathfrak{su}(2)_{1}$
conformal field theory, thereby providing an analytical explanation
for the gapless region observed numerically.

Finally, the fourth part of
Sec.\
\ref{sec:analytical_exploration_of_the_phase_diagrams}
addresses the direct transitions between competing
ordered phases in Hamiltonian
\eqref{eq:def HA and HB c}.
A Landau-Ginzburg analysis
shows that these transitions differ qualitatively depending on the
symmetry of the competing order parameters. When an enlarged
continuous symmetry is present, fluctuations invalidate the mean-field
prediction of a first-order transition and allow a continuous
deconfined quantum critical point to emerge. When only discrete
symmetries are present, the coexistence of competing ordered states
leads instead to a weakly first-order transition. This distinction
accounts for the contrasting behavior of the phase boundaries observed
in the numerical phase diagrams.

\subsection{The U(1)-symmetric line $\theta=\pi/4$}
\label{subsec:The U(1) symmetric lin thetaequalpiover4}

\subsubsection{Introduction}

\begin{widetext}
Section \ref{subsec:The U(1) symmetric lin thetaequalpiover4}
is devoted exclusively to Hamiltonian
\begin{equation}
\begin{split}
\widehat{H}^{(\mathsf{b})}_{\mathrm{U}(1)}(\lambda)\coloneqq
\frac{\cos(\lambda)}{\sqrt{2}}
\sum_{j=1}^{2N-\mathsf{b}}
\left(
\widehat{X}^{\,}_{j}\,
\widehat{X}^{\,}_{j+1}
+
\widehat{Y}^{\,}_{j}\,
\widehat{Y}^{\,}_{j+1}
\right)
+
\frac{\sin(\lambda)}{\sqrt{2}}
\sum_{j=1}^{2N-3\mathsf{b}}
\widehat{Z}^{\,}_{j}\,
\left(
\widehat{X}^{\,}_{j+1}\,
\widehat{X}^{\,}_{j+2}
+
\widehat{Y}^{\,}_{j+1}\,
\widehat{Y}^{\,}_{j+2}
\right)
\widehat{Z}^{\,}_{j+3}
\end{split}
\label{eq:def HbU(1) as function lambda}
\end{equation}
that was defined by Eqs.\
(\ref{eq:def H U1}),
(\ref{eq:def HA and HB c}),
and
(\ref{eq:def HA and HB d})
along the interval $\lambda\in[0,\pi]$.
The domain of definition of $\widehat{H}^{(\mathsf{b})}_{\mathrm{U}(1)}(\lambda)$
is the $2^{2N}$-dimensional Hilbert space
$\mathfrak{H}\cong\mathbb{C}^{2^{2N}}$.
This domain of definition is isomorphic to that of the lattice 
Hamiltonian
\begin{subequations}
\label{eq:def HbU(1) fermionic as function lambda}
\begin{equation}
\widehat{H}^{(\mathsf{b})}_{\mathrm{U(1),f}}(\lambda)\coloneqq
-
\sqrt{2}\cos(\lambda)
\sum_{j=1}^{2N-\mathsf{b}}
\left(
\hat{c}^{\dag}_{j}\,
\hat{c}^{\, }_{j+1}
+
\mathrm{H.c.}
\right)
-
\sqrt{2}\sin(\lambda)
\sum_{j=1}^{2N-3\mathsf{b}}
\left(2\hat{n}^{\,}_{j}-1\right)\,
\left(
\hat{c}^{\dag}_{j+1}\,
\hat{c}^{\, }_{j+2}
+
\mathrm{H.c.}
\right)
\left(2\hat{n}^{\,}_{j+3}-1\right),
\label{eq:def HbU(1) fermionic as function lambda a}
\end{equation}
where
\begin{equation}
\hat{n}^{\,}_{j}\coloneqq
\hat{c}^{\dag}_{j}\,
\hat{c}^{\, }_{j},
\label{eq:def HbU(1) fermionic as function lambda b}
\end{equation}
for the spinless fermions obeying the fermion algebra
\begin{equation}
\left\{
\hat{c}^{\,}_{j},
\hat{c}^{\dag}_{j'}
\right\}=
\delta^{\,}_{j,j'},
\qquad
\left\{
\hat{c}^{\,}_{j},
\hat{c}^{\,\vphantom{\dag}}_{j'}
\right\}=
\left\{
\hat{c}^{\dag}_{j},
\hat{c}^{\dag}_{j'}
\right\}=
0,
\qquad
j,j'=1,\ldots,2N.
\label{eq:def HbU(1) fermionic as function lambda c}
\end{equation}
\end{subequations}
\end{widetext}
We shall explain shortly that
\begin{equation}
\widehat{H}^{(1)}_{\mathrm{U}(1)}(\lambda)=
\widehat{H}^{(1)}_{\mathrm{U}(1),\mathrm{f}}(\lambda).
\label{eq:relation HbU(1) and HbU(10 fermionic for bone}
\end{equation}
The relationship between
$\widehat{H}^{(0)}_{\mathrm{U}(1)}(\lambda)$
and
$\widehat{H}^{(0)}_{\mathrm{U}(1),\mathrm{f}}(\lambda)$
is more involved as will be explained in Sec.\
\ref{subsubsec:Jordan-Wigner transformation along U1 symmetric line}
through the choice of the boundary conditions obeyed by the spinless fermions.
We have argued numerically that the zero-temperature
phase diagram of Hamiltonian
$\widehat{H}^{(\mathsf{b})}_{\mathrm{U}(1)}(\lambda)$
consists of
(see Fig.\ \ref{fig:phase diagram U1})
\begin{itemize}
\item
a line of deconfined quantum critical points for
$
\lambda\in
\left[
0,
\lambda^{\,}_{\mathrm{gFM}^{\,}_{z}}
\right[
$
that ends with the tricritical point
$\lambda^{\,}_{\mathrm{gFM}^{\,}_{z}}$,  
\item
a line of quantum critical points for
$
\lambda\in
\left]
\lambda^{\,}_{\mathrm{gFM}^{\,}_{z}},
\lambda^{\mathrm{SPT}}_{\mathrm{gFM}^{\,}_{z}}
\right[
$
that ends with the tricritical point
$\lambda^{\mathrm{SPT}}_{\mathrm{gFM}^{\,}_{z}}$,  
\item
a line of deconfined quantum critical points for
$
\lambda\in
\left]
\lambda^{\mathrm{SPT}}_{\mathrm{gFM}^{\,}_{z}},
\lambda^{\mathrm{SPT}}_{\mathrm{Neel}^{\,}_{z}}
\right[
$
that ends with the tricritical point
$\lambda^{\mathrm{SPT}}_{\mathrm{Neel}^{\,}_{z}}$,
\item
a line of quantum critical points for
$
\lambda\in
\left]
\lambda^{\mathrm{SPT}}_{\mathrm{Neel}^{\,}_{z}},
\lambda^{\,}_{\mathrm{Neel}^{\,}_{z}}
\right[
$
that ends with the tricritical point
$\lambda^{\,}_{\mathrm{Neel}^{\,}_{z}}$,
\item
and a line of deconfined quantum critical points for
$
\lambda\in
\left]
\lambda^{\,}_{\mathrm{Neel}^{\,}_{z}},
\pi
\right]
$.
\end{itemize}
The goal of Sec.\
\ref{subsec:The U(1) symmetric lin thetaequalpiover4}
is twofold.
First,
the existence of the pair of tricritical points
$\lambda^{\,}_{\mathrm{gFM}^{\,}_{z}}$
and
$\lambda^{\mathrm{SPT}}_{\mathrm{gFM}^{\,}_{z}}$
on the one hand, and the pair of tricritical points
$\lambda^{\mathrm{SPT}}_{\mathrm{Neel}^{\,}_{z}}$
and
$\lambda^{\,}_{\mathrm{Neel}^{\,}_{z}}$
is captured by a mean-field approximation of the fermionic Hamiltonian
$\widehat{H}^{(0)}_{\mathrm{U}(1),\mathrm{f}}(\lambda)$
that we describe in Sec.\
\ref{subsubsec:Mean-field analysis of the transitions to the gFMz and Neelz}.
Second,
the minimal degeneracy of two
of the line of deconfined quantum critical points
$
\lambda\in
\left]
\lambda^{\mathrm{SPT}}_{\mathrm{gFM}^{\,}_{z}},
\lambda^{\mathrm{SPT}}_{\mathrm{Neel}^{\,}_{z}}
\right[
$
when OBCs are selected 
is explained with the help of degenerate perturbation theory of
$\widehat{H}^{(1)}_{\mathrm{U}(1)}(\lambda)$
with respect to $\lambda=\pi/2$
in Sec.\
\ref{subsubsec:effective boundary interaction around lambda piover2}.

\subsubsection{Jordan-Wigner transformation
along the $\mathrm{U}(1)$-symmetric line}
\label{subsubsec:Jordan-Wigner transformation along U1 symmetric line}

In preparation for the mean-field approximation from Sec.\
\ref{subsubsec:Mean-field analysis of the transitions to the gFMz and Neelz},
we follow Lieb, Schultz, and Mattis
in Ref.\ \cite{Lieb61}
who took advantage of the Jordan-Wigner transformation \cite{Jordan28}
to show that the point $\lambda=0$
is integrable, as it is realized by free fermions, and gapless.

\paragraph{Case of OBCs $(\mathsf{b=1})$.}
Define
\begin{subequations}
\label{eq: def Jordan-Wigner trsf U(1) symmetric}
\begin{align}
&
\hat{c}^{\dag}_{j}\coloneqq
\frac{
\widehat{X}^{\,}_{j}
+
\mathrm{i}
\widehat{Y}^{\,}_{j}
   }{2}\,
\widehat{Z}^{\,}_{1}\,
\cdots\,
\widehat{Z}^{\,}_{j-1},
\\
&
\hat{c}^{\,}_{j}\coloneqq
\widehat{Z}^{\,}_{1}\,
\cdots\,
\widehat{Z}^{\,}_{j-1}\,
\frac{
\widehat{X}^{\,}_{j}
-
\mathrm{i}
\widehat{Y}^{\,}_{j}
   }{2},
\end{align}
\end{subequations}
for any site $j=1,\ldots,2N$.
One verifies that:
\begin{enumerate}
\item
The Pauli algebra
(\ref{eq:def Hilbert space})
implies the fermion algebra
(\ref{eq:def HbU(1) fermionic as function lambda c}).
\item
The relation
\begin{equation}
\hat{n}^{\,}_{j}=
\frac{1}{2}
\left(
\mathds{1}^{\,}_{\mathfrak{H}}
+
\widehat{Z}^{\,}_{j}
\right)
\ \Longleftrightarrow\
\widehat{Z}^{\,}_{j}=2\hat{n}^{\,}_{j}-\mathds{1}^{\,}_{\mathfrak{H}}
\label{eq:relation local JW fermion local Z}
\end{equation}
holds for $j=1,\ldots,2N$.
\item
The relations
\begin{subequations}
\label{eq:c's in terms of spins}
\begin{align}
\hat{c}^{\dag}_{j}\,
\hat{c}^{\,}_{j+1}=&\, 
\frac{
\widehat{X}^{\,}_{j}
+
\mathrm{i}
\widehat{Y}^{\,}_{j}
   }{2}\,
\widehat{Z}^{\,}_{j}\,
\frac{
\widehat{X}^{\,}_{j+1}
-
\mathrm{i}
\widehat{Y}^{\,}_{j+1}
   }{2}
\nonumber\\
=&\,
-
\frac{
\widehat{X}^{\,}_{j}
+
\mathrm{i}
\widehat{Y}^{\,}_{j}
   }{2}\,
\frac{
\widehat{X}^{\,}_{j+1}
-
\mathrm{i}
\widehat{Y}^{\,}_{j+1}
   }{2},
\\
\hat{c}^{\dag}_{j+1}\,
\hat{c}^{\,}_{j}=&\, 
\frac{
\widehat{X}^{\,}_{j+1}
+
\mathrm{i}
\widehat{Y}^{\,}_{j+1}
   }{2}\,
\widehat{Z}^{\,}_{j}\,
\frac{
\widehat{X}^{\,}_{j}
-
\mathrm{i}
\widehat{Y}^{\,}_{j}
   }{2}
\nonumber\\
=&\,
-
\frac{
\widehat{X}^{\,}_{j}
-
\mathrm{i}
\widehat{Y}^{\,}_{j}
   }{2}\,
\frac{
\widehat{X}^{\,}_{j+1}
+
\mathrm{i}
\widehat{Y}^{\,}_{j+1}
   }{2},
\end{align}
\end{subequations}
hold for $j=1,\ldots,2N-1$.
\end{enumerate}
Equation (\ref{eq:relation HbU(1) and HbU(10 fermionic for bone})
then follows. Moreover, we may recast
Eq.\ (\ref{eq:relation local JW fermion local Z})
as the global statement that the conserved total fermion-number operator
\begin{subequations}
\label{eq:relationship between N and Mz U(1) symmetric line}
\begin{equation}
\widehat{N}\coloneqq
\sum_{j=1}^{2N}
\hat{n}^{\,}_{j}
\label{eq:relationship between N and Mz U(1) symmetric line a}
\end{equation}
and the conserved total uniform magnetization
\begin{equation}
\widehat{M}^{z}\coloneqq
\sum_{j=1}^{2N}
\frac{\widehat{Z}^{\,}_{j}}{2}
\label{eq:relationship between N and Mz U(1) symmetric line b}
\end{equation}
along the $z$ quantization axis in quantum spin-$1/2$ space
are related by
\begin{equation}
\widehat{N}=
\frac{1}{2}
\left(
2N\mathds{1}^{\,}_{\mathfrak{H}}
+
2
\widehat{M}^{z}
\right)=
N\mathds{1}^{\,}_{\mathfrak{H}}
+
\widehat{M}^{z}.
\label{eq:relationship between N and Mz U(1) symmetric line c}
\end{equation}
\end{subequations}
Because the number $2N$ of sites is even, the fact that
the eigenvalues of $\widehat{M}^{z}$ belong to the set
\begin{subequations}
\begin{equation}
\left\{-N,-N+1,\ldots,-1,0,+1,\ldots,+N-1,+N\right\}
\end{equation}
of cardinality $2N+1$ implies that the eigenvalues of
$\widehat{N}$ belong to the set
\begin{equation}
\left\{0,1,\ldots,2N-1,2N\right\},
\end{equation}
\end{subequations}
of cardinality $2N+1$.
In particular, if the ground state of
$\widehat{H}^{(1)}_{\mathrm{U}(1)}(\lambda)$
defined in Eq.\
(\ref{eq:def HbU(1) as function lambda})
is an eigenstate of $\widehat{M}^{z}$ with vanishing eigenvalue,
then the ground state of
$\widehat{H}^{(1)}_{\mathrm{U(1),f}}$
defined in Eq.\
(\ref{eq:def HbU(1) fermionic as function lambda a})
is an eigenstate of $\widehat{N}$ with eigenvalue $N=(2N)/2$,
the condition for half filling. As we shall see, half-filling
in the fermionic representation at $\lambda=0$
corresponds to a vanishing chemical potential. 
With the same reasoning, we deduce that
\begin{itemize}
\item
the magnetic phase
$\mathrm{gFM}^{\,}_{z}$
corresponds to a fermionic phase away from half-filling,
i.e., with a non-vanishing chemical potential,
\item
the magnetic phase
$\mathrm{Neel}^{\,}_{z}$
corresponds to a fermionic phase at half-filling but with
a staggered chemical potential.
\end{itemize}

\paragraph{Case of PBCs $(\mathsf{b=0})$.}
We select PBCs by choosing $\mathsf{b=0}$
for Hamiltonian
(\ref{eq:def HbU(1) as function lambda})
We decompose the $2^{2N}$-dimensional Hilbert space
(\ref{eq:def Hilbert space d})
on which
$\widehat{H}^{(\mathsf{b})}_{\mathrm{U}(1)}(\lambda)$
acts as the direct sum
\begin{subequations}
\label{sec:def direct sum over Z or PF eigenvalues}
\begin{equation}
\mathfrak{h}=
\mathfrak{h}^{\,}_{+}\oplus\mathfrak{h}^{\,}_{-}
\label{sec:def direct sum over Z or PF eigenvalues a}
\end{equation}
over the orthogonal $2^{2N-1}$-dimensional eigenspaces
\begin{equation}
\mathfrak{h}^{\,}_{\pm}\equiv
\widehat{P}^{\,}_{\pm}\,\mathfrak{h}\coloneqq
\frac{\mathds{1}^{\,}_{\mathfrak{H}}\pm\widehat{Z}}{2}\,\mathfrak{h} 
\label{sec:def direct sum over Z or PF eigenvalues b}
\end{equation}
of
\begin{equation}
\widehat{Z}\coloneqq
\prod_{j=1}^{2N} \widehat{Z}^{\,}_{j}
\label{sec:def direct sum over Z or PF eigenvalues c}
\end{equation}
\end{subequations}
with eigenvalues $\pm1$, respectively.
We then have the identity
\begin{subequations}
\label{eq:case PBC fermion and parity +}
\begin{equation}
\begin{split}
\widehat{P}^{\,}_{+}\,
\widehat{H}^{(0)}_{\mathrm{U(1)}}(\lambda)\,
\widehat{P}^{\,}_{+}=&\,
\widehat{P}^{\,}_{+}\,
\widehat{H}^{(0)}_{\mathrm{U(1),f}}(\lambda)\,
\widehat{P}^{\,}_{+}
\end{split}
\label{eq:case PBC fermion and parity + a}
\end{equation}
if the fermion operators obey the PBCs
\begin{equation}
\hat{c}^{\dag}_{j+2N}=+\hat{c}^{\dag}_{j},
\quad
\hat{c}^{\,}_{j+2N}=+\hat{c}^{\,}_{j},
\quad
j=1,\ldots,2N,
\label{eq:case PBC fermion and parity + b}
\end{equation}
\end{subequations}
and the identity
\begin{subequations}
\label{eq:case APBC fermion and parity -}
\begin{equation}
\begin{split}
\widehat{P}^{\,}_{-}\,
\widehat{H}^{(0)}_{\mathrm{U(1)}}(\lambda)\,
\widehat{P}^{\,}_{-}=&\,
\widehat{P}^{\,}_{-}\,
\widehat{H}^{(0)}_{\mathrm{U(1),f}}(\lambda)\,
\widehat{P}^{\,}_{-}
\end{split}
\label{eq:case APBC fermion and parity - a}
\end{equation}
if the fermion operators obey the antiperiodic boundary conditions
(ABCs)
\begin{equation}
\hat{c}^{\dag}_{j+2N}=-\hat{c}^{\dag}_{j},
\quad
\hat{c}^{\,}_{j+2N}=-\hat{c}^{\,}_{j},
\quad
j=1,\ldots,2N.
\label{eq:case APBC fermion and parity - b}
\end{equation}
\end{subequations}
Here, it is understood that 
\begin{subequations}
\begin{equation}
\widehat{P}^{\,}_{\pm}\coloneqq
\frac{\mathds{1}^{\,}_{\mathfrak{H}}\pm\widehat{P}^{\,}_{\mathrm{F}}}{2},
\end{equation}
where
\begin{equation}
\widehat{P}^{\,}_{\mathrm{F}}\coloneqq
\prod_{j=1}^{2N}
\left(2\hat{n}^{\,}_{j}-\mathds{1}^{\,}_{\mathfrak{H}}\right)=
e^{\mathrm{i}\pi\left(\widehat{N}-2N\mathds{1}^{\,}_{\mathfrak{H}}\right)}=
e^{\mathrm{i}\pi\widehat{N}}
\end{equation}
\end{subequations}
in the fermion representation. Hence, working in the subspace
$\mathfrak{H}^{\,}_{+}$ $(\mathfrak{H}^{\,}_{-})$ 
of the fermionic Fock space $\mathfrak{H}$ 
amounts to demanding that
the total number of fermions is even (odd) and that the fermions
obey PBCs (ABCs).

\begin{figure}
\centering
\includegraphics[width=0.5\columnwidth]{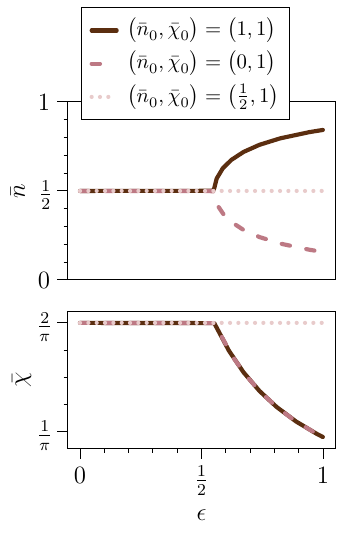}
\caption{Solutions $\bar{n}=\bar{n}(\epsilon)$
and $\bar{\chi}=\bar{\chi}(\epsilon)$
to the mean-field equations \eqref{eq:self-consistency eqs Fourier gFMz}
as a function of $\epsilon$, obtained with the set of initial values
\eqref{eq:gFMz MF intial values}.
}
\label{fig:MF Solutions}
\end{figure}

\subsubsection{Mean-field analysis of the transitions to the
$\mathrm{gFM}^{\,}_{z}$ and $\mathrm{Neel}^{\,}_{z}$ phases
along the $\mathrm{U}(1)$-symmetric line}
\label{subsubsec:Mean-field analysis of the transitions to the gFMz and Neelz}

Our aim is to understand the onset of long-range magnetic order
in the phases $\mathrm{gFM}^{\,}_{z}$ and $\mathrm{Neel}^{\,}_{z}$.
This onset is independent of the choice of boundary conditions.
For this reason, we shall study the instabilities of the
free-fermion point at $\lambda=0$
upon increasing (decreasing) $\lambda$
of the fermionic Hamiltonian
$\widehat{H}^{(0)}_{\mathrm{U(1),f}}(\lambda)$
defined in Eq.\
(\ref{eq:def HbU(1) fermionic as function lambda a})
when $2N/4$ is integer, i.e., when the fermions obey the PBCs
given in Eq.\
(\ref{eq:case PBC fermion and parity + b}).
We are going to perform a mean-field approximation of the interaction
in the fermionic Hamiltonian
$\widehat{H}^{(0)}_{\mathrm{U(1),f}}(\lambda)$.
The onset of the
$\mathrm{gFM}^{\,}_{z}$
($\mathrm{Neel}^{\,}_{z}$)
phase is described by a non-vanishing uniform (staggered) chemical potential.
The uniform mean-field Ansatz predicts the onset of the
$\mathrm{gFM}^{\,}_{z}$ phase above a non-vanishing and positive value
of $\lambda$,
whereas the staggered mean-field Ansatz predicts the onset of the
$\mathrm{Neel}^{\,}_{z}$ phase for any negative value of $\lambda$.

For convenience,
we do the replacement
\begin{equation}
\cos\lambda\mapsto \frac{1}{\sqrt{2}}\,(1-\epsilon),
\qquad
\sin\lambda\mapsto \frac{\sgn(\lambda)}{\sqrt{2}}\,\epsilon,
\label{eq:change parametrization MF}
\end{equation}
with $0\leq\epsilon\leq1$. This choice is suitable since
\begin{itemize}
\item[(i)] the interpolation between the limiting Hamiltonian at $\lambda=0$ and $\lambda=\pm\,\frac{\pi}{2}$ is preserved,
\item[(ii)] we are not interested in quantitatively locating
the onset of long-range magnetic order,
\item[(iii)] the onset of the $\mathrm{gFM}^{\,}_{z}$ phase
is obtained by imposing $0\leq\lambda\leq\tfrac{\pi}{2}$, and
\item[(iv)] the onset of the $\mathrm{Neel}^{\,}_{z}$ phase
is obtained by imposing $-\tfrac{\pi}{2}\leq\lambda\leq0$.
\end{itemize}

\paragraph{Onset of the $\mathrm{gFM}^{\,}_{z}$ phase.}

Under the transformation \eqref{eq:change parametrization MF}
with $0\leq\lambda\leq\tfrac{\pi}{2}$, Hamiltonian 
$\widehat{H}^{(0)}_{\mathrm{U(1),f}}(\lambda)$
becomes
\begin{widetext}
\begin{equation}
\widehat{H}^{\mathrm{gFM}^{\,}_{z}}(\epsilon)
\coloneqq
-\sum_{j=1}^{2N}\left(
\hat{c}^{\dag}_{j}\,
\hat{c}^{\, }_{j+1}
+
\mathrm{H.c.}
\right)
+2\epsilon
\sum_{j=1}^{2N}\left(
\hat{n}^{\,}_{j-1}
+\hat{n}^{\,}_{j+2}
-2\,\hat{n}^{\,}_{j-1}\hat{n}^{\,}_{j+2}
\right)
\left(
\hat{c}^{\dag}_{j}\,
\hat{c}^{\, }_{j+1}
+
\mathrm{H.c.}
\right)
.
\label{eq:H gFMz before MF}
\end{equation}
\end{widetext}
The mean-field approximation
$\widehat{H}^{\mathrm{gFM}^{\,}_{z}}_{\mathrm{MF}}$
to Hamiltonian
\eqref{eq:H gFMz before MF},
follows from the following procedure.

We postulate that for $j=1,\ldots,2N$ the local operators
\begin{equation}
\hat{n}^{\,}_{j}\coloneqq
\hat{c}^{\dag}_{j}\,
\hat{c}^{\,}_{j},
\qquad
\widehat{\chi}^{\,}_{j,j+1}\coloneqq
\hat{c}^{\dag}_{j}\,
\hat{c}^{\,}_{j+1}
+
\mathrm{H.c.},
\label{eq:Local Opeators Fermion U(1)}
\end{equation}
acquire the position-independent
expectation values
\begin{equation}
\bar{n}\coloneqq
\langle\mathrm{FS}|\,
\hat{n}^{\,}_{j}\,
|\mathrm{FS}\rangle,
\qquad
\bar{\chi}\coloneqq
\langle\mathrm{FS}|\,
\widehat{\chi}^{\,}_{j,j+1}\,
|\mathrm{FS}\rangle,
\label{eq:self-consistency eqs lattice gFMz}
\end{equation}
in the ground state
$|\mathrm{FS}\rangle$ (a Fermi sea) of
$\widehat{H}^{\mathrm{gFM}^{\,}_{z}}_{\mathrm{MF}}$.
Here, the self-consistency is a consequence of the fact that
we inserted the expansion
\begin{equation}
\hat{n}^{\,}_{j}\eqqcolon
\bar{n}
+
\widehat{\delta n}^{\,}_{j},
\qquad
\widehat{\chi}^{\,}_{j,j+1}\eqqcolon
\bar{\chi}
+
\widehat{\delta\chi}^{\,}_{j,j+1},
\label{eq:expansion MF parameters gFMz}
\end{equation}
into Hamiltonian \eqref{eq:H gFMz before MF}
and kept only the quadratic terms in powers of the fermion operators
to obtain
\begin{subequations}
\label{eq:H gFMz summary MF}
\begin{align}
\widehat{H}^{\mathrm{gFM}^{\,}_{z}}_{\mathrm{MF}}(\epsilon)
={}&
E^{\,}_{\mathrm{MF}}(\bar{n},\bar{\chi};\epsilon)
\nonumber \\
&{}-
t^{\,}_{\mathrm{MF}}(\bar{n};\epsilon)\,
\sum_{j=1}^{2N}
\left(
\hat{c}^{\dag}_{j}\,\hat{c}^{\,}_{j+1}
+
\mathrm{H.c.}
\right)
\nonumber \\
&{}-
\mu^{\,}_{\mathrm{MF}}(\bar{n},\bar{\chi};\epsilon)
\sum_{j=1}^{2N}
\hat{c}^{\dag}_{j}\,\hat{c}^{\,}_{j},
\label{eq:H gFMz summary MF a}
\end{align}
where
\begin{align}
&
E^{\,}_{\mathrm{MF}}(\bar{n},\bar{\chi};\epsilon)\coloneqq
8\,
\epsilon\,
\bar{n}\,
(2\,\bar{n}-1)\,
\bar{\chi}\,
N,
\\
&
t^{\,}_{\mathrm{MF}}(\bar{n};\epsilon)\coloneqq
1
-
4\,
\epsilon\,
\bar{n}\,
(1-\bar{n}),
\\
&
\mu^{\,}_{\mathrm{MF}}(\bar{n},\bar{\chi};\epsilon)\coloneqq
4\,
\epsilon\,
(2\,\bar{n}-1)\,
\bar{\chi}.
\label{eq:H gFMz summary MF b}
\end{align}
\end{subequations}
The Fermi sea $\ket{\mathrm{FS}}$ of
$\widehat{H}^{\mathrm{gFM}^{\,}_{z}}_{\mathrm{MF}}$
is thus the mean-field approximation to the
ground state of Hamiltonian \eqref{eq:H gFMz before MF}.

In order to construct $\ket{\mathrm{FS}}$,
we insert the Fourier expansions
\begin{subequations}
\label{eq:Fourier MF}
\begin{align}
\hat{c}^{\dag}_{j}
&\eqqcolon
\frac{1}{\sqrt{2N}}
\sum_{k\in\Lambda^{\star}}
e^{+\ii k j}\,
\hat{c}^{\dag}_{k},
\\
\hat{c}^{\,}_{j}
&\eqqcolon
\frac{1}{\sqrt{2N}}
\sum_{k\in\Lambda^{\star}}
e^{-\ii k j}\,
\hat{c}^{\,}_{k},
\end{align}
where
\begin{equation}
\Lambda^{\star}=
\left\{\left.
k^{\,}_{m}=\frac{2\pi}{2N}m\,
\right|
m=-N+1,\ldots,N
\right\},
\end{equation}
\end{subequations}
into Hamiltonian \eqref{eq:H gFMz summary MF a}.
In doing so, we find that the Fermi sea of
\begin{subequations}
\begin{equation}
\widehat{H}^{\mathrm{gFM}^{\,}_{z}}_{\mathrm{MF}}(\epsilon)
=
E^{\,}_{\mathrm{MF}}(\bar{n},\bar{\chi};\epsilon)
+
\sum_{k\in\Lambda^{\star}}
\varepsilon^{\,}_{\mathrm{MF},k}(\bar{n},\bar{\chi};\epsilon)\,
\hat{c}^{\dag}_{k}\,\hat{c}^{\,}_{k},
\label{eq:H gFMz summary MF k}
\end{equation}
where
\begin{equation}
\varepsilon^{\,}_{\mathrm{MF},k}(\bar{n},\bar{\chi};\epsilon)\coloneqq
-
2\,
t^{\,}_{\mathrm{MF}}(\bar{n};\epsilon)\,
\cos(k)
-\mu^{\,}_{\mathrm{MF}}(\bar{n},\bar{\chi};\epsilon),
\end{equation}
is given by
\begin{equation}
\ket{\mathrm{FS}}\coloneqq
\prod_{k\in\Lambda^{\star}}^{\varepsilon^{\,}_{\mathrm{MF},k}(\bar{n},\bar{\chi};\epsilon)<0}
\hat{c}^{\dag}_{k}\ket{0}.
\label{eq:self-consistent Fermi sea}
\end{equation}
\end{subequations}
The pair \eqref{eq:self-consistency eqs lattice gFMz} of
self-consistency equations becomes
\begin{subequations}
\label{eq:self-consistency eqs Fourier gFMz}
\begin{align}
\bar{n}
&=
\frac{1}{2N}
\sum_{k\in\Lambda^{\star}}
\ev{\,\hat{c}^{\dag}_{k}\,\hat{c}^{\,}_{k}\,}{\mathrm{FS}},
\\
\bar{\chi}
&=
\frac{2}{2N}
\sum_{k\in\Lambda^{\star}}
\ev{\,\hat{c}^{\dag}_{k}\,\hat{c}^{\,}_{k}\,}{\mathrm{FS}}\,\cos(k).
\end{align}
\end{subequations}

At the free-fermion point $\epsilon=0$,
the solution $\bar{n}=\bar{n}(\epsilon)$
and $\bar{\chi}=\bar{\chi}(\epsilon)$
to Eq.~\eqref{eq:self-consistency eqs Fourier gFMz}
is exact and given by
\begin{equation}
\bar{n}(0)=1/2,
\qquad
\bar{\chi}(0)=2/\pi,
\end{equation}
respectively, in the thermodynamic limit $2N\to\infty$.

Upon adding interactions with $\epsilon\neq0$,
we numerically solve
the pair \eqref{eq:self-consistency eqs Fourier gFMz}
of coupled non-linear equations
in the limit $2N\to\infty$.
The numerical solver needs to be provided with
the pair of initial values $(\bar{n}^{\,}_{0},\bar{\chi}^{\,}_{0})$
of the solutions $(\bar{n},\bar{\chi})$.
We present in
Fig.~\ref{fig:MF Solutions}
the solutions $(\bar{n},\bar{\chi})$
for the initial values
\begin{equation}
(\bar{n}^{\,}_{0},\bar{\chi}^{\,}_{0})=
(0,1),\ (1/2,1),\ (1,1).
\label{eq:gFMz MF intial values}
\end{equation}
We make the following observations.
\begin{itemize}
\item[(i)]
When $\bar{n}^{\,}_{0}\neq1/2$, the solution
$\bar{n}(\epsilon)$
deviates from its trivial free-fermion value $\bar{n}(0)=1/2$,
decreasing (increasing) for $\epsilon\geq\epsilon^{\,}_{\mathrm{c}}$,
where
\begin{equation}
\epsilon^{\,}_{\mathrm{c}}\simeq 0.55,
\end{equation}
if $\bar{n}^{\,}_{0}<1/2$ ($\bar{n}^{\,}_{0}>1/2$).
Moreover, the solution
$\bar{\chi}(\epsilon)$ also
deviates from its trivial free-fermion value $\bar{\chi}(0)=2/\pi$,
decreasing for $\epsilon\geq\epsilon^{\,}_{\mathrm{c}}$
if $\bar{n}^{\,}_{0}\neq1/2$.
\item[(ii)]
When $\bar{n}^{\,}_{0}=1/2$, the pair
$(\bar{n},\bar{\chi})\equiv(1/2,2/\pi)$
is a solution for any $\epsilon$. 
\end{itemize}

Observation (i) implies the onset of two non-trivial solutions
at $\epsilon=\epsilon^{\,}_{c}$ towards a phase
where $\bar{n}\neq 1/2$.
According to Eqs.~\eqref{eq:relation local JW fermion local Z}
and \eqref{eq:expansion MF parameters gFMz}, this density imbalance
corresponds in the spin language to a non-vanishing uniform magnetization
\begin{equation}
m^{z}_{\mathrm{uni}}=2\bar{n}-1.	
\end{equation}
The two solutions with $\bar{n}\neq1/2$ for
$\epsilon\geq\epsilon^{\,}_{c}$ are therefore interpreted
at the mean-field level as the two symmetry-related
ferromagnetic states with $m^{z}_{\mathrm{uni}}>0$
and $m^{z}_{\mathrm{uni}}<0$.
This mean-field interpretation should, however,
be used with some care.
In one spatial dimension, quantum fluctuations can
invalidate an order parameter obtained at the mean-field level.
In the present case, however, $m^{z}_{\mathrm{uni}}$
breaks only a discrete Ising-like symmetry,
rather than the continuous U(1) symmetry generated by
fermion-number conservation.
There is therefore no Mermin-Wagner obstruction to a
non-zero $m^{z}_{\mathrm{uni}}$.
Quantum fluctuations may renormalize its magnitude,
and eventually suppress it at the phase boundary,
but they are not symmetry-forced to erase it
throughout the phase.
The numerical results of Sec.~\ref{ssub:the_texorpdfstring_mathrm_gfm}
confirm that this density imbalance survives in the full
quantum model.
Indeed, inside the $\mathrm{gFM}^{\,}_{z}$ phase,
the longitudinal correlator approaches the non-zero constant 
\begin{equation}
\lim_{s\to\infty}
C^{z}_{s^{\,}_{0},s^{\,}_{0}+s}=
\big(m^{z}_{\mathrm{uni}}\big)^{2},
\end{equation}
whereas the transverse correlators and the connected
longitudinal correlator decay algebraically with
incommensurate oscillations.
The two solutions $\bar{n}\neq1/2$ are thus understood
as the two stable solutions to the mean-field equations
\eqref{eq:self-consistency eqs Fourier gFMz}, i.e.,
the two global minima of the mean-field free energy
at zero temperature.
Notice that changing the initial value
$\bar{\chi}^{\,}_{0}$ of $\bar{\chi}$ does not alter the results.

Observation (ii) indicates the existence of a solution for
$\epsilon\geq\epsilon^{\,}_{c}$, which is the free-fermion
solution $\bar{n}=1/2$ and $\bar{\chi}=2/\pi$,
selected only by the initial value $\bar{n}^{\,}_{0}=1/2$
for any $\epsilon$.
This solution is the unstable solution to the mean-field equations
\eqref{eq:self-consistency eqs Fourier gFMz}, i.e.,
corresponds to a local maximum of the mean-field
free energy at zero temperature.

\paragraph{Onset of the $\mathrm{Neel}^{\,}_{z}$ phase.}

A staggered mean-field Ansatz designed to capture the
antiferromagnetic order along the $z$ axis predicts that the
free-fermion point $(\pi/4,0)$ is unstable for any $\lambda<0$.
This instability results from the opening of an energy gap that
is exponentially small in $1/|\lambda|$,
owing to a logarithmic divergence in the self-consistency equation.
This Ansatz thus overestimates the instability
toward the $\mathrm{Neel}^{\,}_{z}$ phase
by pushing the value of
$\lambda^{\prime}_{\mathrm{Neel}^{\,}_{z}}$
in Fig.\
\ref{fig:phase diagram U1}
up to the value $\lambda=0$.

\subsubsection{Bosonization along the $\mathrm{U}(1)$-symmetric line}
\label{ssub:bosonization_along_the_mathrm_u}

We complement the mean-field analysis of Sec.%
~\ref{subsubsec:Mean-field analysis of the transitions to the gFMz and Neelz}
with a bosonization of the Hamiltonian
$\widehat{H}^{(0)}_{\mathrm{U}(1)}(\lambda)$
defined in Eq.~\eqref{eq:def H U1}.
The bosonization is performed in the vicinity of the
free-fermion critical point $\lambda=0$.

This exercise serves two purposes.
(i) It provides the demonstration of the equivalence between
Hamiltonian \eqref{eq:def H U1} where
$\lambda^{\,}_{\mathrm{Neel}^{\,}_{z}}
<\lambda
<\lambda^{\,}_{\mathrm{gFM}^{\,}_{z}}$
and Hamiltonian \eqref{eq:def H XXZ} where
$-1<\Delta<1$,
as identified numerically in
Sec.~\ref{subsubsec:The Neelx-Neely phase boundary}.
(ii) It provides the continuum Hamiltonian
on which the discussion of the nature of the tricritical points
$\lambda=\lambda^{\,}_{\mathrm{gFM}^{\,}_{z}}$
and
$\lambda=\lambda^{\prime}_{\mathrm{Neel}^{\,}_{z}}$
will be built in
Secs.%
~\ref{ssub:lifshitz_instability_and_the_tricritical_point}
and%
~\ref{ssub:SU2 symmetric tricritical point},
respectively.

\paragraph{Fermionic Hamiltonian.}

Starting from Hamiltonian
$\widehat{H}^{(0)}_{\mathrm{U}(1)}$
defined in Eq.~\eqref{eq:def H U1},
we perform the Jordan-Wigner 
transformation \eqref{eq: def Jordan-Wigner trsf U(1) symmetric}
to obtain Hamiltonian
$\widehat{H}^{(0)}_{\mathrm{U}(1),\mathrm{f}}$
defined in Eq.~\eqref{eq:def HbU(1) fermionic as function lambda a}.
As we choose to impose PBCs and we demand that $2N/4$ is an integer,
the Jordan-Wigner fermions must obey the PBCs
\eqref{eq:case PBC fermion and parity + b}.
The fermions obey the algebra
\eqref{eq:def HbU(1) fermionic as function lambda c}.

We rewrite Hamiltonian
\eqref{eq:def HbU(1) fermionic as function lambda a} as
\begin{subequations}
\label{eq:bosonization U1 fermionic Hamiltonian}
\begin{align}
\widehat{H}^{(0)}_{\mathrm{U}(1),\mathrm{f}}(\lambda)
=&
-J(\lambda)\,
\sum_{j=1}^{2N}\,
\hat{\chi}^{\,}_{j,j+1}
\nonumber \\
&-4g(\lambda)
\sum_{j=1}^{2N}\,
\hat{\mathsf{n}}^{\,}_{j-1}\,
\hat{\mathsf{n}}^{\,}_{j+2}\,
\hat{\chi}^{\,}_{j,j+1}.
\label{eq:bosonization U1 fermionic Hamiltonian a}
\end{align}
where we denote with
\begin{align}
&\hat{\chi}^{\,}_{j,j'}
\coloneqq
\hat{c}^{\dag}_{j}\,
\hat{c}^{\,}_{j'}
+
\mathrm{H.c.},
\label{eq:bosonization U1 fermionic Hamiltonian b}
\end{align}
the fermionic hopping between sites $j$ and $j'$, and with
\begin{align}
\hat{\mathsf{n}}^{\,}_{j}
\coloneqq
\hat{n}^{\,}_{j}
-
\frac{1}{2},
\qquad
\hat{n}^{\,}_{j}
\coloneqq
\hat{c}^{\dag}_{j}\,
\hat{c}^{\,}_{j},
\label{eq:bosonization U1 fermionic Hamiltonian c}
\end{align}
the on-site fermionic density measured relative to half-filling.
Here, the pair $\hat{c}^{\dag}_{j}$ and $\hat{c}^{\,}_{j}$ of
fermionic operators obey the algebra
\eqref{eq:def HbU(1) fermionic as function lambda c}.
Finally, the pair of couplings
\begin{align}
J(\lambda)
\coloneqq
\sqrt{2}\cos\lambda,
\qquad
g(\lambda)
\coloneqq
\sqrt{2}\sin\lambda,
\label{eq:bosonization U1 fermionic Hamiltonian d}
\end{align}
\end{subequations}
encode the dependence on $\lambda$ of the hopping and
interaction strengths, respectively.

We are going to define
Wick normal ordering with respect to
the half-filled Fermi sea of the Hamiltonian
(\ref{eq:bosonization U1 fermionic Hamiltonian a})
at the free-fermion point $\lambda=0$.

First, the half-filled Fermi sea is constructed as follows.
We insert the Fourier transform \eqref{eq:Fourier MF}
into $\widehat{H}^{(0)}_{\mathrm{U}(1),\mathrm{f}}(0)$.
At $\lambda=0$, the single-particle dispersion is
\begin{subequations}
\begin{equation}
\varepsilon^{\,}_{k}=
-2\,\sqrt{2}\,\cos(k),
\label{eq:bosonization U1 free dispersion}
\end{equation}
and the half-filled non-interacting Fermi sea is
\begin{equation}
\ket{\mathrm{FS}}
\coloneqq
\prod_{k\in\Lambda^{\star}}^{\varepsilon^{\,}_{k}<0}
\hat{c}^{\dag}_{k}\ket{0},
\qquad
\hat{c}^{\,}_{k}\ket{0}=0,
\qquad
k\in\Lambda^{\star}.
\label{eq:bosonization U1 half filled Fermi sea}
\end{equation}
\end{subequations}
In the thermodynamic limit $2N\to\infty$,
the occupied momenta belong to the interval
\begin{equation}
-k^{\,}_{\mathrm{F}}<k<+k^{\,}_{\mathrm{F}},
\qquad
k^{\,}_{\mathrm{F}}=\frac{\pi}{2},
\end{equation}
and the Fermi points are denoted by $\pm\,k^{\,}_{\mathrm{F}}$.
Using the notation
\begin{equation}
\ev{\,\bm{\cdot}\,}{\mathrm{FS}}\equiv\ev{\,\bm{\cdot}\,},
\end{equation}
the reference state obeys
\begin{equation}
\ev*{\hat{n}^{\,}_{j}}=
\frac{1}{2},
\qquad
\ev*{\hat{\chi}^{\,}_{j,j+1}}=
\frac{2}{\pi},
\label{eq:bosonization U1 Fermi sea local averages}
\end{equation}
in the thermodynamic limit. 

Indeed, for any pair of lattice sites $j$ and $j+r$,
we use translational invariance to
define the Fermi-sea contraction
\begin{subequations}
\begin{equation}
C^{\,}_{r}\coloneqq
\ev*{\hat{c}^{\dag}_{j}\,\hat{c}^{\,}_{j+r}}.
\label{eq:bosonization U1 lattice contraction definition}
\end{equation}
In the thermodynamic limit,
\begin{equation}
C^{\,}_{r}=
\int\limits_{-\pi/2}^{+\pi/2}
\frac{\dd k}{2\pi}\,
e^{+\ii k\,r}=
\begin{cases}
\dfrac{1}{2},
&r=0,
\\[3mm]
\dfrac{1}{\pi\,r}
\sin\left(\dfrac{\pi\,r}{2}\right),
&r\neq0.
\end{cases}
\label{eq:bosonization U1 lattice contraction}
\end{equation}
In particular,
\begin{equation}
C^{\,}_{1}=
\frac{1}{\pi},
\qquad
C^{\,}_{2}=
0,
\qquad
C^{\,}_{3}=
-\frac{1}{3\pi}.
\label{eq:bosonization U1 relevant lattice contractions}
\end{equation}
\end{subequations}

Second, let $\hat{a}^{\,}_{1},\ldots,\hat{a}^{\,}_{m}$
denote fermionic creation or
annihilation operators and define the contraction of two such
operators with respect to the half-filled Fermi sea by
$\ev*{\hat{a}^{\,}_{i}\,\hat{a}^{\,}_{j}}$.
The Wick-normal-ordered product
${\bm{:}\,}\hat{a}^{\,}_{1}\cdots\hat{a}^{\,}_{m}{\,\bm{:}}$ is then
defined recursively through Wick's theorem,
\begin{equation}
\hat{a}^{\,}_{1}\cdots\,\hat{a}^{\,}_{m}=
\sum_{P}
(-1)^{\sigma(P)}
\left[
\prod_{(i,j)\in P}
\ev*{\hat{a}^{\,}_{i}\,\hat{a}^{\,}_{j}}\right]
{\bm{:}\,}
\prod_{k\notin P}\hat{a}^{\,}_{k}
{\,\bm{:}},
\label{eq:wick-normal-ordering}
\end{equation}
where the sum runs over all partial pairings $P$ of the operators,
including the empty pairing, the uncontracted operators retain their
original relative order, and $(-1)^{\sigma(P)}$ is the fermionic sign
associated with the permutations required to realize the contractions.
The term corresponding to the empty pairing is
${\bm{:}\,}\hat{a}^{\,}_{1}\cdots\hat{a}^{\,}_{m}{\,\bm{:}}$.

The Wick-normal-ordered lattice Hamiltonian
\eqref{eq:bosonization U1 fermionic Hamiltonian}
is
\begin{widetext}
\begin{subequations}
\label{eq:bosonization U1 Wick decomposition}
\begin{equation}
\widehat{H}^{(0)}_{\mathrm{U}(1),\mathrm{f}}(\lambda)=
\widehat{H}^{\,}_{0}(\lambda)
+
\widehat{H}^{\,}_{2}(\lambda)
+
\widehat{H}^{\,}_{4}(\lambda)
+
\widehat{H}^{\,}_{6}(\lambda),
\label{eq:bosonization U1 Wick decomposition a}
\end{equation}
where
\begin{align}
\widehat{H}^{\,}_{0}(\lambda)\coloneqq&
\left[
-
\frac{4N}{\pi}\,J(\lambda)
+
\frac{64N}{9\pi^{3}}\,g(\lambda)
\right]\,
\widehat{\mathds{1}},
\label{eq:bosonization U1 Wick decomposition b}
\\
\widehat{H}^{\,}_{2}(\lambda)\coloneqq&
\left[
-
J(\lambda)
+
\frac{28}{9\pi^{2}}\,g(\lambda)
\right]
\sum_{j=1}^{2N}\,
{\bm{:}\,}\hat{\chi}^{\,}_{j,j+1}{\,\bm{:}}
-
\frac{20}{3\pi^{2}}\,g(\lambda)\,
\sum_{j=1}^{2N}\,
{\bm{:}\,}\hat{\chi}^{\,}_{j-1,j+2}{\,\bm{:}},
\label{eq:bosonization U1 Wick decomposition c}
\\
\widehat{H}^{\,}_{4}(\lambda)\coloneqq&
-\frac{8}{\pi}\,g(\lambda)\,
\sum_{j=1}^{2N}\,
{\bm{:}\,}
\hat{n}^{\,}_{j-1}\,
\hat{n}^{\,}_{j+2}
{\,\bm{:}}
-
\frac{4}{3\pi}\,g(\lambda)\,
\sum_{j=1}^{2N}\,
{\bm{:}\,}
\hat{\chi}^{\,}_{j-1,j+2}\,
\hat{\chi}^{\,}_{j,j+1}
{\,\bm{:}}
\nonumber\\
&-
\frac{4}{\pi}\,g(\lambda)\,
\sum_{j=1}^{2N}\left(
{\bm{:}\,}
\hat{c}^{\dag}_{j-1}\,
\hat{c}^{\dag}_{j}\,
\hat{c}^{\,}_{j-1}\,
\hat{c}^{\,}_{j+2}
{\,\bm{:}}
+
{\bm{:}\,}
\hat{c}^{\dag}_{j-1}\,
\hat{c}^{\dag}_{j+2}\,
\hat{c}^{\,}_{j-1}\,
\hat{c}^{\,}_{j}
{\,\bm{:}}
-
{\bm{:}\,}
\hat{c}^{\dag}_{j-1}\,
\hat{c}^{\dag}_{j+2}\,
\hat{c}^{\,}_{j+2}\,
\hat{c}^{\,}_{j+1}
{\,\bm{:}}
-
{\bm{:}\,}
\hat{c}^{\dag}_{j+2}\,
\hat{c}^{\dag}_{j+1}\,
\hat{c}^{\,}_{j-1}\,
\hat{c}^{\,}_{j+2}
{\,\bm{:}}
\right),
\label{eq:bosonization U1 Wick decomposition d}
\\
\widehat{H}^{\,}_{6}(\lambda)\coloneqq&
\,4g(\lambda)\,
\sum_{j=1}^{2N}
{\bm{:}\,}
\hat{n}^{\,}_{j-1}\,
\hat{n}^{\,}_{j+2}\,
\hat{\chi}^{\,}_{j,j+1}
{\,\bm{:}}.
\label{eq:bosonization U1 Wick decomposition e}
\end{align}
\end{subequations}
\end{widetext}

\paragraph{Continuum limit.}

We now take the continuum limit of
the Wick-normal-ordered lattice Hamiltonian
\eqref{eq:bosonization U1 Wick decomposition}.
Given the lattice spacing $\mathfrak{a}$ of the chain of length
$L=2N\,\mathfrak{a}$, we introduce the continuum coordinate
\begin{equation}
x\coloneqq
j\,\mathfrak{a},
\qquad
j=1,\ldots,2N.
\label{eq:bosonization U1 continuum coordinate}
\end{equation}
The half-filled Fermi sea has the pair of physical Fermi momenta
$\pm\,k^{\,}_{\mathrm{F}}/\mathfrak{a}$.
We introduce the slowly varying
right- and left-moving fermionic fields
$\widehat{\psi}^{\,}_{R}(x)$ and $\widehat{\psi}^{\,}_{L}(x)$
through the substitution
\begin{equation}
\frac{\hat{c}^{\,}_{j}}{\sqrt{\mathfrak{a}}}
\longrightarrow
e^{+\ii \frac{\pi}{2\mathfrak{a}}x}\,
\widehat{\psi}^{\,}_{R}(x)
+
e^{-\ii \frac{\pi}{2\mathfrak{a}}x}\,
\widehat{\psi}^{\,}_{L}(x).
\label{eq:bosonization U1 continuum decomposition}
\end{equation}
The continuum fermions obey the equal-time algebra
\begin{subequations}
\label{eq:bosonization U1 continuum fermion algebra}
\begin{align}
&\bigl\{
\widehat{\psi}^{\,}_{R}(x),\widehat{\psi}^{\,\dag}_{R}(y)
\bigr\}
=
\bigl\{
\widehat{\psi}^{\,}_{L}(x),\widehat{\psi}^{\,\dag}_{L}(y)
\bigr\}
=
\delta(x-y),
\\
&\bigl\{
\widehat{\psi}^{\,}_{R}(x),\widehat{\psi}^{\,}_{L}(y)
\bigr\}
=
\bigl\{
\widehat{\psi}^{\,}_{R}(x),\widehat{\psi}^{\,\dag}_{L}(y)
\bigr\}
=
0.
\end{align}
\end{subequations}
The lattice sum is replaced by
\begin{equation}
\sum_{j=1}^{2N}
\longrightarrow
\frac{1}{\mathfrak{a}}
\int\dd x,
\end{equation}
and we have the Taylor expansion
\begin{equation}
\widehat{\psi}^{\,}_{M}(x+r\mathfrak{a})
=
\sum_{n=0}^{\infty}
\frac{(r\mathfrak{a})^{n}}{n!}\,
\partial^{n}_{x}\widehat{\psi}^{\,}_{M}(x),
\qquad
r\in\mathbb{Z},
\end{equation}
where $M=L,R$.

Wick normal ordering in the continuum is inherited from
taking the continuum limit of Eq.\
(\ref{eq:bosonization U1 Wick decomposition a})
through (i) the substitution
\eqref{eq:bosonization U1 continuum decomposition},
(ii) performing a Taylor expansion in powers of $\mathfrak{a}$
(a gradient expansion), (iii) ignoring any composite operator
that is multiplied by $\exp(\ii\,n\,\frac{\pi}{2\mathfrak{a}}x)$
for any $n\not\equiv0\bmod 4$, and (iv) truncating the Taylor
expansion so as to keep local composite operators
whose bare scaling dimensions are not larger than four.
The continuum limit of Eq.\
(\ref{eq:bosonization U1 Wick decomposition a})
thus results in the substitution
\begin{equation}
\widehat{H}^{(0)}_{\mathrm{U}(1),\mathrm{f}}(\lambda)
-
\widehat{H}^{\,}_{0}(\lambda)\to
\int\mathrm{d}x\,
\left[
\widehat{\mathcal{H}}^{\,}_{2}(\lambda)
+
\widehat{\mathcal{H}}^{\,}_{4}(\lambda)
+
\widehat{\mathcal{H}}^{\,}_{6}(\lambda)
\right].
\label{eq:bosonization U1 fermionic continuum Hamiltonian}
\end{equation}

The continuum limit of the quadratic contribution
$\widehat{H}^{\,}_{2}(\lambda)$
defined in Eq.\ (\ref{eq:bosonization U1 Wick decomposition c})
is the space integral over the local composite operator
given by, up to total derivatives,
\begin{subequations}
\label{eq:bosonization U1 continuum quadratic Hamiltonian}
\begin{align}
\widehat{\mathcal{H}}^{\,}_{2}(\lambda)=&\,
-
\ii
v^{\,}_{2}(\lambda)
\left(
{\bm{:}\,}
\widehat{\psi}^{\,\dag}_{R}\,\partial^{\,}_{x}\widehat{\psi}^{\,}_{R}
{\,\bm{:}}
-
{\bm{:}\,}
\widehat{\psi}^{\,\dag}_{L}\,\partial^{\,}_{x}\widehat{\psi}^{\,}_{L}
{\,\bm{:}}
\right)
\nonumber\\
&\,
+
\ii
\gamma^{\,}_{2}(\lambda)
\left(
{\bm{:}\,}
\widehat{\psi}^{\,\dag}_{R}\,\partial^{3}_{x}\widehat{\psi}^{\,}_{R}{\,\bm{:}}
-
{\bm{:}\,}
\widehat{\psi}^{\,\dag}_{L}\,\partial^{3}_{x}\widehat{\psi}^{\,}_{L}
{\,\bm{:}}
\right)
\nonumber\\
&\,
+
\cdots,
\label{eq:bosonization U1 continuum quadratic Hamiltonian a}
\end{align}
where
\begin{align}
v^{\,}_{2}(\lambda)
&\coloneqq
\mathfrak{a}
\left[
2\,J(\lambda)
-
\frac{416}{9\pi^{2}}\,g(\lambda)
\right],
\label{eq:bosonization U1 continuum quadratic Hamiltonian b} 
\\
\gamma^{\,}_{2}(\lambda)
&\coloneqq
\mathfrak{a}^{3}
\left[
-\frac{1}{3}\,J(\lambda)
+
\frac{1648}{27\pi^{2}}\,g(\lambda)
\right].
\label{eq:bosonization U1 continuum quadratic Hamiltonian c}
\end{align}
\end{subequations}
The first line on the right-hand side of
Eq.~\eqref{eq:bosonization U1 continuum quadratic Hamiltonian a}
has the bare scaling dimension two,
whereas the second line has the bare scaling dimension four.
The dots refer to local composite operators whose bare scaling
dimensions are larger than four.

The continuum limit of the quartic contribution
$\widehat{H}^{\,}_{4}(\lambda)$
defined in Eq.\ (\ref{eq:bosonization U1 Wick decomposition d})
is the space integral over the local composite operator
given by, up to total derivatives,
\begin{subequations}
\label{eq:bosonization U1 continuum quartic Hamiltonian}
\begin{align}
\widehat{\mathcal{H}}^{\,}_{4}(\lambda)=&\,
-\frac{256\,\mathfrak{a}}{3\pi}\,g(\lambda)\,
{\bm{:}\,}
\widehat{\rho}^{\,}_{R}\,\widehat{\rho}^{\,}_{L}
{\,\bm{:}}
\nonumber\\
&\,
+
\frac{128\,\mathfrak{a}^{3}}{\pi}\,g(\lambda)\,
\Big(
\widehat{\mathcal{K}}^{\,}_{R}
+
\widehat{\mathcal{K}}^{\,}_{L}
\Big)
\nonumber\\
&\,
+
\frac{4\,\mathfrak{a}^{3}}{\pi}\,g(\lambda)\,
\Big(
\widehat{\mathcal{D}}^{\,}_{RL}
+
\widehat{\mathcal{D}}^{\dag}_{RL}
\Big)
\nonumber\\
&\,
-
\frac{320\,\mathfrak{a}^{3}}{3\pi}\,g(\lambda)\,
\Big(
\widehat{\mathcal{U}}
+
\widehat{\mathcal{U}}^{\dag}
\Big)
\nonumber\\
&\,
+
\cdots,
\label{eq:bosonization U1 continuum quartic Hamiltonian a}
\end{align}
where the first line on the right-hand side
is the current-current interaction in the left- and right-mover
representation with the bare scaling dimension two that has been expressed
as a bilinear in the chiral densities
\begin{equation}
\widehat{\rho}^{\,}_{M}(x)\coloneqq
{\bm{:}\,}
\widehat{\psi}^{\,\dag}_{M}(x)\,
\widehat{\psi}^{\,}_{M}(x)
{\,\bm{:}},
\qquad
M=L,R,
\label{eq:bosonization U1 continuum quartic Hamiltonian b}
\end{equation}
the second line on the right-hand side is expressed in terms of
the same-chirality four-fermion derivative operators
\begin{align}
\widehat{\mathcal{K}}^{\,}_{R}\coloneqq&\,
{\bm{:}\,}
\widehat{\psi}^{\,\dag}_{R}\,
\bigl(\partial^{\,}_{x}\widehat{\psi}^{\,\dag}_{R}\bigr)\,
\widehat{\psi}^{\,}_{R}\,
\bigl(\partial^{\,}_{x}\widehat{\psi}^{\,}_{R}\bigr)
{\,\bm{:}},
\label{eq:bosonization U1 continuum quartic Hamiltonian c}
\\
\widehat{\mathcal{K}}^{\,}_{L}\coloneqq&\,
{\bm{:}\,}
\widehat{\psi}^{\,\dag}_{L}\,
\bigl(\partial^{\,}_{x}\widehat{\psi}^{\,\dag}_{L}\bigr)\,
\widehat{\psi}^{\,}_{L}\,
\bigl(\partial^{\,}_{x}\widehat{\psi}^{\,}_{L}\bigr)
{\,\bm{:}},
\label{eq:bosonization U1 continuum quartic Hamiltonian d}
\end{align}
with the bare scaling dimension four,
the third line on the right-hand side is expressed in terms of
the mixed-chirality four-fermion derivative operators
\begin{align}
\widehat{\mathcal{D}}^{\,}_{RL}\coloneqq&\,
+24\
{\bm{:}\,}
\widehat{\psi}^{\,\dag}_{R}\,
\widehat{\psi}^{\,\dag}_{L}\,
\widehat{\psi}^{\,}_{R}\,
\bigl(\partial^{2}_{x}\widehat{\psi}^{\,}_{L}\bigr)
{\,\bm{:}}
\nonumber\\
&
-
\frac{80}{3}\
{\bm{:}\,}
\widehat{\psi}^{\,\dag}_{R}\,
\widehat{\psi}^{\,\dag}_{L}\,
\bigl(\partial^{\,}_{x}\widehat{\psi}^{\,}_{R}\bigr)\,
\bigl(\partial^{\,}_{x}\widehat{\psi}^{\,}_{L}\bigr)
{\,\bm{:}}
\nonumber\\
&
+24\
{\bm{:}\,}
\widehat{\psi}^{\,\dag}_{R}\,
\widehat{\psi}^{\,\dag}_{L}\,
\bigl(\partial^{2}_{x}\widehat{\psi}^{\,}_{R}\bigr)\,
\widehat{\psi}^{\,}_{L}
{\,\bm{:}}
\nonumber\\
&
+
\frac{16}{3}\
{\bm{:}\,}
\widehat{\psi}^{\,\dag}_{R}\,
\bigl(\partial^{\,}_{x}\widehat{\psi}^{\,\dag}_{L}\bigr)\,
\widehat{\psi}^{\,}_{R}\,
\bigl(\partial^{\,}_{x}\widehat{\psi}^{\,}_{L}\bigr)
{\,\bm{:}}
\nonumber\\
&
+
\frac{32}{3}\
{\bm{:}\,}
\widehat{\psi}^{\,\dag}_{R}\,
\bigl(\partial^{\,}_{x}\widehat{\psi}^{\,\dag}_{L}\bigr)\,
\bigl(\partial^{\,}_{x}\widehat{\psi}^{\,}_{R}\bigr)\,
\widehat{\psi}^{\,}_{L}
{\,\bm{:}},
\label{eq:bosonization U1 continuum quartic Hamiltonian e}
\end{align}
with the bare scaling dimension four,
the fourth line on the right-hand side is the umklapp operator
\begin{equation}
\widehat{\mathcal{U}}
\coloneqq
{\bm{:}\,}
\widehat{\psi}^{\,\dag}_{R}
\bigl(\partial^{\,}_{x}\widehat{\psi}^{\,\dag}_{R}\bigr)\,
\widehat{\psi}^{\,}_{L}
\bigl(\partial^{\,}_{x}\widehat{\psi}^{\,}_{L}\bigr)
{\,\bm{:}},
\label{eq:bosonization U1 continuum quartic Hamiltonian f}
\end{equation}
\end{subequations}
with the bare scaling dimension four, and the dots on the last line
refer to local composite operators whose bare scaling
dimensions are larger than four.

Finally, the continuum limit of the sextic contribution
$\widehat{H}^{\,}_{6}(\lambda)$
defined in Eq.\ (\ref{eq:bosonization U1 Wick decomposition e})
is the space integral over the local composite operators
\begin{equation}
\widehat{\mathcal{H}}^{\,}_{6}(\lambda)=\cdots,
\label{eq:bosonization U1 continuum six fermion contribution}
\end{equation}
whose bare scaling dimensions are larger than four.

\paragraph{Bosonic representation.}

Our goal is to bosonize the fermionic continuum Hamiltonian
\eqref{eq:bosonization U1 fermionic continuum Hamiltonian}.
To this end, we introduce a pair of free chiral bosonic fields
$\widehat{\varphi}^{\,}_{R}$ and $\widehat{\varphi}^{\,}_{L}$.
At the free-fermion point $\lambda=0$, they obey
\begin{subequations}
\begin{align}
&
\partial^{\,}_{t}\widehat{\varphi}^{\,}_{M}(t,x)=
-
\sigma^{\,}_{M}\,
v^{\,}_{\mathrm{F}}\,
\partial^{\,}_{x}\widehat{\varphi}^{\,}_{M}(t,x),
\qquad
M=L,R,
\\
&
\sigma^{\,}_{R}=-\sigma^{\,}_{L}=+1,
\end{align}
so that
$\widehat{\varphi}^{\,}_{R}$ and $\widehat{\varphi}^{\,}_{L}$ propagate toward
increasing and decreasing $x$ with the speed
$v^{\,}_{\mathrm{F}}=2\sqrt{2}\,\mathfrak{a}$,
respectively.
Their equal-time algebra is
\begin{align}
&
[\widehat{\varphi}^{\,}_{M}(x),\widehat{\varphi}^{\,}_{M}(y)]=
\ii\pi\sigma^{\,}_{M}\sgn(x-y),
\
M=L,R,
\\
&
[\widehat{\varphi}^{\,}_{R}(x),\widehat{\varphi}^{\,}_{L}(y)]=
\ii\pi.
\end{align}
\end{subequations}
The constant mixed commutator is understood
to reside in the zero-mode sector.
We therefore decompose the full chiral fields as
\begin{equation}
\widehat{\varphi}^{\,}_{M}(x)=
\widehat{\varphi}^{\,0}_{M}
+
\widehat{\varphi}^{\,\mathrm{osc}}_{M}(x),
\qquad
M=L,R,
\end{equation}
where all zero modes commute with all oscillator modes.

With respect to the Fock vacuum of the free chiral Hamiltonian,
we decompose the nonzero-momentum oscillator part as
\begin{subequations}
\label{eq:def-osc-expansion-widehatvarphiM}
\begin{equation}
\widehat{\varphi}^{\,\mathrm{osc}}_{M}(x)=
\widehat{\varphi}^{\,<}_{M}(x)
+
\widehat{\varphi}^{\,>}_{M}(x),
\label{eq:def-osc-expansion-widehatvarphiM-a}
\end{equation}
with
\begin{equation}
\langle 0|\,\widehat{\varphi}^{\,<}_{M}(x)=0,
\qquad
\widehat{\varphi}^{\,>}_{M}(x)\,|0\rangle=0,
\label{eq:def-osc-expansion-widehatvarphiM-b}
\end{equation}
for $M=L,R$.
Oscillator components of opposite chirality commute.
The only independent nonvanishing oscillator commutators are
\begin{equation}
\left[
\widehat{\varphi}^{\,>}_{M}(x),\widehat{\varphi}^{\,<}_{M}(y)
\right]=
-
\log
\left(
\frac{\mathsf{a}-\ii\sigma^{\,}_{M}(x-y)}{\ell}
\right)
\label{eq:def-osc-expansion-widehatvarphiM-c}
\end{equation}
\end{subequations}
for $M=L,R$.
Here, $\mathsf{a}>0$ is a short-distance
regulator and $\ell>0$ is an infrared reference length fixing the
additive constant of the massless contraction.
For $\mathsf{a}>0$,
this prescription gives
\begin{equation}
[\widehat{\varphi}^{\,\mathrm{osc}}_{M}(x),
\widehat{\varphi}^{\,\mathrm{osc}}_{M}(y)]=
\log
\left(
\frac{\mathsf{a}+\ii\sigma^{\,}_{M}(x-y)}
{\mathsf{a}-\ii\sigma^{\,}_{M}(x-y)}
\right)
\end{equation}
for $M=L,R$,
which approaches $\ii\pi\sigma^{\,}_{M}\sgn(x-y)$ as
$\mathsf{a}\to0^+$ provided $x\neq y$
and the principal branch of the logarithm is chosen.

The zero-mode exponentials are denoted by
\begin{equation}
\widehat{U}^{\,}_{M,q}\equiv
e^{\ii q\,\widehat{\varphi}^{\,0}_{M}},
\qquad
q\in\mathbb Z,
\qquad
M=L,R.
\end{equation}
They obey the Weyl algebra
\begin{subequations}
\begin{align}
&
\widehat{U}^{\,}_{M,q}\,
\widehat{U}^{\,}_{M,q'}=
\widehat{U}^{\,}_{M,q+q'},
\\
&
\widehat{U}^{\,}_{R,q}\,
\widehat{U}^{\,}_{L,q'}=
e^{-\ii\pi q\,q'}\,
\widehat{U}^{\,}_{L,q'}\,
\widehat{U}^{\,}_{R,q},
\\
&
\widehat{U}^{\dag}_{M,q}=
\widehat{U}^{\,}_{M,-q},
\end{align}
\end{subequations}
for $M=L,R$. Equivalently, the mixed Weyl relation follows formally from
\begin{equation}
[\widehat{\varphi}^{\,0}_{R},\widehat{\varphi}^{\,0}_{L}]=
\ii\pi.
\end{equation}

Let $\widehat{N}^{\,}_{R}$ and $\widehat{N}^{\,}_{L}$ denote the chiral
charge operators. They commute pairwise
\begin{subequations}
\begin{equation}
[\widehat{N}^{\,}_{M},\widehat{N}^{\,}_{M'}]=0,
\qquad
M,M'=L,R.
\end{equation}
Their algebra with the zero-mode exponentials is
\begin{equation}
[\widehat{N}^{\,}_{M},\widehat{U}^{\,}_{M',q}]=
q\,\delta^{\,}_{MM'}\,\widehat{U}^{\,}_{M',q},
\quad
M,M'=L,R.
\end{equation}
\end{subequations}
The neutral bosonic Fock vacuum obeys
\begin{equation}
\widehat{N}^{\,}_{M}\,|0\rangle=0,
\qquad
\widehat{\varphi}^{\,>}_{M}(x)\,|0\rangle=0,
\qquad
M=L,R.
\end{equation}
We define the vacua in the charged zero-mode sectors by
\begin{equation}
|n^{\,}_{R},n^{\,}_{L}\rangle\coloneqq
\widehat{U}^{\,}_{R,n^{\,}_{R}}\,
\widehat{U}^{\,}_{L,n^{\,}_{L}}\,
|0\rangle,
\qquad
n^{\,}_{R},n^{\,}_{L}\in\mathbb Z.
\end{equation}
They satisfy
\begin{equation}
\widehat{N}^{\,}_{M}\,|n^{\,}_{R},n^{\,}_{L}\rangle=
n^{\,}_{M}\,|n^{\,}_{R},n^{\,}_{L}\rangle,
\qquad
M=L,R.
\end{equation}
With this phase convention, the action of the
zero-mode exponentials on the charged oscillator vacua is
\begin{subequations}
\begin{align}
&
\widehat{U}^{\,}_{R,q}\,|n^{\,}_{R},n^{\,}_{L}\rangle=
|n^{\,}_{R}+q,n^{\,}_{L}\rangle,
\\
&
\widehat{U}^{\,}_{L,q}\,|n^{\,}_{R},n^{\,}_{L}\rangle=
e^{\ii\pi q\,n^{\,}_{R}}\,|n^{\,}_{R},n^{\,}_{L}+q\rangle.
\end{align}
\end{subequations}
In particular,
\begin{subequations}
\begin{align}
&
\widehat{U}^{\,}_{R,q}\,|0\rangle=
|q,0\rangle,
\\
&
\widehat{U}^{\,}_{L,q}\,|0\rangle=
|0,q\rangle.
\end{align}
\end{subequations}

We define the cutoff-normalized oscillator vertex operator by
\begin{equation}
\begin{split}
\widehat{W}^{\,}_{M,q;\mathsf{a}/\ell}(x)\equiv&\,
\left(\frac{\mathsf{a}}{\ell}\right)^{q^{2}/2}\,
{\bm{:}\,} e^{\ii q\,\widehat{\varphi}^{\,\mathrm{osc}}_{M}(x)}{\,\bm{:}}
\\
\coloneqq&\,
\left(\frac{\mathsf{a}}{\ell}\right)^{q^{2}/2}\,
e^{\ii q\,\widehat{\varphi}^{\,<}_{M}(x)}
e^{\ii q\,\widehat{\varphi}^{\,>}_{M}(x)}
\end{split}
\end{equation}
for $M=L,R$.
Equivalently, by the Baker-Campbell-Hausdorff formula,
\begin{subequations}
\begin{equation}
\begin{split}
&
\widehat{W}^{\,}_{M,q;\mathsf{a}/\ell}(x)=
\\
&\quad
\left(\frac{\mathsf{a}}{\ell}\right)^{q^{2}/2}
e^{
\frac{(\ii q)^{2}}{2}\,
\left[\widehat{\varphi}^{\,<}_{M}(x),\widehat{\varphi}^{\,>}_{M}(x)\right]
+
\ii q\,\widehat{\varphi}^{\,<}_{M}(x)
+
\ii q\,\widehat{\varphi}^{\,>}_{M}(x)
  },
\end{split}
\end{equation}
which, owing to Eqs.\
(\ref{eq:def-osc-expansion-widehatvarphiM-a})
and
(\ref{eq:def-osc-expansion-widehatvarphiM-c}),
simplifies to
\begin{align}
\widehat{W}^{\,}_{M,q;\mathsf{a}/\ell}(x)=&\,
e^{\ii q\,\widehat{\varphi}^{\,\mathrm{osc}}_{M}(x)},
\qquad
M=L,R,
\end{align}
\end{subequations}
for the oscillator field regulated at $\mathsf{a}>0$.

The cutoff-normalized full vertex operator is
\begin{equation}
\begin{split}
\widehat{V}^{\,}_{M,q;\mathsf{a}/\ell}(x)\coloneqq&\,
\widehat{U}^{\,}_{M,q}\,
\widehat{W}^{\,}_{M,q;\mathsf{a}/\ell}(x)
\\
=&\,
\left(\frac{\mathsf{a}}{\ell}\right)^{q^{2}/2}
\widehat{U}^{\,}_{M,q}\,
e^{\ii q\,\widehat{\varphi}^{\,<}_{M}(x)}
e^{\ii q\,\widehat{\varphi}^{\,>}_{M}(x)}
\end{split}
\end{equation}
for $M=L,R$. Since the zero modes commute with all oscillator modes,
we may write this formally as
\begin{align}
\widehat{V}^{\,}_{M,q;\mathsf{a}/\ell}(x)=&\,
\left(\frac{\mathsf{a}}{\ell}\right)^{q^{2}/2}
{\bm{:}\,}e^{\ii q\,\widehat{\varphi}^{\,}_{M}(x)}{\,\bm{:}}
\nonumber\\
=&\,
e^{\ii q\,\widehat{\varphi}^{\,}_{M}(x)},
\qquad
M=L,R,
\end{align}
where normal ordering acts on the oscillator sector and the zero-mode
operator $\widehat{U}^{\,}_{M,q}$ is placed to the left in the first
equality. A useful identity needed for bosonization of point-split
composite operators of same chirality is
\begin{equation}
\begin{split}
&
\widehat{V}^{\,}_{M,q;\mathsf{a}/\ell}(x)\,
\widehat{V}^{\,}_{M,q';\mathsf{a}/\ell}(y)=
\\
&\
\left[
\frac{\mathsf{a}+\ii\sigma^{\,}_{M}(x-y)}
{\mathsf{a}-\ii\sigma^{\,}_{M}(x-y)}
\right]^{-\frac{q\,q'}{2}}\,
\widehat{U}^{\,}_{M,q+q'}\,
e^{
\ii q\,\widehat{\varphi}^{\,\mathrm{osc}}_{M}(x)
+
\ii q'\,\widehat{\varphi}^{\,\mathrm{osc}}_{M}(y)
  }
\end{split}
\label{useful-indentity-point-split-composites-for-boso}
\end{equation}
for $M=L,R$.

For opposite chiralities, the zero-mode Weyl algebra gives
\begin{equation}
\widehat{V}^{\,}_{R,q;\mathsf{a}/\ell}(x)\,
\widehat{V}^{\,}_{L,q';\mathsf{a}/\ell}(y)=
e^{-\ii\pi qq'}\,
\widehat{V}^{\,}_{L,q';\mathsf{a}/\ell}(y)\,
\widehat{V}^{\,}_{R,q;\mathsf{a}/\ell}(x).
\end{equation}
Consequently, for $q,q'=\pm1$,
\begin{equation}
\widehat{V}^{\,}_{R,q;\mathsf{a}/\ell}(x)\,
\widehat{V}^{\,}_{L,q';\mathsf{a}/\ell}(y)=
-
\widehat{V}^{\,}_{L,q';\mathsf{a}/\ell}(y)\,
\widehat{V}^{\,}_{R,q;\mathsf{a}/\ell}(x).
\end{equation}
At nonzero $\mathsf{a}$ and $x-y$,
the equal-chirality exchange relation is
\begin{equation}
\begin{split}
&
\widehat{V}^{\,}_{M,q;\mathsf{a}/\ell}(x)\,
\widehat{V}^{\,}_{M,q';\mathsf{a}/\ell}(y)=
\\
&\qquad
\left[
\frac{\mathsf{a}+\ii\sigma^{\,}_{M}(x-y)}
     {\mathsf{a}-\ii\sigma^{\,}_{M}(x-y)}
     \right]^{-q\,q'}\,
\widehat{V}^{\,}_{M,q';\mathsf{a}/\ell}(y)\,
\widehat{V}^{\,}_{M,q;\mathsf{a}/\ell}(x),
\end{split}
\end{equation}
which gives
\begin{equation}
\begin{split}
&
\widehat{V}^{\,}_{M,q;\mathsf{a}/\ell}(x)\,
\widehat{V}^{\,}_{M,q';\mathsf{a}/\ell}(y)=
\\
&\qquad\qquad
e^{-\ii\pi\sigma^{\,}_{M}qq'\sgn(x-y)}\,
\widehat{V}^{\,}_{M,q';\mathsf{a}/\ell}(y)\,
\widehat{V}^{\,}_{M,q;\mathsf{a}/\ell}(x)
\end{split}
\end{equation}
in the limit $\mathsf{a}\to0^+$. Hence, the $q=\pm1$ full vertex
operators anticommute within a given chirality at distinct points and
between opposite chiralities. We choose $\ell$ to be the lattice spacing
$\mathfrak{a}$ and identify
\begin{subequations}
\label{eq:bosonization U1 chiral fermion bosonization}
\begin{align}
\widehat{\psi}^{\,}_{R}(x)
&\coloneqq
\frac{1}{\sqrt{2\pi\mathfrak{a}}}\,
{\bm{:}\,}
e^{+\ii\widehat{\varphi}^{\,}_{R}(x)}
{\,\bm{:}},
\label{eq:bosonization U1 chiral fermion bosonization a}
\\
\widehat{\psi}^{\,}_{L}(x)
&\coloneqq
\frac{1}{\sqrt{2\pi\mathfrak{a}}}\,
{\bm{:}\,}
e^{-\ii\widehat{\varphi}^{\,}_{L}(x)}
{\,\bm{:}},
\label{eq:bosonization U1 chiral fermion bosonization b}
\end{align}
\end{subequations}
with the Fermi fields obeying the fermion algebra
\eqref{eq:def HbU(1) fermionic as function lambda c},
in the limit $\mathfrak{a}\to0^{+}$.

Any local composite operator built out of
$\widehat{\psi}^{\,\dag}_{M}(x)$
or
$\widehat{\psi}^{\,}_{M}(x)$
and their derivatives
that enters the right-hand side of Eq.\
(\ref{eq:bosonization U1 fermionic continuum Hamiltonian})
can be bosonized using (i) point splitting,
(ii) Eq.\
(\ref{eq:bosonization U1 chiral fermion bosonization}),
(iii) and a Taylor expansion in powers of $x-y$ of
(\ref{useful-indentity-point-split-composites-for-boso}).
There follows the local bosonization rules 
\begin{subequations}
\begin{align}
&
-\ii\sigma^{\,}_{M}\,
{\bm{:}\,}
\widehat{\psi}^{\,\dag}_{M}\partial^{\,}_{x}\widehat{\psi}^{\,}_{M}
{\,\bm{:}}\longrightarrow
\pi\,
{\bm{:}\,}
\widehat{\rho}^{2}_{M}
{\,\bm{:}}\,,
\\
&
+\ii\sigma^{\,}_{M}\,
{\bm{:}\,}
\widehat{\psi}^{\,\dag}_{M}\partial^{3}_{x}\widehat{\psi}^{\,}_{M}
{\,\bm{:}}\longrightarrow
\left[
\frac{\pi}{2}\,
{\bm{:}\,}
(\partial^{\,}_{x}\widehat{\rho}^{\,}_{M})^{2}
{\,\bm{:}}
+
2\pi^{3}\,
{\bm{:}\,}
(\widehat{\rho}^{\,}_{M})^{4}
{\,\bm{:}}
\right],
\end{align}
\end{subequations}
for the fermionic composite operators of quadratic order
on the right-hand side of Eq.\
(\ref{eq:bosonization U1 continuum quadratic Hamiltonian a}),
whereby we observe that
\begin{subequations}
\begin{align}
&
{\bm{:}\,}
\widehat{\psi}^{\,\dag}_{M}\,
\widehat{\psi}^{\,}_{M}
{\,\bm{:}}\longrightarrow
\frac{1}{2\pi}\,
\partial^{\,}_{x}\widehat{\varphi}^{\,}_{M}
\equiv
\widehat{\rho}^{\,}_{M},
\\
&
{\bm{:}\,}
\widehat{\psi}^{\,\dag}_{M}\,
\partial^{\,}_{x}\widehat{\psi}^{\,\dag}_{M}
{\,\bm{:}}\longrightarrow
\sigma^{\,}_{M}\,
\frac{\ii}{2\pi\mathfrak{a}^{2}}\,
{\bm{:}\,}
e^{-2\ii\widehat{\varphi}^{\,}_{M}}
{\,\bm{:}}\,.
\end{align}  
\end{subequations}
For the fermionic composite operators of quartic order
on the right-hand side of Eq.\
(\ref{eq:bosonization U1 continuum quartic Hamiltonian a}),
we find the bosonization rules
\begin{subequations}
\begin{align}
&
\widehat{\mathcal{K}}^{\,}_{M}\longrightarrow
+
\frac{5}{12}\,
{\bm{:}\,}
(\partial^{\,}_{x}\widehat{\rho}^{\,}_{M})^{2}
{\,\bm{:}}
-
\frac{\pi^{2}}{3}\,
{\bm{:}\,}
\widehat{\rho}^{4}_{M}
{\,\bm{:}}\,,
\\
&
\widehat{\mathcal{D}}^{\,}_{RL}
+
\widehat{\mathcal{D}}^{\dag}_{RL}\longrightarrow
+
\frac{376}{9}\,
{\bm{:}\,}
(\partial^{\,}_{x}\widehat{\rho}^{\,}_{R})\,
(\partial^{\,}_{x}\widehat{\rho}^{\,}_{L})
{\,\bm{:}}
\nonumber\\
&
\hphantom{
\left(
\widehat{\mathcal{D}}^{\,}_{RL}
+
\widehat{\mathcal{D}}^{\dag}_{RL}
\right)=
         }
+
\frac{224\,\pi^{2}}{3}\,
{\bm{:}\,}
\widehat{\rho}^{2}_{R}\,
\widehat{\rho}^{2}_{L}
{\,\bm{:}}
\nonumber\\
&
\hphantom{
\left(
\widehat{\mathcal{D}}^{\,}_{RL}
+
\widehat{\mathcal{D}}^{\dag}_{RL}
\right)=
         }
+
\frac{512\,\pi^{2}}{9}\,
{\bm{:}\,}
\widehat{\rho}^{\,}_{R}\,
\widehat{\rho}^{\,}_{L}\,
(\widehat{\rho}^{2}_{R}+\widehat{\rho}^{2}_{L})
{\,\bm{:}}\,,
\\
&
\widehat{\mathcal{U}}
+
\widehat{\mathcal{U}}^{\dag}\longrightarrow
\frac{1}{2\pi^{2}\mathfrak{a}^{4}}\,
{\bm{:}\,}
\cos
\left(
2
\left[
\widehat{\varphi}^{\,}_{L}(x)
+
\widehat{\varphi}^{\,}_{R}(x)
\right]
\right)\!
{\,\bm{:}}\,.
\end{align}  
\end{subequations}

At this stage, it is convenient to trade the pair of chiral fields
for the pair of nonchiral bosonic fields
$\widehat{\phi}(x)$ and $\widehat{\theta}(x)$ through
\begin{subequations}
\label{eq:bosonization U1 nonchiral bosonic fields}
\begin{align}
\widehat{\phi}(x)\coloneqq&\,
\frac{1}{\sqrt{2\pi}}\,
[\widehat{\varphi}^{\,}_{L}(x)
+
\widehat{\varphi}^{\,}_{R}(x)],
\label{eq:bosonization U1 nonchiral bosonic fields a}
\\
\widehat{\theta}(x)\coloneqq&\,
\frac{1}{\sqrt{8\pi}}\,
[\widehat{\varphi}^{\,}_{L}(x)
-
\widehat{\varphi}^{\,}_{R}(x)].
\label{eq:bosonization U1 nonchiral bosonic fields b}
\end{align}
\end{subequations}
The pair of non-chiral fields inherits the equal-time algebra
\begin{equation}
[\widehat{\phi}(x),\widehat{\theta}(y)]=
\ii\Theta(y-x),
\label{eq:bosonization U1 dual bosonic field algebra}
\end{equation}
where $\Theta$ is the Heaviside function
with the convention $\Theta(0)=1/2$.
Hence, the operator $\partial^{\,}_{x}\widehat{\theta}(x)$
is canonically conjugate to $\widehat{\phi}(x)$.

Combining the local bosonization rules
expressed in terms of the pair of non-chiral fields
(\ref{eq:bosonization U1 nonchiral bosonic fields}),
bosonization delivers
\begin{equation}
\widehat{H}^{(0)}_{\mathrm{U}(1),\mathrm{f}}(\lambda)\ 
\longmapsto\ 
\widehat{H}^{(0)}_{\mathrm{U}(1),\mathrm{b}}(\lambda)
\end{equation}
with the bosonic Hamiltonian
\begin{equation}
\widehat{H}^{(0)}_{\mathrm{U}(1),\mathrm{b}}(\lambda)
\coloneqq
\widehat{\mathsf{H}}^{\,}_{0}(\lambda)
+
\widehat{\mathsf{H}}^{\,}_{2}(\lambda)
+
\widehat{\mathsf{H}}^{\,}_{4}(\lambda)
+
\widehat{\mathsf{H}}^{\,}_{u}(\lambda).
\label{eq:Bosonized U(1) Final}
\end{equation}
Hamiltonian
$\widehat{\mathsf{H}}^{\,}_{0}(\lambda)$
is proportional to the identity with the
same proportionality constant as in Eq.\
(\ref{eq:bosonization U1 Wick decomposition b}).
The Hamiltonian
\begin{subequations}
\label{eq:bosonization U1 final H2}
\begin{equation}
\widehat{\mathsf{H}}^{\,}_{2}(\lambda)\coloneqq
\int\dd x\,
\left[\,
\rho^{\,}_{\phi}(\lambda)\,
{\bm{:}\,}
(\partial^{\,}_{x}\widehat{\phi})^{2}
{\,\bm{:}}
+
\rho^{\,}_{\theta}(\lambda)\,
{\bm{:}\,}
(\partial^{\,}_{x}\widehat{\theta})^{2}
{\,\bm{:}}\,
\right],
\end{equation}
where
\begin{align}
\rho^{\,}_{\phi}(\lambda)\coloneqq&\,
\mathfrak{a}
\left[
\frac{1}{2}
J(\lambda)
-
\frac{200}{9\pi^{2}}\,
g(\lambda)
\right],
\label{eq:bosonization U1 final bosonic coefficients a}
\\
\rho^{\,}_{\theta}(\lambda)\coloneqq&\,
\mathfrak{a}
\left[
2J(\lambda)
-
\frac{32}{9\pi^{2}}\,
g(\lambda)
\right],
\label{eq:bosonization U1 final bosonic coefficients b}
\end{align}
\end{subequations}
contains operators of scaling dimension two.
The Hamiltonian
\begin{subequations}
\label{eq:bosonization U1 final H4}
\begin{align}
\widehat{\mathsf{H}}^{\,}_{4}(\lambda)\coloneqq
\int\dd x\,
\Bigl[
&\,
\kappa^{\,}_{\phi}(\lambda)\,
{\bm{:}\,}
(\partial^{2}_{x}\widehat{\phi})^{2}
{\,\bm{:}}
+
\kappa^{\,}_{\theta}(\lambda)\,
{\bm{:}\,}
(\partial^{2}_{x}\widehat{\theta})^{2}
{\,\bm{:}}
\nonumber\\
&\,
+
u^{\,}_{\phi}(\lambda)\,
{\bm{:}\,}
(\partial^{\,}_{x}\widehat{\phi})^{4}
{\,\bm{:}}
+
u^{\,}_{\theta}(\lambda)\,
{\bm{:}\,}
(\partial^{\,}_{x}\widehat{\theta})^{4}
{\,\bm{:}}
\nonumber\\
&\,
+
u^{\,}_{\phi,\theta}(\lambda)\,
{\bm{:}\,}
(\partial^{\,}_{x}\widehat{\phi})^{2}
(\partial^{\,}_{x}\widehat{\theta})^{2}
{\,\bm{:}}\,
\Bigr],
\end{align}
where
\begin{align}
&
\kappa^{\,}_{\phi}(\lambda)\coloneqq
\mathfrak{a}^{3}
\left[
-\frac{1}{24}J(\lambda)
+
\frac{1130}{27\pi^{2}}\,
g(\lambda)
\right],
\label{eq:bosonization U1 final bosonic coefficients c}
\\
&
\kappa^{\,}_{\theta}(\lambda)\coloneqq
\mathfrak{a}^{3}
\left[
-\frac{1}{6}J(\lambda)
+
\frac{8}{27\pi^{2}}\,
g(\lambda)
\right],
\label{eq:bosonization U1 final bosonic coefficients d}
\\
&
u^{\,}_{\phi}(\lambda)\coloneqq
\mathfrak{a}^{3}
\left[
-\frac{\pi}{48}J(\lambda)
+
\frac{385}{27\pi}\,
g(\lambda)
\right],
\label{eq:bosonization U1 final bosonic coefficients e}
\\
&
u^{\,}_{\theta}(\lambda)\coloneqq
\mathfrak{a}^{3}
\left[
-\frac{\pi}{3}J(\lambda)
+
\frac{16}{27\pi}\
g(\lambda)
\right],
\label{eq:bosonization U1 final bosonic coefficients f}
\\
&
u^{\,}_{\phi,\theta}(\lambda)\coloneqq
-\pi\mathfrak{a}^{2}\,
\rho^{\,}_{\phi}(\lambda),
\label{eq:bosonization U1 final bosonic coefficients g}
\end{align}
\end{subequations}
contains operators of scaling dimension four.
Finally, the Hamiltonian
\begin{subequations}
\begin{equation}
\widehat{\mathsf{H}}^{\,}_{u}(\lambda)\coloneqq
g^{\,}_{\mathrm{u}}(\lambda)
\int\dd x\
{\bm{:}}
\cos
\big(
\sqrt{8\pi}\,
\widehat{\phi}\,
\big)
{\,\bm{:}},
\label{eq:umklapp Hamiltonian}
\end{equation}
where
\begin{equation}
g^{\,}_{\mathrm{u}}(\lambda)\coloneqq
-
\frac{160}
{3\pi^{3}\mathfrak{a}}\,
g(\lambda),
\end{equation}
\end{subequations}
contains a vertex operator whose
Gaussian scaling dimension depends
on the ratio of $\rho^{\,}_{\phi}(\lambda)$
to $\rho^{\,}_{\theta}(\lambda)$.

It is useful to introduce the speed $v(\lambda)$ and
the Gaussian parameter $\eta(\lambda)$ through
\begin{equation}
v(\lambda)\coloneqq
2\,\sqrt{\rho^{\,}_{\phi}(\lambda)\,\rho^{\,}_{\theta}(\lambda)},
\qquad
\eta(\lambda)\coloneqq
\sqrt{\frac{\rho^{\,}_{\phi}(\lambda)}{\rho^{\,}_{\theta}(\lambda)}}.
\label{eq:bosonization U1 velocity and eta definitions}
\end{equation}
Hamiltonian $\widehat{H}^{\,}_{2}(\lambda)$ can
then be written as
\begin{equation}
\widehat{\mathsf{H}}^{\,}_{2}(\lambda)=
\frac{v(\lambda)}{2}
\int\dd x\,
\left[\,
\eta(\lambda)\,
{\bm{:}\,}
(\partial^{\,}_{x}\widehat{\phi})^{2}
{\,\bm{:}}
+
\frac{1}{\eta(\lambda)}\,
{\bm{:}\,}
(\partial^{\,}_{x}\widehat{\theta})^{2}
{\,\bm{:}}\,
\right].
\label{eq:bosonization U1 Gaussian bosonic Hamiltonian}
\end{equation}
The Gaussian scaling dimension of the umklapp vertex in
Eq.~\eqref{eq:umklapp Hamiltonian} is
\begin{equation}
\Delta^{\,}_{\mathrm{u}}(\lambda)=
\frac{2}{\eta(\lambda)}.
\label{eq:bosonization U1 Gaussian dimension umklapp}
\end{equation}
At the free-fermion fixed point
\begin{equation}
v(0)=2\sqrt{2}\,\mathfrak{a},
\qquad
\eta(0)=\frac{1}{2},
\qquad
\Delta^{\,}_{\mathrm{u}}(0)=4.
\label{eq:v and eta bosonization U(1)}
\end{equation}

\begin{figure}
\centering
\includegraphics[width=0.51\columnwidth]{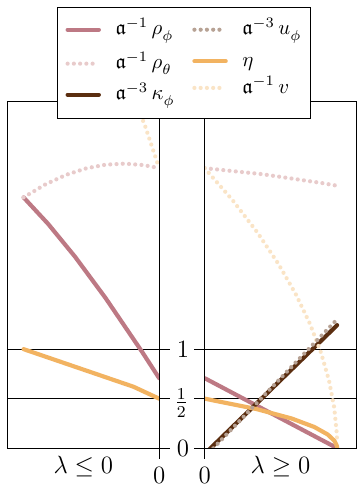}
\caption{The bare couplings of Hamiltonian
\eqref{eq:Bosonized U(1) Final}
given in
Eqs.~\eqref{eq:bosonization U1 final H2},
\eqref{eq:bosonization U1 final H4},
and \eqref{eq:v and eta bosonization U(1)}
are useful to characterize the tricritical point
$\lambda^{\,}_{\mathrm{gFM}^{\,}_{z}}$
in the region $\lambda\geq0$ and
$\lambda^{\prime}_{\mathrm{Neel}^{\,}_{z}}$
in the region $\lambda\leq0$.
}
\label{fig:Boson Couplings}
\end{figure}

\subsubsection{Lifshitz instability and the tricritical point}
\label{ssub:lifshitz_instability_and_the_tricritical_point}

We now use the bosonic continuum Hamiltonian derived in
Sec.~\ref{ssub:bosonization_along_the_mathrm_u}
to describe the transition from the
$\mathrm{U}(1)$-symmetric Luttinger liquid to the
$\mathrm{gFM}^{\,}_{z}$ phase.
The bosonization is performed perturbatively near the free-fermion
point $\lambda=0$.
The coefficients obtained therein are therefore bare couplings defined
at the microscopic matching scale $\mathfrak{a}$.
They determine the operator content of the continuum theory but are not
expected to locate quantitatively the tricritical point
$\lambda=\lambda^{\,}_{\mathrm{gFM}^{\,}_{z}}$.
In particular, the zero of the bare stiffness
$\rho^{\,}_{\phi}(\lambda)$ need not coincide with the numerical value
$\lambda^{\,}_{\mathrm{gFM}^{\,}_{z}}$.

We denote the corresponding long-distance effective couplings by
$\bar{\rho}^{\,}_{\phi}$,
$\bar{\rho}^{\,}_{\theta}$,
$\bar{\kappa}^{\,}_{\phi}$, and
$\bar{u}^{\,}_{\phi}$.
At the matching scale, their bare values obey
\begin{equation}
\left.
\left(
\bar{\rho}^{\,}_{\phi},
\bar{\rho}^{\,}_{\theta},
\bar{\kappa}^{\,}_{\phi},
\bar{u}^{\,}_{\phi}
\right)
\right|_{\mathfrak{a}}
=
\left(
\rho^{\,}_{\phi},
\rho^{\,}_{\theta},
\kappa^{\,}_{\phi},
u^{\,}_{\phi}
\right).
\label{eq:Lifshitz matching conditions}
\end{equation}
We interpret the tricritical point as a
Lifshitz instability at which the renormalized stiffness of the
$\widehat{\phi}$ sector vanishes,
\begin{equation}
\bar{\rho}^{\,}_{\phi}=0,
\qquad
\bar{\rho}^{\,}_{\theta}>0,
\qquad
\bar{\kappa}^{\,}_{\phi}>0,
\qquad
\bar{u}^{\,}_{\phi}>0.
\label{eq:renormalized Lifshitz conditions}
\end{equation}
Their values near the tricritical point
include renormalization effects
that are not determined by the perturbative bosonization.
The assumption \eqref{eq:renormalized Lifshitz conditions}
follows from the qualitative behavior of the bare couplings
for $\lambda\geq0$, as shown in
Fig.~\ref{fig:Boson Couplings}.

We derive below the corresponding effective
Lifshitz field theory and analyze the phases on its two
sides.

\paragraph{Effective Lifshitz theory.}

The minimal Hamiltonian that captures the instability of the
$\widehat{\phi}$ sector is
\begin{align}
\widehat{H}^{\,}_{\mathrm{Lif}}
\coloneqq
\int\dd x\,
\Bigl[
&\bar{\rho}^{\,}_{\theta}\,
{\bm{:}\,}(\partial^{\,}_{x}\widehat{\theta})^{2}{\,\bm{:}}
+
\bar{\rho}^{\,}_{\phi}\,
{\bm{:}\,}(\partial^{\,}_{x}\widehat{\phi})^{2}{\,\bm{:}}
\nonumber\\
&+
\bar{\kappa}^{\,}_{\phi}\,
{\bm{:}\,}(\partial^{2}_{x}\widehat{\phi})^{2}{\,\bm{:}}
+
\bar{u}^{\,}_{\phi}\,
{\bm{:}\,}(\partial^{\,}_{x}\widehat{\phi})^{4}{\,\bm{:}}
\Bigr].
\label{eq:effective Lifshitz Hamiltonian}
\end{align}
The coefficient $\bar{\rho}^{\,}_{\theta}>0$ multiplies the square of
the momentum density
$\partial^{\,}_{x}\widehat{\theta}$.
The coefficient $\bar{\rho}^{\,}_{\phi}$ multiplies
$(\partial^{\,}_{x}\widehat{\phi})^{2}$ and its sign
controls the instability.
The coefficient $\bar{\kappa}^{\,}_{\phi}>0$ multiplying
$(\partial^{2}_{x}\widehat{\phi})^{2}$ suppresses spatial variations
of $\partial^{\,}_{x}\widehat{\phi}$.
The coefficient $\bar{u}^{\,}_{\phi}>0$ multiplying
$(\partial^{\,}_{x}\widehat{\phi})^{4}$ bounds the Hamiltonian from
below when $\bar{\rho}^{\,}_{\phi}<0$.
The other dimension-four operators in the bosonized Hamiltonian do not
change the mean-field structure of this instability and are omitted
from Eq.~\eqref{eq:effective Lifshitz Hamiltonian}.

The equal-time commutation relation
\eqref{eq:bosonization U1 dual bosonic field algebra}
identifies $\partial^{\,}_{x}\widehat{\theta}$ as the momentum
conjugate to $\widehat{\phi}$.
In the phase-space path integral, the corresponding integration fields
are written without hats.
The Lagrangian density associated with
Eq.~\eqref{eq:effective Lifshitz Hamiltonian} is
\begin{align}
\mathcal{L}^{\,}_{\mathrm{Lif}}
={}&
(\partial^{\,}_{x}\theta)(\partial^{\,}_{t}\phi)
-
\bar{\rho}^{\,}_{\theta}(\partial^{\,}_{x}\theta)^{2}
-
\bar{\rho}^{\,}_{\phi}(\partial^{\,}_{x}\phi)^{2}
\nonumber\\
&-
\bar{\kappa}^{\,}_{\phi}(\partial^{2}_{x}\phi)^{2}
-
\bar{u}^{\,}_{\phi}(\partial^{\,}_{x}\phi)^{4}.
\label{eq:effective Lifshitz phase space Lagrangian}
\end{align}
Variation with respect to $\partial^{\,}_{x}\theta$ gives
\begin{equation}
\partial^{\,}_{x}\theta=
\frac{1}{2\bar{\rho}^{\,}_{\theta}}
\partial^{\,}_{t}\phi,
\label{eq:effective Lifshitz theta constraint}
\end{equation}
up to an $x$-independent mode of $\theta$.

The Lifshitz instability occurs when
$\bar{\rho}^{\,}_{\phi}=0$.
Temporarily neglecting the relevant interaction
$\bar{u}^{\,}_{\phi}(\partial^{\,}_{x}\phi)^{4}$ defines
a Gaussian reference theory.
The Gaussian theory is used below to characterize the
quadratic dispersion and the correlations.
However, it need not coincide with the infrared fixed point
of the interacting Lifshitz theory.
We use the real-time Fourier convention
\begin{equation}
\phi(t,x)=
\int\frac{\dd\omega}{\sqrt{2\pi}}
\int\frac{\dd k}{\sqrt{2\pi}}\,
e^{+\ii(kx-\omega
t)}
\phi(\omega,k),
\label{eq:Gaussian Lifshitz real time Fourier convention}
\end{equation}
where $\omega\in\mathbb{R}$ is the real frequency
and $k\in\mathbb{R}$ is the wave number.
The positive-frequency dispersion of the Gaussian reference theory is
\begin{equation}
\omega(k)=
c^{\,}_{\mathrm{Lif}}\,k^{2},
\qquad
c^{\,}_{\mathrm{Lif}}\coloneqq
2\sqrt{\bar{\rho}^{\,}_{\theta}\,\bar{\kappa}^{\,}_{\phi}}>0.
\label{eq:Gaussian Lifshitz dispersion}
\end{equation}
The Gaussian reference theory therefore has the dynamical exponent
$\mathsf{z}=2$.
The interaction
$\bar{u}^{\,}_{\phi}(\partial^{\,}_{x}\phi)^{4}$ remains part of the
full theory and is omitted only in the Gaussian analysis.

\paragraph{Phases adjacent to the Lifshitz point.}

We denote ground-state expectation values in the positive-
and negative-stiffness phases by
$\langle\,\bm{\cdot}\,\rangle^{\,}_{+}$ and
$\langle\,\bm{\cdot}\,\rangle^{\,}_{-}$, respectively.

For $\bar{\rho}^{\,}_{\phi}>0$, the static Hamiltonian
\eqref{eq:effective Lifshitz Hamiltonian}
is minimized by a uniform configuration satisfying
\begin{equation}
\langle
\partial^{\,}_{x}\widehat{\phi}
\rangle^{\,}_{+}=
0.
\label{eq:positive stiffness uniform ground state}
\end{equation}
The terms proportional to $\bar{\rho}^{\,}_{\phi}$ and
$\bar{\kappa}^{\,}_{\phi}$ become comparable at
\begin{equation}
k^{\,}_{\times}
\coloneqq
\sqrt{\frac{\bar{\rho}^{\,}_{\phi}}
           {\bar{\kappa}^{\,}_{\phi}}},
\qquad
\xi^{\,}_{\times}
\coloneqq
\frac{1}{k^{\,}_{\times}}.
\label{eq:Lifshitz crossover scales}
\end{equation}
For $|k|\ll k^{\,}_{\times}$, the dispersion is linear,
\begin{equation}
\omega(k)
\simeq
v^{\,}_{+}|k|,
\qquad
v^{\,}_{+}
\coloneqq
2\sqrt{\bar{\rho}^{\,}_{\theta}\,\bar{\rho}^{\,}_{\phi}}.
\end{equation}
The infrared theory is a Luttinger liquid.
The crossover length $\xi^{\,}_{\times}$ diverges as
$\bar{\rho}^{\,}_{\phi}\to0^{+}$.

For $\bar{\rho}^{\,}_{\phi}<0$, we introduce the slope field
$m\coloneqq\partial^{\,}_{x}\phi$.
The static Hamiltonian density obtained from
Eq.~\eqref{eq:effective Lifshitz Hamiltonian} is
\begin{equation}
\mathcal{H}^{\,}_{\mathrm{stat}}[m]
=
\bar{\kappa}^{\,}_{\phi}(\partial^{\,}_{x}m)^{2}
-
|\bar{\rho}^{\,}_{\phi}|\,m^{2}
+
\bar{u}^{\,}_{\phi}\,m^{4}.
\label{eq:gradient ordered static Hamiltonian}
\end{equation}
The two gradient-ordered ground states therefore satisfy
\begin{equation}
\langle
\partial^{\,}_{x}\widehat{\phi}
\rangle_{-}
=
m^{\,}_{\star},
\qquad
m^{\,}_{\star}
\coloneqq
\pm
\sqrt{\frac{|\bar{\rho}^{\,}_{\phi}|}
{2\bar{u}^{\,}_{\phi}}}.
\label{eq:negative stiffness gradient order parameter}
\end{equation}
The corresponding configurations
$\phi^{\,}_{\star}(x)=\phi^{\,}_{0}+m^{\,}_{\star}x$ are exchanged by
spatial inversion.
Expanding
$\phi=\phi^{\,}_{0}+m^{\,}_{\star}x+\varphi$ to quadratic order gives
\begin{equation}
\mathcal{L}^{\,}_{\mathrm{quad},\varphi}
=
\frac{1}{4\bar{\rho}^{\,}_{\theta}}
(\partial^{\,}_{t}\varphi)^{2}
-
2|\bar{\rho}^{\,}_{\phi}|
(\partial^{\,}_{x}\varphi)^{2}
-
\bar{\kappa}^{\,}_{\phi}
(\partial^{2}_{x}\varphi)^{2}.
\label{eq:gradient ordered quadratic Lagrangian}
\end{equation}
At small wave number, the operator
$(\partial^{2}_{x}\varphi)^{2}$ is subleading.
The gradient-ordered phase is consequently an infrared Luttinger
liquid with speed
\begin{equation}
v^{\,}_{-}\coloneqq
2\sqrt{2\,\bar{\rho}^{\,}_{\theta}\,|\bar{\rho}^{\,}_{\phi}|}.
\end{equation}
The decomposition of $\phi$ implies
\begin{equation}
e^{+\ii\alpha\phi(t,x)}=
e^{+\ii\alpha\phi^{\,}_{0}}
e^{+\ii\alpha m^{\,}_{\star}x}
e^{+\ii\alpha\varphi(t,x)}.
\end{equation}
The algebraic correlations of this vertex are modulated at the wave
number $\alpha\,m^{\,}_{\star}$.
In particular, the umklapp operator
$\cos\big(\sqrt{8\pi}\,\phi\big)$ defined in
Eq.~\eqref{eq:umklapp Hamiltonian} carries the wave number
$q^{\,}_{\mathrm{u}}\coloneqq\sqrt{8\pi}\,m^{\,}_{\star}$.
It averages to zero under coarse graining unless
$q^{\,}_{\mathrm{u}}$ is commensurate with a reciprocal lattice
vector.

Because $m^{\,}_{\star}\propto\sqrt{|\rho^{\,}_{\phi}|}$
we can also interpret the Lifshitz point as a disorder point
that separates a phase with commensurate $\phi$-correlations
and a phase with incommensurate $\phi$-correlations.
More detailed discussion is found in
Sec.~\ref{sec:disorder point} and
Refs.~\cite{schollwock1996,Herviou2025,Fitouchi2026,chitov2025,beradze2023,feng2022,
huang2022,Pixley2014,Chepiga2016,Saito2024}.

\paragraph{Vertex correlations at the Gaussian Lifshitz point.}

We denote by
$\langle\,\bm{\cdot}\,\rangle^{\,}_{0}$
expectation values in the Gaussian reference theory.
At
$\bar{\rho}^{\,}_{\phi}=0$,
the Euclidean Gaussian action is
\begin{equation}
S^{(\mathrm{E})}_{0,\mathrm{Lif},\phi}
=
\int\dd\tau\int\dd x\,
\left[
\frac{1}{4\bar{\rho}^{\,}_{\theta}}
(\partial^{\,}_{\tau}\phi)^{2}
+
\bar{\kappa}^{\,}_{\phi}
(\partial^{2}_{x}\phi)^{2}
\right].
\label{eq:Gaussian Lifshitz Euclidean action}
\end{equation}
We use the Euclidean Fourier convention
\begin{equation}
\chi(\tau,x)
=
\int\frac{\dd\varpi}{\sqrt{2\pi}}
\int\frac{\dd k}{\sqrt{2\pi}}\,
e^{+\ii(\varpi\tau-kx)}
\chi(\varpi,k),
\label{eq:Gaussian Lifshitz Euclidean Fourier convention}
\end{equation}
with $\chi=\phi,\theta$ and where
$\varpi\in\mathbb{R}$
denotes the Euclidean frequency and
$k\in\mathbb{R}$
denotes the wave number.

We define the Euclidean Green function
$G^{(\mathrm{E})}_{\mathrm{Lif}}(\tau,x)$
by
\begin{equation}
\left(
-\partial^{2}_{\tau}
+
c^{2}_{\mathrm{Lif}}\,
\partial^{4}_{x}
\right)
G^{(\mathrm{E})}_{\mathrm{Lif}}(\tau,x)=
\delta(\tau)\,
\delta(x).
\label{eq:Gaussian Lifshitz Green function definition}
\end{equation}
We select the solution that is even under
$x\mapsto-x$
and obeys
\begin{equation}
\lim_{|x|\to\infty}
G^{(\mathrm{E})}_{\mathrm{Lif}}(\tau,x)
=
-\frac{|x|}{4c^{\,}_{\mathrm{Lif}}}.
\label{eq:Gaussian Lifshitz Green function boundary condition}
\end{equation}
These conditions fix the affine homogeneous
ambiguity in the solution of
Eq.~\eqref{eq:Gaussian Lifshitz Green function definition}.
The resulting Green function is
\begin{align}
G^{(\mathrm{E})}_{\mathrm{Lif}}(\tau,x)
={}&
-
\frac{|x|}{4c^{\,}_{\mathrm{Lif}}}
\operatorname{erf}\left(
\frac{|x|}
{2\sqrt{c^{\,}_{\mathrm{Lif}}|\tau|}}
\right)
\nonumber\\
&-
\frac{\sqrt{c^{\,}_{\mathrm{Lif}}|\tau|}}
{2\sqrt{\pi}\,c^{\,}_{\mathrm{Lif}}}\,
e^{
-
\frac{x^{2}}
{4c^{\,}_{\mathrm{Lif}}|\tau|}
}.
\label{eq:Gaussian Lifshitz Green function}
\end{align}
For fixed
$x\neq0$,
its equal-time limit is
\begin{equation}
G^{(\mathrm{E})}_{\mathrm{Lif}}(0,x)
=
-\frac{|x|}{4c^{\,}_{\mathrm{Lif}}}.
\label{eq:Gaussian Lifshitz equal time Green function}
\end{equation}

The two-point function of the field
$\phi$
is
\begin{equation}
\left\langle
\phi(\tau,x)\,\phi(0,0)
\right\rangle^{\,}_{0}=
2\bar{\rho}^{\,}_{\theta}\,
G^{(\mathrm{E})}_{\mathrm{Lif}}(\tau,x).
\label{eq:Gaussian phi correlator}
\end{equation}
It is defined up to an additive contribution from the spatially
uniform mode.
This ambiguity cancels from the field-difference correlator.
Indeed, using translational invariance and
Eq.~\eqref{eq:Gaussian Lifshitz equal time Green function},
we obtain
\begin{equation}
\left\langle
[\phi(0,x)-\phi(0,0)]^{2}
\right\rangle^{\,}_{0}
=
\frac{\bar{\rho}^{\,}_{\theta}}
{c^{\,}_{\mathrm{Lif}}}
|x|.
\label{eq:Gaussian Lifshitz phi difference correlator}
\end{equation}
The mean-square fluctuations of $\phi$
therefore grow linearly with the spatial separation.

We now specify the ultraviolet normalization of the vertex operators.
For $\chi=\phi,\theta$,
we first Wick order the nonzero-mode fluctuations with respect to the
Gaussian vacuum at the short-distance reference scale $\mathfrak{a}$.
The remaining finite multiplicative freedom is fixed by the condition
\begin{equation}
\lim_{\tau\downarrow0}\,
\langle
\mathcal{V}^{\,}_{q,\chi}(\tau,\mathfrak{a})
\mathcal{V}^{\,}_{-q,\chi}(0,0)
\rangle^{\,}_{0}
=1,
\qquad
\chi=\phi,\theta.
\label{eq:Gaussian Lifshitz vertex normalization condition}
\end{equation}
Equivalently, we write
\begin{equation}
\mathcal{V}^{\,}_{q,\chi}(\tau,x)
\coloneqq
Z^{\,}_{q,\chi}(\mathfrak{a})\,
{\bm{:}\,}
e^{+\ii q\,\chi(\tau,x)}
{\,\bm{:}},
\qquad
\chi=\phi,\theta,
\label{eq:Gaussian Lifshitz normalized vertex definition}
\end{equation}
where the finite factor
$Z^{\,}_{q,\chi}(\mathfrak{a})$
is chosen to satisfy
Eq.~\eqref{eq:Gaussian Lifshitz vertex normalization condition}.
This convention fixes the amplitudes of the two-point functions at the
reference separation $\mathfrak{a}$.
It does not modify their dependence on the separation at
$|x|\gg\mathfrak{a}$.

For the normalized $\phi$-vertex operator,
the Gaussian identity gives the ratio
\begin{equation}
\frac{
\langle
\mathcal{V}^{\,}_{\alpha,\phi}(0,x)
\mathcal{V}^{\,}_{-\alpha,\phi}(0,0)
\rangle^{\,}_{0}
}{
\langle
\mathcal{V}^{\,}_{\alpha,\phi}(0,\mathfrak{a})
\mathcal{V}^{\,}_{-\alpha,\phi}(0,0)
\rangle^{\,}_{0}
}
=
e^{
-
\frac{\alpha^{2}}{2}
\left[
D^{\,}_{\phi}(x)-D^{\,}_{\phi}(\mathfrak{a})
\right]
},
\label{eq:Gaussian Lifshitz phi vertex ratio}
\end{equation}
where
\begin{equation}
D^{\,}_{\phi}(x)
\coloneqq
\left\langle
[\phi(0,x)-\phi(0,0)]^{2}
\right\rangle^{\,}_{0}.
\label{eq:Gaussian Lifshitz phi difference definition}
\end{equation}
Using
Eqs.~\eqref{eq:Gaussian Lifshitz vertex normalization condition}
and \eqref{eq:Gaussian Lifshitz phi difference correlator},
we obtain
\begin{align}
\langle
\mathcal{V}^{\,}_{\alpha,\phi}(0,x)
\mathcal{V}^{\,}_{-\alpha,\phi}(0,0)
\rangle^{\,}_{0}
=
e^{
-
\frac{\alpha^{2}\bar{\rho}^{\,}_{\theta}}
{2c^{\,}_{\mathrm{Lif}}}
\left(|x|-\mathfrak{a}\right)
},
\label{eq:Gaussian Lifshitz normalized phi vertex correlator}
\end{align}
for $|x|\geq\mathfrak{a}$.
The normalization condition fixes the value of the correlator at
$|x|=\mathfrak{a}$.
The exponential decay rate is independent of this choice.
Under a different finite normalization of the composite operator, the
right-hand side of
Eq.~\eqref{eq:Gaussian Lifshitz normalized phi vertex correlator}
is multiplied by a separation-independent constant.
The normalized
$\phi$-vertex operators are therefore short-ranged in the Gaussian
reference theory.

The two-point function of the field $\theta$
has a distributional equal-time contribution
and must therefore be treated by point splitting.
For $k\neq0$,
the Euclidean continuation of
Eq.~\eqref{eq:effective Lifshitz theta constraint}
gives
\begin{equation}
\theta(\varpi,k)=
-
\frac{\ii}{2\bar{\rho}^{\,}_{\theta}}
\frac{\varpi}{k}
\phi(\varpi,k).
\label{eq:Gaussian Lifshitz theta phi relation}
\end{equation}
The mode with
$k=0$
is not fixed by
Eq.~\eqref{eq:Gaussian Lifshitz theta phi relation}.
Using the Gaussian propagator associated with
Eq.~\eqref{eq:Gaussian Lifshitz Euclidean action},
we obtain
\begin{subequations}
\begin{align}
\left\langle
\theta(\tau,x)\,
\theta(0,0)
\right\rangle^{\,}_{0}=&\,
-
\frac{1}{2\bar{\rho}^{\,}_{\theta}}\,
\delta(\tau)\,
G^{\,}_{0}(x)
\nonumber\\
&\,
+
\frac{c^{\,}_{\mathrm{Lif}}}{4\bar{\rho}^{\,}_{\theta}}\,
\frac{1}{\sqrt{4\pi c^{\,}_{\mathrm{Lif}}|\tau|}}\,
e^{
-
\frac{x^{2}}{4c^{\,}_{\mathrm{Lif}}|\tau|}
},
\label{eq:Gaussian theta correlator}
\end{align}
where the Green function
$G^{\,}_{0}(x)$
of the one-dimensional Laplacian is defined by
\begin{equation}
-\partial^{2}_{x}G^{\,}_{0}(x)=
\delta(x).
\label{eq:one dimensional Laplacian Green function}
\end{equation}
\end{subequations}
The first term in
Eq.~\eqref{eq:Gaussian theta correlator}
is supported at equal imaginary time.
It is the contact contribution associated
with the canonical relation between
$\partial^{\,}_{x}\theta$ and $\phi$.
We denote the second term in
Eq.~\eqref{eq:Gaussian theta correlator} by
$\left\langle
\theta(\tau,x)\,
\theta(0,0)
\right\rangle^{\,}_{0,\mathrm{reg}}$.
It is the regular dynamical contribution.
For every fixed
$x\neq0$,
\begin{equation}
\lim_{\tau\downarrow0}
\left\langle
\theta(\tau,x)\theta(0,0)
\right\rangle^{\,}_{0,\mathrm{reg}}
=0.
\label{eq:Gaussian regular theta point split limit}
\end{equation}

For the neutral pair of normalized
$\theta$-vertex operators, the Gaussian identity gives the ratio
\begin{align}
&\frac{
\langle
\mathcal{V}^{\,}_{\beta,\theta}(\tau,x)
\mathcal{V}^{\,}_{-\beta,\theta}(0,0)
\rangle^{\,}_{0}
}{
\langle
\mathcal{V}^{\,}_{\beta,\theta}(0^{+},\mathfrak{a})
\mathcal{V}^{\,}_{-\beta,\theta}(0,0)
\rangle^{\,}_{0}
}
\nonumber \\
&\qquad=
e^{
\beta^{2}
\left[
\left\langle
\theta(\tau,x)\theta(0,0)
\right\rangle^{\,}_{0,\mathrm{reg}}
-
\left\langle
\theta(0^{+},\mathfrak{a})\theta(0,0)
\right\rangle^{\,}_{0,\mathrm{reg}}
\right]
}.
\label{eq:Gaussian Lifshitz theta vertex ratio}
\end{align}
Here, $0^{+}$
denotes the point-split limit from positive imaginary time.
Using
Eqs.~\eqref{eq:Gaussian Lifshitz vertex normalization condition}
and
\eqref{eq:Gaussian regular theta point split limit},
we find
\begin{equation}
\lim_{\tau\downarrow0}\,
\langle
\mathcal{V}^{\,}_{\beta,\theta}(\tau,x)
\mathcal{V}^{\,}_{-\beta,\theta}(0,0)
\rangle^{\,}_{0}
=1,
\qquad
|x|\geq\mathfrak{a}.
\label{eq:Gaussian Lifshitz normalized theta vertex correlator}
\end{equation}
The value on the right-hand side follows from the
normalization condition
\eqref{eq:Gaussian Lifshitz vertex normalization condition}.
A different finite normalization multiplies it by a nonzero
separation-independent constant.
The absence of spatial decay is independent of this choice.
The normalized
$\theta$-vertex operators therefore exhibit long-range order in the
Gaussian reference theory.

This result is consistent with the numerical evolution of the
transverse spin correlations upon approaching the tricritical point
from the Luttinger-liquid side. Indeed, along the
$\mathrm{U}(1)$-symmetric critical line, the transverse spin
correlation functions obey
$C^{x}_{s^{\,}_{0},s^{\,}_{0}+s}=
C^{y}_{s^{\,}_{0},s^{\,}_{0}+s}\sim s^{-\eta^{x}}$,
with $\eta^{x}=\eta^{y}$ decreasing continuously to zero as
$\lambda\uparrow\lambda^{\,}_{\mathrm{gFM}^{\,}_{z}}$, as shown in
Fig.~\ref{fig:DQ Neel Eta}.
Simultaneously, the velocity of the Luttinger liquid vanishes upon
approaching the Lifshitz instability, since
$v^{\,}_{+}=2\sqrt{\bar{\rho}^{\,}_{\theta}\,\bar{\rho}^{\,}_{\phi}}\to0$
as
$\bar\rho^{\,}_{\phi}\to0^{+}$.
The limiting value $\eta^{x}=\eta^{y}=0$
inferred numerically is therefore consistent with the absence of spatial decay
of the corresponding vertex correlations at the Gaussian Lifshitz
point. Taken together, the numerical approach from the
Luttinger-liquid side and the analytical result at the Lifshitz point
support the spontaneous breaking of the microscopic
$\mathrm{U}(1)$ symmetry at the tricritical point.
The longitudinal spin correlations remain algebraic.

\paragraph{Stability against defects for $\bar{\rho}^{\,}_{\phi}<0$.}

The two gradient-ordered ground states are separated by
configurations in which the slope $m=\partial^{\,}_{x}\phi$
interpolates between $-m^{\,}_{\star}$ and $+m^{\,}_{\star}$.
A straight static domain wall has a finite tension and
an Euclidean action that grows linearly with
the temporal extent of the system.
Its positional entropy grows only logarithmically
with the system size.
Such domain walls therefore do not proliferate in the thermodynamic limit.

A spatially uniform tunneling configuration that connects
the two gradient-ordered ground states in imaginary time
has an action that grows quadratically with the linear system size.
These global tunneling events are consequently
suppressed even more strongly than the static domain walls.
The gradient order is therefore stable against
the two classes of defects considered here.

\paragraph{Summary.}

The Lifshitz instability is controlled by the sign of
the renormalized stiffness $\bar{\rho}^{\,}_{\phi}$
multiplying $(\partial^{\,}_{x}\widehat{\phi})^{2}$.
The three regimes are
\begin{equation*}
\begin{array}{ccl}
\bar{\rho}^{\,}_{\phi}>0
&:&
\text{uniform Luttinger liquid},
\\[1mm]
\bar{\rho}^{\,}_{\phi}=0
&:&
\text{Lifshitz point with Gaussian exponent }\mathsf{z}=2,
\\[1mm]
\bar{\rho}^{\,}_{\phi}<0
&:&
\text{gradient-ordered Luttinger liquid}.
\end{array}
\label{eq:Lifshitz continuum phase structure}
\end{equation*}
For $\bar{\rho}^{\,}_{\phi}>0$, the ground state has
$\langle\partial^{\,}_{x}\widehat{\phi}\rangle^{\,}_{+}=0$,
and the infrared excitations have a linear dispersion.

For $\bar{\rho}^{\,}_{\phi}=0$, the quadratic dispersion is
$\omega(k)=c^{\,}_{\mathrm{Lif}}\,k^{2}$.
At Gaussian level, the vertex operators constructed from
$\phi$ are short-ranged, whereas the neutral two-point functions
of the normalized $\theta$-vertex operators exhibit long-range order.
This analytical result is consistent with
the numerical evolution along the $\mathrm{U}(1)$-symmetric
Luttinger-liquid line, where both the transverse scaling exponent
$\eta^{x}=\eta^{y}$ and the Luttinger-liquid velocity vanish upon
approaching the tricritical point. We therefore interpret the
Lifshitz tricritical point as realizing spontaneous breaking of the
continuous $\mathrm{U}(1)$ symmetry, with long-range transverse spin
correlations coexisting with algebraic longitudinal correlations.

For $\bar{\rho}^{\,}_{\phi}<0$, the ground state develops one
of the two uniform gradients $m^{\,}_{\star}$ given in
Eq.~\eqref{eq:negative stiffness gradient order parameter}.
The fluctuations around either gradient-ordered configuration
remain gapless and recover a Luttinger liquid in the infrared.
The defects considered (domain walls, tunneling)
do not proliferate in the thermodynamic limit,
and the gradient order is stable against these.
We consequently identify the $\mathrm{gFM}^{\,}_{z}$
phase with the gradient-ordered side of the Lifshitz instability.

\subsubsection{$\mathrm{SU}(2)$-symmetric tricritical point}
\label{ssub:SU2 symmetric tricritical point}

We now use the bosonic continuum Hamiltonian derived in
Sec.~\ref{ssub:bosonization_along_the_mathrm_u}
to describe the transition from the
$\mathrm{U}(1)$-symmetric Luttinger liquid to the gapped
$\mathrm{Neel}^{\,}_{z}$ phase.

As $\lambda\leq0$ is decreased from the free-fermion point,
the Gaussian parameter $\eta$ defined in
Eq.~\eqref{eq:bosonization U1 velocity and eta definitions}
increases from its free-fermion value $\eta=1/2$.
The Gaussian Hamiltonian
\eqref{eq:bosonization U1 Gaussian bosonic Hamiltonian}
reaches the $\mathrm{SU}(2)$-symmetric value when
\begin{equation}
\eta=1,
\qquad
\rho^{\,}_{\phi}
=
\rho^{\,}_{\theta},
\label{eq:bosonization U1 SU2 condition}
\end{equation}
as shown in Fig.~\ref{fig:Boson Couplings}.
The value of $\lambda$ at which
Eq.~\eqref{eq:bosonization U1 SU2 condition}
is satisfied provides only a bare estimate of the transition.
The physical location
$\lambda=\lambda^{\prime}_{\mathrm{Neel}^{\,}_{z}}$
includes renormalization effects that are not captured quantitatively
by the perturbative expansion about $\lambda=0$.

At $\eta=1$, the Gaussian scaling dimension
\eqref{eq:bosonization U1 Gaussian dimension umklapp}
of the umklapp operator
\eqref{eq:umklapp Hamiltonian}
is
\begin{equation}
\Delta^{\,}_{\mathrm{u}}=2.
\label{eq:bosonization U1 marginal umklapp dimension}
\end{equation}
The umklapp perturbation is therefore marginal at tree level.
Its flow is determined beyond tree level by the coupled
renormalization of its amplitude and of the Gaussian parameter
$\eta$.
The continuum Hamiltonian at
Eq.~\eqref{eq:bosonization U1 SU2 condition}
has the same Gaussian parameter and operator content as the bosonized
antiferromagnetic $\mathrm{SU}(2)$ point $\Delta=1$ of the
spin-$1/2$ XXZ chain defined in
Eq.~\eqref{eq:def H XXZ}.
We consequently identify
$\lambda=\lambda^{\prime}_{\mathrm{Neel}^{\,}_{z}}$
with the $\mathrm{SU}(2)$-symmetric endpoint of the
$\mathrm{U}(1)$-symmetric critical line.

On the Luttinger-liquid side, the umklapp perturbation is irrelevant
at tree level when $\eta<1$ and becomes marginally irrelevant upon
approaching the $\mathrm{SU}(2)$-symmetric critical trajectory.
The infrared theory remains gapless and has central charge $c=1$.
At
$\lambda=\lambda^{\prime}_{\mathrm{Neel}^{\,}_{z}}$,
the umklapp perturbation is marginal at tree level.
The renormalization-group trajectory entering the
$\mathrm{Neel}^{\,}_{z}$ phase is marginally relevant and flows to
strong coupling.
Inside the $\mathrm{Neel}^{\,}_{z}$ phase, the renormalized Gaussian
parameter satisfies $\eta>1$, so that
\begin{equation}
\Delta^{\,}_{\mathrm{u}}
=
\frac{2}{\eta}<
2.
\label{eq:bosonization U1 relevant umklapp Neelz}
\end{equation}
The umklapp operator is therefore relevant at tree level in the
$\mathrm{Neel}^{\,}_{z}$ phase.
Its flow to strong coupling pins the field $\widehat{\phi}$ at one of
the minima of the normalized vertex operator
${\bm{:}}
\cos\big(\sqrt{8\pi}\,\widehat{\phi}\,\big)
{\,\bm{:}}$
appearing in
Eq.~\eqref{eq:umklapp Hamiltonian}.
The pinning generates a finite excitation gap and long-range
staggered order in the $z$ axis in spin space.
The transition encountered along the
$\mathrm{U}(1)$-symmetric line at $\lambda^{\prime}_{\mathrm{Neel}^{\,}_{z}}$
is therefore inferred to be of the
BKT type.

\subsubsection{Effective interaction between boundary spins for
$\lambda\leq\pi/2$}
\label{subsubsec:effective boundary interaction around lambda piover2}

We have provided in Sec.\
\ref{subsubsec:The NeelSPTx-NeelSPTy phase boundary}
numerical evidences that the ground-state degeneracy
of the Hamiltonian
$\widehat{H}^{(1)}_{\mathrm{U}(1)}(\lambda)$,
defined in Eq.~\eqref{eq:def H U1},
changes from four at
$\lambda=\pi/2$ to two in the open interval
$\pi/2-\lambda^{\,}_{\mathrm{gFM}^{\,}_{z}}<\lambda<\pi/2$.
The aim of this section is to elucidate the mechanism
responsible for this change of degeneracy.

Using degenerate perturbation theory
\begin{subequations}
in powers of
\begin{equation}
0\leq\epsilon\equiv\frac{\pi}{2}-\lambda\ll1,
\end{equation}
we shall show that
this change originates from the effective interaction
between one quantum spin-$1/2$ localized at the
left boundary of the chain and one quantum spin-$1/2$ localized at the
right boundary of the chain.
In particular, we shall show that for finite system sizes
$2N\in[20,30]$
accessible by exact diagonalization (ED),
the energy difference
$E^{\,}_{1}-E^{\,}_{0}$ is proportional to
\begin{equation}
\epsilon^{2}\,(2N-1)^{\bar{\kappa}^{\,}_{xy}-3},
\qquad
\bar{\kappa}^{\,}_{xy}=-3.080(5),
\label{eq:E1-E0 Propto Analytics}
\end{equation}
and the energy difference
$E^{\,}_{2}-E^{\,}_{1}$ is proportional to
\begin{equation}
\epsilon^{2}\,(2N-1)^{\bar{\kappa}^{\,}_{z}-4},
\qquad
\bar{\kappa}^{\,}_{z}=2.953(3).
\end{equation}
Additionally, we shall show that in the thermodynamic limit
$2N\to\infty$, the energy differences
$E^{\,}_{1}-E^{\,}_{0}$ and $E^{\,}_{2}-E^{\,}_{1}$
vanish as
\begin{equation}
\epsilon^{2}\,(2N)^{-6}
\end{equation}
and
\begin{equation}
\epsilon^{2}\,(2N)^{-1}.
\end{equation}
respectively.
Finally, we shall show that the energy difference
\begin{equation}
E^{\,}_{3}-E^{\,}_{2}=0
\end{equation}
\end{subequations}
is identically vanishing
for any system size $2N$.

The simplest effective Hamiltonian invariant under all the symmetries
of $\widehat{H}^{(1)}_{\mathrm{U}(1)}(\lambda)$
that captures the boundary magnetism observed numerically in
Sec.~\ref{subsubsec:The NeelSPTx-NeelSPTy phase boundary}
along the interval $\pi/2-\lambda^{\,}_{\mathrm{gFM}^{\,}_{z}}<\lambda<\pi/2$
is
\begin{align}
\widehat{H}^{\,}_{\mathrm{eff}}(\lambda)
\coloneqq&\,
+
J^{\,}_{xy}(\lambda)\,
\left(
\widehat{X}^{\,}_{1}
\widehat{X}^{\,}_{2N}
+
\widehat{Y}^{\,}_{1}
\widehat{Y}^{\,}_{2N}
\right)
\nonumber \\
&\,+
J^{\,}_{z}(\lambda)\,
\widehat{Z}^{\,}_{1}
\widehat{Z}^{\,}_{2N},
\label{eq:H Effective Boundary Spins}
\end{align}
with the effective anisotropic couplings
$J^{\,}_{xy}\neq0$ and $J^{\,}_{z}>0$ functions of $\lambda$.
The domain of definition of Hamiltonian
$\widehat{H}^{\,}_{\mathrm{eff}}(\lambda)$
is $\mathfrak{H}^{\,}_{1,2N}\cong\mathbb{C}^{4}$.

Indeed, using the Jordan-Wigner transformation
\eqref{eq: def Jordan-Wigner trsf U(1) symmetric}, 
we may rewrite Hamiltonian \eqref{eq:H Effective Boundary Spins} as
\begin{align}
\widehat{H}^{\,}_{\mathrm{eff,f}}(\lambda)\coloneqq&\,
-
2\,J^{\,}_{xy}(\lambda)\,
\left(
\hat{c}^{\dag}_{1}
\hat{c}^{\,}_{2N}
+
\hat{c}^{\dag}_{2N}
\hat{c}^{\,}_{1}
\right)
\nonumber \\
&\,+
J^{\,}_{z}(\lambda)\,
\left(
2\,
\hat{c}^{\dag}_{1}
\hat{c}^{\,}_{1}
-1
\right)\!\left(
2\,
\hat{c}^{\dag}_{2N}
\hat{c}^{\,}_{2N}
-1
\right)\!,
\label{eq:H Effective Boundary Fermions}
\end{align}
whose domain of definition is isomorphic to $\mathfrak{H}^{\,}_{1,2N}$.
Hamiltonian \eqref{eq:H Effective Boundary Fermions}
can be diagonalized.
One finds the eigenstates
\begin{subequations}
\begin{equation}
\begin{aligned}
&\ket{\Psi^{\,}_{0}}=
\frac{\ket{1}^{\,}_{1}\otimes\ket{0}^{\,}_{2N}-
\ket{0}^{\,}_{1}\otimes\ket{1}^{\,}_{2N}}
{\sqrt{2}},
\\
&\ket{\Psi^{\,}_{1}}=
\frac{\ket{1}^{\,}_{1}\otimes\ket{0}^{\,}_{2N}+
\ket{0}^{\,}_{1}\otimes\ket{1}^{\,}_{2N}}
{\sqrt{2}},
\\
&\ket{\Psi^{\,}_{2}}=
\ket{1}^{\,}_{1}\otimes\ket{1}^{\,}_{2N},
\\
&\ket{\Psi^{\,}_{3}}=
\ket{0}^{\,}_{1}\otimes\ket{0}^{\,}_{2N},
\end{aligned}
\end{equation}
associated to the eigenenergies
\begin{equation}
\begin{aligned}
&E^{\,}_{0}(\lambda)=-J^{\,}_{z}(\lambda)-2\,|J^{\,}_{xy}(\lambda)|,
\\
&E^{\,}_{1}(\lambda)=-J^{\,}_{z}(\lambda)+2\,|J^{\,}_{xy}(\lambda)|,
\\
&E^{\,}_{2}(\lambda)=+J^{\,}_{z}(\lambda),
\\
&E^{\,}_{3}(\lambda)=+J^{\,}_{z}(\lambda),
\end{aligned}
\end{equation}
respectively, in the basis chosen such that
\begin{equation}
\hat{c}^{\dag}_{j}\ket{0}^{\,}_{j}=
\ket{1}^{\,}_{j},\qquad
\hat{c}^{\,}_{j}\ket{0}^{\,}_{j}=
0,\qquad
j=1,\,2N.
\label{eq:Boundary States Fermions}
\end{equation}
\end{subequations}
Hamiltonian \eqref{eq:H Effective Boundary Spins}
captures phenomenologically
the boundary magnetism observed numerically in
Sec.~\ref{subsubsec:The NeelSPTx-NeelSPTy phase boundary} if
\begin{subequations}
\label{eq:Jxy Jz Effective Expected}
\begin{align}
&
|J^{\,}_{xy}(\lambda)|=
\frac{A^{\,}_{\mathrm{p}}(\lambda)}{4}\,
(2N-1)^{\kappa^{\,}_{xy}-3},
&&
\kappa^{\,}_{xy}
=-3.119(5),
\label{eq:Jxy Jz Effective Expected a}
\\
&
J^{\,}_{z}(\lambda)\simeq
\frac{v^{\,}_{2,1}(\lambda)}{2}\,
(2N-1)^{\kappa^{\,}_{z}-4},
&&
\kappa^{\,}_{z}
=+2.950(3),
\label{eq:Jxy Jz Effective Expected b}
\end{align}
\end{subequations}
for system sizes $2N\in[20,30]$ and assuming $\pi/2-\lambda\ll1$.
Here, the $\lambda$ dependence of the effective couplings
is assigned entirely to the finite-size level-spacing amplitudes
$A^{\,}_{\mathrm{p}}$ and $v^{\,}_{2,1}$
that are given in the right column of
Table \ref{tab:Fits DQ Neel SPT OBCs}.
Also, the exponents $\kappa^{\,}_{xy}$ and $\kappa^{\,}_{z}$
are the ones given in the left column of 
Table \ref{tab:Fits DQ Neel SPT OBCs}.
Finally, the approximation in Eq.~\eqref{eq:Jxy Jz Effective Expected b}
follows from neglecting the subleading contribution
$|J^{\,}_{xy}(\lambda)|$.

We are now going to derive the effective Hamiltonian
(\ref{eq:H Effective Boundary Spins})
with the effective couplings
(\ref{eq:Jxy Jz Effective Expected})
with the help of degenerate perturbation theory
about Hamiltonian $\widehat{H}^{(1)}_{\mathrm{U}(1)}(\lambda)$
at $\lambda=\pi/2$.

In order to study the perturbation introduced by
$\pi/2-\lambda^{\,}_{\mathrm{gFM}^{\,}_{z}}<\lambda<\pi/2$
to Hamiltonian $\widehat{H}^{(1)}_{\mathrm{U}(1)}(\lambda)$
at $\lambda=\pi/2$,
we first do the rescaling 
\begin{subequations}
\label{eq:Param PT Boundary Spins}
\begin{align}
&
\cos\lambda\mapsto\frac{1}{\sqrt{2}}\,\cos\lambda
\\
&
\sin\lambda\mapsto\frac{1}{\sqrt{2}}\,\sin\lambda
\end{align}
of the unit of energy. We then insert the expansions
\begin{align}
&
\cos\left(\frac{\pi}{2}-\epsilon\right)=
\epsilon+\mathcal{O}(\epsilon^{3}),
\\
&
\sin\left(\frac{\pi}{2}-\epsilon\right)=
1
-
\frac{\epsilon^{2}}{2}
+
\mathcal{O}(\epsilon^{4}).
\end{align}
In doing so, Hamiltonian $\widehat{H}^{(1)}_{\mathrm{U}(1)}(\lambda)$
becomes
\begin{align}
\widehat{H}(\epsilon)
\coloneqq&\,
\frac{1-\frac{\epsilon^{2}}{2}}{2}
\sum_{j=2}^{2N-2}
\widehat{Z}^{\,}_{j-1}
\left(
\widehat{X}^{\,}_{j}\,
\widehat{X}^{\,}_{j+1}
+
\widehat{Y}^{\,}_{j}\,
\widehat{Y}^{\,}_{j+1}
\right)
\widehat{Z}^{\,}_{j+2}
\nonumber\\
&\,
\mathrlap{{}+\frac{\epsilon}{2}}
\hphantom{\frac{1-\frac{\epsilon^{2}}{2}}{2}}
\sum_{j=1}^{2N-1}
\hphantom{\widehat{Z}^{\,}_{j-1}}
\left(
\widehat{X}^{\,}_{j}\,
\widehat{X}^{\,}_{j+1}
+
\widehat{Y}^{\,}_{j}\,
\widehat{Y}^{\,}_{j+1}
\right)
\nonumber\\
&\,
+
\mathcal{O}(\epsilon^{3}).
\label{eq:H Effective Epsilon Before}
\end{align}
\end{subequations}

Next, we apply the unitary $\widehat{U}^{(1)}_{\mathrm{CZ}}$
defined in Eq.~\eqref{eq:action U_CZ on 3 Paulis}
to Hamiltonian \eqref{eq:H Effective Epsilon Before}.
In doing so, we obtain
\begin{subequations}
\begin{equation}
\widehat{H}'(\epsilon)\coloneqq
\widehat{U}^{(1)}_{\mathrm{CZ}}\,
\widehat{H}(\epsilon)\,
\widehat{U}^{(1)}_{\mathrm{CZ}},
\end{equation}
where
\begin{align}
\widehat{H}'(\epsilon)
=&\,
\frac{1-\frac{\epsilon^{2}}{2}}{2}
\sum_{j=2}^{2N-2}
\hphantom{\widehat{Z}^{\,}_{j-1}}
\left(
\widehat{X}^{\,}_{j}\,
\widehat{X}^{\,}_{j+1}
+
\widehat{Y}^{\,}_{j}\,
\widehat{Y}^{\,}_{j+1}
\right)
\nonumber\\
&\,
\mathrlap{{}+\frac{\epsilon}{2}}
\hphantom{\frac{1-\frac{\epsilon^{2}}{2}}{2}}
\sum_{j=2}^{2N-2}
\widehat{Z}^{\,}_{j-1}
\left(
\widehat{X}^{\,}_{j}\,
\widehat{X}^{\,}_{j+1}
+
\widehat{Y}^{\,}_{j}\,
\widehat{Y}^{\,}_{j+1}
\right)
\widehat{Z}^{\,}_{j+2}
\nonumber \\
&\,
+
\frac{\epsilon}{2}\,
\hphantom{\widehat{Z}^{\,}_{2N-2}}
\left(
\widehat{X}^{\,}_{1}\,
\widehat{X}^{\,}_{2}
+
\widehat{Y}^{\,}_{1}\,
\widehat{Y}^{\,}_{2}
\right)
\widehat{Z}^{\,}_{3}
\nonumber \\
&\,
+
\frac{\epsilon}{2}\,
\widehat{Z}^{\,}_{2N-2}
\left(
\widehat{X}^{\,}_{2N-1}\,
\widehat{X}^{\,}_{2N}
+
\widehat{Y}^{\,}_{2N-1}\,
\widehat{Y}^{\,}_{2N}
\right)
\nonumber\\
&\,
+
\mathcal{O}(\epsilon^{3}).
\label{eq:H Effective Epsilon After}
\end{align}
\end{subequations}
When $\epsilon=0$,
the ground state is four-fold degenerate as 
$\widehat{H}'(\epsilon=0)$ commutes with the pair of triplets
$\widehat{X}^{\,}_{1}$, $\widehat{Y}^{\,}_{1}$, $\widehat{Z}^{\,}_{1}$
and
$\widehat{X}^{\,}_{2N}$, $\widehat{Y}^{\,}_{2N}$, $\widehat{Z}^{\,}_{2N}$.
The term of order $\epsilon^{2}$ on the first line of the right-hand side
yields only a rescaling of the unit energy and thus
does not lift the four-fold degeneracy of the ground state at $\epsilon=0$.
Hence, we may safely ignore it when measuring all energies
with respect to the ground-state energy.

We are going to perform the approximation by which we neglect
the perturbation to the bulk of the chain, i.e.,
all the terms in the second line of \eqref{eq:H Effective Epsilon After}
except the ones for which $j=2$ and $j=2N-2$. In particular, it is
the terms $j=2$ and $j=2N-2$ in the second line of
\eqref{eq:H Effective Epsilon After}
that will contribute to the term proportional to $J^{\,}_{z}$
in Hamiltonian \eqref{eq:H Effective Boundary Spins}.
The third and last lines of \eqref{eq:H Effective Epsilon After}
will contribute to the term proportional to $J^{\,}_{xy}$
in Hamiltonian \eqref{eq:H Effective Boundary Spins}.

At last, we apply the Jordan-Wigner transformation
\eqref{eq: def Jordan-Wigner trsf U(1) symmetric} to the
approximation to Hamiltonian \eqref{eq:H Effective Epsilon After}.
In this way, we obtain the effective perturbative Hamiltonian
\begin{widetext}
\begin{subequations}
\label{eq:H Effective Epsilon Fermion}
\begin{equation}
\widehat{H}^{\,}_{\mathrm{f}}(\epsilon)
\coloneqq
\widehat{H}^{\,}_{0}
+
\widehat{H}^{\,}_{\mathrm{L}}(\epsilon)
+
\widehat{H}^{\,}_{\mathrm{R}}(\epsilon)
+	
\widehat{H}^{\prime}_{\mathrm{L}}(\epsilon)
+
\widehat{H}^{\prime}_{\mathrm{R}}(\epsilon),
\label{eq:H Effective Epsilon Fermion a}
\end{equation}
where
\begin{align}
&
\mathrlap{\widehat{H}^{\,}_{0}}
\hphantom{\widehat{H}^{\,}_{\mathrm{R}}(\epsilon)}=-
\sum_{j=2}^{2N-2}\,
\left(
\hat{c}^{\dag}_{j} 
\hat{c}^{\,}_{j+1}+
\hat{c}^{\dag}_{j+1}
\hat{c}^{\,}_{j}
\right),
\label{eq:H Effective Epsilon Fermion b}
\\
&\widehat{H}^{\,}_{\mathrlap{\mathrm{L}}\hphantom{\mathrm{R}}}(\epsilon)
=-
\epsilon
\left(
\hat{c}^{\dag}_{1}
\hat{c}^{\,}_{2}+
\hat{c}^{\dag}_{2}
\hat{c}^{\,}_{1}
\right)
\left(
2\,\hat{c}^{\dag}_{3}\hat{c}^{\,}_{3}
-
\mathds{1}
\right),
\label{eq:H Effective Epsilon Fermion c}
\\
&\widehat{H}^{\,}_{\mathrm{R}}(\epsilon)
=-
\epsilon
\left(
2\,\hat{c}^{\dag}_{2N-2}\hat{c}^{\,}_{2N-2}
-
\mathds{1}
\right)
\left(
\hat{c}^{\dag}_{2N-1}
\hat{c}^{\,}_{2N}+
\hat{c}^{\dag}_{2N}
\hat{c}^{\,}_{2N-1}
\right),
\label{eq:H Effective Epsilon Fermion d}
\\
&\widehat{H}^{\prime}_{\mathrlap{\mathrm{L}}\hphantom{\mathrm{R}}}(\epsilon)
=-
\epsilon
\left(
2\,\hat{c}^{\dag}_{1}\hat{c}^{\,}_{1}
-
\mathds{1}
\right)
\left(
\hat{c}^{\dag}_{2}
\hat{c}^{\,}_{3}+
\hat{c}^{\dag}_{3}
\hat{c}^{\,}_{2}
\right)
\left(
2\,\hat{c}^{\dag}_{4}\hat{c}^{\,}_{4}
-
\mathds{1}
\right),
\label{eq:H Effective Epsilon Fermion e}
\\
&\widehat{H}^{\prime}_{\mathrm{R}}(\epsilon)
=-
\epsilon
\left(
2\,\hat{c}^{\dag}_{2N-3}\hat{c}^{\,}_{2N-3}
-
\mathds{1}
\right)
\left(
\hat{c}^{\dag}_{2N-2}
\hat{c}^{\,}_{2N-1}+
\hat{c}^{\dag}_{2N-1}
\hat{c}^{\,}_{2N-2}
\right)
\left(
2\,\hat{c}^{\dag}_{2N}\hat{c}^{\,}_{2N}
-
\mathds{1}
\right).
\label{eq:H Effective Epsilon Fermion f}
\end{align}
\end{subequations}
\end{widetext}

In order to obtain the effective Hamiltonian
\eqref{eq:H Effective Boundary Fermions}
together with the couplings \eqref{eq:Jxy Jz Effective Expected},
it is sufficient to perform degenerate perturbation theory
up to second order in powers of $\epsilon$.
Indeed, according to the right column of
Table \ref{tab:Fits DQ Neel SPT OBCs}, we have
\begin{equation}
A^{\,}_{\mathrm{p}}(\epsilon)\propto
\epsilon^{2}+
\mathcal{O}(\epsilon^{4}),
\qquad
v^{\,}_{2,1}(\epsilon)\propto
\epsilon^{2}+
\mathcal{O}(\epsilon^{4}).
\end{equation}
This calculation is summarized in Appendix
\ref{sec:effective_interactions_between_boundary_spin_1_2_degrees_of_freedom}.

\subsection{The free Majorana line $\theta=\pi/2$}
\label{subsec:The free Majorana line theta=piover2}

\subsubsection{Introduction}

Section \ref{subsec:The free Majorana line theta=piover2}
is devoted exclusively to Hamiltonian
(\ref{eq:def HA and HB d})
along the line $\theta=\pi/2$, 
a line that is unitarily equivalent to the line $\theta=0$
according to Eq.\
(\ref{eq:action U_R on Hamiltonians}).
Along the line $\theta=\pi/2$,
Hamiltonian
(\ref{eq:def HA and HB d})
interpolates between the members
$r=-1$ and $r=3$
of the family of quantum spin-$1/2$ cluster-$r$ Hamiltonians
\begin{subequations}
\label{eq:def quantum spin1over2 cluster-r Hamiltonians}
\begin{equation}
\widehat{H}^{(\mathsf{b})}_{r}\coloneqq
\begin{cases}
\sum\limits_{j=1}^{2N-\mathsf{b}|r|}
\widehat{Y}^{\,}_{j}\,
\widehat{X}^{\,}_{j+1}\,
\cdots
\widehat{X}^{\,}_{j+|r|-1}\,
\widehat{Y}^{\,}_{j+|r|},
&
r<-1,
\\
\sum\limits_{j=1}^{\mathrlap{2N-\mathsf{b}}
\phantom{2N-\mathsf{b}|r|}}
\widehat{Y}^{\,}_{j}\,
\widehat{Y}^{\,}_{j+1},
&
r=-1,
\\
\sum\limits_{j=1}^{\mathrlap{2N}
\phantom{2N-\mathsf{b}|r|}}
\widehat{X}^{\,}_{j},
&
r=0,
\\
\sum\limits_{j=1}^{\mathrlap{2N-\mathsf{b}}
\phantom{2N-\mathsf{b}|r|}}
\widehat{Z}^{\,}_{j}\,
\widehat{Z}^{\,}_{j+1},
&
r=+1,
\\
\sum\limits_{j=1}^{2N-\mathsf{b}|r|}
\widehat{Z}^{\,}_{j}\,
\widehat{X}^{\,}_{j+1}\,
\cdots
\widehat{X}^{\,}_{j+|r|-1}\,
\widehat{Z}^{\,}_{j+|r|},
&
r>+1,
\end{cases}
\label{eq:def quantum spin1over2 cluster-r Hamiltonians a}
\end{equation}
labeled by the integer $r\in\mathbb{Z}$
and where, as usual, $\mathsf{b}=0$ ($\mathsf{b}=1$) selects PBCs (OBCs), i.e.,
\begin{equation}
\widehat{H}^{(\mathsf{b})}_{\mathrm{II}}(\pi/2,\lambda)\coloneqq
\cos(\lambda)\,\widehat{H}^{(\mathsf{b})}_{-1}
+
\sin(\lambda)\,\widehat{H}^{(\mathsf{b})}_{+3}.
\label{eq:def quantum spin1over2 cluster-r Hamiltonians b}
\end{equation}
\end{subequations}
The family of quantum spin-$1/2$ cluster-$r$ Hamiltonians
(\ref{eq:def quantum spin1over2 cluster-r Hamiltonians a})
realizes four distinct bosonic SPT phases
that are protected by the symmetry group generated by
\cite{Verresen17,MudryChamon2025}
\begin{subequations}
\begin{equation}
\widehat{U}^{\,}_{X}\coloneqq
\prod_{j=1}^{2N}\widehat{X}_{j} 
\end{equation}
and 
\begin{equation}
\widehat{U}^{\,}_{T'}\coloneqq
\widehat{U}^{\,}_{Y}\,\widehat{U}^{\,}_{T}\equiv
\widehat{U}^{\,}_{Y}\left(\widehat{U}^{\,}_{Y}\,\mathsf{K}\right)=
\mathsf{K}.
\end{equation}
\end{subequations}
On the one hand,
if
\begin{equation}
|\cos(\lambda)|\neq|\sin(\lambda)|
\label{eq: cos lambda neq sin lambda in magnitude}
\end{equation}
and the thermodynamic limit has been taken,
then
$\widehat{H}^{(0)}_{\mathrm{II}}(\pi/2,\lambda)$
is gapped with a two-fold degenerate ground state,
while 
$\widehat{H}^{(1)}_{\mathrm{II}}(\pi/2,\lambda)$
is gapped with either a two-fold or eight-fold degenerate ground state
when
$|\cos(\lambda)|>|\sin(\lambda)|$
or
$|\cos(\lambda)|<|\sin(\lambda)|$,
respectively \cite{Verresen17,MudryChamon2025}.
On the other hand,
if
\begin{equation}
|\cos(\lambda)|=|\sin(\lambda)|
\label{eq: cos lambda eq sin lambda in magnitude}
\end{equation}
and the thermodynamic limit has been taken,
$\widehat{H}^{(\mathsf{b})}_{\mathrm{II}}(\pi/2,\lambda)$
is gapless and realizes a free-fermion conformal field theory with central
charge $\mathsf{c}=2$ \cite{Verresen17,MudryChamon2025}.

The goal of
Sec.\ \ref{subsec:The free Majorana line theta=piover2}
is to study the finite-size spectra of
$\widehat{H}^{(\mathsf{b})}_{\mathrm{II}}(\pi/2,\lambda)$
when $\lambda=\pi/4,3\pi/4$.
In Sec.~\ref{sec:Stability analysis of the critical points}, we will
study the stability against the perturbation induced by
$\theta\neq\pi/2$ of the CFT with central charge $\mathsf{c}=2$
when $\lambda=\pi/4,3\pi/4$.
To this end, we shall first perform a Jordan-Wigner transformation
in Sec.\ \ref{subsubsec:Jordan-Wigner transformation}
under which $\widehat{H}^{(\mathsf{b})}_{\mathrm{II}}(\pi/2,\lambda)$
is mapped to either a free open ($\mathsf{b}=1$) or a
free closed ($\mathsf{b}=0$) Majorana chain.

\subsubsection{Jordan-Wigner transformation}
\label{subsubsec:Jordan-Wigner transformation}

\paragraph{Case of OBCs $(\mathsf{b=1})$.}
Define
\begin{subequations}
\label{eq: def Jordan-Wigner trsf}
\begin{align}
&
\hat{\xi}^{\,}_{j}\coloneqq
\widehat{X}^{\,}_{1}\,
\cdots\,
\widehat{X}^{\,}_{j-1}\,
\widehat{Y}^{\,}_{j}
=
\hat{\xi}^{\dag}_{j}=
-\mathsf{K}\,\hat{\xi}^{\,}_{j}\,\mathsf{K},
\\
&
\hat{\eta}^{\,}_{j}\coloneqq
\widehat{X}^{\,}_{1}\,
\cdots\,
\widehat{X}^{\,}_{j-1}\,
\widehat{Z}^{\,}_{j}=
\hat{\eta}^{\dag}_{j}=
+\mathsf{K}\,\hat{\eta}^{\,}_{j}\,\mathsf{K},
\end{align}
\end{subequations}
for any site $j=1,\ldots,2N$.
It follows that
\begin{subequations}
\begin{align}
&
+\mathrm{i}
\hat{\xi}^{\,}_{j+r}\,
\hat{\eta}^{\,}_{j}=
+
\widehat{Y}^{\,}_{j}\,
\widehat{X}^{\,}_{j+1}\cdots
\widehat{X}^{\,}_{j+r-1}\,
\widehat{Y}^{\,}_{j+r}
\end{align}
for $j=1,\ldots,2N-r$,
\begin{align}
&
+\mathrm{i}
\hat{\xi}^{\,}_{j}\,
\hat{\eta}^{\,}_{j}=
-\widehat{X}^{\,}_{j}
\end{align}
for $j=1,\ldots,2N$,
\begin{align}
&
+\mathrm{i}
\hat{\xi}^{\,}_{j}\,
\hat{\eta}^{\,}_{j+r}=
+
\widehat{Z}^{\,}_{j}\,
\widehat{X}^{\,}_{j+1}\cdots
\widehat{X}^{\,}_{j+r-1}\,
\widehat{Z}^{\,}_{j+r}
\end{align}
for $j=1,\ldots,2N-r$, and 
\begin{equation}
\widehat{X}\equiv
\prod_{j=1}^{2N}\widehat{X}^{\,}_{j}=
\prod_{j=1}^{2N}
\left(
-\mathrm{i}
\hat{\xi}^{\,}_{j}\,
\hat{\eta}^{\,}_{j}
\right)\equiv
\widehat{P}^{\,}_{\mathrm{F}}.
\label{eq:parity operator in Majorana rep}
\end{equation}
Moreover, one verifies that the $\hat{\xi}$'s and $\hat{\eta}$'s
obey the Majorana algebra given by
\begin{equation}
\left\{
\hat{\xi}^{\,}_{j},
\hat{\xi}^{\,}_{j'}
\right\}=
\left\{\vphantom{\hat{\xi}^{\,}_{j}}
\hat{\eta}^{\,}_{j},
\hat{\eta}^{\,}_{j'}
\right\}=2\delta^{\,}_{j,j'},
\qquad
\left\{
\hat{\xi}^{\,}_{j},
\hat{\eta}^{\,}_{j'}
\right\}=
0,
\label{eq:Majorana Algebra}
\end{equation}
\end{subequations}
for $j,j'=1,\ldots,2N$. We may thus rewrite
Hamiltonian (\ref{eq:def quantum spin1over2 cluster-r Hamiltonians b})
as the non-interacting Majorana Hamiltonian
\begin{equation}
\begin{split}
\widehat{H}^{(1)}_{\mathrm{II}}(\pi/2,\lambda)=&\,
\cos(\lambda)
\sum_{j=1}^{2N-1}
\mathrm{i}
\hat{\xi}^{\,}_{j+1}\,
\hat{\eta}^{\,}_{j}
\\
&\,
+
\sin(\lambda)
\sum_{j=1}^{2N-3}
\mathrm{i}
\hat{\xi}^{\,}_{j}\,
\hat{\eta}^{\,}_{j+3}
\end{split}
\label{eq:Majorana-representation-theta=piover2-b=1}
\end{equation}
on an open chain
(see Refs.\ \cite{Verresen17} and \cite{MudryChamon2025}).

The two-fold degeneracy of the gapped ground state
when $\lambda=0$ is attributed to the absence of operators
$\hat{\xi}^{\,}_{1}$ and $\hat{\eta}^{\,}_{2N}$ in
$\widehat{H}^{(1)}_{\mathrm{II}}(\pi/2,0)$.

The eight-fold degeneracy of the gapped ground state
when $\lambda=\pi/2$ is attributed to the absence of operators
$\hat{\eta}^{\,}_{1}$, $\hat{\eta}^{\,}_{2}$, $\hat{\eta}^{\,}_{3}$,
$\hat{\xi}^{\,}_{2N-2}$, $\hat{\xi}^{\,}_{2N-1}$, and $\hat{\xi}^{\,}_{2N}$ in
$\widehat{H}^{(1)}_{\mathrm{II}}(\pi/2,\pi/2)$.

The many-body spectral gap of
$\widehat{H}^{(1)}_{\mathrm{II}}(\pi/2,\lambda)$
on the interval $\lambda\in[0,\pi]$
closes if and only if
$\lambda=\pi/4,3\pi/4$.
We show in Appendix
\ref{appsec:Degeneracies of the ground states H1II piover4 or 3piover3}
that the dependence on the number $2N$ of sites of
the ground state degeneracy (GSD) for Hamiltonian
$\widehat{H}^{(1)}_{\mathrm{II}}(\pi/2,\pi/2\mp\pi/4)$
is given by
\begin{equation}
\mathrm{GSD}^{(1)}_{\mathrm{II}}
\left(\frac{\pi}{2},\frac{\pi}{2}\mp\frac{\pi}{4}\right)=
\begin{cases}
1,&\mathrm{if}\ 2N\,\mathrm{mod}\,4=0,
\\&\\
2,&\mathrm{if}\ 2N\,\mathrm{mod}\,2=2.
\end{cases}
\label{eq:sensitivity to 2Nover4 being integer or half integer of GSD H1II if gapless} 
\end{equation}
This result is reported in Table
\ref{tab:TQ HB Degeneracy}.

\begin{table}
\caption{
Dependence on the value of
$2N \bmod 4$, denoted by $2N^{\,}_{4}$, of the ground-state degeneracy
$d^{\,}_{\lambda,\mathsf{b};2N^{\,}_{4}}$
of
$\widehat{H}^{(\mathsf{b})}_{\mathrm{II}}(\pi/2,\lambda)$
at the critical values $\lambda=\pi/4,3\pi/4$
for both $\mathsf{b}=0$ (PBCs) and $\mathsf{b}=1$ (OBCs).
    }
\label{tab:TQ HB Degeneracy}
\centering
\renewcommand{\arraystretch}{1.1}
\setlength{\tabcolsep}{5pt}
\begin{tabular}{c c cc c cc c cc c cc}
\hline\hline
$\lambda$
& \phantom{ab}
& \multicolumn{5}{c}{$\frac{\pi}{4}$}
& \phantom{ab}
& \multicolumn{5}{c}{$\frac{3\pi}{4}$} \\
\cmidrule{3-7} \cmidrule{9-13}
$\mathsf{b}$ 
& 
& \multicolumn{2}{c}{0} 
&
& \multicolumn{2}{c}{1}
& 
& \multicolumn{2}{c}{0} 
&
& \multicolumn{2}{c}{1} \\
\cmidrule{3-7} \cmidrule{9-13}
$2N^{\,}_{4}$
&
& 0 & 2
&
& 0 & 2 
& 
& 0 & 2
&
& 0 & 2 \\
\midrule
$d^{\,}_{\lambda,\mathsf{b};2N^{\,}_{4}}$ 
&
& 1 & 2
&
& 1 & 2 
& 
& 1 & 4
&
& 1 & 2 \\
\hline\hline
\end{tabular}
\end{table}

\paragraph{Case of PBCs $(\mathsf{b=0})$.}
We select PBCs by choosing $\mathsf{b=0}$
for Hamiltonian
(\ref{eq:def quantum spin1over2 cluster-r Hamiltonians b}).
Following Refs.\ \cite{MudryChamon2025} and
\cite{Aksoy2024},
we decompose the $2^{2N}$-dimensional Hilbert space
(\ref{eq:def Hilbert space d})
on which
$\widehat{H}^{(\mathsf{b})}_{\mathrm{II}}(\pi/2,\lambda)$
acts as the direct sum
\begin{subequations}
\label{sec:def direct sum over X or PF eigenvalues}
\begin{equation}
\mathfrak{h}=
\mathfrak{h}^{\,}_{+}\oplus\mathfrak{h}^{\,}_{-}
\label{sec:def direct sum over X or PF eigenvalues a}
\end{equation}
over the orthogonal $2^{2N-1}$-dimensional eigenspaces
\begin{equation}
\mathfrak{h}^{\,}_{\pm}\coloneqq
\widehat{P}^{\,}_{\pm}\,\mathfrak{h},
\qquad
\widehat{P}^{\,}_{\pm}\coloneqq
\frac{\mathds{1}^{\,}_{\mathfrak{H}}\pm\widehat{X}}{2}
\label{sec:def direct sum over X or PF eigenvalues b}
\end{equation}
\end{subequations}
of $\widehat{X}$ with eigenvalues $\pm1$, respectively.
If the Majorana operators obey the PBCs
\begin{subequations}
\label{eq:case PBC Majorana and parity +}
\begin{equation}
\hat{\xi}^{\,}_{j+2N}=+\hat{\xi}^{\,}_{j},
\quad
\hat{\eta}^{\,}_{j+2N}=+\hat{\eta}^{\,}_{j},
\quad
j=1,\ldots,2N,
\label{eq:case PBC Majorana and parity + b}
\end{equation}
we then have the identity
\begin{equation}
\begin{split}
\widehat{P}^{\,}_{+}\,
\widehat{H}^{(0)}_{\mathrm{II}}(\pi/2,\lambda)\,
\widehat{P}^{\,}_{+}=&\,
\cos(\lambda)
\sum_{j=1}^{2N}
\mathrm{i}
\hat{\xi}^{\,}_{j+1}\,
\hat{\eta}^{\,}_{j}
\\
&\,
+
\sin(\lambda)
\sum_{j=1}^{2N}
\mathrm{i}
\hat{\xi}^{\,}_{j}\,
\hat{\eta}^{\,}_{j+3}.
\end{split}
\label{eq:case PBC Majorana and parity + a}
\end{equation}
\end{subequations}
Similarly,
if the Majorana operators obey the antiperiodic boundary conditions
(ABCs)
\begin{subequations}
\label{eq:case APBC Majorana and parity -}
\begin{equation}
\hat{\xi}^{\,}_{j+2N}=-\hat{\xi}^{\,}_{j},
\quad
\hat{\eta}^{\,}_{j+2N}=-\hat{\eta}^{\,}_{j},
\quad
j=1,\ldots,2N.
\label{eq:case APBC Majorana and parity - b}
\end{equation}
we then have the identity
\begin{equation}
\begin{split}
\widehat{P}^{\,}_{-}\,
\widehat{H}^{(0)}_{\mathrm{II}}(\pi/2,\lambda)\,
\widehat{P}^{\,}_{-}=&\,
\cos(\lambda)
\sum_{j=1}^{2N}
\mathrm{i}
\hat{\xi}^{\,}_{j+1}\,
\hat{\eta}^{\,}_{j}
\\
&\,
+
\sin(\lambda)
\sum_{j=1}^{2N}
\mathrm{i}
\hat{\xi}^{\,}_{j}\,
\hat{\eta}^{\,}_{j+3}.
\end{split}
\label{eq:case APBC Majorana and parity - a}
\end{equation}
\end{subequations}
Here, it is understood that 
\begin{equation}
\widehat{P}^{\,}_{\pm}\coloneqq
\frac{\mathds{1}^{\,}_{\mathfrak{H}}\pm\widehat{P}^{\,}_{\mathrm{F}}}{2}
\label{eq:def-widehatPpm-Majorana-rep}
\end{equation}
in the Majorana representation.

By construction, the ground states of
Hamiltonians
(\ref{eq:case PBC Majorana and parity + a})
and
(\ref{eq:case APBC Majorana and parity - a})
are non-degenerate and gapped whenever condition
(\ref{eq: cos lambda neq sin lambda in magnitude})
holds. We show in Appendix
\ref{appsec:Degeneracies of the ground states H0II piover4 or 3piover3}
that, at the two gap-closing topological transitions
set by $\lambda=\pi/4,3\pi/4$
for which condition
(\ref{eq: cos lambda eq sin lambda in magnitude})
holds, the dependence on the number $2N$ of sites of the
number $Z^{\,}_{\lambda,\mathsf{p};2N}$
of (single-particle) fermionic zero modes for
the Majorana representation of Hamiltonian
\begin{subequations}
\label{eq:dege H0II resolved by parity at criticality}
\begin{equation}
\widehat{H}^{(0)}_{\mathrm{II}}(\pi/2,\lambda;\mathsf{p})\equiv
\widehat{P}^{\,}_{\mathsf{p}}\,
\widehat{H}^{(0)}_{\mathrm{II}}(\pi/2,\lambda)\,
\widehat{P}^{\,}_{\mathsf{p}},
\qquad
\mathsf{p}=\pm,
\label{eq:dege H0II resolved by parity at criticality a}
\end{equation}
is given by
\begin{equation}
Z^{\,}_{\frac{\pi}{4},\mathsf{p};2N}=
\begin{cases}
4\delta^{\,}_{\mathsf{p},+},
&
\mathrm{if}\,2N\,\mathrm{mod}\,8=0,
\\&\\
0,
&
\mathrm{if}\,2N\,\mathrm{mod}\,8=2,6,
\\&\\
4\delta^{\,}_{\mathsf{p},-},
&
\mathrm{if}\,2N\,\mathrm{mod}\,8=4,
\end{cases}
\label{eq:dege H0II resolved by parity at criticality b}
\end{equation}
and
\begin{equation}
Z^{\,}_{\frac{3\pi}{4},\mathsf{p};2N}=
\begin{cases}
4\delta^{\,}_{\mathsf{p},+},
&
\mathrm{if}\,2N\,\mathrm{mod}\,4=0,
\\&\\
2,
&
\mathrm{if}\,2N\,\mathrm{mod}\,4=2,
\end{cases}
\label{eq:dege H0II resolved by parity at criticality c}
\end{equation}
\end{subequations}
respectively. We also show in
Appendix
\ref{appsec:Degeneracies of the ground states H0II piover4 or 3piover3}
how to deduce from Eqs.\
(\ref{eq:dege H0II resolved by parity at criticality b})
and
(\ref{eq:dege H0II resolved by parity at criticality c})
in combination with the evaluation of the four ground-state energy eigenvalues
of $\widehat{H}^{(0)}_{\mathrm{II}}(\pi/2,\lambda;\mathsf{p})$
when $\lambda=\pi/4,3\pi/4$ and $\mathsf{p}=\pm$
the dependence on the number $2N$ of sites of the degeneracy of
\begin{equation}
\widehat{H}^{(0)}_{\mathrm{II}}(\pi/2,\lambda)=
\bigoplus\limits_{\mathsf{p}=\pm} 
\widehat{H}^{(0)}_{\mathrm{II}}(\pi/2,\lambda;\mathsf{p})
\label{eq:Majorana-representation-theta=piover2-b=0} 
\end{equation}
as reported in Table
\ref{tab:TQ HB Degeneracy}.

\subsection{Stability analysis of the critical points
$\widehat{H}^{(\mathsf{b})}_{\mathrm{II}}(\pi/2,\lambda)$
for
$\lambda=\pi/4$ and $3\pi/4$}
\label{sec:Stability analysis of the critical points}

\subsubsection{Introduction}

With the help of the Majorana representations
(\ref{eq:Majorana-representation-theta=piover2-b=1})
and
(\ref{eq:Majorana-representation-theta=piover2-b=0}), 
we are going to show that Hamiltonian
$\widehat{H}^{(\mathsf{b})}_{\mathrm{II}}(\theta,\lambda)$
defined in Eq.~\eqref{eq:def HA and HB d}
supports two quantum critical points at
\begin{equation}
(\pi/2,\pi/4),
\qquad
(\pi/2,3\pi/4).
\label{eq:critical-points-along-piover2-vertical-line}
\end{equation}
We have seen in Sec.\
(\ref{subsec:The free Majorana line theta=piover2})
that these two quantum critical points
are unstable to gap-opening one-body terms
induced by the perturbation
\begin{equation}
(\pi/2,\pi/4+\delta\lambda),
\qquad
(\pi/2,3\pi/4+\delta\lambda),
\end{equation}
for any non-vanishing $\delta\lambda$.
The goal of
Sec.~\ref{sec:Stability analysis of the critical points}
is to investigate the stability of the two critical points
(\ref{eq:critical-points-along-piover2-vertical-line})
to the perturbation
\begin{equation}
(\pi/2+\delta\theta,\pi/4),
\qquad
(\pi/2+\delta\theta,3\pi/4),
\end{equation}
for any non-vanishing $\delta\theta$.
With the help of the Majorana representation
(\ref{eq:Majorana-representation-theta=piover2-b=0}), 
we are going to show that normal-ordering of the quartic interaction at
$(\pi/2+\delta\theta,3\pi/4)$
produces a quadratic term that opens a gap,
whereas normal-ordering of the quartic interaction at
$(\pi/2+\delta\theta,\pi/4)$
produces a quadratic term that does not open a gap.
In the latter case, the remaining normal-ordered
interactions are marginal at tree level and
their stability must therefore
be determined beyond tree level.

When imposing the PBCs
$\mathsf{b}=0$
on
$\widehat{H}^{(\mathsf{b})}_{\mathrm{II}}(\theta,\lambda)$
defined in Eq.~\eqref{eq:def HA and HB d},
we can only perform the Jordan-Wigner 
transformation (\ref{eq: def Jordan-Wigner trsf})
in the subspaces of fixed fermion parity
subject to the conditions
(\ref{eq:case PBC Majorana and parity + b})
and
(\ref{eq:case APBC Majorana and parity - b}),
respectively. The Majorana representation
of $\widehat{H}^{(0)}_{\mathrm{II}}(\theta,\lambda)$
that we shall use is given by
\begin{subequations}
\label{eq:def parity resolved Hamiltonian stability}
\begin{align}
&
\widehat{H}^{(0)}_{\mathrm{II}}(\theta,\lambda)=
\bigoplus_{\mathsf{p}=\pm}
\widehat{H}^{(0)}_{\mathrm{II}}(\theta,\lambda;\mathsf{p}),
\label{eq:def parity resolved Hamiltonian stability a}
\\
&
\widehat{H}^{(0)}_{\mathrm{II}}(\theta,\lambda;\mathsf{p})\coloneqq
\widehat{P}^{\,}_{\mathsf{p}}\,
\widehat{H}^{(0)}_{\mathrm{II}}(\theta,\lambda)\,
\widehat{P}^{\,}_{\mathsf{p}},
\label{eq:def parity resolved Hamiltonian stability b}
\end{align}
where $\widehat{P}^{\,}_{\mathsf{p}}$ is defined in Eq.\
(\ref{eq:def-widehatPpm-Majorana-rep})
and the sector $\mathsf{p}=\pm$ selects the boundary conditions
\begin{equation}
\hat{\xi}^{\,}_{j+2N}=
\mathsf{p}\,
\hat{\xi}^{\,}_{j},
\qquad
\hat{\eta}^{\,}_{j+2N}=
\mathsf{p}\,
\hat{\eta}^{\,}_{j},
\label{eq:def parity resolved Hamiltonian stability c}
\end{equation}
\end{subequations}
for $j=1,\ldots,2N$ obeyed by the Majorana operators.
The sector $\mathsf{p}=+$ ($\mathsf{p}=-$)
imposes periodic (antiperiodic) boundary conditions
for the Majorana operators.

The goal of this section is a stability analysis of the
bulk thermodynamic phase.
Hence, without loss of generality,
we shall perform the stability analysis in the sector
$\mathsf{p}=+$
in which Hamiltonian
(\ref{eq:def HA and HB d})
has the Majorana representation
\begin{widetext}
\begin{subequations}
\label{eq:def Exact Hamiltonian Majorana Stability}
\begin{equation}
\widehat{H}^{(0)}_{\mathrm{II}}(\theta,\lambda;+)=
\cos\lambda\,
\sum_{j=1}^{2N}
\left(
-
\cos\theta\,
\hat{\xi}^{\,}_{j}\,
\hat{\eta}^{\,}_{j}\,
\hat{\xi}^{\,}_{j+1}\,
\hat{\eta}^{\,}_{j+1}
-
\sin\theta\,
\ii
\hat{\eta}^{\,}_{j}\,
\hat{\xi}^{\,}_{j+1}
\right)
+
\sin\lambda\,
\sum_{j=1}^{2N}
\left(
+
\cos\theta\,
\hat{\xi}^{\,}_{j}\,
\hat{\xi}^{\,}_{j+1}\,
\hat{\eta}^{\,}_{j+2}\,
\hat{\eta}^{\,}_{j+3}
+
\sin\theta\,
\ii
\hat{\xi}^{\,}_{j}\,
\hat{\eta}^{\,}_{j+3}
\right)
\label{eq:def Exact Hamiltonian Majorana Stability a}
\end{equation}
with the Majorana operators obeying the PBCs
\begin{equation}
\hat{\xi}^{\,}_{j+2N}=+\hat{\xi}^{\,}_{j},
\qquad
\hat{\eta}^{\,}_{j+2N}=+\hat{\eta}^{\,}_{j},
\qquad
j=1,\ldots,2N.
\label{eq:def Exact Hamiltonian Majorana Stability b}
\end{equation}
\end{subequations}
\end{widetext}
As it should be, we recover the quadratic Hamiltonian
(\ref{eq:case PBC Majorana and parity + a})
when $\theta=\pi/2$.

We introduce the dimensionless parameter
$0\leq\epsilon\ll1$
by expanding Hamiltonian
(\ref{eq:def Exact Hamiltonian Majorana Stability})
up to linear order in powers of
\begin{subequations}
\begin{equation}
0\leq\epsilon\coloneqq
\begin{cases}
\theta
-
\dfrac{\pi}{2}\ll1,
&
\hbox{if $\lambda=\dfrac{3\pi}{4}$,}
\\ \\
\dfrac{\pi}{2}
-
\theta\ll1,
&
\hbox{if $\lambda=\dfrac{\pi}{4}$,}
\end{cases}
\end{equation}
with the help of the expansions 
\begin{align}
&
\cos\left(\frac{\pi}{2}\pm\epsilon\right)=
\mp\,\epsilon+\mathcal{O}(\epsilon^{3}),
\\
&
\sin\left(\frac{\pi}{2}\pm\epsilon\right)=
1
-
\frac{\epsilon^{2}}{2}
+
\mathcal{O}(\epsilon^{4}).
\end{align}
\end{subequations}
With $\widehat{H}^{\,}_{+}(\epsilon)$ defined by
\begin{subequations}
\label{eq:Relation Hamiltonian Original VS H+-}
\begin{equation}
\sqrt{2}\,
\widehat{H}^{(0)}_{\mathrm{II}}(\pi/2+\epsilon,3\pi/4;+)=
\widehat{H}^{\,}_{+}(\epsilon)
+
\mathcal{O}(\epsilon^{2}),
\label{eq:Relation Hamiltonian Original VS H+- a}
\end{equation}
and $\widehat{H}^{\,}_{-}(\epsilon)$ defined by
\begin{equation}
\sqrt{2}\,
\widehat{H}^{(0)}_{\mathrm{II}}(\pi/2-\epsilon,\hphantom{3}\pi/4;+)=
\widehat{H}^{\,}_{-}(\epsilon)
+
\mathcal{O}(\epsilon^{2}),
\label{eq:Relation Hamiltonian Original VS H+- b}
\end{equation}
\end{subequations}
for $0\leq\epsilon<\pi/2$, we find that
\begin{subequations}
\label{eq:def Perturbed Hamiltonian Majorana Stability}
\begin{equation}
\widehat{H}^{\,}_{\pm}(\epsilon)=
\widehat{T}^{\,}_{\pm}
+
\epsilon\,
\widehat{V}^{\,}_{\pm},
\label{eq:def Perturbed Hamiltonian Majorana Stability a}
\end{equation}
where
\begin{align}
&
\widehat{T}^{\,}_{\pm}\coloneqq
\sum_{j=1}^{2N}
\left(
\pm
\ii
\hat{\eta}^{\,}_{j}\,
\hat{\xi}^{\,}_{j+1}
+
\ii
\hat{\xi}^{\,}_{j}\,
\hat{\eta}^{\,}_{j+3}
\right),
\label{eq:def Perturbed Hamiltonian Majorana Stability b}
\\
&
\widehat{V}^{\,}_{\pm}\coloneqq
\sum_{j=1}^{2N}
\left(
-
\hat{\xi}^{\,}_{j}\,
\hat{\eta}^{\,}_{j}\,
\hat{\xi}^{\,}_{j+1}\,
\hat{\eta}^{\,}_{j+1}
\mp
\hat{\xi}^{\,}_{j}\,
\hat{\xi}^{\,}_{j+1}\,
\hat{\eta}^{\,}_{j+2}\,
\hat{\eta}^{\,}_{j+3}
\right),
\label{eq:def Perturbed Hamiltonian Majorana Stability c}
\end{align}
\end{subequations}
with the Majorana operators obeying the PBCs
(\ref{eq:def Exact Hamiltonian Majorana Stability b}).
The kinetic term $\widehat{T}^{\,}_{\pm}$
is quadratic in the Majorana operators.
It defines the Gaussian critical theory.
The interacting term $\widehat{V}^{\,}_{\pm}$
is quartic in the Majorana operators.
It is the leading perturbation generated by the deviation
from $\theta=\pi/2$.

\subsubsection{Diagonalization of the free Hamiltonians $\widehat{T}^{\,}_{\pm}$}

If we interpret the pair of Majorana operators
$\hat{\xi}_{j}^{\,}$
and
$\hat{\eta}_{j}^{\,}$
as twice the imaginary and real parts of the operator
$\hat{c}^{\dag}_{j}$, respectively, i.e.,
\begin{equation}
\hat{\xi}^{\,}_{j}=
-\ii
\left(
\hat{c}^{\dag}_{j}-\hat{c}^{\,}_{j}
\right),
\qquad
\hat{\eta}^{\,}_{j}=
\hat{c}^{\dag}_{j}+\hat{c}^{\,}_{j},
\label{eq:Def Majorana X-basis}
\end{equation}
respectively, it then follows that
the operators $\hat{c}^{\,}_{j},\hat{c}^{\dag}_{j}$
obey the fermionic algebra
\eqref{eq:def HbU(1) fermionic as function lambda c}.
We can rewrite the kinetic Hamiltonian
\eqref{eq:def Perturbed Hamiltonian Majorana Stability b}
in terms of these fermionic operators as
\begin{align}
\widehat{T}^{\,}_{\pm}=
\sum_{j=1}^{2N}
\left[\vphantom{\hat{c}^{\dag}_{j}}\right.
&
\pm\left(
\hat{c}^{\dag}_{j}\,
\hat{c}^{\dag}_{j+1}
-
\hat{c}^{\dag}_{j}\,
\hat{c}^{\,}_{j+1}
+
\mathrm{H.c.}
\right)
\nonumber\\
&
+
\left.\left(
\hat{c}^{\dag}_{j}\,
\hat{c}^{\dag}_{j+3}
+
\hat{c}^{\dag}_{j}\,
\hat{c}^{\,}_{j+3}
+
\mathrm{H.c.}
\right)
\right].
\label{eq:T pm Fermions Lattice}
\end{align}

If we insert the Fourier expansions \eqref{eq:Fourier MF}
in Hamiltonian \eqref{eq:T pm Fermions Lattice},
we obtain the Nambu representation of Hamiltonian
\eqref{eq:T pm Fermions Lattice}
given by
\begin{subequations}
\label{eq:T pm Fermions Fourier}
\begin{equation}
\widehat{T}^{\,}_{\pm}=
\frac{1}{2}
\sum_{k\in\Lambda^{\star}}
\widehat{\Psi}^{\,\dag}_{k}\,
H^{\,}_{\pm,k}\,
\widehat{\Psi}^{\,}_{k},
\label{eq:T pm Fermions Fourier a}
\end{equation}
where the Brillouin zone $\Lambda^{\star}$
was defined in Eq.\ (\ref{eq:Fourier MF})
and we have introduced the operator-valued Nambu spinors
\begin{equation}
\widehat{\Psi}^{\,\dag}_{k}\coloneqq
\pmqty{
\hat{c}^{\dag}_{k}
&
\hat{c}^{\,}_{-k}
      },
\qquad
\widehat{\Psi}^{\,}_{k}\coloneqq
\pmqty{
\hat{c}^{\,}_{k}
\\
\hat{c}^{\dag}_{-k}
   },
\label{eq:T pm Fermions Fourier b}
\end{equation}
together with the $2\times2$
single-particle Bogoliubov-de Gennes (BdG) matrix
\begin{equation}
H^{\,}_{\pm,k}\coloneqq
\pmqty{
t^{\,}_{\pm,k} & \Delta^{*}_{\pm,k}
\\
\Delta^{\,}_{\pm,k} & -t^{\,}_{\pm,k}
   }
\label{eq:T pm Fermions Fourier c}
\end{equation}
whose matrix elements are given by the $k$-even function
\begin{align}
&
t^{\,}_{\pm,k}\coloneqq
2
\left[
\mp\cos(k)+\cos(3k)
\right]=
+
t^{\,}_{\pm,-k}
\in\mathbb{R},
\end{align}
and the $k$-odd function
\begin{align}
&
\Delta^{\,}_{\pm,k}\coloneqq
2\ii
\left[
\mp\sin(k)-\sin(3k)
\right]=
-
\Delta^{\,}_{\pm,-k}
\in\mathbb{C}.
\label{eq:T pm Fermions Fourier d}
\end{align}
\end{subequations}
It appears that the right-hand side of
Eq.\ (\ref{eq:T pm Fermions Fourier a})
is invariant under the transformation
$\widehat{\Psi}^{\,}_{k}\mapsto e^{+\ii\alpha}\,\widehat{\Psi}^{\,}_{k}$
and
$\widehat{\Psi}^{\dag}_{k}\mapsto e^{-\ii\alpha}\,\widehat{\Psi}^{\dag}_{k}$
for any $\alpha\in[0,2\pi[$. However, this is a spurious symmetry, for
the only sensible $\mathrm{U}(1)$ transformation
consists in multiplying
the lattice fermion operator $\hat{c}^{\,}_{j}$ in Eq.\
(\ref{eq:T pm Fermions Lattice})
by a global phase factor,
an operation that changes the lattice Hamiltonian
(\ref{eq:T pm Fermions Lattice}).

For any $k\in\Lambda^{\star}$,
the eigenvalues of the $2\times2$ single-particle matrix
(\ref{eq:T pm Fermions Fourier c})
are obtained from its characteristic equation
\begin{align}
0&=
\det
\pmqty{
t^{\,}_{\pm,k}-\lambda & \Delta^{*}_{\pm,k}
\\
\Delta^{\,}_{\pm,k} & -t^{\,}_{\pm,k}-\lambda
}
\nonumber\\
&=
-\bigl(t^{\,}_{\pm,k}\bigr)^{2}
+\lambda^{2}
-\bigl|\Delta^{\,}_{\pm,k}\bigr|^{2}.
\end{align}
Hence
\begin{equation}
\lambda^{2}=
\bigl(t^{\,}_{\pm,k}\bigr)^{2}
+
\bigl|\Delta^{\,}_{\pm,k}\bigr|^{2}.
\end{equation}
We denote the two eigenvalues by
\begin{subequations}
\begin{equation}
\lambda^{\,}_{\pm,s,k}\coloneqq
s\,E^{\,}_{\pm,k},
\qquad
s=\pm,
\end{equation}
where
\begin{equation}
E^{\,}_{\pm,k}\coloneqq
\sqrt{
\bigl(t^{\,}_{\pm,k}\bigr)^{2}
+
\bigl|\Delta^{\,}_{\pm,k}\bigr|^{2}
     }\geq
0.
\label{eq:def-Epmk}
\end{equation}
\end{subequations}
Using the trigonometric identities
\begin{subequations}
\begin{align}
&
\cos^{2}(x)+\sin^{2}(x)=1,
\\
&
\cos(x)\cos(y)-\sin(x)\sin(y)=\cos(x+y),
\end{align}
together with
\begin{align}
&1-\cos(4k)=2\sin^{2}(2k),
\\
&1+\cos(4k)=2\cos^{2}(2k),
\end{align}
\end{subequations}
we obtain
\begin{subequations}
\label{eq:def-Epmk-final}
\begin{align}
&
E^{\,}_{+,k}=
4\,|\sin(2k)|=
E^{\,}_{+,-k},
\\
&
E^{\,}_{-,k}=
4\,|\cos(2k)|=
E^{\,}_{-,-k}.
\end{align}
\end{subequations}
Equivalently, the complete single-particle BdG spectra are
\begin{subequations}
\begin{align}
&
\lambda^{\,}_{+,s,k}=
4s\,|\sin(2k)|=
\lambda^{\,}_{+,s,-k},
\\
&
\lambda^{\,}_{-,s,k}=
4s\,|\cos(2k)|=
\lambda^{\,}_{-,s,-k},
\end{align}
\end{subequations}
with $s=\pm$.
Interestingly, we observe that
\begin{equation}
E^{\,}_{-,k}=E^{\,}_{+,k+\pi/4}.
\label{eq:Two Spectra Phase Shift}
\end{equation}

The family of BdG matrices $\{H^{\,}_{\pm,k}\}$ labeled by $k\in\Lambda^{\star}$
is gapless over the Brillouin zone $\Lambda^{\star}$. More precisely,
\begin{subequations}
\label{eq:def-4-nodes-gapless-superconductor}
\begin{align}
E^{\,}_{+,k}=0
&\ \Longleftrightarrow\
k\in\Lambda^{\star}_{+,\mathrm{F}}\coloneqq
\biggl\{\,0,\ \pm\,\frac{\pi}{2},\ \pi\biggr\},
\\
E^{\,}_{-,k}=0
&\ \Longleftrightarrow\
k\in\Lambda^{\star}_{-,\mathrm{F}}\coloneqq
\biggl\{\pm\,\frac{\pi}{4},\ \pm\,\frac{3\pi}{4}\biggr\}.
\end{align}
\end{subequations}
At the nodal momenta $k\in\Lambda^{\star}_{\pm,\mathrm{F}}$,
the matrix $H^{\,}_{\pm,k}$ vanishes and the
associated bulk single-particle modes have
exactly zero energy.
These are accidental nodal zero-modes that are not
relevant for the bulk thermodynamic stability analysis
that we aim at.

We now diagonalize the regular part of the kinetic Hamiltonian
(\ref{eq:T pm Fermions Fourier a}), namely the
part supported on momenta away from the nodal zero modes.
To this end, we define
\begin{equation}
\Lambda^{\star}_{\pm,\mathrm{reg}}\coloneqq
\Lambda^{\star}\setminus
\Lambda^{\star}_{\pm,\mathrm{F}},
\end{equation}
by subtracting a subset of $\Lambda^{\star}$
of vanishing measure in the thermodynamic limit.
For all $k\in\Lambda^{\star}_{\pm,\mathrm{reg}}$, one has
$E^{\,}_{\pm,k}>0$. Hence the two eigenvalues
$\pm E^{\,}_{\pm,k}$ of $H^{\,}_{\pm,k}$ are non-degenerate and the
corresponding Bogoliubov rotation can be chosen in the usual way.

For $k\in\Lambda^{\star}_{\pm,\mathrm{reg}}$,
we define the Bogoliubov quasiparticles by
\begin{subequations}
\label{eq:Def Bogoliubov Quasiparticles Regular}
\begin{align}
\hat{\gamma}^{\,}_{\pm,+k}\coloneqq
u^{\,}_{\pm,k}\,\hat{c}^{\,}_{k}
+
v^{\,}_{\pm,k}\,\hat{c}^{\dag}_{-k}.
\end{align}
The coherence factors are chosen as
\begin{align}
&
u^{\,}_{\pm,k}\coloneqq
\sqrt{
\frac{
E^{\,}_{\pm,k}+t^{\,}_{\pm,k}
     }
     {
2E^{\,}_{\pm,k}
     }
     }=
+u^{\,}_{\pm,-k}
\in\mathbb{R},
\\
&
v^{\,}_{\pm,k}\coloneqq
\frac{
\Delta^{*}_{\pm,k}
     }
     {
\left|\Delta^{\,}_{\pm,k}\right|
     }
\sqrt{
\frac{
E^{\,}_{\pm,k}-t^{\,}_{\pm,k}
     }
     {
2E^{\,}_{\pm,k}
     }
     }=
-v^{\,}_{\pm,-k}
\in\mathbb{C},
\end{align}
whenever $\Delta^{\,}_{\pm,k}\neq0$.
\end{subequations}
If $\Delta^{\,}_{\pm,k}=0$ while
$E^{\,}_{\pm,k}>0$, then $H^{\,}_{\pm,k}$ is already diagonal in the
Nambu basis.

The operators $\hat{\gamma}^{\,}_{\pm,k}$ obey the canonical fermionic
algebra
\begin{subequations}
\begin{align}
\Bigl\{
\hat{\gamma}^{\,}_{\pm,k},
\hat{\gamma}^{\dag}_{\pm,k'}
\Bigr\}
&=
\delta^{\,}_{k,k'},
\\
\Bigl\{
\hat{\gamma}^{\,}_{\pm,k},
\hat{\gamma}^{\,}_{\pm,k'}
\Bigr\}
&=
\Bigl\{
\hat{\gamma}^{\dag}_{\pm,k},
\hat{\gamma}^{\dag}_{\pm,k'}
\Bigr\}
=0,
\end{align}
\end{subequations}
for $k,k'\in\Lambda^{\star}_{\pm,\mathrm{reg}}$.

With this choice, the kinetic Hamiltonian
(\ref{eq:T pm Fermions Fourier a})
takes the diagonal form
\begin{equation}
\widehat{T}^{\,}_{\pm}=
\sum_{k\in\Lambda^{\star}_{\pm,\mathrm{reg}}}
E^{\,}_{\pm,k}
\left(
\hat{\gamma}^{\dag}_{\pm,k}
\hat{\gamma}^{\,}_{\pm,k}
-\frac{1}{2}
\right).
\label{eq:diagonal-BdG-kinetic energy}
\end{equation}
Hamiltonian $\widehat{T}^{\,}_{\pm}$
describes a one-dimensional non-interacting gapless superconductor.
Taking into account the particle-hole redundancy,
linearization of the spectrum $E^{\,}_{\pm,k}$
around the four nodes (\ref{eq:def-4-nodes-gapless-superconductor})
provides four independent massless Majorana fields.
Thefore, the Gaussian fixed point $\widehat{T}^{\,}_{\pm}$
has total central charge $\mathsf{c}=2$.

Normal ordering will be red{taken} with respect to the reference state
$\ket{\Omega^{\,}_{\pm,0}}$ defined as
\begin{equation}
\hat{\gamma}^{\,}_{\pm,k}
\ket{\Omega^{\,}_{\pm}}=
0,
\qquad
k\in\Lambda^{\star}_{\pm,\mathrm{reg}},
\label{eq:Def Reference Ground State Normal Ordering}
\end{equation}
together with
the condition that the BdG single-particle levels
corresponding to the nodal zero-modes are unocupied.
We shall use the short-hand notation
\begin{equation}
\ev{\,\bm{\cdot}\,}{\Omega^{\,}_{\pm}}
\equiv
\ev{\,\bm{\cdot}\,}^{\,}_{\pm}.
\end{equation}
For any operator $\widehat{O}$, the normal-ordered operator
${\bm{:}\,}\widehat{O}{\bm{:}}$ is defined by subtracting all
contractions evaluated in the state $\ket{\Omega^{\,}_{\pm}}$.
In particular, for two Majorana operators
\begin{equation}
\widehat{\Xi},\widehat{\Xi}'\in
\left\{
\left.
\hat{\xi}^{\,}_{j},
\hat{\eta}^{\,}_{j}
\,\right|
j=1,\ldots,2N
\right\},
\end{equation}
the definition reduces to
\begin{equation}
{\bm{:}\,}
\widehat{\Xi}\,
\widehat{\Xi}'
{\bm{:}}
\equiv
\widehat{\Xi}\,
\widehat{\Xi}'
-
\ev*{\,
\widehat{\Xi}\,
\widehat{\Xi}'
\,}^{\,}_{\pm},
\end{equation}
so that
\begin{equation}
\ev*{\,
{\bm{:}\,}
\widehat{\Xi}\,
\widehat{\Xi}'
{\bm{:}}
\,}^{\,}_{\pm}
=0.
\end{equation}

Equivalently, in the diagonal fermionic basis, the elementary
contractions are
\begin{subequations}
\begin{align}
\big\langle
\hat{\gamma}^{\dag}_{\pm,k}\,
\hat{\gamma}^{\,}_{\pm,k'}
\big\rangle^{\,}_{\pm}
&=
0,
\\
\big\langle
\hat{\gamma}^{\,}_{\pm,k}\,
\hat{\gamma}^{\dag}_{\pm,k'}
\big\rangle^{\,}_{\pm}
&=
\delta^{\,}_{k,k'},
\\
\big\langle
\hat{z}^{\dag}_{\pm,\alpha}\,
\hat{z}^{\,}_{\pm,\beta}
\big\rangle^{\,}_{\pm}
&=
0,
\\
\big\langle
\hat{z}^{\,}_{\pm,\alpha}\,
\hat{z}^{\dag}_{\pm,\beta}
\big\rangle^{\,}_{\pm}
&=
\delta^{\,}_{\alpha,\beta}.
\end{align}
\end{subequations}
These contractions determine the normal ordering of products of
Majorana operators through the inverse Bogoliubov transformation and
Wick's theorem.

In summary, the single-particle BdG matrix $H^{\,}_{\pm,k}$ has the
particle-hole symmetric eigenvalues
$\lambda^{\,}_{\pm,s,k}=sE^{\,}_{\pm,k}$, with $s=\pm$. Away from
the Fermi momenta $\Lambda^{\star}_{\pm,\mathrm{F}}$, this gives a
well-defined Bogoliubov basis. At the Fermi momenta,
$E^{\,}_{\pm,k}=0$, the two eigenvalues collapse to zero and the
Bogoliubov basis is not uniquely fixed. We therefore choose the
vacuum convention
\eqref{eq:Def Reference Ground State Normal Ordering}
with the absence of occupied nodal zero-modes
in order to define the reference state used for normal ordering.

\medskip
\subsubsection{Perturbation-induced quadratic corrections}

We have constructed the ground state $\ket{\Omega^{\,}_{\pm}}$
of the free Hamiltonians $\widehat{T}^{\,}_{\pm}$, with respect to which
we can normal-order the interactions $\widehat{V}^{\,}_{\pm}$
defined in Eq.~\eqref{eq:def Perturbed Hamiltonian Majorana Stability c}.
The interactions $\widehat{V}^{\,}_{\pm}$ are monomials of order four in
the fermionic operators.
The interactions $\widehat{V}^{\,}_{\pm}$
differ from their normal-ordered forms
${\bm{:}\,}\widehat{V}^{\,}_{\pm}{\,\bm{:}}$
by inconsequential $\mathbb{C}$-numbers (here zero)
and by quadratic terms, namely
\begin{equation}
\epsilon\,
\widehat{V}^{\,}_{\pm}=
\mp
\frac{4\,\epsilon}{\pi}
\sum_{j=1}^{2N}\,
\ii\hat{\xi}^{\,}_{j}\hat{\eta}^{\,}_{j+1}
+
\epsilon\,
{\bm{:}\,}\widehat{V}^{\,}_{\pm}{\,\bm{:}}\,.
\label{eq:Quadratic Corrections Stability}
\end{equation}
\begin{proof}
We normal order the interaction    
\eqref{eq:def Perturbed Hamiltonian Majorana Stability c}
by using Wick's theorem for operators obeying the
algebra \eqref{eq:Majorana Algebra}.
In doing so, we obtain
\begin{widetext}
\begin{subequations}
\label{eq:Normal-Ordering Quartic Majorana Interactions}
\begin{align}
\hat{\xi}^{\,}_{j}\,
\hat{\eta}^{\,}_{j}\,
\hat{\xi}^{\,}_{j+1}\,
\hat{\eta}^{\,}_{j+1}
=&\ 
{\bm{:}\,}
\hat{\xi}^{\,}_{j}\,
\hat{\eta}^{\,}_{j}\,
\hat{\xi}^{\,}_{j+1}\,
\hat{\eta}^{\,}_{j+1}
{\bm{:}}
\nonumber \\
&+
\ev*{\hat{\xi}^{\,}_{j}\,\hat{\eta}^{\,}_{j}}^{\,}_{\pm}
{\bm{:}\,}\hat{\xi}^{\,}_{j+1}\,\hat{\eta}^{\,}_{j+1}{\bm{:}\,}
-
\ev*{\hat{\xi}^{\,}_{j}\,\hat{\xi}^{\,}_{j+1}}^{\,}_{\pm}
{\bm{:}\,}\hat{\eta}^{\,}_{j}\,\hat{\eta}^{\,}_{j+1}{\bm{:}\,}
+
\ev*{\hat{\xi}^{\,}_{j}\,\hat{\eta}^{\,}_{j+1}}^{\,}_{\pm}
{\bm{:}\,}\hat{\eta}^{\,}_{j}\,\hat{\xi}^{\,}_{j+1}{\bm{:}\,}
\nonumber \\
&+
\ev*{\hat{\eta}^{\,}_{j}\,\hat{\xi}^{\,}_{j+1}}^{\,}_{\pm}
{\bm{:}\,}\hat{\xi}^{\,}_{j}\,\hat{\eta}^{\,}_{j+1}{\bm{:}\,}
-
\ev*{\hat{\eta}^{\,}_{j}\,\hat{\eta}^{\,}_{j+1}}^{\,}_{\pm}
{\bm{:}\,}\hat{\xi}^{\,}_{j}\,\hat{\xi}^{\,}_{j+1}{\bm{:}\,}
+
\ev*{\hat{\xi}^{\,}_{j+1}\,\hat{\eta}^{\,}_{j+1}}^{\,}_{\pm}
{\bm{:}\,}\hat{\xi}^{\,}_{j}\,\hat{\eta}^{\,}_{j}{\bm{:}\,}
\nonumber \\
&+
\ev*{\hat{\xi}^{\,}_{j}\,\hat{\eta}^{\,}_{j}}^{\,}_{\pm}
\ev*{\hat{\xi}^{\,}_{j+1}\,\hat{\eta}^{\,}_{j+1}}^{\,}_{\pm}
-
\ev*{\hat{\xi}^{\,}_{j}\,\hat{\xi}^{\,}_{j+1}}^{\,}_{\pm}
\ev*{\hat{\eta}^{\,}_{j}\,\hat{\eta}^{\,}_{j+1}}^{\,}_{\pm}
+
\ev*{\hat{\xi}^{\,}_{j}\,\hat{\eta}^{\,}_{j+1}}^{\,}_{\pm}
\ev*{\hat{\eta}^{\,}_{j}\,\hat{\xi}^{\,}_{j+1}}^{\,}_{\pm},
\label{eq:Normal-Ordering Quartic Majorana Interactions a}
\end{align}
and
\begin{align}
\hat{\xi}^{\,}_{j}\,
\hat{\xi}^{\,}_{j+1}\,
\hat{\eta}^{\,}_{j+2}\,
\hat{\eta}^{\,}_{j+3}
=&\ 
{\bm{:}\,}
\hat{\xi}^{\,}_{j}\,
\hat{\xi}^{\,}_{j+1}\,
\hat{\eta}^{\,}_{j+2}\,
\hat{\eta}^{\,}_{j+3}
{\bm{:}}
\nonumber \\
&+
\ev*{\hat{\xi}^{\,}_{j}\,\hat{\xi}^{\,}_{j+1}}^{\,}_{\pm}
{\bm{:}\,}\hat{\eta}^{\,}_{j+2}\,\hat{\eta}^{\,}_{j+3}{\bm{:}\,}
-
\ev*{\hat{\xi}^{\,}_{j}\,\hat{\eta}^{\,}_{j+2}}^{\,}_{\pm}
{\bm{:}\,}\hat{\xi}^{\,}_{j+1}\,\hat{\eta}^{\,}_{j+3}{\bm{:}\,}
+
\ev*{\hat{\xi}^{\,}_{j}\,\hat{\eta}^{\,}_{j+3}}^{\,}_{\pm}
{\bm{:}\,}\hat{\xi}^{\,}_{j+1}\,\hat{\eta}^{\,}_{j+2}{\bm{:}\,}
\nonumber \\
&+
\ev*{\hat{\xi}^{\,}_{j+1}\,\hat{\eta}^{\,}_{j+2}}^{\,}_{\pm}
{\bm{:}\,}\hat{\xi}^{\,}_{j}\,\hat{\eta}^{\,}_{j+3}{\bm{:}\,}
-
\ev*{\hat{\xi}^{\,}_{j+1}\,\hat{\eta}^{\,}_{j+3}}^{\,}_{\pm}
{\bm{:}\,}\hat{\xi}^{\,}_{j}\,\hat{\eta}^{\,}_{j+2}{\bm{:}\,}
+
\ev*{\hat{\eta}^{\,}_{j+2}\,\hat{\eta}^{\,}_{j+3}}^{\,}_{\pm}
{\bm{:}\,}\hat{\xi}^{\,}_{j}\,\hat{\xi}^{\,}_{j+1}{\bm{:}\,}
\nonumber \\
&+
\ev*{\hat{\xi}^{\,}_{j}\,\hat{\xi}^{\,}_{j+1}}^{\,}_{\pm}
\ev*{\hat{\eta}^{\,}_{j+2}\,\hat{\eta}^{\,}_{j+3}}^{\,}_{\pm}
-
\ev*{\hat{\xi}^{\,}_{j}\,\hat{\eta}^{\,}_{j+2}}^{\,}_{\pm}
\ev*{\hat{\xi}^{\,}_{j+1}\,\hat{\eta}^{\,}_{j+3}}^{\,}_{\pm}
+
\ev*{\hat{\xi}^{\,}_{j}\,\hat{\eta}^{\,}_{j+3}}^{\,}_{\pm}
\ev*{\hat{\xi}^{\,}_{j+1}\,\hat{\eta}^{\,}_{j+2}}^{\,}_{\pm}.
\label{eq:Normal-Ordering Quartic Majorana Interactions b}
\end{align}
\end{subequations}
\end{widetext}
The double contractions are scalar energy shifts. 
The single contractions produce quadratic operators.

\begin{table}[t]
\caption{Short-distance Majorana contractions
defined in Eq.~\eqref{eq:Two-point Functions Majorana pm,r}
in the ground states $\ket{\Omega^{\,}_{\pm}}$
for relevant values of $r=1,2,3$ in the limit $2N\to\infty$.
The value of $\Gamma^{\xi\eta}_{\pm,0}$ is also given.}
\label{tab:Contractions Majorana Stability}
\begin{ruledtabular}
\begin{tabular}{ccccc}
$r$ & $0$ & $1$ & $2$ & $3$ \\
\midrule
$\Gamma^{\eta\eta}_{\pm,r}$ & & $0$ & $0$ & $0$ \\
$\Gamma^{\eta\xi}_{\pm,r}$ & & $\pm\dfrac{2\ii}{\pi}$ & $0$ & $0$ \\
$\Gamma^{\xi\eta}_{\pm,r}$ & $0$ & $0$ & $0$ & $+\dfrac{2\ii}{\pi}$ \\
$\Gamma^{\xi\xi}_{\pm,r}$ & & $0$ & $0$ & $0$ \\
\end{tabular}
\end{ruledtabular}
\end{table}

To evaluate Eq.~\eqref{eq:Normal-Ordering Quartic Majorana Interactions},
it suffices to compute the following two-point functions
\begin{subequations}
\label{eq:Two-point Functions Majorana pm,r}
\begin{align}
&\Gamma^{\eta\eta}_{\pm,r}\coloneqq
\ev*{\hat{\eta}^{\,}_{j}\,\hat{\eta}^{\,}_{j+r}}^{\,}_{\pm},
\qquad
\Gamma^{\eta\xi}_{\pm,r}\coloneqq
\ev*{\hat{\eta}^{\,}_{j}\,\hat{\xi}^{\,}_{j+r}}^{\,}_{\pm},
\\
&\Gamma^{\xi\eta}_{\pm,r}\coloneqq
\ev*{\hat{\xi}^{\,}_{j}\,\hat{\eta}^{\,}_{j+r}}^{\,}_{\pm},
\qquad
\Gamma^{\xi\xi}_{\pm,r}\coloneqq
\ev*{\hat{\xi}^{\,}_{j}\,\hat{\xi}^{\,}_{j+r}}^{\,}_{\pm},
\end{align}
\end{subequations}
for $r=1,2,3$ together with $\Gamma^{\xi\eta}_{\pm,0}$ for $r=0$.
This we do by inserting the definition
\eqref{eq:Def Majorana X-basis},
the Fourier expansions
\eqref{eq:Fourier MF}
and the Bogoliubov transformation
(\ref{eq:Def Bogoliubov Quasiparticles Regular})
into Eq.\,
\eqref{eq:Two-point Functions Majorana pm,r}.
If we first insert Eq.\
\eqref{eq:Def Majorana X-basis}
into Eq.\
\eqref{eq:Two-point Functions Majorana pm,r},
we find that
\begin{subequations}
\begin{align}
&\Gamma^{\eta\eta}_{\pm,r}=
(h^{\,}_{\pm,r}-h^{*}_{\pm,r})
+
(p^{\,}_{\pm,r}-p^{*}_{\pm,r}),
\\
&\Gamma^{\eta\xi}_{\pm,r}=
+\ii\,(h^{\,}_{\pm,r}+h^{*}_{\pm,r})
+
\ii\,(p^{\,}_{\pm,r}+p^{*}_{\pm,r}),
\\
&\Gamma^{\xi\eta}_{\pm,r}=
-\ii\,(h^{\,}_{\pm,r}+h^{*}_{\pm,r})
+
\ii\,(p^{\,}_{\pm,r}+p^{*}_{\pm,r}),
\\
&\Gamma^{\xi\xi}_{\pm,r}=
(h^{\,}_{\pm,r}-h^{*}_{\pm,r})
-
(p^{\,}_{\pm,r}-p^{*}_{\pm,r}),
\end{align}
for $r=1,2,3$ and
\begin{equation}
\Gamma^{\xi\eta}_{\pm,0}=
\ii\,(1-2h^{\,}_{\pm})
\end{equation}
for $r=0$, whereby we have defined
\begin{equation}
h^{\,}_{\pm,r}\coloneqq
\ev*{\hat{c}^{\dag}_{j}\,\hat{c}^{\,}_{j+r}}^{\,}_{\pm},
\qquad
p^{\,}_{\pm,r}\coloneqq
\ev*{\hat{c}^{\,}_{j}\,\hat{c}^{\,}_{j+r}}^{\,}_{\pm},
\label{eq:Def h And p}
\end{equation}
\end{subequations}
for any integer $r$.

Second, with the help of the Bogoliubov transformation
(\ref{eq:Def Bogoliubov Quasiparticles Regular}),
we find that
\begin{equation}
\ev*{\hat{c}^{\dag}_{k}\,\hat{c}^{\,}_{k}}^{\,}_{\pm}=
|v^{\,}_{\pm,k}|^{2},
\qquad
\ev*{\hat{c}^{\,}_{k}\,\hat{c}^{\,}_{-k}}^{\,}_{\pm}=
u^{\,}_{\pm,k}\,v^{\,}_{\pm,k}.
\label{eq:EV Fermion Fourier Stability}
\end{equation}

Third, if we insert the Fourier transformation
\eqref{eq:Fourier MF}
and the result
\eqref{eq:EV Fermion Fourier Stability}
into Eq.\ \eqref{eq:Def h And p},
we obtain
\begin{subequations}
\label{eq:Def hr And pr}
\begin{align}
&h^{\,}_{\pm,r}=
\frac{1}{4N}
\sum_{k\in\Lambda^{\star}_{\pm,\mathrm{reg}}}
e^{\ii kr}\left(
1-\frac{t^{\,}_{\pm,k}}{E^{\,}_{\pm,k}}
\right),
\\
&p^{\,}_{\pm,r}=
-\frac{1}{4N}
\sum_{k\in\Lambda^{\star}_{\pm,\mathrm{reg}}}
e^{-\ii kr}\,
\frac{\Delta^{\,}_{\pm,k}}{E^{\,}_{\pm,k}}.
\end{align}
\end{subequations}

With the help of Eqs.\
(\ref{eq:T pm Fermions Fourier d})
and
(\ref{eq:def-Epmk-final}),
we can write
\begin{align}
\frac{t^{\,}_{\mathsf{s},k}+\Delta^{\,}_{\mathsf{s},k}}
     {E^{\,}_{\mathsf{s},k}}=&\,
\frac{
2\,\bigl(
-
\mathsf{s}\,
e^{+\mathrm{i}k}
+
e^{-3\mathrm{i}k}
\bigr)
     }
     {
E^{\,}_{\mathsf{s},k}
     }
\nonumber\\
=&\,
e^{-\ii k}
\begin{cases}
-\ii\sgn(\sin(2k)), & \mathsf{s}=+, \\
\hphantom{-\ii}\sgn(\cos(2k)), & \mathsf{s}=-.
\end{cases}
\label{eq:Trick t+Delta}
\end{align}
Moreover, as
\begin{subequations}
\begin{equation}
t^{\,}_{\pm,k}\in\mathbb{R},
\qquad
\ii\Delta^{\,}_{\pm,k}\in\mathbb{R},
\end{equation}
we have the pair of identities
\begin{align}
&
\frac{t^{\,}_{\pm,k}}
     {E^{\,}_{\pm,k}}
=
\Re\left(
\frac{t^{\,}_{\pm,k}+\Delta^{\,}_{\pm,k}}
     {E^{\,}_{\pm,k}}
\right),
\\
&
\frac{\Delta^{\,}_{\pm,k}}
     {E^{\,}_{\pm,k}}
=
\ii\Im\left(
\frac{t^{\,}_{\pm,k}+\Delta^{\,}_{\pm,k}}
     {E^{\,}_{\pm,k}}
\right).
\end{align}
\end{subequations}
We can then insert the real or imaginary parts of
Eq.~\eqref{eq:Trick t+Delta}
into Eq.~\eqref{eq:Def hr And pr}.

As an example, we explicitly compute
$h^{\,}_{+,r}$ for $r=1,2,3$.
In the aim of taking the thermodynamic limit $2N\to\infty$,
we will make the replacement
\begin{equation}
\frac{1}{2N}
\sum_{k\in\Lambda^{\star}_{\pm,\mathrm{reg}}}^{k>0}
\longrightarrow
\frac{1}{2\pi}
\int\limits_{0}^{\pi}\dd k.
\end{equation}
Hence, we have
\begin{align}
&
\lim_{2N\to\infty}
h^{\,}_{+,r}=
\nonumber\\
&\qquad
\lim_{2N\to\infty}
\frac{1}{4N}
\sum_{k\in\Lambda^{\star}_{+,\mathrm{reg}}}
e^{\ii kr}
\Re
\left[
\ii
e^{-\ii k}
\sgn(\sin(2k))
\right]=
\nonumber\\
&\qquad
\lim_{2N\to\infty}
\frac{1}{2N}
\sum_{k\in\Lambda^{\star}_{+,\mathrm{reg}}}^{k>0}
\cos(kr)\,\sin(k)\,
\sgn(\sin(2k))=
\nonumber\\
&\qquad
\frac{1}{2\pi}
\int\limits_{0}^{\pi}\dd k
\cos(kr)\,\sin(k)\,
\sgn(\sin(2k)).
\end{align}

We can apply the same procedure to evaluate
$h^{\,}_{+,r}$ for $r=0$,
$h^{\,}_{-,r}$ for $r=0,1,2,3$ and $p^{\,}_{\pm,r}$ for $r=1,2,3$
in the thermodynamic limit $2N\to\infty$.
We obtain
\begin{subequations}
\begin{align}
&\lim_{2N\to\infty}
h^{\,}_{\pm,r}=
\begin{cases}
+\dfrac{1}{2}, & r=0, \\
\pm\,\dfrac{1}{2\pi}, & r=1, \\
0, & r=2, \\
-\,\dfrac{1}{2\pi}, & r=3,
\end{cases}
\\
&\lim_{2N\to\infty}
p^{\,}_{\pm,r}=
\begin{cases}
\pm\,\dfrac{1}{2\pi}, & r=1, \\
0, & r=2, \\
+\,\dfrac{1}{2\pi}, & r=3.
\end{cases}
\end{align}
\end{subequations}
We list in Table \ref{tab:Contractions Majorana Stability}
the values of the two-point functions
\eqref{eq:Two-point Functions Majorana pm,r}
obtained in the limit $2N\to\infty$.
One verifies from Table \ref{tab:Contractions Majorana Stability}
that all the $\mathbb{C}$-numbers
on the right-hand side of Eqs.\
(\ref{eq:Normal-Ordering Quartic Majorana Interactions a})
and
(\ref{eq:Normal-Ordering Quartic Majorana Interactions b})
vanish.

Finally, the insertion of the $\mathbb{C}$-numbers obtained from
\eqref{eq:Two-point Functions Majorana pm,r}
and their values listed in Table \ref{tab:Contractions Majorana Stability}
on the right-hand sides of
Eqs.\
\eqref{eq:Normal-Ordering Quartic Majorana Interactions}
and
(\ref{eq:def Perturbed Hamiltonian Majorana Stability c})
yields Eq.~\eqref{eq:Quadratic Corrections Stability}.
\end{proof}

\subsubsection{Effect of the interaction-induced quadratic terms}
\label{ssub:effect_of_the_interactions}

We analyze the effect of the one-body term induced by normal-ordering perturbation
\eqref{eq:def Perturbed Hamiltonian Majorana Stability c}
for the non-interacting Hamiltonian
\eqref{eq:def Perturbed Hamiltonian Majorana Stability b}
that describes a one-dimensional non-interacting gapless superconductor.
For $\widehat{H}^{\,}_{+}(\epsilon)$
defined in Eq.\
(\ref{eq:Relation Hamiltonian Original VS H+- a}),
we find that it opens the single-quasiparticle BdG gap
\begin{equation}
\Delta^{\,}_{+}(\epsilon)\coloneqq
\frac{8\,\epsilon}{\pi}
\label{eq:Gap H+ lambda 3pi4}
\end{equation}
in the BdG spectrum that is linear in $\epsilon$.
For $\widehat{H}^{\,}_{-}(\epsilon)$
defined in Eq.\
(\ref{eq:Relation Hamiltonian Original VS H+- b}),
it shifts the set of nodal points of $\widehat{H}^{\,}_{-}(0)$ to the pair
\begin{equation}
\Lambda^{\star}_{-,\mathrm{F}}(\epsilon)\coloneqq
\left\{\pm\left(\frac{\pi}{4}+\frac{\epsilon}{\pi}\right),
\ \pm\left(\frac{3\pi}{4}-\frac{\epsilon}{\pi}\right)\right\}
\label{eq:Fermi H- lambda pi4}
\end{equation}
up to first order in $\epsilon$.
Surprisingly, following
Eq.~\eqref{eq:Two Spectra Phase Shift},
the two critical points
$\widehat{H}^{\,}_{+}(0)$ and
$\widehat{H}^{\,}_{-}(0)$
have equivalent-looking spectra, but the
microscopic perturbation embeds differently
into their low-energy theories,
making one unstable through an induced mass
while leaving the other with
a marginal interaction problem.
\begin{proof}
The Hamiltonian \eqref{eq:def Perturbed Hamiltonian Majorana Stability}
can be rewritten as
\begin{subequations}
\label{eq:Hamiltonian Decomposotion H_pm}
\begin{equation}
\widehat{H}^{\,}_{\pm}(\epsilon)=
\widehat{T}^{\,}_{\pm,\mathrm{eff}}(\epsilon)
+
\epsilon\,
{\bm{:}\,}\widehat{V}^{\,}_{\pm}{\,\bm{:}}\,,
\label{eq:Hamiltonian Decomposotion H_pm a}
\end{equation}
where
\begin{equation}
\widehat{T}^{\,}_{\pm,\mathrm{eff}}(\epsilon)\coloneqq
\sum_{j=1}^{2N}\,
\ii\hat{\xi}^{\,}_{j}
\left(
\mp\,
\hat{\eta}^{\,}_{j-1}
\mp\frac{4\epsilon}{\pi}
\hat{\eta}^{\,}_{j+1}
+
\hat{\eta}^{\,}_{j+3}
\right)\!.
\label{eq:Hamiltonian Decomposotion H_pm b}
\end{equation}
\end{subequations}
Insertion of the Fourier expansions
\begin{subequations}
\begin{align}
&
\hat{\xi}^{\,}_{j}=
\frac{1}{\sqrt{2N}}
\sum_{k\in\Lambda^{\star}}
e^{\mathrm{i}kj}\,
\hat{\xi}^{\,}_{k},
\qquad
\hat{\xi}^{\dag}_{k}=\hat{\xi}^{\,}_{-k},
\\
&
\hat{\eta}^{\,}_{j}=
\frac{1}{\sqrt{2N}}
\sum_{k\in\Lambda^{\star}}
e^{\mathrm{i}kj}\,
\hat{\eta}^{\,}_{k},
\qquad
\hat{\eta}^{\dag}_{k}=\hat{\eta}^{\,}_{-k},
\end{align}
\end{subequations}
gives
\begin{subequations}
\label{eq:Teffepsilon-final}
\begin{align}
\widehat{T}^{\,}_{\pm,\mathrm{eff}}(\epsilon)=&\,
\sum_{k\in\Lambda^{\star}}
f^{\,}_{\pm,k}(\epsilon)\,
\hat{\xi}^{\,}_{k}\,\hat{\eta}^{\,}_{-k}
\nonumber\\
=&\,
\sum_{k\in\Lambda^{\star}}^{k>0}
\left[
f^{\,}_{\pm,k}(\epsilon)\,
\hat{\xi}^{\,}_{k}\,\hat{\eta}^{\,}_{-k}
+
f^{*}_{\pm,k}(\epsilon)\,
\hat{\eta}^{\,}_{k}\,\hat{\xi}^{\,}_{-k}
\right],
\label{eq:Teffepsilon-final-a}
\end{align}
where
\begin{equation}
f^{\,}_{\pm,k}(\epsilon)\coloneqq
\ii e^{-\ii k}
\left(
\mp\,e^{+\ii 2k}
+e^{-\ii 2k}
\mp \frac{4\epsilon}{\pi}
\right).
\label{eq:Teffepsilon-final-b}
\end{equation}
\end{subequations}
The Hermiticity of $\widehat{T}^{\,}_{\pm,\mathrm{eff}}(\epsilon)$
is manifest on the right-hand side
of the second equality in Eq.\
(\ref{eq:Teffepsilon-final-a}).

The single-particle spectrum associated to the
Majorana quadratic form 
(\ref{eq:Teffepsilon-final-a})
is
\begin{subequations}
\begin{equation}
\lambda^{\,}_{\mathsf{s},k}(\epsilon)=
\pm
E^{\,}_{\mathsf{s},k}(\epsilon),
\end{equation}
where
\begin{align}
E^{\,}_{\mathsf{s},k}(\epsilon)
&=
2\left|
f^{\,}_{\mathsf{s},k}(\epsilon)
\right|
\nonumber \\
&=
\begin{cases}
\left|4\ii\,\mathrlap{\sin}\phantom{\cos}(2k)
+
\dfrac{8\,\epsilon}{\pi}\right|,
& \mathsf{s}=+,
\\ &\\
\left|
4\phantom{\ii}\cos(2k)
+
\dfrac{8\,\epsilon}{\pi}
\right|,
& \mathsf{s}=-.
\end{cases}
\end{align}
\end{subequations}
The fact that $8\,\epsilon/\pi\in\mathbb{R}$
has important consequences.

In the case of $\widehat{T}^{\,}_{+,\mathrm{eff}}(\epsilon)$,
there exists no real solution to the equation
\begin{equation}
E^{\,}_{+,k}(\epsilon)=0
\end{equation}
for any $\epsilon>0$.
Moreover, we have the bound
\begin{equation}
E^{\,}_{+,k}(\epsilon)\geq\frac{8\,\epsilon}{\pi},
\qquad
\epsilon\geq0,
\end{equation}
with the equality holding only for
$k\in\Lambda^{\star}_{+,\mathrm{F}}$.
Hence, the spectrum of
$\widehat{T}^{\,}_{+,\mathrm{eff}}(\epsilon)$
is gapped for any
$\epsilon>0$.
The magnitude of this gap is the
minimal energy of a BdG quasiparticle excitation, namely
\begin{equation}
\Delta^{\mathrm{(bdg)}}_{+}(\epsilon)\coloneqq
\min_{k\in\Lambda^{\star}}
E^{\,}_{+,k}(\epsilon)=
\frac{8\,\epsilon}{\pi}.
\end{equation}

In the case of $\widehat{T}^{\,}_{-,\mathrm{eff}}(\epsilon)$,
the solutions to the equation
\begin{equation}
E^{\,}_{-,k}(\epsilon)=0,
\end{equation}
for any $0\leq\epsilon<\pi/2$ are given by
\begin{subequations}
\label{eq:Fermi Points Shifted Leading}
\begin{equation}
k^{\,}_{\mathrm{F}}(\epsilon)\in
\Big\{
k^{\,}_{1}(\epsilon),
k^{\,}_{2}(\epsilon)
\Big\},
\label{eq:Fermi Points Shifted Leading a}    
\end{equation}
where
\begin{align}
k^{\,}_{1}(\epsilon)\coloneqq&\,
+
\frac{1}{2}\,
\arccos(-\frac{2\,\epsilon}{\pi})
\nonumber\\
=&\,
\frac{\pi}{4}+\frac{\epsilon}{\pi}+\mathcal{O}\bigl(\epsilon^{3}\bigr),
\label{eq:Fermi Points Shifted Leading b}
\\
k^{\,}_{2}(\epsilon)\coloneqq&\,
\pi
-
\frac{1}{2}\,
\arccos(-\frac{2\,\epsilon}{\pi})
\nonumber\\
=&\,
\frac{3\pi}{4}-\frac{\epsilon}{\pi}+\mathcal{O}\bigl(\epsilon^{3}\bigr).
\label{eq:Fermi Points Shifted Leading c}
\end{align}
\end{subequations}   
Observe that
\begin{subequations}
\begin{align}
&
k^{\,}_{1}(\epsilon)
+
k^{\,}_{2}(\epsilon)=\pi,
\\
&
k^{\,}_{1}(\epsilon)
-
k^{\,}_{2}(\epsilon)=
\arccos(-\frac{2\,\epsilon}{\pi})
-
\pi,
\end{align}
\end{subequations}
\end{proof}

\subsubsection{Effect of the normal-ordered quartic term in $\widehat{H}^{\,}_{+}(\epsilon)$}
\label{Effect of the normal-ordered quartic term in H+}

The quadratic term induced by normal-ordering of the interaction
$\epsilon\,\widehat{V}^{\,}_{+}$
already produces a gap linear in $\epsilon$.
The residual normal-ordered quartic interaction
$\epsilon\,{\bm{:}\,}\widehat{V}^{\,}_{+}{\,\bm{:}}$
can renormalize the value of this gap but does not alter the conclusion
that the critical point $(0,3\pi/4)$
is unstable to the perturbation.

\subsubsection{Effect of the normal-ordered quartic term in $\widehat{H}^{\,}_{-}(\epsilon)$}
\label{Effect of the normal-ordered quartic term in H-}

We are going to show within a one-loop perturbative
renormalization-group (RG) treatment that the
normal-ordered interaction
$\epsilon\,{\bm{:}\,}\widehat{V}^{\,}_{-}{\,\bm{:}}$
defined in Eq.\
(\ref{eq:Quadratic Corrections Stability})
destroys the incommensurate critical phase with central charge
$\mathsf{c}=2$ that is encoded by the lattice hopping Hamiltonian 
$\widehat{T}^{\,}_{-,\mathrm{eff}}(\epsilon)$
defined in Eq.\
(\ref{eq:Hamiltonian Decomposotion H_pm b})
by gapping half of the unperturbed critical modes.
More precisely, we show that (i) the unperturbed critical point
consists of two decoupled theories of free Dirac fermions, each carrying a
$\mathsf{c}=1$ central charge,
(ii) the unperturbed critical point can be interpreted as
two decoupled $\mathsf{c}=1$ sectors, each one realizing an
$\mathfrak{su}(2)^{\,}_{1}$ current algebra,
and (iii) the interaction, under the one-loop RG flow,
approaches a marginally relevant
$\mathrm{SU}(2)$-symmetric current-current interaction
that dynamically gaps one of the two critical sectors 
realizing an $\mathfrak{su}(2)^{\,}_{1}$ current algebra.
The outcome is an interval of critical points, each one of which realizes an
$\mathfrak{su}(2)^{\,}_{1}$  current algebra with
central charge $\mathsf{c}=1$,
but differing by their incommensuration.
This is the scenario (Ca) shown in Fig.\
(\ref{fig:TQ gFMz HB scenarios}).

We seek the continuum limit of the lattice Hamiltonian
$\widehat{H}^{\,}_{-}(\epsilon)$
defined by Eqs.\
(\ref{eq:Hamiltonian Decomposotion H_pm})
and
(\ref{eq:def Perturbed Hamiltonian Majorana Stability})
that describes a gapless superconductor.
To this end, we do a gradient expansion of the BdG quasiparticle
energies $E^{\,}_{-,k}$ in Eq.\
(\ref{eq:diagonal-BdG-kinetic energy})
around the two incommensurate positive nodal points
$k^{\,}_{a}$ with $a=1,2$ defined in
Eq.~\eqref{eq:Fermi Points Shifted Leading}
of the Brillouin zone at which
$E^{\,}_{-,k^{\,}_{a}}=0$.
To leading order in this gradient expansion, we find
\begin{subequations}
\begin{align}
\widehat{H}^{\,}_{-}(\epsilon)\simeq
\widehat{H}^{\,}_{0,-}(\epsilon)
+
\widehat{H}^{\,}_{\mathrm{int},-}(\epsilon),
\label{eq:Def H_ for RG}
\end{align}
where the non-interacting (interacting) contribution
$\widehat{H}^{\,}_{0,-}$
($\widehat{H}^{\,}_{\mathrm{int},-}$)
is quadratic (quartic) in the slowly varying
right- and left-moving BdG fields
$\widehat{\gamma}^{\,}_{a,R}(x)$
and
$\widehat{\gamma}^{\,}_{a,L}(x)$
labeled by the incommensurate nodal points $a=1,2$
from which they originate.
Their equal-time algebra is that of fermions with
\begin{equation}
\left\{
\widehat{\gamma}^{\,}_{a,\mu}(x),
\widehat{\gamma}^{\dag}_{a',\mu'}(x')
\right\}=
\delta^{\,}_{\mu,\mu'}\,
\delta^{\,}_{a,a'}\,
\delta(x-x'),
\end{equation}
\end{subequations}
for $\mu,\mu'=R,L$, $a,a'=1,2$ and $x,x'\in\mathbb{R}$
their only non-vanishing equal-time anticommutators.
Here, it is important to remember that the transformation
by which $\widehat{\gamma}^{\,}_{a,\mu}(x)$
is multiplied by a global phase factor
has no dynamical significance
for the same reason as explained below
Eq.\ (\ref{eq:T pm Fermions Fourier}).

The non-interacting BdG contribution is
\begin{subequations}
\label{eq:H0 RG stability Hminus}
\begin{equation}
\widehat{H}^{\,}_{0,-}(\epsilon)\coloneqq
-\ii
v\,
\int\dd x
\left(
\widehat{\Gamma}^{\dag}_{R}\,
\partial^{\,}_{x}\widehat{\Gamma}^{\,}_{R}
-
\widehat{\Gamma}^{\dag}_{L}\,
\partial^{\,}_{x}\widehat{\Gamma}^{\,}_{L}
\right),
\label{eq:H0 RG stability Hminus a}
\end{equation}
where we have introduced the field-valued chiral spinor
\begin{equation}
\widehat{\Gamma}^{\,}_{\mu}\equiv
\begin{pmatrix}
\widehat{\gamma}^{\,}_{1,\mu}
\\
\widehat{\gamma}^{\,}_{2,\mu}
\end{pmatrix},
\qquad
\mu=L,R,
\label{eq:H0 RG stability Hminus b}
\end{equation}
and the speed is given by 
\begin{equation}
v\coloneqq
8\,
\mathfrak{a}>0
\label{eq:H0 RG stability Hminus c}
\end{equation}
\end{subequations}
to the leading order in powers of $\epsilon$.
There is a hidden dependence on $\epsilon$ on the right-hand side
of Eq.~\eqref{eq:H0 RG stability Hminus a}
through the implicit choice of the incommensurate nodal points
in the Brillouin zone.

The non-interacting Hamiltonian
(\ref{eq:H0 RG stability Hminus a})
defines a conformal field theory
with central charge $\mathsf{c}=2$ whose only non-vanishing
Euclidean propagators are given by
\begin{subequations}
\label{eq:conformal-properties-theta=0lambda=piover4}
\begin{align}
&
\ev*{
\widehat{\gamma}^{\,}_{a,R}(\tau,x)\,
\widehat{\gamma}^{\dag}_{b,R}(0,0)
    }=
\frac{\delta^{\,}_{ab}}{2\pi}\,
\frac{1}{v\,\tau-\ii x},
\label{eq:conformal-properties-theta=0lambda=piover4-a}
\\
&
\ev*{
\widehat{\gamma}^{\,}_{a,L}(\tau,x)\,
\widehat{\gamma}^{\dag}_{b,L}(0,0)
    }=
\frac{\delta^{\,}_{ab}}{2\pi}\,
\frac{1}{v\,\tau+\ii x}.
\label{eq:conformal-properties-theta=0lambda=piover4-b}
\end{align}
\end{subequations}
Owing to the fact that both nodes in the Brillouin zone
share the same speed $v$,
it then follows that there exists a conserved
\begin{subequations}
\begin{equation}
\mathfrak{so}(4)^{\,}_{1}\cong
\mathfrak{su}(2)^{\,}_{1}\oplus\mathfrak{su}(2)^{\,}_{1}
\label{eq:conformal-properties-theta=0lambda=piover4-c}
\end{equation}
chiral current algebra whose twelve chiral currents,
six for each chirality, are given by \cite{Yang1989,Yang1990,Zhang1991}
\begin{equation}
\begin{split}
&
\widehat{J}^{\,1}_{\mu}\coloneqq
\frac{\mathcal{N}}{2}\,
{\bm{:}}
\left(
\widehat{\gamma}^{\dag}_{1,\mu}
\widehat{\gamma}^{\dag}_{2,\mu}
+
\widehat{\gamma}^{\,}_{2,\mu}
\widehat{\gamma}^{\,}_{1,\mu}
\right)
{\bm{:}}\,,
\\
&
\widehat{J}^{\,2}_{\mu}\coloneqq
\frac{\mathcal{N}}{2\ii}\,
{\bm{:}}
\left(
\widehat{\gamma}^{\dag}_{1,\mu}
\widehat{\gamma}^{\dag}_{2,\mu}
-
\widehat{\gamma}^{\,}_{2,\mu}
\widehat{\gamma}^{\,}_{1,\mu}
\right)
{\bm{:}}\,,
\\
&
\widehat{J}^{\,3}_{\mu}\coloneqq
\frac{\mathcal{N}}{2}\,
{\bm{:}}
\left(
\widehat{\gamma}^{\dag}_{1,\mu}
\widehat{\gamma}^{\,}_{1,\mu}
+
\widehat{\gamma}^{\dag}_{2,\mu}
\widehat{\gamma}^{\,}_{2,\mu}
\right)
{\bm{:}}\,,
\end{split}
\label{eq:conformal-properties-theta=0lambda=piover4-d}
\end{equation}
on the one hand, and
\begin{equation}
\begin{split}
&
\widehat{K}^{1}_{\mu}\coloneqq
\sigma^{\,}_{\mu}\,
\frac{\mathcal{N}}{2}\,
{\bm{:}}
\left(
\widehat{\gamma}^{\dag}_{1,\mu}
\widehat{\gamma}^{\,}_{2,\mu}
+
\widehat{\gamma}^{\dag}_{2,\mu}
\widehat{\gamma}^{\,}_{1,\mu}
\right)
{\bm{:}}\,,
\\
&
\widehat{K}^{2}_{\mu}\coloneqq
\sigma^{\,}_{\mu}\,
\frac{\mathcal{N}}{2\ii}\,
{\bm{:}}
\left(
\widehat{\gamma}^{\dag}_{1,\mu}
\widehat{\gamma}^{\,}_{2,\mu}
-
\widehat{\gamma}^{\dag}_{2,\mu}
\widehat{\gamma}^{\,}_{1,\mu}
\right)
{\bm{:}}\,,
\\
&
\widehat{K}^{3}_{\mu}\coloneqq
\frac{\mathcal{N}}{2}\,
{\bm{:}}
\left(
\widehat{\gamma}^{\dag}_{1,\mu}
\widehat{\gamma}^{\,}_{1,\mu}
-
\widehat{\gamma}^{\dag}_{2,\mu}
\widehat{\gamma}^{\,}_{2,\mu}
\right)
{\bm{:}}\,,
\end{split}
\label{eq:conformal-properties-theta=0lambda=piover4-e}
\end{equation}
\end{subequations}
on the other hand, for $\mu=L,R$ and $\sigma^{\,}_{R}=-\sigma^{\,}_{L}=+1$.
Here, Eq.\
(\ref{eq:conformal-properties-theta=0lambda=piover4-e})
follows from Eq.\
(\ref{eq:conformal-properties-theta=0lambda=piover4-d})
by performing the particle-hole transformation
$\widehat{\gamma}^{\,}_{2,\mu}\mapsto \widehat{\gamma}^{\dag}_{2,\mu}$
for $\mu=L,R$ followed by a global rotation by $\pi$ generated by
$\widehat{K}^{3}_{L}$.
With the choice
\begin{subequations}
\begin{equation}
\mathcal{N}=2\pi
\end{equation}
for the multiplicative normalization factor of the currents,
one verifies that the only non-vanishing and singular
operator-product expansions (OPEs) between any two
pairs of these twelve currents are the standard ones for the affine
$\mathfrak{su}(2)^{\,}_{k}\oplus\mathfrak{su}(2)^{\,}_{k}$
current algebra with the level $k=1$,
i.e., if $I$ stands for either $J$ or $K$,
the only singular OPEs are given by
\begin{equation}
\widehat{I}^{a\,}_{R}(z)\,
\widehat{I}^{\,b}_{R}(0)\sim
\frac{k\,\delta^{ab}}{2\,z^{2}}
+
\frac{\ii\epsilon^{abc}}{z}\,
\widehat{I}^{\,c}_{R}(0),
\end{equation}
in the holomorphic sector for which $z:=v\,\tau-\ii x$ and
\begin{equation}
\widehat{I}^{a}_{L}(\bar{z})\,
\widehat{I}^{b}_{L}(0)\sim
\frac{k\,\delta^{ab}}{2\,\bar{z}^{2}}
+
\frac{\ii\epsilon^{abc}}{\bar{z}}\,
\widehat{I}^{c}_{L}(0),
\end{equation}
\end{subequations}
in the anti-holomorphic sector for which $\bar{z}:=v\,\tau+\ii x$,
for $a,b=1,2,3$.
Hence, the non-interacting Hamiltonian
\eqref{eq:H0 RG stability Hminus a}
can be organized into the two
commuting $\mathfrak{su}(2)^{\,}_{k}$ sectors
with $k=1$ as
\cite{Sugawara1968}
\begin{subequations}
\label{eq:free JK Hamiltonian decomposition}
\begin{equation}
\widehat
H^{\,}_{0,-}
=
\widehat
H^{\,}_{0,J}
+
\widehat
H^{\,}_{0,K},
\label{eq:free JK Hamiltonian decomposition a}
\end{equation}
where we have dropped the dependence on $\epsilon$ as it is
implicit through the choice of the nodal points and
\begin{align}
&
\widehat{H}^{\,}_{0,J}\coloneqq
\frac{v}{6\pi}
\int\dd x\
{\bm{:}\,}\!
\left(
\widehat{\bm{J}}^{\,}_{R}
\cdot
\widehat{\bm{J}}^{\,}_{R}
+
\widehat{\bm{J}}^{\,}_{L}
\cdot
\widehat{\bm{J}}^{\,}_{L}
\right)\!
{\,\bm{:}}\,,
\label{eq:free JK Hamiltonian decomposition b}
\\
&
\widehat{H}^{\,}_{0,K}
\coloneqq
\frac{v}{6\pi}
\int\dd
x\
{\bm{:}\,}\!
\left(
\widehat{\bm{K}}^{\,}_{R}
\cdot
\widehat{\bm{K}}^{\,}_{R}
+
\widehat{\bm{K}}^{\,}_{L}
\cdot
\widehat{\bm{K}}^{\,}_{L}
\right)\!
{\,\bm{:}}\,.
\label{eq:free JK Hamiltonian decomposition c}
\end{align}
\end{subequations}

Only processes of bare scaling dimension two and
conserving momentum at the shifted Fermi
points are retained in
$\widehat{H}^{\,}_{\mathrm{int},-}(\epsilon)$.
We introduce the $a=1,2$- and $\mu=L,R$-resolved operators
\begin{subequations}
\label{eq:RG interaction stability Hminus}
\begin{align}
&\widehat{J}^{\,}_{a,R}(x)\coloneqq
{\bm{:}\,}
\widehat{\gamma}^{\dag}_{a,R}(x)\,
\widehat{\gamma}^{\,}_{a,R}(x)
{\,\bm{:}},
\label{eq:RG interaction stability Hminus a'}
\\
&\widehat{J}^{\,}_{a,L}(x)\coloneqq
{\bm{:}\,}
\widehat{\gamma}^{\dag}_{a,L}(x)\,
\widehat{\gamma}^{\,}_{a,L}(x)
{\,\bm{:}},
\label{eq:RG interaction stability Hminus a}
\end{align}
the same-flavor operators
\begin{align}
&
\widehat{J}^{\,}_{a}(x)
\coloneqq
\widehat{J}^{\,}_{a,R}(x)
+
\widehat{J}^{\,}_{a,L}(x),
\label{eq:RG interaction stability Hminus b}
\\
&
\widehat{I}^{\,}_{a}(x)
\coloneqq
\widehat{J}^{\,}_{a,R}(x)
-
\widehat{J}^{\,}_{a,L}(x),
\label{eq:RG interaction stability Hminus c}
\\
&
\widehat{Y}^{\,}_{a}(x)
\coloneqq
{\bm{:}\,}\widehat{\gamma}^{\,}_{a,R}(x)\,\widehat{\gamma}^{\,}_{a,L}(x){\,\bm{:}}
+
{\bm{:}\,}\widehat{\gamma}^{\dag}_{a,L}(x)\,\widehat{\gamma}^{\dag}_{a,R}(x){\,\bm{:}},
\label{eq:RG interaction stability Hminus d}
\\
&
\widehat{X}^{\,}_{a}(x)
\coloneqq
\ii\,
{\bm{:}\,}\widehat{\gamma}^{\,}_{a,R}(x)\,\widehat{\gamma}^{\,}_{a,L}(x){\,\bm{:}}
-\ii\,
{\bm{:}\,}\widehat{\gamma}^{\dag}_{a,L}(x)\,\widehat{\gamma}^{\dag}_{a,R}(x){\,\bm{:}},
\label{eq:RG interaction stability Hminus e}
\end{align}
together with the mixed-flavor operators
\begin{align}
&\widehat{M}^{\,}_{R}
\coloneqq
{\bm{:}\,}\widehat{\gamma}^{\dag}_{1,R}(x)\,\widehat{\gamma}^{\,}_{2,R}(x){\,\bm{:}},
\label{eq:RG interaction stability Hminus f'}
\\
&\widehat{M}^{\,}_{L}
\coloneqq
{\bm{:}\,}\widehat{\gamma}^{\dag}_{2,L}(x)\,\widehat{\gamma}^{\,}_{1,L}(x){\,\bm{:}}.
\label{eq:RG interaction stability Hminus f}
\end{align}
\end{subequations}
The marginal continuum interaction can then
be expressed as the space integral over
the following six local composite operators
\begin{subequations}
\begin{equation}
\widehat{H}^{\,}_{\mathrm{int},-}(\epsilon)
\coloneqq
\sum_{A\in\mathfrak{A}}\,
g^{\,}_{A}(\epsilon)
\int\dd x\ 
\widehat{\mathcal{O}}^{\,}_{A}(x),
\label{eq:RG interaction stability Hminus g}
\end{equation}
with
\begin{equation}
\mathfrak{A}=
\{Y,X,C,M,J,I\},
\label{eq:RG interaction stability Hminus h}
\end{equation}
and where each of the operators
\begin{align}
&\widehat{\mathcal{O}}^{\,}_{Y}(x)
\coloneqq
\widehat{Y}^{\,}_{1}(x)\,\widehat{Y}^{\,}_{2}(x),
\\
&\widehat{\mathcal{O}}^{\,}_{X}(x)
\coloneqq
\widehat{X}^{\,}_{1}(x)\,
\widehat{X}^{\,}_{2}(x),
\\
&\widehat{\mathcal{O}}^{\,}_{C}(x)
\coloneqq
\widehat{J}^{\,}_{1,R}(x)\,
\widehat{J}^{\,}_{1,L}(x)
+
\widehat{J}^{\,}_{2,R}(x)\,
\widehat{J}^{\,}_{2,L}(x),
\\
&\widehat{\mathcal{O}}^{\,}_{M}(x)
\coloneqq
\widehat{M}^{\,}_{R}(x)\,
\widehat{M}^{\,}_{L}(x)
+
\widehat{M}^{\dag}_{R}(x)\,
\widehat{M}^{\dag}_{L}(x),
\\
&\widehat{\mathcal{O}}^{\,}_{J}(x)
\coloneqq
\widehat{J}^{\,}_{1}(x)\,\widehat{J}^{\,}_{2}(x),
\\
&\widehat{\mathcal{O}}^{\,}_{I}(x)
\coloneqq
\widehat{I}^{\,}_{1}(x)\,\widehat{I}^{\,}_{2}(x),
\end{align}
is fully normal ordered by construction%
~\footnote{
For example, the operators in the normal-ordered product
$\protect{\widehat{Y}}^{\,}_{1}(x)$ commute with those
in the normal-ordered product
$\protect{\widehat{Y}}^{\,}_{2}(x)$.
         }.
Their six couplings are dimensionful and given,
to leading order in $\epsilon$, by
\begin{equation}
g^{\,}_{A}(\epsilon)=
\epsilon\,
\mathfrak{a}\,
\mathfrak{g}^{\,}_{A},
\qquad
A\in\mathfrak{A},
\label{eq:RG interaction stability Hminus i}
\end{equation}
with
\begin{equation}
\begin{split}
&
\Big(
\mathfrak{g}^{\,}_{Y},
\mathfrak{g}^{\,}_{X},
\mathfrak{g}^{\,}_{C},
\mathfrak{g}^{\,}_{M},
\mathfrak{g}^{\,}_{J},
\mathfrak{g}^{\,}_{I}
\Big)=
\\
&
\qquad\qquad
\Bigl(
-8,
-8,
+16,
-16,
-8,
+8
\Bigr).
\end{split}
\label{eq:RG interaction stability Hminus j}
\end{equation}
\end{subequations}

We proceed by deriving the one-loop renormalization group (RG)
obeyed by the six couplings
(\ref{eq:RG interaction stability Hminus i}).
We introduce the logarithmic RG scale
$\ell$ through the running short-distance cutoff
\begin{equation}
\mathfrak{a}'(\ell)\coloneqq
\mathfrak{a}\,e^{\ell}.
\label{eq:rescaling mathfrak a}
\end{equation}
Increasing $\ell>0$ thus corresponds to flowing toward
longer length scales and lower energies.
One then verifies that the partition function at zero temperature
with the right-hand side of
Eq.\
(\ref{eq:Def H_ for RG})
as Hamiltonian is form-invariant under the rescaling
(\ref{eq:rescaling mathfrak a})
provided we introduce the six dimensionless running couplings
\begin{subequations}
\label{eq:RG equations stability Hminus}
\begin{equation}
\bar{g}^{\,}_{A}(\epsilon,\ell)\coloneqq
\frac{g^{\,}_{A}(\epsilon,\ell)}{2\pi\,v},
\qquad
A\in\mathfrak{A},
\label{eq:RG equations stability Hminus 0}
\end{equation}
that obey the six coupled one-loop $\beta$-functions
\begin{align}
&
\frac{\dd\bar{g}^{\,}_{Y}}{\dd\ell}=
-
2\,\bar{g}^{\,}_{C}\,\bar{g}^{\,}_{Y}
+
2\,(\bar{g}^{\,}_{J}-\bar{g}^{\,}_{I})\,\bar{g}^{\,}_{X},
\label{eq:RG equations stability Hminus a}
\\
&
\frac{\dd\bar{g}^{\,}_{X}}{\dd\ell}=
-
2\,\bar{g}^{\,}_{C}\,\bar{g}^{\,}_{X}
+
2\,(\bar{g}^{\,}_{J}-\bar{g}^{\,}_{I})\,\bar{g}^{\,}_{Y},
\label{eq:RG equations stability Hminus b}
\\
&
\frac{\dd\bar{g}^{\,}_{C}}{\dd\ell}=
-
2
\left(
\bar{g}^{2}_{Y}
+
\bar{g}^{2}_{X}
\right)
+
\bar{g}^{2}_{M},
\label{eq:RG equations stability Hminus c}
\\
&
\frac{\dd\bar{g}^{\,}_{M}}{\dd\ell}=
+
(
2\,\bar{g}^{\,}_{C}
+
2\,\bar{g}^{\,}_{I}
-
2\,\bar{g}^{\,}_{J}
)\,
\bar{g}^{\,}_{M},
\label{eq:RG equations stability Hminus d}
\\
&
\frac{\dd\bar{g}^{\,}_{J}}{\dd\ell}=
+
2\,\bar{g}^{\,}_{X}\,\bar{g}^{\,}_{Y}
-
\frac{1}{2}\,
\bar{g}^{2}_{M},
\label{eq:RG equations stability Hminus e}
\\
&
\frac{\dd\bar{g}^{\,}_{I}}{\dd\ell}=
-
2\,\bar{g}^{\,}_{X}\,\bar{g}^{\,}_{Y}
+
\frac{1}{2}\,
\bar{g}^{2}_{M},
\label{eq:RG equations stability Hminus f}
\end{align}
with the initial conditions fixed by Eqs.\
(\ref{eq:H0 RG stability Hminus c}),
(\ref{eq:RG interaction stability Hminus i}),
(\ref{eq:RG interaction stability Hminus j}),
and
\begin{equation}
\bar{g}^{\,}_{A}(\epsilon,0)\coloneqq
\frac{\epsilon\,\mathfrak{g}^{\,}_{A}}{16\,\pi},
\qquad
A\in\mathfrak{A}.
\label{eq:RG equations stability Hminus g}
\end{equation}
\end{subequations}

The system of first-order differential equations
\eqref{eq:RG equations stability Hminus}
can be simplified as follows.
First, we observe that the combination
$\bar{g}^{\,}_{X}-\bar{g}^{\,}_{Y}$
obeys the differential equation
\begin{subequations}
\label{eq:RG Invariance X-Y}
\begin{equation}
\frac{\dd}{\dd\ell}
(\bar{g}^{\,}_{X}-\bar{g}^{\,}_{Y})=
-
2\,(\bar{g}^{\,}_{C}+\bar{g}^{\,}_{J}-\bar{g}^{\,}_{I})\,
(\bar{g}^{\,}_{X}-\bar{g}^{\,}_{Y}).
\end{equation}
Thus, it vanishes identically because of the initial condition
\begin{equation}
\bar{g}^{\,}_{X}(\epsilon,0)=\bar{g}^{\,}_{Y}(\epsilon,0).
\end{equation}
\end{subequations}
Second, we observe that the combination
$\bar{g}^{\,}_{I}+\bar{g}^{\,}_{J}$
obeys the differential equation
\begin{subequations}
\begin{equation}
\frac{\dd}{\dd\ell}
(\bar{g}^{\,}_{I}+\bar{g}^{\,}_{J})=0.
\end{equation}
Thus, it vanishes identically because of the initial condition
\begin{equation}
\bar{g}^{\,}_{I}(\epsilon,0)=-\bar{g}^{\,}_{J}(\epsilon,0).
\end{equation}
\end{subequations}
Third, we observe that the combination
$\bar{g}^{\,}_{J}+\bar{g}^{\,}_{C}/2$
obeys the differential equation
\begin{subequations}
\begin{equation}
\frac{\dd}{\dd\ell}
\left(\bar{g}^{\,}_{J}+\frac{\bar{g}^{\,}_{C}}{2}\right)=
-(\bar{g}^{\,}_{X}-\bar{g}^{\,}_{Y})^{2}.
\end{equation}
Thus, it vanishes identically because of
Eq.~\eqref{eq:RG Invariance X-Y}
and the initial condition
\begin{equation}
\bar{g}^{\,}_{J}(\epsilon,0)=-\frac{\bar{g}^{\,}_{C}(\epsilon,0)}{2}.
\end{equation}
\end{subequations}
We conclude that, given the initial conditions
(\ref{eq:RG equations stability Hminus g}),
the flow remains on the invariant submanifold
\begin{equation}
\bar{g}^{\,}_{X}\equiv\bar{g}^{\,}_{Y},
\qquad
\bar{g}^{\,}_{J}
\equiv
-\bar{g}^{\,}_{I}
\equiv
-\frac{\bar{g}^{\,}_{C}}{2}.
\label{eq:invariant RG manifold stability Hminus}
\end{equation}
The equations \eqref{eq:RG equations stability Hminus}
on the submanifold \eqref{eq:invariant RG manifold stability Hminus}
admit the closed solutions
\begin{subequations}
\label{eq:closed RG solution stability Hminus}
\begin{align}
&
\bar{g}^{\,}_{Y}(\epsilon,\ell)=
-
\frac{\gamma(\epsilon)}{2}\,
\frac{1-\sin\bigl(4\,\gamma(\epsilon)\,\ell\bigr)}
     {\cos\bigl(4\,\gamma(\epsilon)\,\ell\bigr)},
\label{eq:closed RG solution stability Hminus a}
\\
&
\bar{g}^{\,}_{X}(\epsilon,\ell)=
\bar{g}^{\,}_{Y}(\epsilon,\ell),
\label{eq:closed RG solution stability Hminus b}
\\
&
\bar{g}^{\,}_{C}(\epsilon,\ell)=
+
\gamma(\epsilon)\,
\frac{1}{\cos\bigl(4\,\gamma(\epsilon)\,\ell\bigr)},
\label{eq:closed RG solution stability Hminus c}
\\
&
\bar{g}^{\,}_{M}(\epsilon,\ell)=
-
\gamma(\epsilon)\,
\frac{1+\sin\bigl(4\,\gamma(\epsilon)\,\ell\bigr)}
     {\cos\bigl(4\,\gamma(\epsilon)\,\ell\bigr)},
\label{eq:closed RG solution stability Hminus d}
\\
&
\bar{g}^{\,}_{J}(\epsilon,\ell)=
-\frac{1}{2}\,\bar{g}^{\,}_{C}(\epsilon,\ell),
\label{eq:closed RG solution stability Hminus e}
\\
&
\bar{g}^{\,}_{I}(\epsilon,\ell)=
+\frac{1}{2}\,\bar{g}^{\,}_{C}(\epsilon,\ell),
\label{eq:closed RG solution stability Hminus f}
\end{align}
where we are using the short-hand notation
\begin{equation}
\gamma(\epsilon)\equiv
\bar{g}^{\,}_{C}(\epsilon,0)=
\frac{\epsilon}{\pi}.
\label{eq:closed RG solution stability Hminus g}
\end{equation}
\end{subequations}
The solution
(\ref{eq:closed RG solution stability Hminus})
to the one-loop RG equations
(\ref{eq:RG equations stability Hminus})
allows us to draw the following insights.

First, for every $\epsilon>0$, the Gaussian fixed point
is marginally unstable.
Indeed, the interaction drives the flow
toward ``infinitely'' strong coupling at the finite RG scale
\begin{equation}
\ell^{\,}_{\star}(\epsilon)=
\frac{\pi}{8\,\gamma(\epsilon)}=
\frac{\pi^{2}}{8\,\epsilon}
\label{eq:strong coupling scale stability Hminus}
\end{equation}
for any $\epsilon>0$.

Second, this instability is associated
to the crossover length 
\begin{equation}
\xi^{\,}_{\star}(\epsilon)
\coloneqq
\mathfrak{a}'\big(\ell^{\,}_{\star}(\epsilon)\big)=
e^{
\frac{\pi^2}{8\epsilon}
  }\,
\mathfrak{a}
\end{equation}
that is exponentially large for small values of $0<\epsilon\ll1$.

Third, the three ratios
\begin{subequations}
\label{eq:strong coupling ray stability Hminus}
\begin{align}
&
\lim_{\ell\uparrow\ell^{\,}_{\star}(\epsilon)}
\frac{\bar{g}^{\,}_{M}(\epsilon,\ell)}
     {\bar{g}^{\,}_{C}(\epsilon,\ell)}=
-2,
\label{eq:strong coupling ray stability Hminus d}
\\
&
\lim_{\ell\uparrow\ell^{\,}_{\star}(\epsilon)}
\frac{\bar{g}^{\,}_{J}(\epsilon,\ell)}
     {\bar{g}^{\,}_{C}(\epsilon,\ell)}=
-\frac{1}{2},
\label{eq:strong coupling ray stability Hminus e}
\\
&
\lim_{\ell\uparrow\ell^{\,}_{\star}(\epsilon)}
\frac{\bar{g}^{\,}_{I}(\epsilon,\ell)}
     {\bar{g}^{\,}_{C}(\epsilon,\ell)}=
+\frac{1}{2},
\label{eq:strong coupling ray stability Hminus f}
\end{align}
have a well-defined limiting value as
$\ell\uparrow\ell^{\,}_{\star}$.
Additionally, the running couplings
$\bar{g}^{\,}_{Y}$ and $\bar{g}^{\,}_{X}$
converge to
\begin{align}
&
\lim_{\ell\uparrow\ell^{\,}_{\star}(\epsilon)}
\bar{g}^{\,}_{Y}(\epsilon,\ell)=
0,
\label{eq:strong coupling ray stability Hminus a}
\\
&
\lim_{\ell\uparrow\ell^{\,}_{\star}(\epsilon)}
\bar{g}^{\,}_{X}(\epsilon,\ell)=
0.
\label{eq:strong coupling ray stability Hminus b}
\end{align}
\end{subequations}
Numerical integration of the one-loop $\beta$-functions
\eqref{eq:RG equations stability Hminus}
for selected small
perturbations of the initial conditions
(\ref{eq:RG equations stability Hminus g})
indicates that the strong-coupling ray
\eqref{eq:strong coupling ray stability Hminus}
is locally attractive.

The divergence at $\ell^{\,}_{\star}$ 
is not to be interpreted literally.
Rather, it signals the breakdown of the one-loop description
and the generation of a nonperturbative infrared scale.
The nature of the phase beyond this scale is
determined by analyzing the strong-coupling operator that is selected by
Eq.~\eqref{eq:strong coupling ray stability Hminus}.

If we factorize the running coupling $\bar{g}^{\,}_{C}(\epsilon,\ell)$
in the interaction
$\widehat{H}^{\,}_{\mathrm{int},-}(\epsilon,\ell)$,
take the limit
$\ell\uparrow\ell^{\,}_{\star}(\epsilon)$,
and make use of the asymptotics
(\ref{eq:strong coupling ray stability Hminus}),
we may write
\begin{subequations}
\begin{equation}
\widehat{H}^{\,}_{\mathrm{int},-}(\epsilon,\ell)\simeq
2\pi v\,
\bar{g}^{\,}_{C}(\epsilon,\ell)
\int\dd x\ 
\widehat{\mathcal{O}}^{\,}_{\star}(x)
\label{eq:strong coupling su2 interaction}
\end{equation}
where
\begin{equation}
\widehat{\mathcal{O}}^{\,}_{\star}
\coloneqq
\widehat{\mathcal{O}}^{\,}_{C}
-
2\,\widehat{\mathcal{O}}^{\,}_{M}
-
\frac{1}{2}\,\widehat{\mathcal{O}}^{\,}_{J}
+
\frac{1}{2}\,\widehat{\mathcal{O}}^{\,}_{I},
\end{equation}
and
\begin{equation}
\lim_{\ell\uparrow\ell^{\,}_{\star}(\epsilon)}
\bar{g}^{\,}_{C}(\epsilon,\ell)=
+\infty,
\qquad
\epsilon>0.
\end{equation}
Remarkably, one verifies that
\begin{equation}
\widehat{\mathcal{O}}^{\,}_{\star}=
\frac{1}{\pi^{2}}\,
\widehat{\bm{K}}^{\,}_{R}
\cdot
\widehat{\bm{K}}^{\,}_{L}
\label{eq:Strong Coupling Operator su2}
\end{equation}
\end{subequations}
with the vector of chiral components
$\widehat{\bm{K}}^{\,}_{L}$
and
$\widehat{\bm{K}}^{\,}_{R}$
defined in Eq.\
(\ref{eq:conformal-properties-theta=0lambda=piover4-e}).
The asymptotically dominant interaction
$\widehat{\mathcal{O}}^{\,}_{\star}$
in Eq.~\eqref{eq:Strong Coupling Operator su2}
is thus the current-current interaction built from the
non-Abelian currents
(\ref{eq:conformal-properties-theta=0lambda=piover4-e})
obeying the $\mathfrak{su}(2)^{\,}_{1}$ current algebra.

We conclude that,
when $\ell\uparrow\ell^{\,}_{\star}(\epsilon)$,
the Hamiltonian \eqref{eq:Def H_ for RG}
is approximated by
\begin{subequations}
\begin{equation}
\widehat{H}^{\,}_{-}\simeq
\widehat{H}^{\,}_{J}
+
\widehat{H}^{\,}_{K},
\end{equation}
where
\begin{align}
&\widehat{H}^{\,}_{J}\coloneqq
\widehat{H}^{\,}_{0,J},
\\
&\widehat{H}^{\,}_{K}\coloneqq
\widehat{H}^{\,}_{0,K}
+
\frac{2v}{\pi}\,
\bar{g}^{\,}_{C}(\epsilon,\ell)
\int\dd x\ 
\widehat{\bm{K}}^{\,}_{R}
\cdot
\widehat{\bm{K}}^{\,}_{L}.
\end{align}
\end{subequations}
The interacting Hamiltonian $\widehat{H}^{\,}_{K}$
is the $\mathrm{SU}(2)^{\,}_{1}$ non-Abelian Thirring model
\cite{Thirring1958,Dashen1973}.
With the positive sign of the coupling constant
$\bar{g}^{\,}_{C}(\epsilon,\ell)$,
the interaction \eqref{eq:strong coupling su2 interaction}
is marginally relevant and generates dynamically
a mass scale in the $\mathfrak{su}(2)^{\,}_{1}$ sector
generated by
$\widehat{\bm{K}}^{\,}_{L}$
and
$\widehat{\bm{K}}^{\,}_{R}$
\cite{Dashen1973,Ludwig2003}.
The currents
$\widehat{\bm{J}}^{\,}_{L}$
and
$\widehat{\bm{J}}^{\,}_{R}$
defined in Eq.\
(\ref{eq:conformal-properties-theta=0lambda=piover4-d})
are absent from the asymptotically dominant interaction.
Hence, the critical sector of the unperturbed Hamiltonian
(\ref{eq:conformal-properties-theta=0lambda=piover4})
that is generated by the currents
$\widehat{\bm{J}}^{\,}_{L}$
and
$\widehat{\bm{J}}^{\,}_{R}$
remains critical. It is a $\mathsf{c}=1$ conformal field theory
with $\mathfrak{su}(2)^{\,}_{1}$ current algebra.

\subsubsection{Comparison with numerics}
\label{ssub:comparison_with_numerics}

The leading effect of the perturbation
$\epsilon\,\widehat{V}^{\,}_{\pm}$
on the free Hamiltonian $\widehat{T}^{\,}_{\pm}$
in Eq.~\eqref{eq:def Perturbed Hamiltonian Majorana Stability}
is to impose the replacement
\begin{equation}
\widehat{T}^{\,}_{\pm}\mapsto
\widehat{T}^{\,}_{\pm,\mathrm{eff}}(\epsilon)
\label{eq:Replacement T - Teff}
\end{equation}
according to Eq.\
(\ref{eq:Hamiltonian Decomposotion H_pm}).
In doing so, we have discarded normal-ordered quartic terms
that are marginal at tree level.
Here, we compare the consequences of the replacement
\eqref{eq:Replacement T - Teff}
analyzed in Sec.~\ref{ssub:effect_of_the_interactions}
to the numerical results of
Sec.~\ref{sec:Numerical exploration of the phase diagrams}.

The comparison between the one-loop	RG analysis in
Sec.~\ref{Effect of the normal-ordered quartic term in H-}
against the numerics is done in
Sec.~\ref{subsubsec:Neelx-NeelSPTx phase boundary}.

\paragraph{Case of $\widehat{H}^{\,}_{+}(\epsilon)$.}

Hamiltonian $\widehat{H}^{\,}_{+}(\epsilon)$ has been studied
numerically in
Sec.~\ref{sec:Quadricritical point}
for $0\leq\epsilon\ll1$.
According to Eq.~\eqref{eq:Ansatz theta NeelZ HB},
the spectral gap of
$\widehat{H}^{(0)}_{\mathrm{II}}(\epsilon,3\pi/4)$
found numerically is
\begin{equation}
\Delta(\epsilon)
\simeq
3.542\,\epsilon,
\qquad
0\leq\epsilon\ll1.
\label{eq:Delta Found Numerically Compared}
\end{equation}

The product of the unitary transformations
$\widehat{U}^{\,}_{\mathrm{R}}$ and
$\widehat{U}^{\,}_{Y,\pi}$
defined in Eqs.~\eqref{eq:U_R} and
\eqref{eq:U_Y even a}, respectively,
realizes the reflection
$\theta\mapsto\pi/2-\theta$
together with the translation $\theta\mapsto\pi-\theta$
in the phase diagram in Fig.~\ref{fig:phase diagram B}.
This implies the identity
\begin{equation}
\widehat{U}\,
\widehat{H}^{(0)}_{\mathrm{II}}(\epsilon,3\pi/4)\,
\widehat{U}^{\dag}=
\widehat{H}^{(0)}_{\mathrm{II}}(\pi/2+\epsilon,3\pi/4),
\end{equation}
with the unitary $\widehat{U}=
\widehat{U}^{\,}_{Y,\pi}\,
\widehat{U}^{\,}_{\mathrm{R}}$
and for $0\leq\epsilon\ll1$.
Therefore, with the definition
\eqref{eq:Relation Hamiltonian Original VS H+- a},
we identify the BdG gap $\widetilde{\Delta}^{\mathrm{(bdg)}}_{+}(\epsilon)$ of
$\widehat{H}^{(0)}_{\mathrm{II}}(\epsilon,3\pi/4)$
with the BdG gap $\Delta^{\mathrm{(bdg)}}_{+}(\epsilon)$
of $\widehat{H}^{\,}_{+}(\epsilon)$ as
\begin{equation}
\widetilde{\Delta}^{\mathrm{(bdg)}}_{+}(\epsilon)\coloneqq
\frac{1}{\sqrt{2}}\,
\Delta^{\mathrm{(bdg)}}_{+}(\epsilon)
\simeq
1.801\,\epsilon,
\qquad
\epsilon\ll1.
\label{eq:Gap H+ lambda 3pi4 rescaled}
\end{equation}
We now compare the analytical result
\eqref{eq:Gap H+ lambda 3pi4 rescaled}
with the numerical result in
\eqref{eq:Delta Found Numerically Compared} as
\begin{equation}
\Delta(\epsilon)\ /\ 
\widetilde{\Delta}^{\mathrm{(bdg)}}_{+}(\epsilon)
\simeq1.967,
\qquad
0<\epsilon\ll1.
\label{eq:Compare Gaps Numerics VS Analytics}
\end{equation}

The factor of approximately $2$ on the right-hand side of
Eq.~\eqref{eq:Compare Gaps Numerics VS Analytics}
originates from the distinction between the minimum
positive single-quasiparticle BdG energy
$\widetilde{\Delta}^{\mathrm{(bdg)}}_{+}(\epsilon)$ and
the many-body level spacing
$\Delta(\epsilon)$ extracted numerically.

The microscopic Majorana chain with PBCs decomposes into
sectors of fixed fermion parity $\mathsf{p}=\pm$.
The parity operator is $\widehat{P}^{\,}_{\mathrm{F}}$
defined in Eq.~\eqref{eq:parity operator in Majorana rep}.
Since the creation of one BdG quasiparticle
reverses fermion parity, the lowest excitation
within a fixed parity sector must contain an
even number of quasiparticles.
The lowest such excitation is therefore expected
to approach the two-quasiparticle threshold
\begin{equation}
\lim_{2N\to\infty}
E^{\,}_{2}-E^{\,}_{1}
=
2\,
\widetilde{\Delta}^{\mathrm{(bdg)}}_{+}(\epsilon).
\label{eq:many body versus BdG gap stability}
\end{equation}

The symmetry-resolved finite-size spectra support
this interpretation.
For $2N=0\bmod4$, the states with energies
$E^{\,}_{1}$ and $E^{\,}_{2}$ belong to the same
$\mathsf{p}=-$ sector and are
therefore separated by a two-quasiparticle excitation that
preserves fermion parity.
For $2N\neq0\bmod4$, the lowest four energies
form two distinct doublets
\begin{equation}
E^{\,}_{0}=E^{\,}_{1}\neq
E^{\,}_{2}=E^{\,}_{3}.
\end{equation}
Both parity sectors are represented in each doublet,
so that parity alone does not determine their
quasiparticle content.
Nevertheless, both finite-size sequences
of system sizes $2N$ extrapolate consistently
with Eq.~\eqref{eq:many body versus BdG gap stability}.

The effective quadratic Hamiltonian consequently predicts
a many-body gap of
$2\widetilde{\Delta}^{\mathrm{(bdg)}}_{+}(\epsilon)$
in good agreement with the numerical result
$\Delta(\epsilon)$,
\begin{equation}
2\,\widetilde{\Delta}^{\mathrm{(bdg)}}_{+}(\epsilon)
-
\Delta(\epsilon)
\simeq
0.059\,\epsilon.
\end{equation}
The two coefficients differ by approximately $1.7\%$.
A complete state-by-state identification would require
treating the two parity sectors away from the critical  point
$(0,3\pi/4)$ in Fig.~\ref{fig:phase diagram B}, including their different
finite-size momentum grids, and is not attempted here.

\paragraph{Case of $\widehat{H}^{\,}_{-}(\epsilon)$.}

Hamiltonian $\widehat{H}^{\,}_{-}(\epsilon)$ has been studied
numerically in
Sec.~\ref{subsubsec:Neelx-NeelSPTx phase boundary}
for $0\leq\epsilon\ll1$.
According to Eq.~\eqref{eq:wave vector for Cx and Cz,c fixed lambda piover4 b},
the wavevector of oscillations that dominates the two-point function
$C^{z,\mathrm{c}}_{s^{\,}_{0},s^{\,}_{0}+s}$ of
$\widehat{H}^{(0)}_{\mathrm{II}}(\epsilon,\pi/4)$
found numerically is
\begin{equation}
q^{z,\mathrm{c}}(\epsilon)
\simeq
\frac{\pi}{2}+
\frac{\pi}{5}\,\epsilon,
\qquad
0\leq\epsilon\ll1.
\label{eq:Qz Found Numerically Compared}
\end{equation}

The unitary transformation
$\widehat{U}^{\,}_{\mathrm{R}}$
defined in Eq.~\eqref{eq:U_R}
realizes the reflection
$\theta\mapsto\pi/2-\theta$
in the phase diagram in Fig.~\ref{fig:phase diagram B}.
This implies the identity
\begin{equation}
\widehat{U}^{\,}_{\mathrm{R}}\,
\widehat{H}^{(0)}_{\mathrm{II}}(\epsilon,\pi/4)\,
\widehat{U}^{\dag}_{\mathrm{R}}=
\widehat{H}^{(0)}_{\mathrm{II}}(\pi/2-\epsilon,\pi/4),
\end{equation}
for $0\leq\epsilon\ll1$.
Therefore, with the definition
\eqref{eq:Relation Hamiltonian Original VS H+- b},
we identify the Fermi points $\widetilde{\Lambda}^{\star}_{-,\mathrm{F}}(\epsilon)$ of
$\widehat{H}^{(0)}_{\mathrm{II}}(\epsilon,\pi/4)$
with the Fermi points $\Lambda^{\star}_{-,\mathrm{F}}(\epsilon)$
of $\widehat{H}^{\,}_{-}(\epsilon)$.

The leading effect of the replacement
\eqref{eq:Replacement T - Teff} is to shift the
nodal points of $\widehat{H}^{\,}_{-}(0)$ to the
incommensurate nodal points of
$\widehat{H}^{\,}_{-}(\epsilon)$ given in
Eq.~\eqref{eq:Fermi H- lambda pi4}.
We may choose
\begin{equation}
k^{\,}_{\mathrm{F}}(\epsilon)=
\frac{\pi}{4}+\frac{\epsilon}{\pi}
\in
\Lambda^{\star}_{-,\mathrm{F}}(\epsilon)
\label{eq:k_F compared to numerics}
\end{equation}
without loss of generality.
The corresponding analytical wavevector is
\begin{equation}
2k^{\,}_{\mathrm{F}}(\epsilon)=
\frac{\pi}{2}+
\frac{2}{\pi}\,\epsilon.
\label{eq:analytical wavevector 2kF}
\end{equation}

We compare this analytical prediction with the
dominant wavevector $q^{z,\mathrm{c}}(\epsilon)$
extracted numerically from the connected
two-point function $C^{z,\mathrm{c}}_{s^{\,}_{0},s^{\,}_{0}+s}$.
Equations~\eqref{eq:Qz Found Numerically Compared} and
\eqref{eq:analytical wavevector 2kF} give
\begin{align}
q^{z,\mathrm{c}}(\epsilon)
-
2k^{\,}_{\mathrm{F}}(\epsilon)&=
\left(
\frac{\pi}{5}
-
\frac{2}{\pi}
\right)\epsilon
%\nonumber\\
\simeq
-0.0083\,\epsilon,
\end{align}
and thus differ by approximately $1.3\%$.
The numerical result is therefore consistent with identifying the
dominant incommensurate wavevector with
$2k^{\,}_{\mathrm F}(\epsilon)$.
In particular, the shift of the nodal points in
Eq.~\eqref{eq:Fermi H- lambda pi4} correctly captures
the deviation of $q^{z,\mathrm{c}}(\epsilon)$
from the commensurate value $\pi/2$ at $\epsilon=0$.

Determining why this wavevector appears and is
dominant in
$C^{z,\mathrm{c}}_{s^{\,}_{0},s^{\,}_{0}+s}$,
rather than in the corresponding $x$- or $y$-spin
correlation function, would require the continuum
expansion of the three microscopic spin components.
Such an operator identification is beyond
the scope of the present analysis.

\subsection{Weakly first-order quantum phase transitions of
$\widehat{H}^{(\mathsf{b})}_{\mathrm{I}}(\pi/2,\lambda)$ at $\lambda=\pi/4,3\pi/4$}
\label{subsec:Weakly first-order quantum phase transitions of}

Section \ref{subsec:Weakly first-order quantum phase transitions of}
is devoted exclusively to Hamiltonian
(\ref{eq:def HA and HB c})
obeying PBCs at any one of the four phase boundaries
\begin{equation}
(\theta,\lambda)=
\left\{\!
\left(0,\frac{\pi}{4}\right)\!,
\left(\frac{\pi}{2},\frac{\pi}{4}\right)\!,
\left(0,\frac{3\pi}{4}\right)\!,
\left(\frac{\pi}{2},\frac{3\pi}{4}\right)\!
\right\}
\label{eq:phase boundary different SPT and order}
\end{equation}
between two distinct SPT phases with competing orders
or at any one of the two phase boundaries
\begin{equation}
(\theta,\lambda)=
\left\{\!
\left(\frac{\pi}{4},0\right)\!,
\left(\frac{\pi}{4},\frac{\pi}{2}\right)\!
\right\}
\label{eq:phase boundary same SPT and different order}
\end{equation}
between the same SPT phases with competing orders.
We are going to use mean-field arguments in support
to the numerical observation from Sec.\
\ref{subsubsec:The Neelx-NeelSPTy phase boundary}
that the phase boundaries
(\ref{eq:phase boundary different SPT and order})
realize the same weakly first-order quantum phase transition,
in contrast to the phase boundaries
(\ref{eq:phase boundary same SPT and different order})
that realize the same deconfined quantum critical point.
Without loss of generality, we only need to compare
the integrable phase boundary $(\pi/4,0)$
from the non-integrable phase boundary $(0,\pi/4)$.

By combining symmetries with a mean-field approximation 
and the Mermin-Wagner theorem,
we are going to show that the point
\begin{equation}
(\theta,\lambda)=(\pi/4,0)
\end{equation}
realizes a continuous quantum phase transition
along the interval $\theta\in[0,\pi/2]$,
whereas the point
\begin{equation}
(\theta,\lambda)=(0,\pi/4)
\end{equation}
realizes a weakly first-order quantum phase transition
along the interval $\lambda\in[0,\pi/2]$.

This we do by considering the Landau-Ginzburg function
\begin{equation}
\begin{split}
V(X,Y)\coloneqq&\,
-
m^{2}
\left(
\cos\alpha\, X^{2}
+
\sin\alpha\, Y^{2}
\right)
\\
&\,
+
\mathrlap{u^{2}}
\hphantom{m^{2}}
\left(
\cos\alpha\, X^{4}
+
\sin\alpha\, Y^{4}
\right)
\\
&\,
+
\frac{4}{\sqrt{2}}\,
v^{2}\,
\cos\alpha\,
\sin\alpha\,
X^{2}\,Y^{2}.
\end{split}
\label{eq:def MF Landau-Ginzburg fct}
\end{equation}
Here, $X\in\mathbb{R}$ ($Y\in\mathbb{R}$)
is the short-hand notation for the staggered magnetization along
the $x$-axis ($y$-axis) in spin-$1/2$ space, while
$m^{2}\geq0$, $u^{2}>0$, $v^{2}\geq0$, and $\alpha\in[0,\pi/2]$
are functions of the microscopic parameters of Hamiltonian
(\ref{eq:def HA and HB c}).
The role of the parameter $v^{2}$ is to forbid the coexistence
of the two Ising order parameters whenever $\alpha\neq\pi/4$,
in agreement with the numerics from Sec.\
\ref{subsubsec:The Neelx-NeelSPTy phase boundary}.
This Landau-Ginzburg potential satisfies
\begin{enumerate}
\item
the Ising symmetry under $X\mapsto-X$,
\item
the Ising symmetry under $Y\mapsto-Y$,
\item
the symmetry under $X\mapsto Y$, $Y\mapsto X$, and
$\alpha\mapsto(\pi/2)-\alpha$,
\item
and the $\mathrm{SO}(2)\cong\mathrm{U}(1)$ symmetry under
$X\mapsto+\cos\beta\,X+\sin\beta\,Y$
and
$Y\mapsto-\sin\beta\,X+\cos\beta\,Y$
for the special case when $\alpha=\pi/4$ and $u^{2}=v^{2}$.
\end{enumerate}
When $\alpha=0$,
\begin{equation}
V(X,Y)=
-
m^{2}\,
X^{2}
+
u^{2}\,
X^{4}
\end{equation}
selects a phase with the pair of non-vanishing mean-field values
\begin{equation}
X^{\,}_{\mathrm{MF}}=\pm\sqrt{\frac{m^{2}}{2u^{2}}}
\end{equation}
of the staggered magnetization along the $x$-axis in spin-$1/2$ space.
When $\alpha=\pi/2$,
\begin{equation}
V(X,Y)=
-
m^{2}\,
Y^{2}
+
u^{2}\,
Y^{4}
\end{equation}
selects a phase with the pair of non-vanishing mean-field values
\begin{equation}
Y^{\,}_{\mathrm{MF}}=\pm\sqrt{\frac{m^{2}}{2u^{2}}}
\end{equation}
of the staggered magnetization along the $y$-axis in spin-$1/2$ space.
For values of $\alpha\in\,]0,\pi/2[$,
the local extrema of the Landau-Ginzburg function
(\ref{eq:def MF Landau-Ginzburg fct})
are located at
\begin{subequations}
\label{eq:extrema LG}
\begin{align}
&
(X,Y)=(0,0),
\label{eq:extrema LG a}
\\
&
(X,Y)=(X^{\,}_{\mathrm{MF}},0),
\label{eq:extrema LG b}
\\
&
(X,Y)=(0,Y^{\,}_{\mathrm{MF}}).
\label{eq:extrema LG c}
\end{align}
\end{subequations}
The energy of the extremum (\ref{eq:extrema LG a}) is zero.
The energy of the extremum (\ref{eq:extrema LG b}) is negative and proportional
to $\cos\alpha$.
The energy of the extremum (\ref{eq:extrema LG c}) is negative and proportional
to $\sin\alpha$.
The energies of the extrema (\ref{eq:extrema LG b}) and (\ref{eq:extrema LG c})
cross at the self-dual point
$\alpha=\pi/4$.
This level crossing implies that both order parameters coexist at
$\alpha=\pi/4$, i.e.,
$\alpha=\pi/4$
is a mean-field first-order quantum phase transition.

This mean-field prediction can be invalidated by fluctuations,
as occurs when
\begin{equation}
u^{2}=v^{2},
\qquad
\alpha=\pi/4,
\end{equation}
values for which
\begin{equation}
V(X,Y)=
-
\frac{m^{2}}{\sqrt{2}} 
\left(
X^{2}
+
Y^{2}
\right)
+
\frac{u^{2}}{\sqrt{2}}
\left(
X^{2}
+
Y^{2}
\right)^{2}.
\end{equation}
This $\mathrm{O}(2)$-symmetric potential is minimized along the circle
parametrized by
\begin{equation}
X^{2}_{\mathrm{MF}}+Y^{2}_{\mathrm{MF}}=
\frac{m^{2}}{2u^{2}}.
\end{equation}
However, this mean-field prediction is invalidated by the Mermin-Wagner
theorem as gapless fluctuations prevent the spontaneous breaking of
a continuous symmetry in one-dimensional space.
This mean-field prediction is not invalidated by
accounting for gapful fluctuations, as occurs when only discrete
microscopic symmetries are to be enforced.

We conclude that at the point
\begin{equation}
(\theta,\lambda)=(\pi/4,0)
\end{equation}
in the phase diagram
\ref{fig:phase diagram A},
microscopic $\mathrm{O}(2)$ symmetry enforces $u^{2}=v^{2}$,
thereby preventing a first-order quantum phase transition.
This is not true of the point
\begin{equation}
(\theta,\lambda)=(0,\pi/4)
\end{equation}
in the phase diagram
\ref{fig:phase diagram A},
for which $u^{2}\neq v^{2}$.
However, if $u^{2}$ and $v^{2}$ are close in values, then the
quantum phase transition is a weakly first-order one due to its proximity
to a continuous quantum phase transition. The numerics from Sec.\
\ref{subsubsec:The Neelx-NeelSPTy phase boundary}
are consistent with a weakly first-order quantum phase transition
separating the
$\mathrm{Neel}^{\,}_{x}$
from the
$\mathrm{Neel}^{\mathrm{SPT}}_{y}$
phases.

\paragraph*{Acknowledgements.}

A.R. was supported by
by SNSF Grant No.\ 219339.
\"O.M.A. was supported in parts 
by SNSF Grant No.\ 214429,
by NSF Grant No.\ DMR2022428,
and by the 
Simons Collaboration on Ultra-Quantum Matter, 
which is a grant from the Simons Foundation (651444, SHS).
C.C. was supported by  DOE Grant No.\ DE-SC0026189.
A.F. was supported by RIKEN Pioneering Project "Mathematical foundation of quantum information".

\bibliography{references}

\appendix

\section{iDMRG}
\label{appsec:iDMRG}

We use infinite-size density-matrix renormalization group
(iDMRG) to study the ground states of the Hamiltonians directly
in the thermodynamic limit.
Although many of the same observables can be computed
using finite-size density-matrix renormalization group
(DMRG), iDMRG avoids boundary and
finite-size effects and provides direct access to the
transfer matrix of the infinite matrix-product state (iMPS).
This is particularly useful for extracting correlation lengths and
the oscillatory components of two-point correlation functions.

\subsection{Notions of iDMRG}

The iDMRG procedure mimics the search of
the ground state in an infinite chain, taking some advantage of
the potential translation invariance of the targeted
ground state, as well as providing a natural
expression of the transfer matrix.
The method is particularly efficient for gapped ground states,
which are often accurately represented by MPSs
of moderate bond dimension.
An exact critical ground state generally requires an
unbounded bond dimension, but it can be approximated by a
finite-bond-dimension MPS with a finite effective correlation length.
For an injective and normalized MPS, the corresponding
transfer matrix has a unique eigenvalue of
largest modulus, which can be normalized to unity.

We follow \cite{schollwock2011} and first review the density
matrix renormalization group (DMRG) and
the matrix product state (MPS) formalism, then
we extend it to iDMRG with the iMPS formalism.

\subsubsection{Definition of MPS.}

We define a MPS as a representation of a pure
quantum state of $L$ sites,
viewed as a rank-$L$ tensor
\begin{equation}
\ket{\psi}=
\sum_{\vb*{\sigma}}
c^{\,}_{\sigma^{\,}_{1},\dotsc,\sigma^{\,}_{L}}
\ket{\vb*{\sigma}}=
\begin{minipage}{2.5cm}
\includegraphics[scale=0.7]{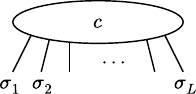}
\end{minipage}
\end{equation}
in terms of a chain of rank-$3$ tensors
\begin{subequations}
\label{eq:MPS definition}
\begin{align}
\ket{\psi}
&=
\sum_{\vb*{\sigma}}\sum_{\vb*{a}}
M^{\sigma^{\,}_{1}}_{1,a^{\,}_{1}}
M^{\sigma^{\,}_{2}}_{a^{\,}_{1},a^{\,}_{2}}\cdots
M^{\sigma^{\,}_{L}}_{a^{\,}_{L-1},1}
\ket{\vb*{\sigma}} \\
&=
\begin{minipage}{5.3cm}
\vspace{9pt}
\includegraphics[scale=0.7]{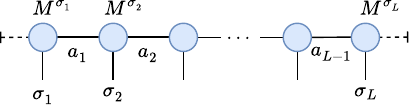}
\end{minipage}
\end{align}
\end{subequations}
and hence can be viewed as a product of matrices $M^{\sigma}$.
The diagrammatic representation of tensor networks
allows one to avoid complex explicit formulations.
A colored circle $M$ is used to represent a rank-$3$ tensor,
with one physical $d$-dimensional leg $\sigma^{\,}_{\ell}$
and two auxiliary legs $a^{\,}_{\ell-1}$ and $a^{\,}_{\ell}$
for the matrix $M^{\sigma^{\,}_{\ell}}$.
Hereto, $d$ is the dimension of the physical Hilbert space
at each site of the chain, e.g.,
$d=2$ for a chain of spin-$1/2$ degrees of freedom.
When two legs are fused, they are contracted, i.e.,
the summation over the corresponding index is performed.
Moreover, if a physical leg $\sigma^{\,}_{\ell}$
is oriented upwards,
the circle labels the complex conjugate matrix
$\bigl(M^{\sigma^{\,}_{\ell}}\bigr)^{*}$.

This MPS formulation will be exploited for ground state search.
First, notice the gauge freedom a matrix product has.
For any invertible matrix $U$,
\begin{equation}
MM'=
MUU^{-1}M'=
\overline{M}\,
\overline{M}',
\label{eq:gaugefree}
\end{equation}
holds. This implies the possibility to express a MPS in
various forms, especially in canonical forms.
The most interesting of them is the
mixed-canonical form
\footnote{A matrix on a site $\ell$ with physical leg
$\sigma^{\,}_{\ell}$ previously denoted by
$M^{\sigma^{\,}_{\ell}}$ is now denoted by
$M^{[\ell]}$ to lighten the notation.}
\begin{equation}
\ket{\psi}=
\begin{minipage}{6.2cm}
\includegraphics[scale=0.7]{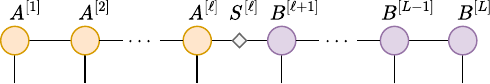}
\end{minipage}
\label{eq:mixed-canonical}	
\end{equation}
with left-normalized matrices $A^{[\ell]}$ and
right-normalized matrices $B^{[\ell]}$.
Left-normalized matrices $A^{[\ell]}$ obey
\begin{align}
\Bigl[
\Bigl(A^{[\ell]}\Bigr)
^{\dag}
A^{[\ell]}
\Bigr]
^{\,}_{a^{\,}_{\ell},a^{\prime}_{\ell}}
&=
\sum_{\sigma^{\,}_{\ell}}
\sum_{a^{\,}_{\ell-1}}
\bigl(A^{\sigma^{\,}_{\ell}}\bigr)
{}^{\dag}_{a^{\prime}_{\ell},a^{\,}_{\ell-1}}
A^{\sigma^{\,}_{\ell}}_{a^{\,}_{\ell-1},a^{\,}_{\ell}}
\nonumber \\
&=
\sum_{\sigma^{\,}_{\ell}}
\sum_{a^{\,}_{\ell-1}}
\bigl(A^{\sigma^{\,}_{\ell}}\bigr)
{}^{*}_{a^{\,}_{\ell-1},a^{\prime}_{\ell}}
A^{\sigma^{\,}_{\ell}}_{a^{\,}_{\ell-1},a^{\,}_{\ell}}
\nonumber \\
&=
\delta^{\,}_{a^{\,}_{\ell},a^{\prime}_{\ell}},
\end{align}
while right-normalized matrices $B^{[\ell]}$ obey
\begin{align}
\Bigl[
B^{[\ell]}
\Bigl(B^{[\ell]}\Bigr)
^{\dag}
\Bigr]
^{\,}_{a^{\,}_{\ell-1},a^{\prime}_{\ell-1}}
&=
\sum_{\sigma^{\,}_{\ell-1}}
\sum_{a^{\,}_{\ell}}
\bigl(B^{\sigma^{\,}_{\ell-1}}\bigr)
{}^{\dag}_{a^{\prime}_{\ell-1},a^{\,}_{\ell}}
B^{\sigma^{\,}_{\ell-1}}_{a^{\,}_{\ell},a^{\,}_{\ell-1}}
\nonumber \\
&=
\sum_{\sigma^{\,}_{\ell-1}}
\sum_{a^{\,}_{\ell}}
\bigl(B^{\sigma^{\,}_{\ell-1}}\bigr)
{\vphantom{B}}^{*}_{a^{\,}_{\ell},a^{\prime}_{\ell-1}}
B^{\sigma^{\,}_{\ell-1}}_{a^{\,}_{\ell},a^{\,}_{\ell-1}}
\nonumber \\
&=
\delta^{\,}_{a^{\,}_{\ell-1},a^{\prime}_{\ell-1}}.
\end{align}
These properties translate into the simplified expressions
\begin{equation}
\Bigl(A^{[\ell]}\Bigr)
^{\dag}
A^{[\ell]}=
\mathds{1},\qquad
B^{[\ell]}
\Bigl(B^{[\ell]}\Bigr)
^{\dag}=
\mathds{1},
\end{equation}
and provides the diagrammatic simplification
\begin{align}
\begin{minipage}{3.3cm}
\includegraphics[scale=0.7]{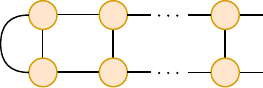}
\end{minipage}%
&=\ \,
\begin{minipage}{3.3cm}
\includegraphics[scale=0.7]{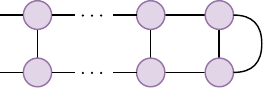}
\end{minipage}
\nonumber \\
&=\mathds{1}.
\end{align}

The expression \eqref{eq:mixed-canonical} is found performing,
for instance, singular value decompositions (SVDs) iteratively.
We briefly recall that the compact SVD
is a decomposition of any $m\times n$ matrix of rank
$r \leq \min(m,n)$ as
\begin{equation}
M=
USV^{\dag},
\end{equation}
with $U$ a left-normalized matrix of dimension $m\times r$,
$S$ a diagonal matrix of dimension $r\times r$ whose entries
are the non-zero singular values of $M$, and
$V^{\dag}$ a right-normalized matrix of dimension
$r\times n$. The form \eqref{eq:mixed-canonical} is
particularly suited to obtain the reduced density matrix
of a state in a bipartite lattice $\mathcal{A} \mathcal{B}$.
Such a state is written in orthonormal bases
$\bigl\{\ket{i_{\mathcal{A}}}\bigr\}$ and
$\bigl\{\ket{j_{\mathcal{B}}}\bigr\}$,
of $\mathcal{A}$ and $\mathcal{B}$, respectively, 
\begin{subequations}
\begin{align}
\ket{\psi}
&=
\sum_{i,j}M^{\,}_{i,j}
\ket{i^{\,}_{\mathcal{A}}}\ket{j^{\,}_{\mathcal{B}}}
\nonumber \\
&=
\sum_{i,j}\sum_{a} U^{\,}_{i,a}
\ket{i^{\,}_{\mathcal{A}}}
S^{\,}_{a,a}
V^{\dag}_{a,j}
\ket{j^{\,}_{\mathcal{B}}} \\
&=
\sum_{a}s^{\,}_{a}
\ket{a^{\,}_{\mathcal{A}}}\ket{a^{\,}_{\mathcal{B}}}
\end{align}
\end{subequations}
which is called the Schmidt decomposition
of $\ket{\psi}$, and with the diagonal elements
$s^{\,}_{a}$ of $S$ being the singular values of $M$.
Decomposing $\{\ket{a^{\,}_{\mathcal{A}}}\}$ into a
left-canonical MPS
\footnote{A MPS entirely made of left-normalized matrices $A$.}
and $\{\ket{a_{\mathcal{B}}}\}$ into a
right-canonical MPS
\footnote{A MPS entirely made of
right-normalized matrices $B$.}, one recovers the MPS
\eqref{eq:mixed-canonical}.
The reduced density matrix of $\mathcal{A}$ is then easily
expressed as
\begin{subequations}
\begin{align}
\rho^{\,}_{\mathcal{A}}
&=
\tr^{\,}_{\mathcal{B}}\rho
\nonumber \\
&=
\tr_{\mathcal{B}}\dyad{\psi} 
\nonumber \\
&=
\sum_{a,b,c}s^{\,}_{a}s^{\,}_{b}
\dyad{a^{\,}_{\mathcal{A}}}{b^{\,}_{\mathcal{A}}}
\braket{c^{\,}_{\mathcal{B}}}{b^{\,}_{\mathcal{B}}}
\braket{a^{\,}_{\mathcal{B}}}{c^{\,}_{\mathcal{B}}} \\
&=
\sum^{\,}_{c}s^{2}_{c}\dyad{c^{\,}_{\mathcal{A}}}.
\end{align}
\end{subequations}
The entanglement entropy of any of the subsystems
$\mathcal{A}$ or $\mathcal{B}$ can be defined as the
von Neumann entropy of the corresponding reduced
density matrix
\footnote{Note that $S^{\,}_{\mathcal{A}}=
S^{\,}_{\mathcal{B}}$ with
$\rho^{\,}_{\mathcal{B}}=
\tr^{\,}_{\mathcal{A}}\dyad{\psi}$.}
\begin{subequations}
\begin{align}
S^{\,}_{\mathcal{A}}
&=
-\tr \rho^{\,}_{\mathcal{A}}
\ln\rho^{\,}_{\mathcal{A}}
\nonumber \\
&=
-\tr \sum_{a}s^{2}_{a}
\dyad{a^{\,}_{\mathcal{A}}}
\ln \sum_{b}s^{2}_{b}
\dyad{b^{\,}_{\mathcal{A}}}
\nonumber \\
&=
-\tr \sum_{a}s^{2}_{a}
\dyad{a^{\,}_{\mathcal{A}}}
\ln s^{2}_{a}
\nonumber \\
&=
-\sum_{c}s^{2}_{c}
\ln s^{2}_{c}.
\end{align}
\label{eq:entropy}
\end{subequations}
For a normalized state, the Schmidt coefficients satisfy
\begin{equation}
\sum_{a}s^{2}_{a}=1
\end{equation}	
so that the eigenvalues $s^{2}_{a}$ of the reduced
density matrix form a normalized probability distribution.

\subsubsection{Brief description of DMRG.}

Having introduced the basic notions of the MPS formulation,
the DMRG algorithm can now be described.
It is a simple iterative ground state search when
expressed into a MPS formalism.
The finite-size DMRG in its $2$-site-update version
amounts to perform iteratively an optimization scheme
of two neighboring tensors of the MPS by minimizing the energy.
As an example, an optimization of a bond $(\ell, \ell+1)$
of a MPS in a mixed-canonical form \eqref{eq:mixed-canonical},
is done as follows.
Isolate the bond and contract the tensors
$A^{[\ell]}$, $S^{[\ell]}$ and $B^{[\ell+1]}$ as
\begin{equation}
\begin{minipage}{2.1cm}
\includegraphics[scale=0.7]{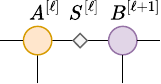}
\end{minipage}
=\ \,
\begin{minipage}{1.8cm}
\includegraphics[scale=0.7]{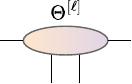}
\end{minipage}
\label{eq:Theta}
\end{equation}
Then to minimize the energy
\begin{equation}
E(\psi)=
\frac{\ev{\mathcal{H}}{\psi}}{\braket{\psi}}
\end{equation}
of a Hamiltonian $\mathcal{H}$,
introduce a Lagrange multiplier $\omega$ to minimize
the quantity
$\ev{\mathcal{H}}{\psi}-\omega\braket{\psi}$ as
\begin{equation}
\pdv{\bigl(\Theta^{[\ell]}\bigr)^{\dag}}
\bigl[\ev{\mathcal{H}}{\psi}-\omega\braket{\psi}\bigr]
=0.	
\end{equation}
The variational optimization of the tensor \eqref{eq:Theta}
translates into solving
\footnote{This simple equation is due to
the mixed-canonical form of the MPS.}
\begin{equation}
\mathcal{H}^{[\ell]}_\mathrm{eff}\,
\Theta^{[\ell]}=
\omega\,\Theta^{[\ell]},
\end{equation}
with
\begin{subequations}
\label{eq:Heff}
\begin{align}
\mathcal{H}^{[\ell]}_\mathrm{eff}
&=
\begin{minipage}{6cm}
\includegraphics[scale=0.7]{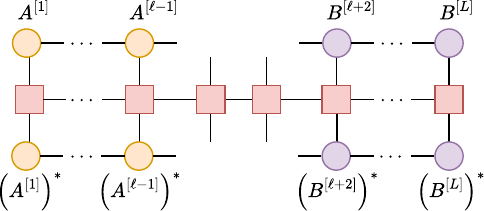}
\end{minipage}
\nonumber \\
&=
\begin{minipage}{3.5cm}
\vspace{10pt}
\includegraphics[scale=0.7]{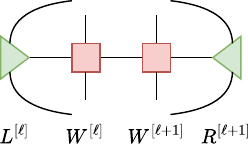}
\end{minipage}
\end{align}
\end{subequations}
and $\Theta^{[\ell]}$ defined in \eqref{eq:Theta}.
To solve it we use, for instance, the Lanczos method
since the effective Hamiltonian is Hermitian,
or a more general algorithm like an implicitly
restarted Arnoldi iteration if some low-lying excited energies
are wanted, to find an updated tensor
$\overline{\Theta}^{[\ell]}$
corresponding to the ground state of the effective
Hamiltonian $\mathcal{H}^{[\ell]}_\mathrm{eff}$.
It is important to use such methods instead of fully
diagonalizing the effective Hamiltonian
$\mathcal{H}^{[\ell]}_\mathrm{eff}$
because it typically is a
$(\chi^{2}\,d^{2})\times(\chi^{2}\,d^{2})$ matrix.
The quantity $\chi$ is the bond dimension, introduced in the next paragraph.
Then we perform a SVD on the reshaped version of
$\overline{\Theta}^{[\ell]}$ obtained by
\begin{equation}
\begin{minipage}{1.75cm}
\vspace{9.5pt}
\includegraphics[scale=0.7]{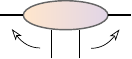}
\end{minipage}
=\ \,
\begin{minipage}{1.8cm}
\vspace{0.5pt}
\includegraphics[scale=0.7]{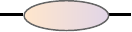}
\end{minipage}
\end{equation}
so that it is now a matrix expressed as
\begin{equation}
\overline{\Theta}^{[\ell]}=
\overline{A}^{[\ell]}\,
\overline{S}^{[\ell]}\,
\overline{B}^{[\ell+1]}.
\end{equation}

One of the most important features of DMRG and
tensor-network methods is the truncation step.
Exploiting the Eckart-Young theorem,
one can approximate very well the matrix
$\overline{\Theta}^{[\ell]}$ of rank $r$ by
$\overline{\Theta}^{[\ell]}_{\mathrm{cut}}$ of rank $r'<r$,
cutting off the lowest singular values of
$\overline{\Theta}^{[\ell]}$ under a precision threshold,
or simply fixing $r'=\chi$,
the usual notation for the maximum bond dimension
allowed for the MPS.
This approximation works well for gapped systems with
short-range correlations, as the singular values of
$\overline{\Theta}^{[\ell]}_\mathrm{cut}$
typically decay rapidly.
For critical systems, they generally decay more slowly,
and larger values of $\chi$ are therefore required to
reach comparable accuracy.
It allows to restrict considerably the dimension of the
Hilbert space, getting rid of the less relevant
states whose weights are typically exponentially small.

The simple DMRG algorithm is performed by the successive
optimization of each bond
\begin{equation}
(1,2),\ (2,3),\ \cdots,\ (L-1,L),
\end{equation}
from the left-most to the
right-most one, and then going back to the left-most one
\begin{equation}
(L-1,L),\ (L-2,L-1),\ \cdots,\ (1,2).
\end{equation}
This is called a sweep.
The sweeps are done from an initial state,
typically a product state with $\chi=1$,
until convergence is reached.
Usual convergence criteria are fixed by 
(i) the variance of the ground-state energy under
some precision threshold,
(ii) a sufficiently small change in the energy
between successive sweeps, and
(iii) a sufficiently small change in the
entanglement entropy or the singular values.

\subsubsection{The extension to iDMRG.}

To extend this method to iDMRG,
one follows \cite{hauschild2018}.
In iDMRG, one optimizes an MPS unit cell that is repeated
indefinitely along the chain.
The tensors inside the unit cell are updated using effective
left and right environments that represent the two infinite half-chains.
During the iterative optimization, these environments are
updated together with the unit-cell tensors.
At convergence, both environments reach fixed points,
and the resulting iMPS is invariant under translations
by one unit cell, up to an MPS gauge transformation.

We denote by $L$ the number of sites in the repeated
iDMRG unit cell, reserving $2N$ for the even
number of sites of the chain in DMRG.

\subsection{The transfer matrix}
\label{sub:the_transfer_matrix}

We now extract the transfer matrix of the iMPS, and
describe how it serves to compute the correlation functions.

We assume the ground state $\ket{\psi}$ of the system
with a unit cell of size $L$
to be in a right-canonical form
\begin{equation}
\ket{\psi}=
\begin{minipage}{3.5cm}
\vspace{9pt}
\includegraphics[scale=0.7]{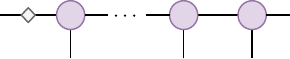}
\end{minipage}
\end{equation}
We define the transfer matrix $T$
of the state $\ket{\psi}$ as
a rank-$4$ tensor
\begin{subequations}
\begin{align}
T
&=
\begin{minipage}{4cm}
\includegraphics[scale=0.7]{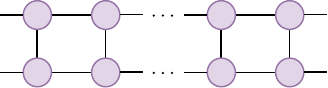}
\end{minipage}
\label{eq:Traw}
\\
&=
\begin{minipage}{2cm}
\includegraphics[scale=0.7]{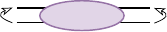}
\end{minipage}
\nonumber \\
&=
\begin{minipage}{1.7cm}
\includegraphics[scale=0.7]{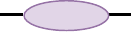}
\end{minipage}
\label{eq:T matrix}
\end{align}
\end{subequations}
reshaped into a non-Hermitian $\chi^{2}\times \chi^{2}$ matrix.

For large $\chi$, explicitly constructing and
storing this $\chi^2\times\chi^2$ matrix $T$ in
Eq.~\eqref{eq:T matrix} is computationally expensive.
In practice, only a small number of its leading eigenvalues
and eigenvectors are required, and the transfer matrix is
therefore applied implicitly using iterative eigensolvers.

We define the left and right eigenvectors
$\bra*{\psi^{\mathrm{L}}_{i}}$ and
$\ket*{\psi^{\mathrm{R}}_{i}}$, respectively,
associated to the eigenvalue $t^{\,}_{i}$ of $T$.
The distinction between the left and right eigenvectors
is necessary since the matrix $T$ is not symmetric.
Therefore, we write
\begin{equation}
T=
\sum_{i=1}^{\chi^{2}}\,
t^{\,}_{i}\,
\dyad{\psi^{\mathrm{R}}_{i}}
{\psi^{\mathrm{L}}_{i}},\qquad
1\eqqcolon t^{\,}_{1}\geq|t^{\,}_{2}|\geq\cdots.
\label{eq:transfer matrix diagonalized}
\end{equation}

We want to understand the relation between the eigenvalues $t^{\,}_{i}$
of the transfer matrix and an arbitrary two-point function
\begin{equation}
\ev*{\widehat{O}^{\,}_{j}\widehat{O}^{\,}_{j'}}
\coloneqq
\ev{\widehat{O}^{\,}_{j}\widehat{O}^{\,}_{j'}}{\psi},\qquad
j,j'\in\mathbb{Z},
\label{eq:def two-point function O}
\end{equation}
where the Hermitian operator $\widehat{O}^{\,}_{j}$ acts on the site $j$
understood on the infinite chain.

By definition $t^{\,}_{1}=1$ and
we assume $t^{\,}_{2}\coloneqq |t^{\,}_{2}|\,e^{\ii \vartheta}\in\mathbb{C}$.
Then we define the correlation length
\begin{equation}
\xi\coloneqq
-\frac{L}{\ln |t^{\,}_{2}|}
\label{eq:xibetter} 
\end{equation}
of the system and compute
\begin{align}
t^{\,}_{2}
&=
e^{\log t^{\,}_{2}}
\nonumber \\
&=
e^{\log \bigl(|t^{\,}_{2}|\,e^{\ii\vartheta}\bigr)}
\nonumber \\
&=
\exp\bigl[\ln |t_2|+\ii(\vartheta + 2\pi m)\bigr]
\nonumber \\
&=
\exp{L\left[\frac{\ln |t_2|}{L}+
\ii\left(\frac{\vartheta}{L}+
\frac{2\pi m}{L}\right)\right]}
\nonumber \\
&\eqqcolon
\exp{L\left[\frac{-1}{\xi}+
\ii\left(q+\frac{2\pi m}{L}\right)\right]},\quad
m\in\mathbb{Z}.
\end{align}
Observe that
\begin{subequations}
\begin{align}
q
&\equiv
q+\frac{2\pi m}{L}\mod\frac{2\pi}{L}
\nonumber \\
&\equiv
\frac{1}{L}\Im\log t^{\,}_{2}\mod\frac{2\pi}{L}.
\end{align}
\label{eq:qmod}
\end{subequations}
This definition is a bit too restrictive, as
$t^{\,}_{2}\in\mathbb{R}$
does not by itself imply the absence of
oscillations in every correlation function.
Therefore, define $q^{\,}_{i}$ for each eigenvalue
$t^{\,}_{i}$ with $i=2,\dotsc,\chi^{2}$ of the transfer matrix
\begin{equation}
q^{\,}_{i}\equiv
\frac{1}{L}\Im\log t^{\,}_{i}\mod\frac{2\pi}{L}.
\label{eq:qimod}
\end{equation}
Similarly, we can define a set of correlation lengths
\begin{equation}
\xi^{\,}_{i}\coloneqq
-\frac{L}{\ln|t^{\,}_{i}|},
\label{eq:xi for t_i}
\end{equation}
for $i=2,\dotsc,\chi^{2}$.

Among the set of $q^{\,}_{i}$ with $i=1,\dotsc,\chi^{2}$,
one wavevector that we label $q^{\,}_{\widehat{O}}$
is associated to the dominant oscillation of
the two-point function \eqref{eq:def two-point function O}.
We define the index $\iota=2,\dotsc,\chi^{2}$ such that
\begin{equation}
q^{\,}_{\iota}=
q^{\,}_{\widehat{O}},\qquad
\xi^{\,}_{\iota}=
\xi^{\,}_{\widehat{O}},
\end{equation}
The wavevector $q^{\,}_{\widehat{O}}$
and associated correlation length $\xi^{\,}_{\widehat{O}}$
enter the Ornstein-Zernicke (OZ) form of
the connected two-point function in a gapped system \cite{Ornstein1914}
\begin{equation}
\ev*{\widehat{O}^{\,}_{s^{\,}_{0}}\widehat{O}^{\,}_{s^{\,}_{0}+s}}
^{\,}_{\mathrm{c}}\sim
\exp(\!-\frac{s}{\xi^{\,}_{\widehat{O}}})\,
\cos\Bigl(q^{\,}_{\widehat{O}}\,s+
\phi^{\,}_{\widehat{O}}\Bigr),
\label{eq:OZ form correlation}
\end{equation}
where $\phi^{\,}_{\widehat{O}}$ is some phase,
$s^{\,}_{0}=1,\dotsc,L$ and $s\in\mathbb{N}$.
More generally, a finite-bond-dimension MPS correlation
function is a sum of such exponentially decaying contributions.

The definition \eqref{eq:qimod}, however, is not sufficient to
distinguish to which correlation $q^{\,}_{i}$ contributes,
and a confusion arises from the modulo.
For instance, in the case of a SU$(2)$-invariant chain,
spin-spin correlation functions in any of the
$x,y,z$-direction are identical.
To remedy this problem, one can introduce the overlaps
\begin{subequations}
\begin{align}
\braket*{\psi^{\mathrm{L}}_{i}}{\widehat{O}^{\,}_{x}}
&=
\begin{minipage}{6cm}
\includegraphics[scale=0.7]{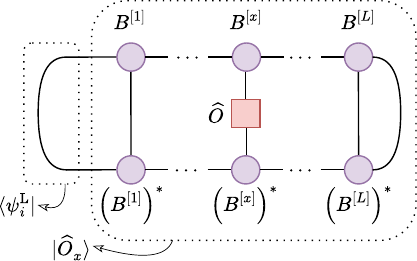}
\end{minipage}
\nonumber \\
&=
\begin{minipage}{4cm}
\includegraphics[scale=0.7]{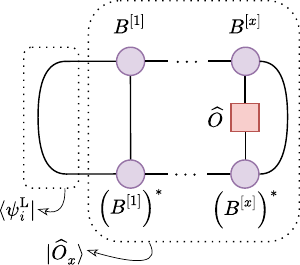}
\end{minipage}
\\
\braket*{\widehat{O}^{\,}_{x}}{\psi^{\mathrm{R}}_{i}}
&=
\begin{minipage}{6cm}
\includegraphics[scale=0.7]{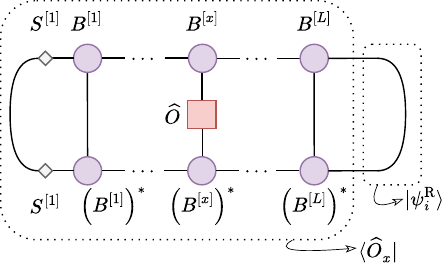}
\end{minipage}
\nonumber \\
&=
\begin{minipage}{4cm}
\includegraphics[scale=0.7]{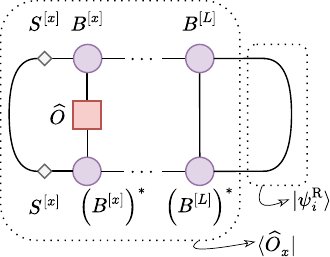}
\end{minipage}
\end{align}
\end{subequations}
where $x=1,\dotsc,L$.
Hence, to verify if a $q^{\,}_{i}$
is associated to a correlation
$\ev*{\widehat{O}^{\,}_{1}\widehat{O}^{\,}_{rL+x}}$,
with $r\in\mathbb N$
and $x=1,\dotsc,L$, compute the weight
\begin{equation}
w^{\,}_{i}(\widehat{O}^{\,}_{1},\widehat{O}^{\,}_{x})=
\braket*{\widehat{O}^{\,}_{1}}{\psi^{\mathrm{R}}_{i}}\!
\braket*{\psi^{\mathrm{L}}_{i}}{\widehat{O}^{\,}_{x}}.
\label{eq:Coverlap}
\end{equation}
If its value is non-zero (or in practice above a chosen
threshold), then there is a non-negligible contribution
from $q^{\,}_{i}$.
This is justified from Eq.~\eqref{eq:transfer matrix diagonalized} as
\begin{align}
\ev*{\widehat{O}^{\,}_{1}\,\widehat{O}^{\,}_{rL+x}}=&\,
\ev*{\widehat{O}^{\,}_{1}\,T^{r}\,\widehat{O}^{\,}_{x}}
\nonumber\\
=&\,
w^{\,}_{1}(\widehat{O}^{\,}_{1},\widehat{O}^{\,}_{x})+
\sum_{k=2}^{\chi^{2}}
t^{r}_{k}\,
w^{\,}_{k}(\widehat{O}^{\,}_{1},\widehat{O}^{\,}_{x}).
\label{eq:correlations from transfer + weights}
\end{align}

We extract two pieces of information from
Eq.~\eqref{eq:correlations from transfer + weights}.
On the one hand, the contribution from the leading eigenvalue is
\begin{equation}
w^{\,}_{1}(\widehat{O}^{\,}_{1},\widehat{O}^{\,}_{x})
=
\ev*{\widehat{O}^{\,}_{1}}\!\ev*{\widehat{O}^{\,}_{x}},
\end{equation}
and therefore gives the disconnected part of the two-point function.
For equivalent sites and a Hermitian operator $\widehat{O}$,
a nonzero limit
\begin{equation}
\lim_{r\to\infty}
\ev*{\widehat{O}^{\,}_{1}\,\widehat{O}^{\,}_{rL+1}}
=
|\!\ev*{\widehat{O}}\!|^{2}
\end{equation}
is consistent with long-range order.
The connected correlation function is obtained
by subtracting this leading contribution.
On the other hand, the eigenvalue $t^{\,}_{k}$
that is the largest in absolute value
among those with a nonzero weight
$w^{\,}_{i}(\widehat{O}^{\,}_{1},\widehat{O}^{\,}_{x})$
controls the decay.
Consequently, the dominant correlation length and wavevector
may depend on the operator.
In the case where
\begin{equation}
|t^{\,}_{k}|= 
|t^{\,}_{k'}|,\qquad
k\neq k',
\end{equation}
e.g., when $t^{\,}_{k}$ and $t^{\,}_{k'}$ have
equal real parts but opposite imaginary parts,
then we have to compute
$w^{\,}_{k}(\widehat{O}^{\,}_{1},\widehat{O}^{\,}_{x})$ and
$w^{\,}_{k'}(\widehat{O}^{\,}_{1},\widehat{O}^{\,}_{x})$.
Moreover, two eigenvalues
$|t^{\,}_{k}|\simeq|t^{\,}_{k'}|$, $k\neq k'$,
can have very close absolute values while associated to
very different
$w^{\,}_{k}(\widehat{O}^{\,}_{1},\widehat{O}^{\,}_{x})$ and
$w^{\,}_{k'}(\widehat{O}^{\,}_{1},\widehat{O}^{\,}_{x})$.
Therefore, to analyze the correlation functions,
we compute the wave vector $q^{\,}_{k}$ from the
eigenvalue $t^{\,}_{k}$ that maximizes the product
\cite{Herviou2025}
\begin{equation}
|t^{\,}_{k}|^{r}\,
|w^{\,}_{k}(\widehat{O}^{\,}_{1},\widehat{O}^{\,}_{x})|,
\qquad
k\geq 2.
\label{eq:weight eigenvalue transfer matrix}
\end{equation}

When the site dependence of the overlaps is dominated by a
single momentum component, the physical wavevector modulo
$2\pi$ can also be extracted from the phase difference
between overlaps on neighboring sites.
With our phase convention, we define
$q^{\,}_{\widehat{O}}$ as
\begin{subequations}
\label{eq:qratio} 
\begin{equation}
q^{\,}_{\widehat{O}}=
\Im\log\frac{\braket*{\widehat{O}^{\,}_{x}}
{\psi^{\mathrm{R}}_{\iota}}}
{\braket*{\widehat{O}^{\,}_{x+1}}
{\psi^{\mathrm{R}}_{\iota}}},
\end{equation}
where
\begin{equation}
\iota=
\argmax_{i\geq 2}
|\!\braket*{\widehat{O}^{\,}_{x}}
{\psi^{\mathrm{R}}_{i}}\!\braket*{\psi^{\mathrm{L}}_{i}}
{\widehat{O}^{\,}_{x}}\!|.
\end{equation}
\end{subequations}
This definition of the wavevector allows to get rid of
the confusing modulo in $q^{\,}_{i}$.
In practice, we use Eq.~\eqref{eq:qratio} when possible
to find the wavevectors $\bmod 2\pi$.

\begin{figure}
\centering
\includegraphics[width=1\columnwidth]{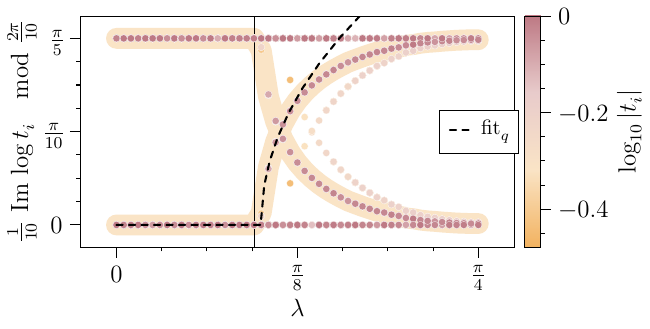}
\caption{
Dependence on $\lambda$ of the wavevector of oscillations
$q^{\,}_{i}$ defined in
Eq.~\eqref{eq:qimod} for the $50$ largest eigenvalues $t^{\,}_{i}$,
$i=1,\ldots,50$ of the transfer matrix of the ground state of
Hamiltonian \eqref{eq:def H U1} obtained with iDMRG ($\chi=256$).
We impose a unit cell of $L=10$ sites.
The color of the points corresponds to the $\log^{\,}_{10}$
of the norm of the $t^{\,}_{i}$.
The light yellow area highlights the dominant
non-trivial wavevector $q^{\,}_{i}$.
The black line indicates the location of the tricritical point
\eqref{eq:tricritical gFMz lambda}.
We have used
$\mathrm{fit}^{\,}_{q}=
(1.50\pm0.05)\times
|\lambda-\lambda^{\,}_{0}|^{0.49\pm0.01}\times
\Theta(\lambda-\lambda^{\,}_{0})$,
where $\lambda^{\,}_{0}=0.3135\pm1\cdot10^{-4}$,
to fit $q^{\,}_{i}$ in the vicinity of
\eqref{eq:tricritical gFMz lambda}.
}
\label{fig:tricritical gFMz Disorder Point}
\end{figure}

\subsection{Example of the coincidence of a disorder point and a
Lifshitz transition separating gapless phases}
\label{sec:disorder point}

In order to investigate the nature of the tricritical point
\eqref{eq:tricritical gFMz lambda}, we diagonalize the transfer matrix
(see Sec.~\ref{sub:the_transfer_matrix})
of the ground state of Hamiltonian \eqref{eq:def H U1}
for $\lambda\in[0,\pi/4]$ obtained using iDMRG.
We then obtain the wavevector associated to each eigenvalue of the
transfer matrix using Eq.~\eqref{eq:qimod}.
Because the transfer matrix is time-reversal symmetric,
its eigenvalues come in complex-conjugate pairs \cite{Herviou2025}.
Hence, the values $q^{\,}_{i}$ for $i=1,\ldots,\chi^{2}$,
defined in Eq.~\eqref{eq:qimod}, can be split into
\begin{equation}
q^{+}_{\iota},\qquad
q^{-}_{\iota}\coloneqq \frac{2\pi}{10}-q^{+}_{\iota},\qquad
\iota=1,\ldots,\frac{\chi^{2}}{2}.
\label{eq:qmod split 2N=10}
\end{equation}
The resulting wavevectors for the $50$ transfer-matrix eigenvalues
of largest modulus are displayed in
Fig.~\ref{fig:tricritical gFMz Disorder Point}.

We now show that the results in
Fig.~\ref{fig:tricritical gFMz Disorder Point}, together with the
correlation-function analysis of
Sec.~\ref{ssub:the_texorpdfstring_mathrm_gfm}, provide evidence that
the tricritical point \eqref{eq:tricritical gFMz lambda}
has the characteristic properties of a disorder point of the first kind.
Here, by a disorder point of the first kind we mean a point that
\begin{itemize}
\item[(I)]
separates a commensurate (C) from an incommensurate (IC) phase,
\item[(II)]
displays either a singularity of a correlation length
or a singularity of an incommensurate wavevector.
\end{itemize}

We first explain how property (I) is diagnosed from the transfer matrix.
Since the iMPS unit cell contains $L=10$ sites,
Eq.~\eqref{eq:qimod} determines every $q^{\,}_{i}$ only modulo
$2\pi/10$.  The vertical axis of
Fig.~\ref{fig:tricritical gFMz Disorder Point} therefore represents a
folded wavevector. Equation~\eqref{eq:qmod split 2N=10} explains the
reflection symmetry of the figure about $q=\pi/10$. The two members of
a complex-conjugate pair satisfy
$q^{-}_{\iota}=2\pi/10-q^{+}_{\iota}$.
Consequently, the branches at $q=0$ and $q=\pi/5$ correspond to the two
ends of the same folded interval.

The connection between the phase of a transfer-matrix eigenvalue and
an oscillation in a correlation function follows from
Eq.~\eqref{eq:correlations from transfer + weights}.
Writing
\begin{equation}
t^{\,}_{i}=|t^{\,}_{i}|\,e^{\ii L q^{\,}_{i}},
\end{equation}
a complex-conjugate pair contributes an oscillatory term to a
two-point function whenever the corresponding eigenvectors have a
nonvanishing overlap with the microscopic operator.
Thus, a continuously varying phase of an operator-relevant
complex-conjugate pair signals an incommensurate contribution to the
corresponding correlation function.

In Sec.~\ref{ssub:the_texorpdfstring_mathrm_gfm},
we have seen that the $\mathrm{gFM}^{\,}_{z}$ phase
possesses nontrivial oscillations, with wavevector
$q^{z,\mathrm{c}}$, in the
$C^{z,\mathrm{c}}_{s^{\,}_{0},s^{\,}_{0}+s}$ two-point function.
In order to find the transfer-matrix wavevector that dominates these
oscillations, one would preferably use Eq.~\eqref{eq:qratio}.
In our case, this method suffers from important numerical imprecision.
Instead, in the region
$\lambda>\lambda^{\,}_{\mathrm{gFM}^{\,}_{z}}$,
we use Eq.~\eqref{eq:qimod} and identify the index $\iota^{\star}$
whose transfer-matrix mode maximizes the quantity
\eqref{eq:weight eigenvalue transfer matrix}.
The corresponding values $q^{+}_{\iota^{\star}}(\lambda)$ and
$q^{-}_{\iota^{\star}}(\lambda)$ are highlighted in light yellow in
Fig.~\ref{fig:tricritical gFMz Disorder Point}.
On the commensurate side, the relevant folded wavevector is pinned at
$q^{+}_{\iota^{\star}}=0$, with its counterpart at
$q^{-}_{\iota^{\star}}=2\pi/10$.
On entering the $\mathrm{gFM}^{\,}_{z}$ phase, the pair departs
continuously from these commensurate values and moves into the interior
of the folded interval. Through
Eq.~\eqref{eq:correlations from transfer + weights}, this departure
corresponds to the appearance of an incommensurate contribution to
$C^{z,\mathrm{c}}_{s^{\,}_{0},s^{\,}_{0}+s}$.
This is the transfer-matrix signature of property (I).

We next test property (II) by studying the onset of the folded
incommensurate wavevector. We use the Ansatz
\begin{subequations}
\label{eq:q zc mod Ansatz}
\begin{equation}
q^{+}_{\iota^{\star}}(\lambda)=
\alpha\times |\lambda-\lambda^{\,}_{0}|^{\sigma}
\times
\Theta(\lambda-\lambda^{\,}_{0}),
\end{equation}
where $\alpha$, $\lambda^{\,}_{0}$, and $\sigma$
are fitting parameters. The fit, denoted by $\mathrm{fit}^{\,}_{q}$ in
Fig.~\ref{fig:tricritical gFMz Disorder Point}, gives
\begin{align}
&\alpha=1.50\pm0.05, \\
&\lambda^{\,}_{0}=0.3135\pm1\cdot10^{-4}, \\
&\sigma=0.49\pm0.01.
\end{align}
\end{subequations}
The exponent $\sigma\approx1/2$ is consistent with a square-root
singularity of the incommensurate wavevector
\cite{schollwock1996,Herviou2025}, and therefore realizes the
wavevector singularity entering property (II).
We note, however, that the fitted onset
$\lambda^{\,}_{0}=0.3135\pm1\cdot10^{-4}$ does not coincide within the
quoted numerical uncertainties with the independently determined
tricritical value \eqref{eq:lambda gFMz}. We therefore interpret the
fit as evidence for a nonanalytic onset of incommensuration in the
vicinity of the tricritical point rather than as an independent precise
determination of its position.

We now explain the relation between
$q^{+}_{\iota^{\star}}(\lambda)$ and the physical wavevector
$q^{z,\mathrm{c}}(\lambda)$. The quantity
$q^{+}_{\iota^{\star}}(\lambda)$ is defined only mod $2\pi/10$,
where $L=10$ is the size of the unit cell chosen here.
Consequently, the physical wavevector $q^{z,\mathrm{c}}(\lambda)$ is
one among the following $20$ quantities
\begin{equation}
q^{+}_{\iota^{\star}},\quad
2\pi-q^{+}_{\iota^{\star}},\quad
\left\{\left.
\frac{2\pi}{10}\,n\pm q^{+}_{\iota^{\star}}\,
\right|
n=1,2,\cdots,9
\right\}.
\label{eq:possible q because of mod}
\end{equation}
We found in Eq.~\eqref{eq:gFMz OBCs HB Fitting Paramters Z} that
$q^{z,\mathrm{c}}(\lambda)$ reaches the minimum value
\cite{Rey2025}
\begin{equation}
q^{z,\mathrm{c}}(\pi/4)=2\pi\times\frac{3}{10}
\end{equation}
at $\lambda=\pi/4$.
Together with
$q^{+}_{\iota^{\star}}(\pi/4)=2\pi/10$ from
Fig.~\ref{fig:tricritical gFMz Disorder Point}, this selects
\begin{equation}
\gamma-q^{+}_{\iota^{\star}}(\lambda),
\qquad
\gamma=2\pi\times\frac{4}{10},
\label{eq:gamma - qi}
\end{equation}
from Eq.~\eqref{eq:possible q because of mod} as the branch that is
consistent with the independently determined value of
$q^{z,\mathrm{c}}(\pi/4)$ on the $\mathrm{gFM}^{\,}_{z}$ side.
We emphasize that this unfolding of the transfer-matrix wavevector uses
information from the real-space correlation function in addition to
Fig.~\ref{fig:tricritical gFMz Disorder Point}.

At the free-fermion point $\lambda=0$, we also know that
\begin{equation}
q^{z,\mathrm{c}}(0)=2\pi\times\frac{5}{10}.
\end{equation}
This value cannot be obtained by continuously extending
Eq.~\eqref{eq:gamma - qi} using the branch visible among the $50$
transfer-matrix eigenvalues shown in
Fig.~\ref{fig:tricritical gFMz Disorder Point}.
One possible explanation is that the transfer-matrix mode associated
with the continuation of $q^{z,\mathrm{c}}(\lambda)$ lies deeper in the
transfer-matrix spectrum on the commensurate side, with
$|t^{\,}_{i}|\ll1$, and is therefore difficult to access accurately in
the present calculation.

Taken together with the independently determined phase boundary and
the continuum analysis below, these transfer-matrix results support the
interpretation of the tricritical point as a disorder point separating
a commensurate critical phase without long-range order from an
incommensurate critical $\mathrm{gFM}^{\,}_{z}$ phase with long-range
order. The characteristic transfer-matrix signature is the formation
of an operator-relevant complex-conjugate pair whose folded wavevector
departs continuously but nonanalytically from its commensurate value.

The same point also admits a Lifshitz description
(see Sec.~\ref{ssub:lifshitz_instability_and_the_tricritical_point}).
In the continuum theory, the coefficient of the quadratic-gradient
term vanishes at this point, so that the leading dispersion becomes
quadratic with dynamical exponent $\mathsf{z}=2$.
On the incommensurate side, the change of sign of the
quadratic-gradient coefficient favors a nonzero average field gradient.
The resulting spatial modulation shifts the wavevector
of the leading algebraically decaying correlations
away from its commensurate value.
This displacement appears continuously but nonanalytically
upon crossing the Lifshitz point.
The disorder-point and Lifshitz-point criteria
therefore describe two complementary aspects of the same transition.
The disorder-point criterion concerns the onset of
incommensurability in the leading connected correlations and
the formation of complex-conjugate transfer-matrix eigenvalues,
whereas the Lifshitz criterion concerns the vanishing of
the quadratic stiffness and the displacement of
the dominant mode in momentum space.
In the present case, the combined numerical and continuum results
support the coincidence of these two criteria with the onset of
long-range order at a tricritical point separating a
commensurate gapless phase from an incommensurate
gapless phase with long-range order.

This mechanism is closely related to a Pokrovsky-Talapov
commensurate-to-incommensurate transition,
for which a quadratic critical mode and
a nonanalytic onset of the incommensurate wavevector
are characteristic \cite{Fitouchi2026}.
The important difference is that the standard
Pokrovsky-Talapov transition separates a commensurate
gapped phase from an incommensurate gapless phase,
whereas the commensurate phase in the present
case is already gapless.
The point identified here should therefore be regarded
as a gapless commensurate-to-incommensurate
Lifshitz transition rather than as a
conventional Pokrovsky-Talapov transition.

Several related phenomena have been reported previously,
but they do not exhibit the complete combination found here.
Lifshitz transitions have been identified in quantum
critical systems and may separate distinct
gapless regimes \cite{chitov2025,beradze2023,feng2022,huang2022}.
These studies, however, do not establish that the
Lifshitz point is also a disorder point, as diagnosed
by the onset of incommensurate asymptotic real-space
correlations and the formation of a complex-conjugate
pair of dominant transfer-matrix eigenvalues.

Coincident disorder and Lifshitz criteria
have also been reported in frustrated spin chains.
In frustrated spin-$1$ models, the disorder line,
defined by the onset of incommensurate real-space correlations,
and the Lifshitz line, defined by the displacement of
the maximum of the static structure factor,
can merge into a single commensurate-to-incommensurate
line inside a dimerized phase \cite{Pixley2014,Chepiga2016}.
In those examples, however,
the merged line characterizes short-range
correlations within a gapped phase and does not represent
a critical point separating two gapless regimes.
Quantum points that are simultaneously tricritical
and Lifshitz have also been reported \cite{Saito2024},
but the simultaneous disorder-point criterion was not
established through the onset of incommensurate asymptotic
correlations or the formation of complex-conjugate
dominant transfer-matrix eigenvalues.

To the best of our knowledge, the point found here is
the first reported example in which a disorder point, a
$\mathsf{z}=2$ Lifshitz point, and a tricritical point
coincide at a transition between commensurate and
incommensurate critical phases.
Its distinguishing features are
(i) the simultaneous onset
of long-range order and of incommensurability
in the leading algebraically decaying correlations,
(ii) the formation of a complex-conjugate pair
among the leading transfer-matrix eigenvalues
governing the connected correlations,
(iii) the vanishing of the quadratic stiffness, and
(iv) the continuous but nonanalytic displacement
of the characteristic wavevector.

\section{Additional evidence for the stability of the $\mathrm{gFM}^{\,}_{z}$ phase}
\label{sec:Additional evidence gFMz}

In Sec.~\ref{ssub:the_texorpdfstring_mathrm_gfm},
we described numerically the properties of the
$\mathrm{gFM}^{\,}_{z}$ phase of Hamiltonians
\eqref{eq:def HA and HB c} and \eqref{eq:def HA and HB d}.
The results presented there establish the existence of critical behavior
and ferromagnetic long-range order along the $z$ axis in spin space
at representative points away from the center
$(\pi/4,\pi/4)$ of the phase.
They do not, however, completely rule out the possibility that the
$\mathrm{gFM}^{\,}_{z}$ phase is restricted to the vertical line
$(\pi/4,\lambda)\in\Omega^{\,}_{\mathrm{Y,B}}$
and the horizontal line
$(\theta,\pi/4)\in\Omega^{\,}_{\mathrm{Y,B}}$.

According to this alternative interpretation,
the extended $\mathrm{gFM}^{\,}_{z}$ regions shown in
Figs.~\ref{fig:phase diagram A} and \ref{fig:phase diagram B}
would correspond to crossover regimes governed by an extremely large,
but finite, physical correlation length.
At length scales shorter than this correlation length,
the system would display the critical properties associated with the
$\mathrm{U}(1)$-symmetric line.
At sufficiently large length scales, however,
a point away from both $\theta=\pi/4$ and $\lambda=\pi/4$
would eventually flow toward one of the neighboring gapped phases.
We refer to this possibility below as the crossover scenario.

In this Appendix, we provide additional numerical evidence against
the crossover scenario.
We focus on representative points satisfying
$\theta\neq\pi/4$ and $\lambda\neq\pi/4$,
at which neither Hamiltonian possesses a microscopic
$\mathrm{U}(1)$ symmetry.
Our analysis combines finite-entanglement scaling,
the scale dependence of the transverse spin correlations,
and the momentum-resolved correlations of the microscopic operators
that break the $\mathrm{U}(1)$ symmetry.
These diagnostics probe complementary consequences of the crossover
scenario.
Taken together, they support the interpretation that the
$\mathrm{gFM}^{\,}_{z}$ regions are genuine extended critical phases.

\subsection{Finite-entanglement scaling}
\label{subsec:Finite entanglement scaling gFMz}

We begin by studying the ground state using iDMRG.
At a finite bond dimension $\chi$,
an MPS approximation to a critical ground state has a finite
effective correlation length $\xi^{\,}_{\chi}$.
This correlation length is not a physical correlation length of the
critical state.
Instead, it is generated by the finite amount of entanglement that
can be represented by an MPS of bond dimension $\chi$.

For a one-dimensional critical ground state described by a CFT,
the finite-entanglement entropy satisfies
Eq.~\eqref{eq:Entropy From Xi}
which is expected to hold in the regime
in which the finite-entanglement correlation length
$\xi^{\,}_{\chi}$ is the dominant infrared cutoff.

The behavior is qualitatively different in a gapped phase.
In that case, increasing $\chi$ eventually allows the MPS to resolve
the finite physical correlation length of the ground state.
Both $\xi^{\,}_{\chi}$ and $S^{\,}_{\chi}$ must then approach finite
values as $\chi$ is increased.
Consequently, the absence or presence of saturation provides a direct
way of distinguishing a critical state from a gapped state,
provided that sufficiently large bond dimensions are reached.

\begin{figure}
\subfigure{\includegraphics[width=0.5\columnwidth]{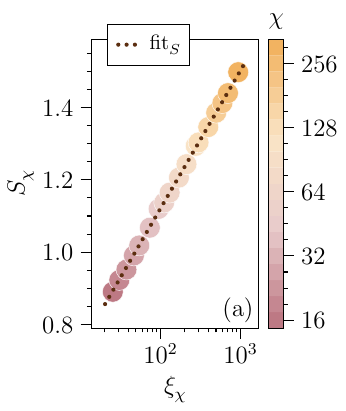}\label{fig:Central Charge gFMz iDMRG a}}%
\subfigure{\includegraphics[width=0.5\columnwidth]{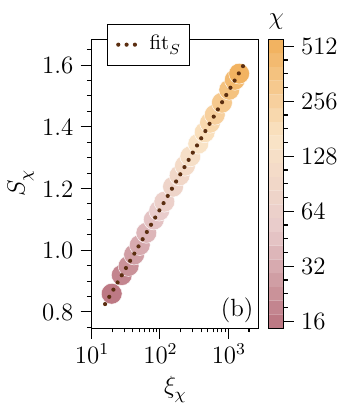}\label{fig:Central Charge gFMz iDMRG b}}%
\caption{
(a)
Dependence of the finite-entanglement entropy
$S^{\,}_{\chi}$ on the MPS correlation length
$\xi^{\,}_{\chi}$ at the point
$(\theta,\lambda)=(\pi/4-\pi/40,\pi/4-\pi/40)$.
The ground states are obtained using iDMRG with several bond dimensions
$\chi$, represented by different colors.
The correlation length $\xi^{\,}_{\chi}$ is defined in
Eq.~\eqref{eq:xibetter} with $L=10$.
The entanglement entropy $S^{\,}_{\chi}$ is evaluated across the first
bond of the repeated unit cell used to approximate the infinite
ground state.
The slope of the linear fit
$\mathrm{fit}^{\,}_{S}$ is related to the central charge through
Eq.~\eqref{eq:Entropy From Xi}, giving
$\mathsf{c}=0.995\pm0.007$.
(b) Same as (a) for Hamiltonian
$\widehat{H}^{(0)}_{\mathrm{II}}$ for which
$\mathsf{c}=0.997\pm0.002$.
}
\label{fig:Central Charge gFMz iDMRG}
\end{figure}

Figure~\ref{fig:Central Charge gFMz iDMRG}
shows $S^{\,}_{\chi}$ as a function of
$\ln\xi^{\,}_{\chi}$ at the representative point
$(\theta,\lambda)=(\pi/4-\pi/40,\pi/4-\pi/40)$,
which lies away from both
$\theta=\pi/4$ and $\lambda=\pi/4$.
This point is therefore not protected by a microscopic
$\mathrm{U}(1)$ symmetry.

For both Hamiltonians
\eqref{eq:def HA and HB c} and \eqref{eq:def HA and HB d},
the numerical results are accurately described by the linear
finite-entanglement relation~\eqref{eq:Entropy From Xi}.
The fitted slopes give
\begin{equation}
\mathsf{c}=
0.995\pm0.007,
\qquad
\mathsf{c}=
0.997\pm0.002,
\end{equation}
for Hamiltonian \eqref{eq:def HA and HB c} and \eqref{eq:def HA and HB d},
respectively.
Both estimates are consistent with
$\mathsf{c}=1$.
Moreover, the correlation length $\xi^{\,}_{\chi}$ continues to
increase over the full range of accessible bond dimensions,
without a visible tendency toward saturation.
The entanglement entropy increases accordingly and continues to follow
Eq.~\eqref{eq:Entropy From Xi}.

This behavior is expected for a Luttinger liquid whose infrared
behavior is regulated by the finite entanglement of the MPS.
It is also consistent with the value
$\mathsf{c}=1$ obtained from the finite-size scaling analysis
presented in Sec.~\ref{ssub:the_texorpdfstring_mathrm_gfm}.

Within the crossover scenario,
the critical relation~\eqref{eq:Entropy From Xi}
could hold only while
\begin{equation}
\xi^{\,}_{\chi}\ll
\xi^{\,}_{\star},
\end{equation}
where $\xi^{\,}_{\star}$ denotes the finite physical crossover length.
Once $\xi^{\,}_{\chi}$ became comparable with or larger than
$\xi^{\,}_{\star}$,
the MPS would resolve the asymptotic gapped regime.
The growth of $\xi^{\,}_{\chi}$ would then stop,
and the entanglement entropy would approach a finite value.
No such departure from the $\mathsf{c}=1$ finite-entanglement scaling
is resolved in Fig.~\ref{fig:Central Charge gFMz iDMRG}.

The results therefore place a lower bound on any hypothetical
crossover length.
If the points studied in
Fig.~\ref{fig:Central Charge gFMz iDMRG}
ultimately flowed toward a gapped phase,
their physical crossover length would have to exceed the largest
finite-entanglement correlation length reached in our calculations.

\subsection{Scale dependence of the transverse exponent}
\label{subsec:Running transverse exponent gFMz}

We next analyze the transverse spin correlation function
$C^{x}_{s^{\,}_{0},s^{\,}_{0}+s}$ using iDMRG.
Inside a critical $\mathrm{gFM}^{\,}_{z}$ phase,
this correlation function is expected to display the algebraic behavior
introduced in Eq.~\eqref{eq:gFMz OBCs algebraic decay},
namely
\begin{equation}
C^{x}_{s^{\,}_{0},s^{\,}_{0}+s}
=
A^{x}_{s}\,
s^{-\eta^{x}},
\end{equation}
where $A^{x}_{s}$ contains the commensurate and incommensurate
oscillations of the correlation function.

To investigate how the apparent exponent evolves with distance,
we define a running exponent
$\eta^{x}_{w^{\,}_{0},w^{\,}_{0}+w}$.
It is obtained by fitting the envelope of
$C^{x}_{s^{\,}_{0},s^{\,}_{0}+s}$
to a power law within a window of size $w$ beginning at
$w^{\,}_{0}$, i.e.,
\begin{equation}
\left|
C^{x}_{s^{\,}_{0},s^{\,}_{0}+s}
\right|
\propto
s^{-\eta^{x}_{w^{\,}_{0},w^{\,}_{0}+w}},
\qquad
s\in
\left\{
w^{\,}_{0},
\ldots,
w^{\,}_{0}+w
\right\}.
\label{eq:Running Exponent}
\end{equation}
This definition allows us to determine whether the algebraic exponent
remains stable as the fitting window is displaced toward larger
separations.

For a critical ground state approximated by an MPS of finite bond
dimension $\chi$,
one expects
\begin{equation}
\eta^{x}_{w^{\,}_{0},w^{\,}_{0}+w}
\simeq
\eta^{x},
\qquad
1
\ll
w^{\,}_{0},
\,
w^{\,}_{0}+w
\ll
\xi^{\,}_{\chi}.
\label{eq:gFMz stability eta plateau}
\end{equation}
When the upper end of the fitting window approaches
$\xi^{\,}_{\chi}$,
the finite-entanglement cutoff modifies the correlation function,
and the running exponent is consequently expected to depart from its
critical value.

\begin{figure}
\subfigure{\includegraphics[width=1\columnwidth]{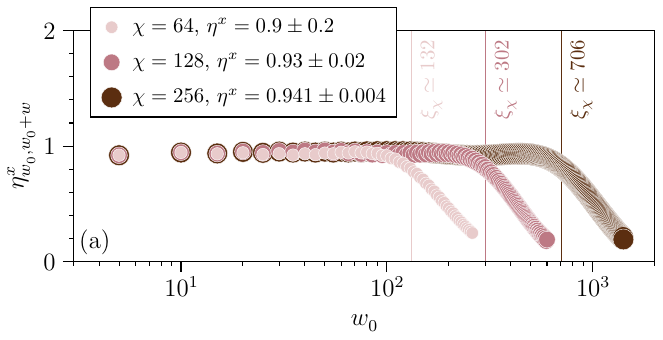}\label{fig:Running Exponent gFMz iDMRG a}}
\subfigure{\includegraphics[width=1\columnwidth]{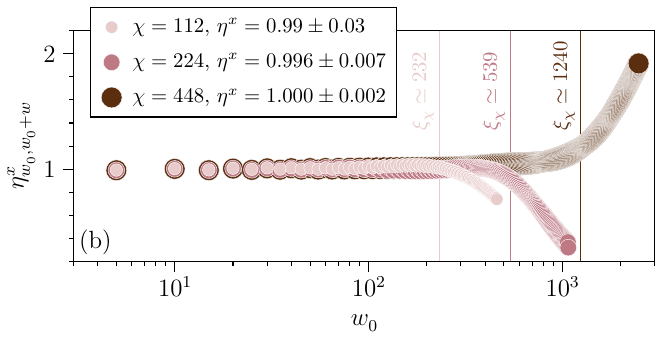}\label{fig:Running Exponent gFMz iDMRG b}}
\caption{
Running exponent
$\eta^{x}_{w^{\,}_{0},w^{\,}_{0}+w}$
defined in Eq.~\eqref{eq:Running Exponent}
as a function of the beginning $w^{\,}_{0}$ of the fitting window.
The window size is fixed to $w=30$.
The ground states of
(a) Hamiltonian \eqref{eq:def HA and HB c} and
(b) Hamiltonian \eqref{eq:def HA and HB d}
are obtained using iDMRG with several bond dimensions
$\chi$ at the point
$(\theta,\lambda)=(\pi/4-\pi/40,\pi/4-\pi/40)$.
The MPS correlation length $\xi^{\,}_{\chi}$,
defined in Eq.~\eqref{eq:xibetter} with $L=10$,
is represented by a vertical line whose color corresponds to the
associated bond dimension.
The estimate of $\eta^{x}$ is obtained from
Eq.~\eqref{eq:Running Exponent} with
$w^{\,}_{0}=8$ and $w=\xi^{\,}_{\chi}/10$.
}
\label{fig:Running Exponent gFMz iDMRG}
\end{figure}

Figure~\ref{fig:Running Exponent gFMz iDMRG}
displays
$\eta^{x}_{w^{\,}_{0},w^{\,}_{0}+w}$
for several values of $\chi$ and for both Hamiltonians at
$(\theta,\lambda)=(\pi/4-\pi/40,\pi/4-\pi/40)$.
For each bond dimension,
the running exponent develops an approximately constant regime at
separations smaller than $\xi^{\,}_{\chi}$.
As $\chi$ is increased,
this regime extends toward larger separations.
At the same time,
the scale at which
$\eta^{x}_{w^{\,}_{0},w^{\,}_{0}+w}$
departs from its plateau moves outward together with
$\xi^{\,}_{\chi}$.
The value of the plateau remains approximately independent of $\chi$.

The residual oscillations visible in
Fig.~\ref{fig:Running Exponent gFMz iDMRG}
originate from subleading incommensurate contributions to
$C^{x}_{s^{\,}_{0},s^{\,}_{0}+s}$.
They appear as an aliasing effect because the running fits are performed
with a fixed window size $w=30$ and with starting points
$w^{\,}_{0}=5n$, where $n\in\mathbb{N}$.
These oscillations do not modify the systematic displacement of the
finite-entanglement distortion toward larger separations as $\chi$ is
increased.

This evolution is incompatible with a crossover for which the physical
correlation length has already been resolved.
If the asymptotic state were a
$\mathrm{Neel}^{\,}_{x}$ state,
the staggered transverse correlation function would eventually satisfy
\begin{equation}
(-1)^{s}
C^{x}_{s^{\,}_{0},s^{\,}_{0}+s}
\longrightarrow
\left(
m^{x}_{\mathrm{sta}}
\right)^{2}
>
0,
\qquad
s\gg1.
\label{eq:gFMz crossover Neel order}
\end{equation}
Its effective algebraic exponent would then approach zero beyond a
physical length scale that becomes independent of $\chi$ once the MPS
has converged.
It is important to notice that
$m^{x}_{\mathrm{sta}}<10^{-6}$
at any point studied in the $\mathrm{gFM}^{\,}_{z}$ phases
in Figs.~\ref{fig:phase diagram A} and \ref{fig:phase diagram B}.

The data show the opposite behavior.
The apparent long-distance distortion of the transverse correlation
function is pushed to increasingly large separations as $\chi$ is
increased.
Its position is controlled by the finite-entanglement correlation
length $\xi^{\,}_{\chi}$ rather than by a fixed physical scale.
The behavior in
Fig.~\ref{fig:Running Exponent gFMz iDMRG}
is therefore naturally interpreted as the finite-entanglement cutoff of
a critical correlation function,
rather than as the onset of physical
$\mathrm{Neel}^{\,}_{x}$ long-range order.
An analogous conclusion applies to a possible crossover toward the
$\mathrm{Neel}^{\,}_{y}$ phase after exchanging the $x$ and $y$
spin directions.

\begin{figure*}
\subfigure{\includegraphics[width=0.221\textwidth]{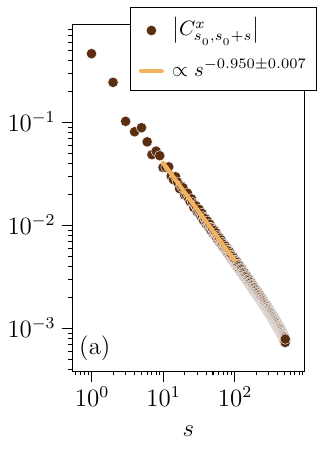}\label{fig:Decay Correlations A gFMz DMRG a}}%
\subfigure{\label{fig:Decay Correlations A gFMz DMRG b}}%
\subfigure{\includegraphics[width=0.389\textwidth]{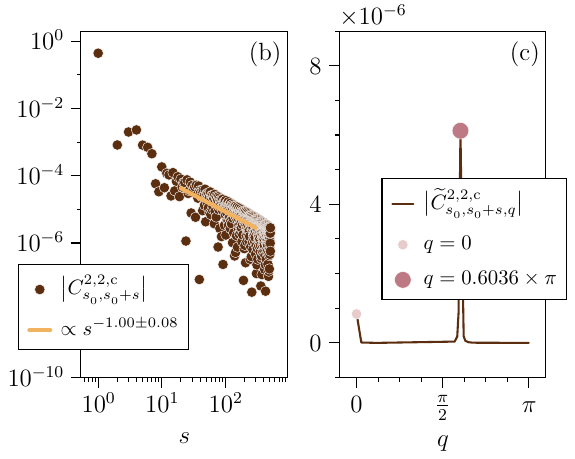}\label{fig:Decay Correlations A gFMz DMRG c}}%
\subfigure{\label{fig:Decay Correlations A gFMz DMRG d}}%
\subfigure{
\includegraphics[width=0.389\textwidth]
{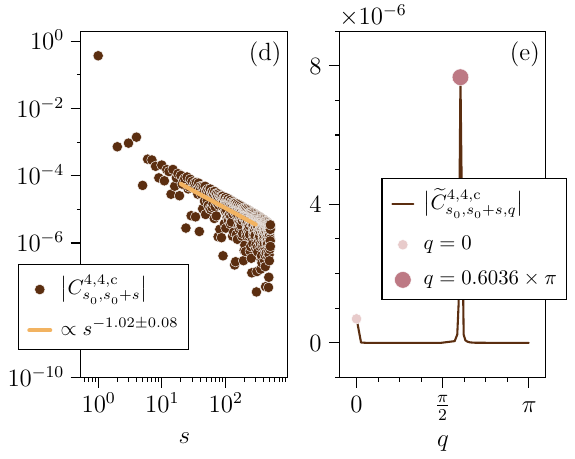}
\label{fig:Decay Correlations A gFMz DMRG e}
}
\subfigure{\label{fig:Decay Correlations A gFMz DMRG f}}%
\caption{
Correlations and momentum-resolved anisotropy diagnostics for
Hamiltonian \eqref{eq:def HA and HB c} at
$(\theta,\lambda)=(\pi/4-\pi/40,\pi/4-\pi/40)$.
The ground state is obtained using DMRG with $2N=1024$,
$\chi=512$, and OBCs.
The reference position is
$s^{\,}_{0}=2N/2-2N/4$.
(a)
Absolute value of the transverse two-point function
$C^{x}_{s^{\,}_{0},s^{\,}_{0}+s}$,
defined in Eq.~\eqref{eq:Calpha},
together with a fit to the algebraic form
\eqref{eq:gFMz OBCs algebraic decay}.
(b)
Absolute value of the connected anisotropy correlation
$C^{2,2,\mathrm{c}}_{s^{\,}_{0},s^{\,}_{0}+s}$,
defined in Eq.~\eqref{eq:gFMz Stability Cab},
and its structure factor, defined in
Eq.~\eqref{eq:gFMz stability SF}.
The fit separates the uniform and finite-momentum algebraic
contributions according to
Eq.~\eqref{eq:gFMz stability anisotropy fit}.
The momenta of the dominant peaks of the structure factor are indicated.
(c)
Same as (b) for the connected dressed-anisotropy correlation
$C^{4,4,\mathrm{c}}_{s^{\,}_{0},s^{\,}_{0}+s}$.
}
\label{fig:Decay Correlations A gFMz DMRG}
\end{figure*}

\begin{figure*}
\subfigure{\includegraphics[width=0.221\textwidth]{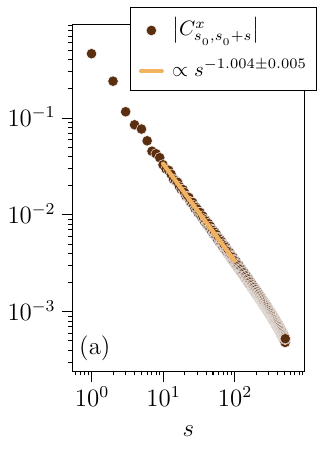}\label{fig:Decay Correlations B gFMz DMRG a}}%
\subfigure{\label{fig:Decay Correlations B gFMz DMRG b}}%
\subfigure{\includegraphics[width=0.389\textwidth]{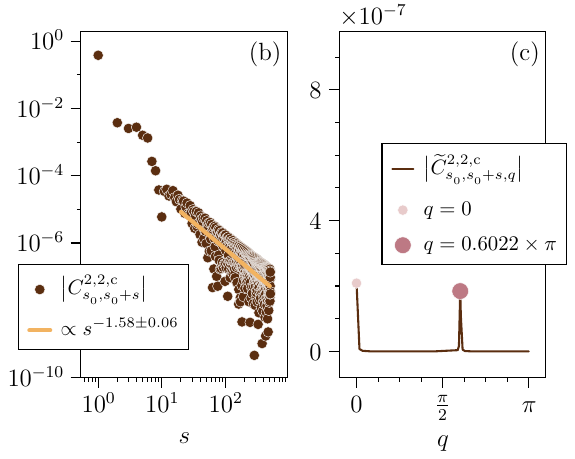}\label{fig:Decay Correlations B gFMz DMRG c}}%
\subfigure{\label{fig:Decay Correlations B gFMz DMRG d}}%
\subfigure{\includegraphics[width=0.389\textwidth]{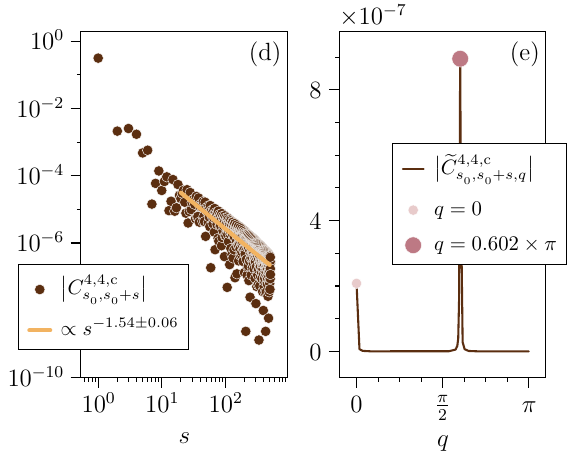}\label{fig:Decay Correlations B gFMz DMRG e}}
\subfigure{\label{fig:Decay Correlations B gFMz DMRG f}}%
\caption{
Correlations and momentum-resolved anisotropy diagnostics for
Hamiltonian \eqref{eq:def HA and HB d} at
$(\theta,\lambda)=(\pi/4-\pi/40,\pi/4-\pi/40)$.
The ground state is obtained using DMRG with $2N=1024$,
$\chi=512$, and OBCs.
The reference position is
$s^{\,}_{0}=2N/2-2N/4$.
(a)
Absolute value of the transverse two-point function
$C^{x}_{s^{\,}_{0},s^{\,}_{0}+s}$,
defined in Eq.~\eqref{eq:Calpha},
together with a fit to the algebraic form
\eqref{eq:gFMz OBCs algebraic decay}.
(b)
Absolute value of the connected anisotropy correlation
$C^{2,2,\mathrm{c}}_{s^{\,}_{0},s^{\,}_{0}+s}$,
defined in Eq.~\eqref{eq:gFMz Stability Cab},
and its structure factor, defined in
Eq.~\eqref{eq:gFMz stability SF}.
The fit separates the uniform and finite-momentum algebraic
contributions according to
Eq.~\eqref{eq:gFMz stability anisotropy fit}.
The momenta of the dominant peaks of the structure factor are indicated.
(c)
Same as (b) for the connected dressed-anisotropy correlation
$C^{4,4,\mathrm{c}}_{s^{\,}_{0},s^{\,}_{0}+s}$.
}
\label{fig:Decay Correlations B gFMz DMRG}
\end{figure*}

\subsection{Momentum content of the anisotropy operators}
\label{subsec:Momentum content anisotropy gFMz}

The preceding subsections provide evidence that the states obtained
away from the $\mathrm{U}(1)$-symmetric lines remain critical on the
largest length scales accessible numerically.
We now examine the operators that break the microscopic
$\mathrm{U}(1)$ symmetry.
This analysis addresses three distinct questions.
First, we determine that their connected correlations decay
algebraically and not exponentially.
Second, we resolve their algebraic decay into uniform and
finite-momentum components.
Third, we compare their decay exponents with the prediction of the
standard bosonization of a magnetized XXZ chain.

These three questions must be considered together.
The decay law determines the scaling dimension of a critical channel,
whereas the structure factor determines the momentum carried by that
channel.
A scaling dimension alone does not establish that a microscopic
perturbation can open a gap,
the corresponding operator must also contribute coherently to the
spatially uniform Hamiltonian.

We introduce the nearest-neighbor anisotropy operator
$\widehat{O}^{\,}_{2,j}$ and its dressed counterpart
$\widehat{O}^{\,}_{4,j}$ as
\begin{subequations}
\label{eq:gFMz Stability Operators}
\begin{align}
\widehat{O}^{\,}_{2,j}\coloneqq&\,
\widehat{X}^{\,}_{j}\widehat{X}^{\,}_{j+1}
-
\widehat{Y}^{\,}_{j}\widehat{Y}^{\,}_{j+1},
\label{eq:gFMz Stability O2}
\\
\widehat{O}^{\,}_{4,j}\coloneqq&\,
\widehat{Z}^{\,}_{j-1}
\left(
\widehat{X}^{\,}_{j}\widehat{X}^{\,}_{j+1}
-
\widehat{Y}^{\,}_{j}\widehat{Y}^{\,}_{j+1}
\right)
\widehat{Z}^{\,}_{j+2}.
\label{eq:gFMz Stability O4}
\end{align}
\end{subequations}
Both operators break the microscopic $\mathrm{U}(1)$ symmetry down to
the same $\mathbb{Z}_{2}$ subgroup.
The operator $\widehat{O}^{\,}_{2,j}$ is the conventional exchange
anisotropy of an XYZ chain.
The operator $\widehat{O}^{\,}_{4,j}$ is its dressed analogue and
appears explicitly in Hamiltonians
\eqref{eq:def HA and HB c} and \eqref{eq:def HA and HB d}.
Comparing them determines whether the dressing merely
changes a nonuniversal amplitude or instead modifies the low-energy
momentum content of the anisotropy.

We denote their connected correlation functions by
\begin{equation}
C^{a,b,\mathrm{c}}_{s^{\,}_{0},s^{\,}_{0}+s}
\coloneqq
\langle
\widehat{O}^{\,}_{a,s^{\,}_{0}}
\widehat{O}^{\,}_{b,s^{\,}_{0}+s}
\rangle
-
\langle
\widehat{O}^{\,}_{a,s^{\,}_{0}}
\rangle
\langle
\widehat{O}^{\,}_{b,s^{\,}_{0}+s}
\rangle
\label{eq:gFMz Stability Cab}
\end{equation}
for $a,b\in\{2,4\}$.
The connected subtraction in
Eq.~\eqref{eq:gFMz Stability Cab}
is required because the one-point functions of the anisotropy operators
need not vanish away from the $\mathrm{U}(1)$-symmetric lines.

The $\mathrm{U}(1)$-breaking parts of the two Hamiltonians are
proportional to the operator combinations
\begin{subequations}
\label{eq:gFMz Stability Physical Perturbations}
\begin{align}
&
\widehat{V}^{\,}_{\mathrm{I},j}\coloneqq
\cos\lambda\,
\widehat{O}^{\,}_{2,j}
+
\sin\lambda\,
\widehat{O}^{\,}_{4,j},
\label{eq:gFMz Stability Physical Perturbation I}
\\
&
\widehat{V}^{\,}_{\mathrm{II,j}}\coloneqq
\cos\lambda\,
\widehat{O}^{\,}_{2,j}
-
\sin\lambda\,
\widehat{O}^{\,}_{4,j}.
\label{eq:gFMz Stability Physical Perturbation II}
\end{align}
\end{subequations}
Equation~\eqref{eq:gFMz Stability Physical Perturbation I}
applies to Hamiltonian \eqref{eq:def HA and HB c}, whereas
Eq.~\eqref{eq:gFMz Stability Physical Perturbation II}
applies to Hamiltonian \eqref{eq:def HA and HB d}.
The correlators of $\widehat{O}^{\,}_{2,j}$ and
$\widehat{O}^{\,}_{4,j}$ identify the microscopic origin of the observed
momentum components.
The correlators of $\widehat{V}^{\,}_{\mathrm{I},j}$ and
$\widehat{V}^{\,}_{\mathrm{II},j}$ are the corresponding direct
diagnostics of the physical symmetry-breaking directions.

For each connected correlator, we compute the finite-window Fourier
transform
\begin{subequations}
\label{eq:gFMz SF}
\begin{equation}
\widetilde{C}^{a,b,\mathrm{c}}_{s^{\,}_{0},s^{\,}_{0}+s,q}
\coloneqq
\sum_{r=0}^{s}
w^{\,}_{s,r}\,
C^{a,b,\mathrm{c}}_{s^{\,}_{0},s^{\,}_{0}+r}\,
e^{-\ii q r},
\label{eq:gFMz stability SF}
\end{equation}
where we use the normalized Hanning window
\begin{equation}
w^{\,}_{s,r}
=
\frac{1}{s}
-
\frac{1}{s}
\cos\left(
\frac{2\pi r}{s}
\right),
\qquad
r\in\{0,\ldots,s\}.
\label{eq:Haning Window}
\end{equation}
The window reduces the spectral leakage caused by the finite real-space
interval.
We refer to
\begin{equation}
S^{a,b}_{s^{\,}_{0},s,q}
\coloneqq
|\widetilde{C}^{a,b,\mathrm{c}}_{s^{\,}_{0},s^{\,}_{0}+s,q}|
\label{eq:gFMz Stability Structure Factor Definition}
\end{equation}
\end{subequations}
as the structure factor on this interval.

We describe the long-distance connected anisotropy correlations by
separating their uniform and finite-momentum components as
\begin{equation}
C^{a,a,\mathrm{c}}_{s^{\,}_{0},s^{\,}_{0}+s}
=
A^{\,}_{a,0}\,s^{-p^{\,}_{a,0}}
+
A^{\,}_{a,Q}\,
\cos\left(Q^{\,}_{a}s\right)\,
s^{-p^{\,}_{a,Q}}.
\label{eq:gFMz stability anisotropy fit}
\end{equation}
Here,
$p^{\,}_{a,0}$ is the decay exponent of the uniform component,
$p^{\,}_{a,Q}$ is the decay exponent of the oscillating component,
and $Q^{\,}_{a}\neq0$ is fixed by the position of the finite-momentum peak
of the corresponding structure factor.
Equation~\eqref{eq:gFMz stability anisotropy fit}
therefore distinguishes two physically different critical channels.
The exponent $p^{\,}_{a,0}$ characterizes a translation-invariant
contribution that can enter a uniform low-energy perturbation directly.
The exponent $p^{\,}_{a,Q}$ characterizes an incommensurate contribution
whose spatial phase must also be considered when assessing its effect
on the infrared theory.

We show the transverse correlation function
$C^{x}_{s^{\,}_{0},s^{\,}_{0}+s}$,
the connected correlations
$C^{2,2,\mathrm{c}}_{s^{\,}_{0},s^{\,}_{0}+s}$ and
$C^{4,4,\mathrm{c}}_{s^{\,}_{0},s^{\,}_{0}+s}$ defined in
Eq.~\eqref{eq:gFMz Stability Cab} and 
and their structure factors
$S^{2,2}_{s^{\,}_{0},s}$ and
$S^{4,4}_{s^{\,}_{0},s}$ 
defined in Eq.~\eqref{eq:gFMz SF}
in Figs.~\ref{fig:Decay Correlations A gFMz DMRG} and
\ref{fig:Decay Correlations B gFMz DMRG}
for Hamiltonians
\eqref{eq:def HA and HB c} and \eqref{eq:def HA and HB d},
respectively.
We chose to study the point
$(\theta,\lambda)=(\pi/4-\pi/40,\pi/4-\pi/40)$
which lies away from the $\mathrm{U}(1)$-symmetric line.
The transverse correlation
$C^{x}_{s^{\,}_{0},s^{\,}_{0}+s}$
and the anisotropy correlations
$C^{2,2,\mathrm{c}}_{s^{\,}_{0},s^{\,}_{0}+s}$ and
$C^{4,4,\mathrm{c}}_{s^{\,}_{0},s^{\,}_{0}+s}$
retain algebraically decaying envelopes over the accessible range.
No rapid exponential cutoff of the type observed in the neighboring
gapped $\mathrm{Neel}^{\,}_{x}$ phase is resolved.

The structure factors demonstrate that the anisotropy correlations away
from the $\mathrm{U}(1)$-symmetric line are not purely uniform.
They contain a robust component at a finite wave number $Q^{\,}_{a}\neq0$.
The same component is visible directly in real space through the
oscillations of the connected correlations.
The finite-momentum peak occurs in the connected correlations of both
$\widehat{O}^{\,}_{2,j}$ and $\widehat{O}^{\,}_{4,j}$,
although its relative weight depends on the Hamiltonian and on the
microscopic operator.

For Hamiltonian \eqref{eq:def HA and HB c},
the operators $\widehat{O}^{\,}_{2,j}$ and
$\widehat{O}^{\,}_{4,j}$ have closely related long-distance behavior.
Their structure factors are dominated by approximately the same finite
momentum 
\begin{equation}
Q^{\,}_{2}\simeq Q^{\,}_{4}\simeq 0.6036\,\pi,
\end{equation}
and their algebraic envelopes are compatible within the
uncertainty associated with the fitting intervals.
This similarity is expected in the presence of ferromagnetic
long-range order along the $z$ axis.
Writing
\begin{equation}
\widehat{Z}^{\,}_{j}
=
m^{z}_{\mathrm{uni}}
+
\delta\widehat{Z}^{\,}_{j},
\end{equation}
one obtains
\begin{align}
\widehat{O}^{\,}_{4,j}
={}&
\left(m^{z}_{\mathrm{uni}}\right)^{2}
\widehat{O}^{\,}_{2,j}
\nonumber\\
&+
m^{z}_{\mathrm{uni}}
\left[
\left(\delta\widehat{Z}^{\,}_{j-1}\right)
\widehat{O}^{\,}_{2,j}
+
\widehat{O}^{\,}_{2,j}
\left(\delta\widehat{Z}^{\,}_{j+2}\right)
\right]
\nonumber\\
&+
\left(\delta\widehat{Z}^{\,}_{j-1}\right)
\widehat{O}^{\,}_{2,j}
\left(\delta\widehat{Z}^{\,}_{j+2}\right).
\label{eq:gFMz stability O4 decomposition}
\end{align}
The data for Hamiltonian \eqref{eq:def HA and HB c}
indicate that this common contribution from the first line
in Eq.~\eqref{eq:gFMz stability O4 decomposition}
dominates the accessible long-distance regime.

For Hamiltonian \eqref{eq:def HA and HB d},
the distinction between the two operators is more pronounced.
The structure factor of $\widehat{O}^{\,}_{2,j}$ contains zero- and
finite-momentum
\begin{equation}
Q^{\,}_{2}\simeq Q^{\,}_{4}\simeq 0.602\,\pi	
\end{equation}
components of comparable magnitude over the analyzed interval.
By contrast, the structure factor of $\widehat{O}^{\,}_{4,j}$
is strongly dominated by its finite-momentum peak.
The $Z$-dressing therefore does not merely rescale
$\widehat{O}^{\,}_{2,j}$ in this Hamiltonian.
It redistributes spectral weight toward the incommensurate sector.

This redistribution has a natural origin because the longitudinal
sector of the $\mathrm{gFM}^{\,}_{z}$ phase is itself incommensurate.
Indeed, the first line of
Eq.~\eqref{eq:gFMz stability O4 decomposition}
reproduces the momentum components already present in
$\widehat{O}^{\,}_{2,j}$.
The last two lines of
Eq.~\eqref{eq:gFMz stability O4 decomposition}
combine the anisotropy operator with the
incommensurate longitudinal fluctuations.
They can therefore enhance finite-momentum components already present
in $\widehat{O}^{\,}_{2,j}$ or generate additional ones.
This explains why the finite-momentum weight of
$\widehat{O}^{\,}_{4,j}$ is particularly stronger in
Fig.~\ref{fig:Decay Correlations B gFMz DMRG f}
than in
Fig.~\ref{fig:Decay Correlations B gFMz DMRG d}.

\begin{figure}
\includegraphics[width=0.55\columnwidth]{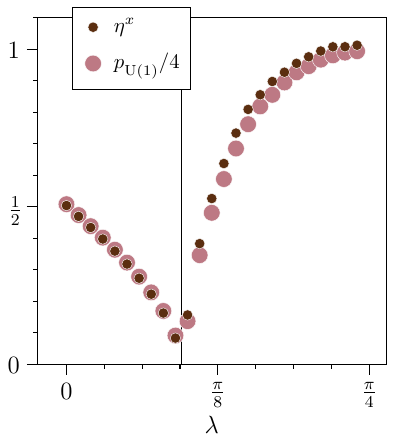}%
\caption{
Decay exponents $\eta^{x}$ defined in
Eq.~\eqref{eq:gFMz OBCs algebraic decay}
and $p^{\,}_{\mathrm{U}(1)}$ defined in
Eq.~\eqref{eq:gFMz Stability Uniform Anisotropy Decay}
as a function of $\lambda$.
They are obtained on the ground state of
Hamiltonian \eqref{eq:def H U1} at $\theta=\pi/4$
using DMRG ($\chi=512$) with OBCs on a chain of length $2N=512$.
The vertical black line indicates the tricritical
point \eqref{eq:lambda gFMz}.
}
\label{fig:Compare Exponents U(1) DMRG}
\end{figure}

We now compare these results with the standard bosonization of the
magnetized XXZ chain.
On the $\mathrm{U}(1)$-symmetric line,
the structure factor of $\widehat{O}^{\,}_{2,j}$ is dominated by the
uniform component.
We denote its algebraic decay exponent on that line by
$p_{\mathrm{U}(1)}$ as defined through
\begin{equation}
C^{2,2,\mathrm{c}}_{s^{\,}_{0},s^{\,}_{0}+s}
\propto
s^{-p^{\,}_{\mathrm{U}(1)}},
\qquad
\theta=\frac{\pi}{4}.
\label{eq:gFMz Stability Uniform Anisotropy Decay}
\end{equation}
In the conventional bosonization dictionary,
$\widehat{O}^{\,}_{2,j}$ is represented by the uniform charge-two
anisotropy vertex.
With the convention used for the transverse exponent
$\eta^{x}$ in Eq.~\eqref{eq:gFMz OBCs algebraic decay},
this identification predicts
\begin{equation}
p^{\,}_{\mathrm{U}(1)}=
4\eta^{x}.
\label{eq:gFMz Stability XXZ Bosonization Relation}
\end{equation}
Equivalently,
the scaling dimension of the uniform anisotropy operator is
\begin{equation}
\Delta^{\,}_{\mathrm{U}(1)}=
\frac{p^{\,}_{\mathrm{U}(1)}}{2}=
2\eta^{x}.
\label{eq:gFMz Stability XXZ Scaling Dimension}
\end{equation}

Figure~\ref{fig:Compare Exponents U(1) DMRG}
compares $p_{\mathrm{U}(1)}$ with $4\eta^{x}$ along the
$\mathrm{U}(1)$-symmetric line.
The agreement
\begin{equation}
p^{\,}_{\mathrm{U}(1)}\simeq
4\eta^{x}
\label{eq:gFMz Stability Exponent Agreement}
\end{equation}
shows that the standard XXZ bosonization correctly describes the
uniform anisotropy channel on this line.
This result provides a controlled reference for interpreting the
correlations away from the $\mathrm{U}(1)$-symmetric line.

The situation changes away from the U(1)-symmetric line.
To make this change explicit,
we analyze each connected anisotropy correlation in two complementary
ways.
First, we fit its absolute value to a single algebraic envelope,
\begin{equation}
|C^{a,a,\mathrm{c}}_{s^{\,}_{0},s^{\,}_{0}+s}|
\propto
s^{-p^{\,}_{a}},
\qquad
a\in\{2,4\},
\label{eq:gFMz Stability Single Power Fit}
\end{equation}
which defines the effective decay exponent $p^{\,}_{a}$ without resolving
the different momentum components.
Second, we fit the correlator to
Eq.~\eqref{eq:gFMz stability anisotropy fit},
which separately defines the exponent $p^{\,}_{a,0}$ of the uniform
component and the exponent $p^{\,}_{a,Q}$ of the component at
$Q^{\,}_{a}\neq0$.
The single-power fit provides the most direct comparison with the
analysis performed on the $\mathrm{U}(1)$-symmetric line,
whereas the momentum-resolved fit determines which part of the
correlation is responsible for the observed decay.

For the anisotropy correlations displayed in
Figs.~\ref{fig:Decay Correlations A gFMz DMRG} and
\ref{fig:Decay Correlations B gFMz DMRG},
we find that
\begin{equation}
p^{\,}_{2}=1.00\pm0.08,\qquad
p^{\,}_{4}=1.02\pm0.08,
\end{equation}
for Hamiltonian \eqref{eq:def HA and HB c} and
\begin{equation}
p^{\,}_{2}=1.58\pm0.06,\qquad
p^{\,}_{4}=1.54\pm0.06,
\end{equation}
for Hamiltonian \eqref{eq:def HA and HB d}.
Because
\begin{subequations}
\begin{align}
&\eta^{x}=0.950\pm0.007,
\\
&\eta^{x}=1.004\pm0.005,
\end{align}
\end{subequations}
for Hamiltonian \eqref{eq:def HA and HB c} and
for Hamiltonian \eqref{eq:def HA and HB d}, respectively,
we find that
\begin{equation}
p^{\,}_{a}
\neq
4\eta^{x},
\qquad
a\in\{2,4\},
\label{eq:gFMz Stability Effective Exponent Mismatch}
\end{equation}
The discrepancy is therefore already present before the correlation is
decomposed into its momentum components.
It cannot be attributed solely to the oscillatory modulation of an
otherwise conventional XXZ envelope.

The momentum-resolved fit defined in
Eq.~\eqref{eq:gFMz stability anisotropy fit}
leads to the stronger result
\begin{subequations}
\label{eq:gFMz Stability Resolved Exponent Mismatch}
\begin{align}
&
p^{\,}_{a,0}\neq
4\eta^{x},
\label{eq:gFMz Stability Uniform Exponent Mismatch}
\\
&
p^{\,}_{a,Q}\neq
4\eta^{x}.
\label{eq:gFMz Stability Finite Momentum Exponent Mismatch}
\end{align}
\end{subequations}
Indeed, based on the data presented in
Fig.~\ref{fig:Decay Correlations A gFMz DMRG}
for Hamiltonian \eqref{eq:def HA and HB c}, we find
\begin{align}
&
p^{\,}_{2,0}=1.015\pm0.001,\qquad
p^{\,}_{2,Q}=1.15\pm0.01,
\\
&
p^{\,}_{4,0}=1.015\pm0.001,\qquad
p^{\,}_{4,Q}=1.16\pm0.02.
\end{align}
Moreover, based on the data presented in
Fig.~\ref{fig:Decay Correlations B gFMz DMRG}
for Hamiltonian \eqref{eq:def HA and HB d}, we find
\begin{align}
&
p^{\,}_{2,0}=1.20\pm0.01,\hphantom{00}\qquad
p^{\,}_{2,Q}=1.929\pm0.007,
\\
&
p^{\,}_{4,0}=1.232\pm0.005,\qquad
p^{\,}_{4,Q}=1.94\pm0.02.
\end{align}
Thus, neither the fitted zero-momentum contribution nor the fitted
finite-momentum contribution is governed by the exponent of the
uniform XXZ anisotropy measured on the $\mathrm{U}(1)$-symmetric line.
The appearance of the incommensurate peak and the change in the decay
exponents are therefore two simultaneous features of the anisotropy
correlations away from that line.

Equations~\eqref{eq:gFMz Stability Exponent Agreement},
\eqref{eq:gFMz Stability Effective Exponent Mismatch}, and
\eqref{eq:gFMz Stability Resolved Exponent Mismatch}
refer to different regimes and are mutually consistent.
On the $\mathrm{U}(1)$-symmetric line,
the dominant anisotropy channel is uniform and obeys the conventional
XXZ relation.
Away from that line,
the microscopic symmetry is reduced to $\mathbb{Z}_{2}$,
the anisotropy correlation contains both zero- and finite-momentum
components, and the exponents of both components no longer obey
any relation of the kind \eqref{eq:gFMz Stability XXZ Bosonization Relation}.
The low-energy projection of the lattice anisotropy is therefore not
obtained by simply continuing the uniform XXZ charge-two vertex away
from the symmetric line while keeping the same relation between its
scaling dimension and $\eta^{x}$.

This conclusion does not imply that the
$\mathsf{c}=1$ low-energy description itself fails.
The finite-entanglement scaling in
Fig.~\ref{fig:Central Charge gFMz iDMRG}
and the running transverse exponent in
Fig.~\ref{fig:Running Exponent gFMz iDMRG}
independently support a critical theory with one gapless mode.
What changes away from the $\mathrm{U}(1)$-symmetric line is the
operator dictionary that relates the microscopic anisotropy to the
fields of this critical theory.
In particular,
the exponents $p^{\,}_{a,0}$ and $p^{\,}_{a,Q}$ define the scaling dimensions
\begin{subequations}
\label{eq:gFMz Stability Resolved Scaling Dimensions}
\begin{align}
&
\Delta^{\,}_{a,0}\coloneqq
\frac{p^{\,}_{a,0}}{2},
\label{eq:gFMz Stability Uniform Scaling Dimension Away}
\\
&
\Delta^{\,}_{a,Q}\coloneqq
\frac{p^{\,}_{a,Q}}{2},
\label{eq:gFMz Stability Finite Momentum Dimension}
\end{align}
\end{subequations}
and the data show that neither is fixed by
$2\eta^{x}$ at the points considered away from the symmetric line.

The zero-momentum result in
Eq.~\eqref{eq:gFMz Stability Uniform Exponent Mismatch}
is especially important.
It shows that the fitted uniform contribution away from the
$\mathrm{U}(1)$-symmetric line should not automatically be identified
with the same uniform charge-two vertex that controls the instability
of the magnetized XXZ chain.
Consequently,
the conventional XXZ relevance criterion
\begin{equation}
\Delta_{\mathrm{U}(1)}=
2\eta^{x}
<
2
\label{eq:gFMz Stability Conventional Relevance Criterion}
\end{equation}
cannot be transferred unchanged to the fitted uniform component away
from the U(1)-symmetric line.
Its relevance must instead be assessed from the exponent
$p^{\,}_{a,0}$ and from its actual coefficient in the complete microscopic
perturbation.

The finite-momentum result in
Eq.~\eqref{eq:gFMz Stability Finite Momentum Exponent Mismatch}
provides an additional and independent distinction from the ordinary
XXZ anisotropy.
The corresponding contribution has the continuum form
\begin{equation}
\widehat{O}^{\,}_{a}(x)
\supset
e^{\ii Q^{\,}_{a}x}\,
\mathcal{O}^{\,}_{a,Q}(x)
+
e^{-\ii Q^{\,}_{a}x}\,
\mathcal{O}^{\,}_{a,-Q}(x),
\qquad
Q^{\,}_{a}\neq0,
\label{eq:gFMz Stability Finite Momentum Operator}
\end{equation}
with local scaling dimension
$\Delta^{a}_{Q}=p^{\,}_{a,Q}/2$.
Even when $\Delta^{a}_{Q}<2$,
the oscillating phase in
Eq.~\eqref{eq:gFMz Stability Finite Momentum Operator}
prevents this component from contributing coherently to a spatially
uniform perturbation when $Q^{\,}_{a}$ is incommensurate.
Its scaling dimension alone therefore does not establish a uniform
gap-opening instability.

The comparison between the two Hamiltonians clarifies how these two
channels enter the physical perturbation.
For Hamiltonian \eqref{eq:def HA and HB c},
$\widehat{O}^{\,}_{2,j}$ and $\widehat{O}^{\,}_{4,j}$ have similar
momentum distributions and predominantly probe the same
incommensurate channel.
For Hamiltonian \eqref{eq:def HA and HB d},
$\widehat{O}^{\,}_{2,j}$ has appreciable weight in both the uniform and
finite-momentum sectors,
whereas the $Z$-dressed operator
$\widehat{O}^{\,}_{4,j}$ strongly enhances the finite-momentum sector.
Since the physical perturbations
\eqref{eq:gFMz Stability Physical Perturbation I} and
\eqref{eq:gFMz Stability Physical Perturbation II}
are linear combinations of these two operators,
their infrared effect cannot be inferred from
$\widehat{O}^{\,}_{2}$,
$\eta^{x}$,
or any one fitted exponent alone.
The uniform and finite-momentum components of the complete perturbing
combination must be considered separately.

\subsection{Summary}

The numerical results can therefore be summarized as follows.
On the exact $\mathrm{U}(1)$-symmetric line,
the anisotropy is dominated by a uniform channel whose exponent obeys
Eq.~\eqref{eq:gFMz Stability Exponent Agreement},
in agreement with the conventional magnetized XXZ theory.
Away from that line,
the anisotropy correlations remain algebraic but their operator content
changes since
\begin{equation}
p^{\,}_{a}
\neq
4\eta^{x},
\quad
p^{\,}_{a,0}
\neq
4\eta^{x},
\quad
p^{\,}_{a,Q}
\neq
4\eta^{x},
\quad
Q^{\,}_{a}\neq0.
\label{eq:gFMz Stability Summary Anisotropy Away}
\end{equation}
The departure from the ordinary XXZ scenario is therefore not limited
to the appearance of an incommensurate peak.
It also includes a change in the algebraic decay of the full
anisotropy correlation and of both momentum-resolved components.

These observations provide further evidence against the crossover
scenario.
If the apparently extended region were already flowing toward a gapped
$\mathrm{Neel}^{\,}_{x}$ or $\mathrm{Neel}^{\,}_{y}$ phase,
the connected anisotropy correlations would eventually develop an
exponential cutoff at a physical length scale independent of the system
size $2N$ and of the MPS bond dimension $\chi$.
No such cutoff is resolved.
Instead,
the full anisotropy correlations and their uniform and incommensurate
components remain algebraic over the accessible distances.
Together with the $\mathsf{c}=1$ finite-entanglement scaling and the
stable running transverse exponent discussed above,
this behavior is consistent with a genuine extended critical
$\mathrm{gFM}^{\,}_{z}$ phase.

We nevertheless emphasize the limitation of this conclusion.
Finite-size and finite-entanglement calculations cannot exclude a
crossover beyond all scales.
The numerical results establish that no such crossover is resolved up
to the largest system sizes and MPS correlation lengths considered here.
If a finite physical crossover length $\xi^{\,}_{\star}$ exists,
it must be larger than these scales.

\section{Effective interactions between boundary spin-$1/2$ degrees of freedom}
\label{sec:effective_interactions_between_boundary_spin_1_2_degrees_of_freedom}

We use degenerate perturbation theory up to second
order to show that Hamiltonian
\eqref{eq:H Effective Boundary Fermions}
is the effective Hamiltonian of $\widehat{H}^{(1)}_{\mathrm{U}(1)}(\lambda)$
for $0\leq\pi/2-\lambda\ll1$ that explains the lifting of the four-fold
degenerate ground state at $\lambda=\pi/2$ for any number $2N$ of sites,
as was observed numerically in
Sec.~\ref{subsubsec:The NeelSPTx-NeelSPTy phase boundary}.

\subsection{Setup of perturbation theory}

We have approximated Hamiltonian $\widehat{H}^{(1)}_{\mathrm{U}(1)}(\lambda)$,
defined in Eq.~\eqref{eq:def H U1},
in the regime for which
\begin{subequations}
\begin{equation}
0\leq\pi/2-\lambda\equiv\epsilon\ll1,
\end{equation}
by the fermionic Hamiltonian
\begin{equation}
\widehat{H}^{\,}_{\mathrm{f}}(\epsilon)\equiv
\widehat{H}^{\,}_{0}
+
\widehat{V}(\epsilon)
\end{equation}
\end{subequations}
defined in Eq.\
\eqref{eq:H Effective Epsilon Fermion}
with the fermionic Fock space
\begin{equation}
\mathfrak{F}^{\,}_{2N}
\cong
\mathbb{C}^{2^{2N}}
\end{equation}
as domain of definition. In doing so, we
(i) made the change of variables
\eqref{eq:Param PT Boundary Spins},
(ii) applied the unitary transformation
\eqref{eq:action U_CZ on 3 Paulis} for $\mathsf{b}=1$,
(iii) neglected the perturbations to the bulk of the chain, i.e.,
any perturbation that does not involve a boundary degree of freedom, and
(iv) fermionized using the Jordan-Wigner transformation
\eqref{eq: def Jordan-Wigner trsf U(1) symmetric}.

If we assume the existence of the expansion
\begin{equation}
\widehat{H}^{\,}_{\mathrm{f}}(\epsilon)=
\sum_{n=0}^{\infty}\widehat{H}^{(n)}(\epsilon)
\end{equation}
in powers of $\widehat{V}(\epsilon)$,
we seek the approximation
\begin{subequations}
\label{eq:Def Effective Ham PT}
\begin{equation}
\widehat{H}^{(2)}_{\mathrm{eff,f}}(\epsilon)\coloneqq
\sum\limits_{n=0}^{2}\,
\widehat{H}^{(n)}(\epsilon),
\label{eq:Def Effective Ham PT a}
\end{equation}
where
\begin{align}
&\widehat{H}^{(0)}(\epsilon)\coloneqq
E^{\,}_{0}\,
\widehat{\mathds{P}}^{\,}_{0},
\label{eq:Def Effective Ham PT b}
\\
&\widehat{H}^{(1)}(\epsilon)\coloneqq
\widehat{\mathds{P}}^{\,}_{0}\,
\widehat{V}(\epsilon)\,
\widehat{\mathds{P}}^{\,}_{0}
\label{eq:Def Effective Ham PT c}
\\
&\widehat{H}^{(2)}(\epsilon)\coloneqq
\widehat{\mathds{P}}^{\,}_{0}\,
\widehat{V}(\epsilon)\,
\widehat{S}\,
\widehat{V}(\epsilon)\,
\widehat{\mathds{P}}^{\,}_{0}.
\label{eq:Def Effective Ham PT d}
\end{align}
Here,
$E^{\,}_{0}$,
$\widehat{\mathds{P}}^{\,}_{0}$,
$\widehat{S}$,
and $\widehat{V}(\epsilon)$ stand for
(i)
the four-fold degenerate ground-state energy
of $\widehat{H}^{\,}_{0}$,
(ii)
the projector on the ground states of
$\widehat{H}^{\,}_{0}$,
(iii)
the propagator
\begin{equation}
\widehat{S}\coloneqq
\bigl(
1-\widehat{\mathds{P}}^{\,}_{0}
\bigr)
\bigl(
E^{\,}_{0}-\widehat{H}^{\,}_{0}
\bigr)^{-1}
\bigl(
1-\widehat{\mathds{P}}^{\,}_{0}
\bigr),
\label{eq:Def Effective Ham PT e}
\end{equation}
and (iv)
the perturbation
\begin{equation}
\widehat{V}(\epsilon)\coloneqq
\widehat{H}^{\,}_{\mathrm{L}}(\epsilon)
+
\widehat{H}^{\,}_{\mathrm{R}}(\epsilon)
+
\widehat{H}^{\prime}_{\mathrm{L}}(\epsilon)
+
\widehat{H}^{\prime}_{\mathrm{R}}(\epsilon)
\label{eq:Def Effective Ham PT f}
\end{equation}
\end{subequations}
to $\widehat{H}^{\,}_{0}$ that was defined in Eq.\
\eqref{eq:H Effective Epsilon Fermion},
respectively.

\subsection{Diagonalization of the unperturbed Hamiltonian}

We now diagonalize $\widehat{H}^{\,}_{0}$ defined in
Eq.~\eqref{eq:H Effective Epsilon Fermion b}.
In doing so, we shall have constructed
the fundamental subspace
\begin{equation}
\mathcal{P}^{\,}_{0}\coloneqq
\widehat{\mathds{P}}^{\,}_{0}\,\mathfrak{F}^{\,}_{2N}
\end{equation}
of $\widehat{H}^{\,}_{0}$.

We introduce the Fourier expansion into $2N-2$ modes of
$\hat{c}^{\,}_{j}$ at $j=2,\cdots,2N-1$,
\begin{subequations}
\label{eq:Fourier Modes PT}
\begin{equation}
\hat{c}^{\,}_{j}=\mathcal{N}\,
\sum_{n=1}^{2N-2}
\hat{f}^{\,}_{n}\,
\sin\!\left(\frac{\pi n(j-1)}{2N-1}\right),
\label{eq:Fourier Modes PT a}
\end{equation}
where $\hat{f}^{\,}_{n}$ is the fermion annihilation
operator of mode $n$ that obeys the fermion algebra
\begin{equation}
\left\{
\hat{f}^{\,}_{n},
\hat{f}^{\dag}_{n'}
\right\}=
\delta^{\,}_{n,n'},
\quad
\left\{
\hat{f}^{\,}_{n},
\hat{f}^{\,\vphantom{\dag}}_{n'}
\right\}=
\left\{
\hat{f}^{\dag}_{n},
\hat{f}^{\dag}_{n'}
\right\}=
0,
\label{eq:Fourier Modes PT b}
\end{equation}
for $n,n'=2,\cdots,2N-1$.
Notice that $\hat{c}^{\,}_{1}$ and $\hat{c}^{\,}_{2N}$ are not involved in this Fourier mode expansion.
The normalization factor
\begin{equation}
\mathcal{N}=\sqrt{\frac{2}{2N-1}}
\label{eq:Fourier Modes PT c}
\end{equation}
\end{subequations}
is a consequence of the fermionic algebras
\eqref{eq:def HbU(1) fermionic as function lambda c}
and
\eqref{eq:Fourier Modes PT b}, and of the identity
\begin{equation}
\frac{2N-1}{2}\,
\delta^{\,}_{n,n'}=
\sum_{j=2}^{2N-1}
\sin\left(\frac{\pi n(j-1)}{2N-1}\right)
\sin\left(\frac{\pi n'(j-1)}{2N-1}\right)
\end{equation}
for $n,n'=2,\cdots,2N-1$.

We then insert the expansion \eqref{eq:Fourier Modes PT}
into Hamiltonian \eqref{eq:H Effective Epsilon Fermion b}
to obtain
\begin{equation}
\widehat{H}^{\,}_{0}=
\sum_{n=1}^{2N-2}
\varepsilon^{\,}_{n}\,
\hat{f}^{\dag}_{n}\hat{f}^{\,}_{n},
\qquad
\varepsilon^{\,}_{n}\coloneqq
2\cos\!\left(\frac{\pi n}{2N-1}\right)\!.
\end{equation}

Because the single-particle energies obey
\begin{subequations}
\begin{align}
&\varepsilon^{\,}_{n}>0,
\qquad
n=1,\ldots,N-1,
\\
&\varepsilon^{\,}_{n}<0,
\qquad
n=N,\cdots,2N-2,
\end{align}
\end{subequations}
we construct the Fermi sea
\begin{subequations}
\label{eq:Fermi Sea and P}
\begin{equation}
\ket{\mathrm{FS}}\coloneqq
\prod_{n=N}^{2N-2}
\hat{f}^{\dag}_{n}\ket{0},
\qquad
\hat{f}^{\,}_{n}\ket{0}=0.
\label{eq:Fermi Sea and P a}
\end{equation}
\end{subequations}
Additionally, because
(i)
the domain of definition
of $\widehat{H}^{\,}_{0}$ is
$\mathfrak{F}^{\,}_{2N}\cong\mathbb{C}^{2^{2N}}$,
(ii)
the state $\ket{\mathrm{FS}}$ is an element of the subspace
\begin{equation}
\mathfrak{F}^{\prime}_{2N}\subset
\mathfrak{F}^{\,}_{2N}
\end{equation}
of dimension $2^{2N-2}$
that is spanned by $\hat{f}^{\dag}_{1},\cdots,\hat{f}^{\dag}_{2N-2}$,
and
(iii)
all fermionic operators
$\hat{f}^{\dag}_{1},\cdots,\hat{f}^{\dag}_{2N-2}$
act trivially on the sites $j=1$ and $j=2N$,
the fundamental subspace of $\widehat{H}^{\,}_{0}$
must be at least four dimensional
[recall Eq.~\eqref{eq:Boundary States Fermions}]
\begin{equation}
\begin{split}
\mathrm{span}\, 
\Bigl\{
&
\hphantom{\hat{c}^{\dag}_{1}}
\ket{0}^{\,}_{1}\otimes
\ket{\mathrm{FS}}\otimes
\hphantom{\hat{c}^{\dag}_{2N}}
\ket{0}^{\,}_{2N},
\\
&
\hphantom{\hat{c}^{\dag}_{1}}
\ket{0}^{\,}_{1}\otimes
\ket{\mathrm{FS}}\otimes
\hat{c}^{\dag}_{2N}
\ket{0}^{\,}_{2N},
\\
&
\hat{c}^{\dag}_{1}
\ket{0}^{\,}_{1}\otimes
\ket{\mathrm{FS}}\otimes
\hphantom{\hat{c}^{\dag}_{2N}}
\ket{0}^{\,}_{2N},
\\
&
\hat{c}^{\dag}_{1}
\ket{0}^{\,}_{1}\otimes
\ket{\mathrm{FS}}\otimes
\hat{c}^{\dag}_{2N}
\ket{0}^{\,}_{2N}
\Bigr\}
\subseteq
\mathcal{P}^{\,}_{0}.
\label{eq:Fermi Sea and P b}
\end{split}
\end{equation}
To show that
\begin{equation}
\begin{split}
\mathcal{P}^{\,}_{0}=
\mathrm{span}\, 
\Bigl\{
&
\hphantom{\hat{c}^{\dag}_{1}}
\ket{0}^{\,}_{1}\otimes
\ket{\mathrm{FS}}\otimes
\hphantom{\hat{c}^{\dag}_{2N}}
\ket{0}^{\,}_{2N},
\\
&
\hphantom{\hat{c}^{\dag}_{1}}
\ket{0}^{\,}_{1}\otimes
\ket{\mathrm{FS}}\otimes
\hat{c}^{\dag}_{2N}
\ket{0}^{\,}_{2N},
\\
&
\hat{c}^{\dag}_{1}
\ket{0}^{\,}_{1}\otimes
\ket{\mathrm{FS}}\otimes
\hphantom{\hat{c}^{\dag}_{2N}}
\ket{0}^{\,}_{2N},
\\
&
\hat{c}^{\dag}_{1}
\ket{0}^{\,}_{1}\otimes
\ket{\mathrm{FS}}\otimes
\hat{c}^{\dag}_{2N}
\ket{0}^{\,}_{2N}
\Bigr\}
\end{split}
\label{eq:def basis mathdsB0}
\end{equation}
it suffices to observe that $\varepsilon^{\,}_{n}\neq 0$
for $n=1,\ldots, 2N-2$. The basis $\mathds{B}^{\,}_{0}$ of $\mathcal{P}^{\,}_{0}$
on the right-hand side of Eq.\ (\ref{eq:def basis mathdsB0})
is orthonormal. We shall reserve the notation $|\zeta\rangle$
for an element of this basis.

\subsection{Degenerate perturbation theory}

We are going to compute the three terms on the right-hand side of
Eq.~\eqref{eq:Def Effective Ham PT}.

\subsubsection{Zeroth order}

We start by deriving Hamiltonian
\eqref{eq:Def Effective Ham PT b}.
This task is done in two steps.

The projector on the fundamental subspace $\mathcal{P}^{\,}_{0}$
defined by Eq.\ (\ref{eq:def basis mathdsB0}) is
\begin{equation}
\widehat{\mathds{P}}^{\,}_{0}\coloneqq
\sum_{\ket{\zeta}\in\mathds{B}^{\,}_{0}}
\dyad{\zeta}.
\end{equation}
The ground-state energy of $\widehat{H}^{\,}_{0}$ is
\begin{align}
E^{\,}_{0}\coloneqq
2
\sum_{n=N}^{2N-2}
\cos\!\left(\frac{\pi n}{2N-1}\right).
\end{align}
The summation on the right-hand side can be performed
in closed form with the help of the finite geometric series.
It delivers
\begin{align}
E^{\,}_{0}=
2
\cos^{2}\left(\frac{\pi N}{2(2N-1)}\right)
\sin^{-1}\left(\frac{\pi}{2(2N-1)}\right).
\end{align}
\medskip

\subsubsection{First order}

We are going to derive Hamiltonian
\eqref{eq:Def Effective Ham PT c}.

To this end, we first define the normal ordering
\begin{subequations}
\begin{equation}
{\bm{:}\,}\hat{c}^{\dag}_{j}\,\hat{c}^{\,}_{j}{\,\bm{:}}\equiv
\hat{c}^{\dag}_{j}\,\hat{c}^{\,}_{j}
-
\frac{1}{2},
\qquad
j=2,\cdots,2N-1,
\end{equation}
for which 
\begin{equation}
\ev{\,{\bm{:}\,}\hat{c}^{\dag}_{j}\,\hat{c}^{\,}_{j}{\,\bm{:}}\,}{\mathrm{FS}}=0,
\end{equation}
since
\begin{equation}
\ev{\,\hat{c}^{\dag}_{j}\,\hat{c}^{\,}_{j}\,}{\mathrm{FS}}=
\frac{1}{2},\qquad
j=2,\cdots,2N-1.
\end{equation}
\end{subequations}

We then insert Eq.~\eqref{eq:Fourier Modes PT}
into the perturbation \eqref{eq:Def Effective Ham PT f}
to obtain
\begin{widetext}
\begin{subequations}
\begin{align}
\widehat{H}^{\,}_{\mathrm{L}}(\epsilon)=&\,
-
2\,\epsilon\,
\mathcal{N}^{\,3}
\sum_{n^{\,}_{1}=1}^{2N-2}
\sum_{n^{\,}_{2}=1}^{2N-2}
\sum_{n^{\,}_{3}=1}^{2N-2}
\left(
\hat{c}^{\dag}_{1}\,\hat{f}^{\,}_{n^{\,}_{1}}
+
\hat{f}^{\dag}_{n^{\,}_{1}}\,\hat{c}^{\,}_{1}
\right)
{\bm{:}\,}\hat{f}^{\dag}_{n^{\,}_{2}}\,\hat{f}^{\,}_{n^{\,}_{3}}{\bm{:}}
\nonumber\\
&\,
\times
\sin\left(\frac{\pi n^{\,}_{1}}{2N-1}\right)
\sin\left(\frac{2\pi n^{\,}_{2}}{2N-1}\right)
\sin\left(\frac{2\pi n^{\,}_{3}}{2N-1}\right),
\\
\widehat{H}^{\,}_{\mathrm{R}}(\epsilon)=&\,
-
2\,\epsilon\,\mathcal{N}^{\,3}
\sum_{n^{\,}_{1}=1}^{2N-2}
\sum_{n^{\,}_{2}=1}^{2N-2}
\sum_{n^{\,}_{3}=1}^{2N-2}
\left(
\hat{c}^{\dag}_{2N}\,\hat{f}^{\,}_{n^{\,}_{1}}
+
\hat{f}^{\dag}_{n^{\,}_{1}}\,\hat{c}^{\,}_{2N}
\right)
{\bm{:}\,}\hat{f}^{\dag}_{n^{\,}_{2}}\,\hat{f}^{\,}_{n^{\,}_{3}}{\bm{:}}
\nonumber\\
&\,
\times
\sin\left(\frac{\pi n^{\,}_{1}(2N-2)}{2N-1}\right)
\sin\left(\frac{\pi n^{\,}_{2}(2N-3)}{2N-1}\right)
\sin\left(\frac{\pi n^{\,}_{3}(2N-3)}{2N-1}\right),
\\
\widehat{H}^{\prime}_{\mathrm{L}}(\epsilon)=&\,
-
2\,\epsilon\,\mathcal{N}^{\,4}
\left(2\,\hat{c}^{\dag}_{1}\,\hat{c}^{\,}_{1}-1\right)
\sum_{n^{\,}_{1}=1}^{2N-2}
\sum_{n^{\,}_{2}=1}^{2N-2}
\sum_{n^{\,}_{3}=1}^{2N-2}
\sum_{n^{\,}_{4}=1}^{2N-2}
\left(
\hat{f}^{\dag}_{n^{\,}_{1}}\,\hat{f}^{\,}_{n^{\,}_{2}}
+
\hat{f}^{\dag}_{n^{\,}_{2}}\,\hat{f}^{\,}_{n^{\,}_{1}}
\right)
{\bm{:}\,}\hat{f}^{\dag}_{n^{\,}_{3}}\,\hat{f}^{\,}_{n^{\,}_{4}}{\bm{:}}
\nonumber\\
&\,
\times
\sin\!\left(\frac{\pi n^{\,}_{1}}{2N-1}\right)
\sin\!\left(\frac{2\pi n^{\,}_{2}}{2N-1}\right)
\sin\!\left(\frac{3\pi n^{\,}_{3}}{2N-1}\right)
\sin\!\left(\frac{3\pi n^{\,}_{4}}{2N-1}\right),
\\
\widehat{H}^{\prime}_{\mathrm{R}}(\epsilon)=&\,
-
2\,\epsilon\,\mathcal{N}^{\,4}
\left(2\,\hat{c}^{\dag}_{2N}\,\hat{c}^{\,}_{2N}-1\right)
\sum_{n^{\,}_{1}=1}^{2N-2}
\sum_{n^{\,}_{2}=1}^{2N-2}
\sum_{n^{\,}_{3}=1}^{2N-2}
\sum_{n^{\,}_{4}=1}^{2N-2}
\left(
\hat{f}^{\dag}_{n^{\,}_{1}}\,\hat{f}^{\,}_{n^{\,}_{2}}
+
\hat{f}^{\dag}_{n^{\,}_{2}}\,\hat{f}^{\,}_{n^{\,}_{1}}
\right)
{\bm{:}\,}\hat{f}^{\dag}_{n^{\,}_{3}}\,\hat{f}^{\,}_{n^{\,}_{4}}{\bm{:}}
\nonumber\\
&\,
\times
\sin\left(\frac{\pi n^{\,}_{1}(2N-2)}{2N-1}\right)
\sin\left(\frac{\pi n^{\,}_{2}(2N-3)}{2N-1}\right)
\sin\left(\frac{\pi n^{\,}_{3}(2N-4)}{2N-1}\right)
\sin\left(\frac{\pi n^{\,}_{4}(2N-4)}{2N-1}\right).
\end{align}
\end{subequations}
\end{widetext}

On the one hand,
because neither
$\widehat{H}^{\,}_{\mathrm{L}}(\epsilon)$
nor
$\widehat{H}^{\,}_{\mathrm{R}}(\epsilon)$
conserve the fermion number
\begin{equation}
\sum_{n=1}^{2N-2}
\hat{f}^{\dag}_{n}\,\hat{f}^{\,}_{n},
\label{eq:Fermion Number Bulk PT}
\end{equation}
they project any state $\ket{\zeta}\in\mathds{B}^{\,}_{0}$
onto a state orthogonal to $\mathcal{P}^{\,}_{0}$, i.e.,
\begin{equation}
\mel{\zeta'}{\widehat{H}^{\,}_{\mathrm{L}}}{\zeta}=
\mel{\zeta'}{\widehat{H}^{\,}_{\mathrm{R}}}{\zeta}=
0,\qquad
\ket{\zeta},\ket{\zeta'}\in\mathds{B}^{\,}_{0}.
\end{equation}
On the other hand, either
$\widehat{H}^{\prime}_{\mathrm{L}}(\epsilon)$
or
$\widehat{H}^{\prime}_{\mathrm{R}}(\epsilon)$
conserve the fermion number \eqref{eq:Fermion Number Bulk PT}.
Nevertheless, we are going to show that
\begin{equation}
\mel{\zeta'}{\widehat{H}^{\prime}_{\mathrm{L}}}{\zeta}=
\mel{\zeta'}{\widehat{H}^{\prime}_{\mathrm{R}}}{\zeta}=
0,\qquad
\ket{\zeta},\ket{\zeta'}\in\mathds{B}^{\,}_{0}.
\end{equation}

One verifies that
\begin{equation}
\mel{\zeta'}{\widehat{H}^{\prime}_{\mathrm{L}}}{\zeta}\equiv
\delta^{\,}_{\zeta,\zeta'}
\ev{\widehat{H}^{\prime}_{\mathrm{L}}}{\mathrm{FS}},
\label{eq:Compute Fisrt Order H'L}
\end{equation}
where we have abused the notation in the following way.
On the left-hand side,
$\widehat{H}^{\prime}_{\mathrm{L}}$ acts on the full Hilbert space
$\mathfrak{F}^{\,}_{2N}$, while on the right-hand side
$\widehat{H}^{\prime}_{\mathrm{L}}$ acts on the subspace
$\mathfrak{F}^{\prime}_{2N}\subset\mathfrak{F}^{\,}_{2N}$ related to the sites
$j=2,\cdots,2N-1$.
We have
\begin{subequations}
\begin{equation}
\ev{\widehat{H}^{\prime}_{\mathrm{L}}}{\mathrm{FS}}=
-2\,\epsilon\,(A+B)\,\mathcal{N}^{\,4}\,
(2\,\hat{c}^{\dag}_{1}\,\hat{c}^{\,}_{1}-1),
\end{equation}
where
\begin{widetext}
\begin{align}
\hspace{-6pt}A
&\coloneqq
\sum_{n^{\,}_{1}=1}^{2N-2}
\sum_{n^{\,}_{2}=1}^{2N-2}
\sum_{n^{\,}_{3}=1}^{N-1}
\sum_{n^{\,}_{4}=N}^{2N-2}
\ev{\,\hat{f}^{\dag}_{n^{\,}_{1}}\,\hat{f}^{\,}_{n^{\,}_{2}}\,
{\bm{:}\,}\hat{f}^{\dag}_{n^{\,}_{3}}\,\hat{f}^{\,}_{n^{\,}_{4}}{\bm{:}}\,}
{\mathrm{FS}}\,
\sin\!\left(\frac{\pi n^{\,}_{1}}{2N-1}\right)
\sin\!\left(\frac{2\pi n^{\,}_{2}}{2N-1}\right)
\sin\!\left(\frac{3\pi n^{\,}_{3}}{2N-1}\right)
\sin\!\left(\frac{3\pi n^{\,}_{4}}{2N-1}\right)\!,
\\
\hspace{-6pt}B
&\coloneqq
\sum_{n^{\,}_{1}=1}^{2N-2}
\sum_{n^{\,}_{2}=1}^{2N-2}
\sum_{n^{\,}_{3}=1}^{N-1}
\sum_{n^{\,}_{4}=N}^{2N-2}
\ev{\,\hat{f}^{\dag}_{n^{\,}_{2}}\,\hat{f}^{\,}_{n^{\,}_{1}}\,
{\bm{:}\,}\hat{f}^{\dag}_{n^{\,}_{3}}\,\hat{f}^{\,}_{n^{\,}_{4}}{\bm{:}}\,}
{\mathrm{FS}}\,
\sin\!\left(\frac{\pi n^{\,}_{1}}{2N-1}\right)
\sin\!\left(\frac{2\pi n^{\,}_{2}}{2N-1}\right)
\sin\!\left(\frac{3\pi n^{\,}_{3}}{2N-1}\right)
\sin\!\left(\frac{3\pi n^{\,}_{4}}{2N-1}\right)\!,
\end{align}
\end{widetext}
\end{subequations}
and where the sums over the indices $n^{\,}_{3}$ and $n^{\,}_{4}$
are restricted to accommodate the normal ordering.
We then perform the Wick contractions
\begin{equation}
\ev{\,\hat{f}^{\dag}_{n}\,\hat{f}^{\,}_{n'}\,
{\bm{:}\,}\hat{f}^{\dag}_{n^{\,}_{3}}\,\hat{f}^{\,}_{n^{\,}_{4}}{\bm{:}}\,}
{\mathrm{FS}}=
\delta^{\,}_{n,n^{\,}_{4}}\delta^{\,}_{n',n^{\,}_{3}},
\end{equation}
and we make the change of variables
\begin{subequations}
\label{eq:Change Variables PT}
\begin{equation}
n^{\,}_{i}\mapsto
2N-1-n^{\,}_{i},\qquad
n^{\,}_{i}=1,\ldots,2N-2,
\label{eq:Change Variables PT a}
\end{equation}
under which
\begin{equation}
\sin\!\left(\frac{\pi n^{\,}_{i}\,m}{2N-1}\right)
\mapsto
(-1)^{m+1}
\sin\!\left(\frac{\pi n^{\,}_{i}\,m}{2N-1}\right).
\label{eq:Change Variables PT b}
\end{equation}
\end{subequations}
In doing so, we find that
\begin{equation}
A=-B
\end{equation}
from which there follows that
\begin{equation}
\ev{\widehat{H}^{\prime}_{\mathrm{L}}(\epsilon)}{\mathrm{FS}}=0.
\end{equation}
The same manipulations deliver
\begin{equation}
\ev{\widehat{H}^{\prime}_{\mathrm{R}}(\epsilon)}{\mathrm{FS}}=0.
\end{equation}

Altogether, we find no contribution at first order,
\begin{equation}
\widehat{H}^{(1)}(\epsilon)\equiv0.
\end{equation}

\subsubsection{Second order}

We turn our attention to the
calculation of the $16$ terms of
Hamiltonian \eqref{eq:Def Effective Ham PT d},
whereby we drop the explicit dependence on $\epsilon$ for conciseness.
This will be done in four steps.
First, we will show that the $4$ homogeneous terms
\begin{subequations}
\label{eq:PT Energy Shifts}
\begin{align}
&\widehat{\mathds{P}}^{\,}_{0}\,
\widehat{H}^{\,}_{\mathrm{L}}\,
\widehat{S}\,
\widehat{H}^{\,}_{\mathrm{L}}\,
\widehat{\mathds{P}}^{\,}_{0},
\qquad
\widehat{\mathds{P}}^{\,}_{0}\,
\widehat{H}^{\,}_{\mathrm{R}}\,
\widehat{S}\,
\widehat{H}^{\,}_{\mathrm{R}}\,
\widehat{\mathds{P}}^{\,}_{0},
\label{eq:PT Energy Shifts a}
\\
&\widehat{\mathds{P}}^{\,}_{0}\,
\widehat{H}^{\prime}_{\mathrm{L}}\,
\widehat{S}\,
\widehat{H}^{\prime}_{\mathrm{L}}\,
\widehat{\mathds{P}}^{\,}_{0},
\qquad
\widehat{\mathds{P}}^{\,}_{0}\,
\widehat{H}^{\prime}_{\mathrm{R}}\,
\widehat{S}\,
\widehat{H}^{\prime}_{\mathrm{R}}\,
\widehat{\mathds{P}}^{\,}_{0}
\label{eq:PT Energy Shifts b}
\end{align}
\end{subequations}
of Hamiltonian \eqref{eq:Def Effective Ham PT d},
only contribute as energy shifts (that we will not compute explicitly).
Second, we will show that the $8$ mixed terms
\begin{subequations}
\label{eq:PT Vanish}
\begin{align}
&\widehat{\mathds{P}}^{\,}_{0}\,
\widehat{H}^{\,}_{\mathrlap{\mathrm{L}}\hphantom{R}}\,
\widehat{S}\,
\widehat{H}^{\prime}_{\mathrlap{\mathrm{L}}\hphantom{R}}\,
\widehat{\mathds{P}}^{\,}_{0}
=
\widehat{\mathds{P}}^{\,}_{0}\,
\widehat{H}^{\prime}_{\mathrlap{\mathrm{L}}\hphantom{R}}\,
\widehat{S}\,
\widehat{H}^{\,}_{\mathrlap{\mathrm{L}}\hphantom{R}}\,
\widehat{\mathds{P}}^{\,}_{0},
\\
&\widehat{\mathds{P}}^{\,}_{0}\,
\widehat{H}^{\,}_{\mathrm{R}}\,
\widehat{S}\,
\widehat{H}^{\prime}_{\mathrm{R}}\,
\widehat{\mathds{P}}^{\,}_{0}
=
\widehat{\mathds{P}}^{\,}_{0}\,
\widehat{H}^{\prime}_{\mathrm{R}}\,
\widehat{S}\,
\widehat{H}^{\,}_{\mathrm{R}}\,
\widehat{\mathds{P}}^{\,}_{0},
\\
&\widehat{\mathds{P}}^{\,}_{0}\,
\widehat{H}^{\,}_{\mathrlap{\mathrm{L}}\hphantom{R}}\,
\widehat{S}\,
\widehat{H}^{\prime}_{\mathrm{R}}\,
\widehat{\mathds{P}}^{\,}_{0}
=
\widehat{\mathds{P}}^{\,}_{0}\,
\widehat{H}^{\prime}_{\mathrm{R}}\,
\widehat{S}\,
\widehat{H}^{\,}_{\mathrlap{\mathrm{L}}\hphantom{R}}\,
\widehat{\mathds{P}}^{\,}_{0},
\\
&\widehat{\mathds{P}}^{\,}_{0}\,
\widehat{H}^{\,}_{\mathrm{R}}\,
\widehat{S}\,
\widehat{H}^{\prime}_{\mathrlap{\mathrm{L}}\hphantom{R}}\,
\widehat{\mathds{P}}^{\,}_{0}
=
\widehat{\mathds{P}}^{\,}_{0}\,
\widehat{H}^{\prime}_{\mathrlap{\mathrm{L}}\hphantom{R}}\,
\widehat{S}\,
\widehat{H}^{\,}_{\mathrm{R}}\,
\widehat{\mathds{P}}^{\,}_{0}
\end{align}
\end{subequations}
vanish. Third, we will show that the $2$ mixed terms
\begin{equation}
\widehat{\mathds{P}}^{\,}_{0}\,
\widehat{H}^{\,}_{\mathrm{R}}\,
\widehat{S}\,
\widehat{H}^{\,}_{\mathrm{L}}\,
\widehat{\mathds{P}}^{\,}_{0}
=
\widehat{\mathds{P}}^{\,}_{0}\,
\widehat{H}^{\,}_{\mathrm{L}}\,
\widehat{S}\,
\widehat{H}^{\,}_{\mathrm{R}}\,
\widehat{\mathds{P}}^{\,}_{0}
\label{eq:PT J xy}
\end{equation}
contribute to the term proportional to $J^{\,}_{xy}$
in Hamiltonian \eqref{eq:H Effective Boundary Spins}.
Finally, we will show that the $2$ mixed terms
\begin{equation}
\widehat{\mathds{P}}^{\,}_{0}\,
\widehat{H}^{\prime}_{\mathrm{R}}\,
\widehat{S}\,
\widehat{H}^{\prime}_{\mathrm{L}}\,
\widehat{\mathds{P}}^{\,}_{0}
=
\widehat{\mathds{P}}^{\,}_{0}\,
\widehat{H}^{\prime}_{\mathrm{L}}\,
\widehat{S}\,
\widehat{H}^{\prime}_{\mathrm{R}}\,
\widehat{\mathds{P}}^{\,}_{0}
\label{eq:PT J z}
\end{equation}
contribute to the term proportional to $J^{\,}_{z}$
in Hamiltonian \eqref{eq:H Effective Boundary Spins}.

\paragraph{Computation of the terms \eqref{eq:PT Energy Shifts}.}

We start by studying the terms
\eqref{eq:PT Energy Shifts a}.
After committing the same abuse of notation as in
Eq.~\eqref{eq:Compute Fisrt Order H'L},
we have to evaluate
\begin{equation}
\ev{\widehat{H}^{\,}_{\mathrm{L}}\,
\widehat{S}\,
\widehat{H}^{\,}_{\mathrm{L}}}
{\mathrm{FS}}=
4\,
\epsilon^{2}\,
\mathcal{N}^{\,6}\,
\left(
\widehat{H}^{\,}_{1}
+
\widehat{H}^{\,}_{2}
\right),
\end{equation}
where
\begin{align}
\widehat{H}^{\,}_{i}\coloneqq&\,
\sum_{l^{\,}_{1}=1}^{2N-2}
\sum_{l^{\,}_{2}=1}^{N-1}
\sum_{l^{\,}_{3}=N}^{2N-2}
\sum_{n^{\,}_{1}=1}^{2N-2}
\sum_{n^{\,}_{2}=1}^{N-1}
\sum_{n^{\,}_{3}=N}^{2N-2}
\nonumber\\
&\,\times
\sin\left(\frac{ \pi l^{\,}_{1}}{2N-1}\right)
\sin\left(\frac{2\pi l^{\,}_{2}}{2N-1}\right)
\sin\left(\frac{2\pi l^{\,}_{3}}{2N-1}\right)
\nonumber\\
&\,\times
\sin\left(\frac{ \pi n^{\,}_{1}}{2N-1}\right)
\sin\left(\frac{2\pi n^{\,}_{2}}{2N-1}\right)
\sin\left(\frac{2\pi n^{\,}_{3}}{2N-1}\right)
\nonumber\\
&\,\times
\hat{h}^{\,}_{i}
\label{eq:Big Sum Hi}
\end{align}
for $i=1,2$.
Here, the operators supported on site $1$ are given by
\begin{subequations}
\label{eq:h1 and h2}
\begin{align}
&
\hat{h}^{\,}_{1}\coloneqq
\frac{
\ev{\,{\bm{:}\,}\hat{f}^{\dag}_{l^{\,}_{3}}\,\hat{f}^{\,}_{l^{\,}_{2}}{\bm{:}}\,
\Bigl(\hat{f}^{\dag}_{l^{\,}_{1}}\,\hat{f}^{\,}_{n^{\,}_{1}}\Bigr)\,
{\bm{:}\,}\hat{f}^{\dag}_{n^{\,}_{2}}\,\hat{f}^{\,}_{n^{\,}_{3}}{\bm{:}}\,}
{\mathrm{FS}}
  }
  {
\varepsilon^{\,}_{n^{\,}_{1}}-\varepsilon^{\,}_{n^{\,}_{2}}+\varepsilon^{\,}_{n^{\,}_{3}}
  }\,
\hat{c}^{\,}_{1}\,\hat{c}^{\dag}_{1},
\\
&
\hat{h}^{\,}_{2}\coloneqq
\frac{
\ev{\,{\bm{:}\,}\hat{f}^{\dag}_{l^{\,}_{3}}\,\hat{f}^{\,}_{l^{\,}_{2}}{\bm{:}}\,
\Bigl(\hat{f}^{\,}_{l^{\,}_{1}}\,\hat{f}^{\dag}_{n^{\,}_{1}}\Bigr)\,
{\bm{:}\,}\hat{f}^{\dag}_{n^{\,}_{2}}\,\hat{f}^{\,}_{n^{\,}_{3}}{\bm{:}}\,}
{\mathrm{FS}}
  }
  {
-\varepsilon^{\,}_{n^{\,}_{1}}-\varepsilon^{\,}_{n^{\,}_{2}}+\varepsilon^{\,}_{n^{\,}_{3}}
  }\,
\hat{c}^{\dag}_{1}\,\hat{c}^{\,}_{1}.
\end{align}
\end{subequations}
We then perform the Wick contractions
\begin{widetext}
\begin{subequations}
\label{eq:Wick Contractions Easy}
\begin{align}
&\ev{\,{\bm{:}\,}\hat{f}^{\dag}_{l^{\,}_{3}}\,\hat{f}^{\,}_{l^{\,}_{2}}{\bm{:}}\,
\Bigl(\hat{f}^{\dag}_{l^{\,}_{1}}\hat{f}^{\,}_{n^{\,}_{1}}\Bigr)\,
{\bm{:}\,}\hat{f}^{\dag}_{n^{\,}_{2}}\,\hat{f}^{\,}_{n^{\,}_{3}}{\bm{:}}\,}
{\mathrm{FS}}
=
\delta^{\,}_{l^{\,}_{1},n^{\,}_{1}}
\delta^{\,}_{l^{\,}_{2},n^{\,}_{2}}
\delta^{\,}_{l^{\,}_{3},n^{\,}_{3}}
\Theta(n^{\,}_{1}-N+1)
+
\delta_{l^{\,}_{1},l^{\,}_{2}}
\delta^{\,}_{n^{\,}_{1},n^{\,}_{2}}
\delta^{\,}_{l^{\,}_{3},n^{\,}_{3}}
-
\delta^{\,}_{l^{\,}_{1},n^{\,}_{3}}
\delta^{\,}_{l^{\,}_{2},n^{\,}_{2}}
\delta^{\,}_{l^{\,}_{3},n^{\,}_{1}},
\label{eq:Wick Contractions Easy a}
\\
&\ev{\,{\bm{:}\,}\hat{f}^{\dag}_{l^{\,}_{3}}\,\hat{f}^{\,}_{l^{\,}_{2}}{\bm{:}}\,
\Bigl(\hat{f}^{\,}_{l^{\,}_{1}}\hat{f}^{\dag}_{n^{\,}_{1}}\Bigr)\,
{\bm{:}\,}\hat{f}^{\dag}_{n^{\,}_{2}}\,\hat{f}^{\,}_{n^{\,}_{3}}{\bm{:}}\,}
{\mathrm{FS}}
=
\delta^{\,}_{l^{\,}_{1},n^{\,}_{1}}
\delta^{\,}_{l^{\,}_{2},n^{\,}_{2}}
\delta^{\,}_{l^{\,}_{3},n^{\,}_{3}}
\Theta(N-n^{\,}_{1})
-
\delta_{l^{\,}_{1},l^{\,}_{2}}
\delta^{\,}_{n^{\,}_{1},n^{\,}_{2}}
\delta^{\,}_{l^{\,}_{3},n^{\,}_{3}}
+
\delta^{\,}_{l^{\,}_{1},n^{\,}_{3}}
\delta^{\,}_{l^{\,}_{2},n^{\,}_{2}}
\delta^{\,}_{l^{\,}_{3},n^{\,}_{1}},
\end{align}
\label{eq:Wick Contractions Easy b}
\end{subequations}
\end{widetext}
where we have introduced the Heaviside function $\Theta$
with the convention $\Theta(0)=0$.
We insert Eq.~\eqref{eq:Wick Contractions Easy}
into Eq.~\eqref{eq:h1 and h2} and then into
Eq.~\eqref{eq:Big Sum Hi}.
Making use of the change of variables
\eqref{eq:Change Variables PT a}
which induces the changes \eqref{eq:Change Variables PT b} and
\begin{equation}
\varepsilon^{\,}_{n^{\,}_{i}}
\mapsto
-\varepsilon^{\,}_{n^{\,}_{i}},
\label{eq:Change Variables PT Epsilon}
\end{equation}
we find that
\begin{equation}
\widehat{H}^{\,}_{1}=
C\,
\hat{c}^{\,}_{1}\,
\hat{c}^{\dag}_{1},
\qquad
\widehat{H}^{\,}_{2}=
C\,
\hat{c}^{\dag}_{1}\,
\hat{c}^{\,}_{1},
\end{equation}
where $C\in\mathbb{R}$.

Therefore, we have found that
\begin{equation}
\widehat{\mathds{P}}^{\,}_{0}\,
\widehat{H}^{\,}_{\mathrm{L}}\,
\widehat{S}\,
\widehat{H}^{\,}_{\mathrm{L}}\,
\widehat{\mathds{P}}^{\,}_{0}
\propto
\epsilon^{2}
\left(
\hat{c}^{\dag}_{1}\,\hat{c}^{\,}_{1}
+
\hat{c}^{\,}_{1}\,\hat{c}^{\dag}_{1}
\right)
\equiv
\epsilon^{2}.
\end{equation}
The same calculation yields
\begin{equation}
\widehat{\mathds{P}}^{\,}_{0}\,
\widehat{H}^{\,}_{\mathrm{R}}\,
\widehat{S}\,
\widehat{H}^{\,}_{\mathrm{R}}\,
\widehat{\mathds{P}}^{\,}_{0}
\propto
\epsilon^{2}.
\end{equation}

The task of evaluating the terms
\eqref{eq:PT Energy Shifts b} is straightforward.
Indeed, it suffices to observe that 
the fermionic operators acting on the boundary sites
$j=1,2N$ are decoupled from the fermionic operators
acting on the bulk sites $j=2,\cdots,2N-1$.
Therefore, we have
\begin{equation}
\widehat{\mathds{P}}^{\,}_{0}\,
\widehat{H}^{\prime}_{\mathrm{L}}\,
\widehat{S}\,
\widehat{H}^{\prime}_{\mathrm{L}}\,
\widehat{\mathds{P}}^{\,}_{0}
\propto
\epsilon^{2}
\left(2\,\hat{c}^{\dag}_{1}\,\hat{c}^{\,}_{1}
-1\right)^{2}
\equiv
\epsilon^{2}.
\end{equation}
Similarly, $\widehat{\mathds{P}}^{\,}_{0}\,
\widehat{H}^{\prime}_{\mathrm{R}}\,
\widehat{S}\,
\widehat{H}^{\prime}_{\mathrm{R}}\,
\widehat{\mathds{P}}^{\,}_{0}
\propto
\epsilon^{2}$.

\paragraph{Computation of the terms \eqref{eq:PT Vanish}.}

We observe that
$\widehat{H}^{\,}_{\mathrlap{\mathrm{L}}\hphantom{R}}\,
\widehat{S}\,
\widehat{H}^{\prime}_{\mathrlap{\mathrm{L}}\hphantom{R}}$
does not conserve the number
\eqref{eq:Fermion Number Bulk PT}, hence
\begin{equation}
\widehat{\mathds{P}}^{\,}_{0}\,
\widehat{H}^{\,}_{\mathrlap{\mathrm{L}}\hphantom{R}}\,
\widehat{S}\,
\widehat{H}^{\prime}_{\mathrlap{\mathrm{L}}\hphantom{R}}\,
\widehat{\mathds{P}}^{\,}_{0}\equiv0.
\end{equation}
This is also so for any of the other terms in
Eq.~\eqref{eq:PT Vanish}.

\paragraph{Computation of the terms \eqref{eq:PT J xy}.}

These terms contribute to the term proportional
to $J^{\,}_{xy}$ in Hamiltonian
\eqref{eq:H Effective Boundary Spins}.
Hence, we compute them explicitly.

\begin{widetext}
We commit the same abuse of notation as in
Eq.~\eqref{eq:Compute Fisrt Order H'L}
and evaluate
\begin{subequations}
\begin{equation}
\ev{\widehat{H}^{\,}_{\mathrm{R}}\,
\widehat{S}\,
\widehat{H}^{\,}_{\mathrm{L}}}
{\mathrm{FS}}=
4\,
\epsilon^{2}\mathcal{N}^{\,6}\,
\left(\widehat{H}^{\prime}_{1}+\widehat{H}^{\prime}_{2}\right),
\label{eq:HR S HL Over FS}
\end{equation}
where
\begin{align}
\widehat{H}^{\prime}_{i}
\coloneqq&\,
\sum_{l^{\,}_{1}=1}^{2N-2}
\sum_{l^{\,}_{2}=1}^{N-1}
\sum_{l^{\,}_{3}=N}^{2N-2}
\sum_{n^{\,}_{1}=1}^{2N-2}
\sum_{n^{\,}_{2}=1}^{N-1}
\sum_{n^{\,}_{3}=N}^{2N-2}
\nonumber\\
&\,\times
\sin\left(\frac{\pi l^{\,}_{1}\,(2N-2)}{2N-1}\right)
\sin\left(\frac{\pi l^{\,}_{2}\,(2N-3)}{2N-1}\right)
\sin\left(\frac{\pi l^{\,}_{3}\,(2N-3)}{2N-1}\right)
\sin\left(\frac{\pi n^{\,}_{1}}{2N-1}\right)
\sin\left(\frac{2\pi n^{\,}_{2}}{2N-1}\right)
\sin\left(\frac{2\pi n^{\,}_{3}}{2N-1}\right)
\nonumber\\
&\,\times
\hat{h}^{\prime}_{i},
\label{eq:Big Sum H'i}
\end{align}
for $i=1,2$.
Here, the operators supported on site $1$ and $2N$ are given by 
\begin{equation}
\hat{h}^{\prime}_{1}\coloneqq
\frac{
\ev{\,{\bm{:}\,}\hat{f}^{\dag}_{l^{\,}_{3}}\,\hat{f}^{\,}_{l^{\,}_{2}}{\bm{:}}\,
\Bigl(\hat{f}^{\dag}_{l^{\,}_{1}}\,\hat{f}^{\,}_{n^{\,}_{1}}\Bigr)\,
{\bm{:}\,}\hat{f}^{\dag}_{n^{\,}_{2}}\,\hat{f}^{\,}_{n^{\,}_{3}}{\bm{:}}\,}
 {\mathrm{FS}}
  }
  {
\varepsilon^{\,}_{n^{\,}_{1}}
-
\varepsilon^{\,}_{n^{\,}_{2}}
+
\varepsilon^{\,}_{n^{\,}_{3}}
  }\,
\hat{c}^{\,}_{2N}\,\hat{c}^{\dag}_{1},
\qquad
\hat{h}^{\prime}_{2}\coloneqq
\frac{
\ev{\,{\bm{:}\,}\hat{f}^{\dag}_{l^{\,}_{3}}\,\hat{f}^{\,}_{l^{\,}_{2}}{\bm{:}}\,
\Bigl(\hat{f}^{\,}_{l^{\,}_{1}}\hat{f}^{\dag}_{n^{\,}_{1}}\Bigr)\,
{\bm{:}\,}\hat{f}^{\dag}_{n^{\,}_{2}}\,\hat{f}^{\,}_{n^{\,}_{3}}{\bm{:}}\,}
{\mathrm{FS}}
  }
  {
-\varepsilon^{\,}_{n^{\,}_{1}}-\varepsilon^{\,}_{n^{\,}_{2}}+\varepsilon^{\,}_{n^{\,}_{3}}
  }\,
\hat{c}^{\dag}_{2N}\,\hat{c}^{\,}_{1}.
\label{eq:h'1 and h'2}
\end{equation}
\end{subequations}
If we insert Eq.~\eqref{eq:Wick Contractions Easy}
into Eq.~\eqref{eq:h'1 and h'2},
make use of the identity
\begin{equation}
\sin\left(\frac{\pi n^{\,}_{i}\,(2N-m-1)}{2N-1}\right)=
(-1)^{n^{\,}_{i}+1}
\sin\!\left(\frac{\pi n^{\,}_{i}\,m}{2N-1}\right)\!
\label{eq:From 2N-m-1 To m}
\end{equation}
valid for any integer $m-1=1,\ldots,2N$,
and make use of the change of variables
\eqref{eq:Change Variables PT a}
that induces the changes
\eqref{eq:Change Variables PT b}
and
\eqref{eq:Change Variables PT Epsilon},
we find
\begin{subequations}
\begin{equation}
\hat{h}^{\prime}_{1}=
-S\,
\hat{c}^{\,}_{2N}\,\hat{c}^{\dag}_{1},
\qquad
\hat{h}^{\prime}_{2}=
+S\,
\hat{c}^{\dag}_{2N}\,\hat{c}^{\,}_{1},
\end{equation}
where
\begin{align}
S\coloneqq&\,
\frac{1}{2}
\sum_{n^{\,}_{1}=1}^{N-1}
\sum_{n^{\,}_{2}=1}^{N-1}
\sum_{n^{\,}_{3}=1}^{N-1}
(-1)^{n^{\,}_{1}+n^{\,}_{2}+n^{\,}_{3}+1}
\nonumber\\
&\,\times
\frac{
\sin\!\left(\frac{\pi n^{\,}_{1}}{2N-1}\right)
\sin\!\left(\frac{2\pi n^{\,}_{2}}{2N-1}\right)
\sin^{2}\!\left(\frac{2\pi n^{\,}_{3}}{2N-1}\right)
\left[
\sin\!\left(\frac{\pi n^{\,}_{1}}{2N-1}\right)
\sin\!\left(\frac{2\pi n^{\,}_{2}}{2N-1}\right)
-
\sin\!\left(\frac{2\pi n^{\,}_{1}}{2N-1}\right)
\sin\!\left(\frac{\pi n^{\,}_{2}}{2N-1}\right)
\right]
}{\cos\!\left(\frac{\pi n^{\,}_{1}}{2N-1}\right)
+\cos\!\left(\frac{\pi n^{\,}_{2}}{2N-1}\right)
+\cos\!\left(\frac{\pi n^{\,}_{3}}{2N-1}\right)}
\label{eq:Awful Sum For J xy}
\end{align}
\end{subequations}
\end{widetext}
Combining everything into
Eq.~\eqref{eq:HR S HL Over FS},
we find that
\begin{equation}
\widehat{\mathds{P}}^{\,}_{0}\,
\widehat{H}^{\,}_{\mathrm{R}}\,
\widehat{S}\,
\widehat{H}^{\,}_{\mathrm{L}}\,
\widehat{\mathds{P}}^{\,}_{0}
=
8\,
\epsilon^{2}\,
\mathcal{N}^{\,6}\,
S\,
\left(
\hat{c}^{\dag}_{1}\,\hat{c}^{\,}_{2N}\,
+
\hat{c}^{\dag}_{2N}\,\hat{c}^{\,}_{1}
\right).
\label{eq:HR S HL Final}
\end{equation}
As the right-hand side is Hermitian,
the identity \eqref{eq:PT J xy} holds.
Moreover, inserting the expressions
\eqref{eq:c's in terms of spins}
into Eq.~\eqref{eq:HR S HL Final}
delivers the term proportional
to $J^{\,}_{xy}$ in Hamiltonian
\eqref{eq:H Effective Boundary Spins}, whereby
the coupling $J^{\,}_{xy}\equiv
\bar{J}^{\,}_{xy}$ with
[the multiplicative $\sqrt{2}$ arises from inverting
the change of parametrization \eqref{eq:Param PT Boundary Spins}]
\begin{equation}
\bar{J}^{\,}_{xy}=
-8\sqrt{2}\,
\epsilon^{2}
\mathcal{N}^{\,6}
S.
\label{eq:J xy Not Yet Estimated}
\end{equation}
We will give in
Sec.~\ref{sub:estimates_of_the_effective_couplings}
an estimate of the dependence of $\bar{J}^{\,}_{xy}$
as a function of $2N$ both
(i) in the range $2N\in[10,30]$ accessible with ED and
(ii) in the limit $2N\gg1$.

\paragraph{Computation of the terms \eqref{eq:PT J z}.}

These terms contribute to the term proportional
to $J^{\,}_{z}$ in Hamiltonian
\eqref{eq:H Effective Boundary Spins}.
Hence, we also compute them explicitly.

We commit the same abuse of notation as in
Eq.~\eqref{eq:Compute Fisrt Order H'L}
and evaluate
\begin{widetext}
\begin{equation}
\ev{\widehat{H}^{\prime}_{\mathrm{R}}\,
\widehat{S}\,
\widehat{H}^{\prime}_{\mathrm{L}}}
{\mathrm{FS}}=
4\,
\epsilon^{2}\,
\mathcal{N}^{8}\,
Z\,
(2\,\hat{c}^{\dag}_{1}\,\hat{c}^{\,}_{1}-1)\,
(2\,\hat{c}^{\dag}_{2N}\,\hat{c}^{\,}_{2N}-1),
\qquad
Z\coloneqq
Z^{\,}_{1}+Z^{\,}_{2}+Z^{\,}_{3}+Z^{\,}_{4},
\label{eq:H'R S H'L Over FS}
\end{equation}
where
\begin{align}
Z^{\,}_{i}
\coloneqq&\,
\sum_{l^{\,}_{1}=1}^{2N-2}
\sum_{l^{\,}_{2}=1}^{2N-2}
\sum_{l^{\,}_{3}=1}^{N-1}
\sum_{l^{\,}_{4}=N}^{2N-2}
\sum_{n^{\,}_{1}=1}^{2N-2}
\sum_{n^{\,}_{2}=1}^{2N-2}
\sum_{n^{\,}_{3}=1}^{N-1}
\sum_{n^{\,}_{4}=N}^{2N-2}
\nonumber\\
&\,\times
\sin\!\left(\frac{\pi n^{\,}_{1}}{2N-1}\right)
\sin\!\left(\frac{2\pi n^{\,}_{2}}{2N-1}\right)
\sin\!\left(\frac{3\pi n^{\,}_{3}}{2N-1}\right)
\sin\!\left(\frac{3\pi n^{\,}_{4}}{2N-1}\right)
\nonumber \\
&\,\times
\sin\!\left[\frac{\pi l^{\,}_{1}\,(2N-2)}{2N-1}\right]
\sin\!\left[\frac{\pi l^{\,}_{2}\,(2N-3)}{2N-1}\right]
\sin\!\left[\frac{\pi l^{\,}_{3}\,(2N-4)}{2N-1}\right]
\sin\!\left[\frac{\pi l^{\,}_{4}\,(2N-4)}{2N-1}\right]
\nonumber\\
&\,\times
z^{\,}_{i},
\label{eq:Big Sum Zi}
\end{align}
with
\begin{subequations}
\label{eq:z1 z2 z3 and z4}
\begin{align}
&z^{\,}_{1}\coloneqq
\ev{\,{\bm{:}\,}\hat{f}^{\dag}_{l^{\,}_{4}}\,
\hat{f}^{\,}_{l^{\,}_{3}}{\bm{:}}\,
\hat{f}^{\dag}_{l^{\,}_{2}}\,
\hat{f}^{\,}_{l^{\,}_{1}}\,
\frac{1-\dyad{\mathrm{FS}}}
{-\varepsilon^{\,}_{n^{\,}_{1}}+
\varepsilon^{\,}_{n^{\,}_{2}}-
\varepsilon^{\,}_{n^{\,}_{3}}+
\varepsilon^{\,}_{n^{\,}_{4}}}\,
\hat{f}^{\dag}_{n^{\,}_{1}}\,
\hat{f}^{\,}_{n^{\,}_{2}}
{\bm{:}\,}\hat{f}^{\dag}_{n^{\,}_{3}}\,
\hat{f}^{\,}_{n^{\,}_{4}}{\bm{:}}\,}
{\mathrm{FS}},
\\
&z^{\,}_{2}\coloneqq
\ev{\,{\bm{:}\,}\hat{f}^{\dag}_{l^{\,}_{4}}\,
\hat{f}^{\,}_{l^{\,}_{3}}{\bm{:}}\,
\hat{f}^{\dag}_{l^{\,}_{2}}\,
\hat{f}^{\,}_{l^{\,}_{1}}\,
\frac{1-\dyad{\mathrm{FS}}}
{\hphantom{-}\varepsilon^{\,}_{n^{\,}_{1}}-
\varepsilon^{\,}_{n^{\,}_{2}}-
\varepsilon^{\,}_{n^{\,}_{3}}+
\varepsilon^{\,}_{n^{\,}_{4}}}\,
\hat{f}^{\dag}_{n^{\,}_{2}}\,
\hat{f}^{\,}_{n^{\,}_{1}}
{\bm{:}\,}\hat{f}^{\dag}_{n^{\,}_{3}}\,
\hat{f}^{\,}_{n^{\,}_{4}}{\bm{:}}\,}
{\mathrm{FS}},
\\
&z^{\,}_{3}\coloneqq
\ev{\,{\bm{:}\,}\hat{f}^{\dag}_{l^{\,}_{4}}\,
\hat{f}^{\,}_{l^{\,}_{3}}{\bm{:}}\,
\hat{f}^{\dag}_{l^{\,}_{1}}\,
\hat{f}^{\,}_{l^{\,}_{2}}\,
\frac{1-\dyad{\mathrm{FS}}}
{-\varepsilon^{\,}_{n^{\,}_{1}}+
\varepsilon^{\,}_{n^{\,}_{2}}-
\varepsilon^{\,}_{n^{\,}_{3}}+
\varepsilon^{\,}_{n^{\,}_{4}}}\,
\hat{f}^{\dag}_{n^{\,}_{1}}\,
\hat{f}^{\,}_{n^{\,}_{2}}
{\bm{:}\,}\hat{f}^{\dag}_{n^{\,}_{3}}\,
\hat{f}^{\,}_{n^{\,}_{4}}{\bm{:}}\,}
{\mathrm{FS}},
\\
&z^{\,}_{4}\coloneqq
\ev{\,{\bm{:}\,}\hat{f}^{\dag}_{l^{\,}_{4}}\,
\hat{f}^{\,}_{l^{\,}_{3}}{\bm{:}}\,
\hat{f}^{\dag}_{l^{\,}_{1}}\,
\hat{f}^{\,}_{l^{\,}_{2}}\,
\frac{1-\dyad{\mathrm{FS}}}
{\hphantom{-}\varepsilon^{\,}_{n^{\,}_{1}}-
\varepsilon^{\,}_{n^{\,}_{2}}-
\varepsilon^{\,}_{n^{\,}_{3}}+
\varepsilon^{\,}_{n^{\,}_{4}}}\,
\hat{f}^{\dag}_{n^{\,}_{2}}\,
\hat{f}^{\,}_{n^{\,}_{1}}
{\bm{:}\,}\hat{f}^{\dag}_{n^{\,}_{3}}\,
\hat{f}^{\,}_{n^{\,}_{4}}{\bm{:}}\,}
{\mathrm{FS}}.
\end{align}
\end{subequations}
Each of the $z^{\,}_{i}$, $i=1,2,3,4$ in
Eq.~\eqref{eq:z1 z2 z3 and z4} is computed using Wick's theorem
and yields a sum of $13$ non-vanishing terms in the form of
products of $4$ Kronecker-$\delta$.
Each of the $z^{\,}_{i}$, $i=1,2,3,4$ is then inserted into
Eq.~\eqref{eq:Big Sum Zi} and in
Eq.~\eqref{eq:H'R S H'L Over FS} so that $Z$
becomes a sum of $4\times 13=52$ terms.

These $52$ terms can be reduced to
\begin{subequations}
\label{eq:Awful Sum For J z}
\begin{equation}
Z=
\sum_{i=1}^{8}\,
S^{\,}_{i},\qquad
S^{\,}_{i}\coloneqq
\sum_{n^{\,}_{1}=1}^{N-1}
\sum_{n^{\,}_{2}=1}^{N-1}
\sum_{n^{\,}_{3}=1}^{N-1}
\sum_{n^{\,}_{4}=1}^{N-1}\,
s^{\,}_{i}\,(n^{\,}_{1},n^{\,}_{2},n^{\,}_{3},n^{\,}_{4}),
\end{equation}
where
\begin{align}
% 2 * A1
s^{\,}_{1}
&\coloneqq
2\times
(-1)^{n^{\,}_{1}+n^{\,}_{2}+n^{\,}_{3}+n^{\,}_{4}+1}\,
\frac{
\sin^{2}\!\left(\frac{\pi n^{\,}_{1}}{2N-1}\right)
\sin^{2}\!\left(\frac{2\pi n^{\,}_{2}}{2N-1}\right)
\sin^{2}\!\left(\frac{3\pi n^{\,}_{3}}{2N-1}\right)
\sin^{2}\!\left(\frac{3\pi n^{\,}_{4}}{2N-1}\right)
}{
2\cos(\frac{\pi n^{\,}_{1}}{2N-1})
+2\cos(\frac{\pi n^{\,}_{2}}{2N-1})
+2\cos(\frac{\pi n^{\,}_{3}}{2N-1})
+2\cos(\frac{\pi n^{\,}_{4}}{2N-1})},
\\
% 4 * A3
s^{\,}_{2}
&\coloneqq
4\times
(-1)^{n^{\,}_{1}+n^{\,}_{4}}\,
\frac{
\sin^{2}\!\left(\frac{\pi n^{\,}_{1}}{2N-1}\right)
\sin\!\left(\frac{2\pi n^{\,}_{2}}{2N-1}\right)
\sin\!\left(\frac{3\pi n^{\,}_{2}}{2N-1}\right)
\sin\!\left(\frac{2\pi n^{\,}_{3}}{2N-1}\right)
\sin\!\left(\frac{3\pi n^{\,}_{3}}{2N-1}\right)
\sin^{2}\!\left(\frac{3\pi n^{\,}_{4}}{2N-1}\right)
}{
2\cos(\frac{\pi n^{\,}_{1}}{2N-1})
+2\cos(\frac{\pi n^{\,}_{4}}{2N-1})},
\\
% 2 * A4
s^{\,}_{3}
&\coloneqq
2\times
(-1)^{n^{\,}_{1}+n^{\,}_{2}+n^{\,}_{3}+n^{\,}_{4}}\,
\frac{
\sin\!\left(\frac{\pi n^{\,}_{1}}{2N-1}\right)
\sin\!\left(\frac{3\pi n^{\,}_{1}}{2N-1}\right)
\sin^{2}\!\left(\frac{2\pi n^{\,}_{2}}{2N-1}\right)
\sin\!\left(\frac{\pi n^{\,}_{3}}{2N-1}\right)
\sin\!\left(\frac{3\pi n^{\,}_{3}}{2N-1}\right)
\sin^{2}\!\left(\frac{3\pi n^{\,}_{4}}{2N-1}\right)
}{
2\cos(\frac{\pi n^{\,}_{1}}{2N-1})
+2\cos(\frac{\pi n^{\,}_{2}}{2N-1})
+2\cos(\frac{\pi n^{\,}_{3}}{2N-1})
+2\cos(\frac{\pi n^{\,}_{4}}{2N-1})},
\\
% 4 * A6
s^{\,}_{4}
&\coloneqq
4\times
(-1)^{n^{\,}_{3}+n^{\,}_{4}}\,
\frac{
\sin\!\left(\frac{\pi n^{\,}_{1}}{2N-1}\right)
\sin\!\left(\frac{2\pi n^{\,}_{1}}{2N-1}\right)
\sin\!\left(\frac{\pi n^{\,}_{2}}{2N-1}\right)
\sin\!\left(\frac{2\pi n^{\,}_{2}}{2N-1}\right)
\sin^{2}\!\left(\frac{3\pi n^{\,}_{3}}{2N-1}\right)
\sin^{2}\!\left(\frac{3\pi n^{\,}_{4}}{2N-1}\right)
}{
2\cos(\frac{\pi n^{\,}_{3}}{2N-1})
+2\cos(\frac{\pi n^{\,}_{4}}{2N-1})},
\\
% 2 * A7
s^{\,}_{5}
&\coloneqq
2\times
(-1)^{n^{\,}_{1}+n^{\,}_{2}+n^{\,}_{3}+n^{\,}_{4}}\,
\frac{
\sin^{2}\!\left(\frac{\pi n^{\,}_{1}}{2N-1}\right)
\sin\!\left(\frac{2\pi n^{\,}_{2}}{2N-1}\right)
\sin\!\left(\frac{3\pi n^{\,}_{2}}{2N-1}\right)
\sin^{2}\!\left(\frac{3\pi n^{\,}_{3}}{2N-1}\right)
\sin\!\left(\frac{2\pi n^{\,}_{4}}{2N-1}\right)
\sin\!\left(\frac{3\pi n^{\,}_{4}}{2N-1}\right)
}{
2\cos(\frac{\pi n^{\,}_{1}}{2N-1})
+2\cos(\frac{\pi n^{\,}_{2}}{2N-1})
+2\cos(\frac{\pi n^{\,}_{3}}{2N-1})
+2\cos(\frac{\pi n^{\,}_{4}}{2N-1})},
\\
% 8 * A8
s^{\,}_{6}
&\coloneqq
8\times
(-1)^{n^{\,}_{1}+n^{\,}_{4}+1}\,
\frac{
\sin\!\left(\frac{\pi n^{\,}_{1}}{2N-1}\right)
\sin\!\left(\frac{3\pi n^{\,}_{1}}{2N-1}\right)
\sin\!\left(\frac{2\pi n^{\,}_{2}}{2N-1}\right)
\sin\!\left(\frac{3\pi n^{\,}_{2}}{2N-1}\right)
\sin\!\left(\frac{\pi n^{\,}_{3}}{2N-1}\right)
\sin\!\left(\frac{2\pi n^{\,}_{3}}{2N-1}\right)
\sin^{2}\!\left(\frac{3\pi n^{\,}_{4}}{2N-1}\right)
}{
2\cos(\frac{\pi n^{\,}_{1}}{2N-1})
+2\cos(\frac{\pi n^{\,}_{4}}{2N-1})},
\\
% 2 * B1
s^{\,}_{7}
&\coloneqq
2\times
(-1)^{n^{\,}_{1}+n^{\,}_{2}+n^{\,}_{3}+n^{\,}_{4}}\,
\frac{
\sin\!\left(\frac{\pi n^{\,}_{1}}{2N-1}\right)
\sin\!\left(\frac{2\pi n^{\,}_{1}}{2N-1}\right)
\sin\!\left(\frac{\pi n^{\,}_{2}}{2N-1}\right)
\sin\!\left(\frac{2\pi n^{\,}_{2}}{2N-1}\right)
\sin^{2}\!\left(\frac{3\pi n^{\,}_{3}}{2N-1}\right)
\sin^{2}\!\left(\frac{3\pi n^{\,}_{4}}{2N-1}\right)
}{
2\cos(\frac{\pi n^{\,}_{1}}{2N-1})
+2\cos(\frac{\pi n^{\,}_{2}}{2N-1})
+2\cos(\frac{\pi n^{\,}_{3}}{2N-1})
+2\cos(\frac{\pi n^{\,}_{4}}{2N-1})},
\\
% 4 * B4
s^{\,}_{8}
&\coloneqq
4\times
(-1)^{n^{\,}_{1}+n^{\,}_{2}+n^{\,}_{3}+n^{\,}_{4}+1}\,
\frac{
\sin\!\left(\frac{\pi n^{\,}_{1}}{2N-1}\right)
\sin\!\left(\frac{2\pi n^{\,}_{1}}{2N-1}\right)
\sin\!\left(\frac{2\pi n^{\,}_{2}}{2N-1}\right)
\sin\!\left(\frac{3\pi n^{\,}_{2}}{2N-1}\right)
\sin\!\left(\frac{\pi n^{\,}_{3}}{2N-1}\right)
\sin\!\left(\frac{3\pi n^{\,}_{3}}{2N-1}\right)
\sin^{2}\!\left(\frac{3\pi n^{\,}_{4}}{2N-1}\right)
}{
2\cos(\frac{\pi n^{\,}_{1}}{2N-1})
+2\cos(\frac{\pi n^{\,}_{2}}{2N-1})
+2\cos(\frac{\pi n^{\,}_{3}}{2N-1})
+2\cos(\frac{\pi n^{\,}_{4}}{2N-1})}.
\end{align}
\end{subequations}
\end{widetext}
This reduction proceeds as follows.
First, we use Eq.~\eqref{eq:From 2N-m-1 To m}.
Second, we make use of the change of variables
\eqref{eq:Change Variables PT a}
which induces the changes \eqref{eq:Change Variables PT b},
\eqref{eq:Change Variables PT Epsilon} and
\begin{equation}
(-1)^{n^{\,}_{i}}\mapsto
(-1)^{n^{\,}_{i}+1}.
\end{equation}
Third, among the $52$ terms, $20$ vanish due to the identity
\begin{equation}
\sum_{n=1}^{N-1}
\sin\!\left(\frac{\pi n}{2N-1}\right)
\sin\!\left(\frac{3\pi n}{2N-1}\right)
=0
\end{equation}
holding for any $2N$.
Fourth, among the remaining $32$ terms, $4$ cancel pairwise
upon direct inspection.
Fifth, the remaining $28$ terms can be grouped into $8$ sets
$\mathcal{S}^{\,}_{i}$, $i=1,\ldots,8$,
within each of which the terms are equal by virtue of suitable
identities and relabeling of the summation indices.
Finally, each of the $S^{\,}_{i}$, $i=1,\ldots,8$
in Eq.~\eqref{eq:Awful Sum For J z} is equal to a term in the set
$\mathcal{S}^{\,}_{i}$
multiplied by the cardinality of the set
$\mathcal{S}^{\,}_{i}$.

Combining everything into
Eq.~\eqref{eq:H'R S H'L Over FS},
we find that
\begin{equation}
\widehat{\mathds{P}}^{\,}_{0}\,
\widehat{H}^{\prime}_{\mathrm{R}}\,
\widehat{S}\,
\widehat{H}^{\prime}_{\mathrm{L}}\,
\widehat{\mathds{P}}^{\,}_{0}
=
4\,
\epsilon^{2}
\mathcal{N}^{8}
Z
\left(2\,\hat{c}^{\dag}_{1}\,\hat{c}^{\,}_{1}-1\right)\!
\left(2\,\hat{c}^{\dag}_{2N}\,\hat{c}^{\,}_{2N}-1\right)\,
\label{eq:H'R S H'L Final}
\end{equation}
As the right-hand side is Hermitian,
the identity \eqref{eq:PT J z} holds.
Moreover, inserting the expressions
\eqref{eq:c's in terms of spins}
into Eq.~\eqref{eq:H'R S H'L Final}
delivers the term proportional
to $J^{\,}_{z}$ in Hamiltonian
\eqref{eq:H Effective Boundary Spins}, whereby
the coupling $J^{\,}_{z}\equiv
\bar{J}^{\,}_{z}$ with
[the multiplicative $\sqrt{2}$ arises from inverting
the change of parametrization \eqref{eq:Param PT Boundary Spins}]
\begin{equation}
\bar{J}^{\,}_{z}=
8\sqrt{2}\,
\epsilon^{2}
\mathcal{N}^{8}
Z.
\label{eq:J z Not Yet Estimated}
\end{equation}
We will give in
Sec.~\ref{sub:estimates_of_the_effective_couplings}
an estimate of the dependence of $\bar{J}^{\,}_{z}$
as a function of $2N$ both
(i) in the range $2N\in[10,30]$ accessible with ED and
(ii) in the limit $2N\gg1$.

\subsection{Estimates of the effective couplings}
\label{sub:estimates_of_the_effective_couplings}

\begin{figure}
\centering
\subfigure[]{\includegraphics[width=1\columnwidth]{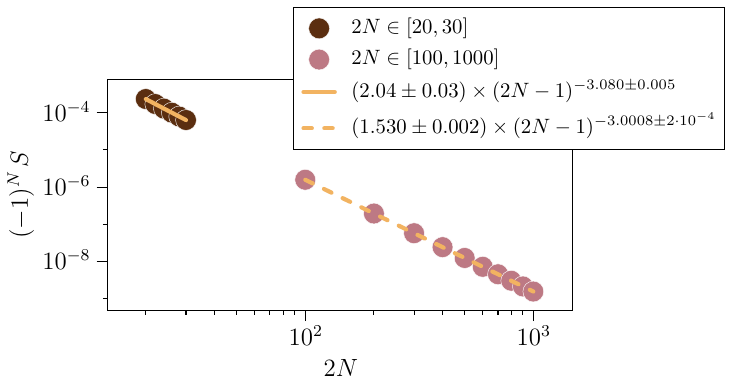} \label{fig:Awful Sums PT a}} \\
\subfigure[]{\includegraphics[width=1\columnwidth]{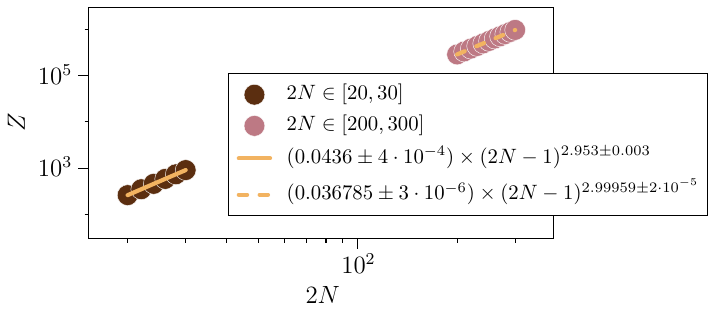} \label{fig:Awful Sums PT b}}
\caption{(a) Dependence on $2N$
of the sum $S$ defined in Eq.~\eqref{eq:Awful Sum For J xy}
in the range $2N\in[20,30]$ of system sizes accessible with ED
and in the range $2N\in[100,1000]$.
(b) Dependence on $2N$
of the sum $Z$ defined in Eq.~\eqref{eq:Awful Sum For J z}
in the range $2N\in[20,30]$ of system sizes accessible with ED
and in the range $2N\in[200,300]$.
}
\label{fig:Awful Sums PT}
\end{figure}

\subsubsection{The coupling $J^{\,}_{xy}$}

We compute the coupling
$J^{\,}_{xy}$ in
Eq.~\eqref{eq:J xy Not Yet Estimated}.
Specifically, we estimate the quantity $S$ defined in
Eq.~\eqref{eq:Awful Sum For J xy}
\begin{itemize}
\item[(I)]
for the range $2N\in[20,30]$ of system sizes accessible with ED
and compare it to \eqref{eq:Jxy Jz Effective Expected a}, 
\item[(II)] and in the thermodynamic limit $2N\to\infty$.
\end{itemize}
On the one hand, we shall compute \eqref{eq:Awful Sum For J xy}
numerically and observe its dependence on $2N-1$
for both (I) and (II).
This is given in Fig.~\ref{fig:Awful Sums PT a}.
On the other hand, only for (II), we shall estimate
analytically the leading
term in $2N-1$ of \eqref{eq:Awful Sum For J xy}.
More precisely, we will show that
\begin{equation}
S=
(-1)^{N}\frac{\,\pi^{4}}{64}\,(2N-1)^{-3}
+\mathcal{O}\bigl(N^{-5}\bigr).
\label{eq:Leading S PT Analytical}
\end{equation}
Notice that it was not possible to obtain a closed form
of \eqref{eq:Awful Sum For J xy} as a function of $2N-1$.

\paragraph{Regime (I).}

In Fig.~\ref{fig:Awful Sums PT a} we find that in the range
$2N\in[20,30]$ of system sizes accessible with ED,
\begin{subequations}
\begin{equation}
S=(-1)^{N}\times(2.04\pm0.03)\times
(2N-1)^{\bar{\kappa}^{\,}_{xy}},
\end{equation}
where
\begin{equation}
\bar{\kappa}^{\,}_{xy}=
-3.080\pm0.005.
\label{eq:Kappa xy Estimate}
\end{equation}
\end{subequations}
This we insert into Eq.~\eqref{eq:J xy Not Yet Estimated} and,
recalling Eq.~\eqref{eq:Fourier Modes PT c},
we obtain
\begin{equation}
|\bar{J}^{\,}_{xy}|=
64\sqrt{2}\,
\epsilon^{2}\times
(2.04\pm0.03)\times
(2N-1)^{\bar{\kappa}^{\,}_{xy}-3}.
\label{eq:J xy Low 2N Analytical}
\end{equation}

We now compare $\bar{J}^{\,}_{xy}$ estimated
in Eq.~\eqref{eq:J xy Low 2N Analytical}
to $J^{\,}_{xy}$ obtained numerically with ED
in Eq.~\eqref{eq:Jxy Jz Effective Expected a}.
On the one hand, we observe that
$\kappa^{\,}_{xy}$ in Eq.~\eqref{eq:Jxy Jz Effective Expected a}
differs from
$\bar{\kappa}^{\,}_{xy}$ in Eq.~\eqref{eq:Kappa xy Estimate}
by $1\%$, i.e., we can write
\begin{equation}
\kappa^{\,}_{xy}
\approx
\bar{\kappa}^{\,}_{xy}.
\end{equation}
On the other hand, we observe that
\begin{subequations}
\begin{align}
&1/4\times A^{\,}_{\mathrm{p}}=
137\pm10,
\label{eq:J xy Compare Prefactors a}
\\
&64\sqrt{2}\times
(2.04\pm0.03)=
185\pm3.
\label{eq:J xy Compare Prefactors b}
\end{align}
\end{subequations}
Comparing Eqs.~\eqref{eq:J xy Compare Prefactors a}
and \eqref{eq:J xy Compare Prefactors b}, we can write
\begin{equation}
|J^{\,}_{xy}|
\approx
|\bar{J}^{\,}_{xy}|.
\end{equation}

\paragraph{Regime (II).}

In Fig.~\ref{fig:Awful Sums PT a} we find that in the
large limit $2N\gg1$ of system sizes,
\begin{subequations}
\begin{equation}
S=(-1)^{N}\times(1.530\pm0.002)\times
(2N-1)^{\bar{\kappa}^{\,}_{xy,\infty}},
\end{equation}
where
\begin{equation}
\bar{\kappa}^{\,}_{xy,\infty}=
-3.0008\pm0.0002.
\end{equation}
\end{subequations}
The fact that $\bar{\kappa}^{\,}_{xy,\infty}$
is close to $-3$, together with the fact that
\begin{equation}
\frac{\pi^{4}}{64}\approx
1.522,
\end{equation}
tend to confirm that \eqref{eq:Leading S PT Analytical}
is the correct leading term of $S$ for $2N\gg1$.

We now prove that Eq.~\eqref{eq:Leading S PT Analytical} holds.
To do this, we compute the leading term of $S$ defined in
Eq.~\eqref{eq:Awful Sum For J xy} as a function of $2N$.
We start by defining
\begin{equation}
2
\mathsf{h}\coloneqq
\frac{\pi}{2N-1},
\qquad
\vartheta^{\,}_{i}\coloneqq
\frac{\pi n^{\,}_{i}}{2N-1},
\label{eq:h, vartheta Large 2N}
\end{equation}
as convenient short-hands. Then, we use
$\sin2\alpha=2\sin\alpha\cos\alpha$ and symmetrize $S$
under the exchange of the summation indices
$n^{\,}_{1}$ and $n^{\,}_{2}$ to obtain
\begin{widetext}
\begin{equation}
S=
4
\sum_{n^{\,}_{1}=1}^{N-1}
\sum_{n^{\,}_{2}=1}^{N-1}
\sum_{n^{\,}_{3}=1}^{N-1}
(-1)^{n^{\,}_{1}+n^{\,}_{2}+n^{\,}_{3}+1}\,
\frac{
\sin^{2}\vartheta^{\,}_{1}
\sin^{2}\vartheta^{\,}_{2}
\sin^{2}\vartheta^{\,}_{3}
\cos^{2}\vartheta^{\,}_{3}
\left(
\cos\vartheta^{\,}_{2}
-
\cos\vartheta^{\,}_{1}
\right)^{2}
}{
\cos\vartheta^{\,}_{1}
+\cos\vartheta^{\,}_{2}
+\cos\vartheta^{\,}_{3}
}.
\label{eq:S as vartheta}
\end{equation}
\end{widetext}
We make the crucial observation that the denominator
is always strictly positive and reaches its minimal value
when $n^{\,}_{1}\sim n^{\,}_{2}\sim n^{\,}_{3}\sim N-1$.

We capture this regime using the variable
\begin{equation}
m^{\,}_{i}\coloneqq
N-1-n^{\,}_{i},\qquad
i=1,2,3,
\label{eq:Change Variables n to m Large 2N}
\end{equation}
which we use to write
\begin{equation}
(-1)^{n^{\,}_{1}+n^{\,}_{2}+n^{\,}_{3}+1}=
(-1)^{N}
(-1)^{m^{\,}_{1}+m^{\,}_{2}+m^{\,}_{3}},
\end{equation}
responsible for the sign $(-1)^{N}$ in
Eq.~\eqref{eq:Leading S PT Analytical}.
Most importantly, we write
\begin{subequations}
\label{eq:Expansion Small h cos sin}
\begin{equation}
\vartheta^{\,}_{i}=
\frac{\pi}{2}-
\frac{1}{2}
\frac{\pi\,(2m^{\,}_{i}+1)}{2N-1},
\end{equation}
that implies
\begin{align}
&
\cos\vartheta^{\,}_{i}=
\mathsf{h}\,
(2m^{\,}_{i}+1)
+\mathcal{O}(\mathsf{h}^{3}),
\\
&\,
\mathrlap{\sin\vartheta^{\,}_{i}}
\hphantom{\cos\vartheta^{\,}_{i}}=
1
+\mathcal{O}(\mathsf{h}^{2}),
\end{align}
\end{subequations}
that is valid in the limit $m^{\,}_{i}\ll\mathsf{h}^{-1}$.
We insert Eq.~\eqref{eq:Expansion Small h cos sin}
into Eq.~\eqref{eq:S as vartheta}.
To accommodate the small $m^{\,}_{i}\ll\mathsf{h}^{-1}$ regime of validity,
we introduce a cutoff through the regulating number $\mathsf{e}>0$ as
\footnote{
We can show that the error that we would commit by
not taking $\lim_{N\to\infty}$ is exponentially
small in $N$.
}
\begin{subequations}
\label{eq:S With Regulator}
\begin{equation}
S=
(-1)^{N}\!\times
16\,
\mathsf{h}^{3}\times
\lim^{\,}_{\mathsf{e}\to0}	
\lim^{\,}_{N\to\infty}	
\mathsf{S}^{\,}_{\mathsf{e},N}
+\mathcal{O}(\mathsf{h}^{5}),
\end{equation}
where
\begin{align}
\mathsf{S}^{\,}_{\mathsf{e},N}
\coloneqq&\,
\sum_{m^{\,}_{1}=0}^{N-2}
\sum_{m^{\,}_{2}=0}^{N-2}
\sum_{m^{\,}_{3}=0}^{N-2}
(-1)^{m^{\,}_{1}+m^{\,}_{2}+m^{\,}_{3}}
\nonumber \\
&
\times
\frac{(m^{\,}_{2}-m^{\,}_{1})^{2}(2m^{\,}_{3}+1)^{2}}
{2(m^{\,}_{1}+m^{\,}_{2}+m^{\,}_{3})+3}
\nonumber \\
&
\times
e^{-\mathsf{e}(m^{\,}_{1}+m^{\,}_{2}+m^{\,}_{3})}.
\end{align}
\end{subequations}
In doing so, we ensure that the limit $N\to\infty$ of
$\mathsf{S}^{\,}_{\mathsf{e,N}}$ is well-defined.

To evaluate $\mathsf{S}^{\,}_{\mathsf{e},N}$, we use
\begin{equation}
\frac{1}{2x+y}=
\int\limits_{0}^{1}\dd t\ t^{2x+y-1}
\label{eq:Denominator As Intergal 0 to 1}
\end{equation}
to rewrite the denominator of the summand of $\mathsf{S}^{\,}_{\mathsf{e},N}$
by choosing $x=m^{\,}_{1}+m^{\,}_{2}+m^{\,}_{3}$ and $y=3$.
Moreover, we use the geometric series
\begin{subequations}
\label{eq:Geometric Series Alternating}
\begin{align}
&\mathcal{S}^{\,}_{0}(s^{\,}_{\mathsf{e}})\coloneqq
\sum_{m=0}^{\infty}
\hphantom{m^{2}}
(-s^{\,}_{\mathsf{e}})^{m}\!=
\frac{1}{1+s^{\,}_{\mathsf{e}}},
\label{eq:Geometric Series Alternating a}
\\
&\mathcal{S}^{\,}_{1}(s^{\,}_{\mathsf{e}})\coloneqq
\sum_{m=0}^{\infty}
\mathrlap{m}
\hphantom{m^{2}}
(-s^{\,}_{\mathsf{e}})^{m}\!=
\frac{-s^{\,}_{\mathsf{e}}}{(1+s^{\,}_{\mathsf{e}})^{2}},
\label{eq:Geometric Series Alternating b}
\\
&\mathcal{S}^{\,}_{2}(s^{\,}_{\mathsf{e}})\coloneqq
\sum_{m=0}^{\infty}
m^{2}
(-s^{\,}_{\mathsf{e}})^{m}\!=
\frac{s^{2}_{\mathsf{e}}-s^{\,}_{\mathsf{e}}}{(1+s^{\,}_{\mathsf{e}})^{3}},
\label{eq:Geometric Series Alternating c}
\end{align}
\end{subequations}
valid for
\begin{equation}
0\leq|s^{\,}_{\mathsf{e}}|<1.
\label{eq:Validity Geometric Series Alternating}
\end{equation}
We choose
\begin{equation}
s^{\,}_{\mathsf{e}}\equiv
e^{-\mathsf{e}}\,t^{2},
\qquad
0\leq t\leq1
\end{equation}
that satisfies Eq.~\eqref{eq:Validity Geometric Series Alternating}
for any $\mathsf{e}>0$,
We can therefore safely write
\footnote{
At fixed $\mathsf{e}>0$, the exponential regulator provides
an absolutely summable majorant, allowing the limit
$N\to\infty$ to be taken inside the integral.
The resulting regulated expression can then be integrated
explicitly before taking the Abel limit $\mathsf{e}\downarrow0$.
}
\begin{align}
\lim_{\mathsf{e}\to0}
\lim_{N\to\infty}
\mathsf{S}^{\,}_{\mathsf{e},N}
=&\,
\int\limits_{0}^{1}\dd t\ 
t^{2}
\nonumber \\
&\times
\Bigl(
2\,\mathcal{S}^{\,}_{0}\bigl(t^{2}\bigr)
\,\mathcal{S}^{\,}_{2}\bigl(t^{2}\bigr)
-2\,\mathcal{S}^{2}_{1}\bigl(t^{2}\bigr)
\Bigr)
\nonumber \\
&\times
\Bigl(
4\,\mathcal{S}^{\,}_{2}\bigl(t^{2}\bigr)
+4\,\mathcal{S}^{\,}_{1}\bigl(t^{2}\bigr)
+\mathcal{S}^{\,}_{0}\bigl(t^{2}\bigr)
\Bigr)
\nonumber \\
=&\,
\int\limits_{0}^{1}\dd t\ 
\frac{-2\,t^{4}\bigl(1-6t^{2}+t^{4}\bigr)}
{\bigl(1+t^{2}\bigr){}^{7}}
\nonumber \\
=&\,
\frac{\pi}{128},
\end{align}
which proves the claim.

\subsubsection{The coupling $J^{\,}_{z}$}

We compute the coupling
$J^{\,}_{z}$ in
Eq.~\eqref{eq:J z Not Yet Estimated}.
Specifically, we estimate the quantity $Z$ defined in
Eq.~\eqref{eq:Awful Sum For J z}
\begin{itemize}
\item[(I)]
for the range $2N\in[20,30]$ of system sizes accessible with ED
and compare it to \eqref{eq:Jxy Jz Effective Expected b}, 
\item[(II)] and in the thermodynamic limit $2N\to\infty$.
\end{itemize}
On the one hand, we shall compute \eqref{eq:Awful Sum For J z}
numerically and observe its dependence on $2N-1$
for both (I) and (II).
This is given in Fig.~\ref{fig:Awful Sums PT b}.
On the other hand, only for (II), we shall estimate
analytically the leading
term in $2N-1$ of \eqref{eq:Awful Sum For J z}.

More precisely, we will show that
\footnote{
The correction $\mathcal{O}\bigl(N^{2}\bigr)$ is likely overestimated.
}
\begin{equation}
Z=
\frac{256}{225\,\pi^{3}}\,(2N-1)^{3}
+\mathcal{O}\bigl(N^{2}\bigr).
\label{eq:Leading Z PT Analytical}
\end{equation}
Notice that it was not possible to obtain a closed form
of \eqref{eq:Awful Sum For J z} as a function of $2N-1$.

\paragraph{Regime (I).}

In Fig.~\ref{fig:Awful Sums PT b} we find that in the range
$2N\in[20,30]$ of system sizes accessible with ED,
\begin{subequations}
\begin{equation}
Z=(0.0436\pm0.0004)\times
(2N-1)^{\bar{\kappa}^{\,}_{z}},
\end{equation}
where
\begin{equation}
\bar{\kappa}^{\,}_{z}=
2.953\pm0.003.
\label{eq:Kappa z Estimate}
\end{equation}
\end{subequations}
This we insert into Eq.~\eqref{eq:J z Not Yet Estimated} and,
recalling Eq.~\eqref{eq:Fourier Modes PT c},
we obtain
\begin{equation}
\bar{J}^{\,}_{z}=
128\sqrt{2}\,
\epsilon^{2}\times
0.0436(4)\times
(2N-1)^{\bar{\kappa}^{\,}_{z}-4}
>0.
\label{eq:J z Low 2N Analytical}
\end{equation}
Notice that the sign $\bar{J}^{\,}_{z}>0$ is the one
needed in Hamiltonian \eqref{eq:H Effective Boundary Spins}.

We now compare $\bar{J}^{\,}_{z}$ estimated
in Eq.~\eqref{eq:J z Low 2N Analytical}
to $J^{\,}_{z}$ obtained numerically with ED
in Eq.~\eqref{eq:Jxy Jz Effective Expected b}.
On the one hand, we observe that
$\kappa^{\,}_{z}$ in Eq.~\eqref{eq:Jxy Jz Effective Expected b}
lies within the uncertainty interval of
$\bar{\kappa}^{\,}_{z}$ in Eq.~\eqref{eq:Kappa z Estimate}
i.e., we can write
\begin{equation}
\kappa^{\,}_{z}
\approx
\bar{\kappa}^{\,}_{z}.
\end{equation}
On the other hand, we observe that
\begin{subequations}
\begin{align}
&1/2\times v^{\,}_{2,1}=
8.45\pm0.03,
\label{eq:J z Compare Prefactors a}
\\
&128\sqrt{2}\times
(0.0436\pm0.0004)=
7.89\pm0.07.
\label{eq:J z Compare Prefactors b}
\end{align}
\end{subequations}
Comparing Eqs.~\eqref{eq:J z Compare Prefactors a}
and \eqref{eq:J z Compare Prefactors b}, we can write
\begin{equation}
J^{\,}_{z}
\approx
\bar{J}^{\,}_{z}.
\end{equation}

\paragraph{Regime (II).}

In Fig.~\ref{fig:Awful Sums PT b} we find that in the
large limit $2N\gg1$ of system sizes,
\begin{subequations}
\begin{equation}
Z=(0.036785\pm0.000006)\times
(2N-1)^{\bar{\kappa}^{\,}_{z,\infty}},
\end{equation}
where
\begin{equation}
\bar{\kappa}^{\,}_{z,\infty}=
2.99959\pm0.00002.
\end{equation}
\end{subequations}
The fact that $\bar{\kappa}^{\,}_{z,\infty}$
is close to $3$, together with the fact that
\begin{equation}
\frac{256}{225\,\pi^{3}}\simeq
0.036695,
\end{equation}
tend to confirm that \eqref{eq:Leading Z PT Analytical}
is the correct leading term of $Z$ for $2N\gg1$.

We now prove that Eq.~\eqref{eq:Leading Z PT Analytical} holds.
To do this, we compute the leading term of $Z$ defined in
Eq.~\eqref{eq:Awful Sum For J z} as a function of $2N\gg1$.
We show that only
\begin{equation}
S^{\,}_{2},\qquad
S^{\,}_{4},\qquad
S^{\,}_{6}
\label{eq:S2, S4, S6}
\end{equation}
contribute to the leading term \eqref{eq:Leading Z PT Analytical}
of order $(2N-1)^{3}$.
We will then show that
\begin{equation}
S^{\,}_{1},\qquad
S^{\,}_{3},\qquad
S^{\,}_{5},\qquad
S^{\,}_{7},\qquad
S^{\,}_{8}
\end{equation}
only contribute to a subleading term of order $(2N-1)^{-1}$.

We start with the derivation of the sums
\eqref{eq:S2, S4, S6}. To do so, we make use of the short-hand notation
\eqref{eq:h, vartheta Large 2N} and obtain
\begin{widetext}
\begin{subequations}
\label{eq:S2, S4, S6 as vartheta}
\begin{align}
&S^{\,}_{2}=
2
\sum_{n^{\,}_{1}=1}^{N-1}
\sum_{n^{\,}_{2}=1}^{N-1}
\sum_{n^{\,}_{3}=1}^{N-1}
\sum_{n^{\,}_{4}=1}^{N-1}
(-1)^{n^{\,}_{1}+n^{\,}_{4}}\,
\frac{
\sin^{2}\vartheta^{\,}_{1}
\sin2\vartheta^{\,}_{2}
\sin3\vartheta^{\,}_{2}
\sin2\vartheta^{\,}_{3}
\sin3\vartheta^{\,}_{3}
\sin^{2}3\vartheta^{\,}_{4}
}{
\cos\vartheta^{\,}_{1}
+\cos\vartheta^{\,}_{4}
},
\\
&S^{\,}_{4}=
2
\sum_{n^{\,}_{1}=1}^{N-1}
\sum_{n^{\,}_{2}=1}^{N-1}
\sum_{n^{\,}_{3}=1}^{N-1}
\sum_{n^{\,}_{4}=1}^{N-1}
(-1)^{n^{\,}_{1}+n^{\,}_{4}}\,
\frac{
\sin^{2}3\vartheta^{\,}_{1}
\sin\vartheta^{\,}_{2}
\sin2\vartheta^{\,}_{2}
\sin\vartheta^{\,}_{3}
\sin2\vartheta^{\,}_{3}
\sin^{2}3\vartheta^{\,}_{4}
}{
\cos\vartheta^{\,}_{1}
+\cos\vartheta^{\,}_{4}
},
\\
&S^{\,}_{6}=
-4
\sum_{n^{\,}_{1}=1}^{N-1}
\sum_{n^{\,}_{2}=1}^{N-1}
\sum_{n^{\,}_{3}=1}^{N-1}
\sum_{n^{\,}_{4}=1}^{N-1}
(-1)^{n^{\,}_{1}+n^{\,}_{4}}\,
\frac{
\sin\vartheta^{\,}_{1}
\sin3\vartheta^{\,}_{1}
\sin2\vartheta^{\,}_{2}
\sin3\vartheta^{\,}_{2}
\sin\vartheta^{\,}_{3}
\sin2\vartheta^{\,}_{3}
\sin^{2}3\vartheta^{\,}_{4}
}{
\cos\vartheta^{\,}_{1}
+\cos\vartheta^{\,}_{4}
}.
\end{align}
\end{subequations}
\end{widetext}
We make the crucial observation that the denominator
of any summand in Eq.~\eqref{eq:S2, S4, S6 as vartheta}
is always strictly positive and reaches its minimal value
when $n^{\,}_{1}\sim n^{\,}_{4}\sim N-1$ for any $n^{\,}_{2},n^{\,}_{3}$.
We capture this regime using the variable
\eqref{eq:Change Variables n to m Large 2N}
for $i=1,4$, that implies Eq.~\eqref{eq:Expansion Small h cos sin}
together with the expansion
\begin{equation}
\label{eq:Change n To m Implies For sin 3x}
\sin3\vartheta^{\,}_{i}=
-1
+\mathcal{O}(\mathsf{h}^{2}).
\end{equation}
Therefore, we obtain
\begin{subequations}
\label{eq:S2, S4, S6 Expanded}
\begin{align}
&S^{\,}_{2}=
2
\mathsf{h}^{-1}\times
\left[
\mathsf{C}
+\mathcal{O}\bigl(\mathsf{h^{2}}\bigr)
\right]\times
\mathsf{A}^{2},
\\
&S^{\,}_{4}=
2
\mathsf{h}^{-1}\times
\left[
\mathsf{C}
+\mathcal{O}\bigl(\mathsf{h^{2}}\bigr)
\right]\times
\mathsf{B}^{2},
\\
&S^{\,}_{6}=
4
\mathsf{h}^{-1}\times
\left[
\mathsf{C}
+\mathcal{O}\bigl(\mathsf{h^{2}}\bigr)
\right]\times
\mathsf{A}\mathsf{B},
\end{align}
where
\begin{align}
&\mathsf{C}\coloneqq
\sum_{m^{\,}_{i}=0}^{N-2}
\sum_{m^{\,}_{j}=0}^{N-2}
\frac{(-1)^{m^{\,}_{i}+m^{\,}_{j}}}
{2(m^{\,}_{i}+m^{\,}_{j})+2},
\\
&\mathsf{A}\coloneqq
\sum_{n^{\,}_{i}=1}^{N-1}
\sin2\vartheta^{\,}_{i}
\sin3\vartheta^{\,}_{i},
\\
&\mathsf{B}\coloneqq
\sum_{n^{\,}_{i}=1}^{N-1}
\sin\vartheta^{\,}_{i}
\sin2\vartheta^{\,}_{i}.
\end{align}
\end{subequations}

We start by deriving $\mathsf{C}$.
We make use of Eq.~\eqref{eq:Denominator As Intergal 0 to 1}
for $x=m^{\,}_{i}+m^{\,}_{j}$ and $y=2$.
Then we use the finite geometric series
\begin{equation}
\sum_{m=0}^{N-2}
\bigl(-t^{2}\bigr)^{m}=
\frac{1-\bigl(-t^{2}\bigr)^{N-1}}{1+t^{2}},
\end{equation}
that we insert into
\begin{align}
\mathsf{C}
&=
\int\limits_{0}^{1}\dd t\ t
\left[\,
\sum_{m=0}^{N-2}\bigl(-t^{2}\bigr)^{m}
\right]^{2}
\nonumber \\
&=
\int\limits_{0}^{1}\dd t\ t
\left[
\frac{1}{1+t^{2}}	
+
(-1)^{N}
\frac{t^{2N-2}}{1+t^{2}}	
\right]^{2}
\nonumber \\
&=
\frac{1}{4}
+\mathcal{O}(\mathsf{h}),
\label{eq:Calculated C}
\end{align}
where passing from the second to the last equality
is done using the inequality
\begin{equation}
\int\limits_{0}^{1}\dd t\
\frac{t^{2N-1}}{(1+t^{2})^{2}}\leq
\int\limits_{0}^{1}\dd t\
t^{2N-1}=
\frac{1}{2N}.
\end{equation}

We are ready to derive $\mathsf{A}$.
We trade the finite series for an integral
\footnote{Specifically, we use the Euler-Maclaurin formula
to obtain the correction to the integral.}
\begin{align}
\mathsf{A}
&=
\sum_{n^{\,}_{i}=1}^{N-1}
\sin2\vartheta^{\,}_{i}
\sin3\vartheta^{\,}_{i}
\nonumber \\
&=
(2\mathsf{h})^{-1}
\int\limits_{0}^{\frac{\pi}{2}}\dd x\ 
\sin 2x\sin 3x
+\mathcal{O}(\mathsf{h})
\nonumber \\
&=
(2\mathsf{h})^{-1}
\times\frac{2}{5}
+\mathcal{O}(\mathsf{h}).
\label{eq:Calculated A}
\end{align}

Finally, we derive $\mathsf{B}$ in the same way as $\mathsf{A}$.
We find
\begin{align}
\mathsf{B}
&=
\sum_{n^{\,}_{i}=1}^{N-1}
\sin\vartheta^{\,}_{i}
\sin2\vartheta^{\,}_{i}
\nonumber \\
&=
(2\mathsf{h})^{-1}
\int\limits_{0}^{\frac{\pi}{2}}\dd x\ 
\sin x\sin 2x
+\mathcal{O}(\mathsf{h})
\nonumber \\
&=
(2\mathsf{h})^{-1}
\times\frac{2}{3}
+\mathcal{O}(\mathsf{h}).
\label{eq:Calculated B}
\end{align}

We insert Eqs.~\eqref{eq:Calculated C},
\eqref{eq:Calculated A} and \eqref{eq:Calculated B}
into Eq.~\eqref{eq:S2, S4, S6 Expanded}
to obtain
\begin{align}
&S^{\,}_{2}=
\frac{4}{25\,\pi^{3}}\,(2N-1)^{3}
+\mathcal{O}\bigl(\mathsf{h}^{-2}\bigr),
\\
&S^{\,}_{4}=
\frac{4}{9\,\pi^{3}}\,(2N-1)^{3}
+\mathcal{O}\bigl(\mathsf{h}^{-2}\bigr),
\\
&S^{\,}_{6}=
\frac{8}{15\,\pi^{3}}\,(2N-1)^{3}
+\mathcal{O}\bigl(\mathsf{h}^{-2}\bigr).
\end{align}

Therefore, we have found that the sum
\begin{equation}
S^{\,}_{2}+S^{\,}_{4}+S^{\,}_{6}=
\frac{256}{225\,\pi^{3}}\,
(2N-1)^{3}
+\mathcal{O}\bigl(N^{2}\bigr)
\end{equation}
is equal to the leading term of $Z$ given by
Eq.~\eqref{eq:Leading Z PT Analytical}.

We now show that the remaining terms
$S^{\,}_{1},S^{\,}_{3},S^{\,}_{5},S^{\,}_{7}$ and $S^{\,}_{8}$
of $Z$ only contribute to a subleading term of order
$(2N-1)^{-1}$.

We make use of the short-hand notation
\eqref{eq:h, vartheta Large 2N}.
We make the crucial observation that the denominator
is always strictly positive and reaches its minimal value
when $n^{\,}_{1}\sim n^{\,}_{2}\sim n^{\,}_{3}\sim n^{\,}_{4}\sim N-1$.
We capture this regime using the variable
\eqref{eq:Change Variables n to m Large 2N}
that implies Eqs.~\eqref{eq:Expansion Small h cos sin}
and \eqref{eq:Change n To m Implies For sin 3x}
together with
\begin{equation}
\sin2\vartheta^{\,}_{i}=
2\mathsf{h}\,
(2m^{\,}_{i}+1)
+\mathcal{O}(\mathsf{h}^{3}).
\end{equation}
To accommodate the small $m^{\,}_{i}\ll\mathsf{h}^{-1}$ regime of validity,
we introduce a cutoff through the regulating number $\mathsf{e}>0$ as
\begin{subequations}
\label{eq:S13 and S578}
\begin{align}
&
S^{\,}_{1}\sim
-S^{\,}_{3}\sim
-4\mathsf{h}\times
\lim_{\mathsf{e}\to0}
\lim_{N\to\infty}
\mathsf{S}^{(1,3)}_{\mathsf{e},N},
\\
&
2S^{\,}_{5}\sim
2S^{\,}_{7}\sim
-S^{\,}_{8}\sim
8\mathsf{h}\times
\lim_{\mathsf{e}\to0}
\lim_{N\to\infty}
\mathsf{S}^{(5,7,8)}_{\mathsf{e},N},
\end{align}
where
\begin{align}
\mathrlap{\mathsf{S}^{(1,3)}_{\mathsf{e},N}}
\hphantom{\mathsf{S}^{(5,7,8)}_{\mathsf{e},N}}
\coloneqq&\,
\sum_{m^{\,}_{i}=0}^{N-2}
\sum_{m^{\,}_{j}=0}^{N-2}
\sum_{m^{\,}_{l}=0}^{N-2}
\sum_{m^{\,}_{k}=0}^{N-2}
(-1)^{m^{\,}_{i}+m^{\,}_{j}+m^{\,}_{l}+m^{\,}_{k}}
\nonumber \\
&\times
\frac{(2m^{\,}_{i}+1)^{2}}
{2(m^{\,}_{i}+m^{\,}_{j}+m^{\,}_{l}+m^{\,}_{k})+4}
\nonumber \\
&\times
e^{-\mathsf{e}(m^{\,}_{i}+m^{\,}_{j}+m^{\,}_{l}+m^{\,}_{k})},
\\
\mathsf{S}^{(5,7,8)}_{\mathsf{e},N}
\coloneqq&\,
\sum_{m^{\,}_{i}=0}^{N-2}
\sum_{m^{\,}_{j}=0}^{N-2}
\sum_{m^{\,}_{l}=0}^{N-2}
\sum_{m^{\,}_{k}=0}^{N-2}
(-1)^{m^{\,}_{i}+m^{\,}_{j}+m^{\,}_{l}+m^{\,}_{k}}
\nonumber \\
&\times
\frac{(2m^{\,}_{i}+1)(2m^{\,}_{j}+1)}
{2(m^{\,}_{i}+m^{\,}_{j}+m^{\,}_{l}+m^{\,}_{k})+4}
\nonumber \\
&\times
e^{-\mathsf{e}(m^{\,}_{i}+m^{\,}_{j}+m^{\,}_{l}+m^{\,}_{k})}.
\end{align}
\end{subequations}

We finally show that the leading terms of
$\mathsf{S}^{(1,3)}_{\mathsf{e},N}$
and $\mathsf{S}^{(5,7,8)}_{\mathsf{e},N}$
are constant functions of $2N-1$.
To do so, we follow the same procedure as for
Eq.~\eqref{eq:S With Regulator} and we rewrite the denominators of
Eq.~\eqref{eq:S13 and S578}
using Eq.~\eqref{eq:Denominator As Intergal 0 to 1}
for $x=m^{\,}_{i}+m^{\,}_{j}+m^{\,}_{l}+m^{\,}_{k}$ and $y=4$.
Then we use the geometric series
\eqref{eq:Geometric Series Alternating} and insert them into
\begin{subequations}
\begin{align}
\lim_{\mathsf{e}\to0}
\lim_{N\to\infty}
\mathsf{S}^{(1,3)}_{\mathsf{e},N}
=&\,
\int\limits_{0}^{1}\dd t\ 
t^{3}
\nonumber \\
&\times
\Bigl(
\mathcal{S}^{\,}_{0}\bigl(t^{2}\bigr)
\Bigr)^{3}
\nonumber \\
&\times
\Bigl(
4\,\mathcal{S}^{\,}_{2}\bigl(t^{2}\bigr)
+4\,\mathcal{S}^{\,}_{1}\bigl(t^{2}\bigr)
+\mathcal{S}^{\,}_{0}\bigl(t^{2}\bigr)
\Bigr)
\nonumber \\
=&\,
\int\limits_{0}^{1}\dd t\ 
\frac{t^{3}\bigl(1-6t^{2}+t^{4}\bigr)}
{\bigl(1+t^{2}\bigr){}^{6}}
\nonumber \\
=&\,-
\frac{1}{40},
\end{align}
and
\begin{align}
\lim_{\mathsf{e}\to0}
\lim_{N\to\infty}
\mathsf{S}^{(5,7,8)}_{\mathsf{e},N}
=&\,
\int\limits_{0}^{1}\dd t\ 
t^{3}
\nonumber \\
&\times
\Bigl(
\mathcal{S}^{\,}_{0}\bigl(t^{2}\bigr)
\Bigr)^{2}
\Bigl(
2\,\mathcal{S}^{\,}_{1}\bigl(t^{2}\bigr)
+\mathcal{S}^{\,}_{0}\bigl(t^{2}\bigr)
\Bigr)^{2}
\nonumber \\
=&\,
\int\limits_{0}^{1}\dd t\ 
\frac{t^{3}\bigl(1-2t^{2}+t^{4}\bigr)}
{\bigl(1+t^{2}\bigr){}^{6}}
\nonumber \\
=&\,
\frac{1}{120}.
\end{align}
\end{subequations}
This closes the proof.

\section{Degeneracies of the ground states of
$\widehat{H}^{(1)}_{\mathrm{II}}(\pi/2,\lambda)$ at $\lambda=\pi/4,3\pi/4$}
\label{appsec:Degeneracies of the ground states H1II piover4 or 3piover3}

We are going to prove assertion
(\ref{eq:sensitivity to 2Nover4 being integer or half integer of GSD H1II if gapless})
for a given value $2N$ of the number of sites.
To this end, we introduce the short-hand notation
\begin{align}
\widehat{H}^{\,}_{\pm}\coloneqq&\,
\sqrt{2}\, \widehat{H}^{(1)}_{\mathrm{II}}(\pi/2,\pi/2\mp\pi/4) 
\nonumber\\
=&\,
\pm
\sum_{j=1}^{2N-1}
\mathrm{i}
\hat{\xi}^{\,}_{j+1}\,\hat{\eta}^{\,}_{j}
+
\sum_{j=1}^{2N-3}
\mathrm{i}
\hat{\xi}^{\,}_{j}\,\hat{\eta}^{\,}_{j+3}.
\label{appeq:def Hpm b=1}
\end{align}
We take advantage of the Majorana algebra to write
\begin{equation}
\begin{split}
\widehat{H}^{\,}_{\pm}=&\,
\pm
\sum_{j=1}^{2N-1}
\frac{\mathrm{i}}{2}
\left(
\hat{\xi}^{\,}_{j+1}\,\hat{\eta}^{\,}_{j}
-
\hat{\eta}^{\,}_{j}\,\hat{\xi}^{\,}_{j+1}
\right)
\\
&\,+
\sum_{j=1}^{2N-3}
\frac{\mathrm{i}}{2}
\left(
\hat{\xi}^{\,}_{j}\,\hat{\eta}^{\,}_{j+3}
-
\hat{\eta}^{\,}_{j+3}\,\hat{\xi}^{\,}_{j}
\right).
\end{split}
\end{equation}

Define the operator-valued row vector
\begin{subequations}
\begin{equation}
\widehat{\Upsilon}^{\mathsf{T}}\coloneqq
\begin{pmatrix}
\hat{\xi}^{\,}_{1}&\cdots&\hat{\xi}^{\,}_{2N}&
\hat{\eta}^{\,}_{1}&\cdots&\hat{\eta}^{\,}_{2N}
\end{pmatrix}.
\end{equation}
Define the $4N\times4N$ real-valued chiral and antisymmetric matrix
\begin{equation}
\mathcal{X}^{\,}_{\pm}\coloneqq
\begin{pmatrix}
0&+A^{\,}_{\pm}\\
-A^{\mathsf{T}}_{\pm}&0\\
\end{pmatrix}=
-
\mathcal{X}^{\mathsf{T}}_{\pm},
\end{equation}
where the real-valued $2N\times2N$ matrix $A$ has the matrix elements
\begin{equation}
\left(A^{\,}_{\pm}\right)^{\,}_{j,j'}\coloneqq
\delta^{\,}_{j',j+3}
\pm
\delta^{\,}_{j',j-1},
\qquad
j,j'=1\cdots,2N.
\end{equation}
We then have the identity
\begin{equation}
\widehat{H}^{\,}_{\pm}=
\frac{\mathrm{i}}{2}\,
\widehat{\Upsilon}^{\mathsf{T}}\,
\mathcal{X}^{\,}_{\pm}\,
\widehat{\Upsilon}.
\end{equation}
\end{subequations}

All non-vanishing
eigenvalues of the single-particle $4N\times 4N$ Hermitian matrix
\begin{subequations}
\begin{equation}
\mathcal{H}^{\,}_{\pm}=
+\frac{\mathrm{i}}{2}\,
\mathcal{X}^{\,}_{\pm}=
\mathcal{H}^{\dag}_{\pm}
\end{equation}
come in pairs with opposite signs owing to the condition
\begin{equation}
\mathcal{H}^{*}_{\pm}=
-\frac{\mathrm{i}}{2}\,
\mathcal{X}^{\,}_{\pm}=
-
\mathcal{H}^{\,}_{\pm}.
\end{equation}
\end{subequations}
Owing to the fact that the dimensionality of the kernel of a square matrix
equals that of its transpose, and that the nonzero eigenvalues of
$\mathcal{H}^{\,}_{\pm}$ come in pairs with opposite signs,
the number $2M^{\,}_{\pm}$
of nonzero single-particle energy eigenvalues
of $\mathcal{H}^{\,}_{\pm}$
is even and smaller or equal to $4N$.
A many-body ground state of
$\widehat{H}^{\,}_{\pm}$
is then the product state of $M^{\,}_{\pm}$ fermionic levels, each one of which
carries a strictly negative single-particle energy eigenvalue of
$\mathcal{H}^{\,}_{\pm}$. If the kernel of
$\mathcal{H}^{\,}_{\pm}$ defined by
\begin{equation}
\mathrm{ker}\,\mathcal{H}^{\,}_{\pm}\coloneqq
\left\{\left.
v\in\mathbb{R}^{4N}\ \right|\
\mathcal{H}^{\,}_{\pm}\,v=0
\right\}
\label{appeq:def kernel mathcal Hpm b=1} 
\end{equation}
is trivial, this ground state is non-degenerate.
Otherwise, the ground state is degenerate with the degeneracy
\begin{subequations}
\label{appeq:def GSDpm b=1}
\begin{equation}
\mathrm{GSD}^{\,}_{\pm}=2^{Z^{\,}_{\pm}},
\label{appeq:def GSDpm b=1 a}
\end{equation}
where the integer
\begin{equation}
2Z^{\,}_{\pm}=4N-2M^{\,}_{\pm}
\label{appeq:def GSDpm b=1 b}
\end{equation}
\end{subequations}
is the dimensionality of
the kernel of $\mathcal{H}^{\,}_{\pm}$, an integer that is necessarily even.
The even integer $2Z^{\,}_{\pm}$ is the number of linearly independent
Majorana zero modes.
Since two linearly independent Majorana zero modes
define a single fermionic zero mode,
it is the occupancy of these $2Z^{\,}_{\pm}/2=Z^{\,}_{\pm}$
fermionic zero modes that fixes the degeneracy (\ref{appeq:def GSDpm b=1 a})
of the many-body ground state.

To compute the degeneracy of the ground state of
Hamiltonian (\ref{appeq:def Hpm b=1}),
we need to construct the kernel
(\ref{appeq:def kernel mathcal Hpm b=1}).
To do so, we need to solve
\begin{subequations}
\label{appeq:kernel}
\begin{equation}
\begin{pmatrix}
0&+A^{\,}_{\pm}\\
-A^{\mathsf{T}}_{\pm}&0\\
\end{pmatrix}
\begin{pmatrix}
0\\
x
\end{pmatrix}=0,
\qquad
x\in\mathbb{R}^{2N},
\label{appeq:kernel a}
\end{equation}
and
\begin{equation}
\begin{pmatrix}
0&+A^{\,}_{\pm}\\
-A^{\mathsf{T}}_{\pm}&0\\
\end{pmatrix}
\begin{pmatrix}
y\\
0
\end{pmatrix}=0,
\qquad
y\in\mathbb{R}^{2N},
\label{appeq:kernel b}
\end{equation}
\end{subequations}
i.e., we need to solve for the kernel of $A^{\,}_{\pm}$ given by 
\begin{subequations}
\label{appeq:kernel bis}
\begin{equation}
\begin{pmatrix}
  0&  0&  0& +1&  0&  0&\cdots\\
 \pm1&  0&  0&  0& +1&  0&\cdots\\
  0& \pm1&  0&  0&  0& +1&\cdots\\
\vdots&\ddots&\ddots&\ddots&\ddots&\ddots&\ddots\\
\end{pmatrix}
\begin{pmatrix}
x^{\,}_{1}\\
x^{\,}_{2}\\
x^{\,}_{3}\\
\vdots
\end{pmatrix}=0
\label{appeq:kernel bis a}
\end{equation}
and we need to solve for the kernel of $A^{\mathsf{T}}_{\pm}$ given by 
\begin{equation}
\begin{pmatrix}
  0& \pm1&  0&  0&  0&  0&\cdots\\
  0&  0& \pm1&  0&  0&  0&\cdots\\
  0&  0&  0& \pm1&  0&  0&\cdots\\
 +1&  0&  0&  0& \pm1&  0&\cdots\\
  0& +1&  0&  0&  0& \pm1&\cdots\\
\vdots&\ddots&\ddots&\ddots&\ddots&\ddots&\ddots\\
\end{pmatrix}
\begin{pmatrix}
y^{\,}_{1}\\
y^{\,}_{2}\\
y^{\,}_{3}\\
y^{\,}_{4}\\
y^{\,}_{5}\\
\vdots
\end{pmatrix}=0,
\label{appeq:kernel bis b}
\end{equation}
\end{subequations}
respectively. Thus, we must solve the coupled linear equations
\begin{subequations}
\label{appeq:kernel bis bis}
\begin{equation}
\begin{split} 
&x^{\,}_{4}=0,\\
&x^{\,}_{j}=\mp x^{\,}_{j+4},\qquad j=1,\ldots,2N-4,\\
&x^{\,}_{2N-3}=x^{\,}_{2N-2}=x^{\,}_{2N-1}=0,
\end{split}
\label{appeq:kernel bis bis a}
\end{equation}
that defines the kernel of $A^{\,}_{\pm}$ 
and the coupled linear equations
\begin{equation}
\begin{split}
&y^{\,}_{2}=y^{\,}_{3}=y^{\,}_{4}=0,\\
&y^{\,}_{j}=\mp y^{\,}_{j+4},\qquad j=4,\cdots,2N-4,\\
&y^{\,}_{2N-3}=0,
\end{split}
\label{appeq:kernel bis bis b}
\end{equation}
\end{subequations}
that defines the kernel of $A^{\mathsf{T}}_{\pm}$, respectively.
When $2N/4$ is an integer, i.e.,
\begin{subequations}
\begin{equation}
2N\ \mathrm{mod}\ 4=0,
\end{equation}
the only solutions
to Eqs.\
(\ref{appeq:kernel bis bis a})
and
(\ref{appeq:kernel bis bis b})
are the trivial ones, namely
\begin{equation}
x^{\,}_{1}=\cdots=x^{\,}_{2N}=0.
\end{equation}
and
\begin{equation}
y^{\,}_{1}=\cdots=y^{\,}_{2N}=0,
\end{equation}
\end{subequations}
respectively.
There are no linearly independent Majorana zero modes.
When $2N/4$ is a half integer, i.e.,
\begin{subequations}
\begin{equation}
2N\ \mathrm{mod}\ 4=2
\Longleftrightarrow
(2N\pm2)/4 \hbox{ is an integer,}
\end{equation}
the only solutions
to Eqs.\
(\ref{appeq:kernel bis bis a})
and
(\ref{appeq:kernel bis bis b})
are, up to a real-valued multiplicative factor,
\begin{equation}
x^{\,}_{\mp}=
\sum_{n=1}^{(N+1)/2}
(\mp1)^{n-1}
\bm{e}^{\,}_{4n-2}
\end{equation}
and
\begin{equation}
y^{\,}_{\mp}=
\sum_{n=0}^{(N-1)/2}
(\mp1)^{n}
\bm{e}^{\,}_{1+4n},
\end{equation}
\end{subequations}
respectively, where
$\bm{e}^{\,}_{1},\cdots,\bm{e}^{\,}_{2N}$ denotes the Cartesian basis
of $\mathbb{R}^{2N}$. It follows that
$2Z^{\,}_{\pm}=2$ is the number of linearly independent Majorana zero modes.
We have proven Eq.\
(\ref{eq:sensitivity to 2Nover4 being integer or half integer of GSD H1II if gapless}).

\section{Degeneracies of the ground states of
$\widehat{H}^{(0)}_{\mathrm{II}}(\pi/2,\lambda)$ at $\lambda=\pi/4,3\pi/4$}
\label{appsec:Degeneracies of the ground states H0II piover4 or 3piover3}

We are first going to prove assertions
(\ref{eq:dege H0II resolved by parity at criticality b})
and
(\ref{eq:dege H0II resolved by parity at criticality c})
for a given value $2N$ of the number of sites.
We will then complete Table
\ref{tab:TQ HB Degeneracy}
when periodic boundary conditions are selected.
To this end, we introduce the short-hand notations
\begin{subequations}
\label{appeq:def Hpm b=0}
\begin{align}
\widehat{H}^{\,}_{\pm}\coloneqq
\pm
\sum_{j=1}^{2N}
\mathrm{i}
\hat{\xi}^{\,}_{j+1}\,\hat{\eta}^{\,}_{j}
+
\sum_{j=1}^{2N}
\mathrm{i}
\hat{\xi}^{\,}_{j}\,\hat{\eta}^{\,}_{j+3}
\label{appeq:def Hpm b=0 a}
\end{align}
and
\begin{align}
\widehat{H}^{\,}_{\pm,\mathsf{p}}\coloneqq
\widehat{P}^{\,}_{\mathsf{p}}\,
\widehat{H}^{\,}_{\pm}\,
\widehat{P}^{\,}_{\mathsf{p}},
\qquad
\widehat{P}^{\,}_{\mathsf{p}}=
\frac{\mathds{1}^{\,}_{\mathfrak{F}^{\,}_{2N}}+\mathsf{p}\widehat{P}^{\,}_{\mathrm{F}}}{2},
\label{appeq:def Hpm b=0 b}
\end{align}
\end{subequations}
where the Majorana parity operator $\widehat{P}^{\,}_{\mathrm{F}}$
was defined in Eq.\
(\ref{eq:parity operator in Majorana rep}),
$\mathsf{p}=\pm$ denotes the sign of the eigenvalues $\pm1$
of $\widehat{P}^{\,}_{\mathrm{F}}$,
and we recall that the Majorana operators obey PBCs
on the subspace
$\mathfrak{h}^{\,}_{+}=\widehat{P}^{\,}_{+}\,\mathfrak{F}^{\,}_{2N}$
and ABCs
on the subspace
$\mathfrak{h}^{\,}_{-}=\widehat{P}^{\,}_{-}\,\mathfrak{F}^{\,}_{2N}$.

Define the operator-valued row vector
\begin{subequations}
\begin{equation}
\widehat{\Upsilon}^{\mathsf{T}}\coloneqq
\begin{pmatrix}
\hat{\xi}^{\,}_{1}&\cdots&\hat{\xi}^{\,}_{2N}&
\hat{\eta}^{\,}_{1}&\cdots&\hat{\eta}^{\,}_{2N}
\end{pmatrix}.
\end{equation}
Define the $4N\times4N$ real-valued chiral and antisymmetric matrix
\begin{equation}
\mathcal{X}^{\,}_{\pm,\mathsf{p}}\coloneqq
\begin{pmatrix}
0&+A^{\,}_{\pm,\mathsf{p}}\\
-A^{\mathsf{T}}_{\pm,\mathsf{p}}&0\\
\end{pmatrix}=
-
\mathcal{X}^{\mathsf{T}}_{\pm,\mathsf{p}},
\end{equation}
where the real-valued $2N\times2N$ matrix $A$ has the matrix elements
\begin{equation}
\begin{split}
\left(A^{\,}_{\pm,\mathsf{p}}\right)^{\,}_{j,j'}\coloneqq&\,
(-1)^{\delta^{\,}_{\mathsf{p},-}\,\Theta(j+3-2N)}\,\delta^{\,}_{j',j+3}
\\
&\,
\pm
(-1)^{\delta^{\,}_{\mathsf{p},-}\,\delta^{\,}_{j,1}}\,\delta^{\,}_{j',j-1}.
\end{split}
\end{equation}
Here, $j,j'\in\mathbb{Z}/\mathbb{Z}^{\,}_{2N}$ and $\Theta(x)$
vanishes when $x\leq0$ while it is the number one otherwise. 
We then have the identity
\begin{equation}
\widehat{H}^{\,}_{\pm,\mathsf{p}}=
\frac{\mathrm{i}}{2}\,
\widehat{\Upsilon}^{\mathsf{T}}\,
\mathcal{X}^{\,}_{\pm,\mathsf{p}}\,
\widehat{\Upsilon}.
\end{equation}
\end{subequations}

All non-vanishing
eigenvalues of the single-particle $4N\times 4N$ Hermitian matrix
\begin{subequations}
\begin{equation}
\mathcal{H}^{\,}_{\pm,\mathsf{p}}=
+\frac{\mathrm{i}}{2}\,
\mathcal{X}^{\,}_{\pm,\mathsf{p}}=
\mathcal{H}^{\dag}_{\pm,\mathsf{p}}
\end{equation}
come in pairs with opposite signs owing to the condition
\begin{equation}
\mathcal{H}^{*}_{\pm,\mathsf{p}}=
-\frac{\mathrm{i}}{2}\,
\mathcal{X}^{\,}_{\pm,\mathsf{p}}=
-
\mathcal{H}^{\,}_{\pm,\mathsf{p}}.
\end{equation}
\end{subequations}
Owing to the fact that the dimensionality of the kernel of a square matrix
equals that of its transpose, and that the nonzero eigenvalues of
$\mathcal{H}^{\,}_{\pm,\mathsf{p}}$ come in pairs with opposite signs,
the number $2M^{\,}_{\pm,\mathsf{p}}$
of nonzero single-particle energy eigenvalues
of $\mathcal{H}^{\,}_{\pm,\mathsf{p}}$
is even and smaller or equal to $4N$.
A many-body ground state of
$\widehat{H}^{\,}_{\pm,\mathsf{p}}$
is obtained by choosing the energy-minimizing occupation
of every of the $M^{\,}_{\pm,\mathsf{p}}$ finite-energy Bogoliubov mode.
If the kernel of
$\mathcal{H}^{\,}_{\pm,\mathsf{p}}$ defined by
\begin{equation}
\mathrm{ker}\,\mathcal{H}^{\,}_{\pm,\mathsf{p}}\coloneqq
\left\{\left.
v\in\mathbb{R}^{4N}\ \right|\
\mathcal{H}^{\,}_{\pm,\mathsf{p}}\,v=0
\right\}
\label{appeq:def kernel mathcal Hpm b=0} 
\end{equation}
is trivial, this ground state is non-degenerate.
Otherwise, the ground state is degenerate with the degeneracy
\begin{subequations}
\label{appeq:def GSDpm b=0}
\begin{equation}
\mathrm{GSD}^{\,}_{\pm,\mathsf{p}}=2^{Z^{\,}_{\pm,\mathsf{p}}-1},
\qquad
Z^{\,}_{\pm,\mathsf{p}}\geq1,
\label{appeq:def GSDpm b=0 a}
\end{equation}
where the integer
\begin{equation}
2Z^{\,}_{\pm,\mathsf{p}}=4N-2M^{\,}_{\pm,\mathsf{p}}
\label{appeq:def GSDpm b=0 b}
\end{equation}
\end{subequations}
is the dimensionality of
the kernel of $\mathcal{H}^{\,}_{\pm,\mathsf{p}}$,
an integer that is necessarily even.
The even integer $2Z^{\,}_{\pm,\mathsf{p}}$
is the number of linearly independent Majorana zero modes.
Since two linearly independent Majorana zero modes
define a single fermionic zero mode,
it is the occupancy of these $2Z^{\,}_{\pm,\mathsf{p}}/2=Z^{\,}_{\pm,\mathsf{p}}$
fermionic zero modes that fixes the degeneracy (\ref{appeq:def GSDpm b=0 a})
of the many-body ground state in a given fermionic parity sector labeled by
$\mathsf{p}$.

The physical fermion parity of a many-body state depends
both on the occupations of the Bogoliubov modes and
on the fermion parity of the Bogoliubov reference vacuum.
It therefore cannot, in general, be inferred solely from
the number of occupied negative-energy single-particle levels.
Occupying one additional complex fermionic zero mode
reverses the total fermion parity.
Consequently, fixing the parity sector $\mathsf{p}$ selects
one of the two parity classes of the
$2^{Z^{\,}_{\pm,\mathsf{p}}}$ zero-mode occupation configurations.
This condition is achieved by subtracting 1 from
$Z^{\,}_{\pm,\mathsf{p}}$ on the right-hand
side of Eq.\ (\ref{appeq:def GSDpm b=0 a}), unlike on the right-hand
side of Eq.\ (\ref{appeq:def GSDpm b=1 a}).

To compute the degeneracy of the ground state of
Hamiltonian (\ref{appeq:def Hpm b=0}),
we need to construct the kernel
(\ref{appeq:def kernel mathcal Hpm b=0}).
This we do by inserting the Fourier transformation
\begin{subequations}
\label{appeq:kernel cal H pm p}
\begin{align}
&
\hat{\xi}^{\,}_{j}=
\frac{1}{\sqrt{2N}}
\sum_{k\in\Lambda^{\star}_{\mathsf{p}}}
e^{+\mathrm{i}kj}
\hat{\xi}^{\,}_{k},
\qquad
\hat{\xi}^{\dag}_{k}=\hat{\xi}^{\,}_{-k},
\label{appeq:kernel cal H pm p a}
\\
&
\hat{\eta}^{\,}_{j}=
\frac{1}{\sqrt{2N}}
\sum_{k\in\Lambda^{\star}_{\mathsf{p}}}
e^{+\mathrm{i}kj}
\hat{\eta}^{\,}_{k},
\qquad
\hat{\eta}^{\dag}_{k}=\hat{\eta}^{\,}_{-k},
\label{appeq:kernel cal H pm p b}
\end{align}
where the cardinality of the Brillouin zone $\Lambda^{\star}_{\mathsf{p}}$
is $2N$, into Eqs.\ (\ref{appeq:def Hpm b=0 a}) and
(\ref{appeq:def Hpm b=0 b}).
We find
\begin{align}
\widehat{H}^{\,}_{\pm,\mathsf{p}}=&\,
\sum_{k\in\Lambda^{\star}_{\mathsf{p}}}
\mathrm{i}
\left(
\pm
e^{+\mathrm{i}k}
+
e^{-3\mathrm{i}k}
\right)
\hat{\xi}^{\,}_{+k}\,
\hat{\eta}^{\,}_{-k}
\nonumber\\
=&\,
\sum_{k\in\Lambda^{\star}_{\mathsf{p}}}
f^{\,}_{\pm,k}\,
\hat{\xi}^{\,}_{+k}\,
\hat{\eta}^{\,}_{-k}
\nonumber\\
=&\,
\sum_{k\in\Lambda^{\star}_{\mathsf{p}}}^{0<k<\pi}
\left(
f^{\,}_{\pm,+k}\,
\hat{\xi}^{\,}_{+k}\,
\hat{\eta}^{\,}_{-k}
+
f^{*}_{\pm,+k}\,
\hat{\eta}^{\,}_{+k}\,
\hat{\xi}^{\,}_{-k}
\right)
\nonumber\\
&\,
+
\delta^{\,}_{\mathsf{p},+}
\left(
f^{\,}_{\pm,0}\,
\hat{\xi}^{\,}_{0}\,
\hat{\eta}^{\,}_{0}
+
f^{\,}_{\pm,\pi}\,
\hat{\xi}^{\,}_{\pi}\,
\hat{\eta}^{\,}_{\pi}
\right).
\label{appeq:kernel cal H pm p c}
\end{align}
where we made the Hermiticity of
$\widehat{H}^{\,}_{\pm,\mathsf{p}}$
manifest in the last equality by taking advantage of
the reality conditions
(\ref{appeq:kernel cal H pm p a})
and
(\ref{appeq:kernel cal H pm p b})
together with
\begin{equation}
\begin{split}
&
f^{\,}_{+,k}=
+2\mathrm{i}\cos(2k)\,e^{-\mathrm{i}k}=
-f^{*}_{+,-k},
\\
&
f^{\,}_{-,k}=
-2\sin(2k)\,e^{-\mathrm{i}k}=
-f^{*}_{-,-k}.
\end{split}
\label{appeq:kernel cal H pm p d}
\end{equation}
The kernel
(\ref{appeq:def kernel mathcal Hpm b=0})
then follows from enumerating all the independent wave numbers
from the Brillouin zone that solve
\begin{equation}
f^{\,}_{\pm,k}=0,
\qquad
k\in\Lambda^{\star}_{\mathsf{p}}.
\label{appeq:kernel cal H pm p e}
\end{equation}
\end{subequations}

The explicit Brillouin zone
\begin{equation}
\Lambda^{\star}_{\mathsf{p}}=
\left\{\left.
k^{\,}_{m}=\frac{2\pi}{2N}m-\frac{\pi}{2N}\,\delta^{\,}_{\mathsf{p},-}
\right|
m=-N+1,\ldots,N
\right\}
\end{equation}
is needed in order to solve for the independent roots of
(\ref{appeq:kernel cal H pm p e}).
To account for the reality conditions
(\ref{appeq:kernel cal H pm p a})
and
(\ref{appeq:kernel cal H pm p b})
obeyed by the Majorana, we seek the positive momenta
from the Brillouin zone that solve condition
(\ref{appeq:kernel cal H pm p e}), i.e., the intersection
\begin{subequations}
\begin{equation}
S^{\,}_{\pm,\mathsf{p}}\equiv 
\Lambda^{\star}_{\mathsf{p}}\cap\Lambda^{\,}_{\pm,>},
\end{equation}
where
\begin{equation}
\Lambda^{\,}_{+,>}=
\left\{\left.
k^{\,}_{n}=\frac{\pi}{4}+n\,\frac{\pi}{2}
\ \right|\
n=0,1,2,\cdots
\right\}
\end{equation}
and
\begin{equation}
\Lambda^{\,}_{-,>}=
\left\{\left.
k^{\,}_{n}=n\,\frac{\pi}{2}
\ \right|\
n=0,1,2,3,\cdots
\right\}.
\end{equation}
\end{subequations}

Care is required when converting the roots of $f^{\,}_{\pm,k}$
into complex fermionic zero modes.
The restriction to $0\leq k\leq\pi$ counts each generic
pair $\{k,-k\}$ only once.
A root at a generic representative momentum
$0<k<\pi$ contributes two complex fermionic zero modes.
By contrast, the momenta $k=0$ and $k=\pi$ are self-conjugate
under $k\mapsto-k$ modulo $2\pi$.
Each zero-energy self-conjugate momentum contributes
one complex fermionic zero mode.

The set of solutions
$S^{\,}_{+,\mathsf{p}}$
is given by the integer
\begin{subequations}
\begin{equation}
1-\delta^{\,}_{\mathsf{p},+}\leq m^{\,}_{+,\mathsf{p},n}\leq N
\end{equation}
such that
(we express the right-hand side as a function of the number $2N$ of sites)
\begin{equation}
m^{\,}_{+,\mathsf{p},n}=
(1+2\,n)\,\frac{2N}{8}
+
\frac{1}{2}\,\delta^{\,}_{\mathsf{p},-}
\end{equation}
\end{subequations}
for some value of $n=0,1,2,\cdots$. When $\mathsf{p}=+$,
the only permissible solutions are
\begin{equation}
 m^{\,}_{+,+,n}=
(1+2n)\,\frac{2N}{8},
\quad 
k^{\,}_{m^{\,}_{+,+,n}}=
(1+2n)\,\frac{\pi}{4},
\end{equation}
for $n=0,1$ if and only if $2N/8$ is an integer.
When $\mathsf{p}=-$,
the only permissible solutions are
\begin{equation}
m^{\,}_{+,-,n}=(1+2n)\,\frac{2N}{8}+\frac{1}{2},
\quad
k^{\,}_{m^{\,}_{+,-,n}}=(1+2n)\,\frac{\pi}{4},
\end{equation}
for $n=0,1$ if and only if $2N/8$ is a half integer.

This completes the proof of Eq.\
(\ref{eq:dege H0II resolved by parity at criticality b}).

The set of solutions
$S^{\,}_{-,\mathsf{p}}$
is given by the integer
\begin{subequations}
\begin{equation}
1-\delta^{\,}_{\mathsf{p},+}\leq m^{\,}_{-,\mathsf{p},n}\leq N
\end{equation}
such that
(we express the right-hand side as a function of the number $2N$ of sites)
\begin{equation}
m^{\,}_{-,\mathsf{p},n}=
n\,\frac{2N}{4}
+
\frac{1}{2}\,\delta^{\,}_{\mathsf{p},-}
\end{equation}
\end{subequations}
for some value of $n=0,1,2,3,\cdots$. When $\mathsf{p}=+$,
the only permissible solutions are
\begin{equation}
 m^{\,}_{-,+,n}=
n\,\frac{2N}{4},
\quad 
k^{\,}_{m^{\,}_{-,+,n}}=
n\,\frac{\pi}{2},
\end{equation}
for $n=0,1,2$ if and only if $2N/4$ is an integer, and
for $n=0,2$ if and only if $2N/4$ is a half integer.
When $\mathsf{p}=-$,
the only permissible solution are
\begin{equation}
m^{\,}_{-,-,n}=n\,\frac{2N}{4}+\frac{1}{2},
\quad
k^{\,}_{m^{\,}_{-,-,n}}=n\,\frac{\pi}{2},
\end{equation}
for $n=1$ if and only if $2N/4$ is a half integer.

This completes the proof of Eq.\
(\ref{eq:dege H0II resolved by parity at criticality c}).

\begin{table}[t!]
\caption{
Dependence on the number $2N$ of sites of the ground-state energy
$E^{(0)}_{\pm,\mathsf{p}}(2N)$
defined by Eq.\
(\ref{appeq:ground state energies as fct 2N}).
As it should be, all ground-state energies scale like
$-(4/\pi)\times 2N$ in the thermodynamic limit $2N\uparrow\infty$.
   }
\label{apptable:ground state energy dependences on 2N}
\begin{ruledtabular}
\begin{tabular}{ccccc}
$2N$ mod $8$ $\vphantom{\Bigg[}$
&
0
&
2
&
4
&
6
\\ \hline
$E^{(0)}_{+,+}(2N)$ $\vphantom{\Bigg[}$
&
$\frac{-8}{\tan\left(\frac{2\pi}{2N}\right)}$
&
$\frac{-4}{\sin\left(\frac{ \pi}{2N}\right)}$ 
&
$\frac{-8}{\sin\left(\frac{2\pi}{2N}\right)}$ 
&
$\frac{-4}{\sin\left(\frac{ \pi}{2N}\right)}$ 
\\
$E^{(0)}_{+,-}(2N)$ $\vphantom{\Bigg[}$
&
$\frac{-8}{\sin\left(\frac{2\pi}{2N}\right)}$ 
&
$\frac{-4}{\sin\left(\frac{ \pi}{2N}\right)}$ 
&
$\frac{-8}{\tan\left(\frac{2\pi}{2N}\right)}$
&
$\frac{-4}{\sin\left(\frac{ \pi}{2N}\right)}$ 
\\ \hline
$E^{(0)}_{-,+}(2N)$ $\vphantom{\Bigg[}$
&
$\frac{-8}{\tan\left(\frac{2\pi}{2N}\right)}$
&
$\frac{-4}{\tan\left(\frac{ \pi}{2N}\right)}$ 
&
$\frac{-8}{\tan\left(\frac{2\pi}{2N}\right)}$
&
$\frac{-4}{\tan\left(\frac{ \pi}{2N}\right)}$ 
\\
$E^{(0)}_{-,-}(2N)$ $\vphantom{\Bigg[}$
&
$\frac{-8}{\sin\left(\frac{2\pi}{2N}\right)}$ 
&
$\frac{-4}{\tan\left(\frac{ \pi}{2N}\right)}$ 
&
$\frac{-8}{\sin\left(\frac{2\pi}{2N}\right)}$ 
&
$\frac{-4}{\tan\left(\frac{ \pi}{2N}\right)}$ 
\end{tabular}
\end{ruledtabular}
\end{table}

\begin{figure}
\centering
\subfigure{\includegraphics[width=1\columnwidth]{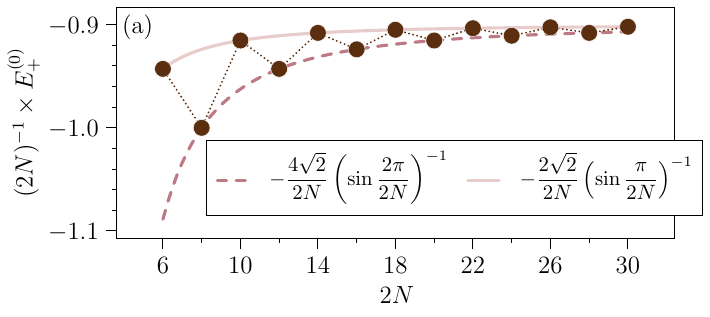}}\\
\subfigure{\includegraphics[width=1\columnwidth]{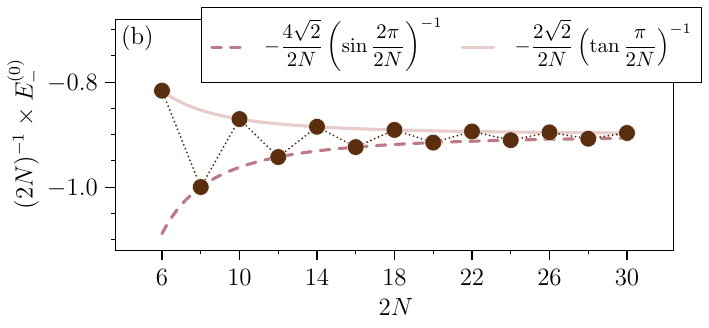}}
\caption{Dependence on $2N\in[6,30]$ of the ground-state energy
$E^{(0)}_{\pm}$ per site of Hamiltonian
$\widehat{H}^{(0)}_{\mathrm{II}}
\left(
\frac{\pi}{2},\frac{\pi}{2}\mp\frac{\pi}{4}
\right)$,
obtained numerically with ED and PBCs.
Comparison with the exact result
\eqref{eq:GS energy H_pm analytical}
is shown as continuous lines.
}
\label{fig:GS Energy HB Analytical vs Numerics}
\end{figure}

We denote the ground-state energy of Hamiltonian
(\ref{appeq:def Hpm b=0 b})
in the parity sector $\mathsf{p}$
by $E^{(0)}_{\pm,\mathsf{p}}(2N)$.
This ground-state energy is given by either
\begin{subequations}
\label{appeq:ground state energies as fct 2N}
\begin{align}
E^{(0)}_{+,\mathsf{p}}(2N)=&\,
-
\sum_{k\in\Lambda^{\star}_{\mathsf{p}}}
\left|f^{\,}_{+,k}\right|
\nonumber\\
=&\,
-2
\sum\limits_{k\in\Lambda^{\star}_{\mathsf{p}}}
|\cos(2k)|
\label{appeq:ground state energies as fct 2N a}
\end{align}
or
\begin{align}
E^{(0)}_{-,\mathsf{p}}(2N)=&\,
-
\sum_{k\in\Lambda^{\star}_{\mathsf{p}}}
\left|f^{\,}_{-,k}\right|
\nonumber\\
=&\,
-
2
\sum\limits_{k\in\Lambda^{\star}_{\mathsf{p}}}
|\sin(2k)|.
\label{appeq:ground state energies as fct 2N b}
\end{align}
\end{subequations}
The summations over the parity-resolved Brillouin zones
in Eqs.\
(\ref{appeq:ground state energies as fct 2N a})
and
(\ref{appeq:ground state energies as fct 2N b})
can be performed in closed forms with the help of the 
finite geometric series. The dependence on $2N$ of the
ground-state energy
$E^{(0)}_{\pm,\mathsf{p}}(2N)$
is reported in Table
\ref{apptable:ground state energy dependences on 2N}.
This dependence has an oscillatory dependence with a period of eight for
$E^{(0)}_{+,\mathsf{p}}(2N)$ and a period of four for
$E^{(0)}_{-,\mathsf{p}}(2N)$. However, the period of the oscillatory
dependence of the energy
\begin{subequations}
\label{eq:GS energy H_pm analytical}
\begin{align}
E^{(0)}_{\pm}(2N)
&\coloneqq
\ev{\,\widehat{H}^{(0)}_{\mathrm{II};\pm}\,}
{\mathrm{GS}}
\nonumber \\
&\,\equiv
\frac{\sqrt{2}}{2}\,
\min\limits_{\mathsf{p}}\,
E^{(0)}_{\pm,\mathsf{p}}(2N)
\label{eq:GS energy H_pm analytical a}
\end{align}
of the ground state $\ket{\mathrm{GS}}$ of
\begin{equation}
\widehat{H}^{(0)}_{\mathrm{II};\pm}\equiv
\widehat{H}^{(0)}_{\mathrm{II}}
\left(
\frac{\pi}{2},\frac{\pi}{2}\mp\frac{\pi}{4}
\right)
\label{eq:GS energy H_pm analytical b}
\end{equation}
\end{subequations}
on the number of sites $2N$ is always four.
We observe in
Fig.~\ref{fig:GS Energy HB Analytical vs Numerics}
that the ground-state energy of the Hamiltonian
$\widehat{H}^{(0)}_{\mathrm{II};\pm}$
computed numerically through exact diagonalization
precisely matches the energy
\eqref{eq:GS energy H_pm analytical b}
for each system size $2N=6,8,\cdots,30$.

We are now in position to deduce the degeneracies of the ground state of
$\widehat{H}^{(0)}_{\mathrm{II};\pm}$
as a function of the number $2N$ of sites.
We shall use the identity $\sin(x)<\tan(x)$ for $0<x<\pi/2$
when comparing the ground-state energies from Table
\ref{apptable:ground state energy dependences on 2N}
in different parity sectors.

\textbf{Case $2N \bmod 8=0$.}
The ground states of
$\widehat{H}^{(0)}_{\mathrm{II};+}$
and
$\widehat{H}^{(0)}_{\mathrm{II};-}$
have parity
$\mathsf{p}=-$ always. According to Eq.\
(\ref{eq:dege H0II resolved by parity at criticality}),
there are no fermionic zero modes in this sector.
Hence the ground states of
$\widehat{H}^{(0)}_{\mathrm{II};+}$
and
$\widehat{H}^{(0)}_{\mathrm{II};-}$
are both non-degenerate.

\textbf{Case $2N \bmod 8=2$.}
The minimal degeneracy of $\widehat{H}^{(0)}_{\mathrm{II};\pm}$ is two
because the ground states in the two parity sectors are degenerate.
However, according to Eq.\
(\ref{eq:dege H0II resolved by parity at criticality}),
$\widehat{H}^{(0)}_{\mathrm{II};+}$
has no fermionic zero modes, whereas for
$\widehat{H}^{(0)}_{\mathrm{II};-}$
the two degenerate parity sectors each contain
two fermionic zero modes.
Hence, the ground state of
$\widehat{H}^{(0)}_{\mathrm{II};+}$
is two-fold degenerate, whereas the ground state of
$\widehat{H}^{(0)}_{\mathrm{II};-}$
is four-fold degenerate.

\textbf{Case $2N \bmod 8=4$.}
The ground state of $\widehat{H}^{(0)}_{\mathrm{II};+}$ has parity $+$.
The ground state of $\widehat{H}^{(0)}_{\mathrm{II};-}$ has parity $-$.
According to Eq.\
(\ref{eq:dege H0II resolved by parity at criticality}),
there are no fermionic zero modes in either sector.
Hence the ground states of
$\widehat{H}^{(0)}_{\mathrm{II};+}$
and
$\widehat{H}^{(0)}_{\mathrm{II};-}$
are both non-degenerate.

\textbf{Case $2N \bmod 8=6$.}
This case is identical to that of $2N \bmod 8=2$.

These degeneracies are reported in Table
\ref{tab:TQ HB Degeneracy}.

\end{document}